%% file: mainflasy.tex
\documentclass[fleqn,openany]{report}
\usepackage{espcrc2}\pdfoutput=1
\usepackage[scaled]{helvet}
\def\vev#1{\left\langle #1\right\rangle}
\DeclareMathAlphabet   {\mathsc}{OT1}{cmr}{m}{sc}
\usepackage{amssymb,amsmath,graphics,graphicx,epsfig}
\usepackage{epigraph}
\usepackage{latexsym}
\usepackage{psfrag}
\usepackage{amsfonts}
\usepackage{multicol}
\usepackage{ifthen}
\usepackage{bm}
\usepackage{bbm}
\usepackage[figuresright]{rotating}
\usepackage[center,footnotesize,hang]{subfigure}
\usepackage{arydshln} 

\def\vev#1{\left\langle #1\right\rangle}
\DeclareMathAlphabet   {\mathsc}{OT1}{cmr}{m}{sc}

\newcommand{\KK}{\mbox{${\bf \underline{3}}$}}
\newcommand{\s}{\mbox{${\bf \underline{1}}$}}
\newcommand{\spr}{\mbox{${\bf \underline{1}'}$}}
\newcommand{\sppr}{\mbox{${\bf {\underline{1}''}}$}}

\newcommand {\beF}{\begin{equation}}
\newcommand {\eeF}{\end{equation}}
\newcommand {\baF}{\begin{eqnarray}}
\newcommand {\eaF}{\end{eqnarray}}

\newcommand\T{\rule{0pt}{2.8ex}} 
\newcommand\B{\rule[-1.2ex]{0pt}{0pt}} 
\newcommand{\ul}{\underline}
\newcommand{\dmsq}{\Delta m^2}
\newcommand{\beqa}{\begin{eqnarray}}
\newcommand{\eeqa}{\end{eqnarray}}

\def\be{\begin{equation}}
\def\ee{\end{equation}}
\def\gs{\mathrel{
   \rlap{\raise 0.511ex \hbox{$>$}}{\lower 0.511ex \hbox{$\sim$}}}}
\def\ls{\mathrel{
   \rlap{\raise 0.511ex \hbox{$<$}}{\lower 0.511ex \hbox{$\sim$}}}}
\newcommand{\obb}{0\mbox{$\nu\beta\beta$}}

\newcommand{\onbb}{neutrino-less double beta decay}
\newcommand{\ba}{\begin{array}{c}}
\newcommand{\baz}{\begin{array}{cc}}
\newcommand{\bad}{\begin{array}{ccc}}
\newcommand{\bav}{\begin{array}{cccc}}
\newcommand{\bea}{\begin{equation} \begin{array}{c}}
\newcommand{\eea}{ \end{array} \end{equation}}
\newcommand{\ea}{\end{array}}

\newcommand{\dms}{\mbox{$\Delta m^2_{\odot}$}}
\newcommand{\dma}{\mbox{$\Delta m^2_{\rm A}$}}

\def\vev#1{\left\langle #1\right\rangle}
\newcommand{\HSdprime}{{\prime\prime}}
\newcommand{\HSpell}{{\ell^\prime}}
\newcommand{\HSBR}{{\text{BR}}}
\newcommand{\HStlll}{\tau \to \overline{\ell}\ell^\prime \ell^\HSdprime}
\newcommand{\HStLeLmLmL}{\tau_L \to \overline{e_L}\mu_L\mu_L}
\newcommand{\HStLmLeLeL}{\tau_L \to \overline{\mu_L}e_Le_L}
\newcommand{\HSmeee}{\mu \to \overline{e}ee}

\newcommand{\HSllg}{\ell \to \ell^\prime \gamma}
\newcommand{\ord}[1]{\mathcal{O}\left({#1}\right)} 

\def\321{SU(3) $\otimes$ SU(2) $\otimes$ U(1)}

\newcommand{\Eqref}[1]{Eq.~\eqref{#1}}
\newcommand{\Figref}[1]{Fig.~\ref{#1}}

\def\gsim{\raise0.3ex\hbox{$\;>$\kern-0.75em\raise-1.1ex\hbox{$\sim\;$}}}
\def\lsim{\raise0.3ex\hbox{$\;<$\kern-0.75em\raise-1.1ex\hbox{$\sim\;$}}}

\def\10{$SO(10)$}

\begin{document}


\title{\Large \textbf{Proceedings of the $1^{st}$ workshop on Flavor Symmetries and consequences in Accelerators and Cosmology (FLASY2011)}}

\author{{\bf Editors:} \\ M. Hirsch$^{a}$, D. Meloni$^{b}$, S. Morisi$^{a}$, S. Pastor$^{a}$, E. Peinado$^{a}$, J.W.F. Valle$^{a}$\\{\bf Authors:}\\ 
Adisorn Adulpravitchai$^{c}$, D. Aristizabal Sierra$^{d}$, F. Bazzocchi$^{e}$, Gautam Bhattacharyya$^{f}$, G. Blankenburg$^{b}$, M. S. Boucenna$^{a}$, I. de Medeiros Varzielas$^{g}$, Marco Aurelio Diaz$^{h}$, Gui-Jun Ding$^{i}$, J. N. Esteves$^{j}$, Yasaman Farzan$^{k}$, Sebastian Garcia Saenz$^{h}$, W. Grimus$^{l}$, Claudia Hagedorn$^{m}$, J. Jones-Perez$^{n}$, Anjan S. Joshipura$^{y}$, Avihay Kadosh$^o$,Kenji Kadota$^{q}$, Sin Kyu Kang$^{p}$, Joern Kersten$^{q}$, Benjamin Koch$^{h}$, Martin B. Krauss$^{s}$, Philipp Leser$^{g}$, Patrick Otto Ludl$^{l}$, Vinzenz Maurer$^{t}$, Luca Merlo$^{u,v}$, Grigoris Panotopoulos$^{w}$, A. Papa$^{x}$, Heinrich Pas$^{g}$, Ketan M. Patel$^{y}$, Werner Rodejohann$^{c}$, U. J. Salda\~na Salazar$^{ab}$, H. Serodio$^{ac}$, Yusuke Shimizu$^{ad}$, Martin Spinrath$^{e}$, Emmanuel Stamou$^{u,v,ae}$, Hiroaki Sugiyama$^{af}$, M. Taoso$^{ag,ah}$, Takashi Toma$^{c,ai}$, Liliana Velasco-Sevilla$^{r}$.
}

\maketitle



{\small \it
\noindent
(a)Departament de F\'{i}sica Te\`{o}rica, IFIC, Universitat de Val\`{e}ncia - CSIC Apt. Correus 22085, E-46071 Val\`{e}ncia, Spain.\\
(b) Dipartimento di Fisica `E.~Amaldi', Universit\`a di Roma Tre.\\
INFN, Sezione di Roma Tre, I-00146 Rome, Italy.\\
(c) Max Planck Institut f\"{u}r Kernphysik, Postfach 10 39 80, 69029 Heidelberg, Germany.\\
(d) IFPA, Dep. AGO, Universite de Liege, Bat B5, Sart Tilman B-4000 \\Liege 1, Belgium.\\
(e) SISSA and INFN, Sezione di Trieste, Via Bonomea 265, 34136 Trieste, Italy.\\
(f) Saha Institute of Nuclear Physics, 1/AF Bidhan Nagar, Kolkata 700064, India.\\
(g) Fakult\"{a}t f\"{u}r Physik, Technische Universit\"{a}t Dortmund D-44221 Dortmund, Germany. \\
(h) Departamento de F\'{\i}sica, Pontificia Universidad Catolica de Chile, Avenida Vicu\~na Mackenna 4860, Santiago, Chile.\\
(i) Department of Modern Physics, University of Science and Technology of China, Hefei, Anhui 230026, China.\\
(j) CAMGSD, Math. Dep., IST, Lisbon. Av. Rovisco Pais 1, 1049-001 Lisboa, Portugal.\\
(k) Institute for research in fundamental sciences (IPM) P.O. Box 19395-5531, Tehran, Iran.\\
(l) University of Vienna, Faculty of Physics, Boltzmanngasse 5, A-1090 Vienna, Austria.\\
(m) Dipartimento di Fisica 'G. Galilei', Universit\`a di Padova \\INFN, Sezione di Padova, Via Marzolo 8, I-35131 Padua, Italy.\\
(n) INFN, Laboratori Nazionali di Frascati, Via E.~Fermi 40, I-00044 Frascati, Italy.\\
(o) Center for Theoretical Physics, University of Groningen, 9747AG, Groningen, The Netherlands.\\
(p) School of Liberal Arts, Seoul National University of Science and Technology, Seoul 139-931, Korea.\\
(q) University of Hamburg, II.\ Institute for Theoretical Physics, Luruper Chaussee 149, 22761 Hamburg, Germany.\\
(r) CINVESTAV-IPN, Apdo.~Postal 14-740, 07000, M\'exico D.F., M\'exico.\\
(s) Institut f{\"u}r Theoretische Physik und Astrophysik, Universit{\"a}t W{\"u}rzburg, Am Hubland, 97074 W{\"u}rzburg, Germany.\\
(t) Department of Physics, University of Basel, Klingelbergstr.\ 82, CH-4056 Basel, Switzerland.\\
(u) Physik-Department, Technische Universit\"at M\"unchen, James-Franck-Strasse, D-85748 Garching, Germany.\\
(v) TUM Institute for Advanced Study, Technische Universit\"at M\"unchen, Lichtenbergstrasse 2a, D-85748 Garching, Germany.\\
(w) Departament de F\'{i}sica Te\`{o}rica, IFIC, Universitat de Val\`{e}ncia - CSIC.\\ 
Apt. Correus 22085, E-46071 Val\`{e}ncia, Spain.\\
(x) Paul Scherrer Institut Villigen, 5232 Switzerland.\\
(y) Physical Research Laboratory, Navarangpura, Ahmedabad-380 009, India.\\
(ab) Instituto de F\'{\i}sica, Universidad Nacional Aut\'onoma de M\'exico, Apdo. Postal 20-364, 01000, M\'exico D.F., M\'exico.\\
(ac) Departamento de F\'{\i}sica and CFTP, Instituto Superior T\'ecnico, Universida de T\'ecnica de Lisboa.\\
(ad) Department of Physics, Niigata University, Niigata 950-2181, Japan.\\
(ae) Excellence Cluster Universe, Technische Universit\"at M\"unchen, Boltzmannstrasse 2, D-85748 Garching, Germany.\\
(af) Ritsumeikan Univ., Shiga 525-8577, Japan.\\
(ag) IFIC, CSIC--Universidad de Valencia, Ed.~Institutos, Apdo.~Correos 22085, E--46071 Valencia, Spain, and MultiDark fellow.\\
(ah) Department of Physics and Astronomy, University of British Columbia, Vancouver, BC V6T 1Z1, Canada.\\
(ai) Institute for Theoretical Physics, Kanazawa University, Kanazawa 920-1192, Japan.\\
}

\onecolumn

\tableofcontents

\chapter[Preface ({\it The editors})]{Preface}
\epigraph{\Large {\em The editors}}{}

\input{preface.tex}


\chapter[ Continuous and Discrete (Flavor)  Symmetries ({\it Hagedorn})]{ Continuous and Discrete (Flavor)  Symmetries}
\epigraph{\Large {\em C. Hagedorn}}{}
%
\input{Author/Hagedorn.tex}

\chapter[Neutrinoless Double Beta Decay and Connections with Flavour Physics ({\it Merlo})]{Neutrinoless Double Beta Decay and Connections with Flavour Physics}
\epigraph{\Large {\em Luca Merlo}}{}
%
\input{Author/LucaMerlo.tex}

\chapter[Deviations and alternatives to tri-bimaximal mixing ({\it  Rodejohann})]{Deviations and alternatives to tri-bimaximal mixing}
\epigraph{\Large {\em Werner Rodejohann}}{}
%
\input{Author/Rodejohann.tex}

\chapter[News on indirect and direct dark matter searches ({\it Taoso})]{News on indirect and direct dark matter searches}
\epigraph{\Large {\em Marco Taoso}}{}
\input{Author/MarcoTaoso.tex}


\chapter[Flavored Orbifold GUT -- $SO(10) \times S_4$ model ({\it Adulpravitchai})]{Flavored Orbifold GUT -- $SO(10) \times S_4$ model}
\epigraph{\Large {\em Adisorn Adulpravitchai}}{}
%
\input{Author/adulpravitchai.tex}

\chapter[Implications of tribimaximal lepton mixing for leptogenesis({\it  Aristizabal Sierra})]{Implications of tribimaximal lepton mixing for leptogenesis}
\epigraph{\Large {\em  D. Aristizabal Sierra}}{}
%
\input{Author/diego-aristizabal.tex}

\chapter[The challenge of low scale flavor symmetry ({\it Bazzocchi})]
{The challenge of low scale flavor symmetry}
\epigraph{\Large {\em Federica Bazzocchi}}{}
%
\input{Author/Bazzocchi.tex}

\chapter[$S_3$ flavor symmetry at the LHC ({\it Leser})]{$S_3$ flavor symmetry at the LHC}
\epigraph{\Large {\em G. Bhattacharyya, \underline{P. Leser}, H. P\"as}}{}
%
\input{Author/Leser.tex}

\chapter[Different $SO(10)$ paths to fermion masses and mixings ({\it Blankenburg})
]{Different $SO(10)$ paths to fermion masses and mixings}
\epigraph{\Large {\em G. Blankenburg}}{}
%
\input{Author/Blankenburg.tex}

\chapter[Stability of dark matter from $A_4$ flavor symmetry({\it Boucenna})
]{Stability of dark matter from $A_4$ flavor symmetry}
\epigraph{\Large {\em M. S. Boucenna}}{}
\input{Author/Boucenna.tex}

\chapter[Aspects of family symmetries ({\it de Medeiros Varzielas})]{Aspects of family symmetries}
\epigraph{\Large {\em Ivo de Medeiros Varzielas}}{}
%
\input{Author/Ivo.tex}

\chapter[ SUSY adjoint $SU(5)$ grand unified model with $S_4$ flavor symmetry({\it Gui-Jun Ding})]{ SUSY adjoint $SU(5)$ grand unified model with $S_4$ flavor symmetry}
\epigraph{\Large {\em Gui-Jun Ding}}{}
\input{Author/dinggj.tex}

\chapter[$A_4$-based neutrino masses with Majoron decaying dark matter ({\it  Esteves})]{$A_4$-based neutrino masses with Majoron decaying dark matter}
\epigraph{\Large {\em Jo\~{a}o N. Esteves}}{}
%
\input{Author/Esteves.tex}

\chapter[Tasting the Flavor of Neutrino Flux from Dark Matter Annihilation {\it (Farzan)}]{Tasting the Flavor of Neutrino Flux from Dark Matter Annihilation }
\epigraph{\Large {\em Yasaman Farzan}}{}
%

\input{Author/YasamanFarzan.tex}

\chapter[Maximal atmospheric neutrino mixing from texture zeros and
quasi-degenerate neutrino masses ({\it Ludl})]
{Maximal atmospheric neutrino mixing from texture zeros and
quasi-degenerate neutrino masses}
\epigraph{\Large {\em W. Grimus and \underline{P.O. Ludl}}}{}
%
\input{Author/Ludl.tex}

\chapter[$U(2)$ and Minimal Flavour Violation in Supersymmetry ({\it Jones P\'erez})]{$U(2)$ and Minimal Flavour Violation in Supersymmetry}
\epigraph{\Large {\em J.~Jones P\'erez}}{}
%
\input{Author/JJonesPerez.tex}

\chapter[ An RS-$A_4$ Flavor model in light of T2K and MEG({\it Kadosh})]{An RS-$A_4$ Flavor model in light of T2K and MEG}
\epigraph{\Large {\em Avihay Kadosh}}{}
\input{Author/AvihayKadosh.tex}

\chapter[Quark-Lepton Complementarity and Triminimal Parametrization of Neutrino Mixing Matrix ({\it Sin Kyu Kang})]{Quark-Lepton Complementarity and Triminimal Parametrization of Neutrino Mixing Matrix}
\epigraph{\Large {\em Sin Kyu Kang}}{}
%
\input{Author/SKKANG.tex}

\chapter[Supersymmetric Musings on the Predictivity of Family Symmetries  ({\it Kersten})]{Supersymmetric Musings on the Predictivity of Family Symmetries}
\epigraph{\Large {\em Kenji Kadota, \underline{J\"orn Kersten}, Liliana Velasco-Sevilla}}{}
%
\input{Author/Kersten.tex}

\chapter[Gravitino Dark Matter and Neutrinos in
Partial Split Supersymmetry ({\it Koch})]{Gravitino Dark Matter and Neutrinos in
Partial Split Supersymmetry }
\epigraph{\Large {\em \underline{Benjamin Koch}, Marco Aurelio D\'\i az$^{a}$, Sebasti\'an Garc\'ia S\'aenz$^{a}$ }}{}
%
\input{Author/BenjaminKoch.tex}

\chapter[LHC tests of neutrino mass from higher dim. eff. operators in SUSY ({\it Krauss})]
{LHC tests of neutrino mass from higher dimensional effective operators in SUSY}
\epigraph{\Large {\em Martin B. Krauss}}{}
%
\input{Author/Krauss.tex}

\chapter[From Flavour to SUSY Flavour Models ({\it Maurer})]{From Flavour to SUSY Flavour Models}
\epigraph{\Large {\em Vinzenz Maurer}}{}
%
\input{Author/Vinzenz_Maurer.tex}

\chapter[The physics of a heavy gauge boson in a Stueckelberg extension ({\it  Panotopoulos})]{The physics of a heavy gauge boson in a Stueckelberg extension of the two-Higgs-doublet model}

\epigraph{\Large {\em Grigoris Panotopoulos}}{}

\input{Author/Panotopoulos.tex}

\chapter[A new limit of the  $\mu^+ \to e^+ \gamma$ decay from the MEG experiment ({\it Papa})]{A new limit of the  $\mu^+ \to e^+ \gamma$ decay from the MEG experiment}
\epigraph{\Large {\em Angela Papa}}{}
\input{Author/Papa.tex}

\chapter[Viability of the exact tri-bimaximal mixing at the GUT scale in $SO(10)$ ({\it Patel})]{Viability of the exact tri-bimaximal mixing at the GUT scale in $SO(10)$}
\epigraph{\Large {\em Anjan S. Joshipura and \underline{Ketan M. Patel}}}{}
\input{Author/Ketan_M_Patel.tex}

\chapter[State of the Art of the M$S_3$IESM ({\it Salda\~na Salazar})]{State of the Art of the M$S_3$IESM}
\epigraph{\Large {\em U. J. Salda\~na Salazar}}{}
%
\input{Author/UJSaldanaSalazar.tex}

\chapter[Leptogenesis and Flavor Models ({\it  Ser\^odio})]{Leptogenesis and Flavor Models}
\epigraph{\Large {\em H. Ser\^odio}}{}
%
\input{Author/HSerodio.tex}

\chapter[Relating Quarks and Leptons
without Grand-Unification ({\it Shimizu})]{Relating Quarks and Leptons
without Grand-Unification}
\epigraph{\Large {\em Yusuke Shimizu}}{}
\input{Author/shimizu.tex}

\chapter[Right Unitarity Triangles and Tri-Bimaximal Mixing from Discrete Symmetries ({\it Spinrath})]{Right Unitarity Triangles and Tri-Bimaximal Mixing from Discrete Symmetries and Unification}
\epigraph{\Large {\em Martin Spinrath}}{}
\input{Author/spinrath.tex}

\chapter[Flavour Changing Neutral Gauge Bosons in $\overline{B}\rightarrow X_s\gamma$ ({\it Stamou})]{Flavour Changing Neutral Gauge Bosons in $\overline{B}\rightarrow X_s\gamma$}
\epigraph{\Large {\em Emmanuel Stamou}}{}
%
\input{Author/EmmanuelStamou.tex}

\chapter[Phenomenology in the Higgs triplet model with $A_4$ symmetry ({\it  Sugiyama})]{Phenomenology in the Higgs triplet model with $A_4$ symmetry}
\epigraph{\Large {\em Hiroaki Sugiyama}}{}
\input{Author/Sugiyama.tex}

\chapter[Indirect and Direct Detection of Dark Matter and Flavor Symmetry ({\it Toma})]{Indirect and Direct Detection of Dark Matter and Flavor Symmetry}
\epigraph{\Large {\em Takashi Toma}}{}
\input{Author/Toma.tex}




\newpage
\noindent
{\bf \Large Acknowledgments}\\
\vskip3.mm
\noindent
The organizers thank IFIC and the MULTIDARK Project, sponsored by the
Spanish Ministry of Economy and Competitiveness's Consolider-Ingenio
2010 Programme under grant CSD2009-00064 for support. We also thank
support from the AHEP group grants: FPA2008-00319/FPA,
FPA2011-22975, Prometeo/2009/091 (Generalitat Valenciana), and the EU
ITN UNILHC PITN-GA-2009-237920.

\vskip3.mm
\noindent
The work of IdMV was supported by DFG grant PA 803/6-1.

\vskip3.mm
\noindent
U.J.S.S. thanks the Local Organizing Committee for supporting his stay in
Valencia, and, the Instituto de F\'isica-UNAM, for supporting his
attendance to this inspiring and interesting workshop. Also he thanks Prof.
A. Mondrag\'on and Dr. M. Mondrag\'on for useful discussions and for
carefully reading the manuscript. For last, the expression of the most
general $S_3$-invariant Higgs potential was found in collaboration with Dr.
M. Mondrag\'on.

\vskip3.mm
\noindent
Y.F. would like to thank the organizers of FLASY 2011 where this
talk was presented for their hospitality and a stimulating
workshop. She is also grateful to ICTP, especially the associate
scheme office, for partially supporting her travel expenses. She
would also like to thank Arman Esmaili, her co-author in papers
that her contribution to this proceedings is based on.

\vskip3.mm
\noindent
CH would like to warmly thank the organizers of this workshop for giving her the 
opportunity to present her talk as well as for the nice and stimulating atmosphere during the 
workshop. 

\vskip3.mm
\noindent
P.L. thanks M.Frigerio for useful discussions. 
His work was supported by DAAD-DST PPP Grant
No.~D/08/04933, and DST-DAAD project
No.~INT/DAAD/P-181/2008. G.B. acknowledges hospitality at
T.U. Dortmund, IST-Lisbon, LPT-Orsay, and ICTP-Trieste, at different
stages of this work. H.P. was supported by DFG Grant No. PA
803/6-1. P.L. and H.P. were also supported by the Physics at the
Terascale Helmholtz Alliance Working Group: Neutrino masses and Lepton
Flavor Violation at the LHC, and they acknowledge hospitality at the
Saha Institute of Nuclear Physics, Kolkata, during a part of this
collaboration.

\vskip3.mm
\noindent
The work of WR was supported by the ERC under the Starting Grant MANITOP and by the Deutsche Forschungsgemeinschaft
in the Transregio 27 ``Neutrinos and beyond -- weakly interacting particles in physics,
astrophysics and cosmology''.

\vskip3.mm
\noindent
The work of S.K.K. is supported by the KRF Grant funded by the Korean Government of the Ministry of Education,
Science and Technology (MEST)(No.2011-0029758).

\vskip3.mm
\noindent
The research of Marco Taoso has been supported by the EC contract UNILHC PITN--GA--2009--237920, by the Spanish grants FPA2008--00319, FPA2011--22975, 
MultiDark CSD2009--00064 (MICINN) and PROMETEO/2009/091 (Generalitat Valenciana) and  by the Institute of Particle Physics 
(IPP) Theory Fellowship and the Natural Sciences and Engineering Research Council (NSERC) of Canada

\vskip3.mm
\noindent
J.~Jones-P\'erez would like to thank R.~Barbieri, G.~Isidori, P.~Lodone and D.~Straub, who collaborated in the original work his talk is based on.

\vskip3.mm
\noindent
MBK would like to thank his collaborators Toshihiko Ota, Werner Porod and Walter Winter, also for their comments on the draft. He acknowledges support from Research Training Group 1147 ``Theoretical astrophysics and particle physics'' of Deutsche Forschungsgemeinschaft.

\vskip3.mm
\noindent
J.K.\ was supported by the Deutsche Forschungsgemeinschaft via the
Junior Research Group `SUSY Phenomenology' within the Collaborative
Research Centre 676 `Particles, Strings and the Early Universe'.

\vskip3.mm
\noindent
V.M.\ acknowledges support by the Swiss National Science Foundation.

\end{document}

%% file: preface.tex
One of the mysteries of theoretical physics is the origin of fermion
masses and mixings. For many years the theory community has attempted
to explain this fascinating problem by assuming an underlying symmetry
acting horizontally between the three known families, namely the {\it
  FLA}vor {\it SY}mmetry. In the last ten years neutrino experiments
have brought substantial improvement in the sensitivity of the
determination of neutrino parameters, which has by now reached the
``precision age''. For the first time neutrino mass and mixing
parameters are measured with great accuracy, opening new expectations
for probing fundamental properties of matter.  In contrast to the
quarks, however, leptons exhibit large mixings, suggesting that
neutrinos can play a special role in the understanding the flavor
problem. For instance it may happen that small quark mixings arise from
a cancellation between the up and down quark sectors or, in contrast,
it may well be that each of these is separately small, of the order of
the CKM angles.

In 2002 Harrison, Perkins and Scott proposed the so called {\it
  tribimaximal ansatz} which, for for many years, has provided a
successful ansatz to theoretical flavor-modelling, as it is very close
to the experimental best fit value.  Recently neutrino oscillation
results from the T2K and Double Chooz collaborations indicate a
nonzero reactor angle. This implies that the tribimaximal ansatz, can
only be a first approximation. In any case different alternative
approaches have also been considered, like bimaximal mixing, golden
ratio and so on.  However tribimaximal can still be a good first step,
corrected either by renormalization effects or by charged sector
diagonalization.

The special structure of tribimaximal suggests a geometrical origin of
neutrino mixing, as in the case of the tetrahedron symmetry
(isomorphic to $A_4$) as flavor symmetry.  Many other non Abelian
discrete groups, like $S_4$, $T'$, $\Delta(27)$, $\Delta(54)$, $D_n$,
$Q_n$ and so on have also been employed in the literature as flavor
symmetry.

Many models have been proposed based on the same hypothesis to
generate a neutrino mass pattern with $\mu-\tau$ invariance and the
property that $m_{11}^\nu + m_{12}^\nu = m_{22}^\nu +m_{23}^\nu$ in
order to yield tribimaximal mixing. But so far a clear way to
distinguish such models is lacking, though some effort in this
direction has been made involving, for example, their predictions
regarding (i) neutrinoless double beta decay, (ii) CP violation, (iii)
accelerator signals, (iv) cosmological issues such as leptogenesis
and dark matter, etc. 

Despite a lot of effort on the part of several groups, an open
challenge remains on how to explain simultaneously quarks within a
flavor-symmetric unified scenario of the fundamental interactions.
 
The goal of the \textit{Workshop on Flavor Symmetries and consequences
  in Accelerators and Cosmology} was to discuss about such important
features of flavor symmetry models, bringing together PhD students,
young post-docs as well as senior scientists.  FLASY has had about 60
participants from about 20 countries.

FLASY has been hosted by the Instituto de Fisica Corpuscular
(IFIC). We thank IFIC and the MULTIDARK Project, sponsored by the
Spanish Ministry of Economy and Competitiveness's Consolider-Ingenio
2010 Programme under grant CSD2009-00064 for support. The main goal of
the MULTIDARK project is to contribute to the identification and
detection of the dark matter. The next FLASY-2012 workshop will take place in Dortmund, hosted by
the University of Dortmund.


%% file: Author/Hagedorn.tex
{\bf Abstract}\\
\vskip5.mm
In this talk I give an overview over continuous and discrete groups and how these are used in the field of model 
building as flavor symmetries. The latter act on the space of the three generations of elementary particles. I mainly 
concentrate on discussing generic mathematical properties of these groups relevant for understanding their possible 
predictive power when applied to explain fermion mass and mixing patterns. I also put emphasis on the 
classification of discrete groups.

\vskip5.mm

\section{Introduction}
The existence of three generations of elementary  particles is well-established. They can be  distinguished according to their mass as well as their mixing. 
The mass hierarchy among charged fermions is known to be strong, especially in the up quark sector
\begin{equation}\nonumber
m_u: m_c: m_t \approx \lambda^8: \lambda^4: 1 \;\; , \;
m_d: m_s: m_b\approx \lambda^4: \lambda^2: 1 \;\; , \;
m_e: m_\mu: m_\tau\approx \lambda^{4\div 5}: \lambda^2: 1 \;\;\; \mbox{with} \;\;\; \lambda \approx 0.22 \, ,
\end{equation}
while it is much milder in the neutrino sector. Neutrino masses are experimentally constrained by the measurements of solar and 
atmospheric mass square differences (at the $2 \sigma$ level) \cite{fogli1}
\begin{equation}\nonumber
\delta m^2 \equiv \Delta m^2_{\mathrm{sol}} \equiv m_2^2 - m_1^2 = (7.58 ^{+0.41} _{-0.42}) \, \times \, 10^{-5} \, \mathrm{eV}^2
\;\; , \;
|\Delta m^2| \equiv \left| m_3^2 - \frac{m_2^2+m_1^2}{2} \right|= (2.35 ^{+0.22} _{-0.18}) \, \times \, 10^{-3} \, \mathrm{eV}^2 \, .
\end{equation} 
The sign of $\Delta m^2$ is undetermined and the absolute neutrino mass scale $m_0$ is bounded, $m_0 \lesssim 0.3 \, \mathrm{eV}$, by cosmological measurements \cite{cosmo}, searches for
neutrinoless double beta \cite{0nubb} and Tritium beta decay \cite{Tritium}. 
The mixing among quarks is small 
\begin{equation}\nonumber
\theta_{12} ^{q} \equiv \theta_{C} \approx 13 ^{\circ} \; , \;\;
\theta_{23} ^{q} \approx 2.4 ^{\circ} \;\;\; \mbox{and} \;\;\; \theta _{13} ^{q} \approx 0.21 ^{\circ} \, .
\end{equation}
At the same time, lepton mixing, measured in neutrino oscillation experiments, involves
two large mixing angles \cite{fogli1}
\begin{equation}\nonumber
\sin^2 \theta_{12}^l=0.306^{+0.036}_{-0.031}  \; , \;\; \sin^2 \theta_{23}^l=0.42^{+0.18}_{-0.06} \;\;\; (2 \sigma \, \mbox{level})
\end{equation}
and a third small angle which is non-zero \cite{fogli1}  - according to recent experimental indications \cite{T2K} 
\begin{equation}
\sin^2 \theta_{13}^l =0.021^{+0.015}_{-0.013}  \;\;\; (2  \sigma \, \mbox{level}) \, .
\end{equation}
This value is obtained using the old estimates for reactor anti-neutrino fluxes.
Neither the hierarchy among the charged fermion masses nor the fermion mixing patterns can be explained in the framework 
of the Standard Model (SM).  The peculiar structure of the lepton mixing is compatible
with special, theoretically motivated, patterns like
\begin{itemize}
\item $\mu\tau$ symmetric mixing \cite{mutau}: $\sin ^{2} \theta _{23}^{l} = \frac{1}{2} \; , \;\;\; \sin ^{2} \theta _{13}^{l} = 0\;\;$ and $\;\; \theta_{12}^l \sim {\cal O}(1)$,
\item tri-bimaximal (TB) mixing \cite{HPS1}:  $\sin^{2} \theta _{12}^{l} = \frac{1}{3} \; , \;\;\; \sin ^{2} \theta _{23}^{l} = \frac{1}{2} \; ,
\;\;\; \sin ^{2} \theta _{13}^{l} = 0$,
\item golden ratio pattern \cite{GR}: $\sin^{2} \theta _{12}^{l} = \frac{1}{\sqrt{5} \phi} \approx 0.276 \; , \;\; \sin ^{2} \theta _{23}^{l} = \frac{1}{2} \; , \;\;\; \sin ^{2} \theta _{13}^{l} = 0$
with $\phi=( 1+ \sqrt{5})/2$,
\item bimaximal mixing \cite{BMorg}:   $\sin^{2} \theta _{12}^{l} = \frac{1}{2} \; , \;\;\; \sin ^{2} \theta _{23}^{l} = \frac{1}{2} \; , \;\;\; \sin ^{2} \theta _{13}^{l} = 0$,
\end{itemize}
up to small corrections. Taking into consideration the success of symmetries describing correctly the gauge interactions of the SM particles,
it is tempting to assume a flavor symmetry $G_f$, i.e. a symmetry acting on the space of generations, to be responsible for the features of fermion masses and mixing.

In section \ref{sec2} I present an overview over general properties of a flavor symmetry $G_f$. Section \ref{sec3} is dedicated to a brief discussion of models with continuous $G_f$
and section \ref{sec4} contains the explanation of the connection between the non-trivial symmetry breaking of a discrete group $G_f$ and the fermion mixing pattern.
I show how TB mixing
arises from the groups $A_4$ and $S_4$ and how predictions for elements of the mixing matrix are derived from (single-valued and double-valued) dihedral groups. In section \ref{sec5} I 
sketch a supersymmetric $D_{14}$ model explaining the Cabibbo angle $\theta_C$ and its extension to the lepton sector which leads to maximal atmospheric mixing,
vanishing $\theta_{13}^l$ and generically large $\theta_{12}^l$. I conclude in section \ref{sec6}.

\mathversion{bold}
\section{General properties of $G_f$}
\label{sec2}
\mathversion{normal}

Several properties of this new symmetry have to be fixed, namely 
\begin{itemize}
\item whether it is abelian or non-abelian,
\item whether it is continuous (like the gauge symmetries) or discrete (like parity), 
\item whether it is local (like the gauge symmetries) or global (like lepton number $U(1)_L$),
\item whether it commutes with the other symmetries (e.g. gauge symmetries) of the theory or not,
\item whether it is broken spontaneously or explicitly,
\item whether it is broken at high energies (like at the seesaw or the GUT scale) or low energies (like the electroweak scale),
\item whether it is broken in an arbitrary way or to one of its subgroups. 
\end{itemize}
Apart from that the size of $G_f$ depends on the gauge group of the model, e.g. in the framework of the SM without right-handed neutrinos
the maximal symmetry of the kinetic terms of the fermions is $U(3)^5$, while it is reduced to $U(3)$ in the case of an $SO(10)$ model in which
all fermions of one generation including the right-handed neutrino are unified into the 16-dimensional representation.

The presence of more than one generation can only be explained with a non-abelian symmetry which has two- and three-dimensional irreducible
representations. An advantage of discrete over continuous symmetries is that the former in general contain several small representations suitable to host
the three fermion generations. Furthermore, it is well-known that the spontaneous breaking of a continuous global (local) symmetry leads to the appearance of Goldstone (gauge) bosons, 
respectively. In order to avoid this I choose the symmetry to be discrete. However, also the breaking of the latter does in general lead to experimentally constrained effects,
namely the generation of domain walls. This does not pose a problem as long as the scale at which the symmetry is broken is comparable with or larger than
the scale of inflation. Thus high energy scales like the seesaw scale or the scale of grand unification seem to be preferred. In this case the flavor symmetry $G_f$ is usually
broken through non-vanishing vacuum expectation values (VEVs) of gauge singlets, called flavons. Models in which $G_f$ is broken at a low energy scale contain instead
several Higgs doublets which also transform non-trivially under $G_f$. Such models seem to be disfavored, because flavor changing neutral
currents and lepton flavor violating processes, like $\mu\to e\gamma$, are induced. Another possibility is to explicitly break $G_f$, for example, at the boundaries of an extra spatial dimension.

A list of possible choices for $G_f$ is
\begin{itemize}
\item continuous symmetries: $SU(2)$, $U(2)$, $SO(3)$, $SU(3)$ and $U(3)$.
\item discrete symmetries:
\begin{itemize}
\item permutation symmetries: symmetric groups $S_N$ and alternating groups $A_N$ 
with $N \in \mathbb{N}$,
\item dihedral symmetries: single-valued groups $D_n$ and double-valued groups $D'_n$ 
with $n \in \mathbb{N}$,
\item further double-valued groups: $T'$, $O'$, $I'$, ...
\item subgroups of $SU(3)$: $\Delta (3 \, n^2)$ and $\Delta (6 \, n^2)$ with $n \in \mathbb{N}$, as well
as $\Sigma$ groups,
\item subgroups of $U(3)$ such as $\Sigma(81)$ or of the listed groups, e.g. $T_7 \cong Z_7 \rtimes Z_3 \subset \Delta (147)$.
\end{itemize}
\end{itemize}

\mathversion{bold}
\section{Overview over models with continuous $G_f$}
\label{sec3}
\mathversion{normal}

In this section I give a brief, and incomplete, survey over models with continuous non-abelian flavor symmetries.
In \cite{U2} supersymmetric models with a $U(2)$ group are discussed in the context of $SU(5)$ and $SO(10)$ grand unified theories (GUTs). Generic features of these models are: 
 the three generations of fermions are assigned to ${\bf 2} + {\bf 1}$ which easily explains the heaviness of the third generation and allows
 to alleviate the so-called supersymmetric flavor problem; the breaking of the symmetry $U(2)$ proceeds in
 two steps, first to a $U(1)$ subgroup and then to nothing. These models predict nine relations among fermion masses and mixing, however,
 only $\theta_{23}^l$ is in general large in the lepton sector, while the two other mixing angles $\theta_{12}^l$ and $\theta_{13}^l$ are small.
 Models with flavor symmetries $SO(3)$ \cite{SO3} and $SU(3)$ \cite{SU3} are more promising, because they allow to unify all three generations of elementary
 particles. Furthermore, the largeness of two mixing angles, $\theta_{12}^l$ and $\theta_{23}^l$, in the lepton sector indicates that all
 three generations are closely related instead of only two. Also in these models the breaking of the flavor symmetry proceeds usually in two
 steps, i.e. $SO(3) \to SO(2) \to \mbox{nothing}$ and $SU(3) \to SU(2) \to \mbox{nothing}$, respectively. The models are supersymmetric and 
the  gauge group is the one of the SM, the Pati-Salam group or, in the case of an extra-dimensional model, $SO(10)$. The masses of the SM fermions arise
usually only from non-renormalizable operators and additional symmetries, such as $Z_n$ or $U(1)$, are imposed in order to forbid certain operators.
Assuming different messenger mass scales allows to differentiate among the expansion parameters present in the up quark, down quark and charged lepton
mass matrices. It has been shown that in such models TB neutrino mixing is predicted through constrained sequential dominance and
a specific structure of the flavons VEVs. The latter can be achieved through F- as well as D-terms. The corrections arising
from the charged lepton sector to the mixing angles are small. Subsequently, models in which the continuous symmetry has been replaced by
a discrete one, like $A_4$, $\Delta(27)$ and $\Delta(108)$, have been discussed \cite{SU3disc}. The results of these models are very similar to those with continuous
flavor groups. In general, the vacuum alignment is achieved in a simpler way; however, in the case of a discrete group D-terms associated with 
the flavor symmetry are obviously absent.

\mathversion{bold}
\section{Non-trivial breaking of discrete $G_f$ and fermion mixing}
\label{sec4}
\mathversion{normal}

In the following I exemplify the idea of the non-trivial breaking of the flavor symmetry in different sectors of the theory, e.g. the charged lepton and the neutrino
one, presenting several models: first, I briefly discuss how $A_4$ and $S_4$ give rise to TB mixing, then dihedral symmetries $D_n$ in general
as well as the group $D_{14}$ in particular which can explain quark and lepton mixing patterns at the same time.

\mathversion{bold}
\subsection{TB mixing and the groups $A_4$ and $S_4$}
\mathversion{normal}

Probably, one of the most famous examples of a flavor symmetry broken in a non-trivial way is the one of the group $A_4$ which leads to the prediction of 
TB mixing. A simple and elegant model can be found in \cite{AF2}. It is an extension of the minimal supersymmetric SM (MSSM) in which the group $A_4$ is spontaneously
broken at high energies through flavon VEVs. The theory is formulated as an effective one with a cutoff $\Lambda$, assumed to be of the order 
of the GUT scale. An elaborate construction of the potential ensures
the correct vacuum alignment. The supermultiplets containing left-handed leptons transform as irreducible triplet of $A_4$ and right-handed charged leptons are
in three inequivalent one-dimensional representations. The two Higgs doublets $h_{u,d}$ do not transform under $A_4$, while the flavons breaking $A_4$ correctly
are two triplets, called $\varphi_T$ and $\varphi_S$ and two singlets $\xi$, $\tilde \xi$ (only one of them acquires a non-zero VEV). Apart from $A_4$
the model has a family independent $Z_3$ symmetry and a family dependent $U(1)_{FN}$ symmetry.
The former allows to separate the two sets of flavons $\{ \varphi_T \}$ and $\{ \varphi_S, \xi, \tilde\xi \}$ and thus
the charged lepton and neutrino sectors in the superpotential at the renormalizable level, while the latter which distinguishes among the right-handed charged leptons
is used to explain the hierarchy among charged lepton masses. This $U(1)_{FN}$ symmetry is also assumed to be spontaneously broken, by a gauge singlet $\theta$, carrying
charge $-1$. 

The group $A_4$ is the symmetry group of the even permutations of four objects and is isomorphic to the symmetry group of a regular tetrahedron. It has 12 elements and
four irreducible representations: three singlets $1$, $1'$, $1''$ and a triplet $3$. The only non-trivial Kronecker product is $3 \times 3 = 1+1'+1''+ 3+3$. The group
can be defined in terms of two generators $S$ and $T$ which satisfy the relations ($E$ is the neutral element of the group) \cite{AF2}
\begin{equation}
\label{A4rel}
S^{2}=E \; , \;\;\; T^{3}=E \; , \;\;\; (ST)^{3} =E \, .
\end{equation}
For the singlets $S$ and $T$ are simply $1$ and the third roots of unity, $1$, $\omega^2$, $\omega$, respectively. For the triplet we use a basis in which $T$ is represented
through a diagonal matrix \cite{AF2}
\begin{equation}
S=\frac{1}{3}\left(\begin{array}{ccc}
       -1 & 2 & 2 \\
       2  & -1 & 2\\
       2  & 2 & -1 \\
       \end{array}\right) \;\;\; , \;\;\; T=\left(\begin{array}{ccc}
                                   1 & 0 & 0 \\
                                   0 & \omega^2 & 0 \\
                                   0 & 0 & \omega \\
                               \end{array}\right) \, .
\end{equation}
With the vacuum alignment (for details see \cite{AF2})
\begin{equation}
\langle \varphi_S \rangle = v_S \left(1,1,1 \right) \;\; , \;\; \langle \xi \rangle = u \;\; , \;\; \langle \tilde\xi\rangle =0  \;\; , \;\; \langle\varphi_T\rangle = v_T \left( 1,0,0 \right) \;\; ,
\end{equation}
the mass matrix ${\cal M}_l$ of the charged leptons is diagonal
\begin{equation}
{\cal M}_l =\frac{v_{T}}{\Lambda} \langle h_d \rangle \; \mbox{diag} 
			\left( y_e \, \frac{\langle \theta \rangle ^2}{\Lambda^2}, 
                         y_\mu \, \frac{\langle \theta \rangle}{\Lambda}, y_\tau \right) \, ,
\end{equation}
while the one of the neutrinos $M_\nu$ arising from the Weinberg operator takes the form
\begin{equation}
\label{MnuA4}
M_\nu=\frac{\langle h_u \rangle^2}{\Lambda}\left(
\begin{array}{ccc}
a+2 b/3& -b/3& -b/3\\
-b/3& 2b/3& a-b/3\\
-b/3& a-b/3& 2 b/3
\end{array}
\right) \, .
\end{equation}
As one immediately sees the latter leads to TB mixing and the neutrino masses are $|a+b|$, $|a|$, $|-a+b|$ in units of $\langle h_u \rangle^2/\Lambda$ with 
$a=x_{a} \, u/\Lambda$, $b=x_{b} \, v_{S}/\Lambda$.
Apart from its simplicity this model is very appealing, because, as has been observed, the group $A_4$ is broken in a specific way. The flavons $\varphi_S$ ($\xi$) break
$A_4$ to a $Z_2$ subgroup in the neutrino sector, since the VEVs of these flavons are eigenvectors to the eigenvalue +1 of the element $S$ of $A_4$. 
From eq.(\ref{A4rel}) it is obvious that $S$ generates
a $Z_2$ group. The VEV of $\varphi_T$ breaks $A_4$ to a $Z_3$ subgroup in the charged lepton sector, because it is an eigenvector to the eigenvalue +1 of the matrix representing the element $T$
of $A_4$ which has order three, see eq.(\ref{A4rel}). Note that in this way $A_4$ is broken completely in the whole theory. The mismatch of the two different subgroups of $A_4$, preserved in 
neutrino and charged lepton sectors, reflects the mismatch between neutrinos and charged leptons  in the flavor space and thus allows for a neat interpretation of lepton mixing.
All this is independent of the particular value of the fermion masses. Similarly, small mixing in the quark sector might be understood as sign that the subgroups preserved in
the up and down quark sectors are the same (for a model realizing this idea with the help of the group $T'$, the double covering of $A_4$, see \cite{FHLM07}).

From the mathematical point of view two things are interesting to notice:
first of all, not only non-zero VEVs of fields forming a $1$ or a $3$ can preserve a $Z_2$ group, but also those of fields transforming as $1'$ or $1''$, and second, apart from 
$Z_2$ and $Z_3$  $D_2$ is a subgroup of $A_4$ (whose preservation is however only compatible with non-zero VEVs of flavons being singlets of $A_4$).
The first aspect can be seen in a different way: the matrix in eq.(\ref{MnuA4}) is not only invariant under the element $S$ of $A_4$, but also under the matrix $P_{23}$ which
is the matrix representing the permutation of second and third rows and columns. The latter generates also a $Z_2$ group and commutes with the element $S$. Thus,
the neutrino mass matrix $M_\nu$ is invariant under $Z_2 \times Z_2$, with one of the two being a subgroup of $A_4$, while the other one is an accidental symmetry which
arises through the particular choice of the flavor symmetry breaking fields. This fact has been used as argument against $A_4$ and 
it has been shown \cite{Lam0708} that indeed from the mathematical view point the group $S_4$ is more appropriate for predicting TB mixing, because it contains the element
$S$ as well as $P_{23}$. The simple idea is then to use $S_4$ as flavor group and break it to $Z_2 \times Z_2$ in the neutrino sector, while the breaking in the charged lepton
sector remains the same, i.e. $S_4 \to Z_3$. Again, the flavor group is broken completely in the whole theory. However, one problem which might arise in $S_4$ models is 
related to the question of how to achieve the mass hierarchy among charged leptons naturally, because $S_4$ only contains two inequivalent singlets in contrast to $A_4$. Thus,
it might not be simple to distinguish the three (right-handed) charged leptons. This problem can however be easily solved \cite{CHMS1} by
extending the flavor group to $S_4 \times Z_3$ and breaking the latter in the charged lepton sector to $Z_3^{(D)}$ which is the diagonal subgroup of the external $Z_3$ factor
and the $Z_3$ contained in $S_4$ and generated by $T$. In order to distinguish the three right-handed charged leptons which transform trivially
under $S_4$, one assigns them to $1$, $\omega^2$ and $\omega$ under the additional $Z_3$ group.

Before discussing other flavor groups let me briefly comment on the question whether or not models predicting TB mixing
are still favored in the light of the recent experimental results \cite{T2K} which indicate a non-zero value for $\theta_{13}^l$. Indeed, in all models the leading order result, in this case
TB mixing, receives corrections from various sources, e.g. higher-dimensional operators. Thus, in all models a non-zero value of $\theta_{13}^l$ is aspected. Concerning its size one can
roughly say: let us assume the size of the corrections to be $\delta$ and that the latter contribute in the same manner to all three mixing angles, then the request to not perturb
too much the result for the solar mixing angle implies that $\delta \lesssim 0.05$. Thus, one might expect $\sin\theta_{13}^l \sim \delta \lesssim 0.05$ which is too small to explain the
recent experimental indication. Since in many models several operators give rise to a correction to $\theta_{13}^l$,  we get in general $\sin\theta_{13}^l \approx |c| \delta$ with $c$ complex.
If $c$ is largish, i.e. the corrections add up, larger values of $\theta_{13}^l$ can be explained. Alternatively, one can consider models in which the corrections to the reactor mixing 
angle are generically larger than to the two other angles. This happens for example if there are two sets of flavor symmetry breaking fields whose VEVs are different in size 
\cite{Lin09}. Yet, another possibility is to assume that the mixing pattern at leading order is bimaximal mixing. Then, the reactor as well as the solar mixing angle have to receive corrections of 
order $\delta \sim \lambda \sim 0.2$ \cite{S4BM}. The crucial issue is then to protect the atmospheric mixing angle $\theta_{23}^l$ from too large corrections.

\mathversion{bold}
\subsection{Dihedral symmetries $D_n$ as $G_f$}
\mathversion{normal}

In order to show the interesting and amusing properties, a non-trivial breaking of a discrete symmetry can have, I discuss the case of a dihedral flavor group $D_n$
which is broken to two distinct $Z_2$ subgroups. If the former is generated by $\rm A$ and $\rm B$, the generating elements of the latter are $\mathrm{B A}^{k_1}$
and $\mathrm{B A}^{k_2}$ ($k_i$ integers), respectively. If two of the left-handed fermion (quark or lepton) generations transform as irreducible two-dimensional representation ${\bf \underline{2}}_{\rm j}$,
one of the elements of the mixing matrix $V$ is of the form
\begin{equation}
|V_{\alpha\beta}|= \left| \cos \left( \frac{\pi \, (k_{1} - k_{2}) \, \mathrm{j}}{n} \right) \right| \, .
\label{Vab}
\end{equation}
As one can nicely see, $|V_{\alpha\beta}|$ only depends on group-theoretical quantities, i.e. the index $n$ of the dihedral group $D_n$, the indices characterizing the two $Z_2$
subgroups $k_{1,2}$ and the index $\rm j$ of the two-dimensional representation. Note that for $k_1=k_2$ the mixing is trivial, as 
expected. The way to reach the result in eq.(\ref{Vab}) can be understood performing the following seven steps: first, one has to choose the group $D_n$ and select the indices $k_{1,2}$ of the two
$Z_2$ subgroups which are associated with the two different fermion sectors 1 and 2. Then, one assigns left-handed fields to ${\bf \underline{1}}_{\rm s} + {\bf \underline{2}}_{\rm j}$, while
the three generations of right-handed fermions can either transform as singlets ${\bf \underline{1}_{\rm i_p}}$ or also as singlet and doublet ${\bf \underline{1}}_{\rm l} + {\bf \underline{2}}_{\rm m}$.
We then consider a generic model in which two sets of scalars exist $\{\Phi_1\}$ and $\{\Phi_2\}$. These sets contain fields in all representations $\mu$ of $D_n$ and we assume their VEVs to be
such that they leave invariant a $Z_2$ group generated by $\mathrm{B A}^{k_{1,2}}$, respectively. As one can check, the most general matrices ${\cal M}_i$ with $i=1,2$ for fermions of sector
1 and 2 are of the form (always given in the left-right basis)
\begin{equation}
\label{Mp1}
\mathcal{M}_i = \left( \begin{array}{ccc}
  0 & A_i & B_i \\
  C_i & D_i & E_i\\
  -C_i \mathrm{e}^{- i \varphi_i \mathrm{j}} & D_i \mathrm{e}^{- i \varphi_i \mathrm{j}}
& E_i \mathrm{e}^{- i \varphi_i \mathrm{j}}
\end{array}
\right) \, ,
\end{equation}
if right-handed fermions are singlets under the group $D_n$, or take the form
\begin{equation} 
\label{Mp2}
\mathcal{M}_i = \left( \begin{array}{ccc}
  A_i & C_i & C_i \mathrm{e}^{- i \varphi_i \mathrm{m}}\\
  B_i & D_i & E_i\\
  B_i \mathrm{e}^{- i \varphi_i \mathrm{j}} & E_i \mathrm{e}^{- i \varphi_i (\mathrm{j}-\mathrm{m})}
& D_i \mathrm{e}^{- i \varphi_i (\mathrm{j}+\mathrm{m})}
\end{array}
\right) \, ,
\end{equation}
if right-handed fermions transform as ${\bf \underline{1}}_{\rm l} + {\bf \underline{2}}_{\rm m}$. The parameters $A_i$,...,$E_i$ are in general complex and not further constrained
by the flavor symmetry $D_n$. The phase $\varphi_i$ which appears in both cases only depends on group-theoretical
quantities
\begin{equation}
\varphi_i = \frac{2 \pi k_i}{n} \, .
\end{equation} 
For both matrices the combination ${\cal M}_i {\cal M}_i^\dagger$ can be written as
\begin{equation}
\label{MMd}
\mathcal{M}_i \mathcal{M}_i^\dagger= \left( \begin{array}{ccc}
  a_i & b_i \mathrm{e}^{i \zeta_i} & b_i \mathrm{e}^{i (\zeta_i + \varphi_i \mathrm{j})}\\
  b_i \mathrm{e}^{- i \zeta_i} & c_i & d_i \mathrm{e}^{i \varphi_i \mathrm{j}}\\
  b_i \mathrm{e}^{- i (\zeta_i+\varphi_i \mathrm{j})} & d_i \mathrm{e}^{- i \varphi_i \mathrm{j}} & c_i 
\end{array}
\right) 
\end{equation}
with $a_i$,...,$d_i$ and $\zeta_i$ depending on the parameters present in the matrices in eqs.(\ref{Mp1})-(\ref{Mp2}).
The matrix $U_i$ diagonalizing the combination ${\cal M}_i {\cal M}_i^\dagger$ is
\begin{equation}
U_i = \left( \begin{array}{ccc}
 0 & \cos \theta_i \, \mathrm{e} ^{i \zeta_i}& \sin \theta_i \, \mathrm{e}^{i \zeta_i}\\   
 - \frac{1}{\sqrt{2}} \mathrm{e}^{i \varphi_i \mathrm{j}} & - \frac{\sin \theta_i}{\sqrt{2}} & \frac{\cos\theta_i}{\sqrt{2}}\\
 \frac{1}{\sqrt{2}}  & - \frac{\sin \theta_i}{\sqrt{2}} \mathrm{e}^{-i \varphi_i \mathrm{j}} & \frac{\cos\theta_i}{\sqrt{2}}
\mathrm{e}^{-i \varphi_i \mathrm{j}}
\end{array}
\right) 
\end{equation}
with $\theta_{i}$ given in terms of the parameters appearing in eq.(\ref{MMd}).
As one clearly sees, one of the eigenvectors only contains the phase $\varphi_i$ and thus only depends on group-theoretical quantities. The physical mixing arises from the misalignment of the
matrices $U_1$ and $U_2$:
\begin{equation}
V=U_1^\dagger U_2 \; .
\end{equation}
Then one of the elements $|V_{\alpha\beta}|$ is of the form
\begin{equation}
|V_{\alpha\beta}| = \frac{1}{2} \left| 1 + \mathrm{e}^{i (\varphi_1 -\varphi_2) \mathrm{j}}\right|
= \left|\cos \left( (\varphi_1 -\varphi_2) \frac{\mathrm{j}}{2} \right) \right|
= \left| \cos \left( \frac{\pi \, (k_{1} - k_{2}) \, \mathrm{j}}{n} \right) \right| \, ,
\end{equation}
while the rest depends also on the angles $\theta_{1,2}$ and the phases $\zeta_{1,2}$.

It has been shown \cite{BHL07}  that $\mu\tau$ symmetric lepton mixing can be realized in this way with the group $D_4$ \cite{myD4}
and that $D_{14}$ can be used to predict the correct size of the Cabibbio angle \cite{BH09}. In the subsequent section a model is outlined
in which $D_{14}$ plays the role of the flavor symmetry and which explains the Cabibbo angle as well as nearly $\mu\tau$
symmetric lepton mixing \cite{HMD14}.

\mathversion{bold}
\section{$D_{14}$ - a symmetry for quarks and leptons}
\label{sec5}
\mathversion{normal}

The group $D_{14}$ belongs to the dihedral symmetries and has 28 elements. As all $D_n$ groups, it only contains one- and two-dimensional representations, called
${\bf \underline{1}}_{{\rm i}}$, ${\rm i}=1,...,4$ and ${\bf \underline{2}}_{{\rm j}}$, ${\rm j}=1,...,6$. It can be described in  terms of two generators $\rm A$ and $\rm B$ which satisfy the relations
\begin{equation}
{\rm A}^{14} = E \; , \;\; {\rm B}^{2} = E \; , \;\; \rm A \, B \, A = B \, .
\end{equation}
The generators $\rm A$ and $\rm B$ for the singlets are just $\pm 1$ and the two-by-two matrices for ${\bf \underline{2}}_{{\rm j}}$ are in a convenient basis
\begin{equation}
\rm A =\left(\begin{array}{cc} 
                           \mathrm{e}^{\left( \frac{\pi i}{7} \right) \, \mathrm{j}} & 0 \\
                            0 & \mathrm{e}^{-\left( \frac{\pi i}{7} \right)
                              \, \mathrm{j}} 
          \end{array}\right) \;\;\;\; \mbox{and} \;\;\;\; \rm B=\left(\begin{array}{cc} 
                                       0 & 1 \\
                                       1 & 0 
                  \end{array}\right) \; .
\end{equation}

The main properties of the original $D_{14}$ model \cite{BH09} which only contains quarks are the following: 
it is an extension of the MSSM formulated as effective theory with a cutoff $\Lambda$, the flavor symmetry $G_f$ is the product $D_{14} \times U(1)_{FN} \times Z_3$, $D_{14}$ and $U(1)_{FN}$ together explain
the quark mass hierarchies and $Z_3$ is used for separating appropriately up and down quark sectors. All flavor symmetries are broken spontaneously at high energies through flavon VEVs. 
The two sets of flavons breaking the group $D_{14} \times Z_3$ are $\{ \psi^{u}_{1,2}, \chi_{1,2}^{u}, \xi^{u}_{1,2}, \eta^{u} \}$ and 
$\{ \psi^{d}_{1,2}, \chi_{1,2}^{d}, \xi^{d}_{1,2}, \eta^{d}, \sigma \}$, while $\theta$ only carries non-trivial $U(1)_{FN}$ charge. The left-handed quark doublets 
$Q_{D}=(Q_1, Q_2)$ and $Q_3$ are in the representations ${\bf \underline{2}}_1$ and ${\bf \underline{1}}_1$, while the right-handed quarks $u^c$, $c^c$, $t^c$ and $d^c$, $s^c$, $b^c$ are singlets under $D_{14}$
and $h_{u,d}$ are trivial singlets of $D_{14}$.

The leading operators contributing to up quark masses are (order one coefficients are omitted in the following and $(\cdots)$ denotes the contraction to a $D_{14}$-invariant)
\begin{eqnarray}
&&Q_3 \, t^c \, h_u +  \frac{1}{\Lambda} (Q_D \psi^u)  t^c h_u + \frac{1}{\Lambda} Q_3 (c^c \eta^u) h_u\\
&+& \frac{\theta^2}{\Lambda^4} (Q_D u^c \chi^u \xi^u) h_u + \frac{\theta^2}{\Lambda^4} \left(Q_D u^c (\xi^u)^2 \right) h_u + \frac{\theta^2}{\Lambda^4} (Q_D  \psi^u \eta^u u^c) h_u\\
&+& \frac{1}{\Lambda^2} (Q_D c^c \chi^u \xi^u)  h_u + \frac{1}{\Lambda^2} \left(Q_D c^c (\xi^u)^2 \right) h_u + \frac{1}{\Lambda^2} (Q_D  \psi^u) (\eta^u c^c) h_u \, ,
\end{eqnarray}
while the ones contributing to the masses of the down quarks read
\begin{equation}
\frac{1}{\Lambda} Q_3 (b^c \eta^d) h_d + \frac{\theta}{\Lambda^2} \, Q_3 \, s^c \sigma h_d + \frac{\theta}{\Lambda^2} \, (Q_D \psi^d) \, s^c h_d \, .
\end{equation}
Note that all these operators respect the separation among the up and down quark sectors, as up-type flavons only contribute to up quark masses and down-type ones only to down quark masses. 
With the appropriate vacuum configuration of the flavons, see \cite{BH09} for details of the superpotential, two distinct $Z_2$ groups
are preserved. One is characterized with an even index, see above,  which can be set without loss of generality to zero, while the second one has an odd index $k$ which remains undetermined.
At the $Z_2$ symmetry preserving level, the up and down quark mass matrices are of the form
\begin{equation}
\mathcal{M}_u = \left(
\begin{array}{ccc}
	- \alpha^u_1 \, t^2 \, \epsilon^2  & \alpha^u_2 \, \epsilon^2 
				& \alpha^u_3 \, \epsilon \\
	\alpha^u_1 \, t^2 \, \epsilon^2 & \alpha^u_2 \, \epsilon^2 & \alpha^u_3 \, \epsilon\\
	0 & \alpha^u_4 \, \epsilon & y_t 
\end{array}
\right) \, \langle h_u \rangle \;\;\; \mbox{and} \;\;\;
\mathcal{M}_d = \left(
\begin{array}{ccc}
	 0 & \alpha^d_1 \, t \, \epsilon  & 0\\
	 0 & \alpha^d_1 \, \mathrm{e} ^{- \pi i k/7} \, t \, \epsilon & 0 \\
	 0 & \alpha^d_2 \, t \, \epsilon & y_b \, \epsilon 
\end{array}
\right) \, \langle h_d \rangle 
\end{equation}
with $\langle\Phi^u\rangle/\Lambda \approx \epsilon$, $\langle\Phi^d\rangle/\Lambda \approx \epsilon$, $t = \langle\theta\rangle/\Lambda \approx \epsilon \approx \lambda^2 \approx 0.04$.
As one easily computes, the quark masses fulfill the relations
\begin{eqnarray}
m_u^2:m_c^2:m_t^2 &\sim& \epsilon^8: \epsilon^4: 1 \; , \;\; m_d^2:m_s^2:m_b^2  \sim 0: \epsilon^2: 1 \; ,\\ \nonumber
m_b^2:m_t^2 &\sim& \epsilon^2:1 \;\;\; \mbox{for} \;\;\; \mbox{small} \;\;\; \tan\beta = \langle h_u \rangle/\langle h_d \rangle
\end{eqnarray}
and the CKM mixing matrix takes the form
\begin{equation}
|V_{CKM}|  = \left( 
\begin{array}{ccc}
	|\cos (\frac{k \, \pi}{14})| & |\sin (\frac{k \, \pi}{14})| & 0 \\
	|\sin (\frac{k \, \pi}{14})| & |\cos (\frac{k \, \pi}{14})| & 0\\
	0   		             & 0			    & 1
\end{array}
\right) +
\left( 
\begin{array}{ccc}
	0  			& \mathcal{O}(\epsilon^4) 	& \mathcal{O}(\epsilon^2)  \\
	\mathcal{O}(\epsilon^2)	& \mathcal{O}(\epsilon^2)	& \mathcal{O}(\epsilon)    \\
	\mathcal{O}(\epsilon) 	& \mathcal{O}(\epsilon)		& \mathcal{O}(\epsilon^2)
\end{array}
\right)  \, .
\end{equation}
As one can see, for the choices $k=1$ and $k=13$ the matrix elements involving the first two generations are very close to their experimental best fit values, e.g.  $|V_{ud}| \approx 0.97493$
should be compared with $|V_{ud}|_{\rm exp} = 0.97419 ^{+0.00022} _{-0.00022}$. Subleading corrections lead to the down quark mass of the correct size as well as allow all
elements of $V_{CKM}$ and the Jarlskog invariant $J_{CP}$ to be accommodated.

In a second step I would like to outline how to extend this model to the lepton sector in a minimalistic way, i.e. to add only a small number of new fields, to not perturb
the results achieved for quarks too much and at the same time to predict $\mu\tau$ symmetric lepton mixing. The setup which fulfills these requirements is the following \cite{HMD14}: the $Z_3$ factor 
of the flavor symmetry is replaced by $Z_7$ and still has the role to segregate the different symmetry breaking sectors, the left-handed leptons $L_1$ and $L_D=(L_2, L_3)$ transform 
as ${\bf \underline{1}}_3$ and ${\bf \underline{2}}_2$, also
the right-handed neutrinos $\nu^c_1$ and $\nu^c_D=(\nu^c_2, \nu^c_3)$ transform 
as ${\bf \underline{1}}_2 + {\bf  \underline{2}}_3$, while all right-handed charged leptons $e^c$, $\mu^c$, $\tau^c$ are trivial singlets of
$D_{14}$. Furthermore, one flavon $\chi^e_{1,2}$ is added which is a doublet ${\bf \underline{2}}_2$ of $D_{14}$.

In the neutrino sector, the leading contributions to the right-handed neutrino mass matrix ${\cal M}_R$ are
\begin{equation}
 \nu^c_1 \nu^c_1 \sigma + (\nu^c_D \nu^c_D) \sigma
\;\;\;\;\;\; \mbox{giving rise to} \;\;\;\;\;\; 
\mathcal{M}_R = \left( \begin{array}{ccc}
 \alpha^M_1 & 0 & 0\\
 0 & 0 & \alpha^M_2\\
 0 & \alpha^M_2 & 0
\end{array}
\right) \, \epsilon \, \Lambda \; ,
\end{equation}
while the Dirac neutrino mass matrix ${\cal M}^D_\nu$ takes the form
\begin{equation}
\mathcal{M}^D_\nu = \left( \begin{array}{ccc}
 0 & \alpha^D_1 & \alpha^D_1\\
\alpha^D_ 2 & 0 & \alpha^D_3\\
-\alpha^D_2 & \alpha^D_3 & 0
\end{array}
\right) \, \epsilon \, \langle h_u \rangle
\end{equation}
and receives at leading level contributions from the operators
\begin{equation}
\frac{1}{\Lambda}  (L_1 \nu_D^c \xi^u) h_u +  \frac{1}{\Lambda} (L_D \nu_1^c \chi^u)  h_u +  \frac{1}{\Lambda} (L_D \nu_D^c \psi^u) h_u \, .
\end{equation}
Consequently, the light neutrino mass matrix $\mathcal{M}_\nu$ can be cast into the form
\begin{equation}
\mathcal{M}_\nu = \left( \begin{array}{ccc}
  2 x^2/v & x & x \\
 x & z & v-z\\
 x & v-z & z
\end{array}
\right) \, \epsilon \, \langle h_u \rangle^2/\Lambda \, .
\end{equation}
The predictions are then: $\mu\tau$ symmetric neutrino mixing, $\tan (\theta_{12}^\nu) = \sqrt{2}\left|\frac{x }{v} \right|$ and
normal ordering with $m_1=0$. 
In addition, one finds $|m_{ee}| = m_2 \, \sin^2 (\theta_{12}^\nu) = \sqrt{\delta m^2} \, \sin^2 (\theta_{12}^\nu)$.
For the charged lepton sector we use the additional flavon $\chi^e_{1,2}$ which acquires a VEV of the form $v^e \, (1,0)$. Such an alignment is easily achieved
with a driving field being a trivial singlet under $D_{14}$ and carrying an appropriate $Z_7$ charge. Interestingly enough, also this alignment preserves a
$Z_2$ subgroup of $D_{14}$ which, however, does not coincide with one of the $Z_2$ symmetries present in the up and down quark sectors. The main contributions
to the charged lepton mass matrix ${\cal M}_l$ originate from ($\alpha$ stands for $e$, $\mu$ and $\tau$)
\begin{equation}
 \frac{1}{\Lambda} (L_D \chi^e) \, \alpha^c \, h_d + \frac{1}{\Lambda^2} (L_D \chi^e \xi^u) \, \alpha^c \, h_d
 + \frac{1}{\Lambda^3} (L_1 \chi^e \psi^u \xi^u) \alpha^c h_d+ \frac{1}{\Lambda^3} (L_1 \eta^u) (\chi^e \chi^u) \alpha^c h_d
\end{equation}
which give rise to
\begin{equation}
\mathcal{M}_l = \left( \begin{array}{ccc}
		\alpha^e_1 \, \epsilon^3 & \alpha^e_2 \, \epsilon^3 & \alpha^e_3 \, \epsilon^3\\
		\alpha^e_4 \, \epsilon^2 & \alpha^e_5 \, \epsilon^2 & \alpha^e_6 \, \epsilon^2\\
		\alpha^e_7 \, \epsilon & \alpha^e_8 \, \epsilon & \alpha^e_9 \, \epsilon
\end{array}
\right) \langle h_d \rangle \;\;\;\;\;\;\;\;\; \mbox{for} \;\;\;\;\; v^e/\Lambda \approx \epsilon \approx \lambda^2 \, .
\end{equation}
As one can see, the charged lepton mass hierarchy is correctly predicted,  $m_e : m_\mu : m_\tau \sim \epsilon^2 : \epsilon : 1$,
and the three mixing angles in the charged lepton sector are small
\begin{equation}
\theta_{12}^e \sim \epsilon \, , \,\, \theta_{13}^e \sim \epsilon^2 \, , \,\, \theta_{23}^e \sim \epsilon \, .
\end{equation}
Thus, the lepton mixing is nearly $\mu\tau$ symmetric, i.e.
\begin{equation}
\sin^2 \theta_{23}^l = \frac{1}{2} + \mathcal{O}(\epsilon) \; , \;\; \sin \theta_{13}^l = \mathcal{O}(\epsilon) \, , \;\; \sin^2 \theta_{12}^l = \mathcal{O}(1) \, .
\end{equation}
As regards the recent indication for $\theta_{13}^l \neq 0$ \cite{T2K}, we note that the model naturally leads only to rather small $\theta_{13}^l$ which might seem to be disfavored
at the moment.

\section{Conclusions}
\label{sec6}

I have presented an overview over flavor symmetries in general and their properties. I commented briefly on the case of a continuous $G_f$ and then focussed on discrete symmetries
playing the role of $G_f$ with special emphasis on those cases in which they are broken in a non-trivial way. I showed a variety of examples: $A_4$ and $S_4$ and their relation to TB mixing,
predictions from dihedral groups $D_n$ in general and from $D_{14}$ in particular. Lastly, I have sketched a $D_{14}$ model in which the Cabibbo angle as well as $\mu\tau$ symmetric lepton
mixing are closely related to how $D_{14}$ is broken and at the same time all fermion mass hierarchies are naturally accommodated.

%% file: Author/LucaMerlo.tex
%


{\bf Abstract}\\

\vskip5.mm
Neutrinoless double beta ($0\nu2\beta$) decay is a fundamental observable to probe the Majorana character of neutrinos and to investigate on their absolute mass scale. The present status of experiments searching for $0\nu2\beta$ decay is reviewed and the most relevant results are discussed. The interplay with flavour physics in general provides clear predictions for $0\nu2\beta$ decay and some major examples are presented.

\vskip5.mm

\section{Introduction}
Non-vanishing masses for neutrinos have been the first evidence of the necessity to go beyond the Standard Model (SM) of Particle Physics. In the last decades, a lot of effort has been put to determine the parameters of the neutrino sector. In the conservative scenario of only 3 active neutrinos, their oscillations can be described by two frequencies, $\Delta m^2_{sol}\equiv m^2_2-m^2_1\approx 7.6\times10^{-5}\;\text{eV}^2$ and $\Delta m^2_{atm}\equiv |m^2_3-m^2_1|\approx 2.5\times 10^{-2}\;\text{eV}^2$, and three mixing angles, the solar $\sin^2\theta_{12}\approx 0.31$, the atmospheric $\sin^2\theta_{23}\approx0.5$ and the reactor $\sin^2\theta_{13}\approx 0.015$ \cite{Schwetz:2011zk}. Furthermore an upper bound on the absolute neutrino mass scale of few $eV$ has been fixed.

On the other hand numerous questions are waiting for an answer: 1) which is the nature of neutrinos, Dirac or Majorana? 2) which is the absolute mass scale? 3) which is the mass ordering (which is the sign of $\Delta m^2_{atm}$)? 4) which are the values of the Dirac and Majorana CP violating phases? The $0\nu2\beta$ decay is a useful observables with this regards, being connected directly or indirectly to the previous questions.

In the following I will first review general notions on the $0\nu2\beta$ decay and the related experimental status. I will then discuss correlations among the $0\nu2\beta$ decay and other observables, with particular emphasis on flavour observables. 

\mathversion{bold}
\section{The $0\nu2\beta$ Decay}
\mathversion{normal}

The $0\nu2\beta$ decay belongs to the class of the $2\beta$ decays, that are spontaneous transitions of an initial nucleus $(A,Z)$ into a nucleus $(A, Z+2)$ with proton number larger by two units, and the emission of two electrons. Such transitions occur in even-even nuclei, i.e. with the same number of protons and neutrons: in such case, the parent and the daughter nuclei are more bound than the intermediate one and as a result the equivalent sequence of two single beta decays is avoided or at least inhibited. 

The most frequent $2\beta$ decays are the $2\nu2\beta$ transitions $(A,Z)\to(A,Z+2)+2e^-+2\bar{\nu}_e$: these are SM allowed decays and occur at the second electroweak order. They provide a natural background for the $0\nu2\beta$ decays $(A,Z)\to(A,Z+2)+2e^-$, that on the contrary are not allowed in the SM. Nine nuclei turn out to be particularly interesting for the $0\nu2\beta$ decay and are actually under investigation: these are $^{48}Ca$, $^{76}Ge$, $^{82}Se$, $^{96}Zr$, $^{100}Mo$, $^{116}Cd$, $^{130}Te$, $^{136}Xe$ and $^{150}Nd$. The reason can be found in the large energy release, the $Q$-factor, that is emitted during the transition: since the $0\nu2\beta$ decay rate goes with $Q^5$, a sufficiently large $Q$-factor allows a better discrimination of the signal over the natural radioactivity, which drops significantly beyond $2.614\;\text{MeV}$. 

It is not clear which is the mechanism that originates the $0\nu2\beta$ decay, but it violates the Lepton number by two units.
This suggests the possibility that the $0\nu2\beta$ decay may be connected to other Lepton number violating (LNV) observables. This is indeed the case as it has been shown in \cite{Schechter:1981bd,Hirsch:2006yk}: following the Schechter--Valle theorem, whatever is the mechanism of the $0\nu2\beta$ decay, it is possible to construct a 4-loop diagram 
that contributes to the Majorana neutrino mass matrix. Even if the corresponding contribution is however tiny $\sim 10^{-23}\;\text{eV}$ and insufficient to explain the observed neutrino masses, the evidence of the $0\nu2\beta$ decay would imply the Majorana nature of neutrinos.

\begin{figure}[h!]
\vspace{-0.5cm}
\centering
\subfigure[Standard mechanism]
{\includegraphics[width=4.5cm]{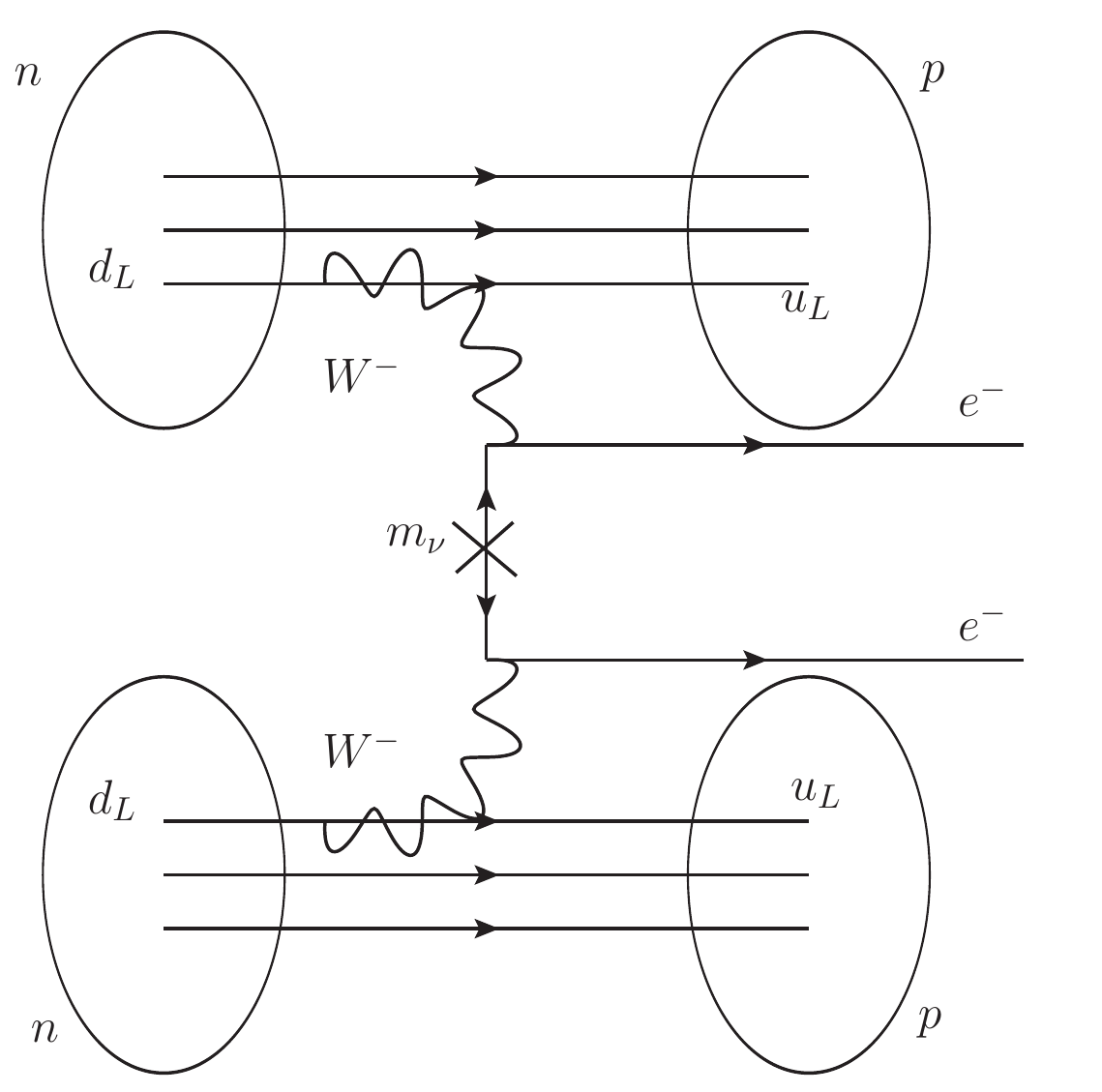}}
\quad
\subfigure[Exotic mechanism: RH neutrinos]
{\includegraphics[width=4.5cm]{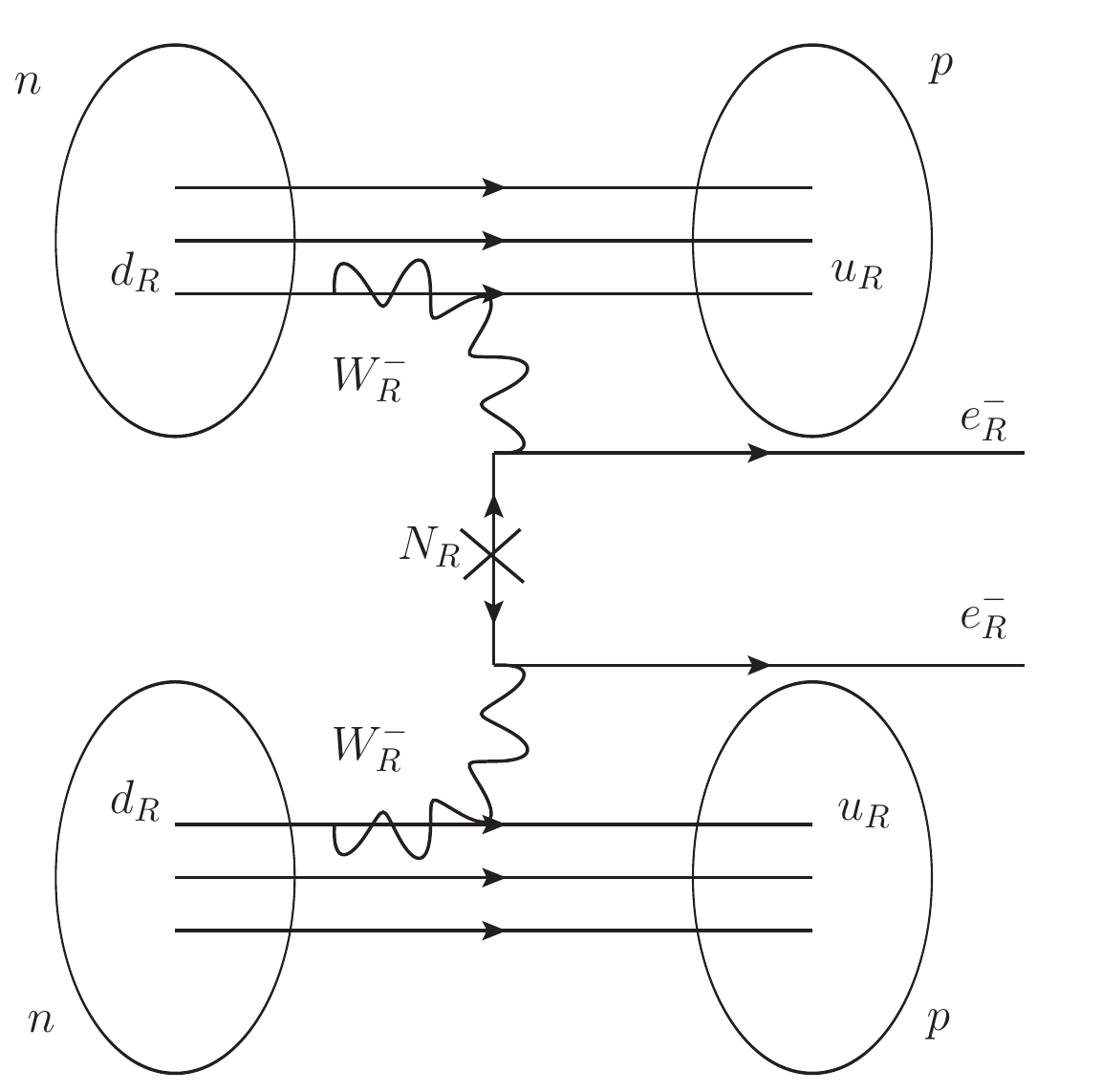}}
\quad
\subfigure[Exotic mechanism: type II See-Saw]
{\includegraphics[width=4.7cm]{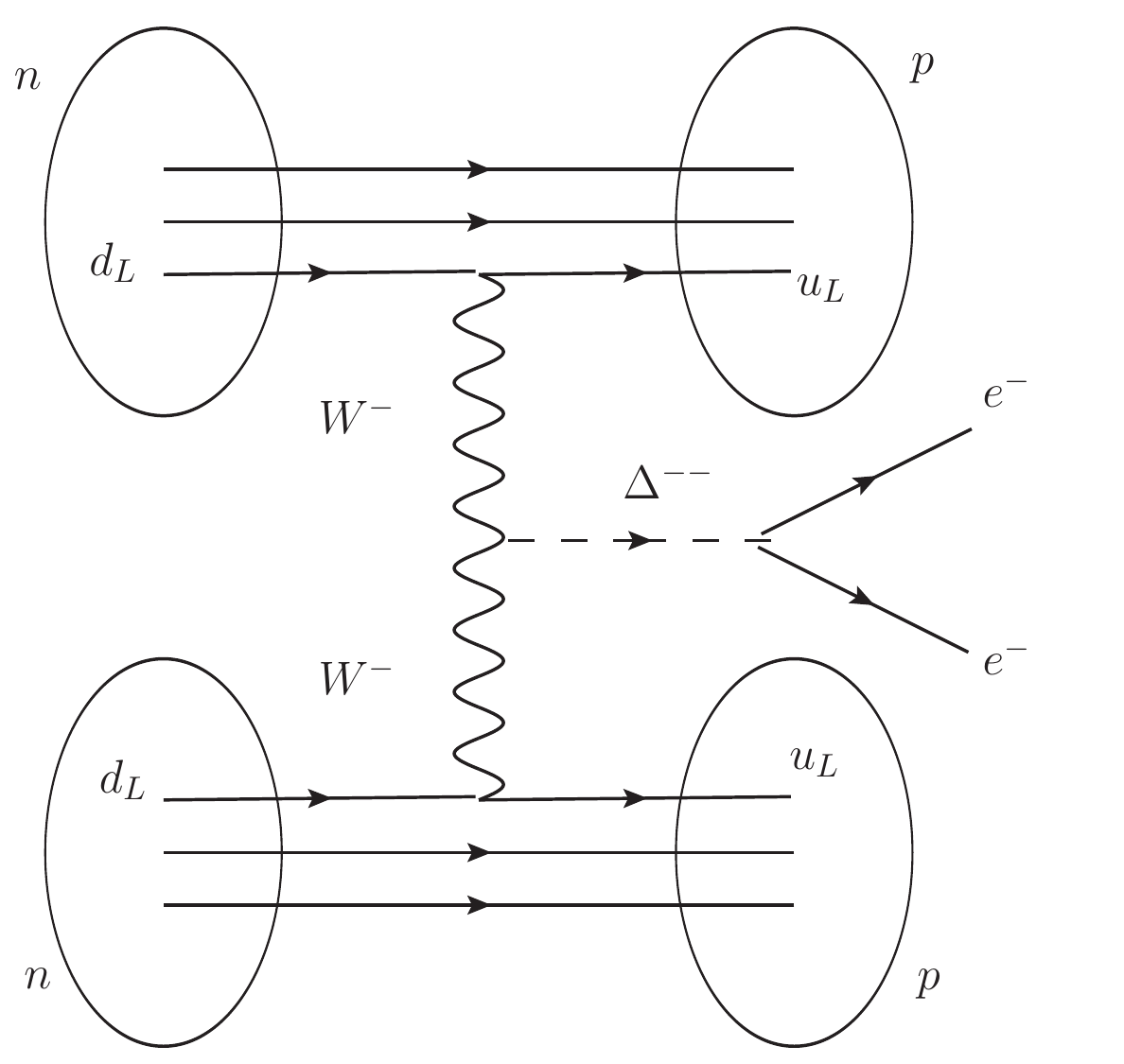}}
\vspace{-1cm}
\caption{\it Standard mechanism and two examples of exotic mechanisms: in (b) the case of RH neutrinos and in (c) the case of the $SU(2)$ Higgs triplet in the type II See-Saw mechanism.}
\vspace{-0.5cm}
\label{fig:mechanisms}
\end{figure}

The mechanisms for the $0\nu2\beta$ decay are usually cataloged into two classes: the standard mechanism and the exotic ones. The first one is illustrated by the  diagram in fig.~\ref{fig:mechanisms}(a), where the mediators are the light oscillating neutrinos. It is possible only if the neutrinos are Majorana particles. On the other hand, when other LNV sources, different from light neutrinos, give the main contribution to the $0\nu2\beta$ decay, such as right-handed (RH) neutrinos in left-right symmetric models (fig.~\ref{fig:mechanisms}(b)) or the $SU(2)$ Higgs triplet in the type II See-Saw mechanism (fig.~\ref{fig:mechanisms}(c)), usually it is referred to exotic mechanisms. A more complete list of exotic mechanisms can be found in \cite{Rodejohann:2011mu}.\\

\begin{figure}[h!]
\vspace{-0.5cm}
\centering
\subfigure[Energy distribution.]
{\includegraphics[width=5cm]{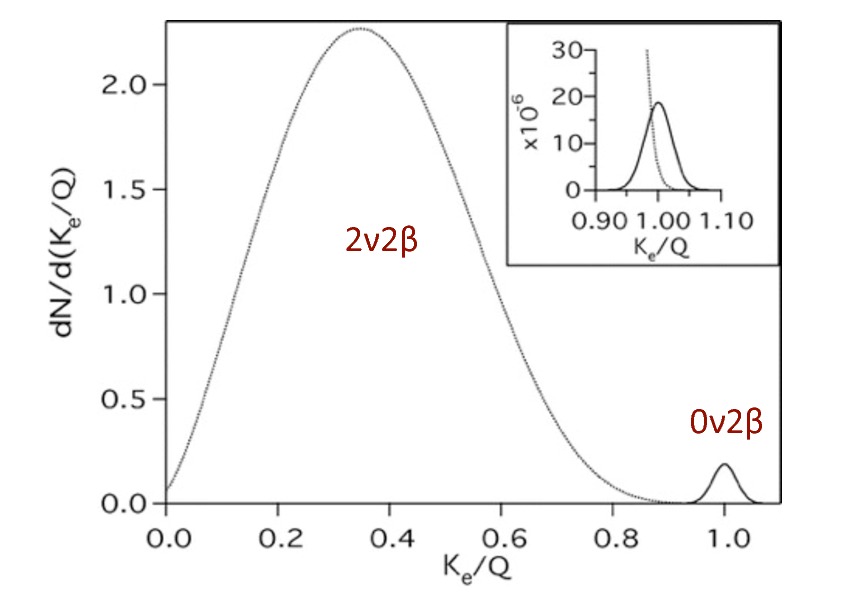}}
\qquad\qquad
\subfigure[Current and future experiments.]
{\includegraphics[width=7.1cm]{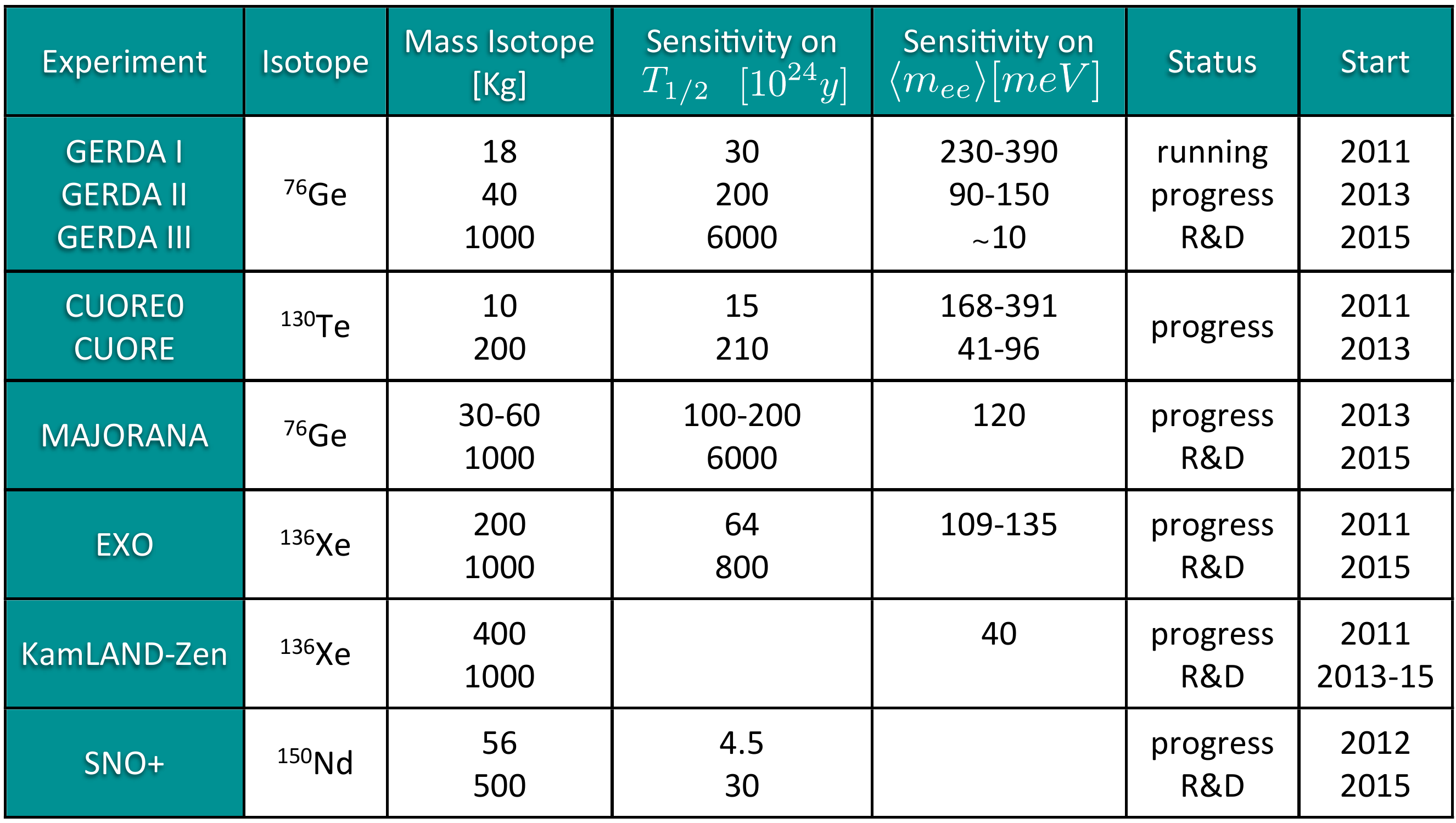}}
\vspace{-1cm}
\caption{\it (a) Energy distribution of the $2\nu2\beta$ and $0\nu2\beta$ decays. (b) Current and future experiments on the $0\nu2\beta$ decay. It is shown, when available, the isotope, the source mass, the sensitivity on the half-time and on the $0\nu2\beta$ effective mass in the case of Standard mechanism, and the status of the experiment.}
\vspace{-0.5cm}
\label{fig:Exp}
\end{figure}

On the experimental side, the present methods to detect the $0\nu2\beta$ decay consists in the direct observation of the two electrons in the final state. Generically the various $2\beta$ decays are separated just on the different distribution of the electron sum energies, as shown in fig.~\ref{fig:Exp}(a): a continuous bell distribution for $2\nu2\beta$ and a sharp line for $0\nu2\beta$. The latter corresponds to the $Q$ value of the nuclear transition. The observables are the electron sum energy, the single electron energy (only in few detectors) and the angular correlation of the two electrons in the final state. 

In all these measurements, a large sensitivity is necessary and can be achieved with: a large source mass with an high isotope abundance; a large $Q$-factor to avoid the environmental background; a good energy resolution to improve signal/background; shields against environmental radioactivity and cosmic rays; minimizing the detector size to lower the background from impurities that scale with the volume of the detector; using pure sources, such as gaseous and liquid sources that can be continuously purified; identifying the daughter isotope in coincidence with the electron pair to reject many backgrounds, possible if the $0\nu2\beta$ decay proceeds into an exited state with the subsequent $\gamma$ ray emission.

The detectors are defined inhomogeneous (homogeneous), if they are distinct  from (coincident with) the sources. Inhomogeneous detectors usually consist in source foils between scintillation detectors and allow a precise reconstruction of the event topology and the study of different type of isotopes. On the other hand only relatively small source masses can be adopted. 

Homogeneous detectors usually consist in semiconductors, liquids or gasses and are based on scintillation techniques. They allow the adoption of large source masses, a very high energy resolution, an high detection efficiency and in some cases the indication of the event topology. On the other hand only a restricted number of isotopes can be studied in such detectors. 

Among the past experiments, the best lower bound on the $0\nu2\beta$ half-time comes from the Hidelberg-Moscow experiment \cite{KlapdorKleingrothaus:2000sn}, based on high pure $^{76}Ge$ semiconductor: at $90\%$ of C.L. $T_{1/2}>19\times10^{24}\;\text{yr}$. Part of the collaboration claimed the observation of the $0\nu2\beta$ decay \cite{KlapdorKleingrothaus:2006ff} corresponding to a value $T_{1/2}=(22.3^{+4.4}_{-3.1})\times10^{24}\;\text{yr}$. The corresponding analysis has been strongly criticized, but only the present and future experiments will be able to confirm or exclude such a result. 

An incomplete list of current and future experiments is presented in fig.~\ref{fig:Exp}(b): all the experiments have been designed to reach and pass the H-M claim and the first results are expected in few years. Furthermore, notice that many different techniques and many different isotopes are involved in these list: this will allow the determination of the systematic errors, a better understanding of signals and backgrounds and possibly the discrimination among different mechanisms originating the $0\nu2\beta$ decay.\\

\begin{figure}[h!]
\vspace{-0.5cm}
\centering
\subfigure[Phase Space.]
{\includegraphics[width=5cm]{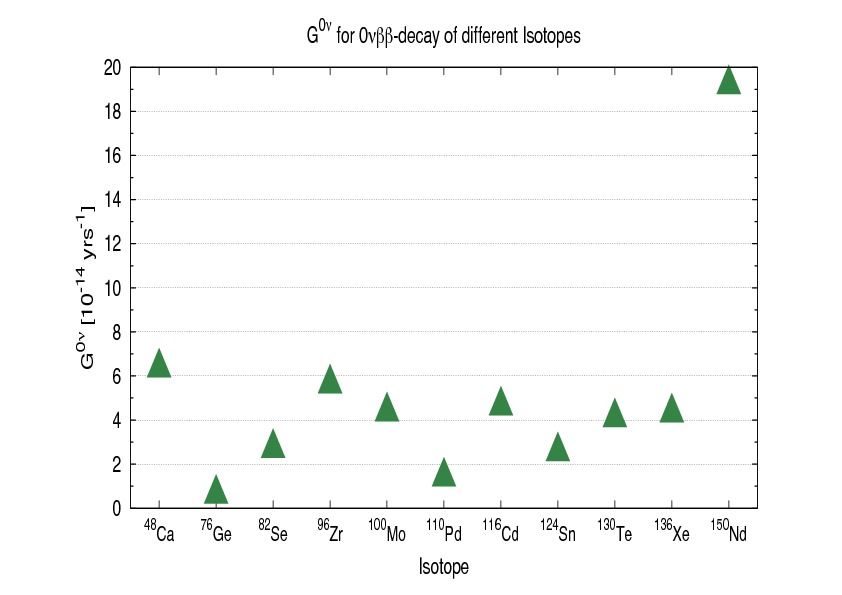}}
\qquad\qquad
\subfigure[Nuclear Matrix Element.]
{\includegraphics[width=6.5cm]{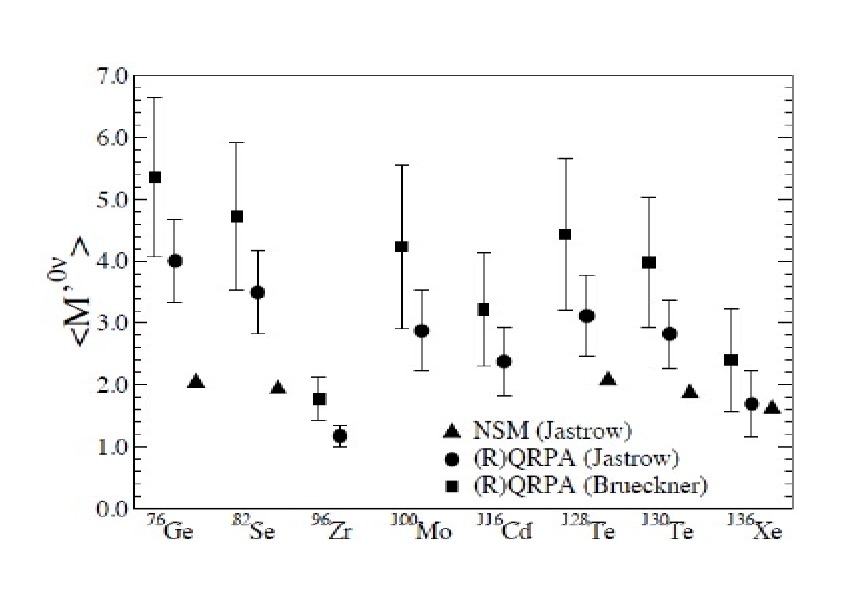}}
\vspace{-1cm}
\caption{\it (a) Phase space $G^{0\nu}(Q^5,Z)$ for different isotopes. (b) Nuclear Matrix Elements for different isotopes in the NSM and QRPS approaches.}
\vspace{-0.5cm}
\label{fig:PhaseAndNuclear}
\end{figure}

The experimental results of the $0\nu2\beta$ half-time is translated into particle physics parameters by the following expression:
\begin{equation}
T^{-1}_{1/2}=G^{0\nu}(Q^5,Z)\left|M^{0\nu}(A,Z)\cdot \Pi\right|^2\,,
\end{equation}
where $G^{0\nu}(Q^5,Z)$ is the phase space, $M^{0\nu}(A,Z)$ the matrix element and $\Pi$ is a function of particle physics parameters. The phase space may depend on the particle physics process that determines the $0\nu2\beta$ decay, but it is almost equal for transitions with only 2 electrons in the final state, as can be noticed in fig.~\ref{fig:PhaseAndNuclear}(a).

The nuclear matrix element $M^{0\nu}(A,Z)$ depends on the particle physics process which determines the transitions and on the isotope considered. Two main approaches are followed: the Nuclear Shell Model (NSM) allows the study of arbitrary complicated configurations, but limited to few single-particle orbitals outside the inert core; the Quasi-particle Random Phase Approximation (QRPS) is able to treat many single-particle states, but only a limited set of configurations. In fig.~\ref{fig:PhaseAndNuclear}(b) we can see the results of the two methods for different isotopes. To be noticed that a lot of effort has been put for the Standard mechanism, but not for the exotic ones. It is from the nuclear matrix element computations that comes the largest theoretical uncertainty in the $0\nu2\beta$ decay rate and unfortunately the precision is not expected to improve more that $20\%$ in the next 10 years. 

The function of the particle physics parameters $\Pi$ is defined by the mechanism that produces the $0\nu2\beta$ transition. In the Standard mechanism, as shown in fig.~\ref{fig:mechanisms}(a), the three light oscillating Majorana neutrinos give the main contribution and the function takes the following form:
\begin{equation}
\Pi \equiv \langle m_{ee}\rangle\,,
\end{equation}
where $\langle m_{ee}\rangle$ is the $0\nu2\beta$ effective mass. We can now translate the experimental present bounds and future sensitivities of the half-time of the $0\nu2\beta$ decay on the equivalent values of the $0\nu2\beta$ effective mass: while the present bound is $0.21-0.53\;\text{eV}$ ($90\%$ C.L.) the H-M claim is $0.32\pm0.03\;\text{eV}$; the future sensitivities can be read from the fifth column of the table in fig.~\ref{fig:Exp}(b).

\mathversion{bold}
\section{The $0\nu2\beta$ Effective Mass}
\mathversion{normal}

The $0\nu2\beta$ effective mass is the function of the particle physics parameters entering in the expression for the $0\nu2\beta$ decay rate. It takes a precise expression when considering the Majorana neutrino mass matrix:
\begin{equation}
\langle m_{ee}\rangle =
\left|\sum_k U^2_{ek}\,m_k\right|
= \left|c^2_{12}\, c^2_{13}\, m_1+
s^2_{12}\, c^2_{13}\, e^{i\alpha_{21}}\,m_2+
s^2_{13}\, e^{i\alpha_{31}}\,m_3\right|\,,
\label{mee}
\end{equation}
where $U_{ek}$ are the elements of the first row of the PMNS matrix, containing the solar and reactor and angles and the Majorana phase, and $m_k$ the neutrino masses. In the expression $c_{ij}$ and $s_{ij}$ refer to cosines and sines of $\theta_{ij}$, while $\alpha_{21,31}$ are the Majorana phases in the usual convention. The r.h.s. of eq.~(\ref{mee}) depends on 7 unknown quantities, while only the quantity $\langle m_{ee}\rangle$ can be read from $T^{-1}_{1/2}$. In order to extract any information from eq.~(\ref{mee}), more constraints and other correlations among the same parameters are necessary. 

\begin{figure}[h!]
\vspace{-0.5cm}
\centering
\subfigure[Cosmological bounds.]
{\includegraphics[width=6.2cm]{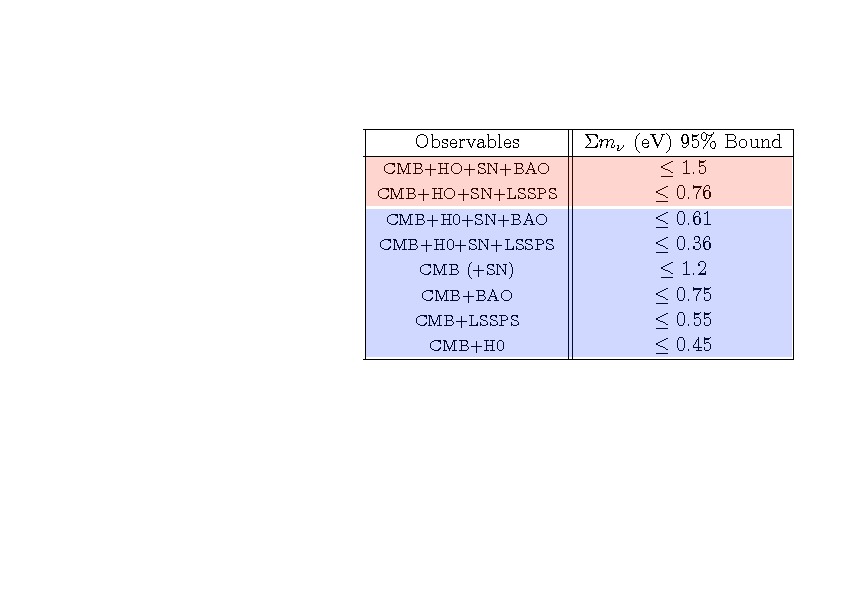}}
\qquad\qquad
\subfigure[Single $\beta$ decay.]
{\includegraphics[width=6.8cm]{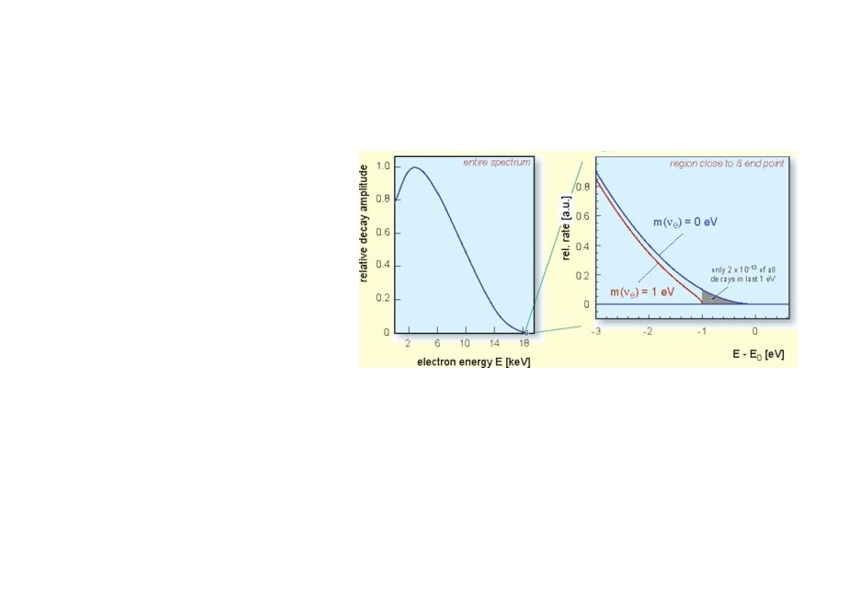}}
\vspace{-1cm}
\caption{\it (a) The cosmological bounds on the sum of the neutrino masses. See the text for details. (b) Energy spectrum of the single $\beta$ decay.}
\vspace{-0.5cm}
\label{fig:CosmoAndSingleBeta}
\end{figure}

Cosmological analyses put bounds on the simple sum of the neutrino masses:
\begin{equation}
\Sigma\,=\,\sum_{k=1}^{3}\,m_k\,.
\end{equation}
In fig.~\ref{fig:CosmoAndSingleBeta}(a), the table shows the recent results \cite{GonzalezGarcia:2010un} considering two contexts and different combination of the cosmological data: the Blue corresponds to the standard cosmological model $\Lambda$CDM with massive neutrinos and the Red the generalization with non-vanishing curvature and with $\omega\neq-1$ in the Dark Matter equation of state. CMB stands for the Cosmic Microwave Background, HO for the Hubble Constant, SN for the high-redshift Type-I SuperNovae, BAO for the Baryon Acoustic Oscillation, LSSPS for Large Scale Structure matter Power Spectrum. In the next few years, the sensitivity should lower down to $0.1\;\text{eV}$ considering the new combined analysis with CMB and Ly-$\alpha$ data.

The single $\beta$ decay experiments are sensible to the following observable:
\begin{equation}
m_{\beta}\,=\,\sqrt{\sum_{k=1}^3\,\left|U_{ek}\right|^2\,m^2_k}\,.
\end{equation}
In fig.~\ref{fig:CosmoAndSingleBeta}(b), the picture shows the region of interest: massive neutrinos produce a tiny effect only in the tail of the curve. The present upper bound on $m_\beta$ is $2.3\;\text{eV}$ ($95\%$ C.L.) from the Mainz \cite{Kraus:2004zw} and Troitsk \cite{Lobashev:2003kt} collaborations. The future sensitivity should be lowered down to $0.2\;\text{eV}$ (0.1\;\text{eV}) by the KATRIN \cite{Host:2007wh} (MARE \cite{Monfardini:2005dk}) experiment.

\begin{figure}[h!]
\vspace{-0.5cm}
\centering
\subfigure[$0\nu2\beta$ effective mass.]
{\includegraphics[width=5.2cm]{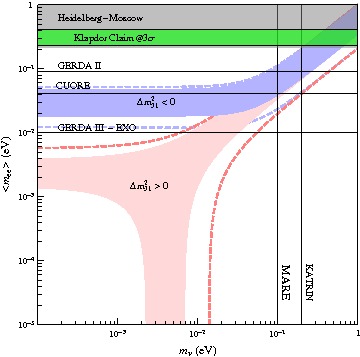}}
\qquad\qquad
\subfigure[Correlations.]
{\includegraphics[width=5cm]{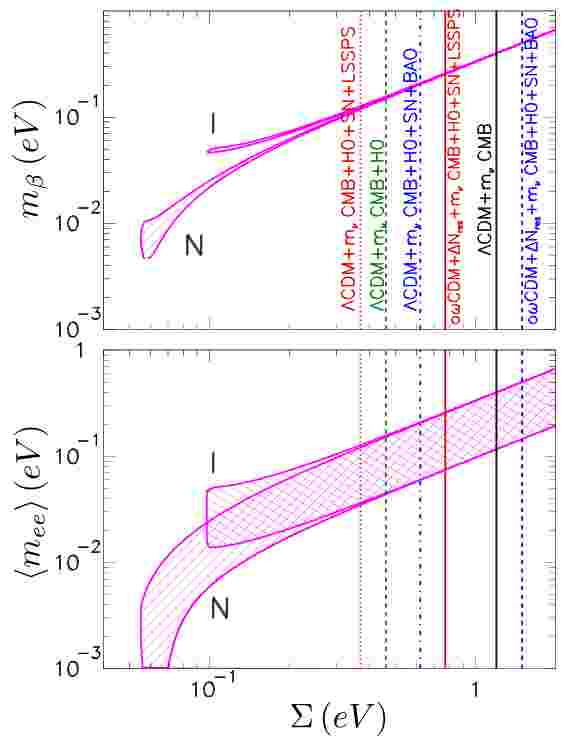}}
\vspace{-1cm}
\caption{\it (a) The $0\nu2\beta$ effective mass as a function of the lightest neutrino mass for both the mass orderings: in Red (Blue) the NO (IO). The coloured areas (dashed lines) correspond to the neutrino oscillation parameters at $1\sigma$ ($3\sigma$) from \cite{Fogli:2011qn}. (b) Correlation between $\Sigma$ and $m_\beta$ and $\langle m_{ee}\rangle$. See details in the text.}
\vspace{-0.5cm}
\label{fig:0nu2betaAndCorrelations}
\end{figure}

Considering the neutrino oscillation experiments, further constraints apply to the expression in eq.~(\ref{mee}): while it is possible to fix the two mixing angles, only two out of the three neutrino masses can be fixed. Indeed in the case of the normal ordering (NO), one can write
\begin{equation}
m_2=\sqrt{m_1^2+\Delta\,m^2_{sol}}\,,\qquad\qquad
m_3=\sqrt{m_1^2+\Delta\,m^2_{atm}}
\label{massesNO}
\end{equation}
while in the case of inverse ordering (IO), they are
\begin{equation}
m_1=\sqrt{m_3^2+\Delta\,m^2_{atm}}\,,\qquad\qquad
m_2=\sqrt{m_3^2+\Delta\,m^2_{sol}+\Delta\,m^2_{atm}}\,.
\label{massesIO}
\end{equation}
The r.h.s. of eq.~(\ref{mee}) now depends only on three parameters: the lightest neutrino mass and the two Majorana phases. Taking these phase in the range $[0,\pi]$, one can generate the plot in fig.~\ref{fig:0nu2betaAndCorrelations}(a), where the $0\nu2\beta$ effective mass is drown as a function of the lightest neutrino mass for both the mass orderings. The horizontal lines corresponds to the past and future sensitivities of the $0\nu2\beta$ decay experiments, while the vertical ones to the KATRIN and MARE future sensitivities.

In fig.~\ref{fig:0nu2betaAndCorrelations}(b), correlations among the sum of the neutrino masses, the single $\beta$ parameter and the $0\nu2\beta$ effective mass are shown. From these plots, one can hope to understand the type of the hierarchy already in the next decade, thank to the improvements in the $0\nu2\beta$ experiments and cosmological analyses.

\mathversion{bold}
\section{Flavour Symmetries in the $0\nu2\beta$ Decay}
\mathversion{normal}

An alternative strategy to increase the predictivity on the parameters entering the $0\nu2\beta$ effective mass is considering flavour symmetries. The main scope of introducing a flavour symmetry is to explain the origin of fermion mass hierarchies and mixings. This leads to specific structures for the neutrino mass matrix, that in some cases benefits of a large predictive power \cite{Rodejohann}. The most known predictive textures are related to the Tri-Bimaximal (TB) pattern \cite{Harrison:2002er,Xing:2002sw}, the Bimaximal (BM) pattern \cite{Barger:1998ta,Altarelli:1998sr} and the Golden Ratio (GR) pattern \cite{Kajiyama:2007gx,Rodejohann:2008ir}. All these three schemes correspond to a maximal atmospheric angle and a vanishing reactor angle:
\begin{equation}
\sin^2\theta_{23}=1/2\,,\qquad\qquad
\sin\theta_{13}=0\,.
\end{equation}
They differ only in the value of the solar angle:
\begin{equation}
\sin^2\theta^{TB}_{12}=1/3\,,\qquad\qquad
\sin^2\theta^{BM}_{12}=1/2\,,\qquad\qquad
\tan\theta^{GR}_{12}=1/\phi\qquad\text{with}\qquad \phi=(1+\sqrt5)/2\,.
\end{equation}
While the TB and GR values of the solar angle are compatible with the experimental data, the BM value is well outside the $3\sigma$ region and needs large corrections \cite{Altarelli:2009gn,Toorop:2010yh,Meloni:2011fx}.

In the following I will concentrate on the TB pattern. The TB mixing matrix is given by 
\begin{equation}
U_{TB}=\left(
         \begin{array}{ccc}
           \sqrt{2/3} & 1/\sqrt3 & 0 \\
           -1/\sqrt6 & 1/\sqrt3 & -1/\sqrt2 \\
           -1/\sqrt6 & 1/\sqrt3 & +1/\sqrt2 \\
         \end{array}
       \right)
\end{equation}
and the most general mass matrix diagonalised by this mixing scheme is given by
\begin{equation}
m^{TB}_\nu= U_{TB}\,\text{diag}\left(m_1,\,m_2\,e^{i\,\alpha_{21}},\,m_3\,e^{i\,\alpha_{31}}\right)\,U_{TB}^T=
\left(
  \begin{array}{ccc}
    x & y & y \\
    y & z & x+y-z \\
    y & x+y-z & z \\
  \end{array}
\right)\,.
\end{equation}
Such mass matrix is $\mu-\tau$ symmetric and magic symmetric, that leads to $(m^{TB}_\nu)_{23}=x+y-z$. Furthermore the $0\nu2\beta$ effective mass is given by
\begin{equation}
\langle m_{ee}^{TB}\rangle=|x|=\dfrac{1}{3}\left|2\,m_1+m_2\, e^{i\alpha_{21}}\right|\,.
\end{equation}
Few comments are in order. Just considering the most general mass matrix of the TB type, the $0\nu2\beta$ effective mass depends only on 3 parameter instead of 7 as in the general case. It is possible to further reduce the dependence on only 2 parameters considering the expressions in eqs.~(\ref{massesNO}) and (\ref{massesIO}): $\langle m_{ee}^{TB}\rangle$ depends now only on the lightest neutrino mass and on the only Majorana phase entering the expression.  

When considering specific TB models, then the lightest neutrino mass and the Majorana phase are not two independent parameters anymore, but they take specific expressions in terms of the parameters of the model, reducing the parameter space of the $0\nu2\beta$ effective mass. In the following I will discuss some examples in which this indeed happens and I will compare the different predictions.

In the low-energy approach, neutrino masses can be described by the Weinberg operator
\begin{equation}
m_{\nu}=Y_{ij}\dfrac{\left(\bar{\ell}^c_i\,H\right)\left(H^T\,\ell_j\right)}{\Lambda_{LN}}\,,
\end{equation}
where $Y_{ij}$ contains the flavour structure and $\Lambda_{LN}$ is the scale of the LNV. In the class of models with discrete flavour symmetries introduced at the high energy scale \cite{HagedornBazzocchi}, $Y_{ij}$ is determined by the vacuum expectations values of new scalar fields, called flavons. The TB mixing naturally arises only when $\ell$ transforms as a triplet of the flavour symmetry, while on the other hand flavons can transform as triplets, doublets and singlets.

\begin{figure}[h!]
\vspace{-0.5cm}
\centering
\subfigure[Effective AF model.]
{\includegraphics[width=5.2cm]{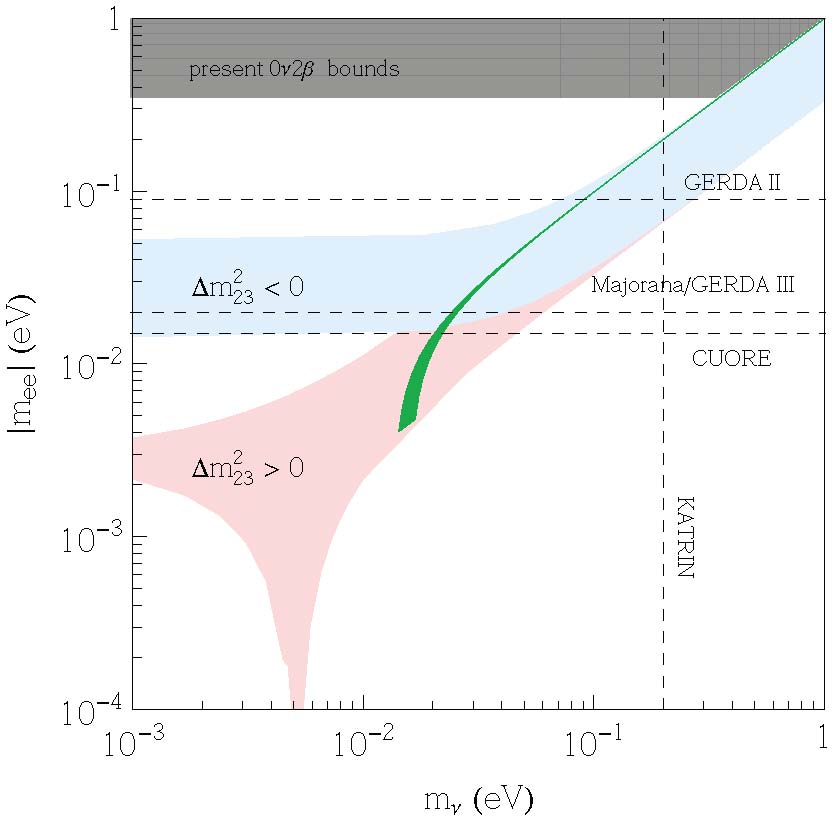}}
\subfigure[See-Saw AF model.]
{\includegraphics[width=5.2cm]{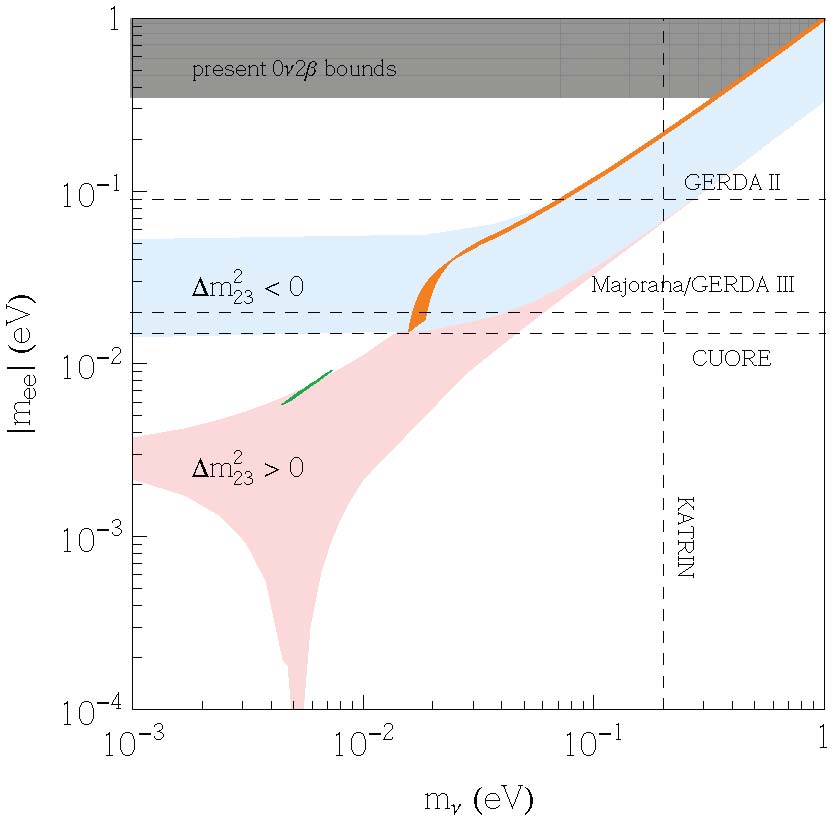}}
\subfigure[Niemeyer model.]
{\includegraphics[width=5.2cm]{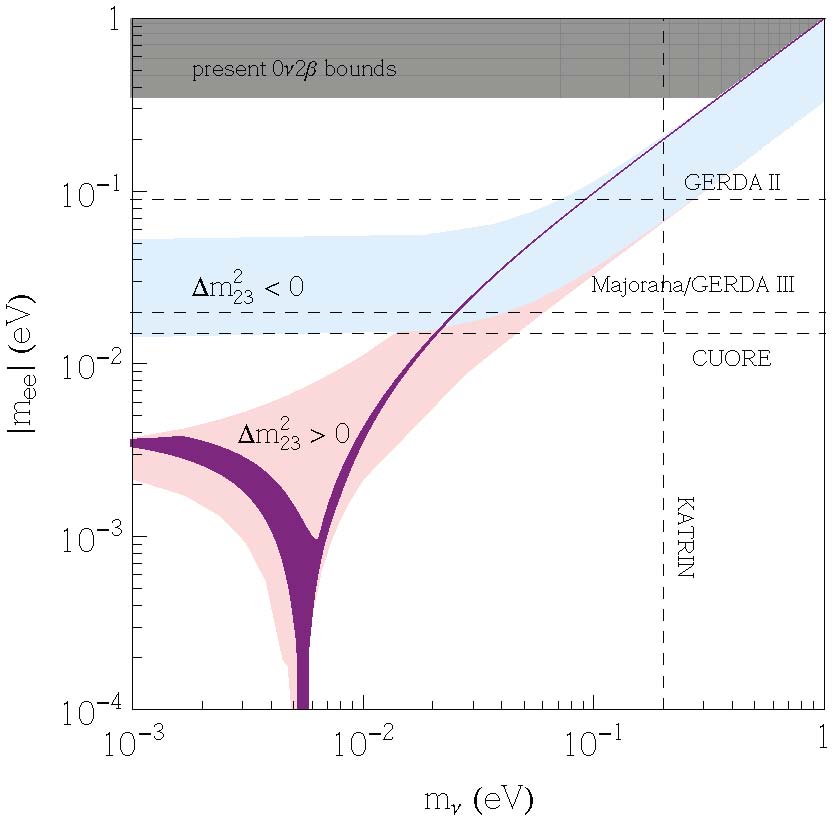}}
\vspace{-1cm}
\caption{\it The $0\nu2\beta$ effective mass as a function of the lightest neutrino mass for both the mass orderings: in Red, Green and Purple (Blue and Orange) the NO (IO). The light coloured areas correspond to the TB parameter space. The dark coloured areas represent the predictions for the AF model for the effective case (a) and the type I See-Saw case (b), and for the Niemeyer model (c).}
\vspace{-0.5cm}
\label{fig:0nu2betaAF}
\end{figure}

The first flavour model I consider is the Altarelli-Feruglio (AF) model \cite{Altarelli:2005yx,Altarelli:2005yp} based on the discrete group $A_4$. Since in this group there are only triplet and singlet representations, only two terms can enter the Weinberg operator:
\begin{equation}
m_{\nu}\supset\left\{\dfrac{\varphi_3}{\Lambda_f}\dfrac{\left(\bar{\ell}^c\,H\right)\left(H^T\,\ell\right)}{\Lambda_{LN}},\,
\dfrac{\varphi_1}{\Lambda_f}\dfrac{\left(\bar{\ell}^c\,H\right)\left(H^T\,\ell\right)}{\Lambda_{LN}}\right\}\,,
\end{equation}
where $\varphi_R$ is the flavon transforming in the $R$ representation, $\Lambda_f$ is the energy scale of the flavour dynamics and the flavour contractions are understood. The resulting Majorana neutrino mass matrix is exactly diagonalized by the TB mixing. A similar analysis can be done considering three RH neutrinos transforming as a triplet of $A_4$ and the same flavons would determine the flavour structure of the neutrino mass matrices. In particular the Majorana mass matrix is of the TB type and the Dirac mass matrix turns out to be proportional to the rotation in the 2-3 sector. Without entering into further details, the results \cite{Feruglio:2007uu} for the $0\nu2\beta$ effective mass are shown in fig.~\ref{fig:0nu2betaAF}. For the effective case, only the NO is allowed and lower bounds for $m_1$ and $\langle m_{ee}\rangle$ can be determined. In the See-Saw case, both the mass orderings are described: for the NO only a narrow parameter space is allowed, while for the IO only a lower bound can be fixed for $m_3$ and $\langle m_{ee}\rangle$.

The second realization I consider is the Niemeyer model \cite{Hirsch:2008rp}, that is also based on the $A_4$ group. In this case only the See-Saw case has been presented. The structure of the neutrino mass matrices is opposite with respect the AF model: the Dirac mass matrix is of the TB type and the Majorana mass matrix turns out to be proportional to the identity. The result for the $0\nu2\beta$ effective mass is shown in fig.~\ref{fig:0nu2betaAF}(c): only the NO is allowed and the lower bound on $m_1$ is below the cancellation point. With respect to the AF model predictions, in the Niemeyer model a vanishing $\langle m_{ee}\rangle$ is possible.

\begin{figure}[h!]
\vspace{-0.5cm}
\centering
\subfigure[Effective $S_4$ model.]
{\includegraphics[width=5.2cm]{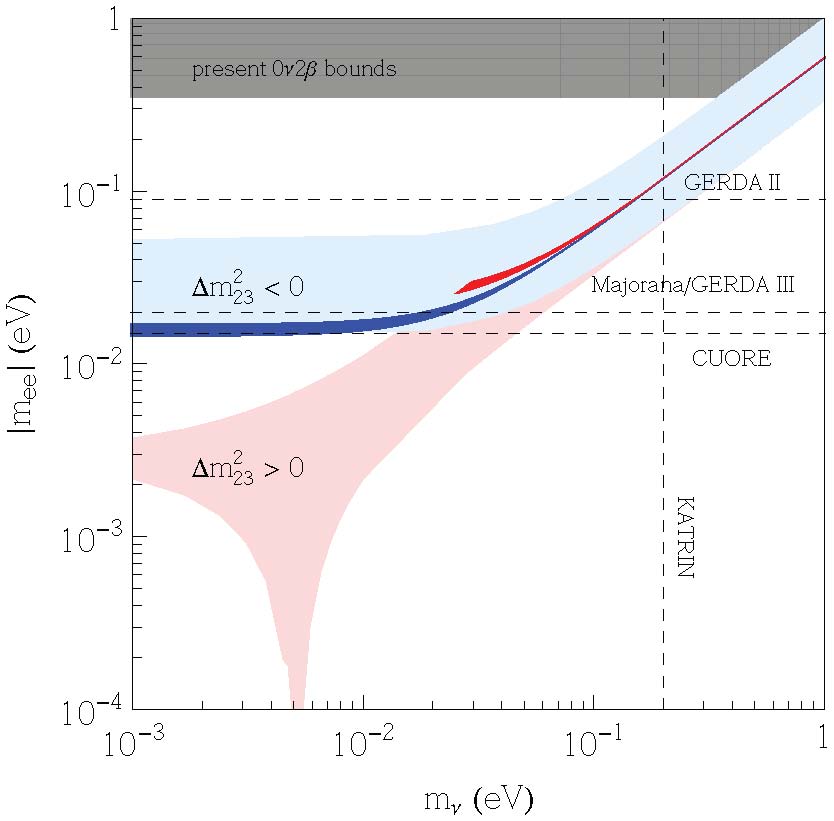}}
\qquad\qquad
\subfigure[See-Saw $S_4$ model.]
{\includegraphics[width=5.2cm]{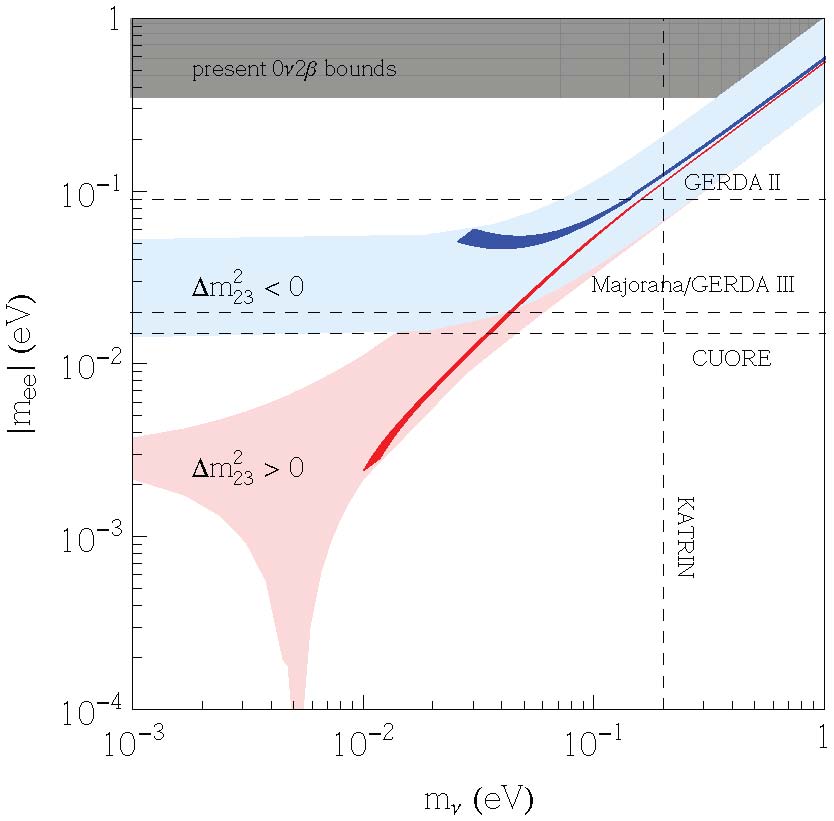}}
\vspace{-1cm}
\caption{\it The $0\nu2\beta$ effective mass as a function of the lightest neutrino mass for both the mass orderings: in Red (Blue) the NO (IO). The light coloured areas correspond to the TB parameter space. The dark coloured areas represent the predictions for the $S_4$ model for the effective case (a) and the type I See-Saw case (b).}
\vspace{-0.5cm}
\label{fig:0nu2betaS4}
\end{figure}

The third model I consider has been presented in \cite{Bazzocchi:2009da,Bazzocchi:2009pv} and it is based on the $S_4$ discrete group. This group is larger than $A_4$ and owns triplet, singlet and doublet representations. In this particular model, in both the effective and See-Saw cases, only triplet and doublet flavons contribute to the neutrino mass matrices. The corresponding prediction for the $0\nu2\beta$ effective mass is shown in fig.~\ref{fig:0nu2betaS4}: both the mass orderings are allowed and lower bounds for the lightest neutrino mass and $\langle m_{ee}\rangle$ are present.

Comparing the plots in figs.~\ref{fig:0nu2betaAF} and \ref{fig:0nu2betaS4}, the predictions from the different models overlap in many points. Only one region allows a discrimination among them: in the soon testable region in which the neutrino mass spectrum is quasi-degenerate, the $A_4$ model lines run along the boundary of the TB allowed region, while the $S_4$ model line stays in the middle of the band. This behaviour reflects the fact that, for almost degenerate masses, in the $A_4$ model the Majorana phase is vanishing, while in the $S_4$ it is not. The reason can be found in the use of single or doublet representations for the flavons contributing to the neutrino mass matrices.

\begin{figure}[h!]
\vspace{-0.5cm}
\centering
\subfigure[Effective perturbed AF model.]
{\includegraphics[width=5.2cm]{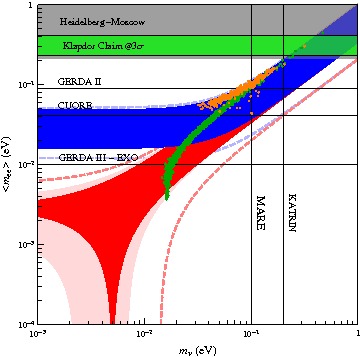}}
\qquad\qquad
\subfigure[See-Saw perturbed AF model.]
{\includegraphics[width=5.2cm]{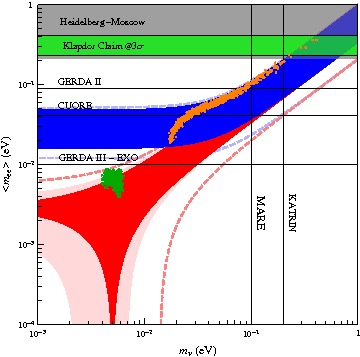}}
\vspace{-1cm}
\caption{\it The $0\nu2\beta$ effective mass as a function of the lightest neutrino mass for the AF model with the higher order corrections. See fig.~\ref{fig:0nu2betaAF} for further details on the plots.}
\vspace{-0.5cm}
\label{fig:0nu2betaAFNLO}
\end{figure}

To be noticed that all of these predictions are strictly valid only considering the first order contributions in the expansion in $\Lambda_f^{-1}$: indeed usually corrections to the mass matrices arise from higher order operators and the predictions are consequently modified. As a title of example, I report in fig.~\ref{fig:0nu2betaAFNLO} the plots of the $0\nu2\beta$ effective mass in the AF model when the higher order corrections are taken in considerations: now in both the cases, both the NO and the IO are allowed and the corresponding points cover a larger area. 

We could expect a similar behaviour also for the other models and as a result the discrimination among the different predictions turns out to be even harder.

\mathversion{bold}
\section{Correlations of the $0\nu2\beta$ Effective Mass}
\mathversion{normal}

\begin{figure}[h!]
\vspace{-0.5cm}
\centering
\subfigure[Correlation with $BR(\tau\to\mu\gamma)$]
{\includegraphics[width=6cm]{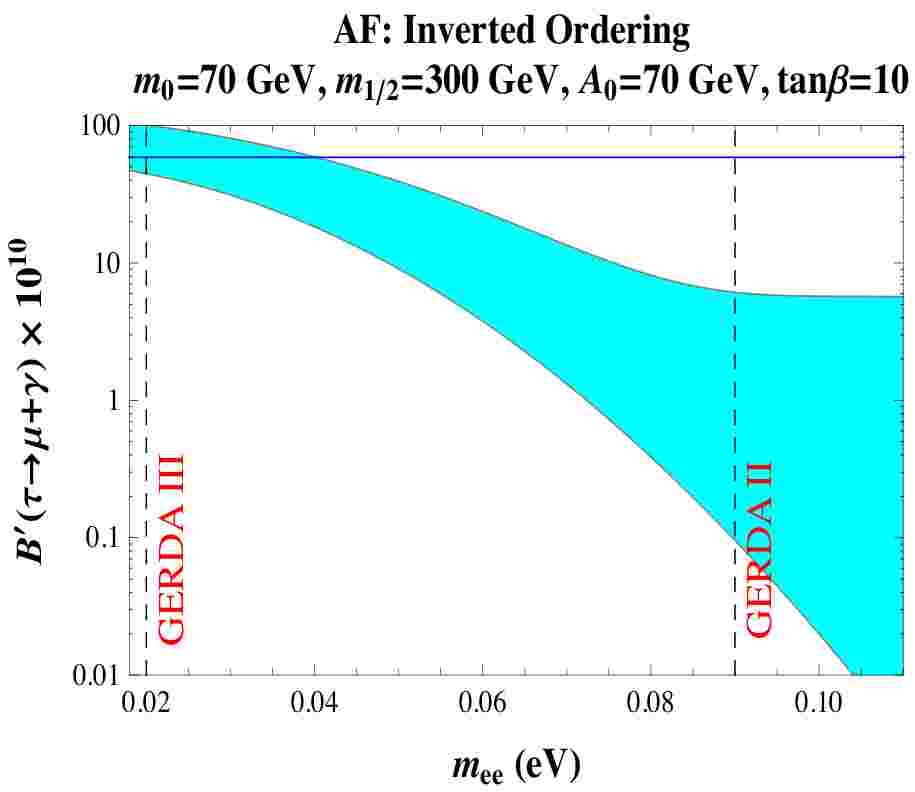}}
\qquad
\subfigure[Correlation with Sum Rules]
{\includegraphics[width=5.5cm]{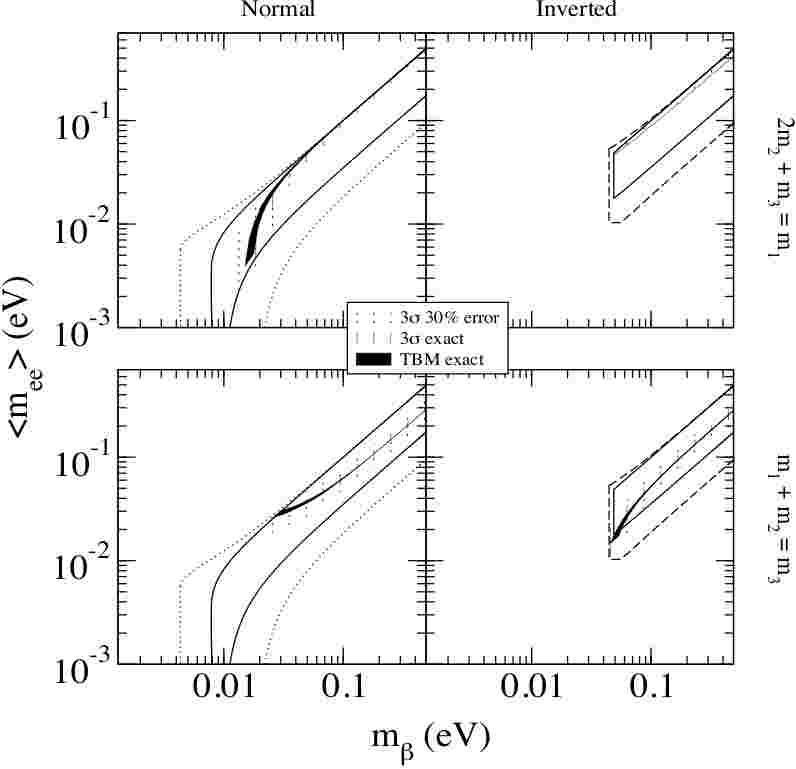}}
\vspace{-1cm}
\caption{\it (a) Correlation between the $0\nu2\beta$ effective mass and the radiative $\tau$ decay for the specific SUSY parameter space specified in the plot and for the IO mass spectrum. Plot from \cite{Hagedorn:2009df}. (b) Correlation between the $0\nu2\beta$ effective mass and neutrino mass sum rules. In the upper row the AF model and in the lower one the $S_4$ model. Plot from \cite{Rodejohann:2011mu}.}
\vspace{-0.5cm}
\label{fig:0nu2betaLFVSumRules}
\end{figure}

In the previous section, I pointed out that the $0\nu2\beta$ effective mass may be expressed in terms of only one single free parameter when considering explicit flavour models. This property may hold also for other observables, such as lepton flavour violating (LFV) decays. In the specific context of the Type I See-Saw mechanism in supersymmetry, the Branching Ratio of the radiative lepton decays $\ell_i\to\ell_j\gamma$ can be approximated by
\begin{equation}
Br(\ell_i\to\ell_j\gamma)\simeq\dfrac{\alpha^3}{G_F^2}\dfrac{\left|\left(m^2_{eLL}\right)_{ij}\right|^2}{m_S^8}\tan^2\beta
\end{equation}
where $m^2_{eLL}$ mainly comes from the RGE effects and takes the form 
\begin{equation}
\left(m^2_{eLL}\right)_{ij}\propto (\hat Y_\nu^\dag\,\hat Y_\nu)_{ij}\log\left(\dfrac{m_1}{m_*}\right)+
(\hat Y_\nu^\dag)_{i2}\,(\hat Y_\nu)_{2j}\log\left(\dfrac{m_2}{m_1}\right)+
(\hat Y_\nu^\dag)_{i3}\,(\hat Y_\nu)_{3j}\log\left(\dfrac{m_3}{m_1}\right)\,.
\end{equation}
Considering the specific case of the AF model, the study of the RG running and of LFV observables turns out to be relevant to constrain the parameter space of the model \cite{Feruglio:2008ht,Feruglio:2009iu,Lin:2009sq,Feruglio:2009hu,Hagedorn:2009df,Merlo:2011hw}. In particular, the previous expression depends only on the lightest neutrino mass and therefore, once a point in the SUSY parameter space has been fixed, the $0\nu2\beta$ effective mass and the $Br(\ell_i\to\ell_j\gamma)$ can be easily correlated: in fig.~\ref{fig:0nu2betaLFVSumRules}(a) there is the correlation between the $0\nu2\beta$ effective mass and the $BR(\tau\to\mu\gamma)$ in the IO mass spectrum case. This particular SUSY parameter space generates a region that could be easily verified in the next years, as the sensitivity on the $0\nu2\beta$ effective mass improves.\\

An other interesting correlation of the $0\nu2\beta$ effective mass is with the neutrino mass sum rules usually present in flavour models. In fig.~\ref{fig:0nu2betaLFVSumRules}(b) the correlations with the sum rules for the AF and the $S_4$ models are shown. These relations are strictly true only at the LO, but the higher order contributions do not introduce large corrections. As a result, these plots are well representative of the models in consideration and allow a good test of such models, as the sensitivities on $\langle m_{ee}\rangle$ and $m_\beta$ will improve.

Further correlations of the $0\nu2\beta$ effective mass with baryogenesis through leptogenesis can also be found in the context of flavour models \cite{Bertuzzo:2009im,AristizabalSierra:2009ex,Hagedorn:2009jy}, but the possibility of measure Majorana phases in the context of $0\nu2\beta$ decay is weak.

\section{Conclusions}
Considering the experimental future improvements on the sensitivity of the $0\nu2\beta$ effective mass, it is possible to outline three possible scenarios: if $\langle m_{ee}\rangle\approx \mathcal{O}(100)\;\text{meV}$ the discovery of the $0\nu2\beta$ is expected in the next $1\div5$ years; if $\langle m_{ee}\rangle\approx 15\div20\;\text{meV}$ in the next $5\div10$ years; if $\langle m_{ee}\rangle\lesssim 2\div5\;\text{meV}$ new experiments must be designed.

However, if $0\nu2\beta$ is discovered, it is possible to conclude that the Lepton number is violated, Neutrinos have Majorana nature (through the Schechter-Valle theorem), but independent different observations are necessary to identify the mechanism that originates the $0\nu2\beta$ decay.

Only if the Standard mechanism turns out to be the main responsible for the $0\nu2\beta$ decay, depending on the sensitivity of the single $\beta$ mass parameter and on the sum of the neutrino masses, it will be possible to determine the type of the neutrino mass spectrum. This would represent a good constraint on flavour models, but it would be still hard to distinguish  among different models only considering $\langle m_{ee}\rangle$, $m_\beta$ and $\Sigma$, mainly due to the theoretical uncertainty on the observables. Only considering the correlation with other observables, such as LFV processes, neutrino mass sum rules, other LNV decays, and Leptogenesis, there would be a hope to rule out or clearly identify some flavour models.



%% file: Author/Rodejohann.tex
{\bf Abstract}\\
\vskip5.mm
The overwhelming majority of flavor symmetry models focusses on 
tri-bimaximal mixing (TBM). Neutrino mass sum-rules are given as one
rather robust example on how to distinguish some of the models from 
each other. We classify mechanisms to deviate from TBM and estimate 
the typical order of magnitude of the corrections. 
Then we present several alternatives to TBM, and outline their
possible origin in flavor symmetries.  Finally, examples on how to to 
accommodate light sterile neutrinos in flavor symmetry model are  
discussed. This works for eV-scale sterile neutrinos to explain the 
reactor anomaly, as well as for keV-scale sterile neutrinos which can 
act as warm dark matter. 


\vskip5.mm

\section{\label{sec:WRintro}Introduction: The Zoo of Models and how to distinguish them}
There is no need here to motivate the need for flavor symmetry models,
or to note the existence of a huge amount of those, for reviews see \cite{revs}. The vast majority
aims to reproduce tri-bimaximal mixing (TBM), 
\be \label{eq:WRU}
U = \left( \bad 
\sqrt{\frac 23} & \sqrt{\frac 13} & 0 \\
-\sqrt{\frac 16} & \sqrt{\frac 13} & -\sqrt{\frac 12} \\
\sqrt{\frac 16} & \sqrt{\frac 13} & \sqrt{\frac 12}
\ea \right) P \,  \Rightarrow \ba 
\sin^2 \theta_{13} = 0 \times \cos^2 \theta_{13} = 0 \\
\sin^2 \theta_{12} = \frac 12 \times \cos^2 \theta_{12} = \frac 13 \\
\sin^2 \theta_{23} = 1 \times \cos^2 \theta_{23} = \frac 12 
\ea \, , 
\ee
see however Section \ref{sec:WRalt} for a discussion of alternative
mixing scenarios. The diagonal matrix $P$ contains the Majorana
phases, $P = {\rm diag}(1, e^{i \alpha_2/2}, e^{i \alpha_3/2})$.  
The TBM form of the mixing matrix originates from
the mass matrix 
\be \label{eq:WRmnu}
\left(
\bad 
A & B & B \\[0.2cm]
\cdot & \frac{1}{2} (A + B + D) & \frac{1}{2} (A + B - D)\\[0.2cm]
\cdot & \cdot & \frac{1}{2} (A + B + D)
\ea 
\right) , \mbox{ where } 
\begin{array}{cl}
A =&  \frac 13 \left(2 \, m_1 + m_2 \, e^{-i\alpha_2} \right) ,\\
B =&  \frac 13 \left(m_2 \, e^{-i\alpha_2} - m_1 \right), \\
D =& m_3 \, e^{-i\alpha_3} \,.
\ea 
\ee
Note that the sum of the elements in each row, and in each column,
equals $A + 2 B = m_2 \,e^{-i\alpha_2}$. 

There are many models. 
Consider the Table in Figure \ref{fig:Rodejohann1}, which introduces a categorization of
$A_4$ models giving rise to tri-bimaximal mixing 
available in the literature \cite{arXiv:1003.2385}. The table illustrates for instance
the simple fact that even after choosing the symmetry group one is far
from done with constructing the model. The question arises how to
distinguish the proposed models from each other. Possibilities are 
lepton flavor violation implied e.g.~by low-lying scalars or 
collider implications of such particles. More aesthetical
considerations are whether leptogenesis is possible, whether the
models are compatible with GUTs, whether dark matter candidates are 
present, etc.

Another method to distinguish the models is made possible by the surprising feature that
neutrino mass sum-rules are often present in such models \cite{arXiv:1007.5217,arXiv:1111.5614}. 
While flavor symmetries cannot predict masses, relations between
masses, such as sum-rules, can very well be predicted. 
Examples are $2 m_2 + m_1 = m_3$, or $1/m_1 + 1/m_2 = 1/m_3$. Here the masses are
understood to be complex, i.e.~including the Majorana phases. The
sum-rules constrain the Majorana phases and masses, and therefore only
certain areas in the parameter space of the different mass observables (effective mass for
\onbb, sum of masses from cosmology and direct mass for direct
searches such as in KATRIN) are allowed \cite{arXiv:1007.5217}. This 
is illustrated for 4 sum-rules in Fig.~\ref{fig:Rodejohann2}. The
predictions of the sum-rules are rather robust. Note
that exotic models such as schizophrenic/bimodal scenarios, in which
some neutrino mass states are Dirac particles
\cite{arXiv:1008.1232}, also have rather distinct 
phenomenology in these observables. 
Another aspect in which \onbb~can say something about flavor symmetry
models has been noted in Ref.~\cite{arXiv:0804.1521}, namely that
often the effective mass is correlated non-trivially with other
neutrino observables, which could also be used to rule out certain
models. 

Inherent to all these arguments is of course that no other of the
countless mechanisms that can mediate \onbb~contributes significantly
to the process, see Ref.~\cite{arXiv:1106.1334} for a recent review.

\begin{figure}[h!]
\begin{center}
\includegraphics[width=10cm,height=8cm]{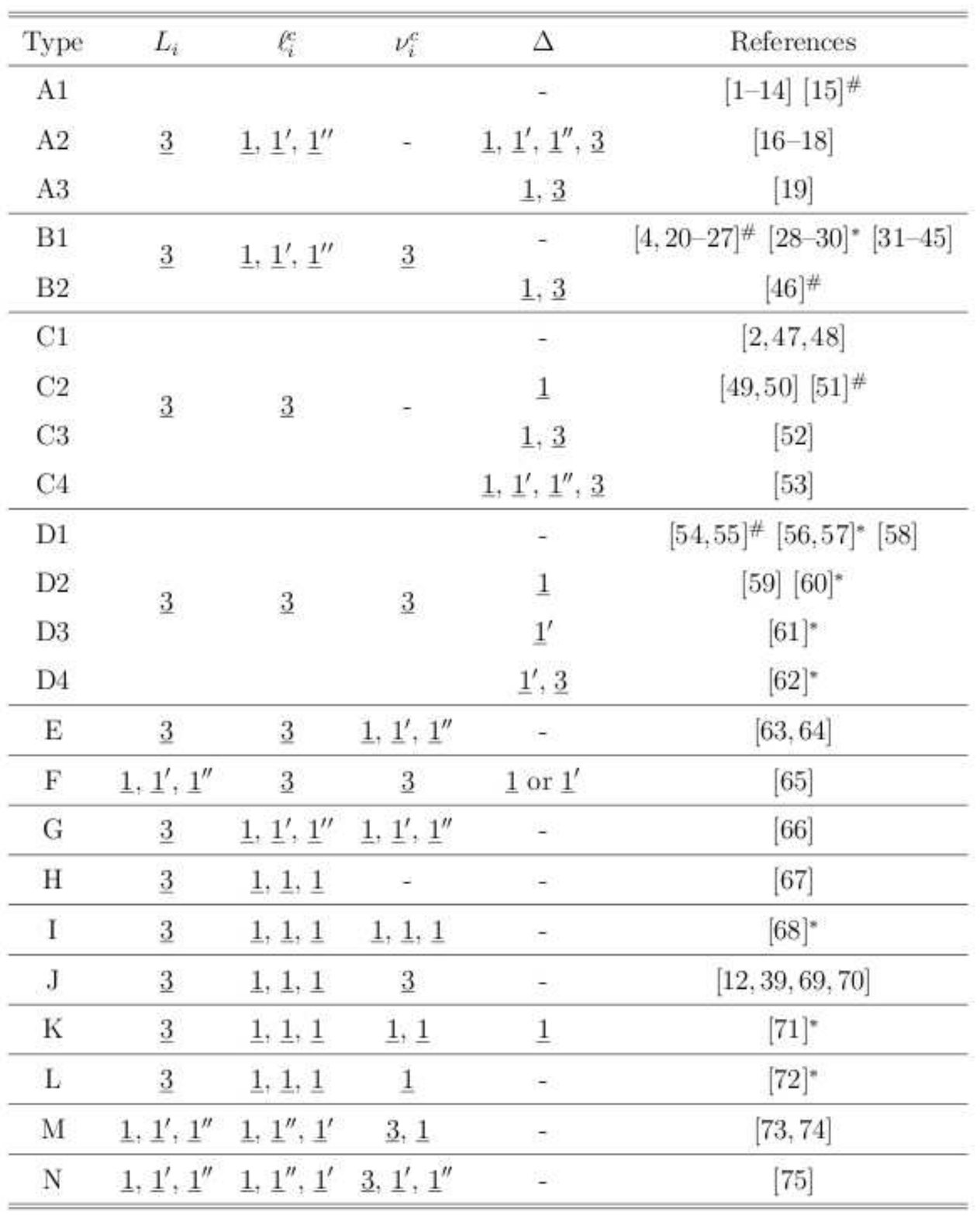}
\caption{\label{fig:Rodejohann1}Particle assignments of $A_4$ models
in the literature. Lepton doublets, charged lepton singlets and
right-handed neutrinos are denoted by $L_i$, $\ell_i^c$ and $\nu_i^c$,
respectively. $\Delta$  denotes the Higgs triplets that gives neutrinos mass in the type II seesaw
mechanism. Models that study the quark sector have the superscript $\#$, those that
embed $A_4$ into a GUT group have the superscript $\ast$. Updated
version of the Table in Ref.~\protect\cite{arXiv:1003.2385}, see the URL 
{\tt http://www.mpi-hd.mpg.de/personalhomes/jamesb/Table$\_$A4.pdf}
for the references and regular updates.}
\end{center}
\end{figure}

\begin{figure}[h!]
\begin{center}
\includegraphics[width=7cm,height=7cm]{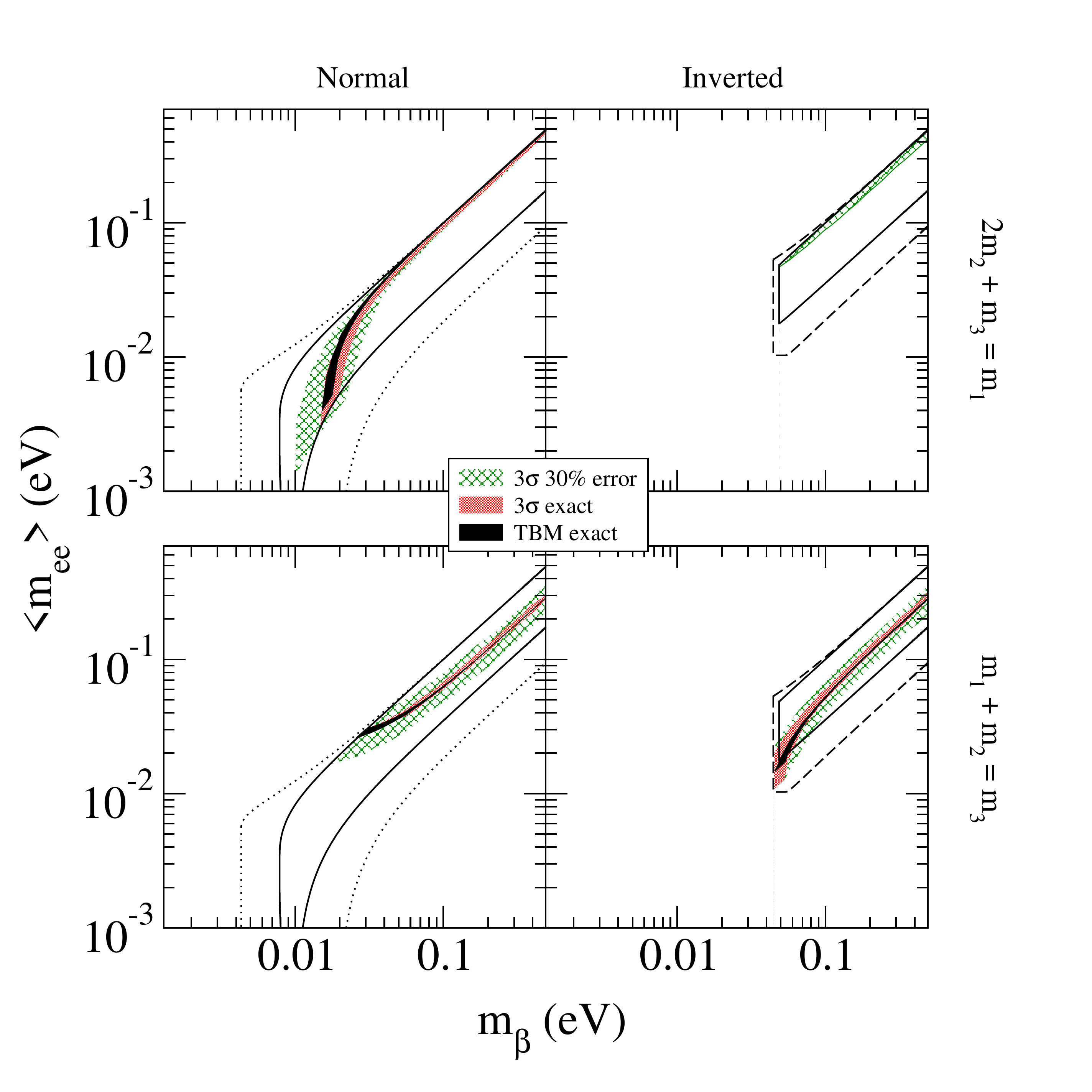}
\includegraphics[width=7cm,height=7cm]{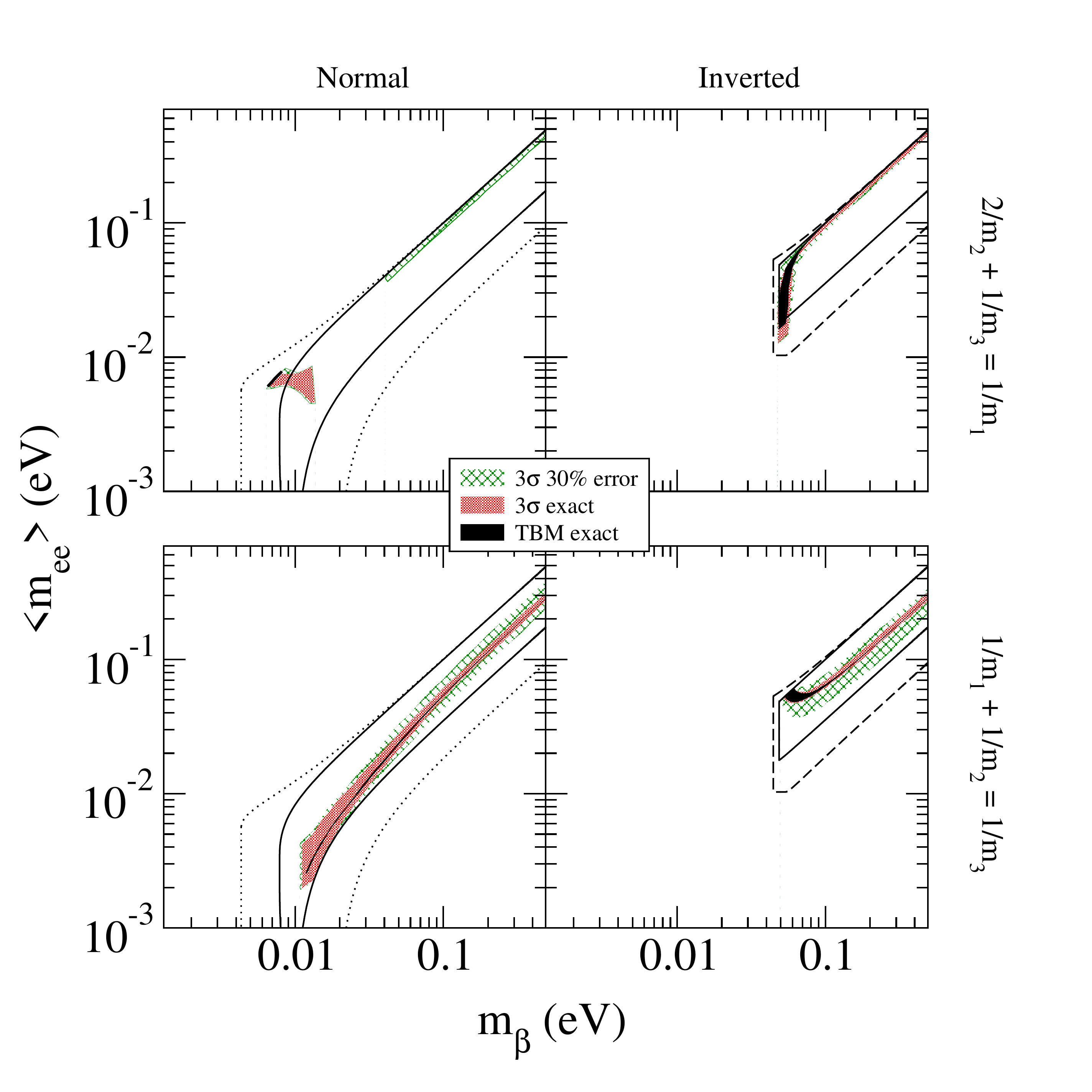}
\caption{\label{fig:Rodejohann2}Neutrino mass sum-rules: Shown in
solid (dashed) lines are the allowed parameter spaces of the KATRIN
observable, $\sqrt{\sum |U_{ei}|^2 m_i^2}$, and the effective mass of
\obb, $|\sum U_{ei}^2 m_i|$, for the best-fit ($3\sigma$) ranges of the oscillation
parameters. The dark areas are the indicated sum-rules and exact TBM,
the red areas for the sum-rules and the current $3\sigma$ ranges of the oscillation
parameters and the green areas for perturbed sum-rules. Taken from \protect\cite{arXiv:1007.5217}. }
\end{center}
\end{figure}

\section{\label{sec:WRpert}Perturbations to Tri-bimaximal Mixing}

While the data on $\theta_{12}$ and $\theta_{23}$ are still compatible
with their tri-bimaximal values, more and more evidence accumulates 
that $U_{e3}$ is non-zero, e.g.~from T2K \cite{arXiv:1106.2822} or Double Chooz
\cite{dc}. In this Section we will show how naive and simple methods
to deviate from TBM can lead to values of $|U_{e3}| \sim 0.1$ \cite{arXiv:0907.2869}. Note
that still the presence of TBM (or any other $\mu$--$\tau$ symmetric 
scheme, or schemes with initially vanishing $\theta_{13}$, for that
matter) is assumed. Alternative approaches are simply too assume that the
resemblance of lepton mixing to the TBM scheme is accidental
\cite{arXiv:1004.0099}, or that initially 
$|U_{e3}|$ is non-zero, see for instance 
Refs.~\cite{arXiv:1107.3486,arXiv:1107.3970,arXiv:1108.2497}. 

Model-independently, it is possible to
describe deviations from TBM in a ``triminimal'' way, via \cite{arXiv:0711.0052}
\be
U = R_{23}(\pi/4) R_{23}(\epsilon_{23}) R_{13}(\epsilon_{13}, \delta)
R_{12}(\epsilon_{12}) R_{12}(\theta_{\rm TBM}) \, , 
\ee
where $R_{ij}(\theta)$ is a rotation in $ij$-space and $\sin^2
\theta_{\rm TBM} = \frac 13$. It is easy to see 
that only one $\epsilon_{ij}$ is responsible for the deviation of 
each mixing angle $\theta_{ij}$ from its TBM value.

\subsection{\label{sec:uell}Charged Lepton Corrections}
One simple method to deviate a mixing scheme is to make use of the
relation $U = U_\ell^\dagger \, U_\nu$, 
where $U_\nu$ $(U_\ell)$ diagonalizes the neutrino (charged lepton)
mass matrix. One of those unitary matrices could give TBM, while the
other leads to deviations. Small to moderate deviations are observed
in experiment, hence the deviating matrix will consist presumably of
small mixing, and one can make the straightforward assumption that it is CKM-like. In
the case of $U_\nu$ giving TBM and $U_\ell$ consisting only of a
12-rotation with angle $\lambda$, one finds  
\be
|U_{e3}| = \frac{\lambda}{\sqrt{2}}~\mbox{ and } ~
\sin^2 \theta_{12} \simeq \frac 13 \left( 1 - 2 \lambda \cos \phi
\right) , 
\ee
where $\phi$ is the Dirac CP phase, and the atmospheric mixing angle
receives only small quadratic corrections. Note that the observed
value of $\theta_{12}$, which lies close to the TBM-value, implies
large to maximal CP violation. Fig.~\ref{fig:Rodejohann3} illustrates
these correlations. Interestingly, in case we would have
started from maximal $\theta_{12}$ (bimaximal mixing, see Section
\ref{sec:WRalt}), then $\phi$ would have to lie
close to zero. This is the Quark-Lepton Complementarity case
\cite{qlc}.  

The opposite case, $U_\ell$ related to TBM and $U_\nu$ CKM-like, leads
to small values of $|U_{e3}| \sim \lambda^2$.

\begin{figure}[h!]
\begin{center}
\includegraphics[width=7cm,height=4cm]{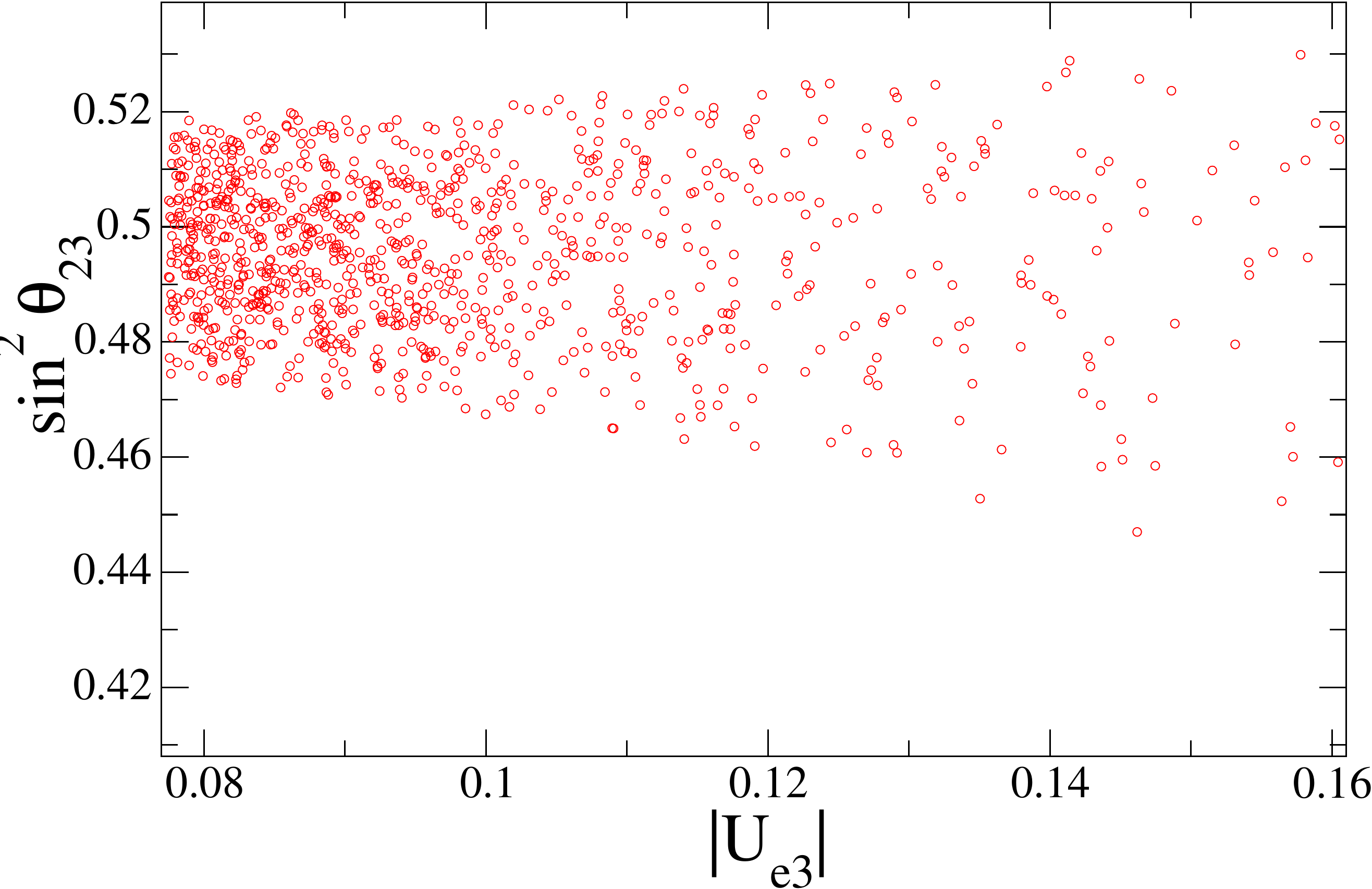}
\includegraphics[width=7cm,height=4cm]{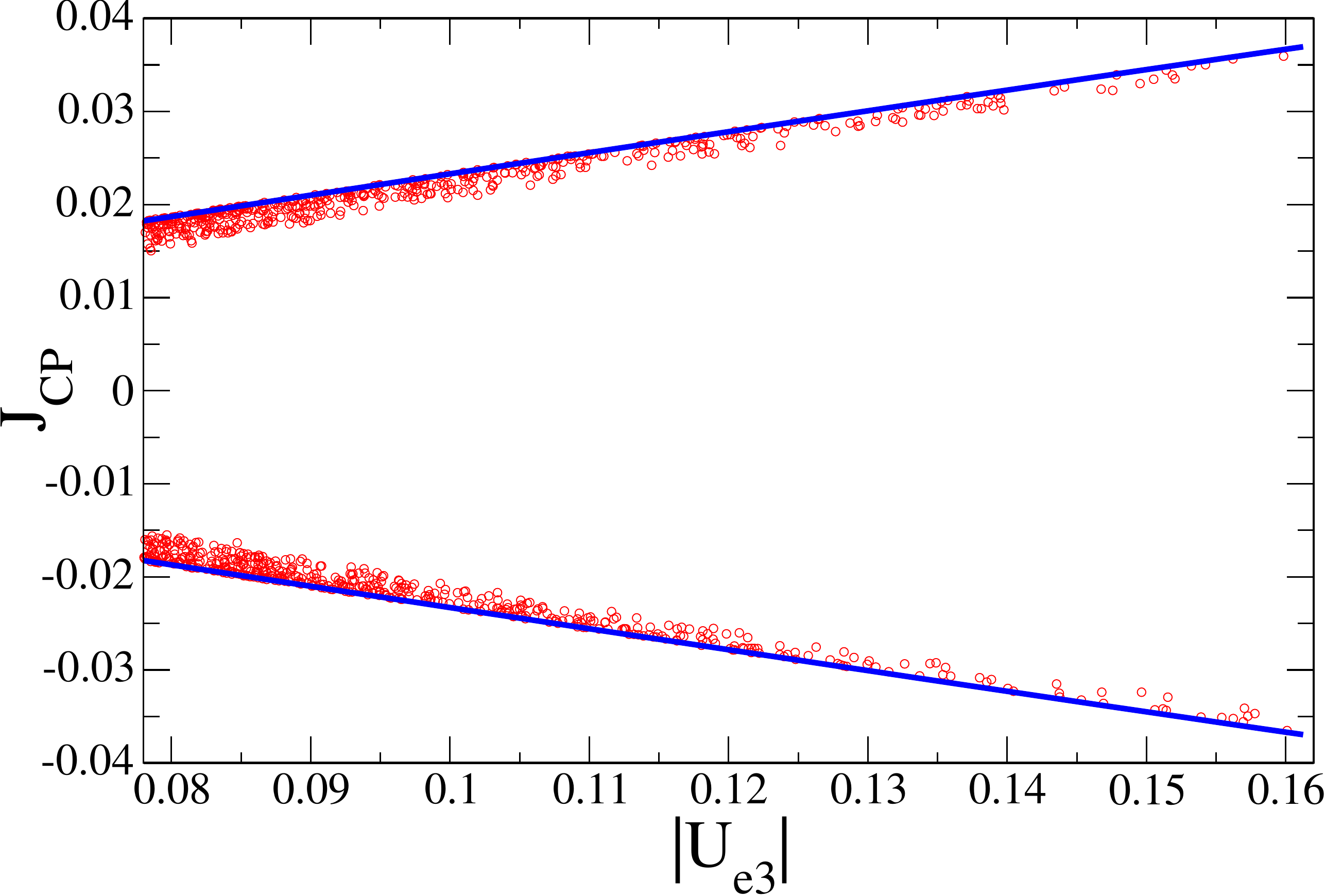}
\caption{\label{fig:Rodejohann3}Charged lepton corrections to tri-bimaximal
mixing. The left plot shows $|U_{e3}|$ against $\sin^2 \theta_{23}$ 
while the right plot gives $|U_{e3}|$ against the Jarlskog invariant $J_{\rm CP} = \frac 18 
\sin 2 \theta_{12} \sin 2 \theta_{23} \sin 2 \theta_{13} \cos
\theta_{13} \sin \delta$. The blue
solid lines display the maximal 
value that $|J_{\rm CP}|$ can take. }
\end{center}
\end{figure}

\subsection{\label{sec:RG}Radiative Corrections}
Renormalization group (RG) effects should be there, and will lead to
deviations from TBM. One can show that \cite{hep-ph/0612328}
\be \label{eq:obs}
\left|\sin \theta_{13}\right| \simeq 
\left| C \, k_{1 3} \, \Delta_\tau \right|~ , ~
\sin^2 \theta_{23} \simeq \frac 12 - C \, k_{23} \, \Delta_\tau~,~
\sin^2 \theta_{12} -\frac{1}{3} \simeq 
\frac{2 \sqrt{2}}{3} \, C \, k_{12} \, \Delta_\tau ~,
\ee
\bigskip
\bigskip

with 
\beqa 
k_{12} \nonumber 
&=& \frac{\sqrt{2}}{6}
\, \frac{\left| m_1 + m_2 \, 
e^{i \alpha_2}\right|^2}{\dmsq_{21}} ,\label{eq:k12} \\
k_{23} 
&=&-\left( \frac{1}{3} \, \frac{\left| m_2 + m_3 \, e^{i(\alpha_3 - \alpha_2)}
\right|^2}{\dmsq_{32}}  
+ \frac{1}{6} \, \frac{\left| m_1 + m_3 \, e^{i\alpha_3}\right|^2}
{\dmsq_{31}} \right)  , \label{eq:kij} \\ 
k_{13} &=& -\frac{\sqrt{2}}{6}\left( 
\frac{\left| m_2 + m_3 \, e^{i(\delta + \alpha_3 - \alpha_2)}\right|^2}{\dmsq_{32}} -  
\frac{\left| m_1 + m_3 \, e^{i(\delta + \alpha_3)}\right|^2}{\dmsq_{31}} 
- \frac{4 \, m_3^2 \, \dmsq_{21}}{\dmsq_{31} \, \dmsq_{32}} \, 
\sin^2{\frac{\delta}{2}} \nonumber 
\right). 
\label{eq:k13}
\eeqa
Here $C = -3/2$ for the SM and $C = +1$ for the MSSM, while 
\beqa
\Delta_\tau &\equiv& \left\{ 
\begin{array}{l} 
\frac{m_\tau^2}{8 \pi^2 \, v^2}\, (1 + \tan^2 \beta) 
\, \ln \frac \Lambda\lambda \simeq 1.4 \cdot 10^{-5}~(1 + \tan^2 \beta)    
\quad \quad {\rm (MSSM)}\, , \\
 \frac{m_\tau^2}{8 \pi^2 \, v^2}\, 
\, \ln \frac \Lambda\lambda \phantom{(1 + \tan^2 \beta)} \simeq  1.5 \cdot 10^{-5} 
\phantom{(1 + \tan^2 \beta)}
\quad \quad {\rm (SM)}\, . \end{array} \right.
\label{delta-tau}
\eeqa
It is easy to see that in the SM it is impossible to generate
$|U_{e3}| \simeq 0.1$ via RG effects, while in the MSSM this is possible for not too
small values of neutrino masses and $\tan \beta$, typically $m_\nu
\tan \beta \simeq 4 - 7$ eV is required for $|U_{e3}| \simeq 0.08 -
0.16$. If such large values of $|U_{e3}|$ are generated, then deviations from maximal 
$\theta_{23}$ are of the same order. Solar neutrino mixing, however,
receives much larger corrections (roughly of order $\dma/\dms \sim
30$), unless the Majorana phase $\alpha_2$ is around $\pi$. This in
turn leads to cancellations in the effective mass for \onbb. 
Fig.~\ref{fig:Rodejohann4} illustrates these correlations.

\begin{figure}[h!]
\begin{center}
\includegraphics[width=7cm,height=4cm]{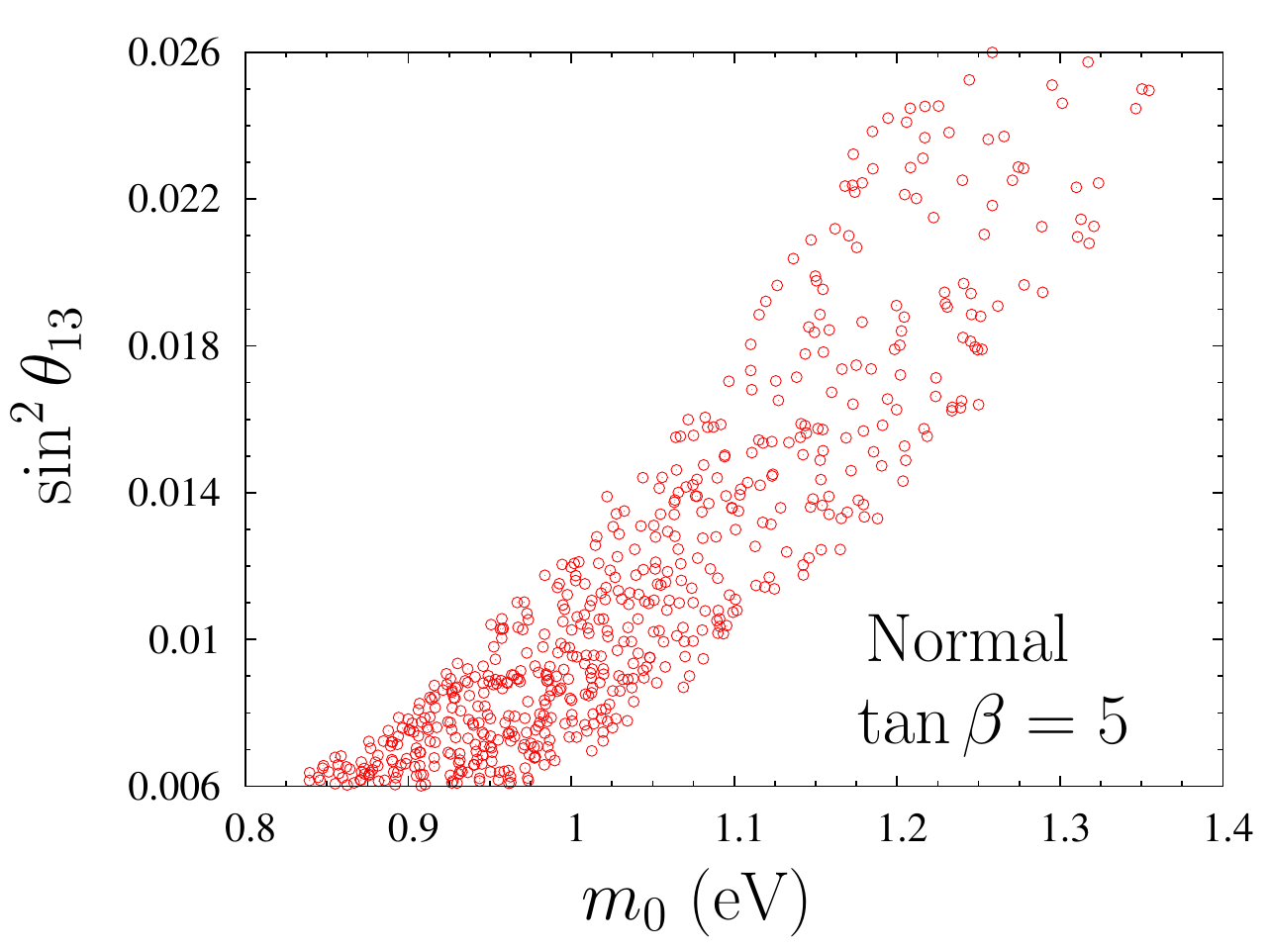}
\includegraphics[width=7cm,height=4cm]{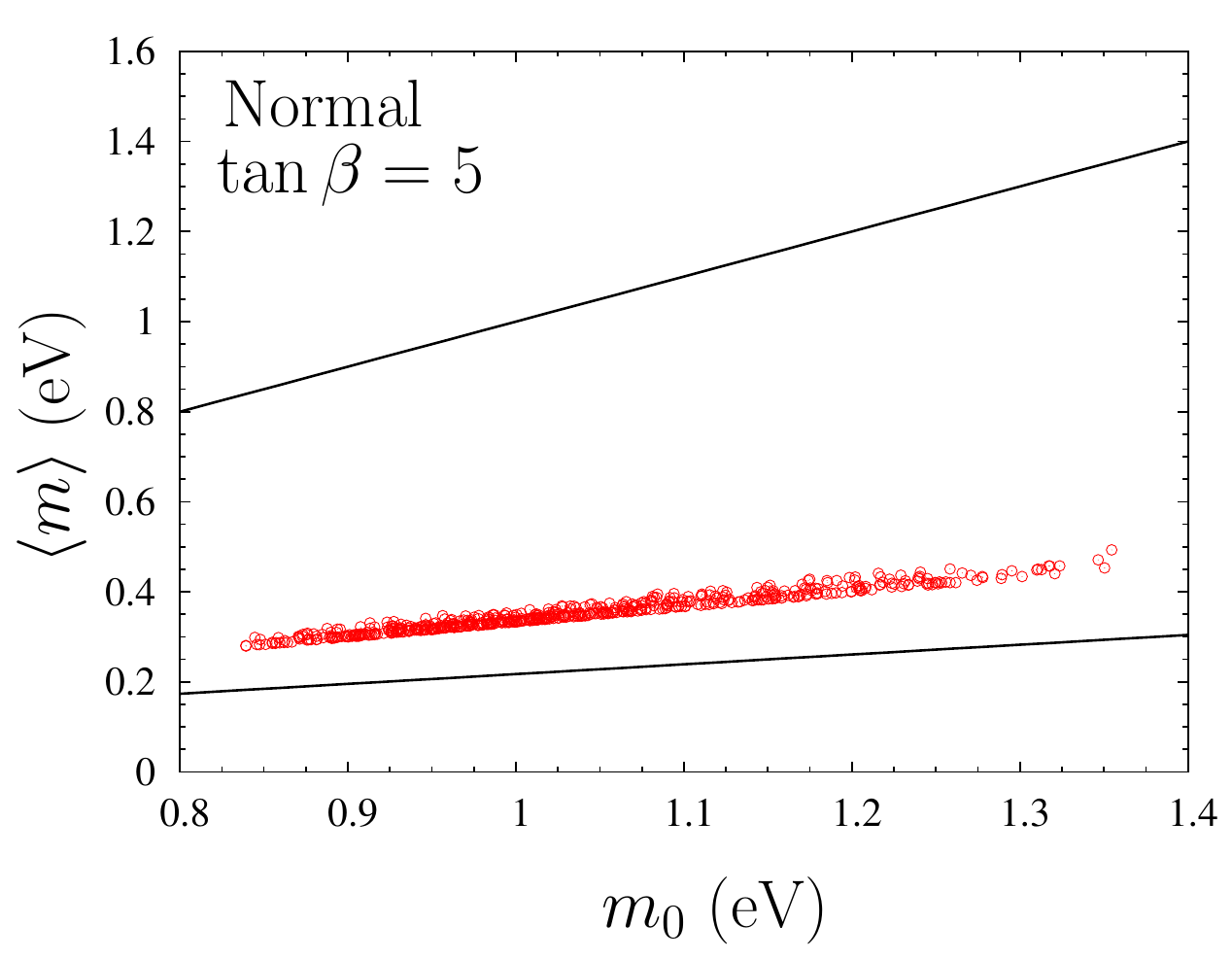}
\includegraphics[width=7cm,height=4cm]{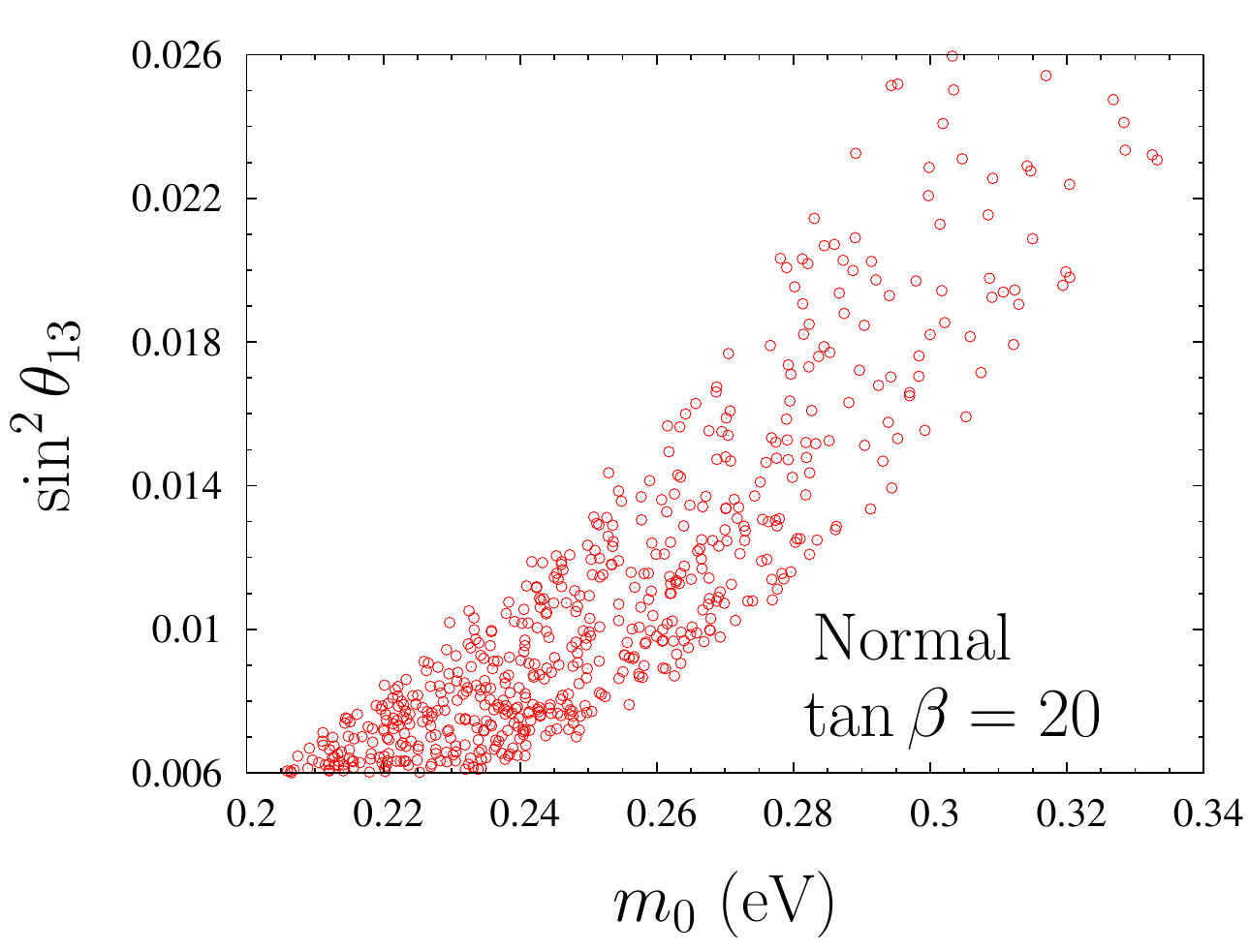}
\includegraphics[width=7cm,height=4cm]{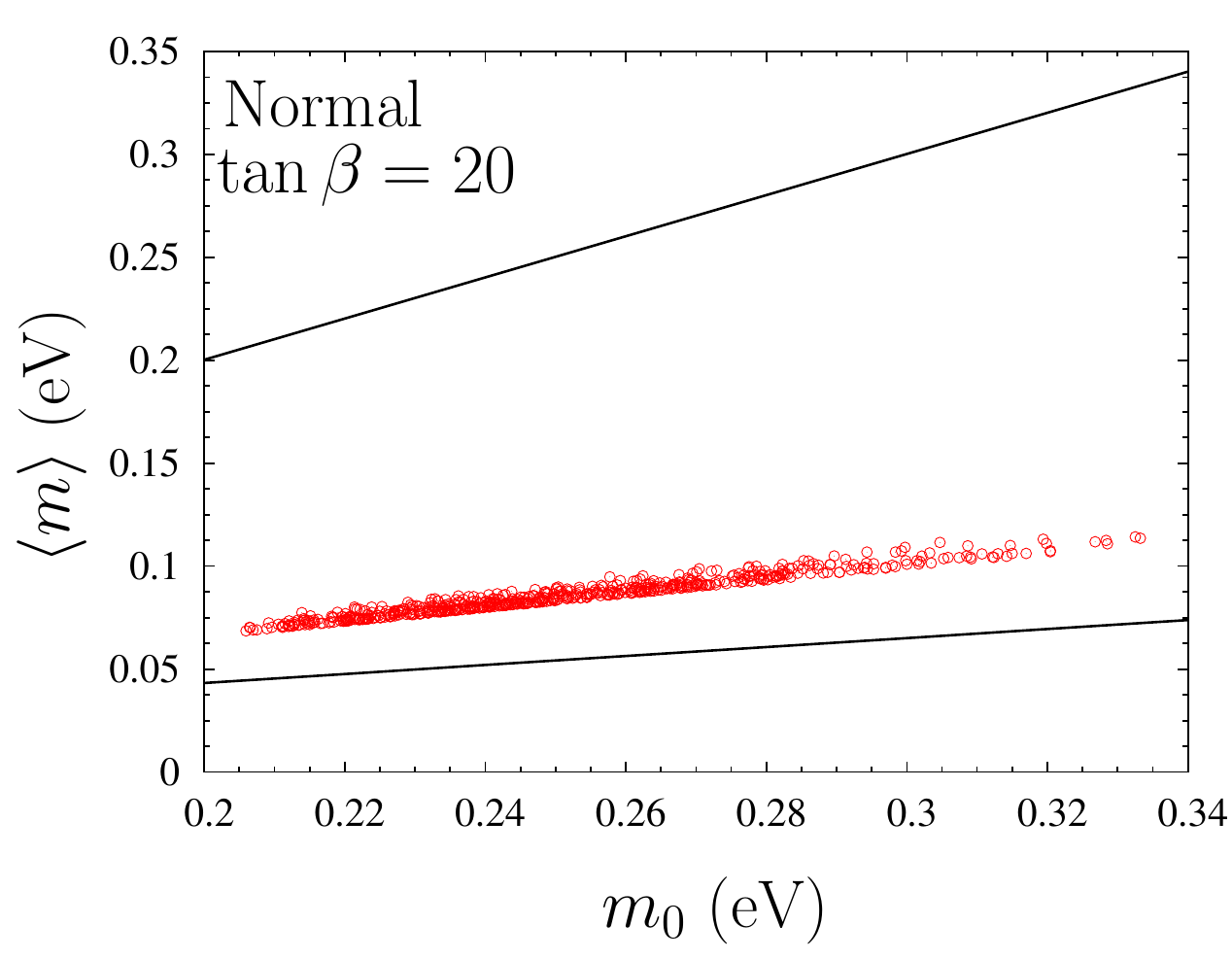}
\caption{\label{fig:Rodejohann4}RG corrections to tri-bimaximal
mixing: the upper plot show the smallest neutrino mass against the
generated $\sin^2 \theta_{13}$ and against the effective mass in the
MSSM for $\tan \beta = 5$. The solid lines show the generally allowed
range of the effective mass. The lower plots show the same for $\tan
\beta = 20$. Taken from \protect\cite{arXiv:0907.2869}.}
\end{center}
\end{figure}

\subsection{\label{sec:expl}Explicit Breaking}
Explicit breaking arises when the mass matrix from
Eq.~(\ref{eq:WRmnu}) is modified to 
\be \label{eq:mnu_dev}
m_\nu =   
\left(
\bad 
A \, (1 + \epsilon_1) & B \, (1 + \epsilon_2) 
& B \, (1 + \epsilon_3)\\[0.2cm]
\cdot & \frac{1}{2} (A + B + D) \, (1 + \epsilon_4) 
& \frac{1}{2} (A + B - D) \, (1 + \epsilon_5)\\[0.2cm]
\cdot & \cdot & \frac{1}{2} (A + B + D) 
\, (1 + \epsilon_6)
\ea 
\right) .
\ee
The complex perturbation parameters $\epsilon_i$ are taken to 
be $|\epsilon_i| \le 0.2$ for $i = 1 - 6$ with their 
phases $\phi_i$ allowed to lie between zero and $2\pi$. 
In case of a normal hierarchy, one finds \cite{arXiv:0804.4581} that 
$|U_{e3}|^2$ is of order $\epsilon^2 \, R$, where 
$\epsilon$ is the magnitude of one of the $\epsilon_i$, and 
$R = \dms/\dma$. 
Hence, a too small value of $|U_{e3}|^2$ is generated in this case. 
It turns out that at least $m_1 \simeq 0.015$ eV is required in 
order to generate $|U_{e3}|^2$ above 0.005.
With the increase of $m_1$ 
starting from 0.015 eV, 
the maximal value of $|U_{e3}|$ grows almost linearly with $m_1$. 
In contrast, in the case of inverted hierarchy (ordering),  
one can generate large values of $|U_{e3}|$ even for 
a vanishing value of the smallest neutrinos mass $m_3$. 
For quasi-degenerate neutrinos,
obviously, 
sizeable values of $|U_{e3}| \simeq 0.1$ can be generated. In
some cases a moderate correlation with Majorana phases and therefore
the effective mass exists, but not as strong as in the RG case
discussed in Section \ref{sec:RG}. 

Fig.~\ref{fig:Rodejohann5} illustrates the results for the normal and
inverted hierarchy, for the sake of comparison also with predictions
of 12 typical $SO(10)$ models, all of which have $|U_{e3}|^2 \gs
10^{-3}$, in contrast to the predictions of the explicitly broken TBM
scenario in case of a normal hierarchy.

\begin{center}
\begin{figure}[ht]
\includegraphics[width=7cm,height=5cm]{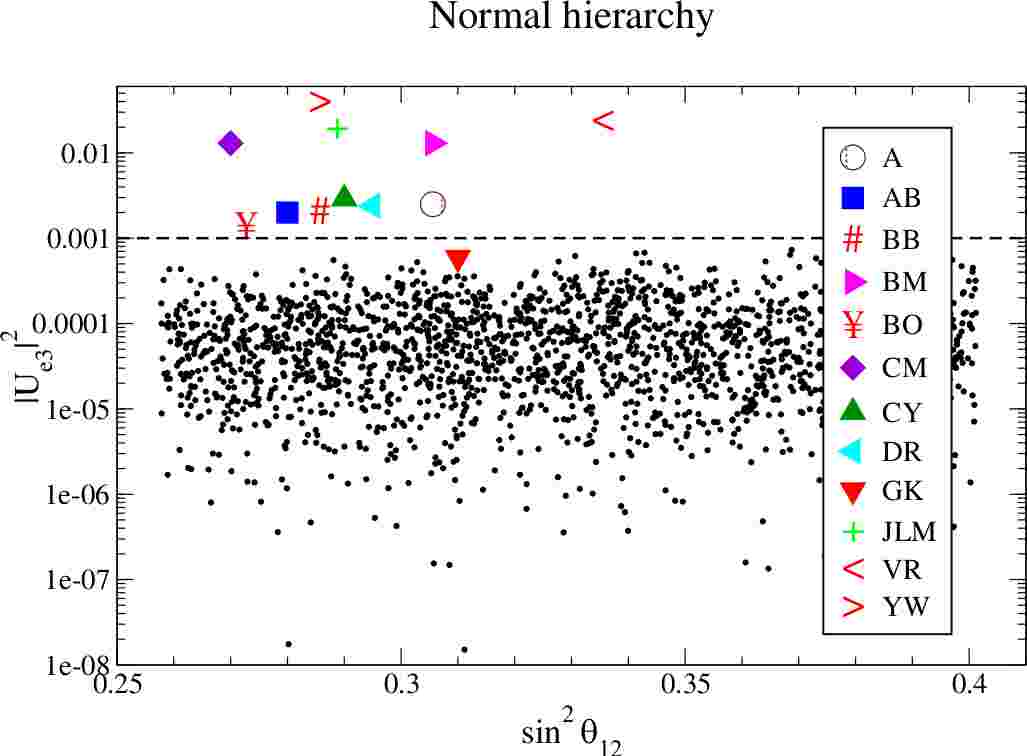}
\includegraphics[width=7cm,height=5cm]{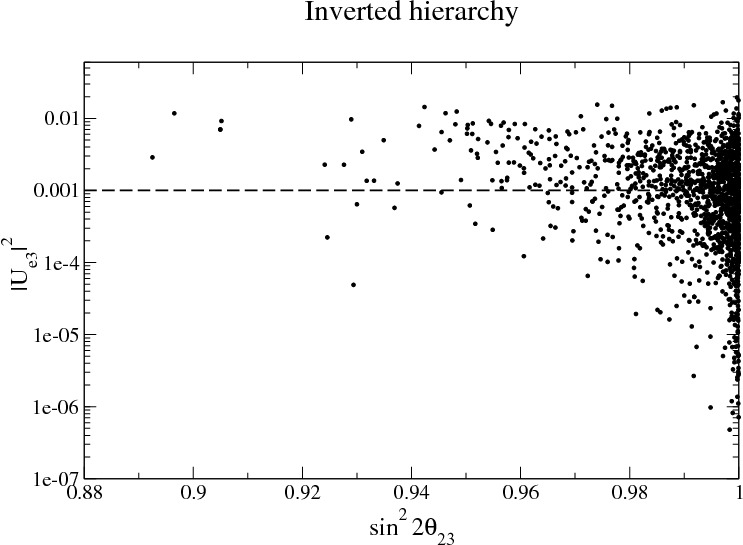}
\caption{\label{fig:Rodejohann5}Explicit corrections to tri-bimaximal
mixing: the left plot shows the result for $\sin^2 \theta_{12}$ and
$|U_{e3}|$ from diagonalizing Eq.~(\ref{eq:mnu_dev}) for the normal
mass hierarchy, together with the predictions of 12 $SO(10)$
models. The right plot is $|U_{e3}|$ vs.~$\sin^2 2 \theta_{23}$ for
the inverted hierarchy. See Ref.~\protect\cite{arXiv:0804.4581} for
references and details.}
\end{figure}
\end{center}

\subsection{\label{sec:vev}Misalignment of VEVs, NLO terms}
Various aspects, such as RG effects or higher dimensional operators,
can be summarized as VEV misalignment, e.g.~if a flavon triplet goes
as $\langle \Phi \rangle = v_\Phi (1,1,1)$ in order to generate TBM,
one studies the effects of $\langle \Phi \rangle = v_\Phi (1 +
\epsilon_1,1 + \epsilon_2,1)$ \cite{arXiv:0801.0181,arXiv:1003.2385}. 
Here $\epsilon_{1,2}$ is typically of order $\langle \Phi
\rangle/\Lambda$, where $\Lambda$ is the cutoff scale of the theory. 
The ratio $\langle \Phi
\rangle/\Lambda$ can be ${\cal O}(0.1)$, ${\cal O}(\lambda)$,
${\cal O}(0.01)$, etc. The outcome is that all mixing angles receive
corrections of the same order, 
$
\theta_{ij} = \theta_{ij}^0 + {\cal O}(\langle \Phi
\rangle/\Lambda) \, , $
with $\theta_{12}$ somewhat more 
sensitive to corrections, because it is connected to the two mass 
eigenstates close in mass. However, large $|U_{e3}|$ in agreement with
recent data can be obtained. 

Somewhat more model-dependent is the careful study of higher
dimensional operators in a specific model, which is nowadays a
necessary part of a paper dealing with flavor symmetry models. A
large variety of corrections is possible here. We illustrate this by a
few examples: in 
the $S_4$ model leading to bimaximal mixing from 
Ref.~\cite{arXiv:0903.1940}, $\sin^2 \theta_{12}$ and $|U_{e3}|$
receive corrections of order $\lambda$ each, whereas $\sin^2
\theta_{23}$ is corrected only quadratically (it is in fact a
realization of the charged lepton correction scenario from Section
\ref{sec:uell}). The $A_4$ model from Ref.~\cite{arXiv:0905.3534} has
large corrections only to $|U_{e3}|$, the $D_4$ model from
Ref.~\cite{arXiv:1007.1888} has $|U_{e3}| \simeq \lambda$, and $\sin^2
\theta_{23} \simeq \frac 12 + {\cal O}(\lambda)$, the $S_4$ model from
Ref.~\cite{arXiv:1004.5004} has $|U_{e3}| \simeq |\frac 13 - \sin^2
\theta_{12}| \simeq \lambda$ and $|\frac 12 - \sin^2 \theta_{23}|
\simeq \lambda^2$, etc.

\section{\label{sec:WRalt}Alternatives to Tri-bimaximal Mixing}
While deviations from TBM have been discussed so far, let us now
focus on alternatives. The following discussion is mostly taken 
from Ref.~\cite{arXiv:1004.2798}. 

Tri-bimaximal mixing is a variant of the more general $\mu$--$\tau$ symmetry, 
which predicts $\sin^2 \theta_{23} = \frac 12$, $|U_{e3}| = 0$, but 
leaves solar neutrino mixing unconstrained. 
From a theoretical point of view, $\theta_{12}$ is unconstrained by 
$\mu$--$\tau$ symmetry and hence can be expected to be a number of
order one. This is indeed in good agreement with data. A simple
$Z_2$ (or $S_2$) 
exchange symmetry acting on the neutrino mass matrix
suffices to generate $\mu$--$\tau$ symmetry. In fact, any symmetry
having $Z_2$ or $S_2$ as a subgroup can be used, for instance, 
$D_4$.  

Other alternatives focus on the value of the solar mixing angle,
particularly interesting are here the two possible golden ratio 
($\varphi = (1 + \sqrt{5})/2$) possibilities: 
\begin{eqnarray}
\varphi_1 : & \cot \theta_{12} = &  
\varphi \Rightarrow \sin^2 \theta_{12} 
= \frac{1}{1 + \varphi^2} \simeq 0.276 \, ,\\
\varphi_2 : & \cos \theta_{12} = & 
\frac{\varphi}{2} 
\Rightarrow \sin^2 \theta_{12} 
= \frac 14 \, (3 - \varphi) \simeq 0.345 \, .
\end{eqnarray}
The first one \cite{hep-ph/0306002,arXiv:0812.1057,arXiv:1101.0393} can be obtained by the flavor symmetry group
$A_5$, which is isomorphic to the symmetry group of the icosahedron,
in whose geometry the golden ratio explicitly shows up. The second
one \cite{arXiv:0810.5239} corresponds to $\theta_{12} = \pi/5$, and this
motivates due to geometrical considerations to use $D_5$ or $D_{10}$,
the symmetry groups of the pentagon or decagon. Mixing angles of $\pi/n$
can be obtained quite generically with dihedral groups $D_n$, as shown in
\cite{arXiv:0710.5061}. In this spirit, $\theta_{12} = \pi/6$ has been
proposed as an Ansatz obtainable from $D_6$ or $D_{12}$, 
and called hexagonal \cite{arXiv:1004.2798} or dodecal
\cite{arXiv:1005.4684} mixing. 

With sizable $|U_{e3}|$ and $\sin^2 \theta_{12} \ls \frac 13$ (as
indicated by global fit results \cite{arXiv:1108.1376}), there
is a simple modification to TBM that can do the job, namely one of the
``tri-maximal'' \cite{trimaximal} variants, in which one row or column of the TBM mixing
matrix is kept constant, and the other two rows or columns are free. This is
equivalent to multiply Eq.~(\ref{eq:WRU}) with a rotation matrix from
the left or right, respectively. If the first row is kept constant,
i.e.~$(|U_{e1}|^2, |U_{\mu 1}|^2, |U_{\tau 1}|^2)^T = (\frac 23 , \frac
16 , \frac 16)^T$, 
the implications of this scheme, denoted TM$_1$, are 
\bea\label{eq:TM_1cond0} \nonumber 
\sin^2 \theta_{12} = \frac 13 \, \frac{1 - 3 \, |U_{e3}|^2}
{1 - |U_{e3}|^2} \simeq 
\frac 13 \, \left( 1 - 2 \, |U_{e3}|^2 \right) \mbox{ and } 
\cos \delta \, \tan 2 \theta_{23} = 
 - \frac{1 - 5 \, |U_{e3}|^2}{2\sqrt{2} \, |U_{e3}| \, 
\sqrt{1 - 3 \, |U_{e3}|^2}} 
. 
\eea
Other proposed 
mixing schemes can be found in 
\cite{arXiv:0805.0416,arXiv:0903.3199,arXiv:1101.2073,arXiv:1107.0696,arXiv:1107.3486,arXiv:1107.3970,arXiv:1108.2497}.
Fig.~\ref{fig:Rodejohann6} illustrates several possible alternatives
to TBM and their correlations in what regards predictions for the
mixing parameters. 

Since some importance is attributed to the value of $|U_{e3}|$, we
attempt in Fig.~\ref{fig:Rodejohann7} to link its magnitude with other
small parameters in flavor physics, such as mass ratios of fermions, 
deviations from $\sin^2 \theta_{23} = \frac 12$ or $\sin^2 \theta_{12}
= \frac 13$, the Cabibbo angle, etc.  As the most simple example on 
how such things may arise, consider a vanishing $ee$ entry of the mass matrix
in case of a strong normal hierarchy. This implies $|U_{e2}|^2 \, m_2
= |U_{e3}|^2 \, m_3$, or 
\be
|U_{e3}|^2 \simeq \sqrt{\frac \dms\dma} \sin^2 \theta_{12} \simeq 0.05\, .
\ee
The deviation
from $\sin^2 \theta_{12} = \frac 13$ is related to $|U_{e3}|$ for
instance in the
scenario TM$_1$ discussed above, etc. 
Values which lie within the 95 \% C.L.~ranges of $|U_{e3}|$ obtained
in Ref.~\cite{arXiv:1111.3330} are $\sqrt{\dms/\dma}$ or
$\lambda/\sqrt{2}$, the latter being a very typical value in
Quark-Lepton Complementarity scenarios. In what regards 
$|U_{e3}| \simeq \sqrt{\dms/\dma}$, this is rather typical for
normal hierarchy scenarios, in which the mass matrix looks
(order-of-magnitude wise) as 
\be
m_\nu = \sqrt{\dma} 
\left(  
\bad 
\epsilon^2 & \epsilon & \epsilon \\
\cdot & 1 & 1 \\
\cdot & \cdot & 1
\ea
\right) , 
\quad \mbox{with}\quad \epsilon \simeq |U_{e3}| \simeq \sqrt{\dms/\dma}.  
\ee

It is rather interesting that
certain mixing schemes can be correlated with simple symmetry
groups, as pointed out in \cite{arXiv:1101.0393}. 
 To see this, one notes that very often the $\mu$--$\tau$
symmetry arises as an accidental symmetry. The underlying flavor
symmetry, say $A_4$, is broken such that in the neutrino mass matrix a
$Z_2$ subgroup is left unbroken and in the charged lepton sector a $Z_3$  subgroup is left unbroken. 
See e.g.~Ref.~\cite{hep-ph/0504165} for an explicit realization of
this scenario within $A_4$. Now one notes that two independent $Z_2$ symmetries are
enough to fully specify the mixing matrix for Majorana neutrinos \cite{Z2}. In this
spirit, consider the most general Majorana mass matrix and their form
after asking for $\mu$--$\tau$ invariance: 
\be
m_\nu = \left( \bad 
a & b & d \\
\cdot & e & f \\
\cdot & \cdot & g 
\ea \right) \stackrel{\mbox{$\mu$--$\tau$}}{\longrightarrow}\left( \bad 
a & b & b \\
\cdot & d & e \\
\cdot & \cdot & d 
\ea \right). 
\ee
The eigenvalue $d-e$ has the eigenvector $(0,-1,1)^T$. To find the second $Z_2$, we
give the most general symmetric matrix $S$ which fulfills $S^2 = \mathbbm{1}$,
det $S = -1$: 
\be
S = \left( \bad
w & x_2 & x_3 \\
x_2 & \frac{x_3^2 - x_2^2 \, w}{x_2^2 + x_3^2} & 
\frac{-x_3 \, x_2 \, (1 + w)}{x_2^2 + x_3^2} \\ 
x_3 &  \frac{-x_3 \, x_2 \, (1 + w)}{x_2^2 + x_3^2} & 
\frac{x_2^2 - x_3^2 \, w}{x_2^2 + x_3^2}
\ea \right) \mbox{ with } w = \sqrt{1 - x_2^2 - x_3^2} \, . 
\ee
Demanding $\left[S,R \right] = 0$ leads to $x_2 = x_3$, and invariance
under $S$ of the $\mu$--$\tau$ symmetric mass matrix
gives\footnote{This $Z_2$ is sometimes called ``hidden $Z_2$'' \cite{hidden}.}
\be
x_3 = \sqrt{2} \, \cos \theta_{12} \, \sin \theta_{12} \, , ~\mbox {
or } 
 S = \left( \bad 
\cos 2 \theta_{12} & \sqrt{\frac 12} \sin 2 \theta_{12} & \sqrt{\frac 12} \sin 2 \theta_{12} \\
\cdot & \sin^2 \theta_{12} & -\cos^2 \theta_{12} \\
\cdot & \cdot & \sin^2 \theta_{12}
\ea \right) .
\ee
Now one assumes that the charged leptons are invariant under a properly
chosen $T$, $T^\dagger  m_\ell^\dagger  m_\ell \, T = m_\ell^\dagger
\, m_\ell$, with $T^n = \mathbbm{1}$. As a result, the tri-bimaximal,
bimaximal and one of the golden ratio possibilities point to simple
groups, as illustrated in Table \ref{tab:rodejohann1}. \\

Let me take this opportunity to enjoy more amusing mixing schemes. For
instance, one could note that vanishing $\theta_{13}$ corresponds to
the fixed point of the sine function, $\sin \theta_{13} = \theta_{13}
\Rightarrow \theta_{13} = 0$. Interestingly, the fixed point
of the cosine function (``Dottie's number''), corresponds to
near-maximal mixing: $\cos \theta_{23} = \theta_{23} \Rightarrow
\theta_{23} = 0.739085\ldots$, or $\sin^2 \theta_{23} = 0.454$. Dottie's number is
irrational, transcendental and a non-trivial example of a fixed point
and an attractor. Other possibilities are to identify $\theta_{12}$ or
$|U_{e2}|$ with the Euler-Mascheroni constant, or identify $\tan 2
\theta_{12} = e$ with Euler's number, etc$\ldots$

\begin{table}[t]
\centering
\begin{tabular}{|c|c|c|c|c|}
\hline
Scenario & $S$ & $T$ & relations &  group      \\ \hline \hline 
bimaximal 
&  
$\sqrt{ \frac 12 } \left(\bad
0  & 1  &  1 \\
\cdot & \sqrt{ \frac 12 }& -\sqrt{ \frac 12 }  \\
\cdot & \cdot & \sqrt{ \frac 12 }  
\ea\right)$ 
& 
${\rm diag}(-1,i,-i) $ 
& 
$T^4 = (S \, T)^3 =  \mathbbm 1$ 
& 
$S_4$
\\ \hline
tri-bimaximal 
&  
$-\frac 13 \left(\bad
1  & 2  &  2 \\
\cdot & 1 & -2  \\
\cdot & \cdot & 1  
\ea\right)$ 
& 
${\rm diag}(e^{-2 i \pi/3},e^{2 i \pi/3},1) $ 
& 
$T^3 = (S \, T)^3 =  \mathbbm 1$ 
& 
$A_4$
\\ \hline
golden ratio $\varphi_1$ 
&  
$\frac{-1}{\sqrt{5}}\left(\bad
1   & \sqrt{2}  & \sqrt{2}  \\
\cdot & 1/\varphi &  -\varphi  \\
\cdot & \cdot &  1/\varphi 
\ea\right)$ 
& 
${\rm diag}(1,e^{-4 i \pi/5},e^{4 i \pi/5}) $ 
& 
$T^5 = (S \, T)^3 =  \mathbbm 1$ 
& 
$A_5$
\\ \hline
\end{tabular}
\caption{\label{tab:rodejohann1}Neutrino mixing scheme and the implied
symmetry group.}
\end{table}%

 \begin{center}
\begin{figure}[ht]
\includegraphics[width=7cm,height=5cm]{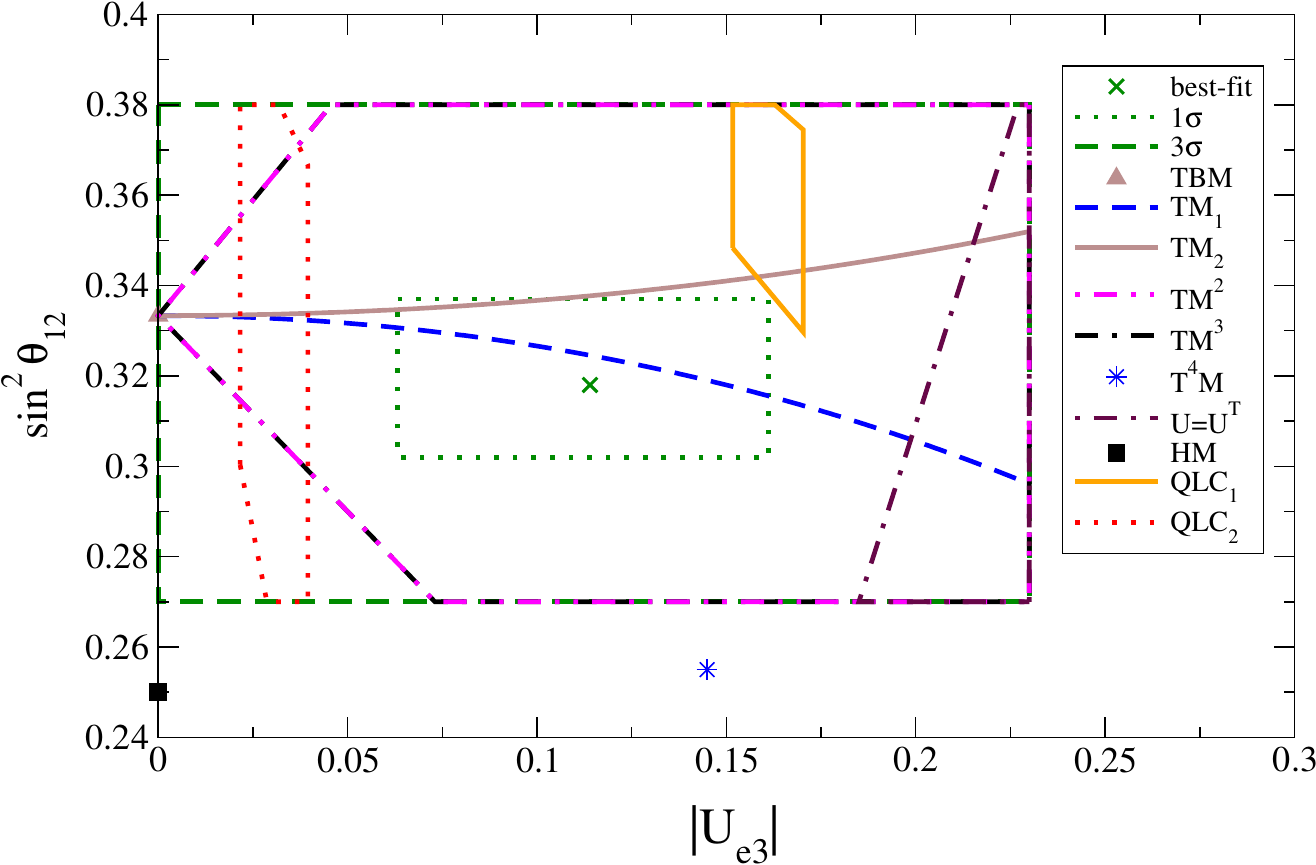}
\includegraphics[width=7cm,height=5cm]{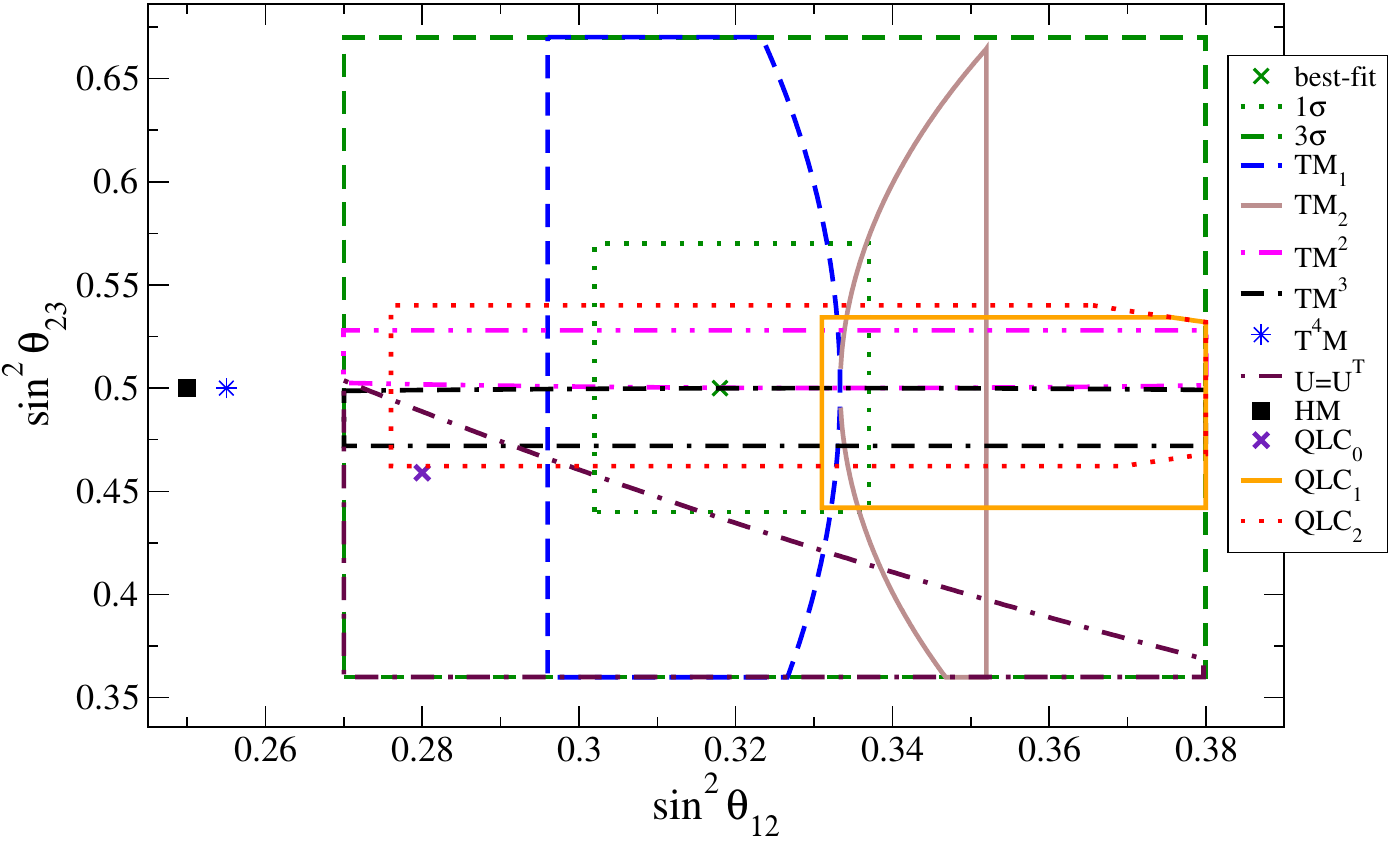}
\caption{\label{fig:Rodejohann6}Alternatives to tri-bimaximal mixing:
Correlations between neutrino mixing parameters for various proposed
mixing schemes. See Ref.~\protect\cite{arXiv:1004.2798} for references and details.}
\end{figure}
\end{center}

 \begin{center}
\begin{figure}[ht]
\includegraphics[width=5cm,height=5.9cm]{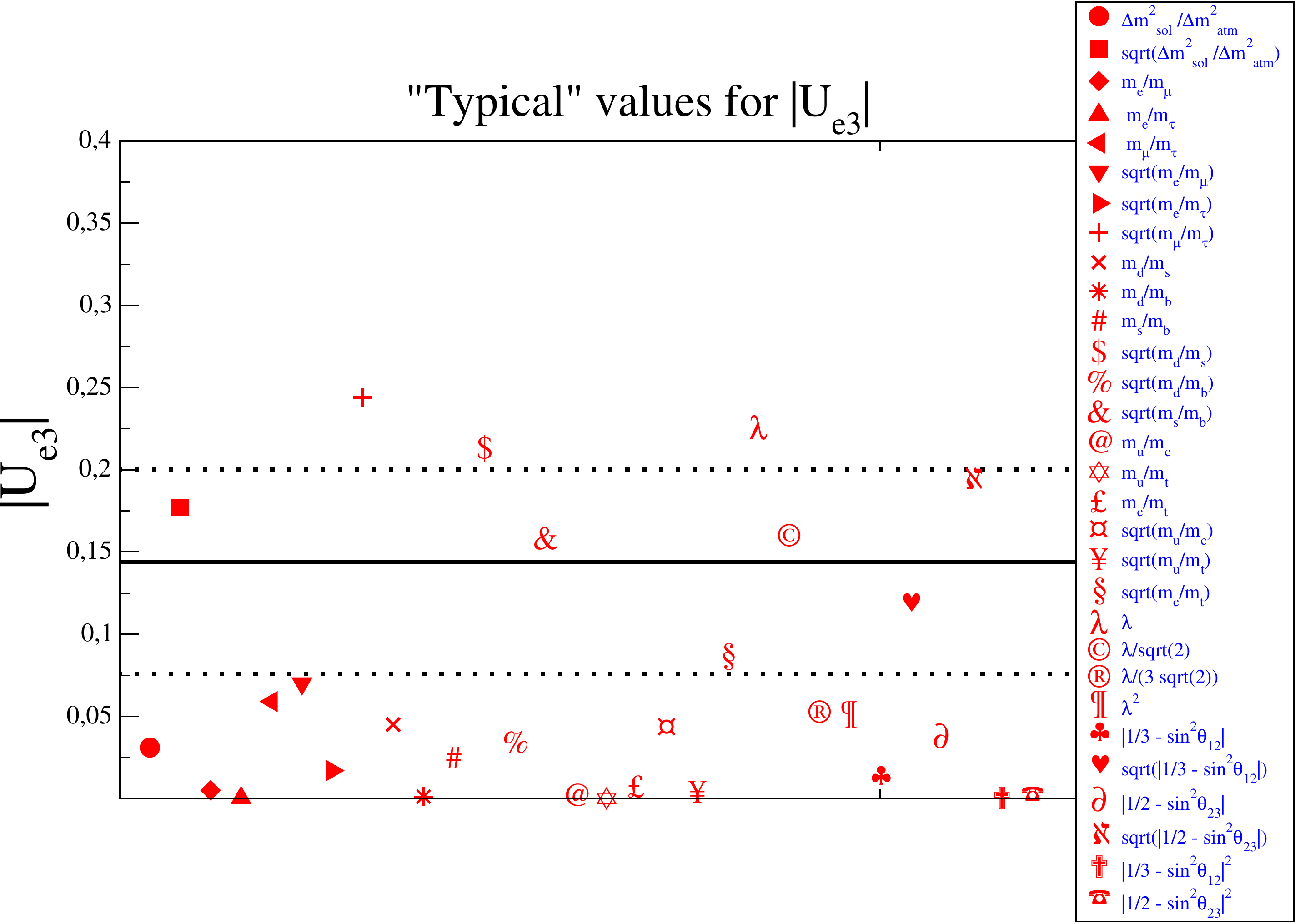}
\includegraphics[width=5cm,height=5.95cm]{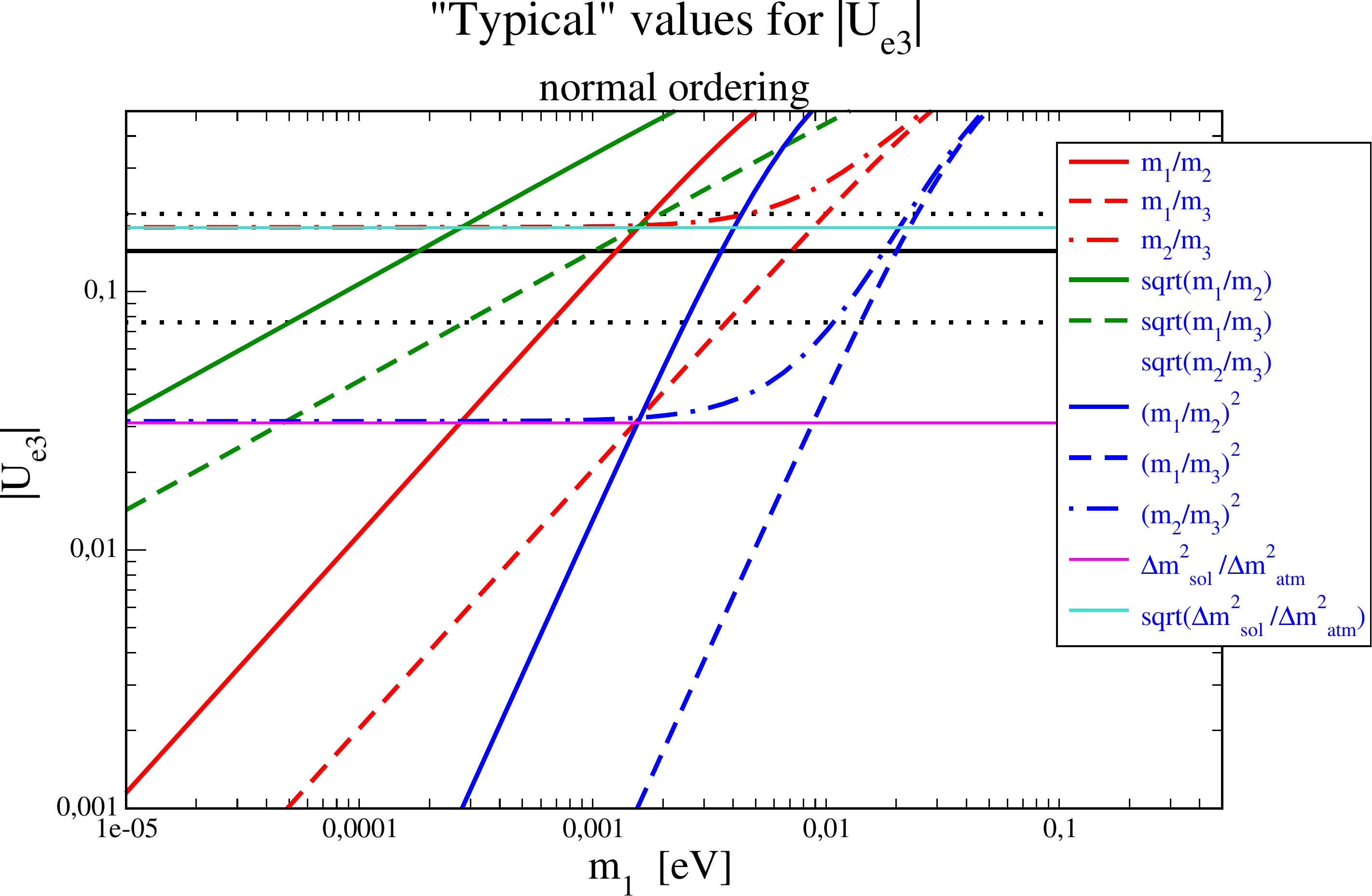}
\includegraphics[width=5cm,height=5.95cm]{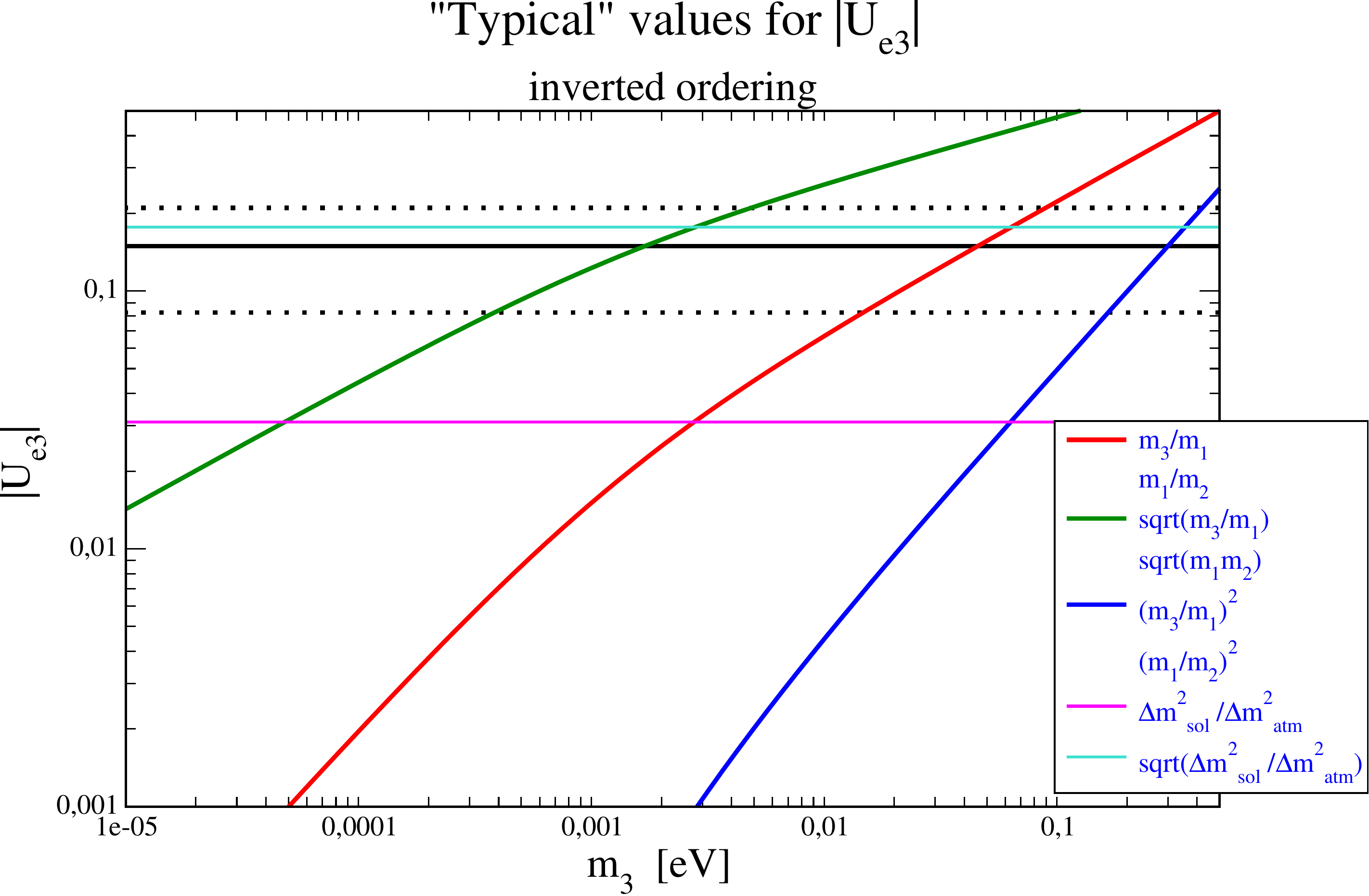}
\caption{\label{fig:Rodejohann7}Left plot: Comparison of $|U_{e3}|$
with typical numbers expected in flavor theories. Central and right plots:
Comparison of $|U_{e3}|$ with mass ratios of neutrinos for the normal
and inverted ordering. Also given are the central and 95\% C.L.~values
of $|U_{e3}|$ from Ref.~\protect\cite{arXiv:1111.3330}.}
\end{figure}
\end{center}

\section{\label{sec:WRster}Sterile Neutrinos and Flavor Symmetry Models}
We conclude this contribution with the increasingly popular sterile
neutrino hypothesis in the context of flavor symmetries. Some 
long-standing issues in particle physics, astrophysics and cosmology
can be solved by the same entity: a light eV-scale sterile neutrinos
with non-negligible mixing with the SM leptons. 
Those issues are the apparent neutrino flavor transitions at LSND and
MiniBooNE, which together with the ``reactor anomaly''
\cite{Mention:2011rkHuber:2011wv} point towards oscillations of eV-scale sterile
neutrinos mixing with strength of order 0.1 with the active ones
(see Refs.~\cite{Kopp:2011qdGiunti:2011gz} for recent global
fits). 
In addition, several hints mildly favoring extra radiation in the Universe have recently emerged from
precision cosmology and Big Bang Nucleosynthesis
\cite{Cyburt:2004ycIzotov:2010caHamann:2010bkGiusarma:2011ex}.
This could be any relativistic degree of freedom or some other New
Physics effect, but has a straightforward interpretation in terms of additional sterile neutrino species.
Although some tension between the neutrino mass scales required by
laboratory experiments and cosmological neutrino mass limits exists 
within the standard $\Lambda$CDM framework,
moderate modifications could arrange for compatibility \cite{Hamann:2011ge}.
Finally, active to sterile oscillations have been proposed to increase
the element yield in $r$-process nucleosynthesis in core collapse
supernovae (which seems to be too low in standard calculations, see 
e.g.~\cite{Beun:2006kaTamborra:2011is}).
It is rather intriguing that indications for the presence of eV sterile
neutrinos come from such fundamentally different probes.

An almost trivial way to incorporate sterile neutrinos to flavor
symmetry models is to simply add them to the particle
spectrum. Consider Table \ref{table:afmodel}, where to the popular $A_4$
model from Ref.~\cite{hep-ph/0504165} a weak, $A_4$ and $Z_3$ singlet
$\nu_s$ is added. Its only non-trivial charge is under a 
Froggatt-Nielsen $U(1)_{\rm FN}$. This additional symmetry is an automatic
ingredient of realistic models in order to generate a hierarchy of
charged lepton masses. We utilize this very $U(1)_{\rm FN}$ in order to control
the magnitude of the sterile neutrino mass. 
\begin{table}[t]
\centering \caption{Particle assignments of the $A_4$ model, modified 
from Ref.~\cite{hep-ph/0504165} to include a sterile
neutrino $\nu_s$. The additional $Z_3$ symmetry decouples the charged
lepton and neutrino sectors; the $U(1)_{\rm FN}$ charge generates the
hierarchy of charged lepton masses and regulates the scale of the
sterile state. Taken from \protect\cite{Barry:2011wb}.} 
\label{table:afmodel} \vspace{8pt}
\begin{tabular}{c|cccccccc|c}
  \hline \hline \T \B Field & $L$ & $e^c$ & $\mu^c$ & $\tau^c$ & $h_{u,d}$ & $\varphi$ & $\varphi'$ & $\xi$ & $\nu_s$ \\
\hline \T $SU(2)_L$ & $2$ & $1$ & $1$ & $1$ & $2$ & $1$ & $1$ & $1$ & $1$ \\
 $A_4$ & $\ul{3}$ & $\ul{1}$ & $\ul{1}''$ & $\ul{1}'$ & $\ul{1}$ & $\ul{3}$ & $\ul{3}$ & $\ul{1}$ & $\ul{1}$ \\
 $Z_3$ & $\omega$ & $\omega^2$ & $\omega^2$ & $\omega^2$ & 1 & 1 & $\omega$ & $\omega$ & 1 \\
 $U(1)_{FN}$ & - & 3 & 1 & 0 & - & - & - & - & 6 \\[1mm] \hline \hline
\end{tabular}
\end{table}
The $A_4$ invariant dimension-5 operator $\frac{1}{\Lambda}(\varphi' L
h_u)\nu_s$ is not allowed by the $Z_3$ symmetry, and we are left with
terms like 
\begin{equation}
 {\cal L}_{{\rm Y}_s} = \frac{x_e}{\Lambda^2}\xi(\varphi' Lh_u)\nu_s 
+ m_s\nu^c_s\nu^c_s + {\rm h.c.},
\label{eq:lagsterile}
\end{equation}
where $m_s$ is a bare Majorana mass.
The model, described in detail in Ref.~\cite{Barry:2011wb}, leads to
diagonal charged leptons and the $4\times4$ neutrino mass matrix 
\begin{equation}
 M^{4\times4}_\nu = \begin{pmatrix} a+\frac{2d}{3} & -\frac{d}{3} & -\frac{d}{3} & e \\ \cdot & \frac{2d}{3} & a-\frac{d}{3} & e \\ \cdot & \cdot & \frac{2d}{3} & e \\ \cdot & \cdot & \cdot & m_s \end{pmatrix},
\label{eq:m4by4}
\end{equation}
which is diagonalized by 
\begin{equation}
   U \simeq \begin{pmatrix} \frac{2}{\sqrt{6}} & \frac{1}{\sqrt{3}} & 0 & 0 \\ -\frac{1}{\sqrt{6}} & \frac{1}{\sqrt{3}} & -\frac{1}{\sqrt{2}} & 0 \\ -\frac{1}{\sqrt{6}} & \frac{1}{\sqrt{3}} & \frac{1}{\sqrt{2}} & 0 \\ 0 & 0 & 0 & 1 \end{pmatrix} + \begin{pmatrix} 0 & 0 & 0 & \frac{e}{m_s} \\ 0 & 0 & 0 & \frac{e}{m_s} \\ 0 & 0 & 0 & \frac{e}{m_s} \\ 0 & -\frac{\sqrt{3}e}{m_s} & 0 & 0 \end{pmatrix} + \begin{pmatrix} 0 & -\frac{\sqrt{3} e^2}{2m_s^2} & 0 & 0 \\ 0 & -\frac{\sqrt{3} e^2}{2m_s^2} & 0 & 0 \\ 0 & -\frac{\sqrt{3} e^2}{2 m_s^2} & 0 & 0 \\ 0 & 0 & 0 & -\frac{3e^2}{2 m_s^2}\end{pmatrix}  .
\label{eq:v4}
\end{equation}
Note that the form-invariance of the mass matrix is lost. Adding a
second singlet is also possible. By suitably 
choosing the parameters, one can also arrange that the sterile state
becomes keV-scale, and acts as a warm dark matter candidate
\cite{arXiv:1110.6382}. See \cite{arXiv:1109.3187} for the current
situation of warm dark matter observations\footnote{Other approaches
for sterile neutrinos in flavor symmetries can be found in
Refs.~\cite{DM_others}, another use of flavor symmetries in 
connection with dark matter in \cite{DM}.}.

More complicated, but also possible, is the task to build a seesaw
model, where we give the particle content and representations in Table
\ref{table:afssmodel_a}, taken from \cite{arXiv:1110.6382}. The right-handed singlet neutrinos of the
type I seesaw are responsible for the light active masses and can be
arranged to be eV-scale, keV-scale, super heavy, etc., depending on
their charge under the very same Froggatt-Nielsen $U(1)_{\rm FN}$
necessary to generate a charged lepton mass hierarchy. A careful study
of several corrections demonstrates that sizable $U_{e3}$ can be
achieved in the model, see Ref.~\cite{arXiv:1110.6382} for details.

\begin{table}[tp]
\centering \caption{Particle assignments of the $A_4$ type I seesaw
model, with three right-handed sterile neutrinos.
Taken from \protect\cite{arXiv:1110.6382}.} \label{table:afssmodel_a} \vspace{8pt}
\begin{tabular}{c|ccccc|ccccccc|ccc}
\hline \hline \T \B Field & $L$ & $e^c$ & $\mu^c$ & $\tau^c$ & $h_{u,d}$ & $\varphi$ & $\varphi'$ & $\varphi''$ & $\xi$ & $\xi'$ & $\xi''$ & $\Theta$ & $\nu^c_{1}$ & $\nu^c_{2}$ & $\nu^c_{3}$ \\
\hline \T $SU(2)_L$ & $2$ & $1$ & $1$ & $1$ & $2$ & $1$ & $1$ & $1$ & $1$ & $1$ & $1$ & $1$ & $1$ & $1$ & $1$ \\
$A_4$ & $\ul{3}$ & $\ul{1}$ & $\ul{1}''$ & $\ul{1}'$ & $\ul{1}$ & $\ul{3}$ & $\ul{3}$ & $\ul{3}$ & $\ul{1}$ & $\ul{1}'$ & $\ul{1}$ & $\ul{1}$ & $\ul{1}$ & $\ul{1}'$ & $\ul{1}$  \\
$Z_3$ & $\omega$ & $\omega^2$ & $\omega^2$ & $\omega^2$ & $1$ & $1$ & $\omega$ & $\omega^2$ & $\omega^2$ & $\omega$ & $1$ & $1$ &  $\omega^2$ & $\omega$ & $1$  \\
$U(1)_{\rm FN}$ & - & $3$ & $1$ & $0$ & - & - & - & - & - & - & - & $-1$ & $F_1$ & $F_2$ & $F_3$ \\[1mm] \hline \hline
\end{tabular}
\end{table}

%% file: Author/MarcoTaoso.tex
\begin{center}
{\bf Abstract}\\
\end{center}
\vskip5.mm
We review the current status of direct and indirect Dark Matter searches,
focusing in particular on those observations which have been interpreted as possible hints of dark matter.
We discuss about the uncertainties affecting these interpretations and highlights the
complementarity between different methods.


\vskip5.mm

\section{Introduction}
\label{Introduction}

Non-baryonic Dark Matter is a fundamental pillar of modern cosmology \cite{896695}.
Astrophysical and cosmological observations have probed its presence from
subgalactic up to cosmological scales and demonstrated that DM constitutes the most
abundant component of the matter budget of the Universe.
Furthermore, the increasing amount of data collected over the last decades have constrained its fundamental properties. 
The emergent picture is that DM is made up of cold (or warm) massive particles, stable over cosmological times and
interacting only (very) weakly with the primordial plasma of the Universe (see Ref.\cite{arXiv:0711.4996} for a review about the properties that a viable dark matter candidate should have). 
Particles fulfilling these requirements arise in many extensions of the Standard Model (SM) of the particle physics,
motivating the connection of the DM problem with searches of new physics beyond the SM.

Presently, the nature of DM remains elusive and its microscopic properties are still undetermined. 
To shed light on this mistery it is needed to go beyond a gravitational detection of DM.
High-energy colliders are ideal tools to search for new particles and interactions.
Since DM should interacts only weakly, it is expected to be invisible to the detectors
exploited at collider experiments, and therefore it should manifest itself as missing energy events.
This is an exciting time for searches of new physics. Indeed, the Large Hadron Collider (LHC) is currently taking data,
probing energies previously unexplored at colliders.
However, in case that a potential DM candidate will be discovered at LHC, additional information will be mandatory
to demonstrate that this new particle constitutes the DM component observed with cosmological and astrophysical
observations. 
Measuring the mass and the interaction cross sections of this particle might allow to reconstruct
its cosmological relic density and to compare this estimate with the DM abundance inferred from cosmlogical
observations.
This will probably be very difficult to do only with the information obtained at LHC \cite{hep-ph/0602187,arXiv:1111.2607}. 
Furthermore, this calculation would
involve cosmological assumptions on the evolution of the Universe.

Direct and Indirect DM searches provide alternative strategies to search for non-gravitational DM signals.
Direct detection experiments looks for the interaction of DM particles with low-background dectors.
These searches are mainly devoted to look for DM in the form of Weakly Interacting Massive
particles (WIMPs), i.e. DM candidates with masses in the GeV-TeV range and interaction strenghts
of the order of those mediated by weak interactions.
WIMPs are predicted in many extensions of the Standard Model, in particular those related
with the electroweak hierarchy problem, like Supersymmetry and extra-dimension models.
These particles were in thermal equilibrium with the plasma in the early Universe and decoupled non-relativistically,
inheriting the correct relic abundance for an annihilation cross-section
$(\sigma v)\sim 10^{-26}\mbox{ cm}^3\mbox{s}^{-1},$ a value typical for weak-scale interactions.
The detection of the WIMPs annihilation products is alternative indirect method to look for WIMPs,
or more in general to search for any DM candidate with sizeable annihilations or decays into SM particles.

In the following, we will focus on direct and indirect DM searches. We will review the present status and
discuss the future prospects for detection.

\section{Direct Detection}

\begin{center}
\begin{figure*}[t]
\centering
\includegraphics[width=0.41\textwidth]{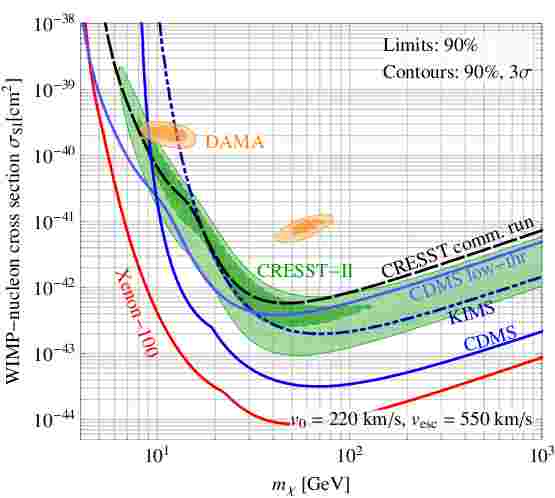}
\hspace{2mm}
\includegraphics[width=0.39\textwidth]{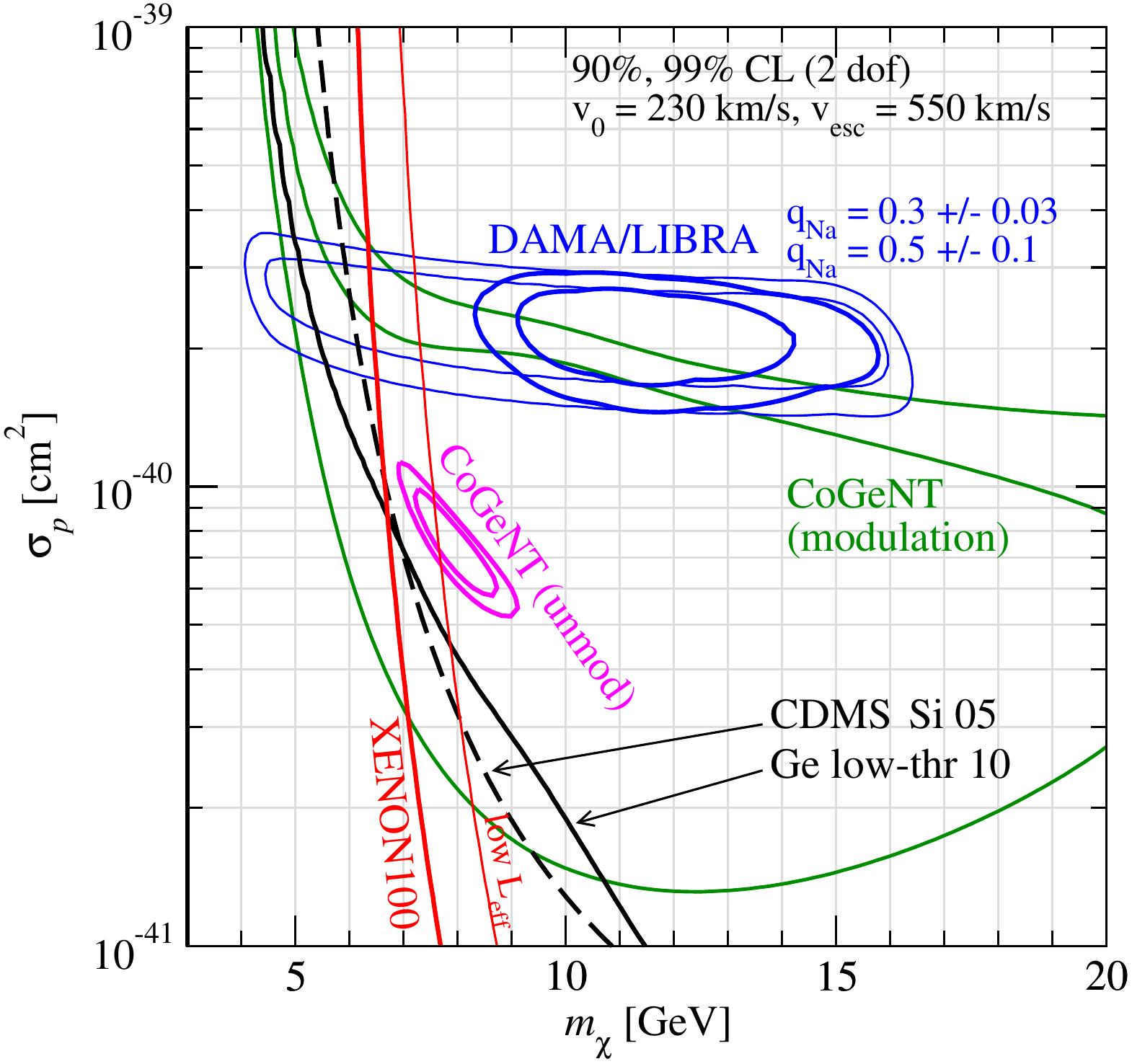}
\caption{Regions of the parameter space preferred by DAMA, CoGent and CRESST-II and constraints from
other experiments. The analysis is for elastic SI interactions. Figure taken from \cite{Schwetz:2011xm,Kopp:2011yr}.}
\label{fig:direct}
\end{figure*}
\end{center}

Direct DM searches aim at detecting DM particles through the measurement of nuclear recoils produced by
DM scattering off target nuclei (see \cite{Cerdeno:2010jj} for a recent review).
The differential rate in events/kg/day/keV is given by: 
\begin{equation}
\frac{dR}{dE_E}=\frac{\rho_{DM}}{m_{DM} m_N}\int_{|\vec{v}|>v_{min}}d^3v \frac{d\sigma}{dE_R} v f(\vec{v})
\label{eq:DD_rate}
\end{equation}
where $m_{DM}$ and $m_{N}$ are respectively the DM and target nucleus masses, $\rho_{DM}$ is the local DM density and $f(\vec{v})$ is the local DM velocity
distribution. The lower limit of the integration $v_{min}$ corresponds to the minimum DM velocity required
for the DM particle to deposit an energy $E_R$ in the detector.
The differential DM-nucleon cross section $d\sigma/dE_R$ depends on the DM particle physics
properties which defines the DM interactions with gluons and quarks, and on nuclear physics
inputs which are needed to promote the fundamental DM-quarks (gluons) cross-section to
an effective DM-nucleus cross section.
In the non-relativistic limit the DM-nucleus interaction can in general be reduced to 
two types of interactions, Spin Indipendent (SI) and Spin Dependent (SD), which respectively
describe the DM coupling with the mass and the spin of the nucleus.
These interactions are often assumed to be elastic and independent on
the momentum exchanged and the DM-nucleus relative velocity.
However, more general scenarios are possible and have extensively been studied during the last years.

\subsection{Hints of DM detection}
Nuclear recoils are measured by current experiments exploiting
scintillation signals, ionization, heat or bubble nucleation in superheated liquids.
Interestingly, several DM direct-detection experiments have reported hints of possible DM signals.
Here we briefly summarize these results and in the next section we 
discuss possible interpretations in terms of DM scatterings.

\begin{itemize}

\item The DAMA/LIBRA experiments has observed a 8.9$\sigma$ evidence
for an annual modulation of their scintillation signal produced by their
highly-purify NaI(Tl) crystals over 13 years annual cycles \cite{Bernabei:2008yi}. The modulation
has been observed in the 2-6 keV range while it is absent at higher energies
and in the 2-6 keV range for multiple-hits events. These features, as well as the phase
and the amplitude of the modulation, are consistent with the DM hypothesis.
Moreover the signal can not be accounted for by any source of background so far considered.

\item The CoGent experiment employs a germanium detector with a very low level
of electronic noise, allowing sensitivity to low nuclear recoil energies.
In their bulk-like events recoil spectrum the CoGent collaboration detected a series of cosmogenic peaks
plus an exponential-like distribution between 0.5 and 3 keV, which is not 
directly attributed to any known source of background \cite{Aalseth:2010vx}.
In addition to that, an analysis of 15-months of cumulative data supports the presence
at 2.8$\sigma$ of an annual modulation of the event rate \cite{Aalseth:2011wp}. The modulation is present
in the energy bins 0.5-3 keV range while there is not statistical evidence at higher energies
and for rejected surface-like events.

\item The CRESST-II detectors is a $\mbox{CaWO}_4$ crystal and exploits phonons and scintillation
measurements to measure the energy deposited in an interaction and to discriminate a possible
signal againts backgrounds events. Recently the CRESST collaboration has reported 67
events in their nuclear-recoil acceptance region \cite{Angloher:2011uu}. The different sources of background considered
so far can only account for a fraction of these events, and the statistical evidence of the excess
has been estimated to be larger than 4$\sigma.$

\end{itemize}

\subsection{Interpretations}
The results summarized in the previous section have triggered a large number of 
works \cite{Schwetz:2011xm,Kopp:2011yr,Farina:2011pw,Kelso:2011gd,DelNobile:2011yb,Fornengo:2011sz,Savage:2010tg,Arina:2011zh}
devoted to study a possible DM interpration of the mentioned signals, in particular
in light of the constraints imposed by the null results from the other direct detection experiments.
The comparison between the results of the various experiments is complicated by issues
of different nature. For instance the interaction of DM with the detector is unkwon and may
differ from the pure elastic SI and SD interactions discussed above. 
Moreover, the event rate induced by DM scattering off the target depends on the poorly known DM velocity distribution
and on experimental uncertanties related to the response of the detector to a scattering event. 
Various analysis in literature have tried to overcome these difficulties considering
several astyophysical and DM scenarios and
studying the impact of the experimental uncertainties on the interpretation of the data.
Here we first consider the case of an elastic momentum-indipendent SI DM-nucleus interaction, under
the hypothesis of an equal coupling of the DM to protons and neutrons. 
The DM velocity distribution is taken to be a Maxwellian.
Then we comment about possible modifications of these assumptions.

Fig.\ref{fig:direct} shows the regions of the DM parameter space preferred by the positive signals discussed before
and the upper bounds imposed by the null results of the other experiments (we refer to \cite{Schwetz:2011xm,Kopp:2011yr} for details
about the analysis).
The results of DAMA, CoGent and CRESST-II all point to a light DM interpretation $M_{DM}\sim 10$ GeV
with a SI cross-section off proton in the same ballpark.
A closer inspection reveals that there is some tension between the different data-sets.
For instance the region preferred by DAMA seems separated from the one which explains
the CoGent unmodulated data. In addition to that, all the regions are in severe tensions
with the bounds imposed by several other experiments, in particular XENON-100 \cite{Aprile:2011hi} and CDMS \cite{Ahmed:2010wy,Akerib:2005kh} experiments.
However, we remind that the upper bounds and the exact position of the regions  are affacted
by several uncertanties.

CoGent has recently estimated an higher fraction of non-rejected low energy
surface events than previously expected. This effects induces a wider CoGent-region shifted
to lower cross-sections, potentially ameliorating the agreement between CoGent and CRESST-II results (see e.g.\cite{Kopp:2011yr,Kelso:2011gd}).
New measurements of the quenching factor for sodium recoils disagree with previous results.
However, the implication of these new estimates would be to exacerbate the tension between the DAMA signal and
the constraints from XENON-100 and CDMS \cite{Kopp:2011yr}.
It is also been proposed that the presence of tidal streams in the local
DM distribution might help to reconcilate the results of DAMA with
those of CRESST-II and the CoGent unmodulated data \cite{Kelso:2011gd}.

The robustness of the constraints derived by XENON and CDMS collaboration
have been recently criticized by some authors, stimulating a long debate (see discussions in \cite{Collar:2011wq,Collar:2010ht,Collaboration:2010er,Plante:2011hw} and references therein).
For instance, some issues have been raised about the determination
of the scintillation light efficiency of the liquid xenon at low
recoil energies, a parameter which is crucial to set the constraints for low mass DM.

In literature there have been considered also different velocity distribution (e.g. \cite{Savage:2009mk}) and DM-nucleus interactions 
than those considered above.
It would be possible for example that DM interacts inelastically with the nucleus \cite{TuckerSmith:2001hy}.
This would change the kinematic of the DM-nucleus interaction and the impact on
the recoil energy spectrum would depend on the target material and the inelastic threshold.
Other possibilities include for instance isospin violating couplings, velocity dependent interactions and
momentum dependent interactions (e.g. \cite{TuckerSmith:2001hy,DelNobile:2011je,Chang:2010yk,Chang:2009yt,Bai:2009cd}).

Recent analysis show that even considering the uncertainties discussed above, in general,
it is difficult to reconcile all the experimental results, even though
the tensions can be alleviated in some scenarios\cite{Schwetz:2011xm,Kopp:2011yr,Farina:2011pw}.
Still, the possibility that these anomalies are produced by dark matter interactions can not
be excluded.
In conclusion, further data are necessary to clarify the situation and understand if DM
is responsible for these excesses or not.

\section{Indirect searches}

Dark matter annihilations or decays into SM particles offer a method
to indirectly detect DM through astrophysical observations.
Indirect searches focus mainly into photons, neutrinos and antimatter cosmic-rays, 
notably positrons, antiprotons and antideuterons.

\subsection{Gamma-rays}
\label{gamma}

\begin{figure*}[t]
\centering
	
\includegraphics[width=0.45\textwidth]{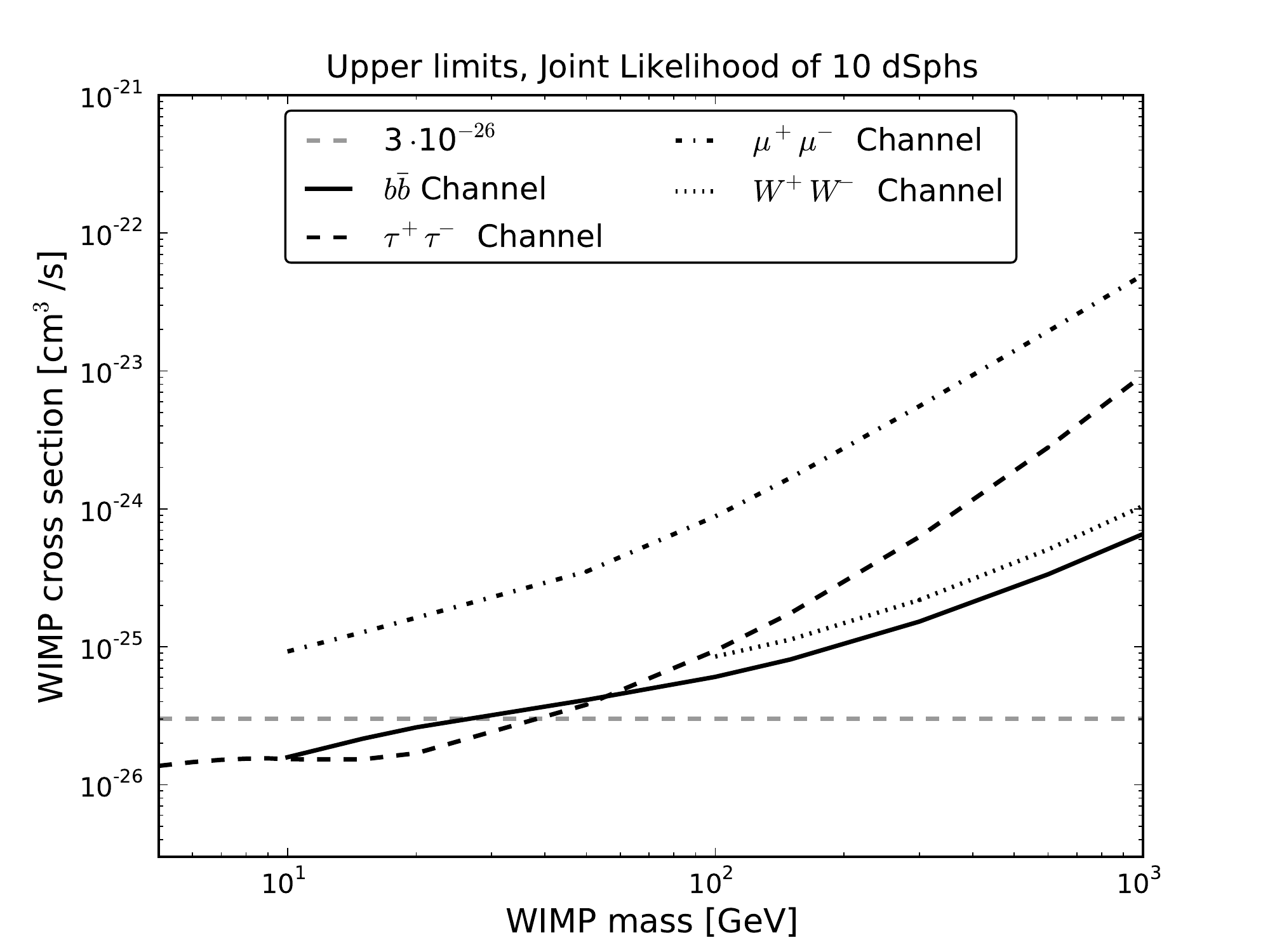}
\hspace{3mm}
\includegraphics[width=0.35\textwidth]{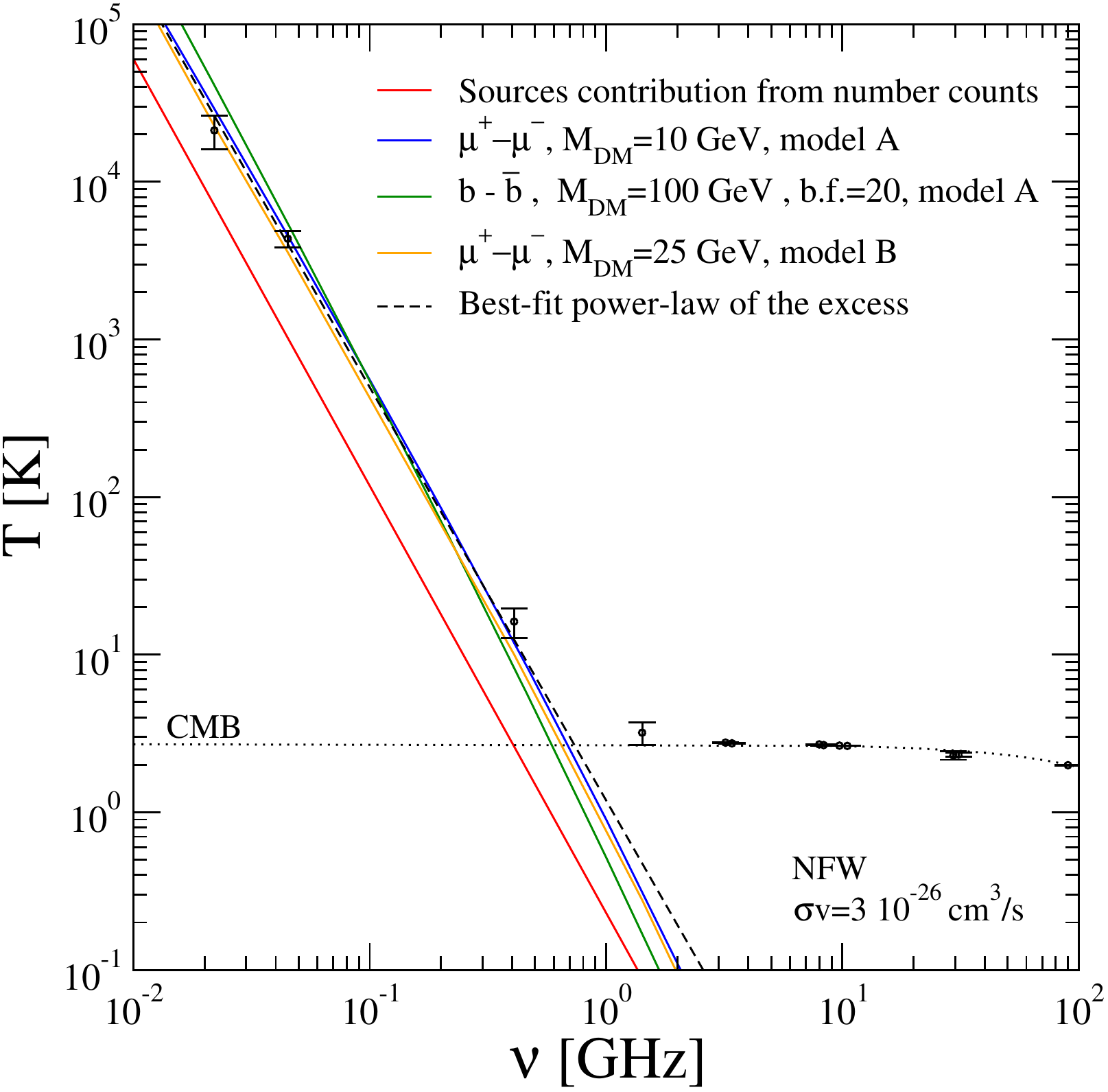} 
\caption{Left: upper bounds on the DM annihilation cross section as a function of the DM mass inferred from
FERMI observations of dwarf galaxies. Figure taken from\cite{collaboration:2011wa}. Right: isotropic diffuse radio background obtained
by the ARCADE collaboration. Lines refer to expectations from astrophysical sources and the contribution from DM annihilations in extragalactic
halos for several DM models. Figure taken from \cite{arXiv:1108.0569}.}
\label{fig:indirect}
\end{figure*}

Gamma-rays have been recognized as particulary interesting targets for indirect DM searches.
Indeed, they can be coupiously produced by annihilations/decays of DM candidates 
with masses in the GeV-TeV range, like WIMPs and many other well motivated DM candidates.
Moreover, contrary to charged particles produced by DM annihilations/decays,
gamma-rays suffer little or null attenuation
when travelling to the Earth and they point directly to the source.

Searches of DM signals with gamma-rays detectors are currently pursued with the
Fermi-LAT satellite and with ground-based Air Cherenkov Telescopes (ACT).
Targets of observations include the galactic center region, the Milky-Way halo,
dwarfs spheroidal galaxies, DM clumps, clusters of galaxies and the extragalactic diffuse emission.
There have not been reported robust evidences for a DM detection so far, even though
some author have claimed the current FERMI data might already indicate the presence of a DM signal.
In the following we will discuss some of the most relevant DM searches with gamma-rays.

\vspace{3mm}
{\bf Dwarfs galaxies} are nearby DM dominated systems with a low gamma-ray background
emission. For this reasons they are ideal targets for DM searches.
Current observations of dwarf galaxies with the FERMI telescope have not reported any
significant gamma-ray flux. These results have been used to constraint possible DM signals,
providing among the most stringent exclusion limits on the DM annihilation cross section $(\sigma v).$
Fig.\ref{fig:indirect} shows the upper bounds on $(\sigma v)$ for different DM annihilation channels\cite{collaboration:2011wa}.

\vspace{3mm}
The energy spectrum of the {\bf Isotropic Diffuse gamma-ray emission} detected by FERMI is
consistent with a power-law between energies of 20 GeV and 100 GeV \cite{Abdo:2010nz}. This implies
that DM should only give a subdominant contribution to this flux. The corresponding constraints on the DM parameter
space are derived in Ref.\cite{Abdo:2010dk}. 
Complementary to the energy spectrum analysis, the study of the small angular anisotropies 
of the isotropic diffuse emission could bring additional information about its origin.
In particular this method could be used to detect a faint DM signal in the data or constraints
the properties of the DM \cite{Ando:2006cr,Fornasa:2009qh,SiegalGaskins:2008ge,Taoso:2008qz,Cuoco:2010jb}.
Analysis of the FERMI data and efforts to model the DM constribution are ongoing \cite{Cuoco:2011ng,Fornasa:2011yb}.

\vspace{3mm}

{\bf The galactic center region}  has been extensively discussed in the context of DM searches with gamma-rays.
This is mainly for two reasons: it is close to us and it is expected to host large DM overdensities. 
However the presence of a strong astrophysical background severely challenges the search of DM  from this region.
Observations pursued with the HESS ACT have not reported any evidence of gamma-ray fluxes from DM annihilations around
the galactic center, implying bounds on $(\sigma v)$ particulary strong for masses around $m_{DM}\sim$ TeV \cite{Abramowski:2011hc}.
At lower energies, a survey of the galactic region is performed by FERMI.
It has been claimed that FERMI data reveal the presence of an unknown extended (up to $50$ degrees in latitude) diffuse emission
towards the galactic center, the so called FERMI Haze (see \cite{Su:2010qj} and references therein). The emission is spatially correlated with
an excess discovered at radio frequencies (WMAP haze), suggesting  a common origin for the two signals.
The properties of the FERMI haze, in particular its elongated bubble-like morphology, disfavour a DM interpration,
even though this possibility can not be excluded \cite{Dobler:2011mk}.

Recently, some authors have analyzed a smaller angular region surrounding the galactic center,
finding an excess in the data with respect to the expected contribution from diffusion emission and point sources\cite{Hooper:2011ti}.
In particular they suggested that this signal could be ascribed to annihilations of
a light WIMP candidate ($M_{DM}\sim$7-45 GeV).
Other analysis disagree arguing that the emission from this region can be instead completely
accounted for by known sources (\cite{Boyarsky:2010dr} and see also \cite{Chernyakova:2010ey} ).

In conclusion, a better understanding of the astrophysical background is necessary
for a reliable detection of a DM signal from the galactic center. This could be reached with more data, in particular
exploiting multi-wavelenght information.

\vspace{3mm}

{\bf Gamma-ray lines} have been proposed as a smoking gun signature for DM annihilations/decays
since ordinary astrophysical processes are not expected to produce monochromatic emissions at these energies.
These signals have been studied in many contexts, both for decaying and 
annihilating dark matter \cite{Mambrini:2009ad,Jackson:2009kg,Bertone:2009cb,Garny:2010eg}.
Unfortunately no line emissions have been detected in the FERMI data so far \cite{Abdo:2010nc,Vertongen:2011mu}.
The bounds on such signals have been used to constraint particle physics models \cite{Huang:2011dq,Restrepo:2011rj,GomezVargas:2011ph}.
Further FERMI and ACTs observations will improve the sensitivity to gamma-ray lines and more in general 
to spectral features generated by DM signals \cite{arXiv:1106.1874}.

\subsection{Cosmic-rays}

The PAMELA satellite has reported an anomalous rise with energy of the positron fraction above $\sim$ 10 GeV \cite{arXiv:0810.4995}.
Recent FERMI results confirm the measurements of PAMELA and suggests that the rise continues at least up to 200 GeV \cite{arXiv:1109.0521}.
These puzzling observations have generated a lot of interest during the last years, expecially in relation
to a possible DM origin of this excess. 
Specific DM properties, pointing to a non-standard DM model, are needed to explain the data.
DM should be i) heavy, with masses above 100 GeV (or perhaps few  TeV)
ii) leptophilic, in order to avoid the constraints from antiprotons measurements
and iii) in the case of an annihilating DM candidate, the annihilation cross section should be 
much larger than the canonical value expected for an s-wave thermal relic.

Soon after the PAMELA measurements of the positron fraction appeared, 
it has been shown that multi-messangers constraints
from photons, antiprotons and neutrinos severly challenge a DM interpretation\cite{arXiv:0811.3744,arXiv:0812.3895,arXiv:0905.0480}.
Recent bounds are futher reducing the viable regions of the DM parameter space \cite{arXiv:1009.5988,arXiv:1012.0588}.
Summarizing, although a DM interpretation of the positron excess is not yet completely ruled out,
it is phenomenologically disfavored.
In addition, standard astrophysical sources can in principle explain all the cosmic-ray data\cite{arXiv:1002.1910}.
The anisotropy in the arrival directions of electrons and positron cosmic-rays
might provide a tool to discriminate the DM solution from an astrophysical scenario\cite{arXiv:1012.0041}.
This could be possible already with upcoming FERMI and AMS-02 data.

Alternative DM searches with cosmic-rays could be pursued looking at
the antideuterons produced by the coalescence of antineutrons and antiprotons generated
by DM annihilations/decays. Promising prospects for detection are expected for the AMS-02 and GAPS experiments\cite{arXiv:0803.2640,arXiv:1006.0983}.

\subsection{Multiwavelenght observations, neutrinos and other astrophysical probes}

The interactions of the electrons and positrons produced by DM annihilations/decays with magnetic fields
and photons produces a continumm multi-wavelength emission ranging from radio/infrared
frequencies to X-ray and gamma-ray bands.
Multi-wavelenght observations are powerful techniques to search for a DM signal and they can provide constraints 
even more restrictive than those inferred from gamma-rays \cite{arXiv:0802.0234,arXiv:1002.0229}.
For instance low radio frequencies are particuary suitable for DM candidates with
low or intermediate masses, $M\lesssim 100$ GeV \cite{arXiv:1110.4337}.

Recently the ARCADE 2 collaboration has measured an isotropic radio emission which is about a factor
5 larger than the expectations from number counts of sources \cite{arXiv:0901.0555}. These observations suggest the presence of an additional population of radio
sources fainter than those current data are probing.
Possible explanations of this excess in terms of known astrophysical sources have shown some difficulties.
Interestingly, the synchrotron signal produced by annihilations of light DM candidates in extragactic halos,
can match the ARCADE measurements, as shown in Fig.\ref{fig:indirect} \cite{arXiv:1108.0569}.

Future radio surveys with improved sensitivities and angular resolutions should help to clarify
the origin of the ARCADE excess and offer the possibility to detect or constraint DM radio signals
induced by DM.

\vspace{3mm}

DM searches through neutrinos are mainly pursued looking at neutrinos from the Sun, the Earth and the galactic center.
Since the neutrino flux from the Sun and the Earth depends on the amounts of DM particles captured
inside these objects, these observations can probe the DM scattering cross-sections off protons.
Moreover, for sizeable annihilation cross sections the DM capture  and  annihilation rates
reach an equilibrium and the resulting DM neutrino flux is independent on the DM annihilation cross section.
Under these assumptions the constraints derived from Super-Kamiokande and IceCube observations
are competitive or even stronger than those from direct detection experiments \cite{arXiv:1107.5227,arXiv:1104.0679}.

\vspace{3mm}

Dark matter particles can accumululate inside the stars as a consequence of their
interactions with the baryons forming the celestial objects.
The cloud of these trapped DM particles can transport or produce energy inside the star, affecting therefore its the properties.
Small modifications of the Sun structure can be tested through measurements of
the solar neutrino fluxes and helioseimology observations \cite{arXiv:1005.5711,arXiv:1005.5102,arXiv:1003.4505}.
Current data are able to constraints the DM scattering cross sections off protons for light DM candidates.
Much dramamtic effects can occour in stars placed in high DM density environments. In this case, ideal targets are first 
stars \cite{arXiv:0705.0521,arXiv:0806.2681,arXiv:0805.4016} or stars close to the galactic center \cite{arXiv:0809.1871}.
Asymmetric dark matter particles captured inside neutron stars and white dwarfs
can form a self-gravitating DM cloud which can eventually collapse to a black hole and destroy the star.
This argument severely constraints the scattering cross sections of asymmetric dark matter candidates with neutrons \cite{arXiv:1103.5472,arXiv:1104.0382}.

\section{Conclusions}

Present DM searches, expecially direct detection experiments, show some hints of dark matter.
These hints could become robust and convicing evidences of DM only after an experimental
high significance confirmation of these excesses will be achieved, as well as a global and consistent theoretical understanding
of all the data.
Interestingly, this might happen soon.
The increasing sensitivity of direct detection experiments will allow in the next years to explore
progressively large portions of the WIMP parameter space.
At the same time, indirect searches will bring complementary informations, testing different DM properties.
LHC is a formidable machine to look for new physics beyond the SM and its results will be extremely valuable
for DM searches, especially if combined with the information from other DM searches\cite{arXiv:1005.4280,arXiv:1111.2607}.

The prospects for DM searches in the forthcoming years are promising and a non-gravitational
detection of DM could be not so far.

%% file: Author/adulpravitchai.tex
{\bf Abstract}\\
\vskip5.mm
In this talk, we briefly discuss the idea of flavored orbifold GUT~\cite{Adulpravitchai:2010na}, that is, the flavor symmetry might emerge due to orbifold compactification of one orbifold and broken by  boundary
conditions of another orbifold where the GUT group is broken. The combination of the orbifold parities in gauge and flavor
space determines the zero modes. An example of the idea is given in a supersymmetric (SUSY) SO(10)$\times
S_4$ orbifold GUT model, which predicts the tribimaximal mixing at leading order in the lepton sector 
as well as the Cabibbo angle in the quark sector.


\vskip5.mm

\section{Introduction}
The leptonic mixing matrix is compatible with the tribimaximal mixing
matrix~\cite{Harrison:2002er} and hints towards non-abelian discrete flavor symmetries. However, the origin of
a potential flavor symmetry is unclear. As the Standard Model of Particle Physics (SM) is constructed based on the two well-known symmetries, namely, continuous 
gauge symmetry and space-time symmetry, it is interesting to investigate their origin from these two symmetries. 
\begin{itemize}
\item {\it Continuous symmetries SO(3), SU(2) or SU(3)}: If one only employs small representations, the only non-Abelian discrete symmetry which can arise is $D_2'$, which is the double covering of $D_2'$~\cite{Adulpravitchai:2009kd}. For larger representations, the other non-Abelian discrete symmetries can be obtained. See also~\cite{largerRep}.
\item {\it Extra-dimensions}: Non-Abelian discrete symmetries can arise as the remnant symmetry of the broken Poincare (Lorentz) via orbifold compactification on two extra-dimensions. The non-Abelian discrete symmetries are $D_3$, $D_4$, $D_6$, $A_4$, and $S_4$~\cite{Altarelli:2006kg}.
\end{itemize}
An orbifold compactification of a GUT can lead to its
breaking and nicely solve e.g.~the doublet-triplet splitting
problem~\cite{Kawamura:2000ev}. Furthermore, the orbifold
compactification can generate the alignment of vacuum expectation values (VEVs)
of flavons~\cite{Kobayashi:2008ih} transmitting the flavor symmetry breaking
into the fermion mass matrices.\\
In this talk, we discuss the combination of the origin of a flavor symmetry as
well as its breaking by the VEV alignment of flavons from an orbifold.  We
assume two orbifolds, where the flavor symmetry originates from the special
geometry of one orbifold and it is broken on another orbifold. We demonstrate it
with a simple model in the context of a SUSY SO(10) orbifold
GUT with an $S_4$ flavor
symmetry~\cite{Adulpravitchai:2010na} (For $SU(5) \times A_4$, see~\cite{Burrows:2010wz}). The model predicts the tribimaximal mixing at leading order in the lepton sector and predicts the Cabibbo angle in quark sector. 

\section{Flavor Symmetries from Orbifolding}
%
Here, we will show briefly how $S_4$ flavor symmetry can be obtained. Let us consider the
$T^2/\mathbb{Z}_2$
orbifold with radii $R=R_5=R_6$, which we
choose as $2\pi R=1$ for simplicity. The discussion of the flavor
symmetry does not change for $2\pi R\neq 1$. It is defined by
\begin{align}
T_1: z \rightarrow& z+1, &
T_2: z \rightarrow& z+ \gamma, &
Z: z \rightarrow& -z.
\end{align}
where $z=x_5+i x_6$ and $\gamma=e^{i \pi/3}$. The shape of this orbifold is
a regular tetrahedron. 
It has been shown in~\cite{Altarelli:2006kg} that the breaking of Poincar\'e
symmetry from 6d to 4d through compactification on the orbifold leads to a
remnant $S_4$ flavor symmetry. Concretely, the orbifold has four fixed points,
\mbox{$(z_1,z_2,z_3,z_4)=(1/2,\,(1+\gamma)/2,\,\gamma/2,\,0)$}, which
are permuted by two translation operations $S_i$, the rotation $T_R$, and two
parity operations $P^{(\prime)}$
\begin{align}
S_1:&z \rightarrow z + 1/2,\;\; &
S_2:&z \rightarrow z + \gamma/2, \;\; &
T_R:&z \rightarrow \gamma^2 z, &
P:& z \rightarrow z^{*}, \;\; &
P^\prime:& z \rightarrow -z^{*}\; .
\end{align}
One can also write these operations explicitly in terms of the interchange of
the fixed points, $S_1[(14)(23)]$, $S_2[(12)(34)]$, $T_R[(123)(4)]$,
$P[(23)(1)(4)]$ and $P^\prime[(23)(1)(4)]$.
From these elements we can define two generators of $S_4$ as $S = S_2 P$ and $T
= T_R$ satisfying the generator relation,
$S^4=T^3=(ST^2)^2=1$. \\
The localization of a brane field defines its
representation of $S_4$. The generators $S$, $T$ can be represented by the matrices,
\begin{align}
 S= \begin{pmatrix} 0 & 0 & 1 & 0 \\ 1 & 0 & 0 & 0 \\ 0 & 0 & 0 & 1 \\ 0 & 1 & 0
& 0 \end{pmatrix}, & \;\;\;\;
 T= \begin{pmatrix} 0 & 0 & 1 & 0 \\ 1 & 0 & 0 & 0 \\ 0 & 1 & 0 & 0 \\ 0 & 0 & 0
& 1 \end{pmatrix},
\end{align}
acting on the brane field $\psi(x_\mu)=(\psi_1,\psi_2,\psi_3,\psi_4)^T$, where
$\psi_{i}=\psi(x_\mu,z_i)$ is a field localized at fixed point $z_i$. We
denote this basis as localization basis. The characters of $S$ and $T$ show that
the four dimensional representation generated by $\left<S,\,T\right>$ can be
decomposed in
$3_1\oplus1_1$. The explicit unitary transformation is
\begin{equation}
 S \rightarrow U^{\dagger} S U = \begin{pmatrix}  S^{fl}_3 &  \\   & 1
\end{pmatrix},\quad
 T \rightarrow U^{\dagger} T U = \begin{pmatrix}   T^{fl}_3 &  \\  & 1
\end{pmatrix}\quad
\end{equation}
with the unitary matrix $U$ (For the explicit form, see \cite{Adulpravitchai:2010na}). 
The transformation of a field $\psi(x)$ is accordingly related to
the flavor basis $\psi^{fl}(x_\mu)=U^\dagger\psi(x_\mu)$ as well as the
Clebsch-Gordan coefficients. The first three components of $\psi^{fl}$ form a
triplet $3_1$ and the last one a singlet $1_1$.
It is possible to remove one of the representations from the
 low-energy spectrum by adding a bulk field transforming as $3_1$ ($1_1$) and oppositely charged to the
brane field, such that they acquire a Dirac mass term. Note that the representations $1_2$ and $3_2$ are analogously obtained by using the
freedom to change the phase of each brane field in a symmetry
transformation (For detail, see \cite{Adulpravitchai:2010na}).
\section{Symmetries Breaking by Boundary Conditions}
In order to demonstrate how the flavor structure can be obtained and broken
appropriately from an orbifold, we implement it in an 8d SUSY SO(10) model. 
The first orbifold is responsible for the origin of the flavor symmetry and the second orbifold is for the symmetry breaking of both flavor and GUT symmetries.
The second orbifold is is also
$T^2/\mathbb{Z}_2$ with two additional boundary condtions at $\hat{z}_1$,
$\hat{z}_3$, i.e. $T^2/(\mathbb{Z}_2^I \times
\mathbb{Z}_2^{PS}\times\mathbb{Z}_2^{GG})$ 
and then we assume that the gauge fields are bulk fields of the two orbifolds,
while all other bulk fields of the second orbifold are brane fields of the first
orbifold. Note that the SO(10) breaking which leads to a splitting of the
doublet and triplet components in the $10$-plet is similar
to the 6d model in~\cite{Asaka:2002nd}. \\
The $N=1$ SUSY in 8d leads to $N=4$ SUSY in
4d~\cite{ArkaniHamed:2001tb}. Therefore, one needs the orbifold parities to break $N=4$ SUSY down to $N=1$ SUSY.
The gauge fields transform under the orbifold parities as
\begin{align}
P_0 V(x_\mu,-z,\hat{z})P_0^{-1} =& \eta_0 V(x_\mu,z,\hat{z})\;, \nonumber \\
P_IV(x_\mu,z,-\hat{z})P_I^{-1} =& \eta_I V(x_\mu,z,\hat{z})\;, \nonumber \\
P_{PS}V(x_\mu,z,-\hat{z}+\hat{z}_1)P_{PS}^{-1} =&
\eta_{PS} V(x_\mu,z,\hat{z}+\hat{z}_1)\;,\nonumber \\
P_{GG}V(x_\mu,z,-\hat{z}+\hat{z}_3)P_{GG}^{-1} =&
\eta_{GG} V(x_\mu,z,\hat{z}+\hat{z}_3)\;,
\end{align}
with $\hat{z}=x_7+i x_8$. The parities are chosen as
$\eta_0=\eta_I=\eta_{PS}=\eta_{GG}=+1$.
The first two parities (corresponding to fixed point $z_4$ and $\hat{z}_4$)
are used to break $N=4$ SUSY to $N=1$
SUSY and the remaining two parities are used
to break the gauge symmetry~\cite{Mirabelli:1997aj}. The parity
assignment of the different components of the bulk fields is given in \cite{Adulpravitchai:2010na}.

Analogously, by assigning the parities to the flavons living in the bulk of
the second orbifold, a zero mode is singled out and consequently the flavor
symmetry is broken. As the flavons are the bulk fields of only one
orbifold, the flavons inherit $N=2$ SUSY in 4d, which can be broken by using one
orbifold parity. If the zero mode acquires a VEV, the breaking of the symmetry
is transmitted to the fermion masses. We assume that flavons transform
non-trivially,
\begin{subequations}
\label{eq:FlavorParities}
\begin{align}
 P_1 \phi(x_\mu,z,-\hat{z}) &= \eta_1\, \phi(x_\mu,z,\hat{z}), \\
 P_2 \phi(x_\mu,z,-\hat{z}+\hat{z}_1) &= \eta_2\,
\phi(x_\mu,z,\hat{z}+\hat{z}_1), \\
 P_3 \phi(x_\mu,z,-\hat{z}+\hat{z}_3) &= \eta_3\,
\phi(x_\mu,z,\hat{z}+\hat{z}_3).
\end{align}
\end{subequations}
The first parity is used to break $N=2$ SUSY to $N=1$, the
remaining two parity operators are used to generate the VEV alignment of the
flavons by singling out the appropriate zero modes.  
 
In order to obtain the VEV alignment for a triplet $\phi\sim 3_1$, we choose
\begin{align}
P_1 =& 1, & P_2 =& TST, &P_3 =& TSTS^2\; .
\end{align}
As $P_1$ is the unit matrix, it does not affect the zero mode. Therefore, the zero mode is entirely determined by $P_{2,3}$.
$P_2$ and $P_3$ are simultaneously diagonalized by the unitary matrix $U$
\begin{equation}
U = \begin{pmatrix}
       \frac{1}{\sqrt{3}} & \frac{2}{\sqrt{6}} & 0 \\
       \frac{1}{\sqrt{3}} & -\frac{1}{\sqrt{6}} & \frac{1}{\sqrt{2}} \\
       \frac{1}{\sqrt{3}} & -\frac{1}{\sqrt{6}} & -\frac{1}{\sqrt{2}}
      \end{pmatrix} \;, \quad\quad
\begin{array}{rl}
U^{\dagger} P_2 U &= \mathrm{diag}(1,\, -1,\, 1) \;,\\
U^{\dagger} P_3 U &=  \mathrm{diag}(1,\, 1,\, -1) \;,
\end{array}
\end{equation}
This leads to the desired VEV aliginment $\left<\phi_{1,1,1}\right>=\left<\phi^0\right>(1,1,1)^T$. (Other VEV alignments, see~\cite{Adulpravitchai:2010na}).
\section{SO(10) $\times S_4$ model}
Here, we briefly show the result of the model. For detail and particle content, see~\cite{Adulpravitchai:2010na}.
Small neutrino masses can be generated by the seesaw mechanism starting from
\begin{equation}
 {\bf W}_\nu = \frac{y_{s}}{\sqrt{\Lambda^2 V}}  \psi \bar{\Delta}_2 S_\nu + \frac{1}{\sqrt{\Lambda^2 V}} \left(y^\nu_\phi\phi_{1,1,1}
+y^{\nu}_\xi \xi\right) S_{\nu} S_{\nu}  \;.
\end{equation}
After flavons obtain their VEV, it leads to neutrino mass matrix which can diagonalized by the tribimaximal mixing matrix, 
\begin{equation}
 M_{\nu} = - \frac{1}{\Lambda^2 V} (y_s \left<\bar\Delta_2\right>) (M_{SS}^{-1}) (y_s \left<\bar\Delta_2\right>)^T
 = - m_0 U^*_\mathrm{tbm}
\begin{pmatrix}
       \frac{1}{3a+b} & 0 & 0 \\
       . & \frac{1}{b} & 0 \\
       . & . & \frac{1}{3a-b}
      \end{pmatrix} U^\dagger_\mathrm{tbm}\;.
\end{equation}
For the quark and charged lepton sectors, we obtain the mass matrices,
\begin{align}
M_{u,d} =&\, m_{t,b} \begin{pmatrix}
       0 & a^{u,d} & 0  \\
       a^{u,d} & b^{u,d} & 0\\
       0 & 0 & 1
      \end{pmatrix}\;,  &
M_{l} =&\, m_b \begin{pmatrix}
       0 & a^d & 0  \\
       a^d & 3\, b^d & 0\\
       0 & 0 & 1
      \end{pmatrix}\label{eq:chargedLepton}\; ,
    \end{align}
leading to the Cabibbo angle $\theta_c$ which is approximately given by
$\sin\theta_c=a^d/b^d-a^u/b^u$. 
\section{Conclusion}
In this talk, we have discussed a model, where an $S_4$ flavor symmetry arises
from an orbifold compactification and it is broken by the boundary conditions
together with the SO(10) gauge symmetry on another orbifold. We gave an example in the context of
SO(10)$\times S_4$ which leads to a phenomenologically viable neutrino mass
matrix as well as enables to fit the masses of quarks and charged leptons. The model predicts the tribimaximal mixing at leading order 
in the lepton sector and a quantitatively correct Cabibbo mixing angle in quark sector.

%% file: Author/diego-aristizabal.tex
  {\bf Abstract}\\
  \vskip5.mm
  In models featuring exact mixing patterns the mass matrices that
  define the effective light neutrino mass matrix -and the light
  neutrino matrix itself- are form-diagonalizable.  We study
  leptogenesis in type I seesaw models in these contexts pointing out
  that the CP asymmetry in right-handed neutrino decays vanishes as a
  consequence of the mass matrices being form-diagonalizable.  A
  non-vanishing CP asymmetry arises once deviations from the exact
  mixing scheme, induced by higher order effective operators, is
  allowed.  Finally we discuss alternative pathways to viable
  leptogenesis in these kind of models.
\vskip5.mm
\section{Introduction}
\label{sec:intro-lepto-FS}
Leptogenesis is a scenario in which the baryon asymmetry of the
Universe is dynamically generated first in the lepton sector and
reprocessed into a baryon asymmetry via standard model electroweak
sphaleron processes \cite{Kuzmin:1985mm}. In order for this mechanism
to take place lepton number must be broken \footnote{The exception
  being scenarios of purely flavored leptogenesis as the one discussed
  in \cite{arXiv:0705.1489}.} thus implying models for Majorana
neutrino masses provide the frameworks for leptogenesis. 

The standard seesaw model (type I seesaw) \cite{seesaw} defines the
scheme for {\it standard leptogenesis} \cite{Davidson:2008bu}. In this
model leptogenesis becomes plausible due to the fact that: (i) the
Yukawa couplings of the fermionic electroweak singlets (right-handed
(RH) for brevity) contain new physical CP phases; (ii) lepton number
violation is provided by the RH neutrino masses; (iii) the expansion
of the Universe guarantees deviations from thermodynamic equilibrium
in RH neutrino decays. With these conditions satisfied the generation
of a net $B-L$ asymmetry proceeds through the decays of the lightest
RH neutrino.

In standard leptogenesis the problem of calculating the baryon
asymmetry depends -in first approximation- on two parameters: the
washout factor $\tilde m$, determined by the contribution of the
lightest RH neutrino to light neutrino masses, and the CP asymmetry
$\epsilon_N$ in RH neutrino decays. Thus, two conditions must be
satisfied in order to produce a net baryon asymmetry: overcome the
washout effects and a non-vanishing $\epsilon_N$. The determination of
both requires the specification of the RH neutrino Yukawa couplings
and mass spectrum, however the former is more involved as it demands
calculating the efficiency factor, which in turn implies solving the
corresponding kinetic equations describing the RH neutrino dynamics.

The seesaw parameter space consist of 18 parameters out of which 9 are
constrained by low-energy data. This implies once these restrictions
are placed there is a remaining arbitrary 9 dimensional parameter
space.  Is indeed partially due to this arbitrariness that
leptogenesis suffers from the lack of testability
\cite{arXiv:0705.1503}. If further restrictions on the parameter space
can be placed the arbitrariness should be reduced, and this is
actually the case if a lepton flavor symmetry is present as some of
these parameters will be either forced to vanish or to be correlated.

Even in the light of recent neutrino data \cite{Schwetz:2011zk} there
is still a strong motivation to believe that the leptonic mixing is a
result of an underlying flavor symmetry operating in the lepton sector
\cite{Altarelli:2010gt}.  The tri-bimaximal mixing (TBM) pattern
\cite{hep-ph/0202074} as an input {\it ansatz} remains as a viable
guideline to construct lepton flavor models accounting for neutrino
masses and mixing angles. We here discuss standard leptogenesis in the
context of the seesaw extended with flavor symmetries \footnote{This
  subject has been recently analyzed in a series of papers
  \cite{Jenkins:2008rb,Hagedorn:2009jy,Bertuzzo:2009im,AristizabalSierra:2009ex,Felipe:2009rr}.}. We
will study the implications that a generic flavor symmetry associated
to a flavor group $G_F$ leading to the TBM pattern may have for the
baryon asymmetry generated via leptogenesis. It will be proven that if
leptogenesis takes place below the scale at which the flavor symmetry
is broken $\epsilon_N$ vanishes in the limit of exact TBM. Viable
leptogenesis becomes possible once departures -induced by higher order
effective operators- are allowed.  We will point out other pathways to
leptogenesis in flavor models.  We will closely follow references
\cite{AristizabalSierra:2009ex}.
\section{General considerations}
\label{sec:general-considerations}
With the addition of three RH neutrinos $N_{R_{i=1,2,3}}$ the standard
model Lagrangian is extended with a new set of interactions that, in a
generic basis in which the charged lepton Yukawa coupling matrix is
diagonal, can be written as
\begin{equation}
  \label{eq:lag}
    -{\cal L}=-i\bar N_{R_i}\,\gamma_\mu\partial^\mu N_{R_i}
  + \bar \ell_{Lj} N_{R_i}\lambda_{ij} \phi
  + \frac{1}{2}\bar N_{R_i} C M_{R_i} \bar N_{R}^T
  + \mbox{h.c.}\,.
\end{equation}
Here $\ell_L$ are the lepton $SU(2)$ doublets, $\phi^T=(\phi^+
\phi^0)$ is the Higgs electroweak doublet, $M_{R_i}$ are the RH
neutrino masses, $C$ is the charge conjugation operator and
$\pmb{\lambda}$ is a $3\times 3$ Yukawa matrix in flavor space (we
will denote matrices in bold-face). In the seesaw limit $M_{R_i}\gg v$
(with $v\simeq 174$ GeV) the effective neutrino mass matrix is
obtained once the heavy fields are integrated out:
\begin{equation}
  \label{eq:seesaw-formula-tree-level}
  \pmb{m_\nu}^{\text{eff}}=
  -\pmb{m_D}\,\pmb{\hat M_R}^{-1}\,\pmb{m_D}^T\,,
\end{equation}
with $\pmb{m_D}=v\,\pmb{\lambda}$.  From now on we will assume the
Lagrangian in (\ref{eq:lag}) to be invariant under a flavor group
$G_F$ that enforces the TBM pattern. This has two
implications. Firstly, the mass matrices $\pmb{m_D}, \pmb{M_R}$ and
$\pmb{m_\nu}^{\text{eff}}$ are form-diagonalizable
\cite{Low:2003dz}. Secondly, the effective neutrino mass matrix is
diagonalized by the TBM leptonic mass matrix, namely
\begin{equation}
  \label{eq:diagonalization}
  \pmb{\hat D}\pmb{U_\text{TB}}^T \pmb{m_\nu}^{\text{eff}}
  \pmb{U_\text{TB}}\pmb{\hat D}=
  \pmb{\hat m_\nu}\qquad\mbox{with}\qquad
  \pmb{U_\text{TB}}=
  \begin{pmatrix}
    \sqrt{2/3}  & 1/\sqrt{3} & 0          \\
    -1/\sqrt{6} & 1/\sqrt{3} & -1/\sqrt{2}\\
    -1/\sqrt{6} & 1/\sqrt{3} & 1/\sqrt{2}
  \end{pmatrix}\,,
\end{equation}
where the matrix
$\pmb{\hat D}=\mbox{diag}(e^{i\varphi_1},e^{i\varphi_2},1)$ contains the
low-energy Majorana CP phases. The matrices $\pmb{M_R}$ and
$\pmb{m_D}$, being in general non-diagonal, can be diagonalized
according to
\begin{equation}
  \label{eq:md-mr-diag}
  \pmb{U_L} \pmb{m_D}\pmb{U_R}^\dagger=\pmb{\hat m_D}
  \qquad\mbox{and}\qquad
  \pmb{V_R}^T \pmb{M_R}\pmb{V_R}=\pmb{\hat M_R}\,,
\end{equation}
with $\pmb{U_{L,R}}$ and $\pmb{V_R}$ unitary $3\times 3$ matrices
characterized in general by three rotation angles and six phases. By
means of eqs. (\ref{eq:diagonalization}) and (\ref{eq:md-mr-diag})
eq. (\ref{eq:seesaw-formula-tree-level}) can be rewritten as
\begin{equation}
  \label{eq:mnueff-rewritten}
  \pmb{\hat m_\nu}=
  -\pmb{\hat D}\,\left(\pmb{U_\text{TB}}^T\,\pmb{U_L}\right)\,\pmb{\hat m_\nu}\,
  \left(\pmb{U_R}^\dagger\,\pmb{V_R}\right)\,
  \pmb{\hat M_R}\,\left(\pmb{V_R}^T\,\pmb{U_R}^*\right)\,
  \pmb{\hat m_D}\,\left(\pmb{U_L}^T\,\pmb{U_\text{TB}}\right)\,\pmb{\hat D}\,.
\end{equation}
Since the mass matrices $\pmb{m_D}$ and $\pmb{M_R}$ are
form-diagonalizable the corresponding diagonalization matrices
$\pmb{U_{L,R}}$ and $\pmb{V_R}$ do not depend upon the corresponding
eigenvalues entering in the diagonal matrices $\pmb{\hat m_D}$ and
$\pmb{\hat M_R}$ \cite{Low:2003dz}. Accordingly,
eq. (\ref{eq:mnueff-rewritten}) is satisfied if and only if the two
following conditions are satisfied:
\begin{equation}
  \label{eq:conditions}
  \pmb{U_\text{TB}}^T\,\pmb{U_L}=\pmb{\hat P_L}\,\pmb{O_{D_i}}
  \qquad\mbox{and}\qquad
  \pmb{U_R}^\dagger\,\pmb{V_R}=\pmb{O_{D_i}}\,\pmb{\hat P_R}\,\pmb{O_{R_m}}\,.
\end{equation}
The matrices $\pmb{O_{D_i}}$ and $\pmb{O_{R_m}}$ are unitary and
complex orthogonal matrices, respectively, that rotate the $i$ and $m$
degenerate eigenvalues in $\pmb{\hat m_D}$ and $\pmb{\hat M_R}$. They
are such that if there is no degeneracy in non of the two mass
matrices $\pmb{O_{D_i}}=\pmb{1}$ and $\pmb{O_{R_m}}=\pmb{1}$. The
matrices $\pmb{\hat P_{L,R}}$ are given by $\pmb{\hat
  P_{L,R}}=\text{diag}(e^{i\alpha_1^{L,R}},e^{i\alpha_2^{L,R}},
e^{i\alpha_3^{L,R}})$ and thus taking $\pmb{\hat M_R}$ to be real
the following constraints on the CP phases must be satisfied:
\begin{equation}
  \label{eq:phases-constraint}
  \varphi_i + \alpha_i^L + \alpha_i^R + \gamma_i=2k\pi
  \quad\mbox{and}\quad
  \alpha_3^L + \alpha_3^R + \gamma_3=2n\pi\,,
\end{equation}
with $\gamma_i$ the CP phases in $\pmb{\hat m_D}$. 

In the basis in which the RH Majorana neutrino mass matrix is diagonal
the Dirac mass matrix is given by
\begin{equation}
  \label{eq:dirac-mm-MRdiag}
  \pmb{m_D^R}=\pmb{m_D}\,\pmb{V_R}\,,
\end{equation}
therefore, taking into account the results in (\ref{eq:conditions})
and (\ref{eq:md-mr-diag}), the Dirac mass matrix can be rewritten
according to
\begin{equation}
  \label{eq:dirac-mm-definitive}
  \pmb{m_D^R}=\pmb{U_\text{TB}}\,\pmb{\hat P_L}\,\pmb{\hat m_D}\,
  \pmb{\hat P_R}\,\pmb{O_{R_m}}\,.
\end{equation}
\begin{figure}
  \centering
  \includegraphics[width=8cm,height=2cm]{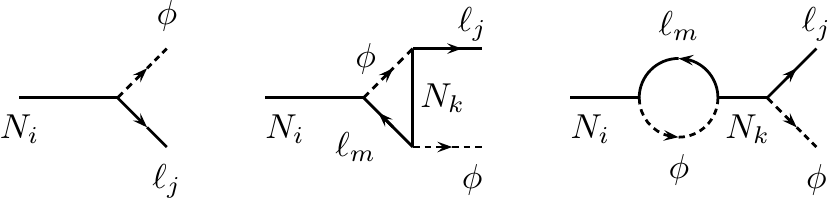}
  \caption{Tree-level and one-loop vertex and wave-function
    corrections responsible for the CP asymmetry in RH neutrino
    decays.}
  \label{fig:cp-asymmetries}
\end{figure}
\section{TBM and leptogenesis}
\label{sec:tbm-leptogenesis}
Depending on the temperature regimen at which leptogenesis takes place
the lepton doublets states that propagate in space-time can be either
a superposition of flavor states or the actual flavor components. For
temperatures above $\sim 10^{13}$ GeV the flavor composition of the
lepton doublets produced in the out-of-equilibrium and CP violating
decays of the lightest RH neutrino can be accurately neglected as all
the standard model lepton Yukawa reactions are slow. In that case the
amount of CP violation generated in RH neutrino decays is entirely
determined by the CP violating asymmetry resulting from the
interference between the tree-level and one-loop vertex and
wave-function Feynman diagrams depicted in figure
\ref{fig:cp-asymmetries}. In the limit of a strongly hierarchical RH
neutrino mass spectrum the result reads \cite{co96}
\begin{equation}
  \label{eq:cp-asymmetry}
  \epsilon_{N_i}=\frac{3}{8v^2\pi}\frac{1}{\left(\pmb{m_D^{R\dagger}}
      \pmb{m_D^R}\right)_{ii}}
  \sum_{i\neq k} \mathfrak{I}\mbox{m}
  \left[\left(\pmb{m_D^{R\dagger}}\pmb{m_D^R}\right)_{ki}^2\right]
  \frac{M_{R_i}}{M_{R_k}}\,.
\end{equation}
From the result in eq. (\ref{eq:dirac-mm-definitive}) the quantity
$\pmb{m_D^{R\dagger}}\pmb{m_D^R}$ can be calculated, namely
\begin{equation}
  \label{eq:mddagger-md}
  \pmb{m_D^{R\dagger}}\pmb{m_D^R}=\pmb{O_{R_m}}^T\pmb{\hat m_D}^2
  \pmb{O_{R_m}}\,.
\end{equation}
From this expression it becomes clear that as long as the $G_F$ flavor
group enforces the mass matrices to be form-diagonalizable the CP
violating asymmetry vanishes thus implying in the limit of TBM
leptogenesis is not viable. Though we have sticked to the concrete TBM
pattern this result remains valid regardless of the mixing scheme. As
it has been stressed it is a consequence of the form-diagonalizable
form of the mass matrices, which as long as we deal with an exact
mixing pattern it is always the case.

A vanishing CP asymmetry, however, can be accommodated in several ways
that we now briefly discuss in turn:
\begin{itemize}
\item \underline{Inclusion of higher order effective operators}
  \cite{Bertuzzo:2009im,AristizabalSierra:2009ex}:\\
  Flavor models involve effective operators that result from
  integrating out heavy fields that account for quark masses and
  mixings. These effective operators, arising from example from a
  quark flavor model {\it \`a la} Froggatt-Nielsen
  \cite{Froggatt:1978nt}, involve different powers of the ratio
  $\delta=\langle S\rangle/M_F$, where $\langle S\rangle$ is the
  vacuum expectation value of an electroweak singlet flavon that
  triggers the breaking of the corresponding quark flavor symmetry (a
  $U(1)_X$ in the case of Froggatt-Nielsen models) and $M_F$ is the
  mass scale of the heavy vectorlike fields.

  Whenever only leading order effective terms are included, that is to
  say order $\delta$ terms, there are no deviations from the exact
  mixing pattern. However, once next-to-leading order terms are
  included departures from this pattern are induced (the mass matrices
  deviate from there form-diagonalizable form) and thus the CP violating 
  asymmetry becomes non-zero. 
\item \underline{Presence of new physical degrees of freedom}
  \cite{arXiv:1101.0602,diego-ivo-federica}:\\
  In models in which the effective neutrino mass matrix receives
  contributions from other degrees of freedom, as for example in
  models featuring an interplay between type I and type II seesaw, the
  CP asymmetry typically contains additional contributions. The
  additional terms can be -in principle- also constrained by the flavor
  symmetry thus leading to a vanishing $\epsilon_N$. However, in
  general, these constraints are not so strong as in models entirely
  based in type I seesaw. Accordingly, realizations exhibiting type I
  and II seesaw models with a non-vanishing $\epsilon_N$ -even at order
  $\delta$- can be constructed.

\item \underline{The role of flavons} \cite{arXiv:1110.3781}:\\
  The analysis leading to the conclusion that the CP asymmetry
  vanishes in the limit of exact TBM has been done assuming
  leptogenesis takes place below the scale at which $G_F$ is broken. A
  new twist occurs if the generation of the lepton asymmetry happens
  at energy scales at which $G_F$ is still an exact symmetry (flavor
  symmetric phase). In that case the conventional contributions to the
  CP asymmetry (those given by the interference of the Feynman
  diagrams shown in fig. \ref{fig:cp-asymmetries} are still zero but
  if: ($a$) some of the RH neutrinos lie in different representations
  of $G_F$; ($b$) the flavons are lighter than one of the RH neutrino
  representations, new non-zero contributions to the CP asymmetry can
  be built thus allowing leptogenesis to proceed even in the flavor
  symmetric phase.
\end{itemize}
\section{Conclusions}
We have analyzed the viability of leptogenesis in seesaw I models
extended with flavor symmetries and assuming the lepton asymmetry is
generated in the flavor broken phase. For concreteness we have taken
the TBM pattern and have shown that in the limit in which this pattern
is exact the CP asymmetry in RH neutrino decays vanishes. We have also
discussed several pathways to leptogenesis in this stage (flavor
broken phase) which include the addition of next-to-leading order
corrections (higher order effective operators) or the addition of new
degrees of freedom, as would be the case in models with an interplay
between type I and II seesaw. Finally we have commented on scenarios
for leptogenesis taking place in the flavor symmetric phase pointing
out the viability of leptogenesis relies -in these cases- on the role
played by the scalar flavons.

%% file: Author/Bazzocchi.tex
{\bf Abstract}\\
\vskip5.mm
A relatively  low scale flavor symmetry is deeply challenging from a phenomenological point of view. We discuss the constraints  that have to be satisfied  in presence of a   unified description of electroweak and flavor  symmetry  breaking and we show how they apply in the context of  specific models.

\vskip5.mm

\section{Introduction}

Much effort in theoretical physics is currently devoted to  many open questions  related
to flavor physics issues, like the 
generation of neutrino masses and the origin of fermion mass hierarchies  and mixing 
matrices.
The present picture offered by the Standard Model (SM) particles is actually  like a 
\emph{puzzle}. There are  three families of quarks and leptons   that present mass hierarchies and mixing completely different. 

One of the most promising idea to explain flavor physics is the use of 
 horizontal \emph{flavor} symmetries that act on the three families.  A crucial point is  identifying the correct symmetry and the scale it should be broken. Flavor changing neutral current (FCNC) processes put a  rough lower bound according if the flavor symmetry is local or global. By comparing the SM contribution to FCNC processes such as $K-\bar{K}$ oscillation we deduce that  a global (local) flavor symmetry cannot be broken at a scale lower than $1$ ($10^6$) TeV. The lower bound corresponding to a global flavor symmetry is particularly interesting because it is close to two other  important scales: the electroweak (EW) scale and the typical scale associated to a weakly interacting cold Dark Matter (DM) candidate. May the new physics responsible of fermion masses and mixings be related to the EW symmetry breaking? or to the origin of  DM? In this talk we address the constraints that such a low scale flavor symmetry has to satisfy considering both constraints arising by the Higgs physics as well as processes involving the  fermion sector \cite{Toorop:2010ex,Toorop:2010kt,Toorop:2011ad}.
 
 \section{Low energy flavor symmetry}
 
The current data on neutrino oscillations seem to point at one small and two large angles in the neutrino mixing matrix \cite{Schwetz:2011zk}. The most recent data indicate that while the solar and atmospheric mixing angles are still in striking agreement  with the tri-bimaximal (TBM) \cite{HPS:TBM,Xing:TBM} mixing pattern the reactor angle deviates from a vanishing value at more than $3\sigma$. Nevertheless TBM mixing is still a good approximation of leptonic mixing.

As long as TBM mixing has been proposed as perfect--good-- approximation of  experimental data,
the use of non-Abelian discrete flavor symmetries has been analyzed in different models (for a review see \cite{AF:ReviewDiscreteSyms}) to generate both the mentioned lepton mixing patterns and the quark ones. In general, in those models, one introduces so called flavons,  scalar fields charged in the flavor space, usually very heavy. Once the flavons develop specific vacuum expectation values (VEVs), this translates to structures in the masses and mixings of the fermions.
However, imposing the correct symmetry breaking patterns on the flavons is highly non-trivial. This holds in particular if two or more flavons are used and the flavor group is broken in different directions. So far, only a few techniques have been developed, all of which need a supersymmetric context or the existence of extra dimensions \cite{AF:ReviewDiscreteSyms}.

Alternatively, one can look at models that require only one flavor symmetry  breaking  direction. In this case the scalar potential that implements the breaking can be non supersymmetric and does not require extra dimensions. Of particular interest is the possibility that one set of fields simultaneously takes the role of the flavons and SM Higgs fields, identifying the breaking scales of the electroweak and the flavor symmetries\cite{MR:A4EWscale,LK:A4EWscale,MP:A4EWscale}. In addition the flavor symmetry breaking may allow a residual  global symmetry that could give rise to a  DM candidate\cite{HMPV:DiscreteDM,MMP:DarkMatterA4,Toorop:2011ad}.

\section{The {$A_4$} Higgs Potential}

As an example of  EW-flavor symmetry unification  we discuss a setup where the flavor symmetry group is $A_4$ and the field scalar content is given by three copies of the SM Higgs field, $\Phi_i$ belonging to the triplet representation of the discrete group. The scalar potential  $A_4\times SM$ is given by
\begin{equation}
\label{A4pot}
\begin{split}
V[\Phi_a]=&\mu^2 (\Phi_1^\dag \Phi_1+ \Phi_2^\dag \Phi_2+ \Phi_3^\dag \Phi_3)+ \lambda_1 (\Phi_1^\dag \Phi_1+ \Phi_2^\dag \Phi_2+ \Phi_3^\dag \Phi_3)^2+\\
&+ \lambda_3 (\Phi_1^\dag\Phi_1 \Phi_2^\dag\Phi_2+ \Phi_1^\dag\Phi_1\Phi_3^\dag \Phi_3+ \Phi_2^\dag\Phi_2\Phi_3^\dag \Phi_3)+ \\
&+ \lambda_4 (\Phi_1^\dag \Phi_2 \Phi_2^\dag \Phi_1 +\Phi_1^\dag \Phi_3 \Phi_3^\dag \Phi_1+ \Phi_2^\dag \Phi_3 \Phi_3^\dag \Phi_2)+\\
&+\frac{\lambda_5}{2}\bigg[ e^{i \epsilon} \left[ (\Phi_1^\dag \Phi_2)^2+(\Phi_2^\dag \Phi_3)^2+(\Phi_3^\dag \Phi_1)^2\right]+  e^{-i \epsilon}   \left[(\Phi_2^\dag \Phi_1)^2+(\Phi_3^\dag \Phi_2)^2+(\Phi_1^\dag \Phi_3)^2\right] \bigg] \,,
\end{split}
\end{equation}
So far  our scenario  is nothing more that a special case of a multi higgs (MH) model with three   scalar doublets. 

\begin{figure}[ht!]
\centering
\includegraphics[width=7cm]{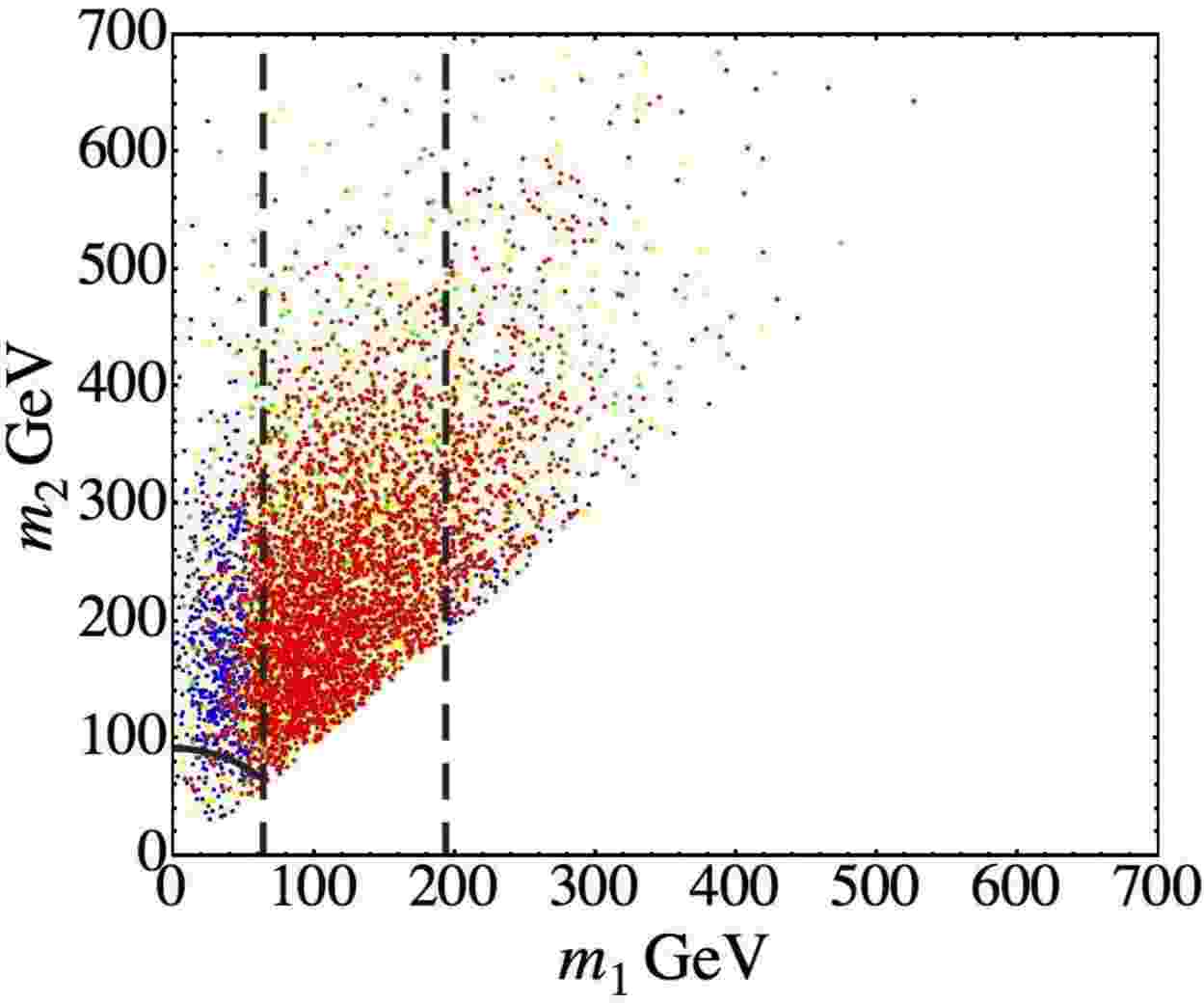}
\includegraphics[width=7cm]{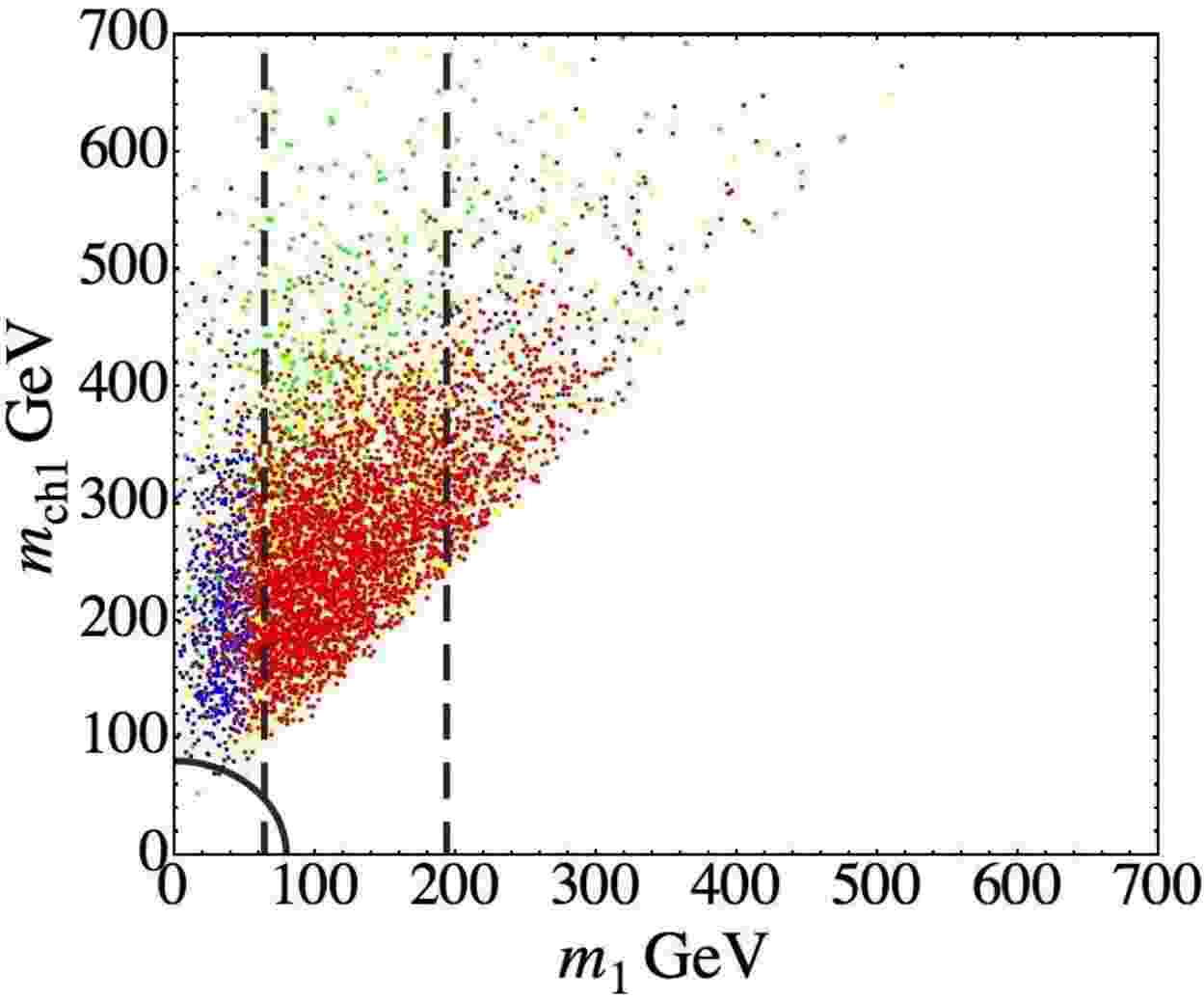}
\caption{CP conserving alignment $(v,v,v)$: \it the upper panels show the lightest neutral mass $m_1$ versus the second lightest neutral mass $m_2$ and the  lightest  charged one $m_{ch_1}$ respectively. The gray arc delimits the region below which the $Z$ ( $W$) decay channel opens. On the left plot the arc is only of $45^\circ$ because $m_2\geq m_1$. For points below the arc the $ Z$ ( $W$) decay may happens.  The dashed vertical lines indicates the approximated cuts that occur at $m_1\sim m_Z/\sqrt{2}$ and $m_1\sim 194 $ GeV arising by bounds on the decays of/into gauge bosons. Only the red points satisfy all the constraints, included thus arising by the oblique parameters $TSU$.}
\label{fig.vvv}
\end{figure}

We want to be as much as possible model independent. Thus we consider all the possible vacuum configurations $(v_1,v_2,v_3)$ of  the scalar potential  given in eq.(\ref{A4pot}). 
and we discuss all the  constraints related only  to the Higgs-gauge boson Lagrangian. We postpone the model dependent analysis to next section. The first model independent constraint comes from the partial wave unitarity for the neutral two-particle amplitudes, which puts upper bounds on the scalar masses\cite{Veltman:Unitarity}. Then  the light scalar mass region can be constrained considering the gauge boson decays. Moreover we may  put an upper bound on the lightest neutral state mass considering the Higgs decay channel $h\rightarrow W^+W^-$\cite{PDG2010,LEP2}. Finally the most stringent bounds arise by the oblique parameters $TSU$\cite{HT:TSUparameters}. The results for the alignment $(v,v,v)$ are shown in fig. \ref{fig.vvv}.

\begin{figure}[ht!]
  \centering
  \includegraphics[width=7cm]{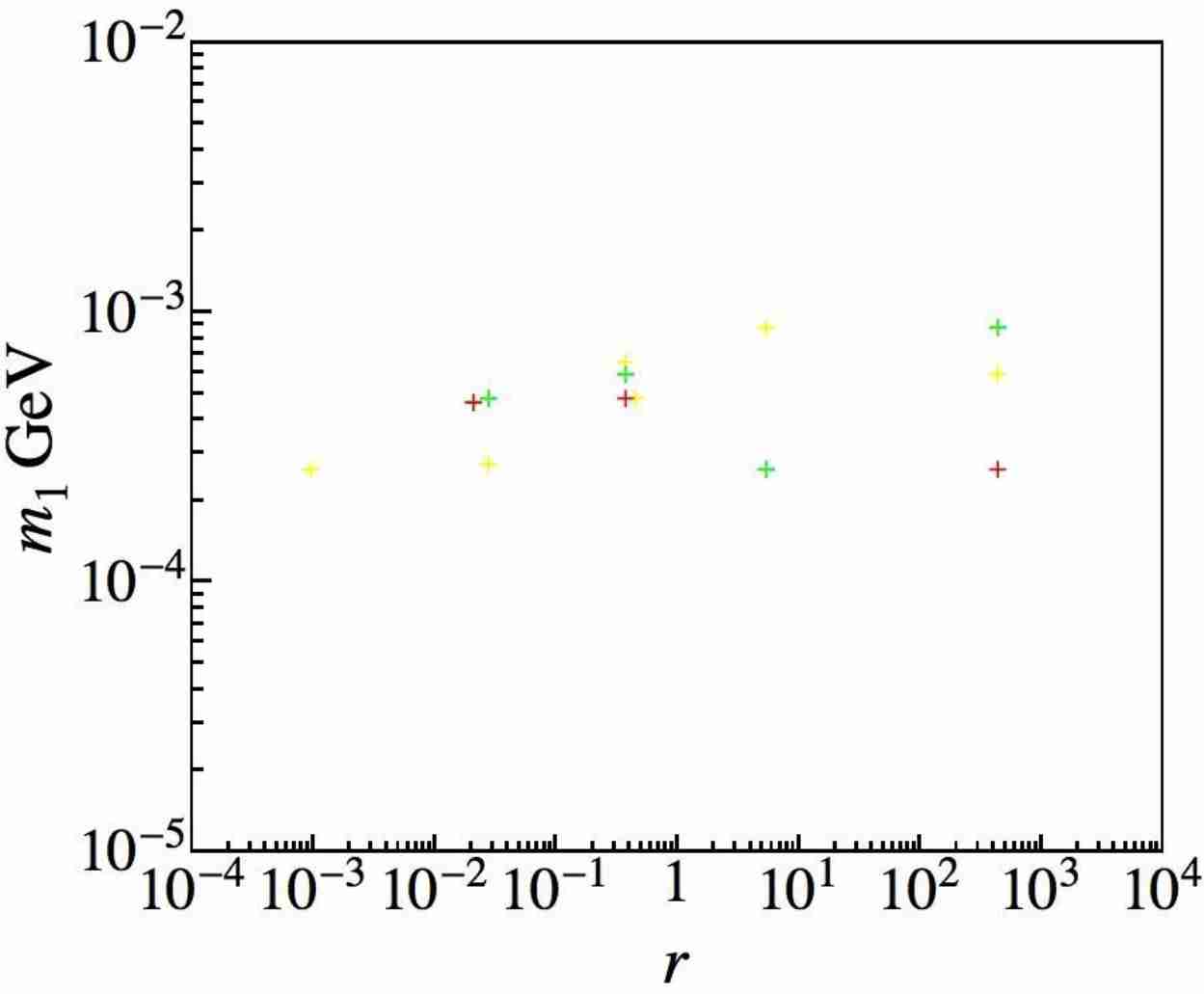}
   \includegraphics[width=7cm]{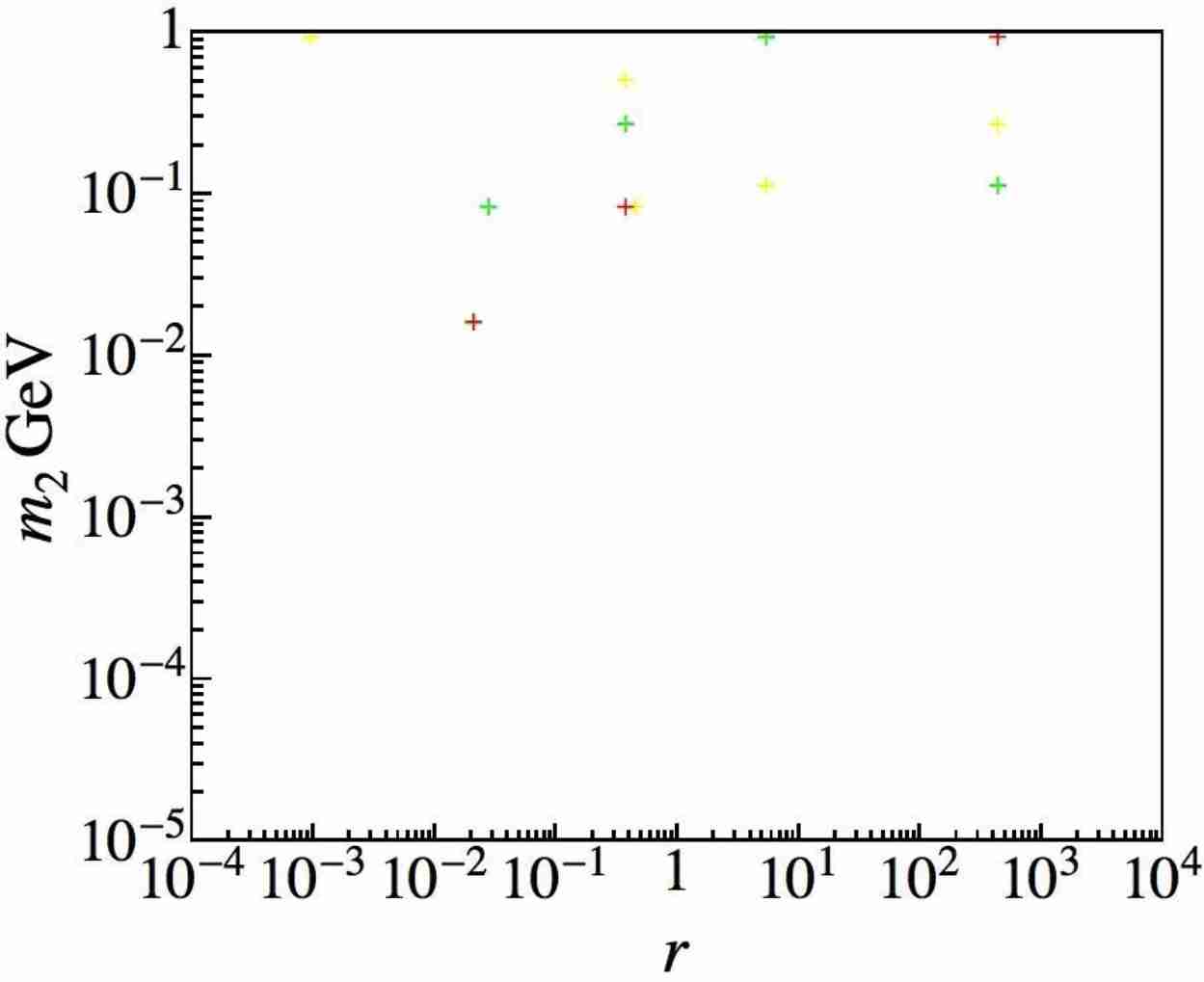}
    \caption{The  alignment $(v e^{i \omega_1}, v e^{-i \omega_1},r v)$: \it  the panels show $m_1$ (on the left) and $m_2$ (on the right) versus $r$. The number of points is small because no tachyonic masses are obtained only by a large tuning of the potential parameter. Moreover the order of magnitude of the masses is extremely low.}
  \label{fig.vvuomega2}
\end{figure}

Some configurations may be obtained only by tuning  the potential parameters, giving rise to scalar spectrums characterized by very light or even massless particles as shown in fig. \ref{fig.vvuomega2}. In this case the $A_4$ scalar potential as to be adequately modified including soft breaking terms in order to be phenomenologically acceptable.

\section{The Fermion Processes}

In this section we  discuss the constraints arising  by  the interactions between the new Higgs sector  and the SM fermions. When more than one Higgs boson couples to all the fermions, the fermion-Higgs interaction matrix generally is no longer diagonal and flavor violating processes can be mediated by the Higgs scalars and pseudoscalars. In this case, there are more channels available for rare fermion decays and meson oscillations. Experimental data place stringent bounds on the masses of the Higgses. We note that these bounds are dependent on details of the model, such as the $A_4$ representations of the fermions.

\begin{figure}[th!]
\begin{center}
\includegraphics[width=7cm]{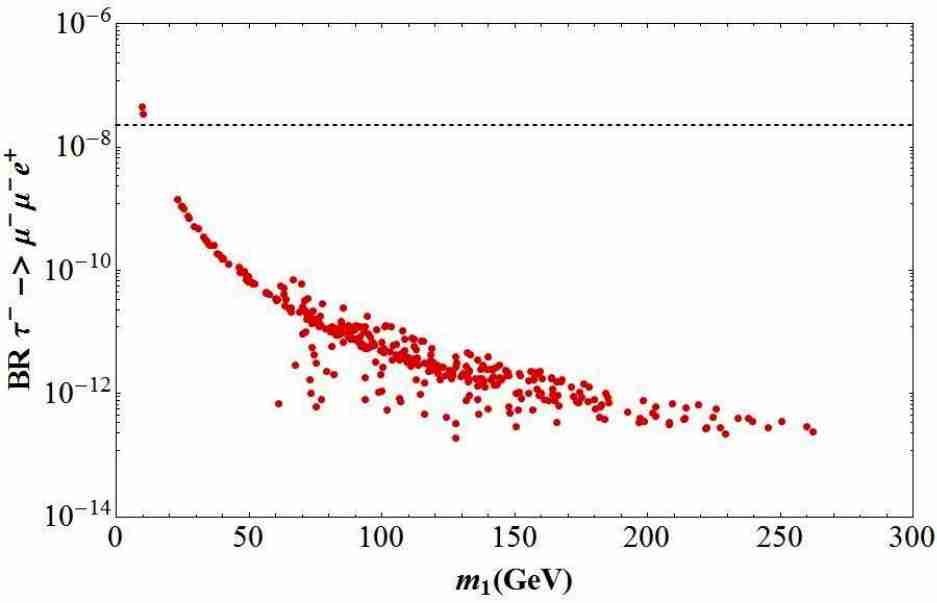}
\caption{\label{figMR} Model 1: \it the branching ratio for the decay $\tau^- \rightarrow \mu^- \mu^- e^+$ as a function of the smallest mass $m_1$. The horizontal line corresponds to the experimental upper bound.}
\end{center}
\end{figure}

\begin{figure}[th!]
\begin{center}
\includegraphics[width=7cm]{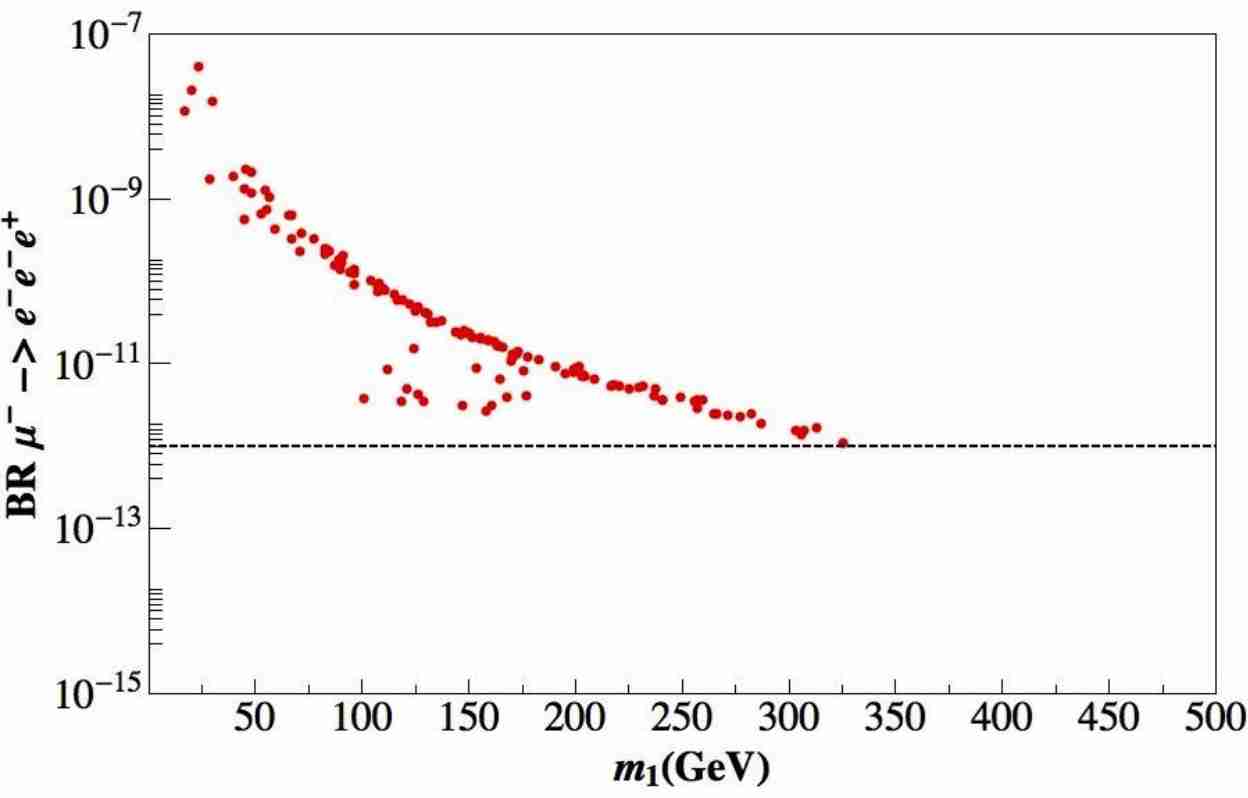}\hskip5.mm
\includegraphics[width=7cm]{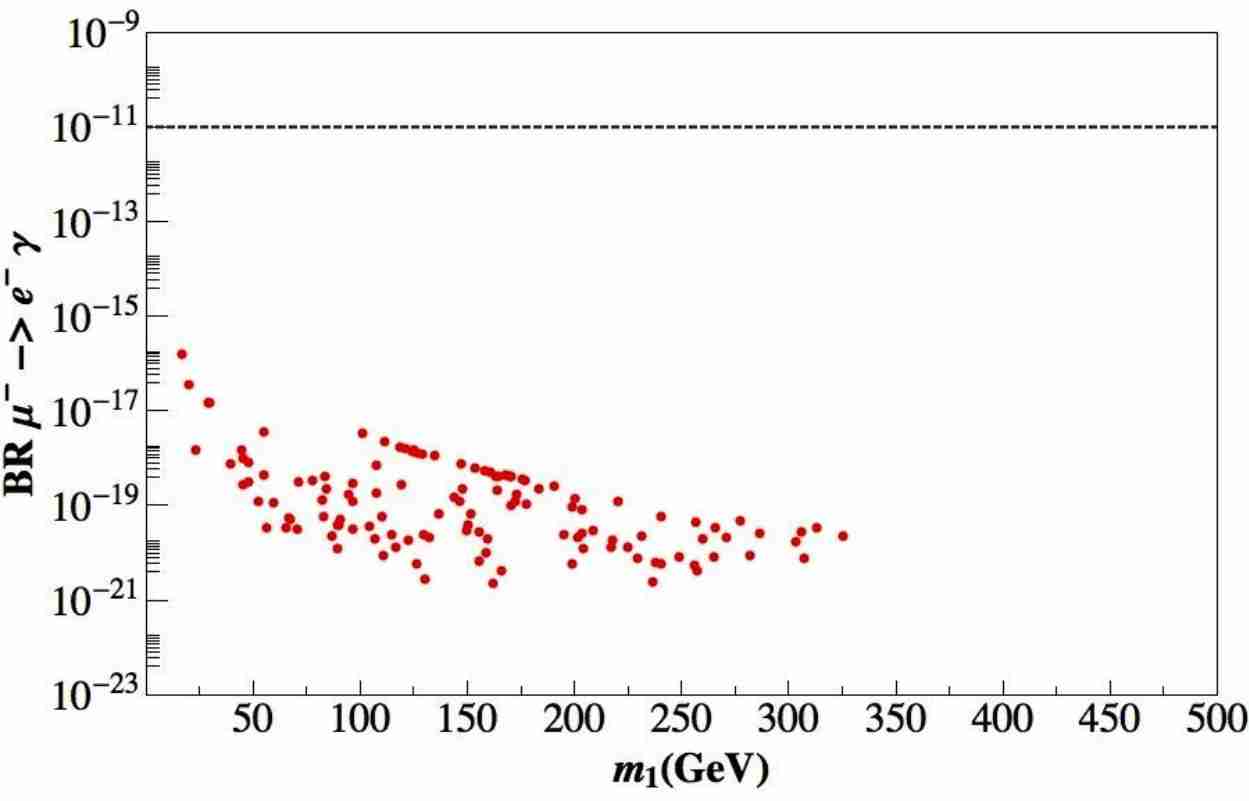}
\caption{\label{figMP}Model 2: \it on the left (right) side, the branching ratio of the decay of a $\mu^-\rightarrow e^-e^-e^+$ ($\mu^- \rightarrow e^- \gamma$) versus the lightest Higgs mass. The horizontal band is the experimental limit \cite{PDG2010}.}
\end{center}
\end{figure}

\begin{center}
\begin{figure}[th!]
\includegraphics[width=7cm]{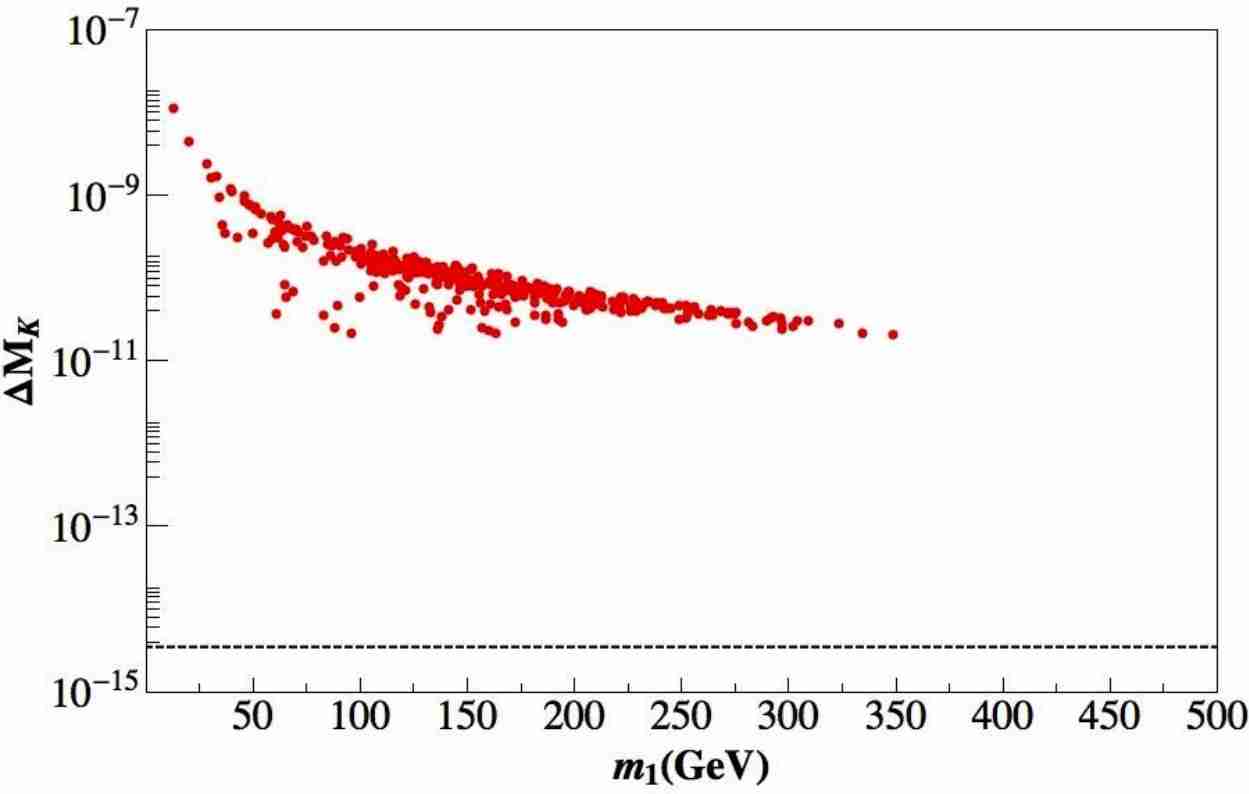}
\includegraphics[width=7cm]{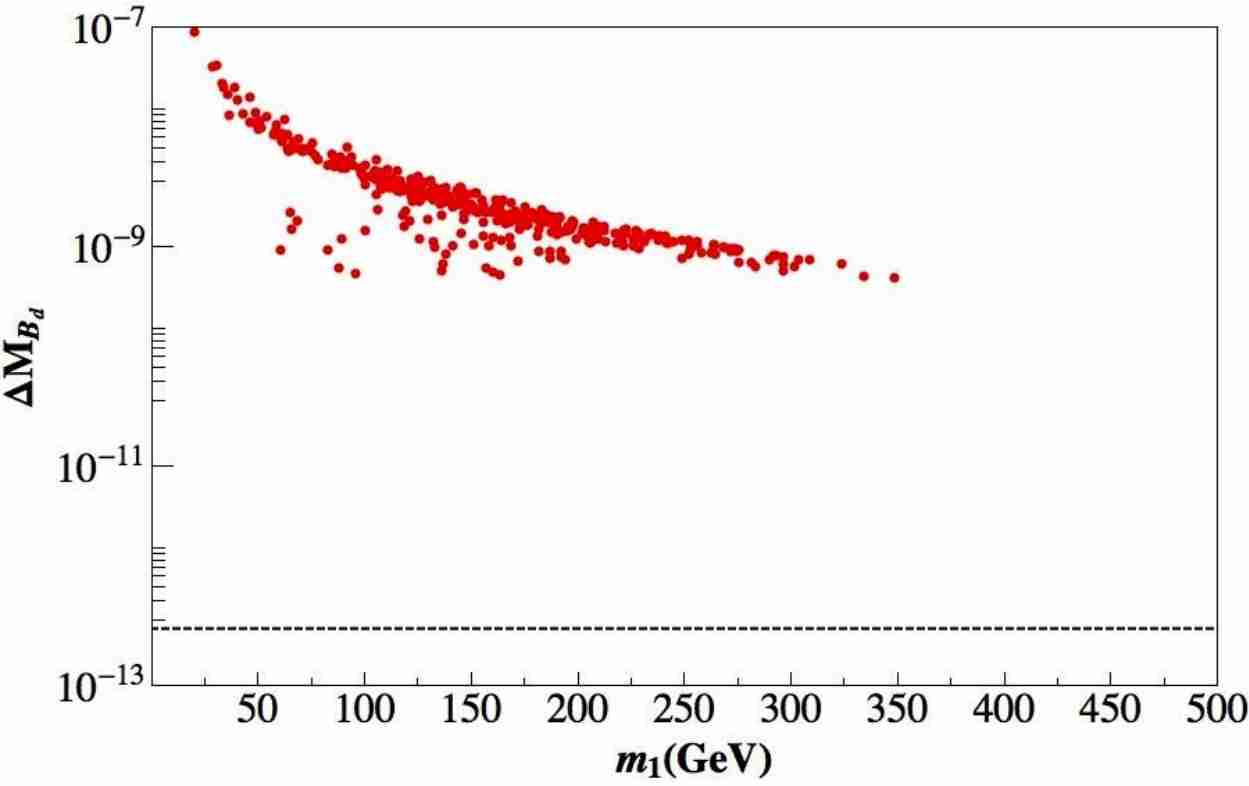}
\caption{\label{figLK}Model 3: \it $\Delta M_{K}$ and $\Delta M_{B_d}$  mass splittings versus the lightest Higgs mass. The horizontal lines correspond to the experimental values as reported in \cite{BCGI:HiggsFCNC}.}
\end{figure}
\end{center}

\begin{center}
\begin{figure}[th!]
\includegraphics[width=7cm]{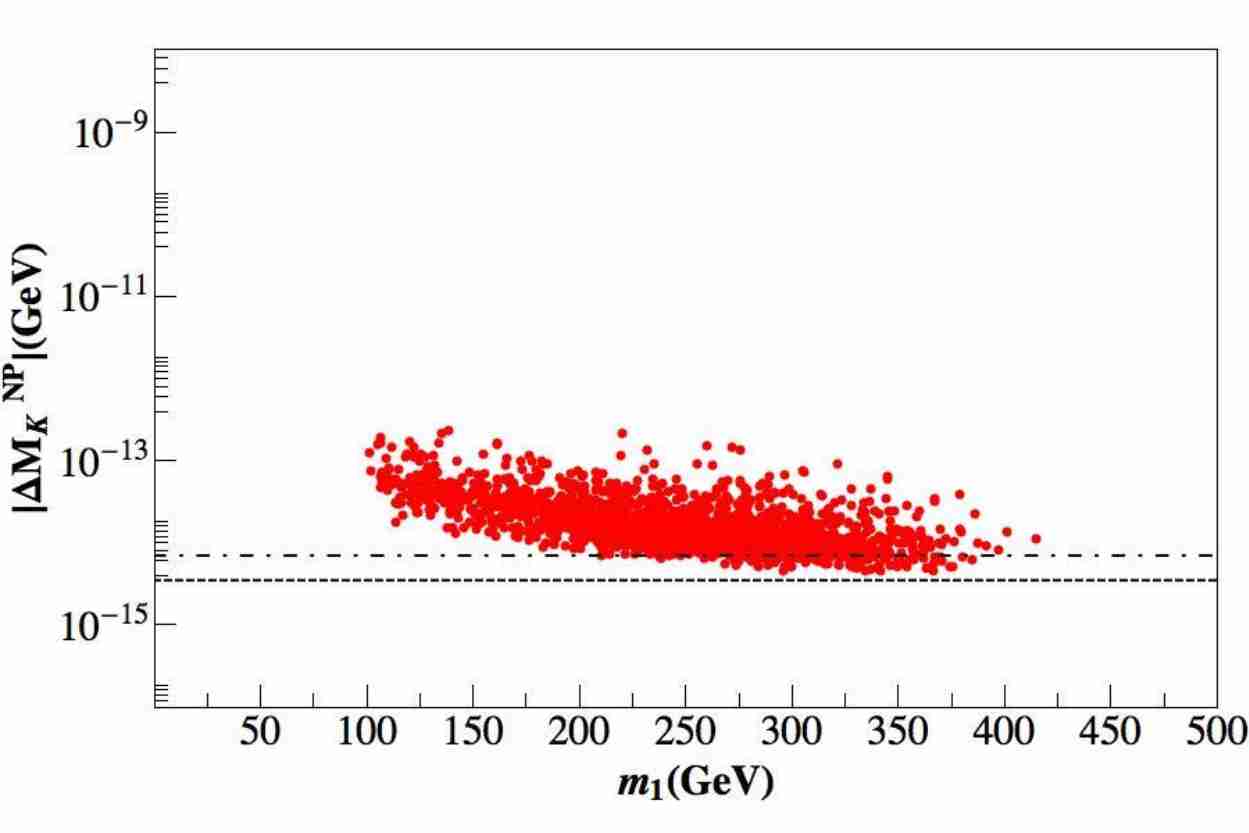}
\includegraphics[width=7cm]{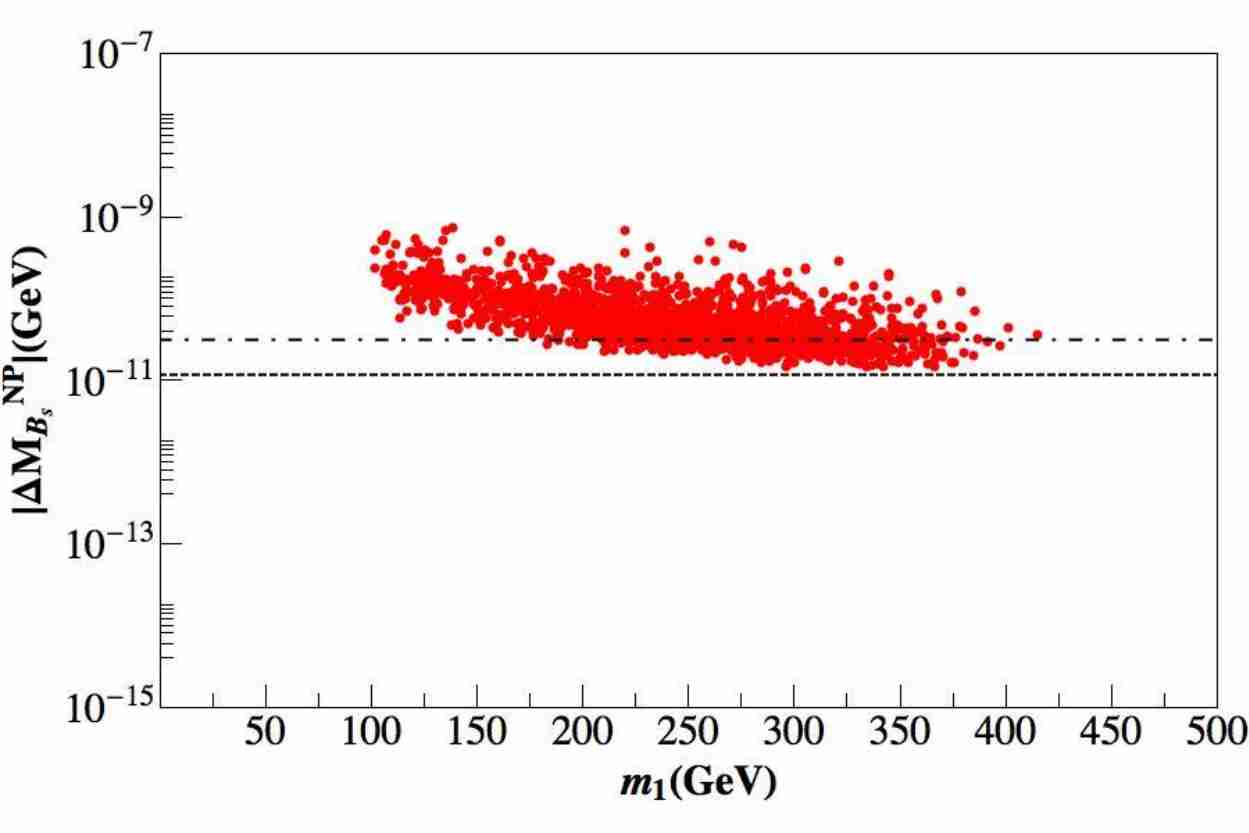}
\caption{\label{figDM}Model 4: \it $K$ and $B_s$ mesons oscillations: the contribution of the new physics to $\Delta M_{K}$  and to $\Delta M_{B_s}$ as function of the lightest Higgs mass. The short-dashed line  indicates the current experimental value \cite{BCGI:HiggsFCNC} that is rather well described by the Standard Model box diagrams. Naively, this is interpreted as an exclusion of the model, which is true in most of parameter space, but not in points where the Standard Model and new physics contributions partially cancel. These values  correspond to the longer-dashed lines. We see that a small number of points is allowed by the data under the assumption of partial cancellation.}
\end{figure}
\end{center}

We consider four model: the model of Ma and Rajasekaran (Model 1) \cite{MR:A4EWscale}, the model of Morisi and Peinado  (Model 2) \cite{MP:A4EWscale}, the model of  Lavoura and Kuhbock (Model 3)\cite{LK:A4EWscale} and the Discrete DM  model (Model 4) \cite{HMPV:DiscreteDM,MMP:DarkMatterA4,Toorop:2011ad}. It turns out that Model 1 is quite robust under the constraints from the Higgs sector and flavor violating processes as can be seen in fig.\ref{figMR}, while the Models  2-3-4 are strongly affected by the experimental constraints, as can be seen in  figs.\ref{figMP}-\ref{figLK}-\ref{figDM}.

\section{Conclusions}

Low scale flavor symmetry allows a unified description of flavor and electroweak symmetry breaking.  This realization is extremely   challenging  but 
 a deep and careful analysis of the phenomenology of this class of  flavor models is fundamental to test their validity beyond the
prediction of the mixing patterns and in addition it  is a powerful tool to discriminate among them.

%% file: Author/Leser.tex
{\bf Abstract}\\
\vskip5.mm
Discrete symmetries employed to explain neutrino mixing and mass hierarchies are often associated with an enlarged scalar sector which might lead to exotic Higgs decay modes. We explore such a possibility in a scenario with $S_3$ flavor symmetry which requires three scalar $SU(2)$ doublets. The spectrum is fixed by minimizing the scalar potential, and we observe that the symmetry of the model leads to tantalizing Higgs decay models potentially observable at the CERN Large Hadron Collider (LHC).

\vskip5.mm

\setcounter{footnote}{0}
\renewcommand{\thefootnote}{\arabic{footnote}}

\section{Introduction}
The permutation group $S_3$ is an attractive candidate to address the flavor puzzle. It was
introduced in \cite{Pakvasa:1977in} and explored further in the article this work is based upon
\cite{Chen:2004rr}.  In this contribution we study the exciting
prospect that such flavor models can predict {\em enlarged Higgs
  sectors with non-standard couplings to fermions and gauge bosons}. A full list of references can be found in the article this contribution is based upon\cite{Bhattacharyya:2010hp}.

The motivation for choosing $S_3$ is that it is the smallest
non-abelian discrete symmetry group that contains a
\mbox{2-dimensional} irreducible representation which can connect two
maximally mixed generations.  It has three irreducible representations:
$\mathbf{1},\mathbf{1'}$ and $\mathbf{2}$, with multiplication rules:
$\mathbf{2}\times\mathbf{2}=\mathbf{1}+\mathbf{1'}+\mathbf{2}$ and
$\mathbf{1'}\times \mathbf{1'}=\mathbf{1}$.  Besides facilitating
maximal mixing through its doublet representation, $S_3$ provides two
inequivalent singlet representations which play a crucial role in
reproducing fermion masses and mixing. To accomplish the latter, three
scalar $SU(2)$ doublets are introduced, which couple to the fermions
as dictated by $S_3$ symmetry. It so happens that large mixing among
up- and down-type quarks cancel each other in the Cabibbo
matrix. Neutrino masses are separately generated by a type-II see-saw
mechanism using scalar $SU(2)$ triplets, so
that the mismatch between the large mixing of the charged leptons and
the diagonal neutrino masses translates directly into the
Pontecorvo-Maki-Nakagawa-Sakata matrix. In this paper we do not deal
with those triplets, but explore the following avenues: ($i$) the mass spectrum and mixing of the scalars ($ii$) the gauge and Yukawa interactions of the neutral
scalars, and ($iii$) different nonstandard production and decay modes
of the neutral CP-even scalars leading to the possibility of their
detection at the LHC.

For definiteness, we study the $S_3$ model pursued in
\cite{Chen:2004rr} to explain the leptonic flavor structure.  We
concentrate on the complementary aspects by exploring the scalar
sector.  The assignments of the fermion and scalar fields are as
follows:
\begin{align}
	\label{eq:multipletassignments}
	(L_\mu,L_\tau)&\in \mathbf{2} & L_e, e^c,\mu^c
&\in \mathbf{1}
        & \tau^c&\in \mathbf{1'} \, , \nonumber \\
	(Q_2,Q_3)&\in\mathbf{2} & Q_1, u^c,c^c,d^c,s^c
&\in\mathbf{1}
        & b^c,t^c&\in\mathbf{1'} \, , \\
	(\phi_1,\phi_2)&\in \mathbf{2} & \phi_3&\in\mathbf{1} \, \nonumber, 
\end{align}
where the notations are standard and self-explanatory.  The vacuum
expectation values (VEVs) of the three scalar doublets $\phi_{1,2,3}$
induce spontaneous electroweak symmetry breaking (SSB).

\section{Scalar potential and  spectrum} 
We use the most general $S_3$ invariant scalar potential involving three scalar doublet fields given by \cite{Chen:2004rr}. It depends on two mass-like parameters $m,m_3$ and eight couplings $\lambda_{\{1,\ldots,8\}}$. After SSB, nine degrees of freedom are left: three neutral scalars,
two neutral pseudoscalars and two charged scalars with two degrees of
freedom each.  We denote the VEVs of $\phi_i$ by $v_i$ and assume the
$\lambda_i$'s to be real.  For the purpose of generating maximal
mixing in the lepton sector, we choose the vacuum alignment $v_1 = v_2
= v$. To ensure that the chosen vacuum alignment actually corresponds to a
minimum of the potential, we adjust parameters to make sure that the
determinant of the Hessian matrix is positive. In this
case, the function is the scalar potential and the Hessian is just the
mass matrix of the scalars.  The positivity of the eigenvalues guarantees that the potential
is minimized. 

We now set out to find the spectrum of the three CP-even neutral
scalars.  We insert the expansion $\phi_i^0=v_i+h_i$ into the scalar potential to obtain the mass matrix. After its
diagonalization the weak basis scalars $h_{1,2,3}$ are expressed in
terms of the physical scalars $h_{a,b,c}$ as
\begin{align}
	\label{eqn:scalarmixing}
	h_1&= U_{1b}~h_b + U_{1c}~h_c - \frac{1}{\sqrt{2}} ~h_a \, , 
        \nonumber\\
	h_2&= U_{2b}~h_b + U_{2c}~h_c + \frac{1}{\sqrt{2}} ~h_a \, ,\\
	h_3&= U_{3b}~h_b + U_{3c}~h_c \, , \nonumber 
\end{align}
where $U_{ib}$ and $U_{ic}$ are analytically tractable but complicated
functions of $\lambda_i$s, $v$ and $v_3$, which we do not display.
The condition $v_1 = v_2$ immediately leads to $U_{1b}=U_{2b}$ and
$U_{1c}=U_{2c}$.  The masses $m_{a,b,c}$ of the three CP-even neutral scalars are a result of the diagonalization of the mass matrix.

A few things are worth noting at this stage:
$(i)$ Since $\phi_{1,2,3}$ are all weak $SU(2)$ doublets, their VEVs
  are related as: $2 v^2 + v_3^2 = v_{\text{SM}}^2$, where $v_{\text{SM}} \approx 246~
  \text{GeV}$.

$(ii)$ One of the physical scalars is given by $h_a=(h_2-h_1)/\sqrt{2}$,
  i.e. there is no dependence on $\lambda_{\{1,\ldots,8\}}$ or on the
  VEVs. 

$(iii)$ We randomly vary the {\em
    magnitude} of the $\lambda_i$'s in the range $[0,1]$, although
  slightly larger (but $<4\pi$) values of $|\lambda_i|$ would have
  still kept the couplings perturbative.  We {\em accept} a given set
  of $\{\lambda_1, \ldots, \lambda_8, v\}$ only if it satisfies the
  minimization conditions.

\begin{figure}[htb]
	\begin{center}
		\includegraphics[width=5cm]{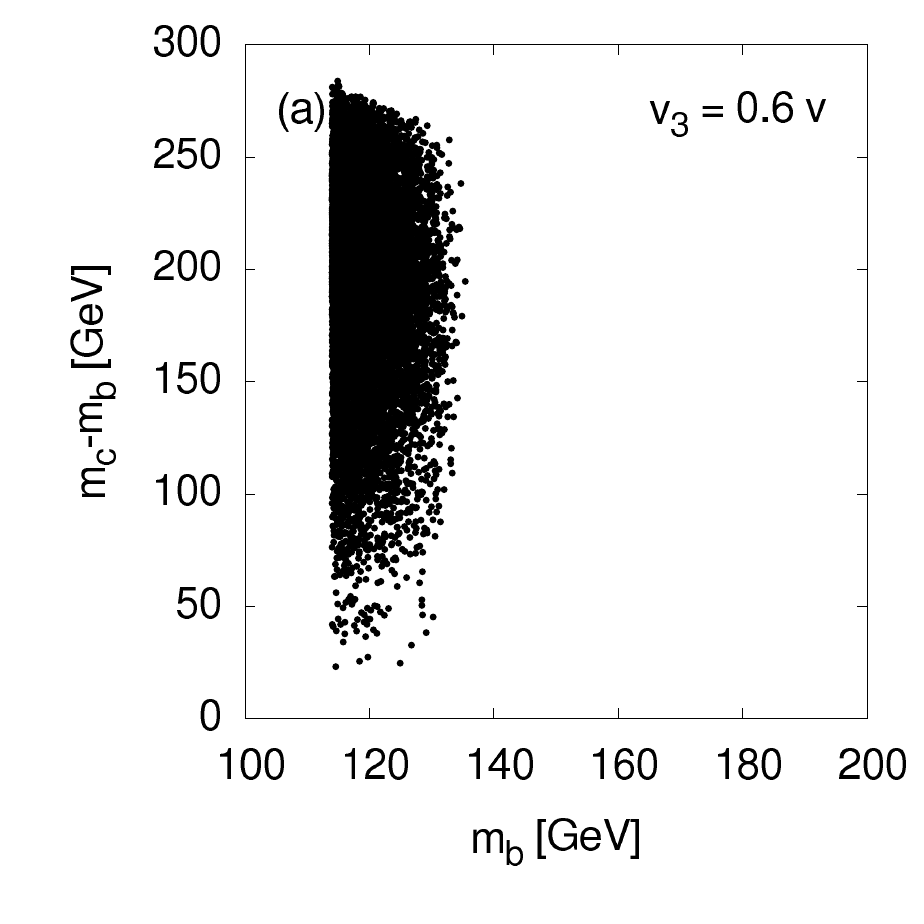}
			\includegraphics[width=5cm]{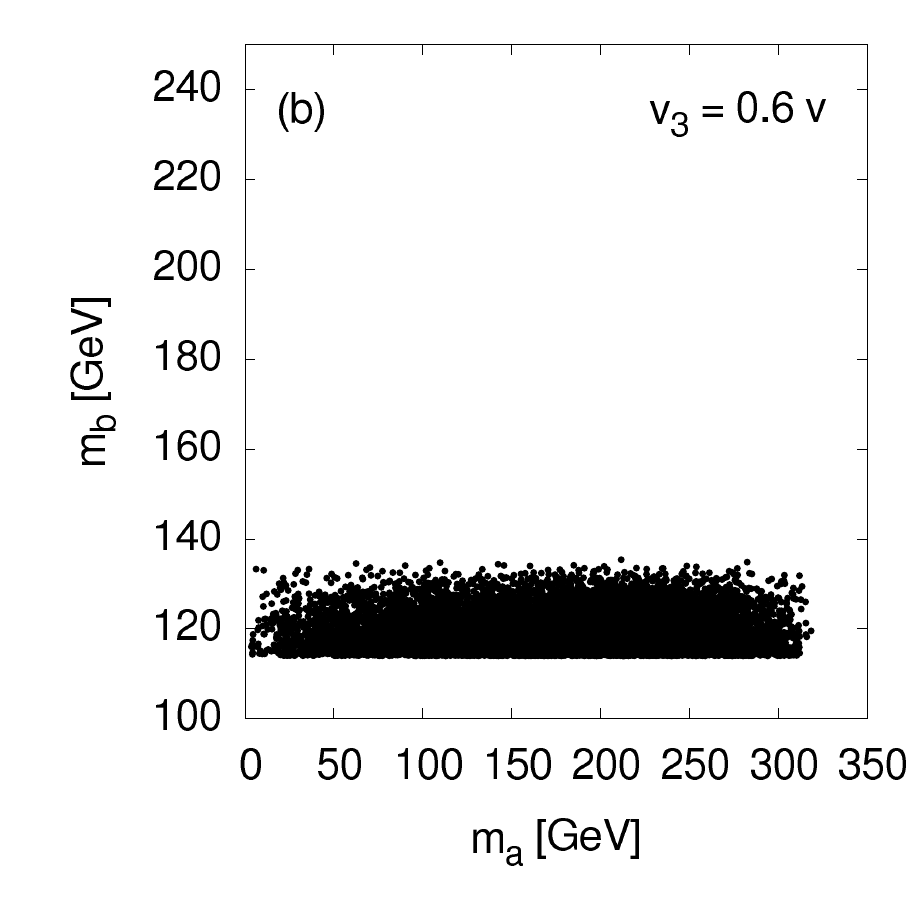}
                        \caption{\small{\sf Results of a random search for
                            allowed scalar masses for a fixed $v_3/v=0.6$. In
                            the left panel (a), we exhibit the splitting
                            $(m_c-m_b)$ for different choices of $m_b$. In the
                            right panel (b), we show the allowed range of
                            $m_a$.}}
	\label{fig:scalar_massrelations}
\end{center}
\end{figure}

$(iv)$ It has been suggested in \cite{Chen:2004rr} that with
  {\em order one} Yukawa couplings, the ratio $v_3/v \sim 0.1$
  reproduces the correct Cabibbo angle in the quark sector. We require
  $v_3/v \geq 0.6$ to ensure that $m_b^2$ stays above the accepted
  limit.  Since $h_b$ and $h_c$ have similar gauge and Yukawa
  properties, quite different from those of $h_a$ (see discussions
  later), we show the mass splitting $(m_c - m_b)$ against $m_b$ in
  Fig.~\ref{fig:scalar_massrelations}(a), and the relation between
  $m_b$ and $m_a$ in Fig.~\ref{fig:scalar_massrelations}(b).

$(v)$ The two CP-odd neutral scalars $\chi_a$ and $\chi_b$ can be light, with $\chi_a$ being lighter than $100$\,GeV and $\chi_b$ being lighter than $250$\,GeV. The charged scalars $h_a^+$ and $h_b^+$ are less restricted: They can be lighter than $250$\,GeV and $300$\,GeV respectively.

\section{Scalar couplings to gauge and matter fields}
The couplings of the symmetry basis $h_i$ to $W^\pm$ and $Z$ are modified by a factor of $v_{i}/v_\text{SM} < 1$
compared to their SM expressions.  In terms of the mass basis scalars,
we observe the following: ($i$) The coupling of $h_b$ to $W^+W^-$ (or,
$ZZ$) is the corresponding SM coupling multiplied by $\left(2v U_{1b}
  + v_3 U_{3b}\right)/v_\text{SM}$ and the corresponding factor for
$h_c$ is $\left(2v U_{1c} + v_3 U_{3c}\right)/v_\text{SM}$.  ($ii$)
The scalar $h_a$ does not have $h_aZZ$ or $h_aWW$ couplings, unlike
the other two scalars. This can be understood as follows. The gauge
couplings of $h_i$ arise from the linear expansion $\phi_i^0 = v_i +
h_i$ in the kinetic term. Since $v_1 = v_2 = v$, the combination
$(h_1+h_2)$ will couple to gauge bosons as proportional to $v$. The
orthogonal combination $(h_2 - h_1)$ that represents the physical
scalar $h_a$ -- see Eq.~(\ref{eqn:scalarmixing}) and point $ii$
following it -- will not have the usual
scalar-gauge-gauge vertex. The four-point $h_a^2ZZ$ and $h_a^2WW$
couplings will, however, exist.

The couplings of $h_{b,c}$ to the quarks and leptons depend on the parameters
$v,v_3, \lambda_i$ and $f_i$ (or, $g_i^{u/d}$), while the couplings of $h_a$
to fermions depend only on $f_i$ (or, $g_i^{u/d}$).  The {\em physical}
scalar couplings to the {\em mass basis} fermions are given by the following Yukawa
matrices, displayed for the charged leptons as an example (the structures for
the quark sector are similar {\em modulo} Cabibbo mixing):
\begin{align}
	\label{eqn:yukawas}
	Y_{h_a} &= 
	\begin{pmatrix}
 0        &	0     &Y_{e_L\tau_R}^a \\
 0        &	0    &Y_{\mu_L\tau_R}^a \\
 Y_{\tau_L e_R}^a &	Y_{\tau_L\mu_R}^a&0        \\ 
	\end{pmatrix}, & 
	Y_{h_{b,c}} &= 
	\begin{pmatrix}
 Y_{e_Le_R}^{b,c}        &	Y_{e_L\mu_R}^{b,c}     &0 \\
 Y_{\mu_L e_R}^{b,c}        &	Y_{\mu_L\mu_R}^{b,c}    &0 \\
 0 &	0&Y_{\tau_L\tau_R}^{b,c}        \\ 
	\end{pmatrix} \, .
\end{align}
The position of the zeros in the matrices deserves some attention.  It turns
out that $h_{a,b,c}$ have off-diagonal fermion couplings at tree level due to
the absence of any natural flavor conservation. $h_a$ couples {\em only} off-diagonally and one of the two
  fermions has to be from the third generation. $h_{b,c}$ couple diagonally as in the SM, {\em but also} possess
  small, numerically insignificant, off-diagonal couplings involving the first
  two generations.  

In the present case, $S_3$ symmetry, under which both scalars and
fermions transform nontrivially, is instrumental in suppressing the
off-diagonal couplings. To provide intuitive understanding, we take,
as an example, only the two-flavor $\mu$--$\tau$ sector together with
two neutral scalars $h_1$ and $h_2$. It is not difficult to see that
the combination $(h_2-h_1)$, which corresponds to $h_a$, couples only
off-diagonally, as mentioned earlier. But the other combination
$(h_2+h_1)$, which corresponds to $h_{b,c}$ following
Eq.~(\ref{eqn:scalarmixing}), couples only diagonally to physical
$\mu$ or $\tau$. When we consider the quark sector, $\mu$ and $\tau$
would be replaced by second and third generation quarks which will
have CKM mixing.  This will yield off-diagonal entries for $h_{b,c}$
couplings to quarks suppressed by the off-diagonal CKM elements.  The
same happens for off-diagonal couplings involving the first two
generations as well. In some
setups where the fermion transformations under $S_3$ are not
appropriately adjusted, the off-diagonal Yukawa couplings may become
order one which induce sizable neutral scalar mediated rare processes,
like $K_L \to \mu e$ or $K_L \to 2\pi$, at tree level.  This requires
those neutral scalars to lie beyond several TeV. But in our case, once we adjust
the couplings to reproduce the fermion masses and mixing, the
off-diagonal Yukawa couplings are determined too.  The largest of them
corresponds to $\bar{c}_L t_R h_a$, which is about 0.8. The second
largest off-diagonal coupling is that for $\bar{s}_L b_R h_a$, and is
about 0.02. The next in line is $\bar{\mu}_L \tau_R h_a$, whose
coefficient is about 0.008.  The others are orders of magnitude
smaller, and are of no numerical significance. Although FCNC processes
like $B_d$--$\bar{B}_d$ and $B_s$--$\bar{B}_s$ mixings proceed at tree
level, the contributions are adequately suppressed even for light
scalar mediators.

\section{Collider signatures}
The perturbativity condition $\left|\lambda_i\right| \leq 1$ and the
requirement $m_{b/c} \geq 114$\,GeV (for which we set $v_3/v \simeq
0.6$) yields $m_b$ in the neighbourhood of 120\,GeV and $m_c$ within
400\,GeV -- see the scatter plots in
Fig.~\ref{fig:scalar_massrelations}.  Both $h_b$ and $h_c$ would decay
into the {\em usual} $ZZ$, $WW$, $b\bar b$, $\gamma\gamma$, $\cdots$
modes, but the dominant decay mode of $h_b$ (or $h_c$) for the case of
$m_a< m_b/2$ (or $m_a < m_c/2$) would be into $h_a h_a$. Recall that
the existing limits on the Higgs mass depend crucially on the gauge
coupling of the Higgs.  Since $h_aZZ$ or $h_aWW$ couplings are
nonexistent, the mass of $h_a$ is unconstrained, i.e. $m_a$ can be
lower than 114\,GeV or larger than 200\,GeV.  We numerically calculate
the strength of the $h_bh_ah_a$ coupling from the set of {\em
  acceptable} parameters characterizing the potential, and introduce a
parameter $k$ which is the ratio of the $h_bh_ah_a$ coupling and the
$h_bWW$ coupling. The magnitude of $k$ depends on the choice of
$\lambda_i$ and $v_3$. Assuming $m_a = 50$\,GeV, we obtain $k$ in the
range of $(5-30)$. Just to compare with a 2HDM
for illustration, the corresponding $k$ value, when the heavier Higgs
weighing around 400\,GeV decays into two lighter Higgs weighing
114\,GeV each, is about 10.

\begin{figure}[tb]
\begin{center}
\includegraphics[width=5cm]{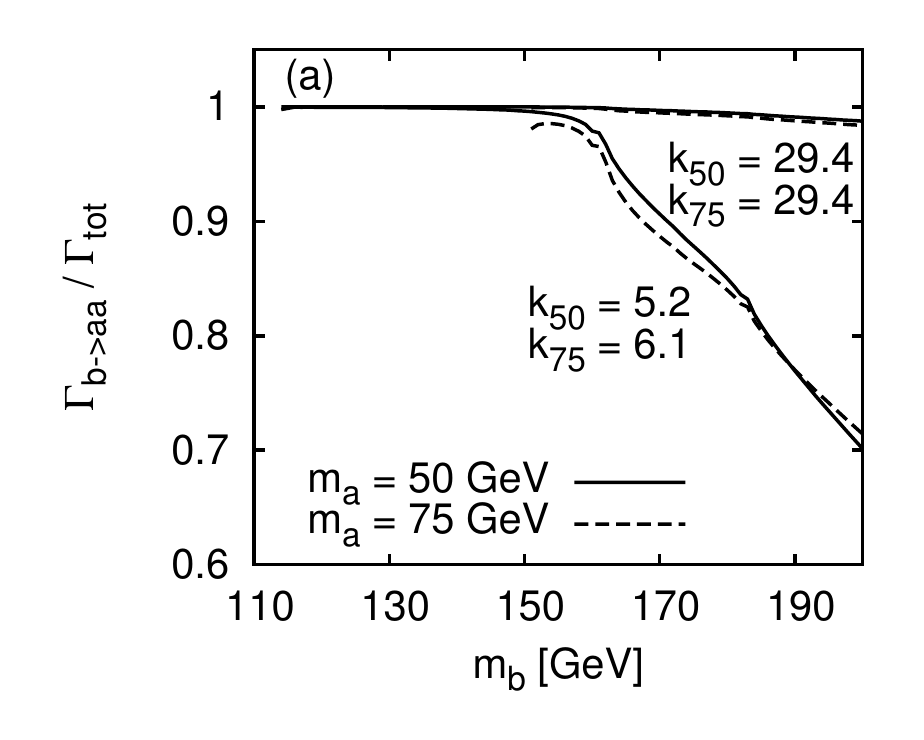}
\includegraphics[width=5cm]{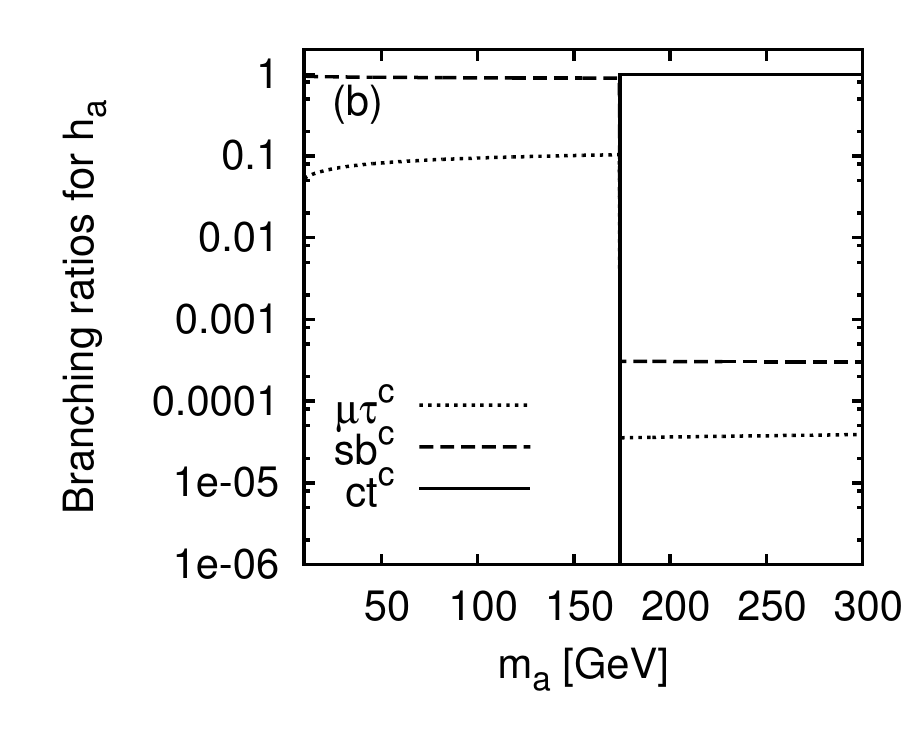}
\includegraphics[width=5cm]{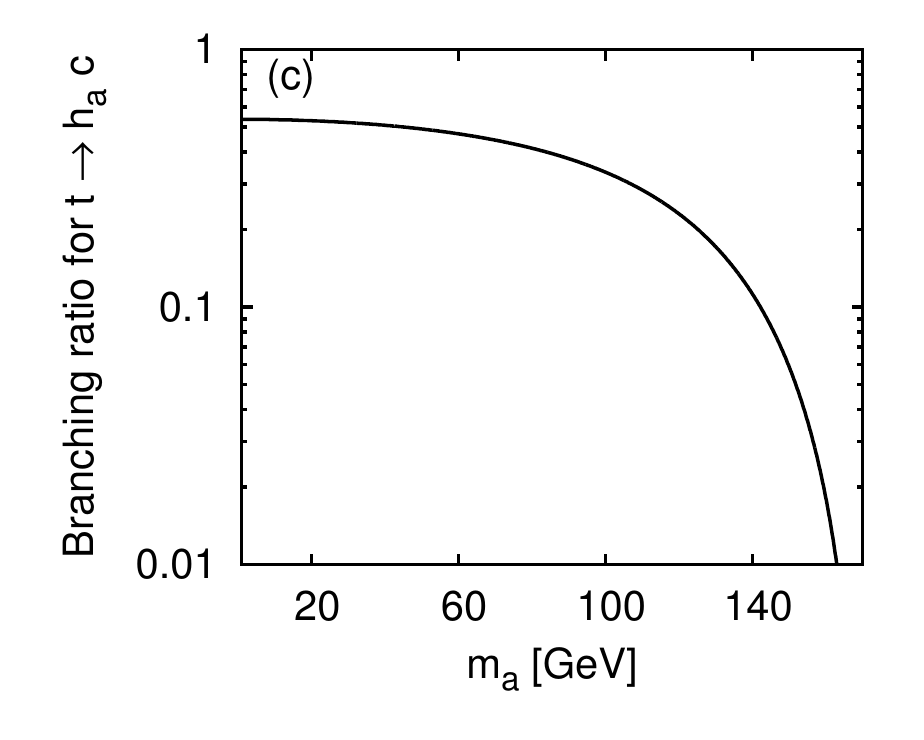}
\caption{\small{\sf (a) Branching ratio of $h_b\to h_ah_a$ for two
    representative values of $m_a$, and in each case for smallest / largest
    values of $k$ in the set of accepted scalar parameters which compares the
    strength of the $h_bh_ah_a$ coupling to the strength of $h_bWW$ coupling;
    (b) branching ratios for the decay of $h_a$, and (c) branching ratio of
    the top quark decay into $h_a$ and charm quark.}}
\label{fig:decays}
\end{center}
\end{figure}

In Fig.~{\ref{fig:decays}}(a) we have plotted the branching ratio of
$h_b \to h_a h_a$ as a function of $m_b$ for two representative values
$m_a = 50$,$75$\,GeV, and for $k \sim$ 5 and 30, which correspond to
the smallest and largest $k$ obtained from the set of
\textit{accepted} scalar parameters.  We observe that till the $WW$ or
$ZZ$ decay modes open up, the branching ratio $h_b \to h_ah_a$ is
almost 100\%. 

As Fig.~\ref{fig:decays}(b) suggests, as long as $m_a < m_t$, $h_a$
will dominantly decay into jets, and one of them can be identified as
the $b$-jet.  The branching ratio of $h_a \to \mu\bar\tau$ is,
nevertheless, not negligible (about 0.1). As shown in
Fig.~\ref{fig:decays}(c), for $m_a \ll m_t$, the branching ratio of $t
\to h_a c$ is quite sizable, which decreases with increasing $m_a$. It may
be possible to reconstruct $h_a$ from $h_a \to \mu \bar\tau$.  In
fact, a light $h_a$ would be copiously produced from the top decay at
the LHC.  On the other hand, if $m_a > m_t$, as can be seen again from
Fig.~\ref{fig:decays}(b), $h_a$ decays to $t\bar c$ with an almost
100\% branching ratio.

If $k$ is large, then there is an interesting twist to the failed
Higgs search at LEP-2. In this case, $h_b \to h_a h_a$ would overwhelm
$h_b \to b\bar b$, and hence the conventional search for the SM-like
scalar ($h_b$, as the lighter between $h_b$ and $h_c$) would fail.
This is similar to what happens in the next-to-minimal supersymmetric
models, when the lightest scalar would dominantly decay into two
pseudoscalars, and each pseudoscalar would then decay into $2b$ or
$2\tau$ final states. In view of these possible $4b$ or $4\tau$ Higgs
signals, LEP data have been reanalyzed putting constraints on the
Higgs production cross section times the decay branching ratios.  The possibility of the Higgs
cascade decays into $4j$ ($j=$ quark/gluon), $2j+2$ photons and 4
photons has been studied too. From a
study of $4b$ final states, a limit $m_h > 110$ GeV (for a SM-like
Higgs) has been obtained (references can be found in\cite{Bhattacharyya:2010hp}).  From all other cascade
decays the limit on $m_h$ will be considerably weaker.  Our $h_a$ has
the special feature that it has only off-diagonal Yukawa couplings
involving one third-family fermion. If $h_b$ is lighter than the top
quark, it would decay as $h_b \to h_a h_a \to 2b+2j$, and into
$b+1j+\mu+\tau$, the latter constituting a spectacular signal with two
different lepton flavors $\mu$ and $\tau$. The standard $2b$ and
cascade $4b$ decay searches are not sensitive to our final states, and
so a value of $m_b$ much lighter than 110 GeV is not ruled out.

\section{Conclusions}
The discrete flavor symmetry $S_3$, besides successfully reproducing
fermion masses and mixing, provides an extended Higgs sector having
unconventional decay properties. We assume all the couplings to be
real, and do not deal with the possibility of CP violation in this
paper.  The potential has been minimized requiring maximal mixing for
the atmospheric neutrinos. In our setup, there are two scalars which
are SM Higgs like, {\em except} that each of them can have a dominant
decay into the third ($h_{b,c} \to h_a h_a$). The latter, i.e. $h_a$,
has no $h_a VV$-type gauge interactions, and has {\em only} flavor
off-diagonal Yukawa couplings with one fermion from the third
generation. It is not unlikely that by evading the conventional search
strategies, both $h_b$ and $h_a$ are already buried in the existing
LEP and Tevatron data.

%% file: Author/Blankenburg.tex
{\bf Abstract}\\
\vskip5.mm
 We make a general study of $SO(10)$ models with type-II see-saw dominance and show that an excellent fit can be obtained for fermion masses and mixings, also in comparison with other realistic $SO(10)$ models.

\vskip5.mm
\section{Introduction}
In the last twenty years we achieved a rather precise knowledge of the leptonic mixing angles, which, within the experimental accuracy, are consistent with the Tri-Bimaximal (TB) pattern \cite{tbm} and, as such, are very different from the quark mixing angles. In fact the quark flavour structure is characterized by hierarchical masses and small mixing angles, while the lepton sector presents a milder hierarchy in the neutrino masses and two large and one small mixing anlges.

It is well known that with the see-saw mechanism the very small neutrino masses point to a very high energy theory of lepton flavour, such as a Grand Unified Theory (GUT).
In particular in this context, among the possible unified groups, $SO(10)$ is very interesting because the right-handed neutrinos are naturally introduced and are not gauge singlets, unlike the Standard Model or $SU(5)$. 
A still open and challenging problem is that of formulating a natural $SO(10)$ Grand Unified model leading to a good description of quark masses and mixing and, at the same time, with a TB lepton mixing structure built-in in a well defined first approximation, due, for example, to an underlying (broken) flavour symmetry.
In $SO(10)$ the main added difficulty with respect to $SU(5)$ is clearly that all fermions in one generation belong to a single 16-dimensional representation, so that one cannot separately play with the properties of the $SU(5)$-singlet right-handed neutrinos in order to explain the striking difference between quark and neutrino mixing.

\section{A class of models}
A promising strategy in order to separate charged fermions and neutrinos is to assume a renormalizable $SO(10)$ model with dominance of type-II see-saw \cite{type2}(with respect to type-I see-saw) for the light neutrino mass matrix.
In renormalizable $SO(10)$ models the fermion
masses are generated by Yukawa couplings with Higgs fields transforming as {\bf 10}, ${\bf \overline{126}}$ (both symmetric) and {\bf 120} (antisymmetric) \cite{ren}
\begin{equation}
W_Y = h\, \psi\psi H_{10} + f\, \psi\psi H_{120}+h'\,\psi\psi H_{\overline{126}},
\end{equation}
where the symbol $\psi$ stands for the {\bf 16} dimensional representation of SO(10) and $H_i$ are the Higgs fields.  I note that in this analysis we assume an underlying "parity" symmetry (justified by the fact that, as we shall see, the resulting fit is very good) that implies that all mass matrices obtained from $h$, $h^\prime$ and $f$ are hermitian \cite{parity}.
The resulting Yukawa mass matrices for the different fermions are:
\begin{eqnarray}
M_u &=& (h + r_2 f +r_3 h^\prime)v_u, \qquad M_d = r_1 (h+ f + h^\prime)v_d, \\ \nonumber
M_e &=& r_1 (h-3f + c_e h^\prime)v_d, \qquad M_{\nu^D} = (h-3 r_2 f + c_\nu h^\prime)v_u,
\end{eqnarray}
and with type-II see-saw dominance the neutrino mass matrix is:
\begin{equation}
m_\nu = fv_L.
\end{equation}
So if type-II see-saw is responsible for neutrino masses, then the neutrino mass matrix (proportional to $f$) is separated from the dominant contributions to the charged fermion masses ($h$ for example) and can therefore show a completely different pattern. This is to be compared with the case of type-I see-saw where the neutrino mass matrix depends on the neutrino Dirac and Majorana matrices and, in $SO(10)$, the relation with the charged fermion mass matrices is tighter.

An important observation is that, without loss of generality, we can always go to a basis where the matrix $f$ is of the TB type. In fact, if we start from a complex symmetric matrix $f'$ not of the TB type, it is sufficient to diagonalise it by a unitary transformation $U$: $f'_{diag}=U^Tf'U$ and then take the matrix 
\begin{equation}
f=U_{TB}^*f'_{diag} U_{TB}^\dagger=U_{TB}^* U^T f' U U_{TB}^\dagger.
\label{efefpr}
\end{equation} 
As a result the matrices $f$ and $f'$ are related by a change of the charged lepton basis induced by the unitary matrix $O= U U_{TB}^\dagger$ (in $SO(10)$ the matrix $O$ rotates the  whole fermion representations $\bf 16_i$).
Since TB mixing is a good approximation to the data we argue that this basis is a good starting point. In fact in this basis the deviations from TB mixing will be generated by the mixing angles from the diagonalisation of $M_e$ which in $SO(10)$ are strongly related to the CKM angles and so are automatically small, while  in general could be large.

\section{The analysis }

An interesting question is to see to which extent the data are compatible with the constraints implied by this interconnected structure. So here we do not consider the problem of formulating a flavour symmetry or another dynamical principle that can lead to approximate TB mixing, but rather study the performance of the type-II see-saw $SO(10)$ model in fitting the data on fermion masses in comparison with other models architectures.

As comparison models we use a set of realistic $SO(10)$ theories with different features: renormalizable or not, with lopsided or with symmetric mass matrices, with various assumed flavour symmetries, with different types of see-saw and so on. Of course in these models TB mixing appears as accidental, and some dedicated parameters are available to fit the observed neutrino masses and mixing angles without a specific TB structure implemented.
The models considered are \cite{gb}: Dermisek, Raby (DR); Albright, Babu, Barr (ABB); Ji, Li, Mohapatra (JLM); Bajc, Senjanovic, Vissani (BSV); Joshipura, Kodrani (JK2); Grimus, Kuhbock (GK).
\begin{table}
\begin{tabular}{llllll}
\hline
Model & d.o.f. & $\chi^2$ &  $\chi^2$/d.o.f. & $d_{FT}$ & $d_{Data}$ \\
\hline
DR & 4 &0.41 & 0.10 &7.0 ~$\times 10^3$&1.3~ $\times 10^3$\\
\hline
ABB   &6 & 2.8 & 0.47 &8.1~$\times 10^3$& 3.8~$\times 10^3$\\
JLM    &4 &2.9 & 0.74 & 9.4~$\times 10^3$& 3.8~$\times 10^3$ \\
\hline
BSV    &$< 0$ &6.9 &-& 2.0~$\times 10^5$& 3.8~$\times 10^3$  \\
JK2   &3 &3.4 & 1.1 &4.7~$\times 10^5$& 3.8~$\times 10^3$  \\
GK    &0 &0.15 & -& 1.5~$\times 10^5$& 3.8~$\times 10^3$  \\
T-IID    &1 & 0.13 & 0.13 & 4.7~$\times 10^5$& 3.8~$\times 10^3$ \\
\hline
\end{tabular}
\caption{Fit results for each model as explained in the text}
\label{tab}
\end{table}

Each model is compared with the same set of data on masses and mixing given at the GUT scale (except for DR that requires a large value of $\tan{\beta}$)\cite{gb}. In this part of the analysis we do not consider the new result from the T2K experiment on the lepton angle $\theta_{13}$.
The results of the analysis are shown in Tab. \ref{tab}, where it is shown the $\chi^2$ and the $\chi^2$/d.o.f. obtained from the fit for each model. We also introduce as addictional quality factor a parameter $d_{FT}$ for a quantitative measure of the amount of fine-tuning of parameters which is needed in each model.  This adimensional quantity is obtained as the sum of the absolute values of the ratios between each parameter and its error (defined for this purpose as the shift from the best fit value that changes $\chi^2$ by one unit with all other parameters fixed at their best fit values), $d_{FT} = \sum \mid \frac{par_i}{err_i} \mid$. It has to be compared with a similar number $d_{Data}$ based on the data (i.e. the sum of the absolute values of the ratios between each observable and its error as derived from the input data), $d_{Data} = \sum \mid \frac{obs_i}{err_i} \mid$.

Moreover we analyse also the model T-IID in the light of the recent results from the T2K collaboration, which give indications of a non zero reactor angle. In particular we compared the model with the old set of data for charged leptons masses and CKM mixings and with the new global neutrino fit \cite{valle}, giving for the reactor angle at 3$\sigma$:
\begin{equation}
\sin^2\theta_{13}=0.013^{+0.022}_{-0.012}
\end{equation}
We show our results in Tab. \ref{tab2}. We note that the goodness of the fit is substantially unchanged compared with the old analysis, showing that in this class of models it is possible to obtain the desired (very small before T2K or more sizeable now) corrections to zero $\theta_{13}$ from the charged lepton sector, taking into account an appreciable amount of finetuning.

\begin{table}[hb]
\begin{tabular}{llllll}
\hline
Model & d.o.f. & $\chi^2$ &  $\chi^2$/d.o.f. & $d_{FT}$ & $d_{Data}$ \\
\hline
T-IID    &1 & 0.14 & 0.14 & 4.6~$\times 10^5$& 3.8~$\times 10^3$ \\
\hline
\end{tabular}
\caption{Fit results for the model T-IID after T2K data}
\label{tab2}
\end{table}

In conclusion  we have shown that a $SO(10)$ model with type-II see-saw dominance can achieve a very good fit of fermion masses and mixings also including the neutrino sector (provided that the representations {\bf 10}, ${\bf\overline{126}}$ and {\bf 120} are all included) and also after the T2K results.  The quality of the fit in terms of $\chi^2$ and $\chi^2$/d.o.f. is better than or comparable with any other realistic $SO(10)$ model that we have tested. However, the tight structure of the T-IID model implies a significantly larger amount of fine tuning with respect to more conventional models like the DR or the ABB and JLM models. But those models have no built-in TB mixing and in fact could accommodate a wide range of mixing angle values.

%% file: Author/Boucenna.tex
\def\321{SU(3) $\otimes$ SU(2) $\otimes$ U(1)}
\def\gsim{\raise0.3ex\hbox{$\;>$\kern-0.75em\raise-1.1ex\hbox{$\sim\;$}}}
\def\lsim{\raise0.3ex\hbox{$\;<$\kern-0.75em\raise-1.1ex\hbox{$\sim\;$}}}

\begin{center}
{\bf Abstract}\\
\end{center}

  We investigate a model in which Dark Matter is stabilized by means
  of a $Z_2$ parity remnant of a non-abelian discrete
  flavor symmetry that accounts for the observed patterns of neutrino
  mixing.
  In this $A_4$ example the standard model is extended by three extra
  Higgs doublets and the $Z_2$ parity emerges from the
  spontaneous breaking of $A_4$ after electroweak symmetry breaking.
  We perform an analysis of the parameter space of the model
  consistent with electroweak precision tests, collider searches and
  perturbativity.
  We determine the regions compatible with the observed relic dark
  matter density and we present prospects for detection in direct as
  well as indirect Dark Matter search experiments.

\vskip5.mm
\section{Introduction}

The existence of Dark Matter (DM) is by now well established, owing
to various cosmological and astrophysical observations. The nature of this non-baryonic
component of the total mass of the Universe is still elusive though, despite great
experimental and theoretical efforts over many years. Elucidating the nature of DM constitutes
one of the most important challenges modern cosmology and particle physics are facing.

Still, we do have some clues about how it \textit{should} be \cite{Bertone:2004pz,Taoso:2007qk}.
Among the most important conditions a DM candidate is required to satisfy are
neutrality, stability over cosmological time scales, and agreement
with the observed relic density.
While the neutrality of the particle is usually easy to accomodate theoretically,
its stability is assumed {\it ad-hoc} in most cases.

A new mechanism of stabilizing DM has been proposed in,Ref.~\cite{Hirsch:2010ru} in which
the stability originates from the flavor structure of the standard model, linking DM to neutrino physics.
Indeed the same discrete flavor symmetry that explains neutrino
mixing patterns~\cite{Schwetz:2008er} is the origin of the DM candidate stability 
\footnote{For models based on non-Abelian discrete symmetries but with a
  decaying dark matter candidate we refer the reader to Ref.\cite{Kajiyama:2010sb} for instance.}.
This link between two sectors that show a striking need for physics beyond the standard model
is attractive and has come under further scrutiny in a series of works, see \cite{Eby:2011qa,Adulpravitchai:2011ei}
for instance.

%

\section{The Model}
\label{model}
  The model considered in~\cite{Hirsch:2010ru} is based on the $A_4$ discrete group,
the group of even permutations of four objects. Matter fields are assigned to its irreducible
representations in the following way: The standard model Higgs doublet $H$ is assigned to the
singlet representation, while the three additional Higgs doublets
$\eta=(\eta_1,\eta_2,\eta_3)$ transform as an $A_4$ triplet, namely
$\eta\sim 3$. The model has in total four Higgs doublets, implying the
existence of four CP even neutral scalars, three physical
pseudoscalars, and three physical charged scalar bosons.
In the fermion sector we have four right-handed neutrinos; three
transforming as an $A_4$ triplet $N_T=(N_1,N_2,N_3)$, and one singlet
$N_4$.  Quarks are $A_4$-blind hence no prediction on their mixing matrix is given (cf.~\cite{Toorop:2011ad}
for the case of a non-trivial charge assignment).

The lepton and Higgs assignments are summarized in table\,\ref{tab1}.
\begin{table}[h!]
\begin{center}
\begin{tabular}{|c|c|c|c|c|c|c|c|c||c|c|}
\hline
&$\,L_e\,$&$\,L_{\mu}\,$&$\,L_{\tau}\,$&$\,\,l_{e}^c\,\,$&$\,\,l_{{\mu}}^c\,\,$&$\,\,l_{{\tau}}^c\,\,$&$N_{T}\,$&$\,N_4\,$&$\,\hat{H}\,$&$\,\eta\,$\\
\hline
$SU(2)$&2&2&2&1&1&1&1&1&2&2\\
\hline
$A_4$ &$1$ &$1^\prime$&$1^{\prime \prime}$&$1$&$1^{\prime \prime}$&$1^\prime$&$3$ &$1$ &$1$&$3$\\
\hline
\end{tabular}\caption{Summary of the relevant quantum numbers}\label{tab1}
\end{center}
\end{table}

The resulting leptonic Yukawa Lagrangian is:
\begin{eqnarray}\label{lag}
\mathcal{L}&=&y_e L_el_{_e}^c \hat{H}+y_\mu L_\mu l_{_\mu}^c \hat{H}+y_\tau L_\tau l_{_\tau}^c \hat{H}+\nonumber\\
&&+y_1^\nu L_e(N_T\eta)_{1}+y_2^\nu L_\mu(N_T\eta)_{1''}+y_3^\nu L_\tau(N_T\eta)_{1'}+\nonumber\\
&&+y_4^\nu L_e N_4 \hat{H}+ M_1 N_TN_T+M_2 N_4N_4+\mbox{h.c.}
\end{eqnarray}
This way the field $\hat{H}$ is responsible for quark and charged
lepton masses, the latter automatically diagonal.  The scalar
potential is:
\begin{eqnarray}\label{potential}
V&=&\mu_\eta^2\eta^\dagger\eta+\mu_{\hat{H}}^2 \hat{H}^\dagger \hat{H} 
+\lambda_1 [\hat{H}^\dagger \hat{H}]_1^2+\lambda_2 [\eta^\dagger\eta]_1^2
+\lambda_3 [\eta^\dagger\eta]_{1^\prime}[\eta^\dagger\eta]_{1^{{\prime}{\prime}}}\nonumber \\
&+&\lambda_4 [\eta^\dagger\eta^\dagger]_{1^\prime}[\eta\eta]_{1^{\prime\prime}}
+\lambda_{4^\prime}[\eta^\dagger\eta^\dagger]_{1^{\prime\prime}}[\eta\eta]_{1^\prime}
+\lambda_5[\eta^\dagger\eta^\dagger]_{1}[\eta\eta]_{1}
+\lambda_6([\eta^\dagger \eta]_{3_{1}}[\eta^\dagger \eta]_{3_{1}}+h.c.)\nonumber \\
&+&\lambda_7 [\eta^\dagger \eta]_{3_{1}}[\eta^\dagger \eta]_{3_{2}} 
+\lambda_8 [\eta^\dagger \eta^\dagger]_{3_{1}}[\eta \eta]_{3_{2}} 
+\lambda_9 [\eta^\dagger \eta]_{1^\prime}[\hat{H}^\dagger \hat{H}] 
+\lambda_{10}[\eta^\dagger \hat{H}]_{3_1}[\hat{H}^\dagger \eta]_{3_1} \nonumber\\
&+&\lambda_{11}([\eta^\dagger\eta^\dagger]_{1}\hat{H} \hat{H}+h.c.) 
+\lambda_{12}([\eta^\dagger\eta^\dagger]_{3_{1}}[\eta \hat{H}]_{3_1}+h.c.) 
+\lambda_{13}([\eta^\dagger\eta^\dagger]_{3_{2}}[\eta \hat{H}]]_{3_1}+h.c.) 
+\lambda_{14}([\eta^\dagger \eta]_{3_{1}}\eta^\dagger \hat{H}+h.c.) \nonumber\\
&+&\lambda_{15}([\eta^\dagger \eta]_{3_{2}}\eta^\dagger \hat{H}+h.c.) 
\end{eqnarray}
where $[...]_{3_{1,2}}$ is the product of two triplets contracted into
one of the two triplet representations of $A_4$, and $[...]_{1,1^\prime,1^\prime\prime}$ is the product
of two triplets contracted into a singlet representation of $A_4$.
In what follows we assume, for simplicity, CP conservation so that
all the couplings in the potential are real.
For convenience we also assume $\lambda_4 = \lambda_4^\prime$ in order
to have manifest conservation of CP in our chosen
$A_4$ basis.

The minimization of the scalar potential results in :

\begin{equation}
  \vev{ H^0}=v_H\ne 0,~~~~ \vev{ \eta^0_1}=v_\eta \ne 0,~~~~
\vev{ \eta^0_{2,3}}=0\,,\label{vevs}
\end{equation}
where all vevs are real.  This vev alignment breaks the group $A_4$ to
its subgroup $Z_2$ responsible for the stability of the DM as well as
for the neutrino phenomenology~\cite{Hirsch:2010ru}.

\begin{center}{\it The stability of the DM}\end{center}

As a consequence of the fields assignments there are no couplings with charged fermions
and quarks. The only Yukawa interactions of the lightest neutral component
of $\eta_{2,3}$ are with the heavy \321 singlet right-handed neutrinos.
This state is charged under the $Z_2$ parity that survives
after the spontaneous breaking of the flavor symmetry.

We now show the origin of such a parity symmetry.  \vskip5.mm

The group $A_4$, has two generators: $S$, and $T$, that satisfy the relations
$S^2=T^3=(ST)^3=I$. In the three dimensional basis $S$ is given by 
\begin{equation}S=\left(
\begin{array}{ccc}
1&0&0\\
0&-1&0\\
0&0&-1\\
\end{array}
\right).\label{z2sim}\end{equation}
$S$ is the generator of the $Z_2$ subgroup of
$A_4$. The alignment $\vev{ \eta} \sim (1,0,0)$ breaks spontaneously $A_4$ to $Z_2$ since $(1,0,0)$ is 
manifestly invariant under the $S$ generator.
The $Z_2$ residual symmetry is defined as
\begin{equation}\label{residualZ2}
\begin{array}{lcrlcr}
N_1 &\to& +N_1\,,\quad& \eta_1 &\to& +\eta_1 ,\\  
N_2 &\to& -N_2\,,\quad& \eta_2 &\to& -\eta_2 ,\\
N_3 &\to& -N_3\,,\quad& \eta_3 &\to& -\eta_3 ,
\end{array}
\end{equation}
The rest of the matter fields being $Z_2$ even.
The DM candidate of the model corresponds to the lightest $Z_2$-odd
neutral spin zero particle which, for the sake of definiteness, we
take as the CP-even state.
The parameters of the model relevant for the DM phenomenology are 8 (scalar) masses, 7 couplings
  and the ratio of the vacuum expectation values $\tan \beta= v_{H}/v_{\eta}.$

\begin{center}
{\it Neutrino phenomenology}  
\end{center}
We present here the main results  obtained in Ref.\cite{Hirsch:2010ru} concerning the neutrino phenomenology.

The model contains four heavy right-handed neutrinos. It is therefore a special
case, called (3,4), of the general type-I seesaw mechanism~\cite{schechter:1980gr}.  
Light neutrinos get Majorana masses by means of the type-I seesaw relation.
The reactor mixing angle $ \theta_{13}$ is null and the hierarchy is inverse.
 The atmospheric angle, the solar angle and the two square mass differences can be fitted.  
The model implies a neutrinoless double beta decay effective mass parameter in the range 0.03 to 0.05~eV at
3~$\sigma$, within reach of upcoming experiments.

Before moving to the calculation of the relic abundance in the next
section we consider the phenomenological constraints on the parameter
space of the model.

\section{Phenomenological constraints}
\label{constraints}

In order to find the viable regions in parameter space where to
perform our study of dark matter, we must impose the following
constraints to the model :

\begin{itemize}
\item {\it Electroweak precision tests} 

  \noindent 
  We compute the effect on $T$ induced by the scalars
  following \cite{Grimus:2007if} and we impose the bounds from
  electroweak measurements \cite{Nakamura:2010zzi}: $-0.08\leq T \leq0.14.$

\item {\it Collider bounds}

  \noindent 
  The bounds imposed by LEP II on the masses of the neutral scalars in
  our model are similar to those constraining the Inert Doublet Model,
  given in Ref.\cite{Lundstrom:2008ai}.
  The excluded region that we adopt is taken from Fig.7 of
  Ref.\cite{Lundstrom:2008ai}.
  We impose a lower limit on the masses of the $Z_2$ even neutral
  and charged scalars of 114 GeV and 100 GeV respectively. 

\item {\it Perturbativity and vacuum stability}

  \noindent The requirement of perturbativity  imposes the following bounds on the Yukawa couplings of the model
  $\lambda_i \lsim \sqrt{4 \pi}$ and $\tan\beta >0.5$.

Finally we must impose the stability of the vacuum which translate as a set of inequalities to be satisfied by the couplings.
of the models.

\end{itemize}

\begin{center}
\begin{figure*}[t]
\includegraphics[width=0.4\textwidth]{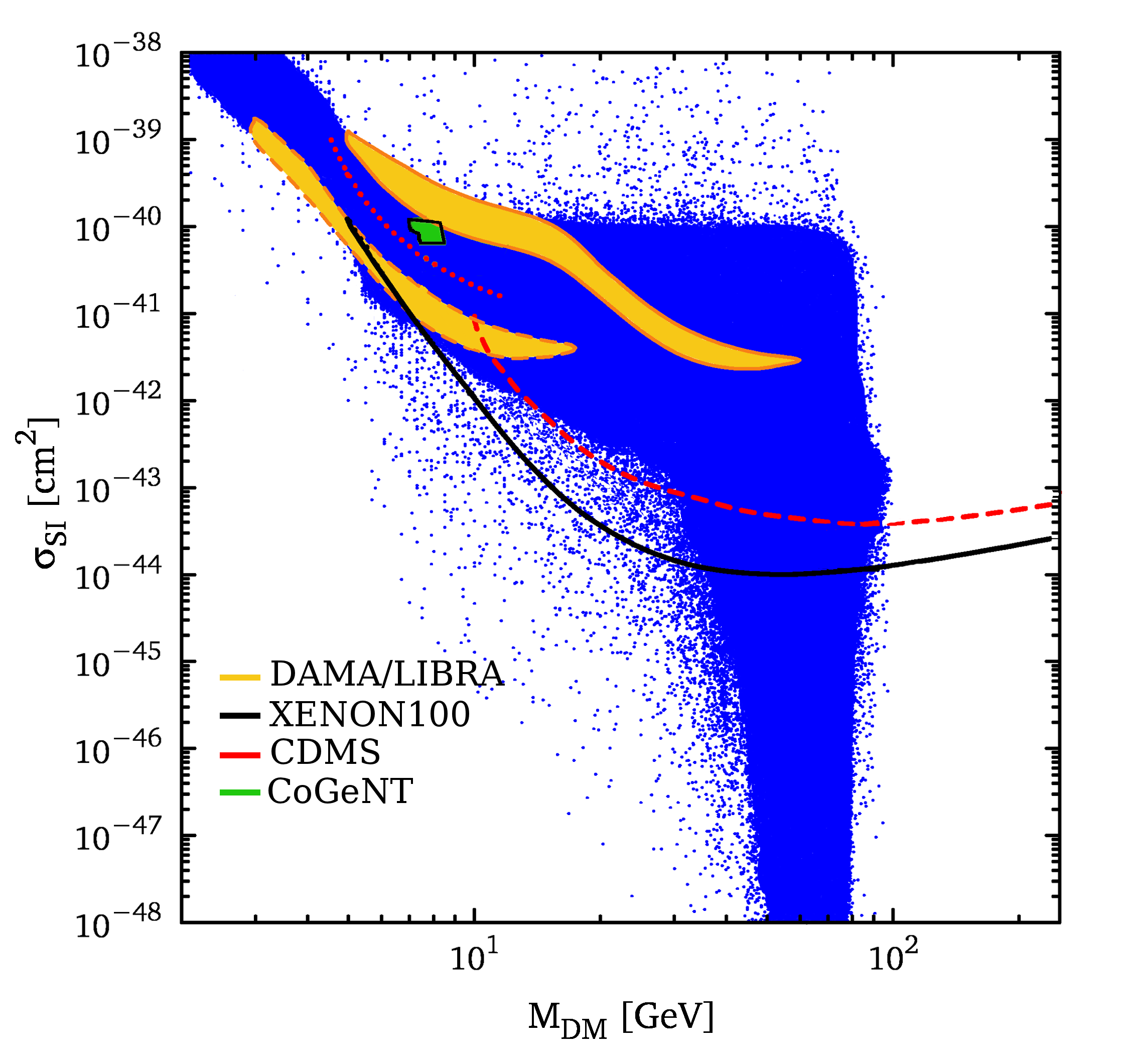}
\includegraphics[width=0.4\textwidth]{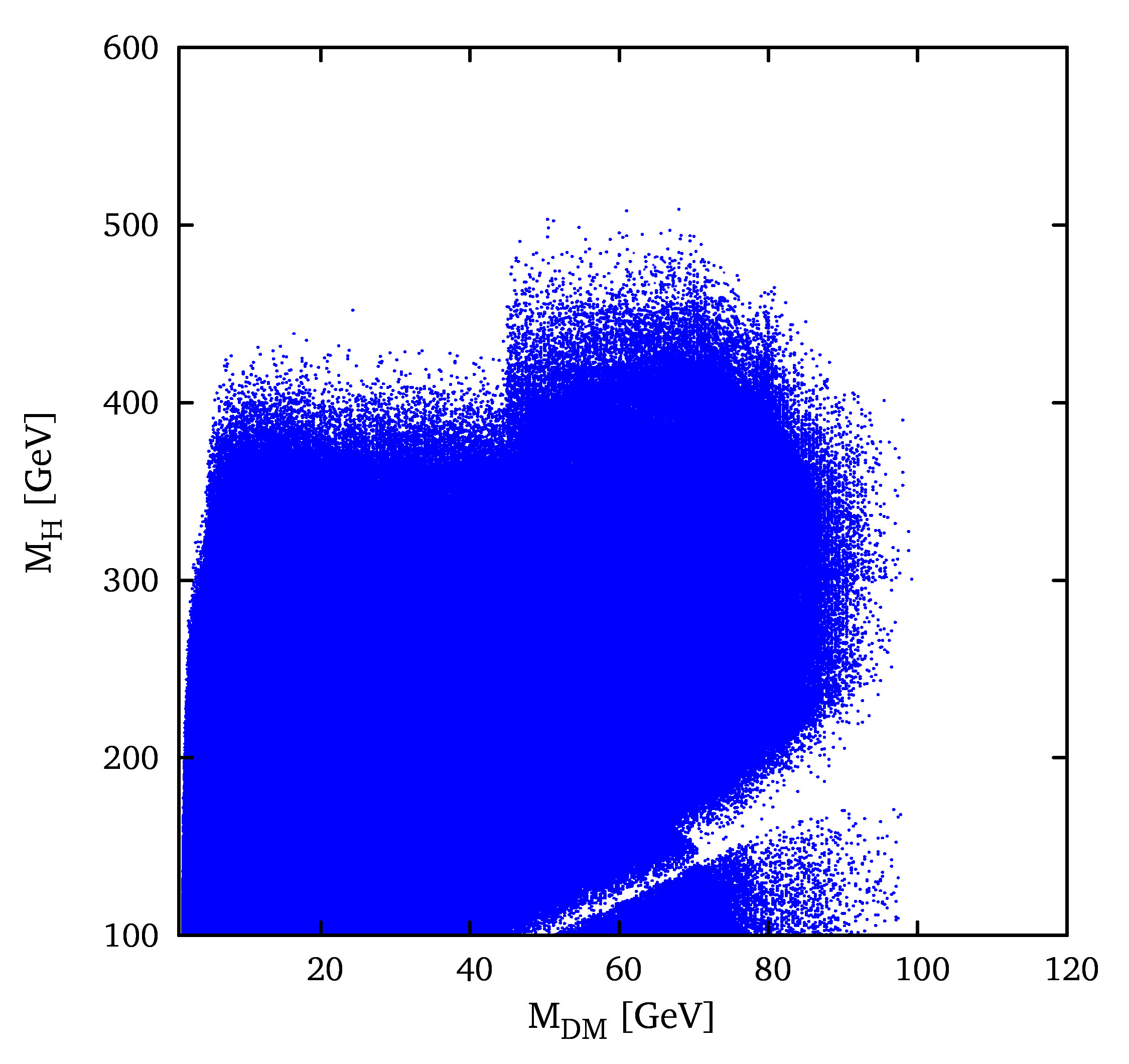}
  \caption{Left plot: Spin-independent DM scattering cross section 
off-protons as a function of the dark matter mass. The orange regions 
delimited by the dashed (solid) line show the DAMA/LIBRA annual modulation 
regions including (neglecting) the channeling effect~\cite{Fornengo:2010mk}.
The green region corresponds to the COGENT data~ \cite{Aalseth:2011wp}.
Dashed and dotted red lines correspond to the upper bound from CDMS
(respectively from \cite{Ahmed:2009zw} and \cite{Ahmed:2010wy}).
XENON100 bounds \cite{Aprile:2011hi} are shown as a solid black line.
Right plot: Regions in the plane DM mass ($M_{DM}$) - lightest Higgs boson $M_H$
    allowed by collider constraints and leading to a DM relic abundance
  compatible with WMAP measurements.}
  \label{omega}
\end{figure*}
\end{center}

\section{Relic Density and (In)Direct Detection}
\label{relic}

The thermal relic abundance of $H_2$ is controlled by its annihilation
cross section into SM particles. In order to study the viable regions of the model we
perform a random scan over the 16-dimensional parameter space
($\lambda_i$ and $\tan \beta$ ) and compute the dark matter relic
abundance using the micrOMEGAs package\cite{Belanger:2010gh,Belanger:2008sj} for the points satisfying 
the constraints discussed in Sec.\ref{constraints}. 

In Fig.~\ref{omega} we show the regions with a correct relic abundance
in the plane DM mass ($M_{DM}$) and the lightest Higgs boson mass
$M_{H}$.  For dark matter masses well below the $W^{\pm}$ threshold,
dark matter annihilations into fermions are driven by the s-channel
exchange of the Higgs scalars of the model.

For $M_{H}\lesssim 400$ GeV the annihilation cross-sections are large
enough to obtain the correct relic density for all DM masses up to the
the $W^{\pm}$ threshold. At larger Higgs boson masses, annihilations
into light fermions are suppressed so that the relic abundance is
typically too large unless efficient co-annihilations with the
pseudoscalar $A_2$ are present. 
For lighter dark matter particles, strong co-annihilations cannot be
reconciled with LEPII constraints as these require $M_{A_2}\gsim 100$
GeV \cite{Lundstrom:2008ai}.
On the other hand, the absence of points on the strip corresponding to the
line $M_{DM}\sim M_H/2$ is associated with the $H$
resonance that enhances the DM annihilation cross section
giving a too small Dark Matter abundance.

For dark matter masses larger than $M_W$ the annihilation cross
section into gauge bosons is typically large enough so that $H_2$ cannot account for
the desired relic density.

The prospects for direct dark matter detection in underground experiments are of particular interest.
We see in Fig.\ref{omega} that the model can potentially explain the DAMA annual modulation data
\cite{Bernabei:2010mq,Bernabei:2000qi} as well as the excess recently
found in the COGENT experiment \cite{Aalseth:2011wp}.  The present upper limits on the spin independent
DM scattering cross section off nucleons can already exclude a large region of the parameter space of the model but
do not constrain the mass of the DM candidate. Note that the direct detection plot of Fig.\ref{omega} has
been updated from \cite{Boucenna:2011tj} to include recent results of
XENON100\cite{Aprile:2011hi} and COGENT\cite{Aalseth:2011wp}.

Indirect dark matter searches through astrophysical observations are
not currently probing the model apart from some small regions of the
parameter space where the dark matter annihilation cross section is
enhanced via a Breit-Wigner resonance.
For further details about this study we refer the reader to \cite{Boucenna:2011tj}.

\section{ Conclusions and discussion}
\label{conclusions}

We have studied a model where the stability of the dark matter
particle arises from a flavor symmetry.  The $A_4$ non-abelian
discrete group accounts both for the observed pattern of neutrino
mixing as well as for DM stability.
We have analyzed the constraints that follow from electroweak
precision tests, collider searches and perturbativity.
Relic dark matter density constraints exclude the region of the
parameter space where simultaneously $M_{DM}\lesssim 40$ GeV and
$M_{H}\gsim 400$ GeV because of the resulting over-abundance of dark
matter.
We have also analyzed the prospects for direct  dark
matter detection and found that, although they already exclude
a large region of the parameter space, we cannot constrain the mass of the
DM candidate.
In contrast, indirect DM detection is not yet sensitive enough to
probe our predictions.

All of the above relies mainly on the properties of the scalar sector
responsible for the breaking of the gauge and flavour symmetry.
A basic idea of our approach is to link the origin of dark matter  to
the origin of neutrino mass and the understanding of the pattern of
neutrino mixing, two of the most oustanding challanges in particle
physics today. Note however that the connection of dark matter to neutrino properties
depends strongly on how the symmetry breaking sector couples to the
leptons.

\section{ Acknowledgments}
This work has been done in collaboration with Martin Hirsch, Stefano Morisi, Eduardo Peinado, Marco Taoso
and Jose W.F. Valle.

%% file: Author/Ivo.tex
{\bf Abstract}\\
\vskip5.mm

We take tri-bi-maximal mixing and decompose the effective neutrino matrix to derive predictions in terms of their masses. We extend this phenomenological analysis to other mass-independent mixing schemes. We classify models through the group structure of their symmetries to point out connections between the groups and the phenomenological analysis.

We study the UV completion of family symmetry models, which in general improves the predictivity over effective models - we illustrate important features by minimally completing an $A_4$ model. We also show that family symmetries can provide Yukawa alignment for multi-Higgs doublet models. 

\vskip5.mm

\section{Phenomenology}

The work summarised here is based in \cite{Pheno} (which also includes a more complete list of references).

We consider tri-bi-maximal mixing (TB) to be a good description of leptonic mixing. The effective neutrino mass matrix with TB mixing can be written without loss of generality in the form
\begin{align}\label{mnuTB}
\begin{split}
m_\text{TB}=U_\text{TB} \, d_\nu  \, U_\text{TB}^T = \frac{1}{3}
{\begin{pmatrix}
2x^\prime+3y^\prime+z^\prime&-x^\prime+z^\prime&-x^\prime+z^\prime\\
-x^\prime+z^\prime&2x^\prime+z^\prime&-x^\prime+3y^\prime+z^\prime\\
-x^\prime+z^\prime&-x^\prime+3y^\prime+z^\prime&2x^\prime+z^\prime
\end{pmatrix}}\,,
\end{split}
\end{align}
where $d_\nu=\text{diag}(x^\prime+y^\prime,y^\prime+z^\prime,x^\prime-y^\prime)$, and $U_\text{TB}$ is the TB mixing matrix.
The mass matrix $m_\text{TB}$ can be separated into three components,
\begin{align}\label{mnu3}
m_\text{TB}&=x^\prime C +y^\prime P+z^\prime D,
\end{align}
\begin{align}\label{CPDstruct}
C=\frac{1}{3}{ \begin{pmatrix}
2&-1&-1\\
-1&2&-1\\
-1&-1&2
\end{pmatrix}}, P={  \begin{pmatrix}
1&0&0\\
0&0&1\\
0&1&0
\end{pmatrix}}, D=\frac{1}{3}{ 
\begin{pmatrix}
1&1&1\\
1&1&1\\
1&1&1
\end{pmatrix}}
\end{align}
$C$, $P$, $D$ respectively denote the well-known magic, $\mu$-$\tau$ symmetric, and democratic matrices.
This decomposition is useful to classify models according to the parameters $x^\prime$, $y^\prime$, and particularly $z^\prime$. If either $x^\prime$ or $y^\prime$ vanishes the neutrino mass spectrum is degenerate, which is excluded by the experimental data and so only $z^\prime$ can be absent in eq.~\eqref{mnu3}.
In the basis where the charged leptons are diagonal and real, the effective low-energy leptonic mixing can be written
\begin{align}
U_\nu=e^{-i \sigma_1/2}\,U_\text{TB}\,
\begin{pmatrix}
1&&\\
&e^{i\gamma_1}&\\
&&e^{i\gamma_2}
\end{pmatrix},
\end{align}
with $\gamma_1=(\sigma_1-\sigma_2)/2,\,\gamma_2=(\sigma_1-\sigma_3)/2$, being the Majorana phases and $\sigma_{1,3}=\text{arg}(x^\prime \pm y^\prime),\,\sigma_2=\text{arg}(y^\prime+z^\prime)$. In turn, the neutrino masses read as
\begin{align}\label{numasses}
\begin{split}
m_1&=\left|xe^{i\alpha_1}+y\right|=\left(x^2+y^2+2xy\cos\alpha_1\right)^{1/2},
\\
m_2&=\left|y+ze^{i\alpha_2}\right|=\left(y^2+z^2+2yz\cos\alpha_2\right)^{1/2},
\\
m_3&=\left|xe^{i\alpha_1}-y\right|=\left(x^2+y^2-2xy\cos\alpha_1\right)^{1/2},
\end{split}
\end{align}
where $x=|x^\prime|$, $y=|y^\prime|$, $z=|z^\prime|$ ,and
$\alpha_1=\text{arg}\,x^\prime-\text{arg}\,y^\prime$,
$\alpha_2=\text{arg}\,z^\prime-\text{arg}\,y^\prime$.
The sign of $(m_3 - m_2)$ is dictated by the ordering of the masses, positive for normal and negative for inverted ordering. In terms of the parameters $\alpha_{1,2}, x, y, z$, from eq.~\eqref{numasses} we obtain:
\begin{align}\label{m2relations}
\begin{split}
\Delta
m^2_{21}=z\left(z+2y\cos\alpha_2\right)-x\left(x+2y\cos\alpha_1\right)\,,\quad \Delta m_{31}^2=-4xy\cos\alpha_1\,.
\end{split}
\end{align}
As we will discuss, the democratic component in eq.~\eqref{mnu3} is naturally absent or suppressed in many FS models so we study what are the phenomenological implications. When $z=0$ we have $\Delta m^2_{21}=-x\left(x+2y\cos\alpha_1\right)$. By definition $\Delta m^2_{21}>0$, so $\pi/2 <\alpha_1 <3\pi/2$ and thus from eq.~\eqref{m2relations} $\Delta m^2_{31}>0$ enforces a normal hierarchy.
Choosing $\alpha_1$ as the free parameter, we have
\begin{align}\label{massesxy}
\begin{split}
m_1=\left(y^2-\Delta m^2_{21}\right)^{1/2},\quad m_2=y, \quad m_3=\left(y^2+\Delta m^2_{31}-\Delta m^2_{21}\right)^{1/2},
\end{split}
\end{align}
\begin{align}\label{xy}
\begin{split}
x=\frac{\left(\Delta m^2_{31}-2\Delta m^2_{21}\right)^{1/2}}{\sqrt{2}}\,, \quad y=-\frac{1}{2\sqrt{2}\cos\alpha_1}\frac{\Delta m^2_{31}}{\left(\Delta
m^2_{31}-2\Delta m^2_{21}\right)^{1/2}}\,.
\end{split}
\end{align}
The Majorana phases are $\gamma_1=\text{arg}(xe^{i\alpha_1}+y)/2$ and $\gamma_2=\gamma_1-\text{arg}(xe^{i\alpha_1}-y)/2$.
From eqs.~\eqref{massesxy} and~\eqref{xy}, we conclude that the lightest neutrino mass has a lower bound, $m^\text{low}_1\simeq 1.56\times 10^{-2}$~eV, for $\alpha_1=\pi$. Moreover, the effective mass parameter $m_{ee}$ that governs $0\nu\beta\beta$ decay is approximately given by
\begin{equation}
\begin{split}
m_{ee}=\frac{m_2}{3} \left[2\left(2-\frac{\Delta
m_{12}^2}{m_2^2}\right)(1+\cos{2\gamma_1})+1-2\frac{\Delta
m_{12}^2}{m_2^2}\right]^{1/2}
\end{split}
\end{equation}
and attains its lowest value $m_{ee}^\text{low}$ at $\alpha_1=\pi$, when $m_2$ is also minimal:
\begin{equation}
m_{ee}^\text{low}\simeq m_2^\text{low}\sqrt{1-\frac{2}{3}\frac{\Delta
m_{12}^2}{\left(m_2^\text{low}\right)^2}}\simeq 1.64\times 10^{-2}\,\text{eV}.
\end{equation}

\begin{figure*}[h!]
\begin{tabular}{cc}
\includegraphics[width=8cm]{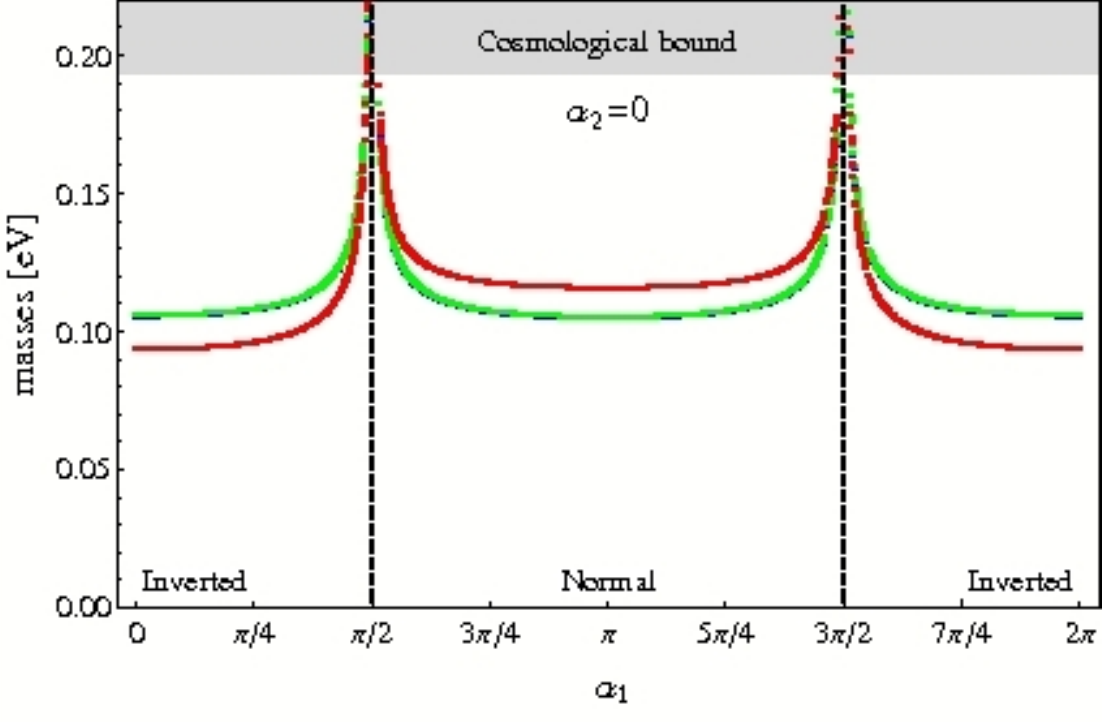}&
\includegraphics[width=7.6cm]{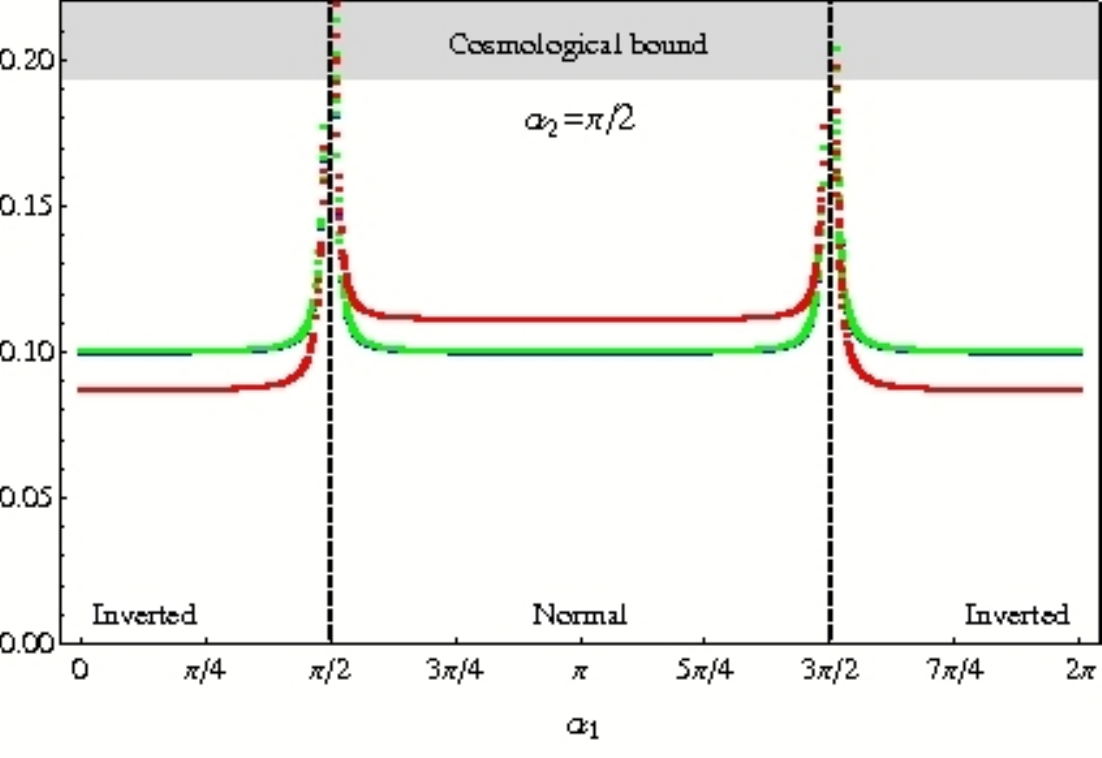}\\
\includegraphics[width=8cm]{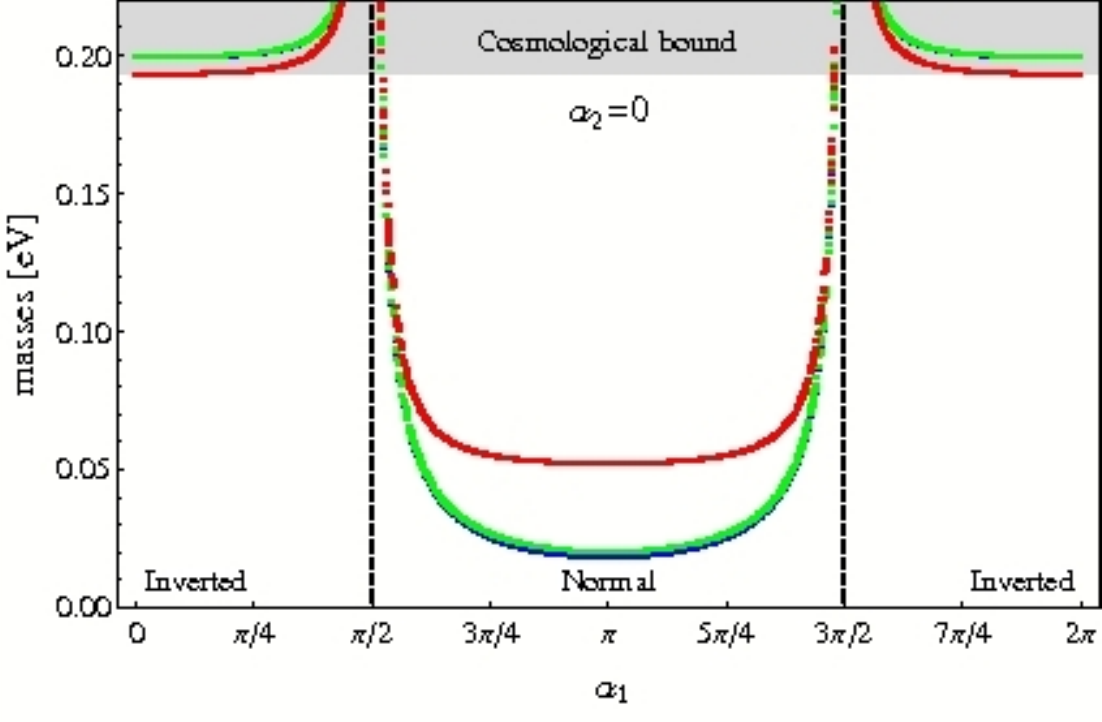}&
\includegraphics[width=7.6cm]{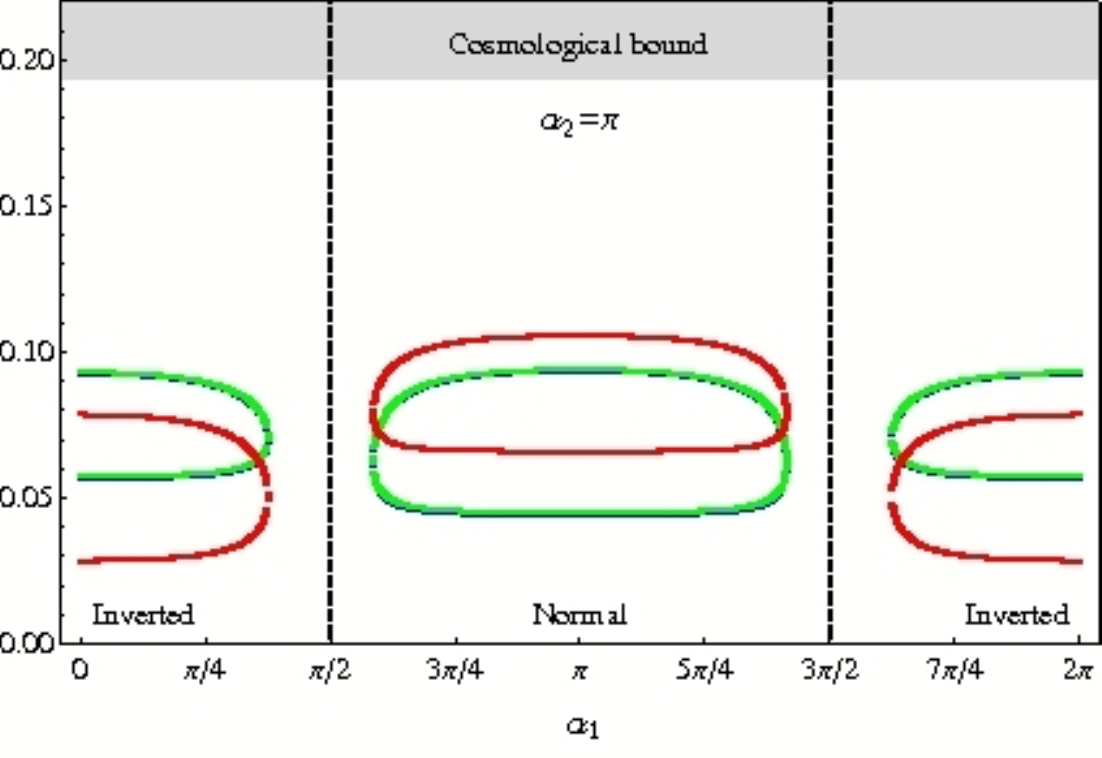}
\end{tabular}
\caption{\label{fig1} Curves correspond to $m_i$ ordered according to the hierarchy as function of $\alpha_1$ for $z=0.1$~eV and $z=z_\text{lim} \simeq 3.3 \times 10^{-3}$~eV with $\alpha_2=0$ (left) and for $z=0.1$~eV and $\alpha_2=\pi/2, \pi$ (right).}
\end{figure*}

\begin{figure*}[h!]
\begin{tabular}{cc}
\includegraphics[width=8cm]{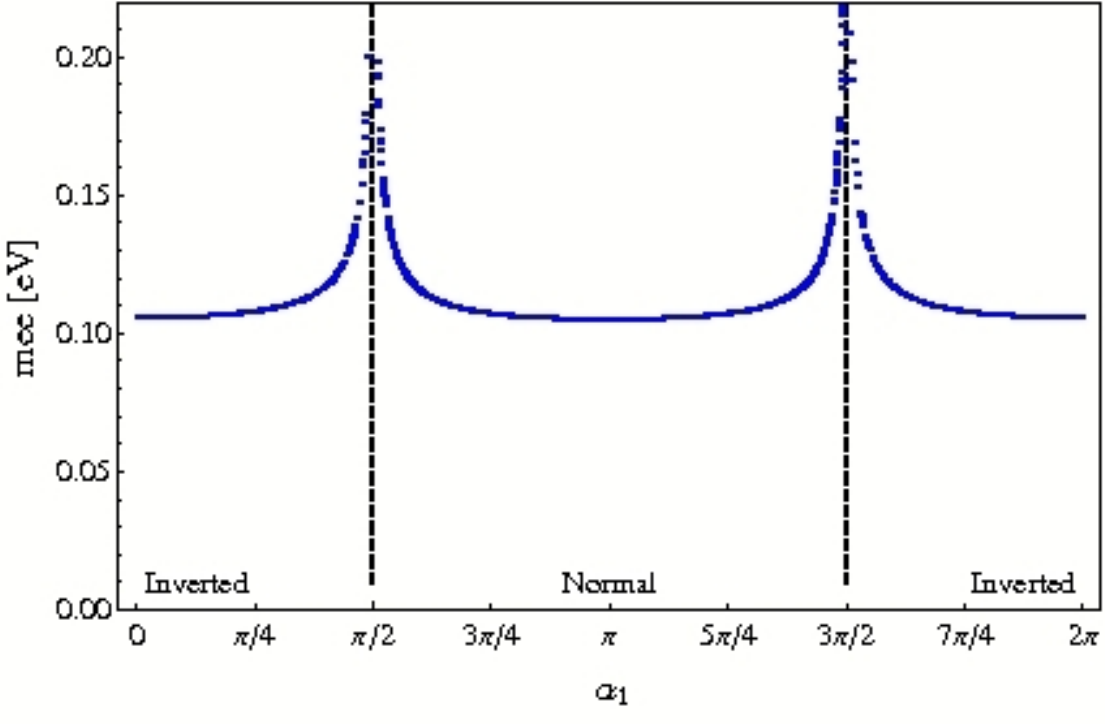}&
\includegraphics[width=7.6cm]{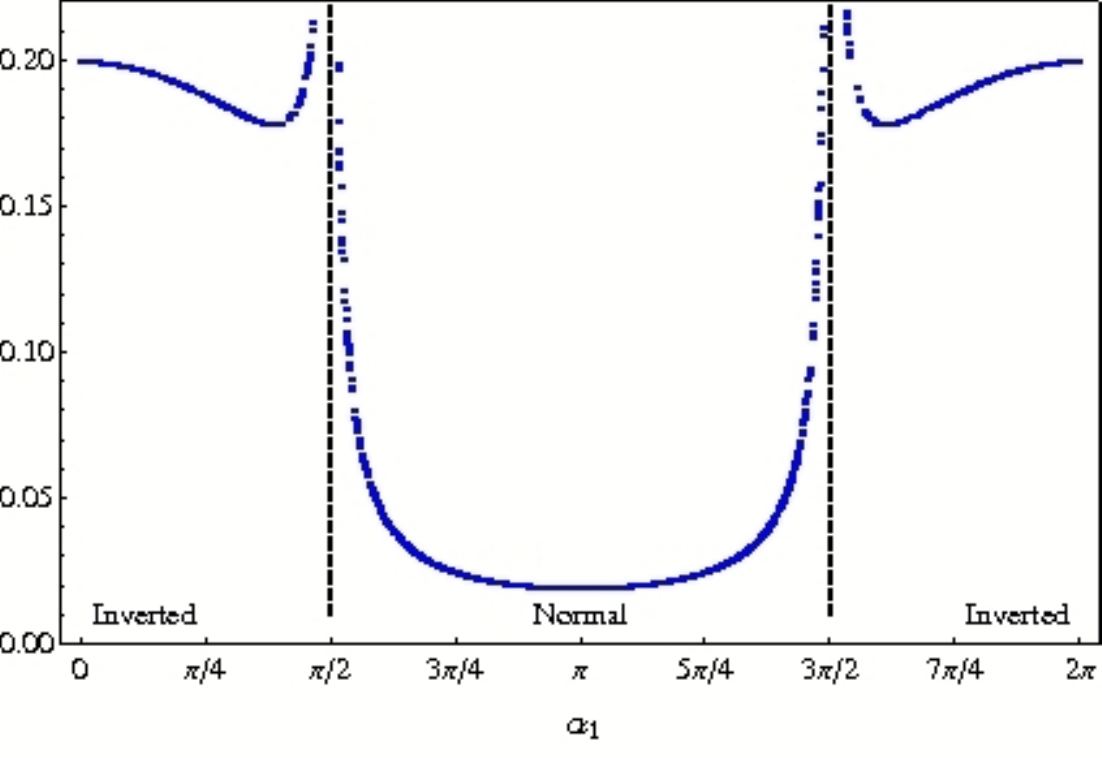}
\end{tabular}
\caption{\label{fig2} $m_{ee}$ for $\alpha_2=0$, $z=0.1$~eV (left) and $z=z_\text{lim}\simeq 3.3 \times 10^{-3}$~eV (right).}
\end{figure*}

We now consider that a small democratic contribution is present. It can be seen from eq.~(\ref{m2relations}) that an inverted hierarchy is now allowed for small values of $z$. Furthermore, such a hierarchy is easier to achieve when $\alpha_2=0$; for other values of $\alpha_2$, the inverted hierarchy is, in general, excluded for $z \lesssim 0.01$~eV. Assuming small $z$ and $\alpha_2=0$, the mass spectrum is
\begin{align}
\begin{split}
m_1\simeq\left(y^2-\Delta m_{21}^2+2yz\right)^{1/2}, m_2\simeq\left(y^2+2yz\right)^{1/2}, m_3\simeq\left(y^2+\Delta m_{31}^2-\Delta m_{21}^2+2yz\right)^{1/2},
\end{split}
\end{align}
so the solar mass-squared difference in eq.~\eqref{m2relations} can be approximated by $\Delta m_{21}^2 \simeq 2yz-x(x+2y\cos\alpha_1)$, with the $2yz$ term enabling the inverted spectrum. The parameters $x$ and $y$ no longer have a closed form (as in eq.~\eqref{xy}) but one can solve numerically for the mass spectrum. For illustration, in Fig.~\ref{fig1} we present the neutrino mass spectrum as a function of the phase $\alpha_1$ for $z=0.1$~eV and $\alpha_2=0,\, \pi/2,\, \pi$. We find that for $z \ge z_\text{lim} \simeq 3.3 \times 10^{-3}$~eV, an inverted mass hierarchy is allowed and this limiting case is shown in the lower left plot of Fig.~\ref{fig1} ($z=z_\text{lim}$ and $\alpha_2=0$). In Fig.~\ref{fig2}, we present the $0\nu\beta\beta$ parameter $m_{ee}$ for $\alpha_2=0$ and $z=0.1$~eV or $z=z_\text{lim}$.

In addition to TB mixing, there are other mass-independent structures that can reproduce the observed leptonic mixing angles at leading order.
The entire phenomenological analysis discussed so far is straightforwardly generalised to other mass-independent structures. If $m_\nu$ is exactly diagonalised by the unitary matrix $U_X$ in a given mass-independent mixing scheme, then $m_\nu = U_{X}\, d_\nu\, U_{X}^T$. Expressing the mixing in terms of $U_\text{TB}$ and an appropriate rotation $K_X$, then $U_X = K_X\, U_\text{TB}$ and we rewrite
\begin{equation}
\label{eq:X}
m_\nu = K_{X}\, m_\text{TB}\, K_{X}^T.
\end{equation}
The decomposition of the neutrino mass matrix in eq.~\eqref{mnu3} is maintained with each component rotated appropriately by $K_X$ and the analysis in terms of $x', y', z'$ holds. Although the mixing matrix is different, many conclusions drawn for TB mixing remain valid for any mass-independent mixing structure. In particular, results that depend only on the neutrino mass spectrum are unchanged. On the other hand since $m_{ee}$ depends directly on the first row of the mixing matrix, it depends on the mixing scheme - although as the mixing angles are constrained by the experimentally allowed ranges, all viable mass-independent schemes should lead to very similar predictions for $m_{ee}$.

The effective neutrino mass term is written as the operator $\ell_i H \ell_j H (...)$, where $(...)$ denotes additional fields that may be present. To establish what kind of family invariants lead to desired structures, it is important to consider the mechanism responsible for the effective term.
In the type-I (III) seesaw, heavy right-handed fermion singlets (triplets) are added to the Standard Model (SM). The type-I seesaw Lagrangian is $\mathcal{L}_\nu=Y_D^{ij}\,\overline{\ell}_iH N_{j}+M_{R}^{ij}\,\overline{N}^c_{i}N_{j}$,
where $N_i$ are the right-handed neutrino fields, $Y_D$ is the Dirac-neutrino Yukawa coupling matrix, and $M_R$ is the heavy Majorana neutrino mass matrix. The type-III Lagrangian is similar, with the right-handed neutrino $N$ replaced by the fermion triplet and the appropriate $SU(2)$ contractions.
In type-II seesaw, heavy scalar triplets $\Delta_a$ are added and the Lagrangian has $Y_a^{ij}\overline{\ell}^c_i\ell_j\Delta_a$.
The neutrino mass matrix structure arising from any of these is controlled by the allowed contractions, and in order to obtain TB mixing from FS invariants it is necessary to specify the representations. Within type-I seesaw, $M_R$ is constructed from invariants with repeated representations and if $N$ belongs to singlet representations there are too many parameters for mass-independent textures to arise without fine-tuning. Thus $N$ must be a family triplet, and the invariant contractions could be $NN$ (if allowed) or $NN\phi$ - depending on the group and the Vacuum Expectation Values (VEVs) of the $\phi$ fields, the $P$, $C$ and/or $D$ structures may appear.
Within type-II seesaw, the effective neutrino mass matrix is obtained directly from repeated representations: $\ell$ must be a triplet representation with family invariant contractions $\ell\ell$ or $\ell\ell \phi$. Notice also that in general $H$ and $\Delta$ are considered as FS singlets but it may be possible to replace $\phi$ by assigning the $H$ and $\Delta$ as FS multiplets.

Considering in detail the representations, we can formulate general arguments to justify the absence or suppression of the democratic contribution to $m_\text{TB}$.
We start with $A_4$ ($\Delta(3 2^2)$) as the FS and assume the VEV $\langle \phi \rangle \propto (1,1,1)$ (VEV alignment should be addressed in actual models). With just this VEV we can get $P$ from $\ell\ell$ and $C$ from $\ell\ell\phi$ to obtain TB with $z'=0$. In order to produce $D$ one must use higher order terms such as $(\ell \phi)(\ell \phi)$. It is therefore legitimate to consider that in $A_4$ models $z'$ naturally vanishes or is small in comparison with other contributions. Now with $\Delta(3 n^2)$ and $n>2$ consider $n=3$, i.e., $\Delta(27)$. There are only 2 triplet irreps. In this case, the $\ell\ell$ term is not allowed and the two choices for three triplet invariants repeating the same irrep. are equivalent and are not readily useful to obtain TB.
For $n=4$, i.e., $\Delta(48)$ there are 5 triplet irreps which can be labeled as $(0,1)$, $(0,2)$, $(0,3)$, $(1,1)$, and $(3,3)$. A similar three-triplet invariant that produces $C$ in $A_4$ can result from one of the outcomes of e.g. $(0,2) \times (0,1) \times (0,1)$, but never available with a repeated $(0,2)$ irrep which would be required to get the invariant $\ell\ell$ like in $A_4$. If we assign $\ell$ to e.g. $(0,1)$ it is possible to obtain all three structures at the cost of an extra VEV. From a $(0,3)$ scalar in the $(1,1,1)$ direction, the product $(0,3)\times(0,1)\times(0,1)$ allows the invariant necessary for the $C$ matrix, while the product $(0,2)\times(0,1)\times(0,1)$ allows an invariant from which both $P$ and $D$ can be constructed with VEVs in the $(1,0,0)$ and $(1,1,1)$ directions respectively. In this sense $\Delta(48)$ is the smallest $\Delta(3n^2)$ group for which the effective neutrinos naturally contain the democratic structure that allows an inverted mass spectrum.

The overarching conclusion is that it proves useful to decompose mass-independent leptonic mixing in particular ways, as doing so can reveal interesting phenomenological results that may even be linked directly with the group structure of the FS generating the mixing.

\section{UV completions}

The work described very briefly here is based in \cite{UV}.
We intend to demonstrate the benefits that UV completions of FS models can provide.
$A_4$ models giving TB are usually constructed at the non-renormalisable level. One such model has charged leptons given by:
\begin{equation}
\begin{split}
{w_\ell}=&\hspace{5mm}\dfrac{y_\tau}{\Lambda} \tau^c (\varphi_T \ell ) \, h_d + \dfrac{y_\mu}{\Lambda^2} \mu^c (\varphi_T \varphi_T\ell ) \, h_d +
\dfrac{y_\mu'}{\Lambda^2} \mu^c (\varphi_T\ell )^{''} \xi' \, h_d +\\
&+\dfrac{y_e}{\Lambda^3} e^c (\varphi_T\varphi_T\ell )^{''} \xi' \, h_d +
\dfrac{y_e'}{\Lambda^3} e^c (\varphi_T\ell )' \xi^{'2} \, h_d +
\dfrac{y_e''}{\Lambda^3} e^c (\varphi_T\ell )' (\varphi_T \varphi_T )'' \, h_d + \\
&+\dfrac{y_e'''}{\Lambda^3} e^c (\varphi_T\ell )'' (\varphi_T \varphi_T )' \, h_d +
\dfrac{y_e^{\rm iv}}{\Lambda^3} e^c (\varphi_T\ell ) (\varphi_T \varphi_T ) \, h_d\;.
\end{split}
\label{AM_cl}
\end{equation}
We explicitly introduce messenger fields. The content and charge assignment are listed in table \ref{table:AMtransformationsMessengers}.

\begin{table}[ht]
\begin{center}
\begin{tabular}{|c||cccc||cccc|}
\hline
&&&&&&&&\\[-4mm]
 & $\chi_\tau$ & $\chi_1$ & $\chi_2$ & $\chi_3$ & $\chi^c_\tau$ & $\chi^c_1$ & $\chi^c_2$ & $\chi^c_3$ \\[2mm]
\hline
&&&&&&&&\\[-4mm]
$A_4$ & $\bf3$ & $\bf1''$ & $\bf1'$ & $\bf1''$ & $\bf3$ & $\bf1'$ & $\bf1''$ & $\bf1'$ \\[2mm]
$Z_4$ & $i$ & $-1$ & $-1$ & $-i$ & $-i$ & $-1$ & $-1$ & $i$ \\[2mm]
\hline
\end{tabular}
\caption{\label{table:AMtransformationsMessengers} Transformation properties of the messengers under $A_4$, $Z_3$.}
\end{center}
\end{table}

With the chosen field content the renormalisable terms of completed model are
\begin{equation}
\label{eq:AM_ren}
w_\ell =M_{\chi_A} (\chi^c_A \chi_A) + h_d(\ell \chi^c_\tau) + \tau^c (\varphi_T \chi_\tau) + \mu^c \xi' \chi_1+ e^c \xi' \chi_3 +(\varphi_T \chi_\tau)'' \chi^c_1 +(\varphi_T \chi_\tau)'  \chi^c_2 +  \chi^c_3 \xi' \chi_2 \,.
\end{equation}

\begin{figure*}[h!]
\begin{center}
\includegraphics[width=8cm]{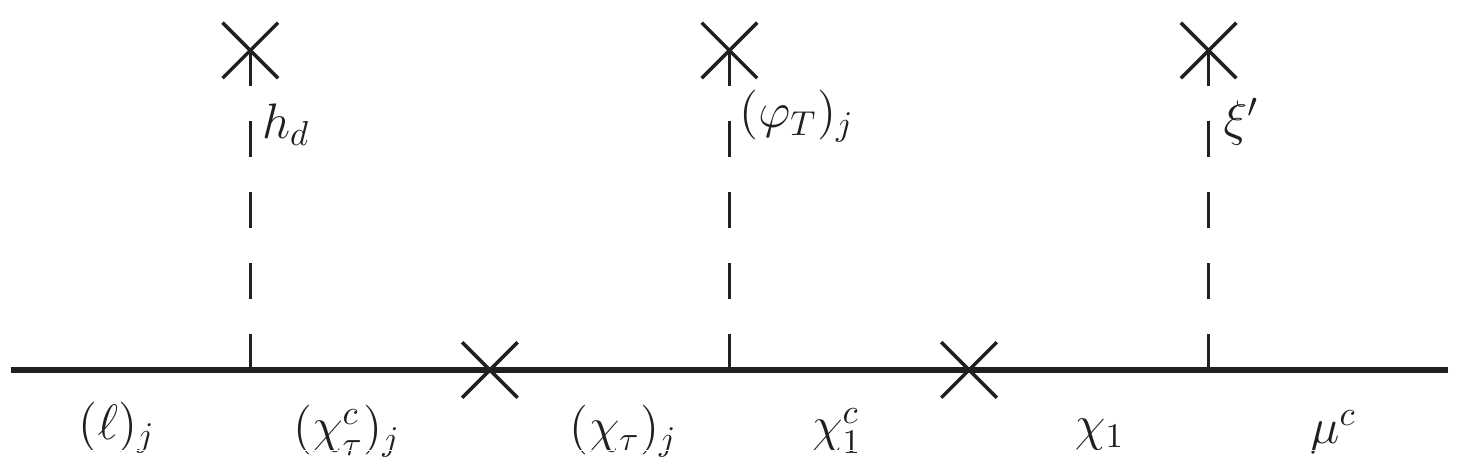}
\caption{\label{muon}Mass term of the muon in the UV complete model.}
\end{center}
\end{figure*}

In the complete model several superfluous effective terms of the original model are absent (due to the messenger content). Focusing on the terms that produce the $\mu, e$ masses we see that the completed model only has one of each, with e.g. the $\mu$ mass arising through the diagram in figure \ref{muon}. The conclusion is that UV completions can readily avoid terms that are present at the non-renormalisable level, enabling more control over the model and therefore increasing predictivity.

\section{Multi-Higgs alignment}

The work described extremely briefly here is based in \cite{MHDM}.
Exact Yukawa alignment in multi-Higgs double models (MHDM) can be achieved through FSs through two requirements: a single FS Invariant Combination (FSIC) is allowed for each family; all Higgs $SU(2)$ doublets FS singlets. The single allowed FSIC can then be invariant under the SM by coupling to any of the doublets. Symbolically (or explicitly, depending on the familon structure) these requirement can be summarised as:
\begin{equation}
\label{eq:Lf}
\mathcal{L}_{f} =  \sum^N_{A=1} c^f_A H_A (F_j \hat{\chi}_f^{jl} f^c_l) \,.
\end{equation}
Where $f$ denotes the family (fermion doublet $F$ and singlet $f^c$), $A$ runs over the $N$ doublets $H_A$. $c^f_A$ is explicitly the only Yukawa coefficient for that family and doublet: each Yukawa has the same structure, given by the spurion (or actual familon) so the Yukawas of each $H_A$ are aligned.
As a generalisation, dropping the constraining single FSIC requirement while maintaining the Higgs as FS singlets means that FSs with a single dominant FSIC can achieve approximate Yukawa alignment - conveniently in line with the requirements of the charged fermion structures (strongly hierarchical third generation). The conclusion: FSs are a potential solution to the flavor problem associated with MHDM.

%% file: Author/dinggj.tex
\begin{center}
{\bf Abstract}
\end{center}
\vskip5.mm

We construct a supersymmetric (SUSY) $SU(5)$ model with the flavor symmetry $S_4\times Z_3\times Z_4$. Three generations of adjoint matter fields are introduced to generate the neutrino masses via the combined type I and type III see-saw mechanism. The first two generations of the the $\mathbf{10}$ dimensional representation in $SU(5)$ are assigned to be a doublet of $S_4$, the second family $\mathbf{10}$ is chose as the first component of the doublet, and the first family as the second component. Tri-bimaximal mixing in the neutrino sector is predicted exactly at leading order, charged
lepton mixing leads to small deviation from the tri-bimaximal mixing pattern. Subleading contributions introduce corrections of order $\lambda^2_c$ to all three lepton mixing angles. The model also reproduces a realistic pattern of quark and charged lepton masses and quark mixings.

\vskip5.mm

\section{Introduction}
So far there is convincing evidence that the so-called solar and atmospheric anomaly can be well explained through the neutrino
oscillation. A very good approximation for the lepton mixing matrix is provided by the so-called Tri-bimaximal (TB) pattern \cite{TBmix}, which suggests the following values of the mixing angles
\begin{equation}
\label{1}\sin^2\theta^{TB}_{12}=\frac{1}{3},~~~\sin^2\theta^{TB}_{23}=\frac{1}{2},~~\sin\theta^{TB}_{13}=0
\end{equation}
which agrees at about the $1\sigma$ level with the data. We note that recently new data from T2K collaboration \cite{Abe:2011sj} and corresponding fits \cite{Fogli:2011qn,Schwetz:2011qt} indicate a $3\sigma$ evidence of a non-vanishing $\theta_{13}$ with a relatively "large" central value. The simple structure of the TB mixing matrix suggests that there may be some symmetry underlying the lepton sector. In the past years, it is found that the flavor symmetry based on finite discrete groups particularly $A_4$ is particularly suitable to reproduce this constant texture. $S_4$ is claimed to be the minimal group which can predict the TB mixing without ad hoc assumptions, from the group theory point of
view \cite{Lam:2008rs}. In this work, we shall present a model combining the $S_4$ flavor symmetry with the $SU(5)$ grand unified theory (GUT) \cite{Ding:2010pc}.

\section{The model}
\begin{table}[hptb]
\begin{center}
\begin{tabular}{|c|c|c|c|c||c|c|c|c|c||c|c|c|c|c|c|c|}\hline\hline
Fields& $T_3$  & $(T_2,T_1)^{T}$ & $F$ &  $A$ &$H_5$ & $H_{45}$ &
$H_{\overline{5}}$ &  $H_{\overline{45}}$&$H_{24}$&$\chi$ &
$\varphi$ & $\zeta$& $\phi$ & $\eta$ & $\Delta$& $\xi$\\ \hline

$SU(5)$&$\mathbf{10}$&$\mathbf{10}$&$\overline{\mathbf{5}}$
&$\mathbf{24}$ & $\mathbf{5}$& $\mathbf{45}$&$\overline{\mathbf{5}}$
&$\overline{\mathbf{45}}$ &$\mathbf{24}$ & $\mathbf{1}$&
$\mathbf{1}$ & $\mathbf{1}$ &$\mathbf{1}$ & $\mathbf{1}$& $\mathbf{1}$&
$\mathbf{1}$\\ \hline

$\rm{S_4}$& $\mathbf{1_1}$& $\mathbf{2}$ &
$\mathbf{3_1}$&$\mathbf{3_1}$ & $\mathbf{1_1}$ & $\mathbf{1_1}$ &
$\mathbf{1_1}$ & $\mathbf{1_1}$&$\mathbf{1_1}$ & $\mathbf{3_1}$& $\mathbf{2}$ & $\mathbf{1_2}$& $\mathbf{3_1}$ & $\mathbf{2}$ & $\mathbf{3_1}$ & $\mathbf{1_1}$ \\
\hline

$\rm{Z_{3}}$& 1 & $\omega$& 1& 1  & 1 & 1 & $\omega$ & 1& 1& 1  &1 & $\mathbf{1}$&
$\omega^2$ & $\omega^2$ & $\omega$& $\omega$\\ \hline

$\rm{Z_{4}}$&1  & $i$& $-i$ & $i$  & 1 & -1 & $1$ & -1&  1 &-1 &-1 & $-1$&
$i$ & $i$ & $i$ & $i$ \\ \hline

$U(1)_R$&1  & 1& 1 & 1  & 0 &0 & $0$ &0&  0 &0 &0 &0 &0  & 0 & 0 &0
\\\hline\hline
\end{tabular}
\end{center}
\caption{\label{tab:trans}Fields and their transformation properties
under the symmetry groups $SU(5)$, $S_4$, $Z_3$ and $Z_4$, where
$\omega=e^{i2\pi/3}=(-1+i\sqrt{3})/2$.}
\end{table}
The flavor symmetry group of the model is $S_4\times Z_3\times Z_4$, where the auxiliary symmetry $Z_3\times Z_4$ plays an important role in
eliminating unwanted couplings, ensuring the needed vacuum alignment and reproducing the observed fermion mass hierarchies.
We introduce three generations chiral superfields $A$ in the adjoint $\mathbf{24}$ representation in addition to $\overline{\mathbf 5}$ matter fields denoted by $F$ and the tenplet $\mathbf{10}$ dimensional matter fields denoted by $T_{1,2,3}$. The neutrino masses are generated through the combination of type I and type III see-saw mechanism in the present model. In the Higgs sector $H_{24}$, $H_5$, $H_{\overline{5}}$, $H_{45}$ and
$H_{\overline{45}}$ are included. Moreover, flavon fields are introduced to spontaneously break the $S_4$ flavor symmetry. The transformation properties of all the fields under $SU(5)$, $S_4$, $Z_3$ and $Z_4$ are summarized in Table \ref{tab:trans}. We note that the first two generations of the tenplet are assigned to be a doublet of $S_4$, the second family $\mathbf{10}$ is taken to be the first component of the doublet, and the first family as the second component. If we reverse the assignment, the down quark and strange quark masses would be of the same order without fine tuning unless some special mechanisms are introduced such as Ref. \cite{Hagedorn:2010th}. Using the standard driving field method, we can show that the scalar components of the flavon
fields acquire vacuum expectation values (VEV) according to the following scheme \cite{Ding:2010pc},
\begin{eqnarray}
\nonumber&&\langle\chi\rangle=\left(\begin{array}{c}v_{\chi}\\v_{\chi}\\v_{\chi}\end{array}\right),~~~\langle\varphi\rangle=\left(\begin{array}{c}v_{\varphi}\\
v_{\varphi}\end{array}\right),~~~\langle\zeta\rangle=0,~~~\langle\phi\rangle=\left(\begin{array}{c}0\\
v_{\phi}\\0\end{array}\right),~~~\langle\eta\rangle=\left(\begin{array}{c}0\\v_{\eta}\end{array}\right),\\
\label{2}&&\langle\Delta\rangle=\left(\begin{array}{c}v_{\Delta}\\0\\0\end{array}\right),~~~~~~~\langle\xi\rangle=v_{\xi}
\end{eqnarray}
In order to produce the observed ratios of up quarks and down quarks and charged lepton masses, the VEVs (scaled by the cutoff $\Lambda$) $v_{\chi}/\Lambda$, $v_{\varphi}/\Lambda$, $v_{\phi}/\Lambda$, $v_{\eta}/\Lambda$, $v_{\Delta}/\Lambda$ and $v_{\xi}/\Lambda$ should be of the same order of magnitude about $\lambda^2_c$ with $\lambda_c\simeq0.22$ being the Cabibbo angle, and we will parameterize the ratio ${\rm VEV}/\Lambda$ by the parameter $\varepsilon$.
\subsection{Neutrino sector}
The LO superpotential which contributes to the neutrino masses is given by
\begin{equation}
\label{22}w_{\nu}=y_{\nu}(FA)_{1_1}H_5+\lambda_1(AA)_{3_1}\chi+\lambda_2(AA)_2\varphi
\end{equation}
It is well-known that there are both $SU(2)$ triplet $\rho_3$ and singlet $\rho_0$ with hypercharge $Y=0$ in the decomposition of the adjoint matter field $A$ with respect to the standard model. From Eq.(\ref{22}) the neutrino Dirac mass matrices read as
\begin{equation}
\label{23}M^D_{\rho_3}=\frac{1}{2}y_{\nu}v_5\left(\begin{array}{ccc}1&0&0\\
0&0&1\\
0&1&0
\end{array}\right),~~~~M^{D}_{\rho_0}=\frac{\sqrt{15}}{10}y_{\nu}v_5\left(\begin{array}{ccc}
1&0&0\\
0&0&1\\
0&1&0
\end{array}\right)
\end{equation}
The last two terms in Eq.(\ref{22}) lead to the Majorana mass matrices of
$\rho_3$ and $\rho_0$
\begin{eqnarray}
\label{24}&&M^{M}_{\rho_3}=\left(\begin{array}{ccc}2\lambda_1v_{\chi}&-\lambda_1v_{\chi}+\lambda_2v_{\varphi}&-\lambda_1v_{\chi}+\lambda_2v_{\varphi}\\
-\lambda_1v_{\chi}+\lambda_2v_{\varphi}&2\lambda_1v_{\chi}+\lambda_2v_{\varphi}&-\lambda_1v_{\chi}\\
-\lambda_1v_{\chi}+\lambda_2v_{\varphi}&-\lambda_1v_{\chi}&2\lambda_1v_{\chi}+\lambda_2v_{\varphi}
\end{array}\right),~~~~M^{M}_{\rho_0}=M^{M}_{\rho_3}
\end{eqnarray}
The light neutrino mass matrix is the sum of type I and type
III see-saw contributions
\begin{eqnarray}
\nonumber M_{\nu}&=&-(M^{D}_{\rho3})^T(M^{M}_{\rho_3})^{-1}M^D_{\rho_3}-(M^{D}_{\rho0})^T(M^{M}_{\rho_0})^{-1}M^D_{\rho_0}\\
\label{25}&=&\left(\begin{array}{ccc}\frac{-a-b}{5b(3a-b)}&\frac{-a+b}{5b(3a-b)}&\frac{-a+b}{5b(3a-b)}\\
\frac{-a+b}{5b(3a-b)}&\frac{-3a^2-4ab+b^2}{5b(9a^2-b^2)}&\frac{-3a^2+2ab-b^2}{5b(9a^2-b^2)}\\
\frac{-a+b}{5b(3a-b)}&\frac{-3a^2+2ab-b^2}{5b(9a^2-b^2)}&\frac{-3a^2-4ab+b^2}{5b(9a^2-b^2)}
\end{array}\right)y^2_{\nu}v^2_5
\end{eqnarray}
where $a\equiv\lambda_1v_{\chi}$ and $b\equiv\lambda_2v_{\varphi}$. This light neutrino mass matrix $M_{\nu}$ is exactly diagonalized by the TB mixing matrix
\begin{equation}
\label{27}U^{T}_{\nu}M_{\nu}U_{\nu}={\rm diag}(m_1,m_2,m_3)
\end{equation}
where $m_{1,2,3}$ are the light neutrino masses, in unit of
$\frac{2}{5}y^2_{\nu}v^2_5$ they are
\begin{eqnarray}
\label{28}&&m_1=\frac{1}{|3a-b|},~~~m_2=\frac{1}{2|b|},~~~m_3=\frac{1}{|3a+b|}
\end{eqnarray}
 The unitary matrix $U_{\nu}$ is given by
\begin{equation}
\label{29}U_{\nu}=U_{TB}\,{\rm
diag}(e^{-i\alpha_1/2},e^{-i\alpha_2/2},e^{-i\alpha_3/2})
\end{equation}
$U_{TB}$ is the well-known TB mixing matrix
\begin{equation}
\label{30}U_{TB}=\left(\begin{array}{ccc}\sqrt{\frac{2}{3}}&\frac{1}{\sqrt{3}}&0\\
-\frac{1}{\sqrt{6}}&\frac{1}{\sqrt{3}}&\frac{1}{\sqrt{2}}\\
-\frac{1}{\sqrt{6}}&\frac{1}{\sqrt{3}}&-\frac{1}{\sqrt{2}}
\end{array}\right)
\end{equation}
The phases $\alpha_1$, $\alpha_2$ and $\alpha_3$ are
\begin{eqnarray}
\label{31}\alpha_1={\rm
arg}(-{y^2_{\nu}v^2_5}/{(3a-b)}),~~~\alpha_2={\rm arg}(-{y^2_{\nu}v^2_5}/{b}),~~~\alpha_3={\rm arg}(-{y^2_{\nu}v^2_5}/{(3a+b)})
\end{eqnarray}
The light neutrino mass spectrum can be both normal hierarchy and inverted hierarchy. Taking into account the experimentally measured mass square differences $\Delta m^2_{sol}$ and $\Delta m^2_{atm}$, we obtain the following constraints on the lightest neutrino mass
\begin{eqnarray}
\nonumber&& m_1\geq0.011{\rm eV},~~~ {\rm for~ normal~ hierarchy}\\
\label{36}&& m_3\geq0.028{\rm eV},~~~ {\rm for ~inverted~ hierarchy}
\end{eqnarray}
\subsection{Charged leptons and quark sector}
The LO superpotential giving rise to the masses of the up type quarks after $S_4$ and $SU(5)$ symmetry breaking, is given by
\begin{eqnarray}
\nonumber
w_u&=&y_tT_3T_3H_5+\sum^4_{i=1}\frac{y_{ci}}{\Lambda^2}TT{\cal
O}^{(1)}_iH_5+\frac{y_{ut1}}{\Lambda^2}TT_3(\phi\chi)_2H_5+
\frac{y_{ut2}}{\Lambda^2}TT_3(\eta\varphi)_2H_5\\
\label{3}&&+\frac{y_{ut3}}{\Lambda^2}TT_3\eta\zeta H_5+\frac{y_{ct}}{\Lambda}TT_3\eta
H_{45}
\end{eqnarray}
with ${\cal
O}^{(1)}=\{(\phi\phi)_{1_1},(\phi\phi)_2,(\eta\eta)_{1_1},(\eta\eta)_2\}$.
The superpotential generating the masses of down quarks and charged
lepton is
\begin{eqnarray}
\nonumber w_d&=&\frac{y_b}{\Lambda}T_3F\phi
H_{\overline{5}}+\frac{y_{s1}}{\Lambda^2}(TF)_{3_1}(\Delta\Delta)_{3_1}H_{\overline{45}}+\frac{y_{s2}}{\Lambda^2}(TF)_{3_1}\Delta\xi
H_{\overline{45}}+\sum^9_{i=1}\frac{y_{di}}{\Lambda^3}T_3F{\cal
O}^{(2)}_iH_{\overline{5}}\\
\label{9}&&+\sum^{6}_{i=1}\frac{x_{di}}{\Lambda^3}T_3F{\cal
O}^{(3)}_iH_{\overline{45}}+\sum^{7}_{i=1}\frac{z_{di}}{\Lambda^3}TF{\cal
O}^{(4)}_iH_{\overline{5}}+...
\end{eqnarray}
where dots stand for higher dimensional operators.
\begin{eqnarray}
\nonumber &&{\cal
O}^{(2)}=\{\chi^2\phi,\chi^2\eta,\varphi\chi\phi,\varphi\chi\eta,\varphi^2\phi,\chi\phi\zeta,\chi\eta\zeta,\varphi\phi\zeta,\phi\zeta^2\}\\
\nonumber&&{\cal
O}^{(3)}=\{\phi^3,\phi^2\eta,\phi\eta^2,\Delta^3,\Delta^2\xi,\Delta\xi^2\}\\
\label{10}&&{\cal
O}^{(4)}=\{\phi^2\chi,\phi^2\varphi,\phi^2\zeta,\eta\phi\chi,\eta\phi\varphi,\eta\phi\zeta,\eta^2\chi\}
\end{eqnarray}
With the vacuum alignment in Eq.(\ref{2}), we can straightforwardly derive the mass matrix as follows
\begin{eqnarray}
\nonumber&&M_u=\left(\begin{array}{ccc}
0&0&4(y_{ut1}\frac{v_{\phi}v_{\chi}}{\Lambda^2}+y_{ut2}\frac{v_{\eta}v_{\varphi}}{\Lambda^2})v_5\\
0&8(y_{c2}\frac{v^2_{\phi}}{\Lambda^2}+y_{c4}\frac{v^2_{\eta}}{\Lambda^2})v_5&8y_{ct}\frac{v_{\eta}}{\Lambda}v_{45}+4y_{ut1}\frac{v_{\phi}v_{\chi}}{\Lambda^2}v_5\\
4(y_{ut1}\frac{v_{\phi}v_{\chi}}{\Lambda^2}+y_{ut2}\frac{v_{\eta}v_{\varphi}}{\Lambda^2})v_5&-8y_{ct}\frac{v_{\eta}}{\Lambda}v_{45}+4y_{ut1}\frac{v_{\phi}v_{\chi}}{\Lambda^2}v_5&8y_tv_5
\end{array}\right)\\
\nonumber&&M_d=\left(\begin{array}{ccc}y^{d}_{11}\varepsilon^3v_{\overline{5}}&y^d_{12}\varepsilon^3v_{\overline{5}}&y^d_{13}\varepsilon^3v_{\overline{5}}+2y^{d'}_{13}\varepsilon^3v_{\overline{45}}\\
y^d_{21}\varepsilon^3v_{\overline{5}}&2y^d_{22}\varepsilon^2v_{\overline{45}}+y^{d'}_{22}\varepsilon^3v_{\overline{5}}&y^d_{23}\varepsilon^3v_{\overline{5}}\\
2y^d_{22}\varepsilon^2v_{\overline{45}}+y^{d'}_{31}\varepsilon^3v_{\overline{5}}&y^d_{32}\varepsilon^3v_{\overline{5}}&y^d_{33}\varepsilon
v_{\overline{5}}
\end{array}\right)\\
\label{12}&&M_{\ell}=\left(\begin{array}{ccc}y^d_{11}\varepsilon^3v_{\overline{5}}&y^{d}_{21}\varepsilon^3v_{\overline{5}}&-6y^d_{22}\varepsilon^2v_{\overline{45}}+y^{d'}_{31}\varepsilon^3v_{\overline{5}}\\
y^d_{12}\varepsilon^3v_{\overline{5}}&-6y^d_{22}\varepsilon^2v_{\overline{45}}+y^{d'}_{22}\varepsilon^3v_{\overline{5}}&y^d_{32}\varepsilon^3v_{\overline{5}}\\
y^d_{13}\varepsilon^3v_{\overline{5}}-6y^{d'}_{13}\varepsilon^3v_{\overline{45}}&y^d_{23}\varepsilon^3v_{\overline{5}}&y^d_{33}\varepsilon
v_{\overline{5}}
\end{array}\right)
\end{eqnarray}
where the factor of 3 difference in the $(13)$, $(22)$ and $(31)$ elements between $M_d$ and $M_{\ell}$ is the so-called Georgi-Jarlskog factor
\cite{Georgi:1979df}, which is induced by the Higgs $H_{\overline{45}}$. Diagonalizing these mass matrices, we find that the CKM matrix elements are as follows
\begin{eqnarray}
\nonumber&&V_{ud}\simeq V_{cs}\simeq V_{tb}\simeq1\\
\nonumber&&V^{*}_{us}\simeq
-V_{cd}\simeq\frac{y^d_{21}}{2y^d_{22}}\frac{v_{\overline{5}}}{v_{\overline{45}}}\varepsilon+\frac{1}{2}\frac{y_{ct}(y_{ut1}v_{\phi}v_{\chi}+y_{ut2}v_{\eta}v_{\varphi})v_5v_{45}}{y_t(y_{c2}v^2_{\phi}+y_{c4}v^2_{\eta})v^2_5+y^2_{ct}v^2_{\eta}v^2_{45}}\frac{v_{\eta}}{\Lambda}\\
\nonumber&&V^*_{ub}=2\frac{y^d_{22}}{y^d_{33}}\frac{v_{\overline{45}}}{v_{\overline{5}}}\varepsilon+\frac{y^{d'}_{31}}{y^d_{33}}\varepsilon^2-\frac{y_{ut1}}{2y_t}\frac{v_{\phi}v_{\chi}}{\Lambda^2}-\frac{y_{ut2}}{2y_t}\frac{v_{\eta}v_{\varphi}}{\Lambda^2}+\frac{1}{2}\frac{y^2_{ct}(y_{ut1}v_{\phi}v_{\chi}+y_{ut2}v_{\eta}v_{\varphi})v^2_{45}}{y^2_t(y_{c2}v^2_{\phi}+y_{c4}v^2_{\eta})v^2_5+y_ty^2_{ct}v^2_{\eta}v^2_{45}}\frac{v^2_{\eta}}{\Lambda^2}\\
\nonumber&&V^{*}_{cb}\simeq-V_{ts}\simeq\frac{y_{ct}v_{45}}{y_tv_5}\frac{v_{\eta}}{\Lambda}\\
\label{21}&&V_{td}=-2\frac{y^d_{22}}{y^d_{33}}\frac{v_{\overline{45}}}{v_{\overline{5}}}\varepsilon-\frac{y^{d'}_{31}}{y^d_{33}}\varepsilon^2+\frac{y_{ut1}}{2y_t}\frac{v_{\phi}v_{\chi}}{\Lambda^2}+\frac{y_{ut2}}{2y_t}\frac{v_{\eta}v_{\varphi}}{\Lambda^2}+\frac{y_{ct}y^d_{21}}{2y_ty^d_{22}}\frac{v_{45}}{v_5}\frac{v_{\overline{5}}}{v_{\overline{45}}}\frac{v_{\eta}}{\Lambda}\varepsilon
\end{eqnarray}
In order to produce the Cabibbo mixing angle between the first and the second family, for the parameters $y^d_{21}$ and $y^d_{22}$ of order ${\cal O}(1)$ we could choose $v_{\overline{45}}\sim\lambda_cv_{\overline{5}}$.
We note that the observed mass hierarchies of quarks and charged lepton are produced. Moreover, we find the following relations between down quarks and charged lepton masses
\begin{equation}
\label{18}m_{\tau}\simeq m_{b},~~~m_{\mu}\simeq3m_{s}
\end{equation}
These are the well-known bottom-tau unification and the Georgi-Jarlskog relation \cite{Georgi:1979df} respectively. Taking into the non-trivial mixing in the charged lepton sector, the lepton mixing angles are given by
\begin{eqnarray}
\nonumber&&\sin\theta_{13}=\simeq\Big|\frac{y^d_{12}}{6\sqrt{2}y^d_{22}}\frac{v_{\overline{5}}}{v_{\overline{45}}}\varepsilon\Big|,~\sin^2\theta_{12}\simeq\frac{1}{3}+\frac{1}{18}\Big[\frac{y^d_{12}}{y^d_{22}}\frac{v_{\overline{5}}}{v_{\overline{45}}}\varepsilon+(\frac{y^d_{12}}{y^d_{22}}\frac{v_{\overline{5}}}{v_{\overline{45}}}\varepsilon)^{*}\Big],~\sin\theta^2_{23}\simeq\frac{1}{2}+\frac{1}{144}\Big|\frac{y^d_{12}}{y^d_{22}}\frac{v_{\overline{5}}}{v_{\overline{45}}}\varepsilon\Big|^2
\end{eqnarray}
Taking into account the results for quark mixing shown in Eq.(\ref{21}), we have
$|V_{us}|\simeq|\frac{y^d_{21}}{2y^d_{22}}\frac{v_{\overline{5}}}{v_{\overline{45}}}\varepsilon|\sim\lambda_c$.
As a result, the model predicts the deviation of the lepton mixing
from the TB pattern as follows
\begin{eqnarray}
\label{40}&&\sin\theta_{13}\sim\frac{\lambda_c}{3\sqrt{2}}\simeq2.97^{\circ},~~~|\sin^2\theta_{12}-\frac{1}{3}|\sim\frac{2}{9}\lambda_c,~~~|\sin^2\theta_{23}-\frac{1}{2}|\sim\frac{\lambda^2_c}{36}
\end{eqnarray}
The lepton mixing angles are predicted to be in agreement at $3\sigma$ error with the experimental data. It is remarkable that Eq.(\ref{40}) belongs to a set of  well-known leptonic mixing sum rules \cite{King:2005bj}, and the same results have been obtained in Ref.\cite{Hagedorn:2010th}. The above LO predictions for the fermion masses and flavor mixing patterns are correction by the next to leading order high dimensional operators allowed by the symmetry of the model. Detailed analysis has shown that the successful LO predictions for the order of magnitudes of both the CKM matrix elements and the quark masses are not spoiled by the subleading corrections, and all the three leptonic mixing angles receive corrections of $\lambda^2_c$ \cite{Ding:2010pc}, they are still compatible with the current experimental data.

%% file: Author/Esteves.tex
{\bf Abstract}\\
\vskip5.mm
We propose an $A_4$ flavor-symmetric \321 seesaw model where lepton
  number is broken spontaneously.  A consistent two-zero texture
  pattern of neutrino masses and mixing emerges from the interplay of
  type-I and type-II seesaw contributions, with important
  phenomenological predictions. We also discuss the possibility of
the decaying but long-lived Majoron to be a good candidate for dark matter.

\vskip5.mm

We suggest \cite{Esteves:2010sh} a version of the seesaw mechanism containing both
type-I~\cite{Minkowski:1977sc,GellMann:1980vs,Yanagida:1979as,Glashow:1979nm,Mohapatra:1979ia,chikashige:1981ui,Schechter:1980gr,Schechter:1981cv}
and type-II
contributions~\cite{Schechter:1980gr,Schechter:1981cv,Magg:1980ut,Lazarides:1980nt,Mohapatra:1980yp,cheng:1980qt} 
in which we implement an $A_4$ flavor symmetry
with spontaneous violation of lepton
number~\cite{chikashige:1981ui,Schechter:1981cv}. We study the
resulting pattern of vacuum expectation values (vevs) and show that
the model reproduces the phenomenologically consistent and
predictive two-zero texture proposed in Ref.~\cite{Hirsch:2007kh}.

In the presence of explicit global symmetry breaking effects, as
might follow from gravitational interactions, the resulting
pseudo-Goldstone boson - Majoron - may constitute a viable candidate
for decaying dark matter if it acquires mass in the keV-MeV range.
Indeed, this is not in conflict with the lifetime constraints which
follow from current cosmic microwave background (CMB) observations
provided by the Wilkinson Microwave Anisotropy Probe
(WMAP)~\cite{Komatsu:2008hk}.
We also show how the corresponding mono-energetic emission line
arising from the sub-leading one-loop induced electromagnetic decay
of the Majoron may be observed in future X-ray
missions~\cite{herder:2009im}.

\section{model}
\label{sec:A4model}
Our model is described by the multiplet content specified in Table
\ref{tab:QuantumNumbers} where the transformation properties under
the SM and $A_4$ groups are shown (as well as the corresponding
lepton number $L$).  The $L_i$ and $l_{Ri}$ fields are the usual SM
lepton doublets and singlets and $\nu_R$ the right-handed neutrinos.
The scalar sector contains an SU(2) triplet $\Delta$, three Higgs
doublets $\Phi_i$ (which transform as a triplet of $A_4$) and a
scalar singlet $\sigma$. Three additional fermion singlets $S_i$ are
also included.
\begin{table}[ht!]
  \centering
  \caption{Lepton multiplet structure ($Q=T_3+Y/2$)}
  \begin{tabular}{ccccccccccc}\hline\hline
 \hskip 10mm
&\phantom{\hskip 2mm}$L_1$\phantom{\hskip 2mm} &\phantom{\hskip
2mm}$L_2$\phantom{\hskip 2mm} &\phantom{\hskip
2mm}$L_3$\phantom{\hskip 2mm} &\phantom{\hskip
2mm}$l_{Ri}$\phantom{\hskip 2mm} &\phantom{\hskip
2mm}$\nu_{iR}$\phantom{\hskip 2mm} &\phantom{\hskip
2mm}$\Phi_i$\phantom{\hskip 2mm} &\phantom{\hskip
2mm}$\Delta$\phantom{\hskip 2mm}
&\phantom{\hskip 2mm}$\sigma$\phantom{\hskip 2mm}
&\phantom{\hskip 2mm}$S_i$\phantom{\hskip 2mm}\\[+2pt] \hline
$SU(2)$&$2$&$2$&$2$&$1$&$1$&$2$&$3$ &$1$ &$1$ \\[+2pt]
$U(1)_Y$&$-1$&$-1$&$-1$&$-2$&$0$&$-1$&$2$ &$0$ &$0$ \\[+2pt]
\hline
$A_4$&$1'$&$1$&$1''$&$3$&$3$&$3$&$1''$ & $1''$ & $3 $\\[+2pt]
\hline
$L$&1&$1$&$1$&$1$&$1$&$0$&$-2$ & $-2$ & $1$\\[+2pt]\hline\hline
  \end{tabular}
  \label{tab:QuantumNumbers}
\end{table}

Taking into account the information displayed in
Table~\ref{tab:QuantumNumbers}, and imposing lepton number
conservation, the Lagrangian responsible for neutrino masses reads
\begin{align}
  \label{eq:Lagneutrmasses}
 -\mathcal{L}_{L}&= h_{1} \overline{L}_1 \left(\nu_R
    \Phi\right)'_1
+h_{2} \overline{L}_2 \left(\nu_R \Phi\right)_1
+h_{3} \overline{L}_3 \left(\nu_R \Phi\right)''_1
+\lambda L_{1}^{T}C\Delta L_{2}+ \lambda L_{2}^{T}C\Delta L_{1}\nonumber\\
 &+ \lambda^{\prime}
L_{3}^{T}C\Delta L_{3}
+ M_R \left(\overline{S_L} \nu_R\right)_1 + h \left(S^T_L C
S_L\right)'_1\sigma+\hbox{h.c.}\,,
\end{align}
where $h$ and $\lambda$ are adimensional couplings, $M_R$ is a mass
scale and
\begin{align}
\label{eq:PhiDelta}
    \Delta=\left(
  \begin{array}{cc}
    \Delta_0 & -\Delta^+/\sqrt{2}  \\
    -\Delta^+/\sqrt{2} &\Delta^{++}
  \end{array}
\right)\,,\,\Phi_i=\left(
  \begin{array}{c}
    \phi_i^0  \\
    \phi_i^{-}
  \end{array}
\right)\,.
\end{align}
Note that the term ($\nu_R^T C \nu_R)'_1\sigma$ is
 allowed by the imposed symmetry. This term however does not
   contribute to the light neutrino masses to the leading order in the
  seesaw expansion and we omit it.
  Alternatively, such term may be forbidden by holomorphy in a
  supersymmetric framework with the following superpotential terms
\begin{equation}\label{eq:SusyPot}
    \mathcal{W}=\dots+\lambda
    \epsilon_{ab}h_i^\nu\hat{L}^a_i\hat{\nu}^c\hat{H}_u^b+ M_R \hat{\nu}^c\hat{S}
    +\frac{1}{2} h \hat{S}\hat{S}\hat{\sigma}  \nonumber
\end{equation}
where the hats denote superfields and the last term replaces the
corresponding bilinear employed in
Ref.~\cite{mohapatra:1986bd,gonzalez-garcia:1989rw}.
Assuming that the Higgs bosons $\Phi_i$, $\Delta^0$ and $\sigma$
acquire the following vevs
\begin{equation}
\label{eq:2} \vev{\phi^0_1}=  \vev{\phi^0_2}=  \vev{\phi^0_3}=
\frac{v}{\sqrt{3}}, \quad \vev{\Delta^0} = u_{\Delta}, \quad
\vev{\sigma} = u_{\sigma}\,,
\end{equation}
we obtain an extended seesaw neutrino mass matrix
$\mathcal{M}$~\cite{mohapatra:1986bd,gonzalez-garcia:1989rw,deppisch:2004fa}
in the ($\nu_L$, $\nu^c$, S) basis
\begin{equation}\label{eq:neutrmassmatrix}
\mathcal{M}= \left(\begin{array}{ccc}
0 & m_D & 0 \\
m_D^T & 0 & M \\
0 & M^T & \mu \\
\end{array}\right)\;,\;m_D= v\ \hbox{diag}(h_{1},h_{2},h_{3})\ U, \quad
U=\frac{1}{\sqrt{3}}\left(
\begin{array}{ccc}
1 &\omega^2 & \omega\\[+2mm]
1  & 1 & 1\\[+2mm]
1 & \omega &\omega^2
\end{array}
\right)\,,
\end{equation}
with $\omega=e^{2\pi i/3}$, $M=M_R\hspace{0.1cm}\text{diag}(1,1,1)$
and $\mu=u_\sigma h\hspace{0.1cm} \text{diag}(1,w^2,w)$.
This leads to an effective light neutrino mass matrix
$\mathcal{M}_\nu^{\rm I}$ given by
\begin{equation}
\label{eq:invseesaw} \mathcal{M}^{\rm I}_{\nu}=m_D{M^T}^{-1}\mu
M^{-1}m_D^T=
\frac{h v^2 u_{\sigma}}{M_R^2} \left(
\begin{array}{ccc}
 h_{1}^2 & 0 & 0 \\[+2mm]
0 & 0 & h_{2} h_{3} \\[+2mm]
0 & h_{2} h_{3}& 0
\end{array}
\right)\,.
\end{equation}
On the other hand the vev of the triplet, $u_{\Delta}$,  will induce
an effective mass matrix for the light neutrinos from type-II seesaw
mechanism
\begin{equation}
\label{eq:II}
 \mathcal{M}^{\rm II}_{\nu}=2u_{\Delta}\left(
    \begin{array}{ccc}
      0 &\lambda  & 0\\[+2mm]
      \lambda & 0 & 0\\[+2mm]
      0 & 0 & \lambda^{\prime}
    \end{array}
\right)\,,
\end{equation}
and the total effective light neutrino mass matrix will then be
\begin{equation}
  \label{eq:T}
  \mathcal{M}_{\nu}=\mathcal{M}^{\rm I}_{\nu}+\mathcal{M}^{\rm
    II}_{\nu}\,.
\end{equation}

In Ref.\cite{Hirsch:2007kh} it was shown that the neutrino mass
matrix given by Eq.~(\ref{eq:T}) could explain the currently
available neutrino data. In section~\ref{sec:neutrino} we will
present an update of that analysis taking into account the latest
neutrino oscillation data. In \cite{Esteves:2010sh} we show that the minimization of the Higgs potential 
is consistent with the $A_4$ symmetry.
\section{neutrino parameter analysis}
\label{sec:neutrino}

Given the two contributions to the light neutrino mass matrix
discussed in Eqs.~(\ref{eq:invseesaw}) and (\ref{eq:II}) one finds
that the total neutrino mass matrix has the following structure:
\begin{equation}
 \mathcal{M}_{\nu}=\left(
    \begin{array}{ccc}
      a & b & 0\\[+2mm]
      b & 0 & c\\[+2mm]
      0 & c & d
    \end{array}
\right). \label{eq:B1}
\end{equation}
This matrix with two-zero texture has been classified as B1 in
\cite{Frampton:2002yf}. One can show that considering the
$(L_1,L_2,L_3)$ transformation properties under $A_4$ as being
$(1^\prime,1^{\prime\prime},1)$ or $(1^{\prime\prime},1^{\prime},1)$
an effective neutrino mass matrix with
$\mathcal{M}_{\nu}(1,2)=\mathcal{M}_{\nu}(3,3)=0$ is obtained (type
B2 in~\cite{Frampton:2002yf}). Moreover, by choosing
$\Delta,\sigma\sim {\bf 1^\prime}$ and appropriate transformation
properties of the $L_i$ doublets, we could obtain the textures B1
and B2 as well. Still, the configuration $\Delta,\sigma\sim {\bf 1}$
would lead to textures which are incompatible with neutrino data
since, in this case, both type I and type II contributions to the
effective neutrino mass matrix would have the same form. Since the
textures of the type B1 and B2 are very similar in what concerns to
neutrino parameter predictions, we will restrict our analysis to B1,
shown in (\ref{eq:B1}).

In general, the neutrino mass matrix is described by nine
parameters: three masses, three mixing angles and three phases (one
Dirac + two Majorana). From neutrino oscillation experiments we have
good determinations for two of the mass parameters (mass squared
differences) and for two of the mixing angles ($\theta_{12}$ and
$\theta_{23}$) as well as an upper-bound on the third mixing angle
$\theta_{13}$.  Using the 3$\sigma$ allowed ranges for these five
parameters and the structure of the mass matrix in Eq.~(\ref{eq:B1})
we can determine the remaining four parameters.
The phenomenological implications of this kind of mass matrix have
been analysed in Refs. \cite{Hirsch:2007kh} and \cite{Dev:2006qe}.
Here we will update the results in light of the recently determined
neutrino oscillation parameters~\cite{Schwetz:2008er}.

\begin{figure}[!h]
  \centering
  \begin{tabular}{ccc}
    \includegraphics[height=3.6cm]{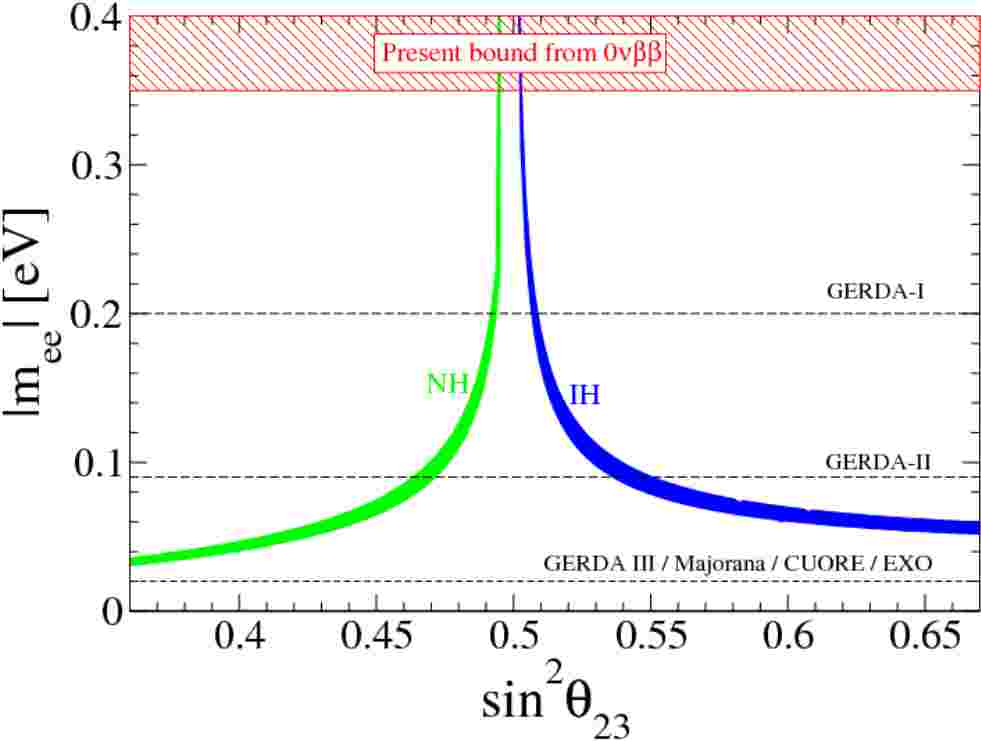}&
    \includegraphics[height=3.6cm]{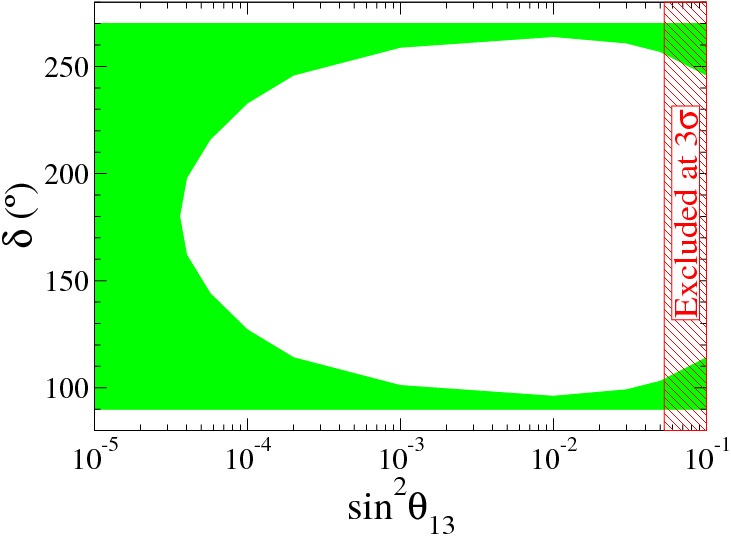}&
    \includegraphics[height=3.6cm]{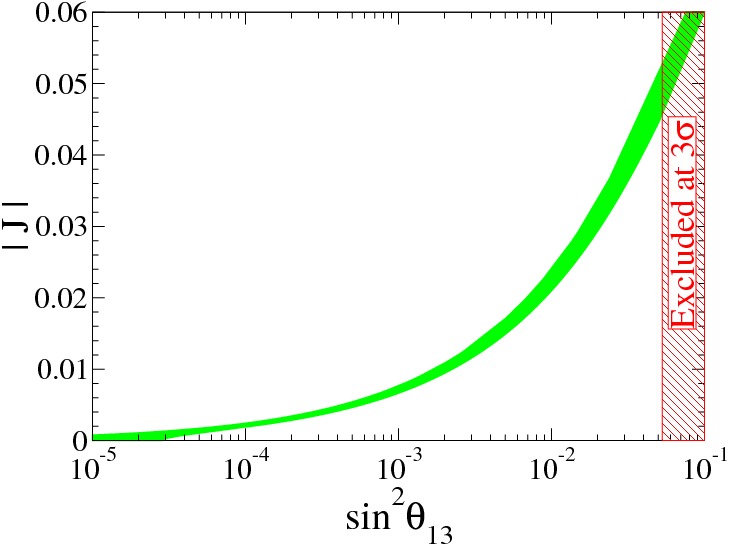}
  \end{tabular}
  \caption{Left panel: correlation between the neutrinoless double beta decay
    amplitude parameter $|m_{ee}|$ and the atmospheric mixing
    parameter.  Experimental sensitivities are also given for
    comparison. Center and right panels: CP violating phase $\delta$ and CP-invariant $J$ in terms
    of the reactor mixing parameter. The 3~$\sigma$-excluded range
    for  $\sin^2\theta_{ij}$ is given for comparison.}
  \label{fig:texture2}
\end{figure}

The main results are shown in \ref{fig:texture2}.  On the left of
figure~\ref{fig:texture2} we plot the
correlation of the mass parameter characterizing the neutrinoless
double beta decay amplitude:
\begin{equation}
\left| m_{ee} \right| = \left| c^2_{13} c^2_{12} m_1 +  c^2_{13}
  s^2_{12} m_2 e^{2i\alpha} + s^2_{13} m_3 e^{2i\beta} \right|,
\end{equation}
with the atmospheric mixing angle $\theta_{23}$.  Here $c_{ij}$ and
$s_{ij}$ stand for $\cos\theta_{ij}$ and $\sin\theta_{ij}$
respectively.  At the zeroth order approximation $m_1/m_3 =
\tan^2\theta_{23}$, and therefore $\theta_{23} < 45^\circ$ for
normal hierarchy (NH), while $\theta_{23} > 45^\circ$ for inverted
hierarchy (IH).  The main result from this plot is a lower bound on
the effective neutrino mass:$\left| m_{ee} \right| > 0.03$ eV.  For
comparison the range of sensitivities of planned experiments as well
as current bounds is also given.
Note that the lower bound we obtain lies within reach of the future
generation of neutrinoless double beta decay experiments.


The center and right panels in Fig.~\ref{fig:texture2} show the CP-violating phase
$\delta$ and the corresponding CP-violating invariant $J$ in neutrino
oscillations
%
%
versus $\sin^2\theta_{13}$.
Note that these hold both for normal and inverted hierarchy spectra.
In the middle panel one sees that $\cos\delta < 0$ since, at first
order in $\sin^2\theta_{13}$, $m_1/m_2 = 1 +
\frac{\cos\theta_{23}}{\cos\theta_{12}\sin\theta_{12}\sin^2\theta_{23}}\sin\theta_{13}\cos\delta$,
and the ratio of masses should satisfy: $m_1/m_2 < 1$. Moreover, for
large $\theta_{13}$ values, where CP violation is likely to be
probed in neutrino oscillations, one can see that our model predicts
maximal violation of CP. Quantitatively, from the right panel one
sees that the 3$\sigma$ bound on $\theta_{13}$: $\sin^2\theta_{13} <
0.053$ implies an upper bound: $|J| \lesssim 0.06$ on the
CP-invariant.

In addition, the two-zero texture structure of our neutrino mass
matrix may have other implications, for example for the expected
pattern of lepton flavor violating decays. In fact, thanks to the
strong renormalization effects due to the presence of the triplet
states, the latter are quite sizeable in sypersymmetric
models~\cite{Rossi:2002zb,Joaquim:2006uz,Esteves:2009vg}.

\section{Majoron dark matter}
\label{sec:MajoronDM}

In models where neutrinos acquire mass through spontaneous breaking
of an ungauged lepton
number~\cite{chikashige:1981ui,Schechter:1981cv} one expects that,
due to non-perturbative effects, the Nambu-Goldstone boson (Majoron)
may pick up a mass that we assume to lie in the kilovolt
range~\cite{berezinsky:1993fm}. This implies that the Majorons will
decay, mainly in neutrinos. As the coupling $g_{J\nu\nu}$ is
proportional to  $\frac{m_\nu}{u_\sigma}$~\cite{Schechter:1981cv},
the corresponding mean lifetime can be extremely long, even longer
than the age of the Universe. As a result the Majoron can, in
principle, account for the observed cosmological dark matter (DM).

This possibility was explored in
Refs.~\cite{lattanzi:2007ux,bazzocchi:2008fh} in a general context.
Here, we just summarize the results. It was found that the relic
Majorons can account for the observed cosmological dark matter
abundance provided
\begin{equation}
  \label{eq:7}
  \Gamma_{J\nu\nu} < 1.3 \times 10^{-19}\ \hbox{s}^{-1}\,\,,\,\,
0.12\ \hbox{keV} < \beta \,m_J < 0.17\ \hbox{keV}\,,
\end{equation}
where $\Gamma_{J\nu\nu}$ is the decay width of $J\rightarrow \nu\nu$
and $m_J$ is the Majoron mass. The parameter $\beta$ encodes our
ignorance about the number density of Majorons, being normalized to
$\beta=1$ if the Majoron was in thermal equilibrium in the early
Universe decoupling sufficiently early, when all other degrees of
freedom of the standard model were excited~\cite{bazzocchi:2008fh}.
In \cite{Esteves:2010sh} we follow their choice and take
\begin{equation}
  \label{eq:9}
  10^{-5} < \beta < 1,
\end{equation}
and calculate both the width into neutrinos as well as the
subleading one-loop induced decay into photons.
%
%
\subsection{numerical results}

In this section we discuss some numerical results regarding the
implementation of the decaying Majoron dark matter hypothesis in our
scenario.  In Ref.~\cite{bazzocchi:2008fh} it was shown that the
experimental limit in the Majoron decay rate into photons is of the
order of $10^{-30}\ \mathrm{s}^{-1}$. It was also shown that, in a
generic seesaw model, a sizeable triplet vev plays a crucial role in
bringing the decay rate close to this experimental bound. Here we
have computed the width of the Majoron into neutrinos and photons in
our extended seesaw model which incorporates the $A_4$ flavor
symmetry, generalizing the models of Ref.~\cite{Hirsch:2007kh}.
\begin{figure}[!htb]
  \centering
  \begin{tabular}{cc}
    \includegraphics[height=3.6cm]{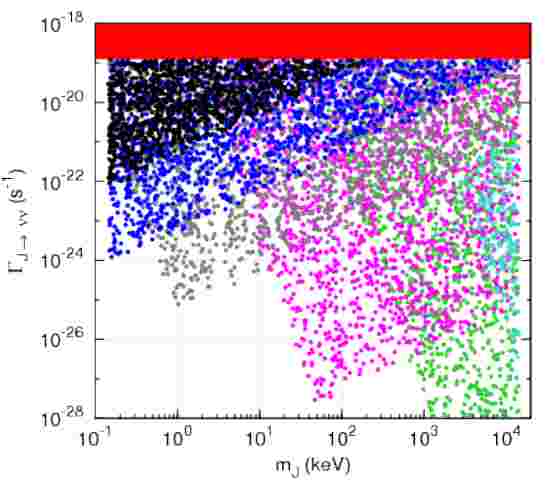}&
    \includegraphics[height=3.6cm]{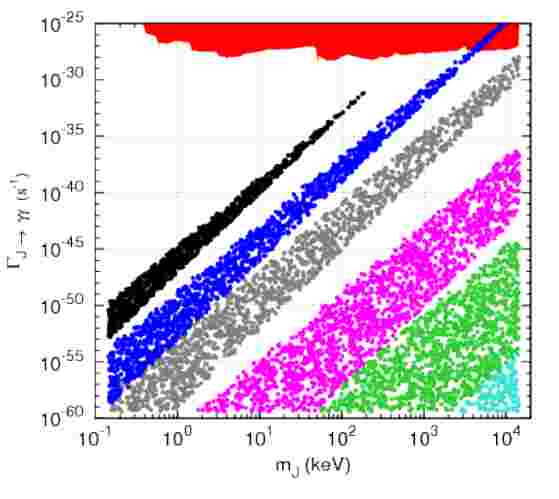}
  \end{tabular}
  \caption{Left panel: $\Gamma_{J\nu\nu}$ as function of the Majoron
    mass respecting Eq.~(\ref{eq:7}) for $u_{\Delta}=$1 eV (turquoise),
    100 eV (dark green), 10keV (magenta), 1MeV (grey), 10MeV (dark
    blue) and 100 MeV (black).  Right panel: $\Gamma_{J\gamma\gamma}$ as
    function of the Majoron mass for the same values of the triplet
    vev as in the left panel. The upper orange shaded region is the
    excluded region from X-ray observations taken from
    Ref.\cite{herder:2009im}. }
\label{fig:JNuNuJGG}
\end{figure}
The results are shown in Fig.~\ref{fig:JNuNuJGG}.  These take into
account the current neutrino oscillation data, discussed in section
\ref{sec:neutrino}.
We chose five values for the triplet vev, $u_{\Delta}=$1~eV
(turquoise), 100~eV (dark green), 10~keV (magenta), 1~MeV (grey) and
10~MeV (dark blue) and 100~MeV (black). For the right panel we
consider only points that satisfy the WMAP constraint (\ref{eq:7})
indicated by the red horizontal band on the top of the left plot.

In order to be able to probe our decaying Majoron dark matter
scenario through the mono-energetic emission line one must be close
to the present experimental limits on the photon decay channel,
discussed in Ref.~\cite{bazzocchi:2008fh} and references therein. As
mentioned, this requires the triplet vev to be sizeable, as shown on
the right panel of Fig.~\ref{fig:JNuNuJGG} for the same choices of
$u_{\Delta}$. In principle there is an additional lower bound on the
Majoron mass coming from the Tremaine-Gunn
argument~\cite{tremaine:1979we}, which, for fermionic dark matter
would be around 500~eV. Under certain assumptions this bound could
be extended to bosons, and is expected to be somewhat
weaker~\cite{Madsen:1990pe}.  The upper orange shaded region is the
excluded region from X-ray observations given in
Ref.~\cite{herder:2009im}. One should point out that, in this model,
because of the vev seesaw relation $u_\Delta u_\sigma\sim v^2$ one
cannot arbitrarily take large values for $u_{\Delta}$ to enhance
$\Gamma_{J\gamma\gamma}$ because then the singlet vev gets
correspondingly smaller values, hence reducing the lifetime of the
Majoron to values in conflict with the WMAP constraint. This
interplay between the CMB bounds and the detectability of the gamma
line is illustrated in Fig.~\ref{fig:JNuNuJGG}, where the dark-blue
points corresponding to $u_{\Delta}=10$ MeV illustrate the
experimental sensitivity to our signal.

%% file: Author/YasamanFarzan.tex
{\bf Abstract}\\
 \vskip5.mm
When the dark matter particles that are trapped inside the Sun
directly annihilate into a pair of neutrinos, the spectrum of the
produced neutrinos will be monochromatic. In this case, the
oscillatory terms in the oscillation probability do not average to
zero. As a result, for a general flavor composition of the
neutrino flux, the number of $\mu$-track events at a neutrino
telescope such as ICECUBE varies during a year when the distance
between the Sun and the Earth changes due to the eccentricity of
the Earth orbit. Information on the dark matter mass as well as on
the flavor structure of its coupling can be derived by studying
this variation.

\vskip5.mm

\section{Introduction}
The local density of Dark Matter (DM) is derived to be of order of
0.39 GeV/cm$^3$ \cite{Catena}. These particles in the vicinity of
the solar system have velocities of order of 200 km/sec relative
to our solar system which is of the same order as the velocity of
the Sun relative to the galaxy center. In principle, when the DM
particles enter the Sun, they can lose kinetic energy and settle
around the Sun center. As a result, the DM density inside the Sun
during its lifetime could have been increased to values much
greater than the universal average. Since the DM annihilation rate
is proportional to the square of the number density, this means
the annihilation rate inside the Sun is much larger than the world
average. If the direct annihilation or the subsequent decay of the
annihilation products creates neutrinos, we will expect a neutrino
flux from the Sun center detectable at neutrino telescopes. This
is the basis of the famous ``indirect DM search'' method which has
been extensively studied in the literature. Considering that no
alternative mechanism is known to give rise to more than $\sim 10$
events at a km$^3$-size neutrino telescope with $E_\nu \gg 10$~MeV
pointing towards the Sun, registering a statistically significant
excess will be a conclusive way of establishing DM particles and
their annihilation.

The neutrinos produced in the Sun center oscillate on their way to
the neutrino telescopes. If the flux at the production is
democratic ({\it i.e.,}
$F_{\nu_e}^0:F_{\nu_\mu}^0:F_{\nu_\tau}^0=1:1:1$), despite the
oscillation, the flavor ratio will remain democratic at Earth:
$$ F_{\nu_e}^{Earth}:F_{\nu_\mu}^{Earth}:F_{\nu_\tau}^{Earth}
=\sum_\alpha F^0_{\nu_\alpha} P(\nu_\alpha \to \nu_e): \sum_\alpha
F^0_{\nu_\alpha} P(\nu_\alpha \to \nu_\mu): \sum_\alpha
F^0_{\nu_\alpha} P(\nu_\alpha \to \nu_\tau)=1:1:1\ $$ where we
have used $\sum_\alpha P(\nu_\alpha \to \nu_\beta)=1$. However,
$F^0_{\nu_e}:F_{\nu_\mu}^0:F_{\nu_\tau}^0$ can in general deviate
from $1:1:1$ so, the oscillation can change the flavor
composition. That is  for
$F^0_{\nu_e}:F_{\nu_\mu}^0:F_{\nu_\tau}^0\ne 1:1:1$,
$\left[\sum_\alpha F_{\nu_\alpha}^0 P(\nu_\alpha \to
\nu_\beta)\right]/\left[\sum_\alpha F_{\nu_\alpha}^0 P(\nu_\alpha
\to \nu_\gamma)\right]$ can be different from
$F_{\nu_\beta}^0/F^0_{\nu_\gamma}$.

The variation of the distance between the Sun and Earth during a
year is of order of 5 million kilometers which is of the same
order as the oscillation length $L_{osc}=4\pi E_\nu/\Delta
m_{12}^2$ for $E_\nu\sim 100$~GeV. This suggests a seasonal
variation of the number of $\mu$-track events due to the
oscillatory terms.
Let us take $L(t)$ to be the distance between the Sun and the
Earth which varies with time. Since $L/L_{osc}\gg 1$,   the phase
of the oscillatory terms will be very large leading to averaging
of the oscillatory terms for a continuous energy spectrum of
neutrinos to zero. That is why in majority of papers devoted to
the study of the oscillation of neutrinos from DM pair
annihilation inside the Sun, the oscillatory terms are dropped.
However, as shown and emphasized in \cite{Esmaili:2009ks}, if the
neutrinos are produced directly by the DM pair annihilation, the
spectrum will be composed of a sharp line corresponding to
un-scattered neutrinos plus  a continuous tail corresponding to
the neutrinos that have gone through scattering off the nuclei
inside the Sun. Although the oscillatory terms have been taken
into account in the study of the oscillation of monochromatic
neutrinos  \cite{Blennow}, no emphasis is put on their oscillatory
behavior as it is discussed that in famous models that the dark
matter particles annihilate directly to a neutrino pair, the
mechanism of annihilation is flavor blind and the oscillatory
terms will therefore have no impact. However, it is possible to
build models that lead to a monochromatic neutrino spectrum with a
non-trivial flavor composition
\cite{Farzan:2011ck,Lindner,McCabe}. As shown in
\cite{Esmaili:2010wa}, studying the oscillatory behavior is a
powerful tool to extract information on the DM parameters.

This letter is organized as follows: In section \ref{first}, we
discuss the impact of the oscillatory terms on the number of
events at ICECUBE. In section \ref{second}, we discuss how one can
derive information on DM properties by measuring this variation.
Conclusions are summarized in sect \ref{conclusion}.

\section{The impact of the oscillatory terms \label{first}}

The DM particles trapped by the gravitational potential of the Sun
can come to thermal equilibrium with the nuclei inside the Sun.
 Setting the kinetic energy equal to $3k_BT/2$,
we find the average velocity of the DM particles to be $v\sim
10^{-4} (100~{\rm GeV}/m_{DM})^{1/2}$. That is these particles are
non-relativistic. Their annihilation will practically take place
at rest so the spectrum of the produced neutrinos will be
monochromatic with $E_\nu=m_{DM}$. To be more precise, the
spectrum will be a narrow Gaussian with a width of $\Delta E/E
\sim v\sim 10^{-4}(100 ~{\rm GeV}/m_{DM})^{1/2}$. This width is
too small to lead to averaging of the oscillatory terms to zero:
$$\frac{\Delta m_{12}^2 L}{2E_\nu}\frac{\Delta E_\nu}{E_\nu}<
1\ .$$ In \cite{Esmaili:2009ks}, various effects that can
potentially widen the spectrum have been discussed and found that
they are all negligible except the effect of scattering of the
neutrinos off the nuclei inside the Sun.

The scattering  can be either of the form of Charged Current (CC)
interaction or of the form of Neutral Current (NC) interaction.
The CC scattering will convert  $\nu_e$ ($\bar{\nu}_e$) to an
electron (positron) which will be absorbed inside the Sun. The CC
scattering of $\nu_\mu$ or ($\bar{\nu}_\mu$) will produce a
charged muon (anti-muon) that will be stopped before decay inside
the Sun. Since the subsequent decay will be at rest, the energy of
$\nu_\mu$ ($\bar{\nu}_\mu$) will be too low to trigger a signal at
neutrino telescopes. However, the tau lepton generated by the CC
scattering of the tau neutrino, having a larger mass and a lower
lifetime, will  be relativistic at the decay time, regenerating a
tau-neutrino with high enough energy to be detectable. In sum, the
CC interactions will reduce the height of the sharp line in the
spectrum but on the other hand, they will result in a continuous
tail with $E_\nu<m_{DM}$ corresponding to the regenerated
neutrinos. The NC interactions convert a neutrino to a neutrino of
a lower energy but of the same flavor. The mediator of NC is a
massive $Z$ boson so, unlike the case of massless photon, there is
no singularity in the propagator at forward scattering. Since the
scattering cross section is finite in the forward direction, the
sharp feature in the spectrum will remain sharp. The effect of
scattering is only to reduce the height of the sharp line and add
a lower energy tail. The oscillatory terms in the oscillation
probability will average out for the neutrinos in the tail of the
spectrum. However, for  the un-scattered neutrinos the oscillatory
terms are important and lead to a seasonal variation of the number
of events during a year. In \cite{Esmaili:2009ks}, it is
demonstrated that the integration over the neutrino production
point does not lead to the average out of the oscillatory terms.

The effect of the oscillatory terms on the seasonal variation of
the $\mu$-track events is proportional to the ratio of the number
of neutrinos that leave the Sun without being scattered. If this
ratio is too low, the effect will be too small to be disentangled
from the statistical fluctuation. Increasing $m_{DM}$, the energy
of neutrinos and their scattering cross section will be increased.
At $m_{DM}=500$~GeV, 65 \% of the initial neutrinos will remain
un-scattered when they leave the Sun. On the other hand, if
$m_{DM}$ is below the detection threshold of the neutrino
telescopes, no event can be registered. In our analysis, we assume
that $100~{\rm GeV}<m_{DM}<500~{\rm GeV}$. To leading order, we
expect a variation due to the oscillatory terms both in the number
of $\mu$-track events as well as the cascade-like events. However,
since the detection threshold of the cascade-like events is
higher, we mainly focus on the $\mu$-track events.

The rate of $\mu$-track events in  neutrino telescope,
$dN_\mu/dt$, is proportional to $A_{eff}(\theta[t])/{ [L(t)]^2}$
where
$A_{eff}(\theta[t])$ is the effective area of the detector. The
effective area depends on the angle between  the neutrino momentum
and the axis of the array of PMTs in detectors, $\theta$. Because
of the tilt of the rotation axis of the Earth, this angle changes
as the Earth moves in its orbit around the Sun. $L(t)$ is the
distance between the Sun and the Earth which varies about $3 \%$
during a year. 
 Apparently, even in the absence of the
oscillatory terms, the rate of events changes due to the seasonal
variation in $L(t)$ and $\theta(t)$. To account for this variation
let us define $\tilde{N}$ as follows \begin{equation}
\label{Ntilde} \tilde{N}(t_0,\Delta t)\equiv
{\int_{t_0}^{t_0+\Delta t} ({d N_\mu }/{dt}) ~dt \over
\int_{t_0}^{t_0+\Delta t} A_{eff}(\theta[t])/[L(t)]^2~dt}.
\end{equation} The variation of $\tilde{N}$ is a measure of the effect of
the oscillatory terms. In \cite{Esmaili:2009ks}, using the
numerical codes, it is shown that the seasonal variation of
$\tilde{N} $ can reach as high as 60\%. Thus, registering a few
hundred events will be statistically enough to establish such a
variation. In the next section, we will discuss how one can derive
information from measuring this seasonal variation.

\section{Information from seasonal variation\label{second}}
Suppose the DM pair annihilates to a neutrino pair as ${\rm
DM}+{\rm DM} \to \nu_\alpha +\stackrel{(-)}{{\nu}_\beta}$ with
amplitude ${\mathcal{M}}_{\alpha \beta}$. The emerging
two-particle state can be written as $|\psi\rangle=\sum_{\alpha
\beta} {\mathcal{M}}_{\alpha \beta}|\nu_\alpha(\vec{p}_1)
\stackrel{(-)}{\nu_\beta}\!\!\!(\vec{p}_2) \rangle$. The neutrinos
are emitted back to back so only one of them can reach us. The
density matrix coming towards us will be the following reduced
matrix
$$ {\rho}_{\alpha \beta}|\nu_\alpha \rangle \langle
\nu_{\beta}|~~~~{\rm where}~~~~{\rho}_{\alpha
\beta}=({\mathcal{M}}{\mathcal{M}}^\dagger)_{\alpha \beta}\ .$$
Notice that although the two-particle state $|\psi\rangle$ is a
pure state, the reduced density matrix $\rho$ is not in general
pure: {\it i.e.,} $\rho \log \rho \ne 0$. Annihilation might be
lepton number conserving ${\rm DM}+{\rm DM}\to \nu_\alpha
+\nu_\beta$ or lepton number violating ${\rm DM}+{\rm DM}\to
\nu_\alpha +\bar\nu_\beta$. In the former case,
${\mathcal{M}}_{\alpha \beta}$ is symmetric and if all the
components of $\rho$ are known then ${\mathcal{M}}_{\alpha \beta}$
can be reconstructed. However, there is unfortunately no
observational way to find out if the annihilation is lepton number
conserving or not.

Consider the  density matrix  of the neutrino flux at production,
$\rho$. Let us show its eigenstates   with
$|\nu_\alpha^\prime\rangle$. These eigenstates can be different
both from the mass eigenstates and flavor eigenstates. The density
matrix can be written as
$$\rho =\sum_\alpha n_\alpha|\nu_\alpha^\prime \rangle \langle
\nu_\alpha^\prime|$$ where $\alpha=1,2,3$, $\langle
\nu_\alpha^\prime|\nu_\beta^\prime\rangle=\delta_{\alpha \beta}$
and $n_\alpha >0$.

As $|\nu_\alpha^\prime\rangle$ propagates, it will evolve in time.
At the Sun surface,
$$|\nu_\alpha^\prime;surface\rangle=a_{\alpha 1}|1\rangle+a_{\alpha 2}|2\rangle+
a_{\alpha 3}|3\rangle\ .$$
When the state $|\nu_\alpha^\prime \rangle$ arrives at the
detector it can be written as \beF |\nu_\alpha^\prime;{\rm
detector}\rangle=a_{\alpha 1} |1\rangle+a_{\alpha 2}
e^{i\Delta_{12}}|2\rangle+a_{\alpha 3} e^{i\Delta_{13}} |3\rangle
\ , \label{detstate}\eeF  where $\Delta_{ij}\equiv \Delta m_{ij}^2
L/(2E)$. The number of $\mu$-track events is given by $\sum_\alpha
n_\alpha P(\nu_\alpha^\prime \to \nu_\mu)$. Using
Eq.~(\ref{detstate}), we can write \beF \label{ppp} P(\nu_\alpha
\to \nu_\mu)=\sum_i |a_{\alpha i}|^2|U_{\mu i}|^2 + \ee
$$ 2\Re [ a_{\alpha 1}^* a_{\alpha 2} U_{\mu 1} U_{\mu 2}^* e^{i
\Delta_{12}}]+2\Re [ a_{\alpha 1}^* a_{\alpha 3} U_{\mu 1} U_{\mu
3}^* e^{i \Delta_{13}}]+2 \Re[a_{\alpha 2}^* a_{\alpha 3} U_{\mu
2} U_{\mu 3}^* e^{i (\Delta_{13}-\Delta_{12})}]\ .$$ As discussed
in \cite{Esmaili:2009ks}, not only the effects of the phase
$\Delta_{12}$ do not average to zero, averaging the effects of the
phase $\Delta_{13}$ is not complete, either. To show this, the
following averages are defined
 \be
\label{ooo12}{ O}_{12}(t, \Delta t)\equiv{\int_t^{t+\Delta t} e^{i
\Delta_{12}(t)} A_{eff}(t)L^{-2}(t) dt \over\int_t^{t+\Delta t}
A_{eff}(t)L^{-2}(t) dt}, \eeF and \beF \label{ooo13}{ O}_{13}(t,
\Delta t)\equiv{\int_t^{t+\Delta t} e^{i \Delta_{13}(t)}
A_{eff}(t)L^{-2}(t) dt \over\int_t^{t+\Delta t}
A_{eff}(t)L^{-2}(t) dt} .\eeF For $E_\nu \sim 100~$GeV, numerical
calculation taking into account the varying speed of the Earth
along its orbit shows that $|{O}_{12}|\sim 1$ and $| O_{13}|\sim
0.1$. With naive estimates overlooking the varying speed of the
Earth, $O_{13}$ turns out to be smaller. The average probability
over the time interval $(t,t+\Delta t)$ is \beF \nonumber \langle
P(\nu_\alpha \to \nu_\mu)\rangle|_{t}^{t+\Delta
t}\equiv{\int_{t}^{t+ \Delta t}P(\nu_\alpha \to \nu_\mu) A_{eff}
L^{-2}(t) dt \over\int_{t}^{t+ \Delta t} A_{eff} L^{-2}(t)}=\ee
\beF \label{average}\sum_i |a_{\alpha i}|^2|U_{\mu i}|^2+2
\Re[a_{\alpha 1}^*a_{\alpha 2} U_{\mu 1}U_{\mu 2}^* O_{12}] +{\cal
O}(a_{\alpha i}^2U_{\mu i}^2O_{13}).\ee

A similar discussion holds for anti-neutrinos
\cite{Esmaili:2009ks}.
Within $(t,t+\Delta t)$, the un-scattered neutrinos induce
muon-track events proportional to \beF \label{KT} K(t,\Delta
t)\equiv \sum_\alpha n_\alpha\left(\langle P(\nu_\alpha \to
\nu_\mu)\rangle|_{t}^{t+\Delta
t}+\frac{\sigma(\bar{\nu})}{\sigma({\nu})}\langle
P(\bar{\nu}_\alpha \to \bar{\nu}_\mu)\rangle|_{t}^{t+\Delta
t}\right)\ , \eeF where $\langle P(\nu_\alpha \to
\nu_\mu)\rangle|_{t}^{t+\Delta t}$ and $\langle P(\bar{\nu}_\alpha
\to \bar{\nu}_\mu)\rangle|_{t}^{t+\Delta t}$ are the probabilites
averaged over time. Adding up the contribution from the continuous
tail of the spectrum, ${\mathcal{A}}$, the muon track events can
be written as \beF \label{ABK}{\mathcal{A}}+{\mathcal{B}}K(t,\Delta
t), \eeF where ${\mathcal{A}}$ and ${\mathcal{B}}$  are unknown. As
discussed in \cite{Esmaili:2009ks}, measuring the muon-track
events over four time intervals yields $\Delta m_{12}^2/m_{DM}$
and the following combinations
 \beF \label{cte}
{\mathcal{A}}+{\mathcal{B}}\sum_\alpha n_\alpha (|a_{\alpha
i}|^2|U_{\mu i}|^2
+\frac{\sigma(\bar{\nu})}{\sigma(\nu)}|\bar{a}_{\alpha
i}|^2|U_{\mu i}|^2)\eeF and \beF \label{amp}
{\mathcal{B}}\left|\sum_\alpha n_\alpha \left(a_{\alpha 1}^*
a_{\alpha 2} U_{\mu 1}U_{\mu
2}^*+\frac{\sigma(\bar{\nu})}{\sigma(\nu)}\bar{a}_{\alpha 1}^*
\bar{a}_{\alpha 2}U_{\mu 1}^*U_{\mu 2}\right)\right|\ . \eeF As
expected, for a democratic composition with $n_\alpha$ independent
of $\alpha$, using $\sum_\alpha a_{\alpha i}^*a_{\alpha
j}=\delta_{ij}$, the dependence on $a_{\alpha i}$ drops from these
combinations.
\section{Summary of conclusions \label{conclusion}}

Observing a variation of  $\tilde{N}$ [see eq.~(\ref{Ntilde}) for
definition] over $t_0$ or time interval $\Delta t$ will indicate
that the spectrum contains a sharp line coming from the direct
annihilation of the DM pair to a neutrino pair. Measuring the
number of $\mu$-track events over four different time intervals,
the value of $\Delta m_{12}^2/m_{DM}$  and therefore $m_{DM}$  can
be derived. It is also possible to discriminate the democratic
flavor composition against a general composition.


%% file: Author/Ludl.tex
{\bf Abstract}\\
\vskip5.mm
Assuming neutrinos to be Majorana particles, in the basis where the charged lepton mass matrix is diagonal,
there are two texture zeros which, in the limit of a quasi-degenerate neutrino mass spectrum, lead to nearly-maximal
atmospheric neutrino mixing irrespective of the values of the solar and reactor mixing angles. In the same limit
the aforementioned cases of texture zeros also lead to maximal CP violation.
Since texture zeros may always be implemented by the use of Abelian symmetries this scenario could
serve as an alternative to non-Abelian family symmetries.

\vskip5.mm

\section{Introduction}

The commonly used parameterization
for the lepton mixing matrix $U$ is given by
\begin{equation}\label{Udefinition}
U=e^{i\hat{\alpha}} V e^{i\hat{\sigma}},
\end{equation}
where~\cite{Nakamura:2010zzi}
\begin{subequations}
\begin{equation}\label{V}
V = \left( \begin{array}{ccc}
c_{13} c_{12} &
c_{13} s_{12} &
s_{13} e^{-i \delta} \\
- c_{23} s_{12} - s_{23} s_{13} c_{12} e^{i \delta} &
c_{23} c_{12} - s_{23} s_{13} s_{12} e^{i \delta} &
s_{23} c_{13} \\
s_{23} s_{12} - c_{23} s_{13} c_{12} e^{i \delta} &
-s_{23} c_{12} - c_{23} s_{13} s_{12} e^{i \delta} &
c_{23} c_{13}
\end{array} \right),
\end{equation}

\begin{equation}
c_{ij} \equiv \cos{\theta_{ij}},\enspace s_{ij} \equiv \sin{\theta_{ij}},\enspace
\theta_{ij}\in[0,\pi/2],\enspace \delta\in[0,2\pi),
\end{equation}
\end{subequations}
and 
\begin{equation}
e^{i\hat \alpha} = \mbox{diag} \left( 
e^{i\alpha_1},\,  e^{i\alpha_2},\,  e^{i\alpha_3} \right) 
\quad \mbox{and} \quad 
e^{i\hat \sigma}  = \mbox{diag} \left( 
e^{i\sigma_1},\,  e^{i\sigma_2},\,  e^{i\sigma_3} \right).
\end{equation}
In stark contrast to the quark sector, where all three mixing angles are rather small,
in the lepton sector there are two large mixing angles $\theta_{12}\sim 34^{\circ}$
and $\theta_{23}\sim 45^{\circ}$ and the smaller reactor angle $\theta_{13}$. For a long
time a vanishing reactor angle was compatible with the experimental data at the $2\sigma$-level.
However, the latest global fits~\cite{Schwetz:2011zk}, which also take into account the
results of the T2K experiment~\cite{Abe:2011sj}, indicate a non-zero $\theta_{13}$ at the
$3\sigma$-level.
This recent result excludes exact realization of the famous 
Harrison-Perkins-Scott (tribimaximal) mixing matrix~\cite{Harrison:2002er} 
\begin{equation}
V_\mathrm{HPS} \equiv \left( \begin{array}{rrc}
2/\sqrt{6} & 1/\sqrt{3} & 0 \\
-1/\sqrt{6} & 1/\sqrt{3} & 1/\sqrt{2} \\
1/\sqrt{6} & -1/\sqrt{3} & 1/\sqrt{2}
\end{array} \right)
\label{HPS}
\end{equation}
at the $3\sigma$-level. However, as a guide line, tribimaximal mixing is
still very popular. Equation~(\ref{HPS}) has lead to the speculation that
there could be a \textit{non-Abelian} family symmetry
which enforces the mixing matrix of equation~(\ref{HPS}), especially $s_{23}^2=1/2$.
The choice of non-Abelian groups is supported by the fact that the only extremal mixing
angle which can be obtained by means of an Abelian symmetry is $\theta_{13}=0^\circ$~\cite{Low:2005yc}. 
However, if we do not insist on an exact realization of $s_{23}^2=1/2$ but
instead claim $s_{23}^2\approx 1/2$ only, Abelian symmetries may still be good candidates
for horizontal symmetries. The purpose of this report is to present a setting in which
$s_{23}^2\approx 1/2$ is achieved through imposing texture zeros in the neutrino mass
matrix and assuming a quasi-degenerate neutrino mass spectrum~\cite{Grimus:2011sf}.
Since in models with an extended scalar sector texture zeros in mass matrices may always
be explained by Abelian symmetries~\cite{Grimus:2004hf}, Abelian family groups may be used
to implement near maximal atmospheric neutrino mixing in the way described in~\cite{Grimus:2011sf}.
						
\section{The framework}

In this section we shortly want to motivate the assumptions on which the analysis
of~\cite{Grimus:2011sf} is based.
Suppose we are given a model in which the charged lepton mass matrix is sufficiently
close to a diagonal matrix, such that in good approximation its diagonalization does not
contribute to the lepton mixing matrix, i.e.
	\begin{equation}
	U_{\ell}\approx \mathbbm{1} \Rightarrow U\approx U_{\nu}. 
	\end{equation}
Then the lepton mixing matrix almost solely stems
from the diagonalization of the neutrino mass matrix. In this situation it could in
principle be possible to have a model in which the lepton mixing matrix is a function
of the neutrino mass ratios only. If this is the case, in the limit of a quasi-degenerate
neutrino mass spectrum the lepton mixing matrix might look like a matrix of ``pure numbers,''
leading for example to the desired relation $s_{23}^2\approx 1/2$.

Assuming that neutrinos are Majorana particles, their mass matrix $\mathcal{M}_\nu$ is symmetric.
In the basis where the charged lepton mass matrix is diagonal--which is the case we are interested
in--there are seven possibilities for a symmetric neutrino mass matrix with two texture zeros which
are compatible with all experimental data~\cite{Frampton:2002yf}. For a list of references to works on the
phenomenology and realization of these seven texture zeros in models we refer the reader to~\cite{Grimus:2011sf}. 
The main result of~\cite{Grimus:2011sf} is that among the seven types of two texture zeros
described in~\cite{Frampton:2002yf}
the two types\footnote{The symbol $\times$ denotes non-zero matrix elements. Since $\mathcal{M}_\nu$
is symmetric, only two of the three zeros of the mass matrices~(\ref{B3B4}) are independent--hence the name ``two
texture zeros''.}
\begin{equation}\label{B3B4}
\mbox{B}_3: \quad
\mathcal{M}_\nu \sim 
\left( \begin{array}{ccc} 
\times & 0 & \times \\ 0 & 0 & \times \\ \times & \times & \times 
\end{array} \right),
\qquad
\mbox{B}_4: \quad
\mathcal{M}_\nu \sim 
\left( \begin{array}{ccc} 
\times & \times & 0 \\ \times & \times & \times \\ 0 & \times & 0 
\end{array} \right)
\end{equation}
lead to $s_{23}^2\approx 1/2$ in the limit $m_k\gg \sqrt{|\Delta m_{ij}^2|}$ of a quasi-degenerate
neutrino mass spectrum. Since the order of magnitude of the cosmological bounds on the sum
of the neutrino masses is~\cite{Nakamura:2010zzi}
\begin{equation}
\sum_{k=1}^3 m_k<1\,\mbox{eV},
\end{equation}
such a spectrum is not in conflict with the current bounds.

\section{Analysis of the cases B$_3$ and B$_4$}

In the following we will explain the method for the analysis of case $B_3$, the analysis of $B_4$ works
completely analogous.

\begin{figure}
\begin{center}
\includegraphics[width=0.5\textwidth,angle=90]{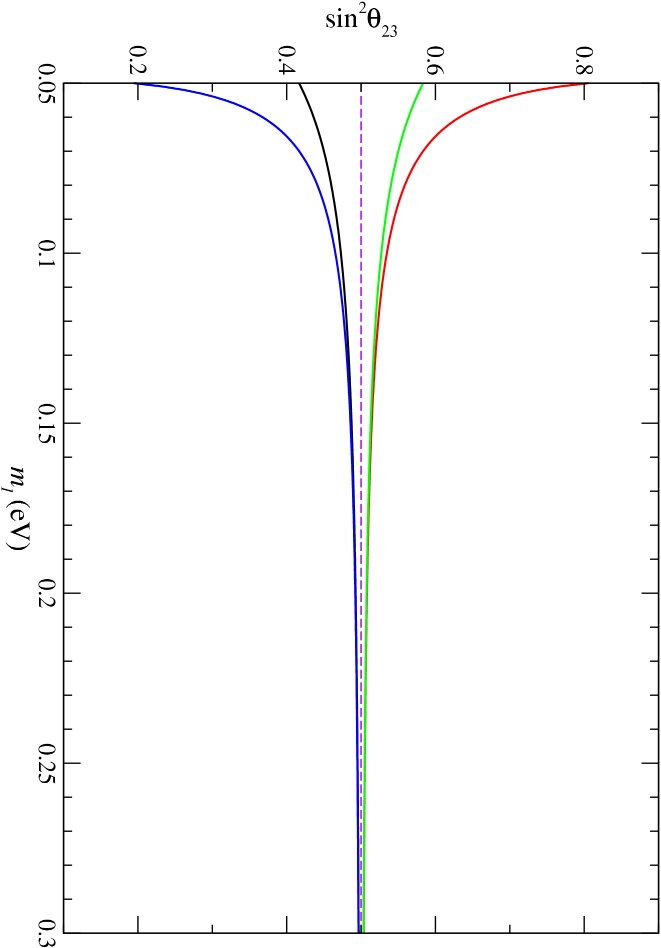}
\end{center}
\caption{$s_{23}^2$ as a function of $m_1$~\cite{Grimus:2011sf}. In descending order the full curves
  refer to case B$_3$ (inverted spectrum), case B$_4$ (normal spectrum), 
  case B$_3$ (normal spectrum), and case B$_4$ (inverted spectrum). The dashed
  line indicates the value 0.5, \textit{i.e.}, maximal atmospheric mixing.
  In this plot, for $s_{12}^2$, $s_{13}^2$, $\Delta m^2_{21}$ and
  $\Delta m^2_{31}$ the best-fit values of~\cite{Schwetz:2011qt} have been used.
  \label{fig-theta23}}
\end{figure}

\begin{figure}
\begin{center}
\includegraphics[width=0.5\textwidth,angle=90]{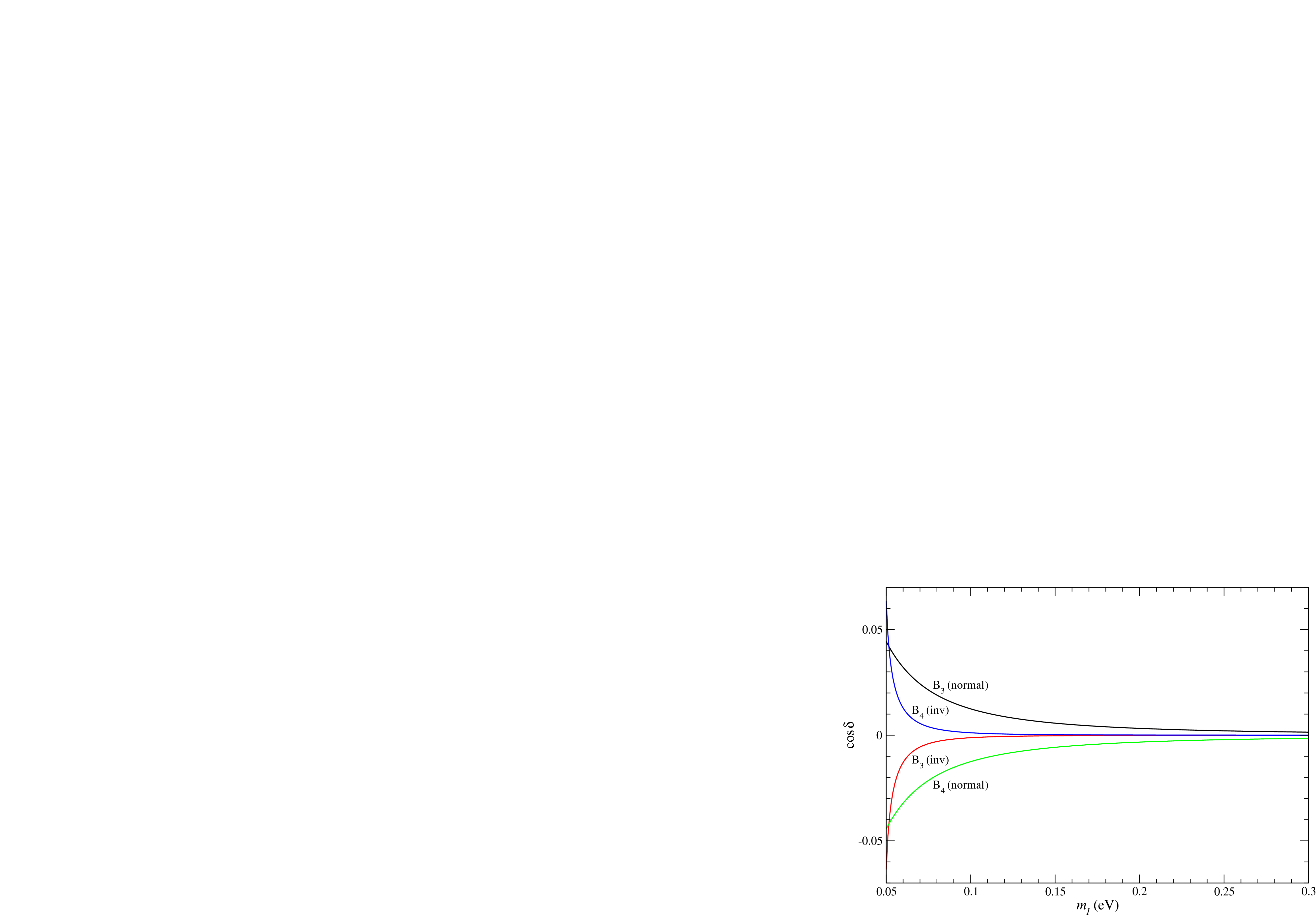}
\end{center}
\caption{$\cos \delta$ as a function of $m_1$~\cite{Grimus:2011sf}. For further details see the
  legend of figure~\ref{fig-theta23}.
  \label{fig-cosd}}
\end{figure}

Inserting $U_{\nu}=U=e^{i\hat{\alpha}} V e^{i\hat{\sigma}}$ into
	\begin{equation}
	\mathcal{M}_{\nu}=U_{\nu}^\ast\,\mathrm{diag}(m_1,\, m_2,\, m_3)\, U_{\nu}^\dagger
	\end{equation}
we find that the condition $(\mathcal{M}_\nu)_{ij}=0$ can be reformulated as
	\begin{equation}
	\sum_{k=1}^3 \mu_k V_{ik} V_{jk}=0 \quad \mbox{with} \quad \mu_k\equiv m_k e^{2i\sigma_k}.
	\end{equation}
Thus, for each of the textures~(\ref{B3B4}), we obtain two linear equations for the complex masses $\mu_k$,
which can be solved for $\mu_1/\mu_3$ and $\mu_2/\mu_3$, respectively~\cite{hep-ph/0201151}.
Using the convention $e^{i\sigma_3}=1$ and defining  
$\epsilon = s_{13} e^{i\delta}$, $t_{12} = s_{12}/c_{12}$ and $t_{23} = s_{23}/c_{23}$
we find
\begin{equation}\label{basic3}
\mathrm{B}_3: \quad
\frac{\mu_1}{m_3} = 
-\frac{t_{12} t_{23} - \epsilon^*}{t_{12} + t_{23} \epsilon}\, t_{23},
\quad
\frac{\mu_2}{m_3} = 
-\frac{t_{23} + t_{12} \epsilon^*}{1 - t_{12} t_{23} \epsilon}\, t_{23}.
\end{equation}
By taking the absolute values of the expressions in equation~(\ref{basic3}) we obtain
$\rho_i=(m_i/m_3)^2$ ($i=1,2$) as a function of the three mixing angles and $\delta$. From these
expressions we can eliminate 
\begin{equation}\label{zeta}
\zeta \equiv 2 t_{12} t_{23} s_{13} \cos\delta =
\frac{t_{12}^2 t_{23}^2 + s_{13}^2 - \rho_1 (t_{12}^2 + t_{23}^2
s_{13}^2)/t_{23}^2}{1 + \rho_1/t_{23}^2}.
\end{equation}
and we finally end up with a cubic equation for $t_{23}^2$:
\begin{equation}\label{t23}
t_{23}^6 + t_{23}^4 \left[ s_{13}^2 + c_{13}^2 \left( c_{12}^2 \rho_1
  + s_{12}^2 \rho_2 \right) \right] -
t_{23}^2 \left[ s_{13}^2 \rho_1 \rho_2 + c_{13}^2 \left( s_{12}^2 \rho_1
  + c_{12}^2 \rho_2 \right) \right] - \rho_1 \rho_2 = 0.
\end{equation}
The resulting equations for the case $B_4$ are similar to the ones for
$B_3$--for further details we refer the reader to~\cite{Grimus:2011sf}.
Using $m_1,\,m_2,\,m_3,\,\theta_{12}$ and $\theta_{13}$ as an input,
equations~(\ref{basic3}), (\ref{zeta}) and (\ref{t23}) allow to determine $\theta_{23}$,
$\delta$ and the Majorana phases $2\sigma_{1,2}$. The numerical results for
$\mathrm{sin}^2\theta_{23}$ and $\mathrm{cos}\,\delta$ as a function of the neutrino mass $m_1$
are shown in figures~\ref{fig-theta23} and~\ref{fig-cosd}. Figure~\ref{fig-theta23} impressively shows
that $\mathrm{sin}^2\theta_{23}$ approaches $1/2$ for $m_1\gg \sqrt{\vert\Delta m_{ij}^2\vert}$.
Taking the limit of quasi-degeneracy $\rho_i\rightarrow 1$ of equation~(\ref{t23}), it becomes clear
that the prediction $\mathrm{sin}^2 \theta_{23}\rightarrow 1/2$ is
independent of the values of $s_{12}^2$ and $s_{13}^2$.
In the same limit $\mathrm{cos}\,\delta$ goes to zero, because $s_{13}^2$ is not
too small\footnote{Note also that $s_{13}^2=0$ is forbidden within the framework of
$B_3$ and $B_4$, because this would lead to $m_1=m_2$--see
equation~(\ref{basic3}).}~\cite{Grimus:2011sf}. Since we consider the limit of a quasi-degenerate
neutrino mass spectrum keeping $\Delta m_{ij}^2$ \textit{fixed}, the limit
$\mathrm{cos}\,\delta\rightarrow 0$ corresponds to maximal CP violation. 
For the Majorana phases one finds the limit $2\sigma_{1,2}\rightarrow \pi$ and the effective mass
$m_{\beta\beta}$ in neutrinoless double beta decay fulfills $m_{\beta\beta}\simeq m_1$ in the
limit of quasi-degeneracy~\cite{Grimus:2011sf}.
%
%

\section{Conclusions}

In this report we have presented two neutrino mass matrices with two texture zeros--namely $B_3$ and $B_4$ in the
nomenclature of~\cite{Frampton:2002yf}--which lead
to near maximal atmospheric neutrino mixing in the limit of a quasi-degenerate neutrino mass
spectrum, provided the charged lepton mass matrix is diagonal.
It can be shown~\cite{Grimus:2011sf} that these two cases
do not automatically lead to a quasi-degenerate neutrino mass spectrum, so the quasi-degeneracy of the neutrinos
has to be \textit{postulated}.
Therefore, texture zeros (which may always be enforced by Abelian family symmetries) together with
the claim for a quasi-degenerate neutrino mass spectrum may serve as an alternative to non-Abelian
symmetries leading to maximal atmospheric neutrino mixing.

%% file: Author/JJonesPerez.tex
{\bf Abstract}\\
\vskip5.mm
In SUSY, the MFV framework is usually called upon in order to ameliorate the New Physics contribution to FCNC. However, this framework, based on a $U(3)^3$ flavour symmetry, is insufficient to solve current tensions in $\Delta F = 2$ processes related to CP Violation. In this work, we analyze the consequences of reducing the symmetry down to a $U(2)^3$ acting on the two lighter generations. We shall outline the $U(2)^3$ framework, and show how it can resolve the current tension between $K^0\to \bar K^0$ and $B^0\to\bar B^0$ mixing, predicting at the same time a larger phase in $B^0_s\to\bar B^0_s$ mixing.

\vskip5.mm

\section{Introduction}
One of the most popular frameworks for low-scale SUSY studies is the so-called CMSSM. Here, sfermions have universal masses at the unification scale, and the RGE-generated flavour structures are of Minimal Flavour Violation (MFV) type~\cite{hep-ph/0207036}. However, there currently exist many reasons to consider other frameworks for the MSSM. To begin with, the lack of new physics signals at the LHC~\cite{arXiv:1109.2352,arXiv:1110.6189} indicate that first generation squark masses should be large, putting at risk the solution of the Higgs mass hierarchy problem, which requires a low mass scale. Moreover, there exists a tension between CP violation observables in the $K$, $B$ and $B_s$ sectors, which MFV structures cannot solve~\cite{arXiv:0805.3887}.

Nevertheless, the lack of significant new physics signals in flavour changing neutral currents (FCNCs) requires the approximate degeneracy of the first two generation squark masses, as well as a MFV-like flavour structure. Furthermore, in addition to LHC data, electric dipole moment experiments also suggest that these squark masses should be very high~\cite{320409}. Thus, all of these facts motivate us to consider a new framework for the MSSM, which should provide some sort of large-mass universality for the first two generations as well as a MFV-like structure, with the third generation being able to have lower masses and structures somewhat different from MFV.

To this end, in this work we describe a $U(2)^3$ flavour symmetry framework acting on the first two quark superfield generations, and briefly show its phenomenological consequences on the $K$, $B$ and $B_s$ sectors. The full details of the framework can be found in~\cite{arXiv:1105.2296}.

\section{Framework}
We consider the quark superfields to transform under a $U(2)_{Q}\times U(2)_{u}\times U(2)_{d}$ group, following:
\begin{align}
  Q  \equiv ( Q_{1}  , Q_{2} )^{\phantom{T}}  \sim (\bar 2,1,1)~, & &
  u_R^{c}\equiv ( u_{1}^c , u_{2}^c )^T    \sim (1,2,1)~, & &
  d_R^{c}\equiv ( d_{1}^c , d_{2}^c )^T    \sim (1,1,2)~, 
\end{align}
with singlet third generation superfields $q_{3}$, $t^c_R$, and $b^c_R$. Considering also a $U(1)_b$ symmetry, under which only $d^c_R$ and $b^c_R$ transform, the only term in the Superpotential in the limit of unbroken symmetry is:
\begin{equation}
 W=y_t ~ q_{3} t^c_R ~H_u~,
\end{equation}
where $y_t$ is the $\ord{1}$ top Yukawa coupling. 

The first step in the construction of the Yukawa matrices lies on the introduction of a spurion $V$ transforming as a (2,1,1), which breaks the symmetry in the $U(2)_Q$ direction. In addition, the $U(1)_b$ symmetry can be broken by the introduction of a $y_b$ spurion. With this, we can write:
\begin{align}
Y_u =y_t \left(\begin{array}{c:c}
 0 & x_t\,V \\\hdashline
 0 & 1
\end{array}\right) & &
Y_d= y_b \left(\begin{array}{c:c}
 0 & x_b\,V \\\hdashline
 0 & 1
\end{array}\right).
\end{align}
In $Y_u$ ($Y_d$), everything above the horizontal line is subject to the $U(2)_Q$ symmetry, while everything to the left of the vertical line is subject to the $U(2)_u$ ($U(2)_d$) symmetry. The parameters $x_t$, $x_b$ are complex, of $\ord{1}$. The size of $y_b$ depends on the value of $\tan\beta$, and in the following we shall assume $\tan\beta\sim10$.

In order to build the masses and mixing of the first two generations we introduce two additional spurions, $\Delta Y_u$ and $\Delta Y_d$, transforming as $(2,\bar 2,1)$ and $(2,1,\bar 2)$, respectively. Combining the various symmetry breaking terms, the Yukawa matrices end up with the following pattern:
\begin{align}
Y_u= y_t \left(\begin{array}{c:c}
 \Delta Y_u & x_t\,V \\\hdashline
 0 & 1
\end{array}\right), & &
Y_d= y_b \left(\begin{array}{c:c}
 \Delta Y_d & x_b\,V \\\hdashline
 0 & 1
\end{array}\right),
\end{align}
where we have absorbed $\ord{1}$ couplings by redefining $\Delta Y_u$ and $\Delta Y_d$.

In order to understand how this structure generates the mass hierarchy and mixings, we shall now go to an explicit parametrization. The leading spurion can always be decomposed as:
\begin{equation}
 V=\epsilon\,U_V  \hat s_2~, \qquad
 \hat s_2 = \left(\begin{array}{c} 0 \\ 1 \end{array}\right)~,
\end{equation}
where $U_V$ is a $2\times 2$ unitary matrix and $\epsilon$ is a real suppression parameter that we require to be of $\ord{\lambda^2_{\rm CKM}}$. Next, the $\Delta Y_u$ and $\Delta Y_d$ spurions can be written in the following way:
\begin{align}
 \Delta Y_u = U_{Q_u}^{\prime\dagger} \Delta Y_u^d\,U^\prime_{u}, & &
 \Delta Y_d = U_{Q_d}^{\prime\dagger} \Delta Y_d^d\,U^\prime_{d},
\end{align}
where $\Delta Y_u^d=\textrm{diag}(\lambda_{u1}, \lambda_{u2})$, and $\Delta Y_d^d=\textrm{diag}(\lambda_{d1}, \lambda_{d2})$, and $U^\prime_i$ are $2\times2$ unitary matrices. We require the $\lambda_i$ to be real, and their size is such that the largest entry is $\lambda_{d2} \approx m_s/m_b =\ord{\epsilon}$. It is now possible to absorb $U^\prime_u$ and $U^\prime_d$ through a redefinition of $u_R^c$ and $d_R^c$, respectively. In addition, we can absorb $U_V$ through a redefinition of $Q_L$, $U^\prime_{Q_u}$ and $U^\prime_{Q_d}$. In such base, the Yukawa matrices assume the explicit form:
\begin{align}
Y_u=\left(\begin{array}{c:c}
 U_{Q_u}^\dagger \Delta Y_u^d & \epsilon\, x_t\hat s_2 \\\hdashline
 0 & 1
\end{array}\right)y_t, & & 
Y_d=\left(\begin{array}{c:c}
 U_{Q_d}^\dagger\Delta Y_d^d & \epsilon\, x_b\hat s_2 \\\hdashline
 0 & 1
\end{array}\right)y_b.
\end{align}

We shall now address the relevant CP phases. We can parametrize each unitary matrix as:
\begin{equation}
 U_{Q_f}=\left(\begin{array}{cc}
 e^{i\omega_{f1}} & 0 \\
0 & e^{i\omega_{f2}}
\end{array}\right)\cdot
\left(\begin{array}{cc}
c_f & s_f\,e^{i\alpha_f} \\
-s_f\,e^{-i\alpha_f} & c_f
\end{array}\right),
\end{equation}
where $c_f$ and $s_f$ are, respectively, the cosine and sine of a mixing angle $\theta_f$. The $\omega_{fi}$ phases can be removed through the rephasing of the components of $u_R^c$ and $d_R^c$, leaving only the phases $\alpha_u$ and $\alpha_d$.

In addition, we also have the phases $\arg(y_t)$, $\arg(y_b)$, $\arg(x_t)$ and $\arg(x_b)$. With a suitable rephasing of the superfields we can remove all of these phases but one. However, in order to maintain a symmetric notation for both Yukawas, we keep the latter two, denoting $x_{f} e^{i\phi_{f}}$, with $x_{f}$ real and positive. Thus, we are left with four phases: $\alpha_u$, $\alpha_d$, $\phi_t$ and $\phi_b$, where only three combinations of them shall be physical.

The Yukawas can now be diagonalized using:
\begin{align}
 Y_u &= U_{uL} Y_u^d U_{uR}^\dagger, & Y_d &= U_{dL} Y_d^d U_{dR}^\dagger,
\end{align}
where, to a very good approximation,the left-handed up-type diagonalization matrices are:
\begin{equation}
 U^\dagger_{uL} =
\left(\begin{array}{ccc}
 c_u &
 s_u\,c_t\,e^{i\alpha_u}  & -s_u s_t e^{i (\alpha_u +\phi_t)}  \\
-s_u\,e^{-i\alpha_u} &  c_u c_t & -c_u s_t  e^{i\phi_{t}}   \\
 0   & s_t  e^{-i\phi_{t}} & c_t
\end{array}\right).
\end{equation}
Here we have $s_t/c_t = \epsilon\, x_t$. An analogous matrix diagonalizes the down-type sector (with $s_u,c_u \to s_d,c_d$, $x_t e^{i\phi_t} \to x_b e^{i\phi_b}$). For both sectors, the right-handed diagonalization matrices are approximately diagonal. The main corrections to left-handed matrices are ${\cal O}(\lambda_{f2}^2)$, while the main deviation from diagonality in the right-handed matrices happens on the $(2-3)$ block, of $\ord{\lambda_{f2}}$.

We now define the CKM matrix $V_{CKM}=(U_{uL}^\dagger\cdot U_{dL})^*$, which acquires the following structure:
\begin{equation}
\label{eq:ckm} 
V_{\rm CKM} \approx \left(\begin{array}{ccc}
 c_u c_d + s_us_d\,e^{i(\alpha_d-\alpha_u)}  &
 -c_u s_d \,e^{-i\alpha_d}  +s_u c_d \,e^{-i\alpha_u}    & s_u s e^{- i (\alpha_u-\xi) } \\
 c_u s_d \,e^{i\alpha_d}  - s_u c_d \,e^{i\alpha_u}    &  c_u c_d + s_us_d\,e^{i(\alpha_u-\alpha_d)}  &
 c_u s  e^{i\xi} \\
-s_d s \,e^{i (\alpha_d-\xi)} & -s c_d  e^{-i\xi} & 1 \\
\end{array}\right)~.
\end{equation}
Here we have $(s/c) e^{i\xi} = \epsilon x_b e^{-i\phi_{b}} -\epsilon x_t e^{-i\phi_{t}}$. In Eq.~(\ref{eq:ckm}) we have set $c=1$. To match this structure  with the standard CKM parametrization, we rephase it imposing real $V_{ud}$, $V_{us}$, $V_{cb}$, $V_{tb}$, and  $V_{cs}$ (which is real at the level of approximation we are working), obtaining 
\begin{equation}
 V_{\rm CKM}=\left(\begin{array}{ccc}
 1- \lambda^2/2 &  \lambda & s_u s e^{-i \delta}  \\
-\lambda & 1- \lambda^2/2   & c_u s  \\
-s_d s \,e^{i (\phi+\delta)} & -s c_d & 1 \\
\end{array}\right),
\label{eq:CKMstand}
\end{equation}
where $\phi = \alpha_d - \alpha_u$, while the phase $\delta$ and the  real and positive parameter $\lambda$ are defined by $s_uc_d - c_u s_d e^{-i\phi}  = \lambda e^{i \delta}$. The $\xi$ phase does not appear on the CKM, but shall be important when we discuss the soft SUSY masses. This means that only the $\alpha_f$ phases are relevant for the CKM phase.

The four parameters $s_u$, $s_d$, $s$, and $\phi$ can be determined completely (up to discrete ambiguities) in terms of the four independent measurements of CKM elements. In particular, using tree-level inputs we get:
\begin{eqnarray}
s  &=& |V_{cb}| = 0.0411 \pm 0.0005~, \\
\frac{s_u}{c_u} &=& \frac{|V_{ub}|}{|V_{cb}|} =  0.095  \pm 0.008~, \\
s_d &=&  - 0.22  \pm 0.01 \qquad {\rm or} \qquad -0.27 \pm 0.01~.
\end{eqnarray}
As a consequence of the $U(2)_Q$ symmetry, $|V_{td}/V_{ts}|$ is naturally of $\ord{\lambda_{\rm CKM}}$ and the smallness of $|V_{ub}/V_{td}|$
is attributed to the smallness of $s_u/s_d$.

\section{Consequences}

Following the pattern outlined above for the Yukawa matrices, we can build the respective soft SUSY masses. If the symmetry is unbroken, all mass matrices shall have the following structure:
\begin{equation}
m^2_{\tilde f}=\left(\begin{array}{ccc}
m_{f_h}^2 & 0 & 0 \\ 
0 & m_{f_h}^2 & 0 \\ 
0 & 0 & m_{f_l}^2 \end{array}\right) 
\end{equation}
where the $m_{f_i}^2$ are real parameters. This structure is modified by the inclusion of spurions, giving:
\begin{eqnarray}
m^2_{\tilde Q} &=&  m_{Q_h}^2
\left(\begin{array}{c:c} 
1  + c_{Qv} V^* V^T   +  c_{Qu}  \Delta Y_u^* \Delta Y_u^{T} 
+  c_{Qd}  \Delta Y_d^* \Delta Y_d^{T} 
&   x_{Q} e^{-i\phi_Q} V^* \\ \hdashline
    x_{Q} e^{i\phi_Q} V^T  \phantom{A^{A^{A^A}}}  &   m_{Q_l}^2/m_{Q_h}^2 
\end{array}\right)~,   \\
m^2_{\tilde d} &=&  m_{d_h}^2
\left(\begin{array}{c:c}
 1  +  c_{dd}  \Delta Y_d^T \Delta Y_d^{*}
&  x_{d} e^{-i\phi_d} \Delta Y_d^T V^* \\ \hdashline
   x_{d} e^{i\phi_d} V^T \Delta Y_d^{*}  \phantom{A^{A^{A^A}}}  &   m_{d_l}^2/m_{d_h}^2 
\end{array}\right)~,  \\
m^2_{\tilde u} &=& m_{u_h}^2
\left(\begin{array}{c:c} 
 1  +  c_{uu}  \Delta Y_u^T \Delta Y_u^{*}
&    x_{u} e^{-i\phi_u} \Delta Y_u^T V^* \\ \hdashline
     x_{u} e^{i\phi_u} V^T \Delta Y_u^{*}  \phantom{A^{A^{A^A}}}  &   m_{u_l}^2/m_{u_h}^2 
\end{array}\right)~,
\end{eqnarray}
where the $c_i$ and the  $x_i$ are real $\ord{1}$ parameters. This structure again gives almost diagonal right-handed squark mixings. On the basis where $Y_d$ is diagonal, the left-handed mixing matrix is given by $W_L^{d\dagger} ~m^2_{\tilde Q}~ W^d_L =  {\rm diag}( m_{Q_1}^2, m_{Q_2}^2, m_{Q_3}^2 )$, resulting in:
\begin{eqnarray}
W^d_L \approx \left(\begin{array}{ccc}
 c_d &  s_d  e^{-i(\delta +\phi)}  & -s_d s_L e^{i\gamma} e^{-i(\delta +\phi)}  \\
-s_d e^{i(\delta +\phi)}  &  c_d & -c_d s_L e^{i\gamma}   \\
  0  &  s_L e^{-i\gamma} & 1 \\
\end{array}\right) 
= \left(\begin{array}{ccc}
 c_d &  \kappa^* & - \kappa^*  s_L e^{i\gamma}  \\
- \kappa  &  c_d & -c_d s_L e^{i\gamma}   \\
  0  &  s_L e^{-i\gamma} & 1 \\
\end{array}\right),
\end{eqnarray}
where  $\kappa =  c_d V_{td}/V_{ts}$ and $s_L e^{i\gamma}  =   e^{-i\xi} (s_{x_b} e^{ - i \phi_b}  + s_Q e^{ - i \phi_Q})$. Notice that the phase $\gamma$ does not involve the phases $\alpha_f$, related to the CKM phase, meaning that it can be treated as a new physics phase.

From this matrix, we can derive simple expressions for the SUSY contribution to FCNC observables in $\Delta F=2$ processes. We find:
\begin{align}
\epsilon_K&=\epsilon_K^\text{SM(tt)}\times\left(1+x^2F_0\right) +\epsilon_K^\text{SM(tc+cc)} &  
\Delta M_d =&\Delta M_d^\text{SM}\times\left|1+xF_0 e^{-2i\gamma}\right| \\
S_{\psi K_S} &=\sin\left(2\beta + \text{arg}\left(1+xF_0 e^{-2i\gamma}\right)\right) & 
S_{\psi\phi} =&\sin\left(2|\beta_s| - \text{arg}\left(1+xF_0 e^{-2i\gamma}\right)\right) \\
\frac{\Delta M_d}{\Delta M_s} &= \frac{\Delta M_d^\text{SM}}{\Delta M_s^\text{SM}},
\end{align}
where $x={s_L^2 c^2_d/|V_{ts}|^2}$. The function $F_0$ is:
\begin{eqnarray} 
F_0 &=& \frac{2}{3} \left(\frac{g_s}{g} \right)^4 \frac{ m_W^2 }{ m^2_{Q_3} } \frac{1}{S_0(x_t)}
\left[ f_{0}(x_g) + \ord{\frac{m_{Q_l}^2}{m_{Q_h}^2}}\right]~, 
\qquad x_g = \frac{ m^2_{\tilde g} }{ m_{Q_3}^2 }~,
\label{eq:F0}
\\
f_{0}(x) & = & \frac{11 + 8 x -19x^2 +26 x\log(x)+4x^2\log(x)}{3(1-x)^3}~, 
\qquad\quad\  f_{0}(1)=1~,
\end{eqnarray}
where $S_0(x_t = m_t^2/m_W^2)\approx 2.4$ is the SM one-loop electroweak coefficient function.

\begin{figure}[h!]
\begin{center}
\includegraphics[width=8cm]{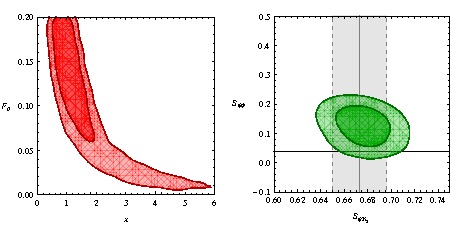}
\end{center}
\caption{Correlation among the preferred values of $x$ and $F_0$ (left) and prediction of $S_{\psi\phi}$ as a function of $S_{\psi K_S}$ (right) as determined from the supersymmetric fit.}
\label{fig:spsiphi}
\end{figure}

In~\cite{arXiv:1105.2296}, we showed that the modifications to these observables allowed us to solve the tensions between $\epsilon_K$, $S_{\psi K_S}$ and $\Delta M_d/\Delta M_s$. This would require a non-zero value for $\gamma$, and would restrict the values of $F_0$ and $x$ to those shown on the left panel of Figure~\ref{fig:spsiphi}. This implies that, to solve the flavour tension, we require values of squark and gluino masses under $1\sim1.5$ TeV.

Moreover, a non-zero value for $\gamma$ means that our framework also provides a contribution to $S_{\psi\phi}$. The requirement of having the correct $S_{\psi K_s}$ restricts the values of $S_{\psi\phi}$ to those shown on the right panel of Figure~\ref{fig:spsiphi}. Notice that having a large value of this observable is favoured by the dimuon anomaly observed by Tevatron~\cite{arXiv:1005.2757}, and shall be accurately probed by the LHCb experiment in the near future.

\section{Conclusions}

We have described a $U(2)^3$ flavour symmetry framework acting on the quark superfields, and shown that it can succesfully accomodate the observed masses and mixings. The framework mantains the degeneracy between the first two generation squark masses,  allowing a light third generation squark mass. This is consistent with the LHC and FCNC bounds, and in principle allows the solution of the Higgs mass hierarchy problem.

In addition, the framework presents deviations from MFV, such that the tensions in the flavour sector can be solved. This requires squarks and gluinos to have masses of a value at most $1\sim1.5$ TeV, and predicts a large $S_{\psi\phi}$. Thus, the framework proves testable, and a good alternative for MFV.

%% file: Author/AvihayKadosh.tex
{\bf Abstract}\\
\vskip5.mm I discuss a model based on an A$_4$ bulk flavor
symmetry in the Randall-Sundrum (RS) setup. After discussing the
setup and leading order results for  the masses and mixings of
quarks and leptons, I elaborate on the effect of higher order
"cross-talk" corrections, their contributions to flavor violating
processes and the resulting constraints on the model parameter
space and the KK mass scale. As a sequel to previous works I focus
on a systematic study of higher order corrections to the PMNS
matrix in light of the new global fits based on recent indications
for $\theta_{13}>0$ from T2K, MINOS and more. In addition, I also
comment on the model new physics contributions to  $\mu\rightarrow
e\gamma$, in light of the new upper bound recently set by the MEG
experiment.

\vskip5.mm

\section{introduction}
Recently  we have proposed a model \cite{A4Warped}  based on a
bulk A$_4$ flavor symmetry \cite{a4} in a warped extra dimension
\,\cite{RS}, in an attempt to account for the  hierarchy of
charged charged fermion masses, the hierarchical mixing pattern in
the quark sector and the large mixing angles and  mild hierarchy
of masses in the neutrino sector. In analogy with a previous RS
realization of A$_{4}$ for the lepton sector \cite{Csaki:2008qq},
the three generations of left-handed quark doublets are unified
into a triplet of A$_4$; this assignment forbids tree level  FCNCs
driven by the exchange of KK gauge bosons. The scalar sector of
the RS-A${}_4$ model consists of two bulk flavon fields, in
addition to a bulk Higgs field. The bulk flavons transform as
triplets of A$_{4}$, and allow for a complete
 "cross-talk" \cite{Volkas} between the A$_{4}\to Z_{2}$
spontaneous symmetry breaking (SSB) pattern associated with the
heavy neutrino sector - with scalar mediator  peaked towards the
UV brane - and the A$_{4}\to Z_{3}$ SSB pattern associated with
the quark and charged lepton sectors - with scalar mediator peaked
towards the IR brane - and allows to obtain realistic masses and
almost realistic mixing angles in the quark sector. A bulk
custodial symmetry \cite{Agashe:2003zs}, broken differently at the
two branes, guarantees the suppression of large contributions to
electroweak precision observables \cite{Carena:2007}, such as the
Peskin-Takeuchi $S$, $T$ parameters. However, the mixing  between
zero modes of the 5D theory and their Kaluza-Klein (KK)
excitations -- after 4D reduction -- may still cause significant
new physics (NP) contributions to SM suppressed flavor changing
neutral current (FCNC) processes (see \cite{A4CPV} and references
therein).

\section{The RS-A$_4$ model}
\begin{figure}
\centering
\includegraphics[width=10truecm]{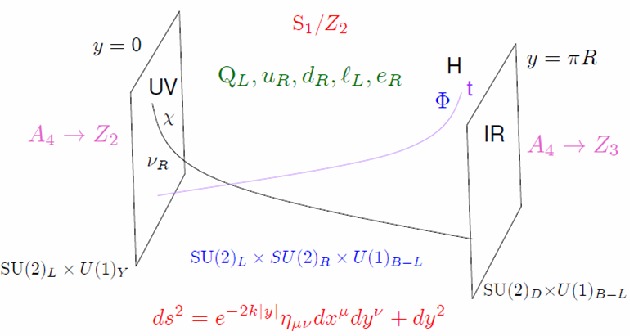}
\caption{A pictorial description of the RS-A$_4$ setup. The bulk
geometry is described by the metric at the bottom and $k\simeq
M_{Pl}$ is the $AdS_5$ curvature scale. All fields propagate in
the bulk and the UV(IR) peaked nature of the heavy RH neutrinos,
the Higgs field, the $t$ quark and the A$_4$ flavons, $\Phi$ and
$\chi$, is emphasized. The SSB patterns of the bulk symmetries on
the UV and IR branes are specified on the side (for A$_4$) and on
the bottom (for $SU(2)_L\times SU(2)_R\times U(1)_{B-L}$) of each
brane. }\label{ModelScheme}
\end{figure}
 The RS-A$_4$ setup  \cite{A4Warped} is illustrated  in
Fig. \ref{ModelScheme}. The bulk geometry is that of a slice of
$AdS_5$ compactified on an orbifold $S_1/Z_2$ \cite{RS}. All 5D
fermionic fields propagate in the bulk and transform under the
following representations of $G=\left(SU(3)_c\times SU(2)_L\times
SU(2)_R\times U(1)_{B-L}\right)\times {\rm A}_4$\cite{A4Warped}:
\begin{equation}
\begin{array}{c}
Q_L \sim \left( 3,2,1,\frac{1}{3} \right) \left( \KK \right) \\
\\\ell_L \sim \left( 1,2,1,-1 \right) \left( \KK \right) \\
\\
\nu_R \sim \left( 1,1,2,0 \right)\left( \KK \right)
\end{array}
\quad\
\begin{array}{c}
 u_R \oplus u'_R
\oplus u''_R \sim \left( 3,1,2,\frac{1}{3} \right)\left(\s \oplus
\spr \oplus \sppr \right) \\
\\
d_R \oplus d'_R \oplus d''_R \sim \left( 3,1,2,\frac{1}{3}
\right)\left(\s \oplus \spr \oplus \sppr \right)\\
\\
\,\,e_R \oplus e'_R \oplus e''_R \sim \left( 1,1,2,-1
\right)\left(\s \oplus \spr \oplus \sppr \right)\, .
\end{array}
\end{equation}
The SM fermions (including RH neutrinos) are identified with the
zero modes of the 5D fermions above. The zero (and KK) mode
profiles are determined by the bulk mass of the corresponding 5D
fermion, denoted by $c_{q_L,u_i,d_i,e_i}k$ and its boundary
conditions \cite{A4CPV}. The scalar sector contains the IR peaked
Higgs field and the UV and IR peaked flavons, $\chi$ and $\Phi$,
respectively. They transform as:
\begin{equation}
\Phi \sim \left( 1,1,1,0 \right) \left( \KK \right),\quad \chi \sim
\left( 1,1,1,0 \right) \left( \KK \right),\quad H\left(1,2,2,0
\right) \left( \s \right)\, .
\end{equation}
The SM Higgs field is identified with the first KK mode of $H$.
All fermionic zero modes acquire masses through Yukawa
interactions with the Higgs field and the A$_4$ flavons after SSB.
The 5D $G$ invariant Yukawa Lagrangian will consist of leading
order(LO) UV/IR peaked interactions and next to leading order
(NLO) "cross-talk" and "cross-brane" interactions\cite{A4Warped}.
 The
LO interactions in the neutrino sector are shown in
\cite{A4Warped} using the see-saw I mechanism, to induce a
tribimaximal (TBM)\cite{TBM} pattern for neutrino mixing while NLO
"cross brane" and "cross talk" interactions, induce small
deviations of $\mathcal{O}(0.04)$, which are still in good
agreement with the current experimental bounds
\cite{Fogli,Tortola}.
 The relevant terms
of the 5D Yukawa lagrangian are of the following form:

\begin{equation}
\mathcal{L}_{5D}^{Y\!uk.}\supset
k^{-2}\,\overline{Q}_L(\overline{\ell}_L)\Phi H
(u_R^{(\prime,\,\prime\prime)},d_R^{(\prime,\,\prime\prime)},(e_R^{(\prime,\,\prime\prime)}))+
k^{-7/2}\,\overline{Q}_L(\overline{\ell}_L)\Phi\chi
H(u_R^{(\prime,\,\prime\prime)},d_R^{(\prime,\,\prime\prime)},(e_R^{(\prime,\,\prime\prime)}))\,,
\label{LYuk5D}\end{equation}
\begin{equation}
\mathcal{L}_{5D}^{\nu}\supset
y_\nu\overline{\ell}_LH\nu_R+(k^{-1/2}y_\nu^\chi\chi+M)\bar{\nu}_R^c\nu_R+
y_\nu^{H\chi}\,k^{-3/2}\,\overline{\ell}_LH\chi\nu_R+y_{\nu}^{\chi^2}k^{-2}\chi^2\bar{\nu}_R^c\nu_R\,.
\label{LNu5D}\end{equation} \noindent Notice that the LO
interactions are peaked towards the UV/IR branes while the NLO
interactions mediate between the two branes due to the presence of
both $\Phi$ (or $H$) and $\chi$.

 \noindent The VEV and physical profiles
for the bulk scalars are obtained by solving the corresponding
equations of motion with a UV/IR localized quartic potential term
and an IR/UV localized  mass term \cite{WiseScalar}. In this way
one can obtain  UV or IR peaked and flat profiles depending on the
bulk mass and the choice of boundary conditions. The resulting VEV
profiles of the RS-A$_4$ scalar sector are:
\begin{equation}
v_{H(\Phi)}^{5D}=H_0(\phi_0)e^{(2+\beta_{H(\phi)})k(|y|-\pi
R)}\qquad
v_\chi^{5D}=\chi_0e^{(2-\beta_\chi)k|y|}(1-e^{(2\beta_\chi)k(|y|-\pi
R)})\,,\label{VEVprofile}
\end{equation}
where $\beta_{H\Phi,\chi}=\sqrt{4+\mu_{H,\Phi,\chi}^2}$, and
$\mu_{H,\Phi,\chi}$ is the bulk mass of the corresponding scalar
in units of $k$, the cutoff of the 5D theory. The following vacua
for the Higgs and the A$_4$ flavons $\Phi$ and $\chi$
\begin{equation}
\langle
\Phi\rangle=(v_\phi,v_\phi,v_\phi)\qquad\langle\chi\rangle=(0,v_\chi,0)\qquad
\langle H\rangle=v_{\footnotesize
H}\left(\begin{array}{cc}1&0\\0&1\end{array}\right)\label{VEValignment},\end{equation}
provide at LO TBM neutrino mixing and zero quark mixing
\cite{a4,Volkas}. The stability of the above vacuum alignment is
discussed in \cite{A4Warped}. The VEV of $\Phi$ induces an A$_4\to
Z_3$ SSB pattern, which in turn induces no quark mixing and is
peaked towards the IR brane. Similarly, the VEV of $\chi$ induces
an A$_4\to Z_2$ SSB pattern peaked towards the UV brane and is in
charge of the TBM mixing pattern in the neutrino sector.
Subsequently, NLO interactions break A$_4$ completely and induce
quark mixing and deviations from TBM, both in good agreement with
experimental data. The Higgs VEV is in charge of the SSB pattern
$SU(2)_L\times SU(2)_R\rightarrow SU(2)_D$, which is peaked
towards the IR brane. The (gauge) SSB pattern on the UV brane is
driven by orbifold BC and a planckian UV localized VEV, which is
effectively decoupled from the model.

\noindent To summarize the implications of the NLO interactions in
the quark and charged lepton sectors, we provide the structure of
the LO+NLO quark mass matrices in the ZMA \cite{A4Warped}:
\begin{equation} \frac{1}{v}(M+\Delta M)_{u,d,e}=\underbrace{\left(\begin{array}{ccc}y_{u,d,e}^{4D}&
 y_{c,s,\mu}^{4D} & y_{t,b,\tau}^{4D}
\\y_{u,d,e}^{4D}& \omega y_{c,s,\mu}^{4D} & \omega^2y_{t,b,\tau}^{4D}\\y_{u,d,e}^{4D}& \omega^2y_{c,s,\mu}^{4D} & \omega
y_{t,b.\tau}^{4D}\end{array}\right)}_{\sqrt{3}U(\omega)
diag(y_{u_i,d_i,e_i}^{4D})}+\left(\begin{array}{ccc}
\hat{f}_\chi^{u,d,e}\tilde{x}_1^{u,d,e}&
\hat{f}_\chi^{c,s,\mu}\tilde{x}_2^{u,d,e}&
\hat{f}_\chi^{t,b,\tau}\tilde{x}_3^{u,d,e}\\0&0&0\\\hat{f}_\chi^{u,d,e}\tilde{y}_1^{u,d,e}&
\hat{f}_\chi^{c,s,\mu}\tilde{y}_2^{u,d,e}&
\hat{f}_\chi^{t,b,\tau}\tilde{y}_3^{u,d,e}\end{array}\right)\,,\label{MDeltaM}
\end{equation}
where $\omega=e^{2\pi i/3}$, $v=174$GeV is the 4D Higgs VEV,
$y^{4D}_{u,c,t,d,s,b,e,\mu,\tau}$ are the effective 4D LO  Yukawa
couplings and $\tilde{x}_i^{u,d,e}$, $\tilde{y}_i^{u,d,e}$ are the
dimensionless coefficients of the 5D NLO Yukawa interactions. The
function $\hat{f}_\chi^{u_i,d_i,e_i}\equiv
y^{4D}_{u_i,d_i,e_i}f_\chi^{u_i,d_i,e_i}\simeq
2y^{4D}_{u_i,d_i,e_i}\beta_\chi
C_\chi/(12-c_{q_L,\ell_L}-c_{u_i,d_i,e_i})\simeq0.05y^{4D}_{u_i,d_i,e_i}$
describes the characteristic suppression of the 4D effective NLO
Yukawa interactions and $C_\chi=\chi_0/M_{Pl}^{3/2}\simeq0.155$.
Finally, the unitary matrix, $U(\omega)$ is the LO left
diagonalization matrix in both the up and down sectors and the
charged lepton sector, $(V_L^{u,d,e})_{LO}$, which is independent
of the LO Yukawa couplings, while $(V_R^{u,d,e})_{LO}=\mathbbm{1}$
(see \cite{A4Warped}). The NLO interactions in the neutrino sector
induce corrections to the (13) and (31) elements of
$M_\nu^{Dirac}$ ($y_\nu^{H\chi}$) and the diagonal elements  of
$M_\nu^{Maj.}$ ($y_\nu^{\chi^2}$) \cite{A4Warped}.

\section{Phenomenology of RS-A$_4$ and constraints on the KK scale}
The main difference between the RS-A$_4$ setup and an anarchic RS
flavor scheme \cite{Agashe:2004cp} lies in the degeneracy of
fermionic LH bulk mass parameters, which implies the universality
of LH zero mode profiles and hence forbids gauge mediated FCNC
processes at tree level, including the KK gluon exchange
contribution to $\epsilon_K$. The latter provides the most
stringent constraint on flavor anarchic models, together with the
neutron EDM \cite{Agashe:2004cp,IsidoriPLB}. However, the choice
of the common LH bulk mass parameter, $c_q^L$ is strongly
constrained by the matching of the top quark mass
($m_t(1.8$\,TeV)$\approx140$\,TeV) and the perturbativity bound of
the 5D top Yukawa coupling, $y_t$. Most importantly, when
considering the tree level corrections to the $Zb\bar{b}$ coupling
 against the stringent EWPM at the Z pole, we
realize \cite{A4Warped} that for an IR scale,
$\Lambda_{IR}\simeq1.8\,$TeV and $m_h\approx 200$\,GeV, $c_q^L$ is
constrained to be larger than 0.35. Assigning $c_q^L=0.4$ and
matching with $m_t$ we obtain $y_t<3$, which easily satisfies the
5D Yukawa perturbativity bound. The constraint on $c_q^L$ from
$Zb\bar{b}$ has a moderate dependence on the Higgs mass, such that
the constraint $\Lambda_{IR}>1.8$\,TeV for $c_q^L=0.35$ and
$m_h\approx200 GeV$, is relaxed to $\Lambda_{IR}>1.3$\,TeV for
$m_h\approx 1$TeV \cite{A4CPV}. The second most significant
constraint comes from 1-loop Higgs mediated dipole operator NP
contributions to the $b\rightarrow s\gamma$ process,
$M_{KK}^{b\rightarrow s\gamma}\gtrsim 1.3\,Y$\,TeV, where $Y$ is
the overall scale of the dimensionless 5D Yukawa coefficients. The
constraints coming from $\epsilon'/\epsilon_K$ and the neutrino
EDM were shown to be weaker by at least factor of 2 (See
\cite{A4CPV}).

Recently the MEG collaboration has established a new upper bound
for the $\mu^+\rightarrow e^+\gamma$ decay, given by
$BR(\mu^+\rightarrow e^+\gamma)<2.4\times10^{-12}\, (90\% {\rm
C.L.})$ \cite{MEG}. The most dominant NP contributions to
$\mu\rightarrow e\gamma$ in the RS-A$_4$ setup will come from
one-loop (dipole)  Higgs diagrams analogous to the quark sector
contributions estimated in \cite{A4CPV}. Using the same formalism
for obtaining a conservative estimation on $BR(\mu^+\rightarrow
e^+\gamma)_{RS-A_4}$, we find that the resulting constraint on $Y$
and $M_{KK}$ is less stringent than the ones mentioned above. A
full analysis of the NP contributions to $\mu\rightarrow e\gamma$,
including (currently neglected) one loop $Z$ mediated
contributions, will be the issue of a separate publication.

\section{Higher order corrections to the PMNS matrix and
$\theta_{13}$}

\begin{center}
\begin{figure}[h!]
\includegraphics[width=6cm]{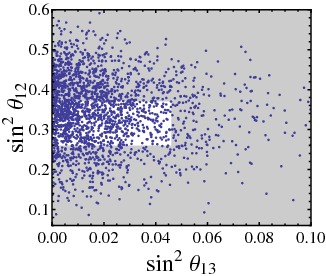}\qquad\qquad
\includegraphics[width=6cm]{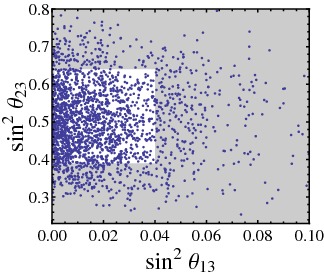}\caption{\it
Model predictions for $\theta_{13}$ vs. $\theta_{12}$ (left) and
$\theta_{23}$ (right) including all dominant higher order and
cross talk effects. The white rectangles represent the $3\sigma$
allowed regions from the global fit of
\cite{Fogli,Tortola}.}\label{NUAngles}
\end{figure}
\end{center}

 The global fits based on the recent indications of $\nu_\mu\rightarrow\nu_e$
appearance in the T2K and MINOS experiments, allow one to obtain a
significance of $3\sigma$ for $\theta_{13}>0$, with best fit
points at around $\theta_{13}\simeq0.15$, depending on the precise
treatment of reactor fluxes  \cite{Fogli,Tortola}. As a
consequence, any flavor model, which predicts TBM at LO, has to be
tested thoroughly for all possible higher order corrections to the
PMNS matrix, such that the new fits are still "accessible" by a
significant portion of the model parameter space.

In our case we are able to obtain analytic expressions for the
corrected diagonalization matrices of both charged leptons and
neutrinos, considering all dominant NLO effects. The resulting
expressions are incredibly long and depend on the $\tilde{x}_i^e,
\tilde{y}_i^e$, $y_\nu^{H\chi}$ and $y_\nu^{\chi^2}$  parameters
and $C_\chi$, which is constrained by the quark sector. Most
importantly, these results do not depend on the LO Yukawa
couplings (Form diagonalizable LO rotation matrices). We performed
a scan over all NLO Yukawa couplings in the range $[0.3,3]$ and
with random complex phases. In Fig.~\ref{NUAngles} we present the
model predictions for $\sin^2\theta_{13}$ vs. $\sin^2\theta_{12}$
(left) and $\sin^2\theta_{23}$ (right) for a set of 3000 randomly
generated points, with the $3\sigma$ allowed ranges of
\cite{Fogli,Tortola} represented by the white rectangles.\\ It can
be seen that the RS-A$_4$ model predictions significantly overlap
with the allowed ranges for the neutrino mixing angles, which
(re-)demonstrates the viability of models predicting TBM at LO.

%% file: Author/SKKANG.tex
{\bf Abstract}\\
\vskip5.mm
We discuss how a parametrization of neutrino mixing matrix reflecting quark-lepton complementarity can be probed by considering phase-averaged oscillation probabilities, flavor composition of neutrino fluxes coming from atmospheric and astrophysical neutrinos and lepton flavor violating radiative decays.
We study about some distinct features of the parametrization by comparing with the triminimal parametrization of perturbations to tri-bimaximal neutrino mixing matrix.

\vskip5.mm

\section{introduction}
The enormous progress made in solar, atmospheric and terrestrial neutrino
experiments \cite{PDG} provides us with very robust evidence for the existence of neutrino
oscillations, a new window to physics beyond the standard model.
The current global fits of the neutrino mixing angles are given at the $1(3)\sigma$ level by
\cite{global1}
\begin{eqnarray}
\theta_{12} &=& 34.4\pm 1.0~\left(^{+3.2}_{-2.9}\right)^{\circ} \nonumber \\
\theta_{23} &=& 42.8^{+4.7}_{-2.9}~\left(^{+10.7}_{-7.3}\right)^{\circ}  \label{fit1} \\
\theta_{13} &=& 5.6^{+3.0}_{-2.7}~(\leq 12.5)^{\circ}. \nonumber
\end{eqnarray}
Those results are well consistent with the so-called tri-bimaximal (TB) neutrino
mixing pattern \cite{tribi}
\begin{eqnarray}
U_0=\left(
\begin{array}{ccc}
\frac{2}{\sqrt{6}} & \frac{1}{\sqrt{3}} & 0 \\
\frac{-1}{\sqrt{6}} & \frac{1}{\sqrt{3}} & \frac{1}{\sqrt{2}}  \\
\frac{1}{\sqrt{6}} & \frac{-1}{\sqrt{3}} & \frac{1}{\sqrt{2}}  \\
\end{array}
\right).
\end{eqnarray}
It corresponds to $\sin2\theta_{12} = 1/3, ~~\sin2\theta_{23} = 1/2$ and $\sin2 \theta_{13} = 0$.
Although TB mixing can be achieved by imposing some flavor symmetries, it is widely accepted that
TB is a good zeroth order approximation to reality and there will be deviations from TB in general.
With this in mind, it is meaningful to parameterize the lepton mixing matrix in such a way that deviations from
TB are manifest.
A useful parametrization of the lepton mixing matrix, so-called ``triminimal" parametrization,  has been proposed such that a mixing angle
in the mixing matrix is given by the sum of a zeroth order angle $\theta_{ij}^0$ and a small perturbation
$\epsilon_{ij}$ with $\theta_{ij}=\theta_{ij}^0+\epsilon_{ij}$.
A merit of this parametrization is that it leads to simple formulas for neutrino flavor mixing
so that the effects of deviations from the TB mixing could be easily probed
as shown in \cite{trimin}, which is not shared by other parameterizations.

In this talk \cite{talk}, we will discuss how the Quark-Lepton Complementarity (QLC) parametrization reflecting a possible hint of the grand unification or quark-lepton symmetry can be probed by considering phase-averaged oscillation probabilities which can be measured from neutrino experiments, flavor composition of neutrino fluxes coming from atmospheric and astrophysical neutrino sources and lepton flavor violating radiative decays.
We note that while consideration of the lepton flavor violating
radiative processes can only be applied to the particular model such as the supersymmetric standard model, that of phase-averaged oscillation
probabilities as well as flavor composition of neutrino fluxes are model independent.
We will also discuss about some distinct features of the QLC parametrization by comparing with the so-called triminimal parametrization of perturbations to tri-bimaximal neutrino mixing matrix \cite{trimin} which has been proposed so that the effects of deviations from the tri-bimaximal mixing could be easily probed.

\section{quark-lepton complementarity}
It has been noted that the solar and atmospheric neutrino mixing angles
$\theta_{12}$ and $\theta_{23}$ measured from neutrino oscillation experiments and the quark mixing angles $\theta_{q_{12}}$
and $\theta_{q_{23}}$ reveal a surprising
relation
\begin{eqnarray}
\theta_{12}+\theta_{q_{12}} \simeq \theta_{23}+\theta_{q_{23}} \simeq 45^{\circ}, \label{qlc}
\end{eqnarray}
which is satisfied by the experimental results
$\theta_{12}+\theta_{q_{12}}=47.4\pm 1.1\left(^{+3.3}_{-3.0}\right)^{\circ} $  and
$\theta_{23}+\theta_{q_{23}}=45.2^{+4.2}_{-2.9}\left(^{+10.8}_{-7.4}\right)^{\circ} $ to within a
few percent accuracy \cite{global1, PDG}.
This quark-lepton complementarity (QLC) relation (\ref{qlc}) has been interpreted  as an
evidence for certain quark-lepton symmetry or quark-lepton unification as shown in Refs. \cite{raidal}.
In the light of the QLC, it is still experimentally allowed for the neutrino mixing matrix
to be composed of a CKM-like matrix and maximal mixing matrices as shown in \cite{skkang2, cabibbo}.
Among possible compositions, in this paper, we consider the following parametrization:
\begin{eqnarray}
U_{PMNS}=R_{32}\left(\frac{\pi}{4}\right)U^{\dagger}_{CKM}R_{21}\left(\frac{\pi}{4}\right), \label{qlc-ckm}
\end{eqnarray}
where $U_{CKM}$ denotes the CKM mixing matrix.
The reason why we consider this particular parametrization for the QLC relations is that it is well compared
with the triminimal parametrization and thus we can simply examine if the effects of deviations from the TB mixing
can reflect the QLC relations by investigating a few observables presented by simple formulas not shared by other parameterizations.
This parametrization can be obtained from the grand unification or quark-lepton symmetry as shown in \cite{skkang2}.
In some unified gauge group such as $SO(10)$, there exist some relations among the Yukawa matrices:
$Y_{\nu}=Y_u=Y_u^T$ and  $Y_e=Y^T_d$, where $Y_{\nu}, Y_{u}, Y_{d}$ and $Y_e$ denote
the Dirac neutrino, up-type quark, down-type quark and charged lepton Yukawa matrices, respectively.
Then, the so-called PMNS neutrino mixing matrix $U_{PMNS}$ is given by $ U_{\rm PMNS} =
V^T_d U_d U_{\rm CKM}^{\dagger}V_M, $ and thus we obtain Eq.(\ref{qlc-ckm}) by taking
$V_d^TU_d=R_{23}(\pi/4)$ and  $V_M=R_{21}(\pi/4)$, where $U_d, V_d$ correspond to the left-handed and right-handed rotation matrices of down-type quark Yukawa matrix, respectively, and
the mixing matrix $V_M$ represents the diagonalizing matrix of
$Y_{\nu}^{diag}V^{\dagger}_0 M_R^{-1}V^{\ast}_0 Y_{\nu}^{diag}$ with a rotation matrix $V_0$ and right-handed heavy Majorana mass matrix $M_R$.
Thus, this parametrization can be used to probe a signal of the grand unification or quark-lepton symmetry.
From the analysis, one can easily prove that this parametrization leads to QLC with an accuracy of order $O(\lambda^2)$.
From now on, we call the parametrization of neutrino mixing matrix given by Eq.(\ref{qlc-ckm}) ``QLC parametrization".

\section{Neutrino Observables}
Let us first take  $U_{CKM}$ as the Wolfenstein parametrization \cite{ckm1}
as follows:
\begin{eqnarray}
U_{CKM}=\left (
\begin{array}{ccc}
1-\frac{1}{2}\lambda^{2}	& \lambda	& A\lambda^{3}(\rho
-i\eta) \\
-\lambda &
1-\frac{1}{2}\lambda^{2}&
A\lambda^{2} \\
A\lambda^{3}(1-\rho-i\eta)	& -A\lambda^{2}& 1
\end{array}
\right ) \; \label{ckm}
\end{eqnarray}
For our numerical calculation, we use the following inputs given by the Particle Data Group \cite{PDG}:
\begin{eqnarray}
\lambda = 0.2257^{+0.0009}_{-0.0010},~~~~A=0.814^{+0.021}_{-0.022}, \nonumber \\
\bar{\rho}=0.135^{+0.031}_{-0.016},~~~~~~\bar{\eta}=0.349^{+0.015}_{-0.017} \label{ckm-fit}
\end{eqnarray}
where $
\bar{\rho}=\rho - \frac{1}{2}\rho \lambda^2 +O(\lambda^4), ~~
\bar{\eta}=\eta-\frac{1}{2}\eta \lambda^2+O(\lambda^4)$.
Inserting Eq.(\ref{ckm}) into Eq.(\ref{qlc-ckm}), we can present the deviations from maximal mixing of the solar and atmospheric mixing angles in powers of $\lambda$: $\theta_{sol}\simeq \pi/4-\lambda$ and $\theta_{atm}\simeq \pi/4-A\lambda^2$.
We also obtain the mixing angle $\theta_{13}$ which is of order $\lambda^3$.

To evaluate the neutrino mixing probabilities for phased-averaged propagation, $P_{\nu_\alpha \leftrightarrow \nu_\beta}$, which is appropriate when the oscillation phase $\Delta m^2 L/4E$ is very large, we need to know $|U_{\alpha i}|^2$ as well discussed in \cite{trimin}.
From Eq.(\ref{qlc-ckm}), we get the matrix form of $|U_{\alpha i}|^2$ which is defined by $\underline{U}_{\alpha i}\equiv |U_{\alpha i}|^2$,
\begin{eqnarray}
\underline{U}=&&\frac{1}{4}\left\{ \left(
\begin{array}{ccc}
2 & 2 & 0 \\
1 & 1 & 2 \\
1 & 1 & 2
\end{array} \right)
+\lambda\left(
\begin{array}{rrr}
4 & -4 & 0 \\
-2 & 2 & 0 \\
-2 & 2 & 0
\end{array} \right) \right.  \label{matrix2} \\
&&+ \left. A\lambda^2\left(
\begin{array}{rrr}
0 & 0 & 0\\
2 & 2 & -1 \\
-2 & -2 & 1
\end{array} \right)
+\lambda^3\left(
\begin{array}{ccc}
-2 & 2 & 0 \\
1-S & -1+S & 0 \\
1+S & -1-S & 0
\end{array} \right) \right\},  \nonumber
\end{eqnarray}
where $S=2A+2A\rho $.
In fact, $\underline{U}_{e3}$ is of order of $\lambda^6$, so we have ignored it.
The neutrino mixing probabilities for phased-averaged propagation is given by $P_{\nu_\alpha \leftrightarrow \nu_\beta}=\sum_{i}
\underline{U}_{\alpha i}\underline{U}_{\beta i}$.
Using  Eq.(\ref{matrix2}), we obtain
\begin{eqnarray}
P_{\nu_e \leftrightarrow \nu_e} & \simeq & \frac{1}{2}+2\lambda^2 \nonumber \\
P_{\nu_e \leftrightarrow \nu_\mu} & \simeq & \frac{1}{4}-(1-\frac{1}{2}A)\lambda^2 \nonumber \\
P_{\nu_e \leftrightarrow \nu_\tau} & \simeq & \frac{1}{4}-(1+\frac{1}{2}A)\lambda^2  \label {prob} \\
P_{\nu_\mu \leftrightarrow \nu_\mu} & \simeq & \frac{3}{8} + \frac{1}{2}(1-A)\lambda^2 \nonumber \\
P_{\nu_\mu \leftrightarrow \nu_\tau } & \simeq & \frac{3}{8} + \frac{1}{2}\lambda^2 \nonumber \\
P_{\nu_\tau \leftrightarrow \nu_\tau} & \simeq & \frac{3}{8} + \frac{1}{2}(1+A)\lambda^2 \nonumber .
\end{eqnarray}
Here, it is interesting to observe that the only terms proportional to $\lambda^2$ survive in each
$P_{\nu_\alpha \leftrightarrow \nu_\beta}$. We have observed that the contributions at the next next leading order are of order $\lambda^4$.
Imposing the experimental results for $\lambda$ and $A$, we predict the values of $P_{\nu_\alpha \leftrightarrow \nu_\beta}$ corresponding
to the best fit values in Eq.(\ref{ckm-fit}) as follows;
\begin{eqnarray}
P_{\nu_e \leftrightarrow \nu_e} &\simeq & 0.6019, ~~~
P_{\nu_e \leftrightarrow \nu_\mu} \simeq  0.2198,  \nonumber \\
P_{\nu_e \leftrightarrow \nu_\tau} &\simeq & 0.1783, ~~~
P_{\nu_\mu \leftrightarrow \nu_\mu} \simeq  0.3797,  \\
P_{\nu_\mu \leftrightarrow \nu_\tau }&\simeq & 0.4005,~~~
P_{\nu_\tau \leftrightarrow \nu_\tau} \simeq  0.4212. \nonumber
\end{eqnarray}

As discussed in \cite{trimin}, it is also interesting to examine how the phase-averaged mixing matrix
in Eq.~(\ref{matrix2}) modifies the flavor composition of the neutrino fluxes.
The most common source for atmospheric and astrophysical neutrinos is thought to be
pion production and decay.
The pion decay chain generates an initial neutrino flux with flavor composition
given approximately~\cite{lipari} by
$\Phi_e^0 : \Phi_\mu^0 : \Phi_\tau^0 = 1 : 2 : 0$ for the neutrino fluxes.
According to Eq.~(\ref{matrix2}), the fluxes $\Phi_\alpha$ arriving at earth
have a flavor ratio of
\begin{small}
\begin{eqnarray}
\Phi_e : \Phi_\mu : \Phi_\tau  &=&
1+4A\lambda^2: 1-\frac{1}{2}A\lambda^2: 1-\frac{1}{2}A\lambda^2 \nonumber \\
&\simeq & 1.2 : 1 : 1.
 \label{flux}
\end{eqnarray}
\end{small}
This result shows that $\nu_\mu\leftrightarrow\nu_\tau$ symmetry is kept in the sense that
$ \Phi_\mu/\Phi_\tau = 1$, which is mainly due to the smallness of $U_{e3}$.
The effects of breaking $\nu_\mu\leftrightarrow\nu_\tau$ symmetry appear at order of $\lambda^4$.

\section{lepton flavor violation}
Now, let us study the implication of the parametrization given by Eq.(\ref{qlc-ckm})
reflecting quark-lepton unification by considering the lepton flavor violating (LFV)
decays particularly in the context of supersymmetric standard model (SSM).
As is well known, the LFV decays in SSM can be caused by the misalignment of lepton and
slepton mass matrices \cite{masiero} and the branching ratios of the LFV decays depend on
the specific structure of the neutrino Dirac Yukawa matrix $Y_{\nu}$ \cite{skkang2}.
It is well known that the RG running induces off-diagonal terms in the slepton mass
matrix even for the case of universal slepton masses at GUT scale
\footnote{We note that the RG-induced off-diagonal terms in the slepton mass
matrix is more precisely given by \cite{ellis} $
m^2_{\tilde{l}_{ij}}\simeq
-\frac{1}{8\pi^2}(3m_0^2+A_0^2)\left(Y^{\dagger}_{\nu
ik}\log\frac{M_G}{M_{R_k}}Y_{\nu kj}\right).$
But for the sake of simplicity we assume the log term to be universal in our study.
}
\cite{casas}:
\begin{eqnarray}
m^2_{\tilde{l}_{ij}}\simeq
-\frac{1}{8\pi^2}(3m_0^2+A_0^2)(Y^{\prime}_{\nu}Y^{\prime\dagger}_{\nu})_{ij}
\log\frac{M_G}{M_X}, \label{slept}
\end{eqnarray}
where $m_0, A_0$ are universal soft scalar mass and soft trilinear $A$ parameter, and
$M_G$ and $M_X$ denote the GUT scale and the characteristic scale of the right-handed neutrinos at which
off-diagonal contributions are decoupled \cite{casas}, respectively.
Here, the Dirac neutrino Yukawa matrix, $Y^{\prime}_{\nu}$, is defined in the basis where the charged lepton Yukawa matrix and
the heavy Majorana mass matrix are real and diagonal, and thus the term $Y^{\prime}_{\nu}Y^{\prime\dagger}_{\nu}$ can be written as
\begin{eqnarray}
Y^{\prime}_{\nu} Y^{\prime\dagger}_{\nu}=
R_{23}\left(\frac{\pi}{4}\right) U^{\dagger}_{\rm CKM}(Y^D_{\nu})^2 U_{\rm CKM}R^{\dagger}_{23} \left(\frac{\pi}{4}\right),
\end{eqnarray}
where  $Y^D_{\nu}$ stands for the diagonal form of the Dirac neutrino Yukawa matrix.
For quark-lepton unification, $Y^D_{\nu} = Y^D_u = y_t \mbox{Diag}[\lambda^{8}, \lambda^{4}, 1]$
where $y_t$ is top quark Yukawa coupling \cite{haba}.
Imposing the above form of $Y^D_{\nu}$, we obtain $(Y^{\prime}_{\nu} Y^{\prime\dagger}_{\nu})_{12}\simeq (Y^{\prime}_{\nu} Y^{\prime\dagger}_{\nu})_{13} \simeq \lambda^3$, which leads to  $Br(\mu\rightarrow e \gamma)/Br(\tau\rightarrow e\gamma)\simeq 1$
and it reflects the $\mu-\tau$ symmetry. Also, one can get $(Y^{\prime}_{\nu} Y^{\prime\dagger}_{\nu})_{12})_{23}\simeq 1$,
so that  $Br(\mu\rightarrow e \gamma)/Br(\tau\rightarrow \mu \gamma)\simeq \lambda^6$.
These results indicate that the branching ratio of the LFV decay $\mu(\tau) \rightarrow e \gamma$ is negligibly small
compared with that of $\tau \rightarrow  \mu \gamma$.
\section{Implications of QLC parametrization}
Now, let us discuss about the implication of the results obtained from the QLC parametrization
by comparing with the triminimal parametrization of perturbations to tri-bimaximal neutrino mixing matrix.
To accommodate the expected deviations from the TB mixing form studied in the literatures \cite{dev1},
the triminimal parametrization of perturbations to tri-bimaximal neutrino mixing matrix has been
proposed \cite{trimin} as follows:
\begin{eqnarray}
U_{TMin}=R_{32}\left(\frac{\pi}{4}\right)U_{\epsilon}(\epsilon_{32};\epsilon_{13},\delta; \epsilon_{21})
         R_{21}\left(\sin^{-1}\frac{1}{\sqrt{3}}\right) \label{trimin}
\end{eqnarray}
where $R_{ij}(\theta)$ describes a rotation in the $ij-$plane through angle $\theta$,
$U_{\delta}=\mbox{diag}(e^{i\delta/2}, 1, e^{-i\delta/2})$ and $U_{\epsilon}=R_{32}(\epsilon_{32})U^{\dagger}_{\delta}R_{13}(\epsilon_{13}(\epsilon_{13})U_{\delta}R_{21}(\epsilon_{21})$.
From the analysis, we obtain the following relations between both parameterizations.
\begin{eqnarray}
\sin\epsilon_{13}e^{-i\delta} & \simeq & \epsilon_{13}e^{-i\delta} \simeq A\lambda^3 [ 1-\rho+i\eta ] \\
\sin\epsilon_{32} & \simeq & \epsilon_{32} \simeq -A\lambda^2 \\
\sin\epsilon_{21} & \simeq & \epsilon_{21} \simeq \frac{s}{c}(1-\frac{\lambda}{cs}+\frac{s^2}{c^2}\lambda^2)
\end{eqnarray}
where $s=(\sqrt{2}-1)/\sqrt{6}, c=(\sqrt{2}+1)/\sqrt{6}$.
The first result of the above relations indicates that the QLC parametrization predicts the size of the neutrino mixing angle $\theta_{13}=\epsilon_{13}$ which is at most of order  $\lambda^3$.
The on-going reactor experiments designed to measure $\theta_{13}$ will test whether the QLC parametrization
is ruled out or not.
The precise measurements of the mixing angles $\theta_{23}$ and $\theta_{12}$ would also be useful to probe the QLC parametrization.
 The determination of $\sin^2 \theta_{12}$ to $2\%$ level which is comparable to that of the Cabibbo angle ($\simeq 1.4\%$) can be
 achievable in the reactor neutrino experiments as shown in \cite{reactor2} and that of $\sin^2 2\theta_{23}$
 to $1\%$ is expected to reach in the JPARC-SK experiment \cite{lbl} .
The QLC parametrization is very predictable because the deviations from two maximal mixing angles can be presented in terms of
the well measured parameters in the CKM matrix.
Although the QLC parametrization looks like leading to similar results from the triminimal parametrization, the results presented in Eqs.(\ref{prob},\ref{flux}) show that the parameter $C$ defined in \cite{trimin} is particularly zero in the QLC parametrization, which is
a distinctive feature of the QLC parametrization.
If future experiments confirm our results obtained from the QLC parametrization, it would be difficult to differentiate
between the QLC parametrization reflecting deviations from the bi-maximal mixing \cite{bimax} and the triminimal parametrization reflecting
deviations from the tri-bimaximal mixing.
Confirmation of our results obtained above may also serve as a possible hint of the grand unification or quark-lepton symmetry.

\section{Conclusion}
In conclusion, we have examined how the QLC parametrization reflecting a possible hint of the grand unification or quark-lepton symmetry can be probed by considering phase-averaged oscillation probabilities which can be measured from neutrino experiments, flavor composition of neutrino fluxes coming from atmospheric and astrophysical neutrino sources and the ratios of the branching fractions of lepton flavor violating radiative decays.
We have found that those observables are predicted in terms of the well measured parameters of CKM matrix.
We have discussed about some distinct features of the QLC parametrization by comparing with the triminimal parametrization which has been proposed so that the effects of deviations from the tri-bimaximal mixing could be probed.

%% file: Author/Kersten.tex
{\bf Abstract}\\
\vskip5.mm
We discuss the predictivity of family symmetries for the soft
supersymmetry breaking parameters in the framework of supergravity.
Unknown details of the messenger sector and the supersymmetry breaking
hidden sector enter into the soft parameters, making it difficult to
obtain robust predictions.  However, specific choices of messenger
fields can improve the predictivity.

\vskip5.mm

\section{Introduction}
Models with family symmetries aim to explain the masses and mixings of
fermions, often assigning the quantum numbers in such a way that
non-zero Yukawa couplings are forbidden.  Instead, the matter fields
couple to a number of flavon and vector-like messenger fields.
Integrating out these heavy messengers yields an effective theory with
non-renormalizable couplings between matter fields and flavons.  When
the latter develop vacuum expectation values, the family symmetry is
broken spontaneously and non-zero Yukawa couplings arise, which are
suppressed by powers of the ratio of flavon vacuum expectation values
(vevs) and messenger masses.
Since the level of suppression is different for different elements of
the Yukawa matrices, one can naturally obtain hierarchical fermion
masses \cite{Froggatt:1978nt}.

Family symmetries also restrict the soft scalar masses and trilinear
couplings in supersymmetric theories, provided that SUSY breaking is
mediated to the visible sector at a scale where the family symmetry is
unbroken \cite{Abel:2001cv,Ross:2002mr}, which is the case for gravity
mediation, for example.  If all matter fields transform under the
three-dimensional representation of a non-Abelian symmetry, only
flavor-diagonal sfermion mass terms $\tilde\psi_i^* \delta_{ij} m_0^2
\tilde\psi_j$ are allowed.  Non-zero trilinear scalar couplings are
forbidden like the Yukawa couplings.  Thus, the SUSY flavor problem is
solved because flavor-changing neutral currents (FCNCs) are suppressed.

The breaking of the family symmetry leads to corrections to the soft
SUSY breaking parameters.  Like the Yukawa couplings, they are
suppressed by the ratio of flavon vevs and messenger masses, so FCNCs
remain under control.  Even better, they are in principle determined by
the model.  Consequently, we can possibly test family symmetries
not only by measuring fermion masses and mixings but also by observing
flavor-changing processes \cite{Dine:1993np}.  In the following, we
shall discuss to which extent this is indeed feasible, summarizing the
results of \cite{Kadota:2010cz}.

\section{Sfermion Masses from an SU(3) Family Symmetry}
\label{sec:NonPred}

\begin{figure}[h!]
\begin{center}
\includegraphics{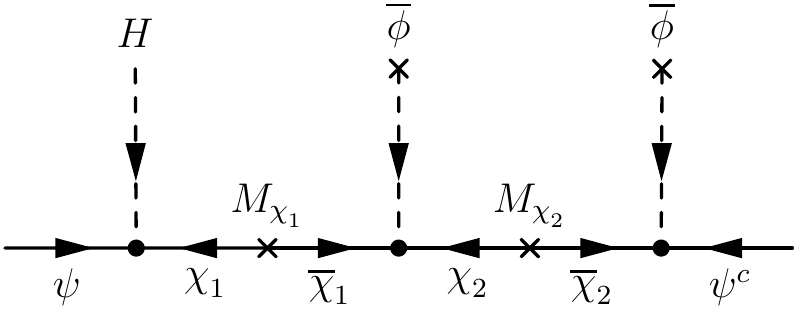} 
\caption{Feynman diagram responsible for Yukawa couplings in the
 renormalizable theory}
\label{fig:Messengers33}
\end{center}
\end{figure}

As an example, let us consider a model with an SU(3) family symmetry
\cite{deMedeirosVarzielas:2005ax}.  The matter superfields $\psi$ and
$\psi^c$ transform as triplets under SU(3), while the flavons transform
as anti-triplets.  For our purposes it will be sufficient to consider a
single flavon $\overline\phi$.  In the renormalizable theory, the diagrams
relevant for Yukawa couplings have the form shown in
\Figref{fig:Messengers33} \cite{Varzielas:2008kr}.
It involves two different messengers $\chi_1$ and $\chi_2$, which
transform as anti-triplet and singlet, respectively.  After integrating
out the messengers and breaking the family symmetry, we obtain Yukawa
couplings
\begin{equation} 
\label{eq:Yd33}
Y \sim	\frac{\vev{\overline\phi}^2}{M_{\chi_1^{}} M_{\chi_2^{}}} ;.
\end{equation}
They contain the product of the two messenger masses.  Thus, only this
product is determined by the observed fermion masses.

\begin{figure}[h!]
\begin{center}
\includegraphics{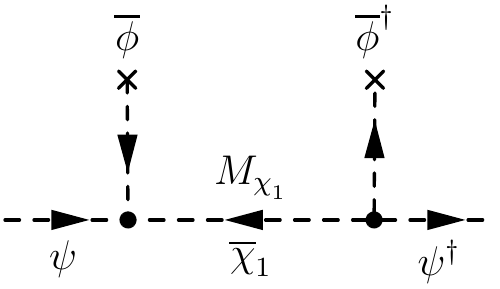} \quad
\includegraphics{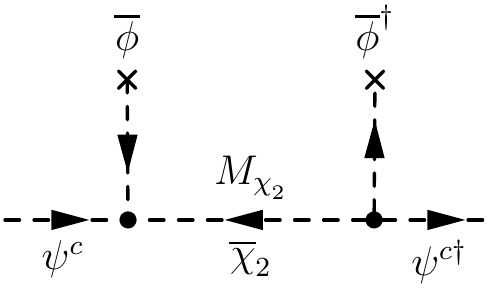}
\caption{Diagrams relevant for the K\"ahler potential.  Only the right
 one is allowed by the family symmetry in the model discussed in
 Sec.~\ref{sec:NonPred}.}
\label{fig:Kaehler}
\end{center}
\end{figure}

As to the soft SUSY breaking parameters, the family symmetry enforces
flavor-conserving scalar masses and vanishing trilinear couplings while
it is unbroken.  In order to find the changes after symmetry breaking,
we consider corrections to the K\"ahler potential, which can be
visualized by diagrams like those shown in \Figref{fig:Kaehler}.
For a rigorous calculation, one starts with the superpotential and
K\"ahler potential of the renormalizable theory and integrates out the
messengers by solving the equations
$\partial W/\partial\chi_i=\partial W/\partial\overline\chi_i=0$
($i=1,2$) for the messenger fields \cite{Brizi:2009nn}.  Plugging the
results into the potentials yields an effective theory, in which one
calculates the soft SUSY breaking parameters using the usual
supergravity formalism \cite{Brignole:1997dp}.  A rough estimate for the
soft scalar mass matrices is
$(\widetilde m_\psi^2)^{}_{ij} \sim m_0^2 \,
 \frac{\partial^2 K_\text{eff}}{\partial \psi^*_i \partial \psi_j}$
and
$(\widetilde m_{\psi^c}^2)^{}_{ij} \sim m_0^2 \,
 \frac{\partial^2 K_\text{eff}}{\partial \psi^{c*}_i \partial \psi^c_j}$.

Considering the diagrams in \Figref{fig:Kaehler}, we see that the left
one is forbidden by the family symmetry since $\overline\chi_1$ is a
triplet.  This has two important consequences.  First, the soft mass
matrix of the SU(2)$_\text{L}$ doublet sfermions remains universal,
$(\widetilde m_\psi^2)^{}_{ij} = m_0^2 \, \delta_{ij}$.  Second, the
messenger mass $M_{\chi_1^{}}$ does not enter into the soft scalar masses.
They depend only on the mass $M_{\chi_2^{}}$ of the family symmetry singlet
messenger, which appears in contributions to the SU(2)$_\text{L}$
singlet scalar masses,
\begin{equation} \label{eq:m2psic}
	\widetilde m^2_{\psi^c} \sim m_0^2
	\left( 1 + \frac{\vev{\overline\phi}^2}{M_{\chi_2^{}}^2} \right) .
\end{equation}
These contributions correspond to the right diagram in
\Figref{fig:Kaehler}.  Including the family indices of the fields and
their vevs \cite{deMedeirosVarzielas:2005ax}, one can easily verify that
the corrections from family symmetry breaking violate flavor.

In summary, the Yukawa couplings depend on a different combination of
messenger masses than the soft scalar masses.  Consequently, the latter
are \emph{not} determined by the fermion masses.  The same
conclusion holds for the trilinear couplings \cite{Kadota:2010cz}.  The
predictivity of the model in the SUSY breaking sector is constrained to
the SU(2)$_\text{L}$ doublet scalar masses.
Further uncertainties stem from a dependence on the hidden sector where
SUSY is broken, although they affect only overall mass scales but not
the flavor structure \cite{Kadota:2010cz}.

\section{A More Predictive Messenger Sector}

\begin{figure}[h!]
\begin{center}
\includegraphics{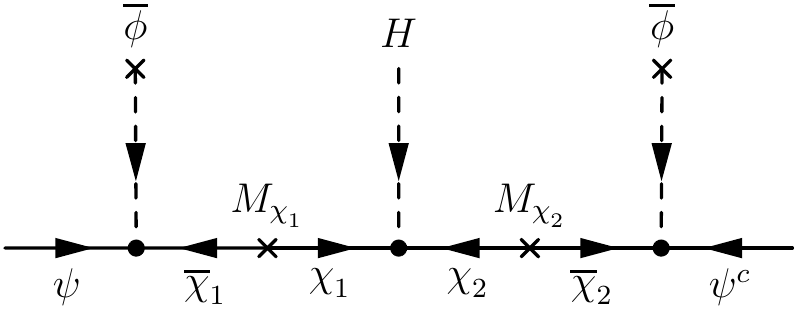} 
\caption{Generating Yukawa couplings with SU(3) singlet messengers only}
\label{fig:Messengers23}
\end{center}
\end{figure}

The situation changes if one generates Yukawa couplings via diagrams
of the type shown in \Figref{fig:Messengers23}.
Both messengers are SU(3) singlets now.
Equation \eqref{eq:Yd33} for the Yukawa couplings is unchanged,
and the right diagram of \Figref{fig:Kaehler} still yields
\Eqref{eq:m2psic} for the SU(2)$_\text{L}$ singlet soft masses.
The crucial difference to the case considered previously is that now
the left diagram of \Figref{fig:Kaehler} is allowed, too.  It yields
\begin{equation}
	\widetilde m^2_\psi \sim m_0^2
	\left( 1 + \frac{\vev{\overline\phi}^2}{M_{\chi_1^{}}^2} \right) .
\end{equation}
Consequently, flavor violation appears in \emph{all} soft scalar masses
in this case.  Even more importantly, the soft SUSY breaking parameters
depend on \emph{all} messenger masses.  This leads to predictions in the
form of correlations between parameters appearing in different
observables.  Introducing the small parameters
$\epsilon_{u,d}$, $\widetilde\epsilon_{Q,L}$, and
$\widetilde\epsilon_{u,d,e}$, which appear in the Yukawa
couplings, SU(2)$_\text{L}$ doublet scalar masses, and SU(2)$_\text{L}$
singlet scalar masses, respectively,%
\footnote{Each parameter is defined by an equation of the form
$\epsilon^2 = \vev{\overline\phi}^2\!/M_\chi^2$.
The large number of expansion parameters is due to the fact that the
model actually contains different messengers for different sectors,
e.g., for up- and down-type quarks.
}
we find
\begin{equation} \label{eq:ConsistencySquarks}
	\widetilde\epsilon_Q \, \widetilde\epsilon_u \sim \epsilon_u^2
	\quad,\quad
	\widetilde\epsilon_Q \, \widetilde\epsilon_d \sim \epsilon_d^2
	\quad,\quad
	\widetilde\epsilon_L \, \widetilde\epsilon_e \sim \epsilon_d^2 \;.
\end{equation}
In particular, no expansion parameter can be smaller than about $0.01$,
since otherwise another parameter would become larger than $1$.

The complete soft scalar mass squared matrices are
\begin{equation}
	\widetilde m^2_f \sim m^2_0 \begin{pmatrix}
	1 & \widetilde\epsilon_f^{\,2} \, \epsilon_d^2 &
	 \widetilde\epsilon_f^{\,2} \, \epsilon_d^2 \\
	\cdot & 1 + \widetilde\epsilon_f^{\,2} & \widetilde\epsilon_f^{\,2} \\
	\cdot & \cdot & 1
	\end{pmatrix}
	\quad,\quad
	f = u,\, d,\, Q,\, e,\, L
\end{equation}
at the family symmetry breaking scale.  The dots stand for elements
determined by hermiticity.  We have omitted complex phases and factors
of order one.

A very simple example consistent with the conditions 
\eqref{eq:ConsistencySquarks} is
\begin{equation}
\widetilde\epsilon_Q = \widetilde\epsilon_d = \widetilde\epsilon_L =
 \widetilde\epsilon_e = \epsilon_d \approx 0.15
\quad,\quad
\widetilde\epsilon_u = \epsilon_u^2/\epsilon_d \approx 0.02 \;.
\end{equation}
In this case, a rough estimate shows that most predicted FCNCs are well
below the experimental limits.  However, the contributions to the mass
difference of the neutral kaons and to the branching ratio of the decay
$\mu \to e\gamma$ can reach their upper bounds.  This demonstrates that
an experimental test of family symmetries via observing FCNCs may indeed
be feasible.

\section{Conclusions}

Non-Abelian family symmetries can solve the SUSY flavor and CP problems.
They also have the potential to make predictions for the soft SUSY
breaking parameters, but this is not possible in all cases.  The
predictivity of a model depends on details of its messenger sector.  We
have illustrated these issues using one particular fermion
mass model with an SU(3) family symmetry as an example.

%% file: Author/BenjaminKoch.tex



{\bf Abstract}\\
We discuss the decays of
gravitinos in the model of partial split supersymmetry (PSS),
finding that for a large range of gravitino masses, the
two body decay channel is dominant.
Taking the gravitino as dark matter candidate 
and demanding that the model also describes the established
neutrino masses and mixings we show that one can derive
a theoretical upper limit for the gravitino mass within
this model. Finally we address questions that have been
raised during this workshop.

\vskip5.mm

\section{introduction}
Split supersymmetry (SS) was originally proposed to address two of the most 
conspicuous 
problems of supersymmetric models, which are fast proton decay and excessive 
flavor changing neutral currents and CP violation \cite{ArkaniHamed:2004fb}. In
SS the 
solution to these problems is 
accomplished by considering all squarks and sleptons very massive, with a mass
scale 
$\widetilde m$ somewhere between the supersymmetric scale $M_{susy}$ and the
Grand 
Unification scale $M_{GUT}$. One of the Higgs bosons remains light, as usual in 
supersymmetric models, as well as the gauginos and higgsinos, with all these
particles 
having a mass accessible to the LHC \cite{Aad:2009wy}.
In addition to this,
the introduction of bilinear R-parity violation into this kind of
models allows for a rich phenomenology, especially in the neutrino sector
\cite{Barbier:2004ez,Nowakowski:1995dx,Hirsch:2000ef,Chun:2004mu,Diaz:2006ee,Diaz:2009yz,Sundrum:2009gv}.

A variation of this model is 
Partial Split Supersymmetry (PSS). In PSS all squarks and sleptons have a mass
of the order of
the SS mass scale $\widetilde m$, but both Higgs doublets remain with a mass at
the electroweak scale \cite{Diaz:2006ee,Sundrum:2009gv}. The addition of RpV to
this
model was introduced to be able to generate a solar neutrino 
mass \cite{Diaz:2006ee}. Loop contributions from neutral CP-even and CP-odd 
Higgs bosons are indeed able to do the job, producing not only the atmospheric
and solar masses, but also the atmospheric, solar, and reactor neutrino mixing 
angles \cite{Diaz:2009gf}.

In R-parity violating PSS the usual dark matter candidates
of supersymmetric models (the neutralinos) are unstable. 
However, the gravitino is an alternative to the neutralino.
Also the gravitino can decay into standard model particles,
but its decay is expected to be further suppressed by the
Planck mass $M_P$. This proceedings paper is essentially
a summary of \cite{Diaz:2011pc}.

\section{Partial Split Supersymmetry}

In PSS both Higgs doublets remain with a mass at the electroweak scale. As
it happens in SS, higgsinos, gauginos, and Higgs bosons interact via induced 
R-parity conserving couplings of the type,
\begin{equation}
{\cal L}_{PSS}^{RpC}\owns 
-\textstyle{\frac{1}{\sqrt{2}}} H_u^\dagger
(\tilde g_u \sigma\widetilde W + \tilde g'_u\widetilde B)\widetilde H_u
-\textstyle{\frac{1}{\sqrt{2}}} H_d^\dagger
(\tilde g_d \sigma \widetilde W - \tilde g'_d \widetilde B)\widetilde H_d 
+\mathrm{h.c.},
\label{RpCterms}
\end{equation}
where $\tilde g_u$, $\tilde g'_u$, $\tilde g_d$, and $\tilde g'_d$ are
couplings induced in the effective low energy lagrangian. At the SS scale
$\widetilde m$ they satisfy the boundary conditions,
$\tilde g_u=\tilde g_d=g\,,\qquad \tilde g'_u=\tilde g'_d=g'$,
evolving with independent RGE down to the electroweak scale. Similarly to the
MSSM, both Higgs fields acquire a vacuum expectation value $\langle H_u\rangle=v_u$
and $\langle H_d\rangle=v_d$, with the constraint 
$v^2=v_u^2+v_d^2=246\,{\mathrm{GeV}}^2$
and the definition $\tan\beta=v_u/v_d$. Gauginos and higgsinos mix forming the
neutralinos, with a mass matrix very similar to the one in the MSSM.
The relevant bi-linear R-parity violating couplings are
\begin{equation}
{\cal L}_{PSS}^{RpV} =
-\epsilon_i \widetilde H_u^T \epsilon L_i  
\ -\ 
\textstyle{\frac{1}{\sqrt{2}}} b_i H_u^T\epsilon
(\tilde g_d \sigma\widetilde W-\tilde g'_d\widetilde B)L_i 
\ + \ h.c., 
\label{LSS2HDMRpV}
\end{equation}
where the $b_i$ are the parameters of the lepton-higgs-gaugino interactions,
$\epsilon=i\sigma_2$, and the $\epsilon_i$ are the supersymmetric BRpV
parameters.
The effective neutrino mass matrix in this approach was found to be \cite{Diaz:2009gf}
\begin{equation}
{\bf M}_\nu|_{ij}=A\Lambda_i\Lambda_j+
C\epsilon_i\epsilon_j,
\quad  \mbox{with}\quad 
\Lambda_i=\mu b_i v_u+\epsilon_i v_d\quad,
\label{DpiH2HDM}
\end{equation}
where
\begin{equation}
A^{(0)}= \frac{M_1 \tilde g^2_d + M_2 \tilde g'^2_d}{4\det{M_{\chi^0}}}.
\label{A0}
\end{equation}
and where the coefficient $C$ arises from loop diagrams with virtual
mixed Higgs fields and neutralinos.
The theoretical background model will be further discussed in  a later section.
In order to make a connection with neutrino physics we will now turn
to the photino-neutrino mixing in this model $U_{\tilde \gamma \nu}$.
This mixing is governed by the same parameters as neutrino mass matrix.
Thus, imposing that the model fits neutrino data reduces the 
parameter space of PSS which translates into an allowed range for the
mixing $U_{\tilde \gamma \nu}$. The determination of this range
was done numerically. For example for $M_1=300\,$GeV one finds
\begin{equation}
2 \cdot 10^{-17}<U_{\tilde \gamma \nu}<3 \cdot 10^{-14}\quad.
\end{equation}
This range will be of importance in the context of gravitino decay.

\section{Gravitino Decay and Induced Photon flux}

In R-parity violating models the gravitino can decay completely
into ordinary particles which should lead to an observable
signature from gravitino dark matter. In our calculations
of this signature we asume that$m_{3/2}<m_W$.

The first coupling that is relevant for gravitino decay is
\begin{equation} \label{coupleGravi}
{\cal L} \owns -\frac{1}{4M_P} \overline\psi_\mu \sigma^{\nu\rho}
\gamma^\mu \lambda_\gamma F_{\nu\rho}
\end{equation}
where $M_P$ is the Planck mass, $\psi_\mu$ is the spin-$3/2$ gravitino 
field, $\lambda_\gamma$ is the spin-$1/2$ photino field, 
$F_{\nu\rho}=\partial_\nu A_\rho-\partial_\rho A_\nu$ is the photon field
strength, and $A_\mu$ is the photon field.
This coupling gives rise to a vertex gravitino photon, photino.
Due to a violation of R-parity the photino can further transform into a 
neutrino. This tranforation is proportional to the mixingo between
photino and neutrino $U_{\tilde \gamma \nu}$.
The resulting two body decay rate is then
\begin{equation} \label{eq:2body}
\Gamma(\widetilde{G}\rightarrow\gamma\nu) = 
\frac{m_{3/2}^3}{32\pi M_P^2}|{U}_{\widetilde{\gamma}\nu}|^2.
\end{equation}
In PSS the mixing between photon and photino is
\begin{equation} \label{eq:gammamixing}
U_{\widetilde{\gamma}\nu_i}\simeq\frac{\mu}{2({\rm det}M_{\chi^0})} 
\left(\tilde{g}_dM_1s_W-\tilde{g}'_dM_2c_W\right)\Lambda_i.
\end{equation}
The constants here are parameters of PSS which are also relevant
for the generation of neutrino masses and mixings within this model.

In addition to this two body decay one has to take into account
also the possible three body decays.
When studying the three body decay a more general part
of the interaction Lagrangian comes into play
\begin{eqnarray}\label{coupleGravi2}
{\mathcal{L}}&\owns&
-\frac{i}{\sqrt{2}M_P}\left[
(D^*_\mu\phi^{i*})\bar \psi_\nu \gamma^\mu \gamma^\nu P_L \chi^i-
(D_\mu \phi^i)\bar \chi^i P_R \gamma^\nu \gamma^\mu \psi_\nu
\right]\\ \nonumber
&&-\frac{i}{8 M_P}\bar \psi_\mu 
\left[ \gamma^\nu,\gamma^\rho\right]
\gamma^\mu \lambda^{(\alpha)a}F_{\nu \rho}^{(\alpha)a}\quad,
\end{eqnarray}
where the second line is in analogy to (\ref{coupleGravi}) and the first line
introduces additional couplings with scalar fields $\phi$.
The 3-body decays of the gravitino were studied in detail for the first time in 
\cite{Choi:2010jt,Choi:2010xn}, where explicit formulae are given. Nevertheless,
our calculations 
have yielded that the three-body results in \cite{Choi:2010jt} have to be
corrected.
We agree, however, with the conclusion that the 3-body decays are indeed
important, 
and cannot be neglected in general.
The interaction (\ref{coupleGravi2}) allows for various amplitdes that
contribute to the decay of a gravitino into a neutrino and two leptons with
opposite charge. The details of this calculation were given in
\cite{arXiv:1106.0308}. 
We find 3-body decay branching ratios for small gravitino
masses $m_{3/2}<20\,$GeV of the
order of $<15\%$.

If dark matter consists of those gravitinos, their
decays should contribute to the photon spectra observed 
by astrophysical experiments.
This contribution consists of a 
mono-energetic line of energy $m_{3/2}/2$ from the two-body decay, plus a 
continuum distribution from the three-body decays. The exact form of the 
spectrum, which depends on $m_{3/2}$ and $M_1$, was studied in detail in 
\cite{Choi:2010jt,Choi:2010xn} using an event generator. 
Since the first contribution is the dominant feature \cite{Buchmuller:2007ui}
only this will be considered.
Here we are interested
in obtaining constraints on the gravitino parameters, for which it suffices 
as an approximation to consider only the photon line from the two-body 
decay, as this is the most prominent feature of the spectrum for values 
of $M_1$ up to 1 TeV \cite{Choi:2010jt}.
Following established models of dark matter distribution
and photon flux propagation
\cite{Navarro:1995iw,Bertone:2007aw,Buchmuller:2007ui}
one obtains the diffenctial flux in dependence of the photon energy
due to (\ref{eq:2body}). This flux, which is supposed
to be strongly peaked at the gravitino mass, can be compared
to the photon flux from the same region of the sky, which was
measured by the Fermi LAT collaboration \cite{Abdo:2010nz}.
Since the observed flux does not show any strong peak,
this non-observation can be used to constrain the parameters 
of the decay rate (\ref{eq:2body}). This method is essentially analogous
to   \cite{Restrepo:2011rj,Huang:2011xr,GomezVargas:2011ph}.
One obtains
\begin{equation} \label{eq:constraint1}
\left(\frac{\tau_{3/2}}{10^{27}~{\rm s}}\right)>\frac{0.851}{p(m_{3/2})}
B(\widetilde{G}\rightarrow\gamma\nu) 
\left(\frac{m_{3/2}}{1~{\rm GeV}}\right)^{\gamma-2},
\end{equation}
where we parametrized the energy dependent resolution for the Fermi LAT
experiment
$\sigma=m_{3/2}p(m_{3/2})/2$ by
\begin{equation}
p(m_{3/2}) = 0.349-0.142\log\left(\frac{m_{3/2}}{2~
{\rm MeV}}\right)+0.019\log^2\left(\frac{m_{3/2}}{2~{\rm MeV}}\right).
\end{equation}
In equation (\ref{eq:constraint1}) further appears
the two body branching ratio of the gravitino decay $B$ and
the lifetime of the  gravitino $\tau_{3/2}$.

\begin{center}
\begin{figure}[h!]
\includegraphics[width=8cm]{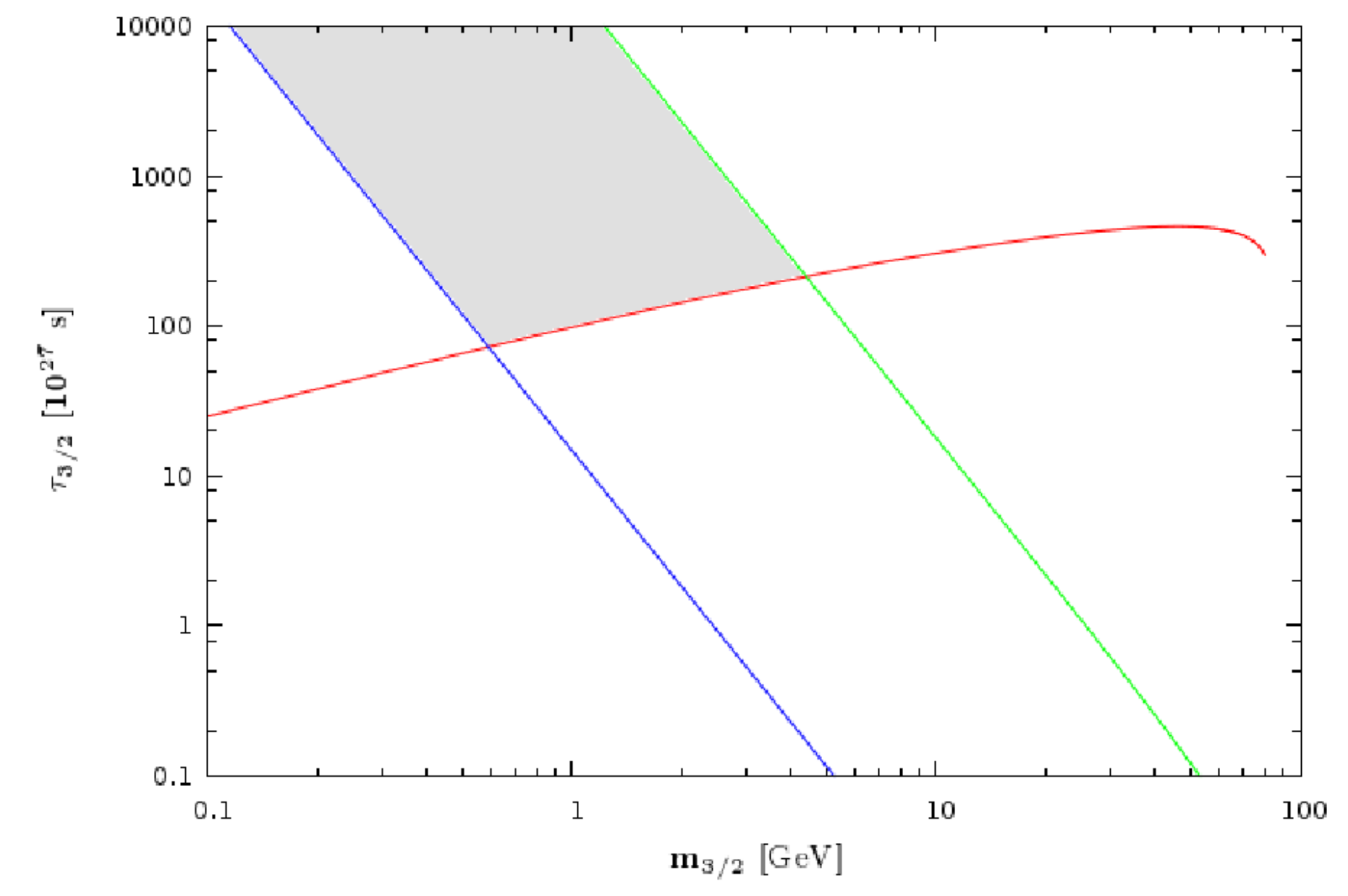}
\caption{\it Allowed region for $M_1=300$ GeV}
\label{fig1ben}
\end{figure}
\end{center}

Now we are in place to combine the constraints from the neutrino
sector and from the observed photon spectrum.
Both constraints can be visualized in a $m_{3/2}-\tau_{3/2}$ plane as it is 
shown in figure \ref{fig1ben}.
From this figure one can also read off the essential conclusion
of this study: PSS as a model for neutrino mass and the possibility
of gravitino dark matter leads to the prediction of an upper limit
of the gravitino mass in the range of $10\,$GeV.

\section{Questions and answers }
We would like use the remaining space to address some questions 
that were raised during the workshop. We appreciate those comments
because they lead to a better understanding of the basic mechanism
at work.
\begin{itemize}
\item {\it``A dimension five operator generating $\nu$-Majorana masses
needs to have (i) lepton number violation and (ii) it needs
to be proportional to the square of a $SU(2)_L$ breaking vacuum expectation value \cite{Weinberg:1979sa}. Are both conditions fulfilled?''}

Both conditions are fulfilled. Condition (i) is rather obvious due to the Majorana nature
of the term in the effective Lagrangian. The vacuum expectation value of condition (ii)
is somewhat hidden in the $C$-term of the equation \ref{DpiH2HDM} which
can be shown to be proportional to $m_Z^2$.
\item  {\it``In m-sugra a vanishing splitting between the real and imaginary s-neutrino
masses drives two of the three neutrino masses to zero. The same should
happen in this model where the s-neutrinos are integrated out''}

The splitting is controlled by the value of $\epsilon(i)B(i)\rightarrow B_{\epsilon}^i$
which in general does not necessarily have to be as small as in m-sugra.
Actually in PSS it is not small.
The reason for this to happen can be seen in equation (B7)
of the paper \cite{Diaz:2006ee}:
If one assumes very large s-neutrino masses $M_{Li}$
this tadpole equation demands equally large values of
$B_{\epsilon}^i \sim v_i M_{Li}^2/v_u$. Thus the mixing parameter
grows in such a way that the Higgses and the s-neutrinos
always mix. As it can be seen in equations (44,45) of the same
paper: The matching conditions between PSS and MSSM at the
large scale $\tilde{m}$ imply that this mixing has a definite value
$s_s^i\sim c_{\alpha}v_i/v_u$ .
There are also no problems from the side of the mass bounds
since the charged scalar/sbottom loops are suppressed by
the large mass scale of the sbottom. 

\item {\it There is a theorem in \cite{hep-ph/9701253} that states:
  If the real part and the imaginary part of the s-neutrino
have equal values $m_{sn Re}=m_{sn Im}$ then the
Majorana like B-L violating mass term has to vanish
which again implies neutrinos have to have zero mass $m_nu=0$. 
Is there a contradiction?}

No there is no contradiction.
The $m_H$ and the $m_A$ Higgses have a fixed
part of s-neutrino (as explained above), thus the general
statement of that paper predicts for the case of PSS: 
If $m_H=m_A$, then $m_\nu =0$;
That this statement is true can be seen from
\cite{Diaz:2009gf}:
Equations (12,13) of this paper show that
$m_H^2\approx m_A^2 +m_Z^2 sin^2(2 \beta)$,
thus the splitting is $\sim m_Z^2 sin^2(2 \beta)$.
However as one can see from equation (23) of the
same paper: If this splitting is zero, then also the
contribution to $m_\nu$ is zero.
This is the realization of the general statement in PSS. 

\item 
{\it It has been shown that: If there is no s-neutrino 
then there is no neutrino Majorana mass \cite{hep-ph/9701253}.
Is there a contradiction?}

No there is no contradiction.
Since the s-neutrinos mix with the Higgses due to
the possibly large parameter $B_\epsilon^i$ there is always
a part of the s-neutrino interaction eigenstates in PSS.
Thus, there is no contradiction with this theorem.
\end{itemize}

%% file: Author/Krauss.tex
{\bf Abstract}\\
\vskip5.mm
We study the generation of neutrino masses by higher than $d=5$ effective operators in supersymmetric models. We will show how these operators can be implemented in the MSSM or NMSSM. As an example we study a decomposition of the $d=7$ operator $LLH_uH_uH_dH_u$. In this model one uses heavy SU(2) doublet fermions as mediators, which can be produced in Drell-Yan processes. Since displaced vertices appear in their decays, there is the possibility to observe them at the LHC. Lepton number violating processes, however, are at the limit of observation.

\vskip5.mm

\section{Introduction}
The following discussion will be based on Ref.~\cite{Krauss:2011ur}. Due to the observation of neutrino oscillation, we know that neutrinos must have small but non-zero masses. This is a hint to new physics beyond the Standard Model. Not knowing the nature of the underlying theory, one can parameterize the effect of this new physics by a tower of effective operators which is added to the SM Lagrangian:
\begin{align}
 \mathcal{L} = \mathcal{L}_{\rm SM} + \mathcal{L}^{d=5}_{\text{eff}} 
+ \mathcal{L}^{d=6}_{\text{eff}} + \cdots
\, , \quad \textrm{with} \quad \mathcal{L}^{d}_{\text{eff}} \propto \frac{1}{\Lambda_{\mathrm{NP}}^{d-4}} \, \mathcal{O}^{d}\,.
\end{align}
The standard type-I seesaw \cite{Minkowski:1977sc,Yanagida:1979as,GellMann:1980vs,Mohapatra:1979ia}, for example, leads, after integrating out the heavy right-handed neutrinos, to the famous Weinberg operator $\mathcal{O}_W = (\overline{L^{c}} {\rm i} \tau^{2} H)\, (H {\rm i} \tau^2 L)$~\cite{Weinberg:1979sa}. One can easily see that also the type-II and type-III seesaw generate the Weinberg operator after integrating out the heavy mediator fields, since an effective operator can have several possible decompositions. After electroweak symmetry breaking, the Higgs field obtains a VEV and the Weinberg operator becomes a Majorana mass term for the neutrino with an effective mass $m_\text{eff} \propto v^2/\Lambda_\text{NP}$. To obtain a neutrino mass in agreement with experimental observations, one has to assume a new physics scale close to the GUT scale.

Models with new physics at $\mathcal{O}(\text{TeV})$ have been discussed recently which can potentially be observed at the LHC. In these scenarios further suppression mechanisms are required. In the following, we will discuss scenarios where the $d=5$ operator is forbidden and hence the leading contribution to neutrino mass comes from higher dimensional operators~\cite{Babu:1999me,Babu:2001ex,Chen:2006hn,Gogoladze:2008wz,Giudice:2008uua,Babu:2009aq,Gu:2009hu,Bonnet:2009ej,Picek:2009is,Liao:2010rx,Liao:2010ku,Liao:2010ny,Kanemura:2010bq}.
In this case we have to satisfy the following conditions:
\begin{itemize}
 \item The operators with dimensions below the leading contribution to neutrino masses are forbidden by a U(1) or a discrete symmetry.
 \item New (scalar) fields are required to construct these operators.\footnote{This is due to the fact that in the SM the only gauge invariant higher-dimensional operators contributing to neutrino mass are of the type $(LLHH)(H^\dagger H)^n$. Since $H^\dagger H$ is a singlet under any of the symmetries mentioned above, operators of this kind will always contradict the first condition. }
\end{itemize}
The simplest scenario fulfilling the second condition is the SM extended by a Higgs singlet~\cite{Chen:2006hn,Gogoladze:2008wz}
\begin{align}
  \mathcal{L}^{d=n+5}_{\text{eff}} =  \frac{1}{\Lambda_{\mathrm{NP}}^{d-4}} (LLHH) (S)^{n} \, , \quad n=1,2,3, \hdots 
\end{align}
or by an additional Higgs doublet as in Two Higgs Doublet Models (THDMs)~\cite{Gunion:1989we,Babu:1999me,Giudice:2008uua,Bonnet:2009ej}
\begin{align}
 \mathcal{L}^{d=2n+5}_{\text{eff}} =  \frac{1}{\Lambda_{\mathrm{NP}}^{d-4}}
	(LLH_{u} H_{u}) (H_{d} H_{u})^{n} \, , \quad n=1,2,3, \hdots \, .
\end{align}
In Ref.~\cite{Bonnet:2009ej} the second case has been studied in detail. 

\section{Neutrino masses from higher dimensional operators in the MSSM and NMSSM}
The Higgs sector of the minimal supersymmetric extension of the SM, the MSSM (Minimal Supersymmetric Standard Model), corresponds to a type II-THDM. All possible operators leading to neutrino mass terms up to $d=9$ are listed in Tab. \ref{tab:opOverviewMSSM}. Compared to the THDM scenario here we have less possibilities, due to the fact that the operators must be invariant under SUSY transformations. For the same reason, a $\mathbb{Z}_3$ symmetry is sufficient to forbid the $d=5$ operator in the MSSM. 
\begin{table}[t!]
\begin{center}
\begin{small}
\begin{tabular}{ccll}
\hline \hline
& Op.\#
&Effective interaction
& Charge \\
\hline
$d=5$
& 1
& $LLH_{u} H_{u}$
& $2q_{L} + 2q_{H_{u}}$
\\
$d=7$
& 3
& $LLH_{u} H_{u} H_{d} H_{u}$
& $2 q_{L} + 3 q_{H_{u}} + q_{H_{d}} $
\\
$d=9$
& 7
&  $LLH_{u} H_{u} H_{d} H_{u} H_{d} H_{u}$
& $2q_{L} + 4 q_{H_{u}} + 2 q_{H_{d}} $
\\
\hline \hline
\end{tabular}
\end{small}
\caption{\label{tab:opOverviewMSSM} Effective operators generating neutrino mass in the MSSM up to $d=9$. The operator numbers have been chosen in consistency with Tab.~\ref{tab:opOverviewNMSSM}. Taken from Ref.~\cite{Krauss:2011ur}.}
\end{center}
\end{table}

To avoid some problems of the MSSM\footnote{In the MSSM the superpotential includes the term $\mu \hat H_u \hat H_d$, which has to break the discrete symmetry explicitly. Otherwise all operators of the type $(LLH_uH_u)(H_u H_d)^n$ would have the same charge for all $n$, which implies that the $d=5$ operator ($n=0$) would always be the leading contribution to neutrino mass.} one can also consider the Next to Minimal Supersymmetric Standard Model (NMSSM), which has an additional singlet Higgs $S$. Its superpotential reads
\begin{align} \label{equ:nmssm}
  W_\text{NMSSM} = W_\text{Yuk} + \lambda \hat S \hat H_u \hat H_d + \kappa \hat S^3\,,
\end{align}
where the superpotential for the Yukawa couplings $W_\text{Yuk}$ is identical to the MSSM. The operators generating neutrino mass in the NMSSM case are listed in Tab.~\ref{tab:opOverviewNMSSM}.
\begin{table}[t!]
\begin{center}
\begin{small}
\begin{tabular}{ccllc}
\hline \hline
& Op.\#
&Effective interaction
& Charge & Same as \\
\hline
$d=5$
& 1
& $LLH_{u} H_{u}$
& $2q_{L} + 2q_{H_{u}}$
\\
\hline
$d=6$
& 2
&$LLH_{u} H_{u} S$ 
& $2q_{L} + q_{H_{u}} - q_{H_d}$
\\
\hline
$d=7$
& 3
& $LLH_{u} H_{u} H_{d} H_{u}$
& $2 q_{L} + 3 q_{H_{u}} + q_{H_{d}} $
\\
& 4
& $LLH_{u} H_{u} S S$
& $2 q_{L} - 2 q_{H_{d}}$
\\
\hline
$d=8$
& 5
& $LLH_{u} H_{u} H_{d} H_{u} S$ 
& $2q_{L} + 2q_{H_{u}}$
& \#1
\\
& 6
& $LLH_{u} H_{u} S S S$ 
& $2q_{L} + 2q_{H_{u}}$
& \#1
\\
\hline
$d=9$
& 7
&  $LLH_{u} H_{u} H_{d} H_{u} H_{d} H_{u}$
& $2q_{L} + 4 q_{H_{u}}+ 2 q_{H_{d}} $
\\
& 8
&  $LLH_{u} H_{u} H_{d} H_{u} S S$
& $2q_{L} + q_{H_{u}} - q_{H_{d}}$
& \#2
\\
& 9
&  $LLH_{u} H_{u} S S S S$
& $2q_{L} + q_{H_{u}} - q_{H_{d}}$
& \#2 
\\
\hline \hline
\end{tabular}
\end{small}
\caption{\label{tab:opOverviewNMSSM} Effective operators generating neutrino mass in the NMSSM up to $d=9$. Taken from Ref.~\cite{Krauss:2011ur}.}
\end{center}
\end{table}
A study for operators \#1, \#2 and \#4 can be found in~\cite{Gogoladze:2008wz}. In the following we want to discuss operator \#3 (for more details see~\cite{Krauss:2011ur}). A possible charge assignment that has this operator as leading contribution to neutrino mass is $ q_{H_u}=0,\ q_{H_d}=1,\ q_L=1,\ (q_S = 2)$ under a $\mathbb{Z}_3$ symmetry. A list of decompositions of this operators can be found in App. A of Ref.~\cite{Krauss:2011ur}. Here we will focus on one example that has possible phenomenological implications at the LHC.

\section{An example with a linear or inverse seesaw structure}
\begin{figure}[h!]
\begin{center}
\includegraphics[width=.45\linewidth]{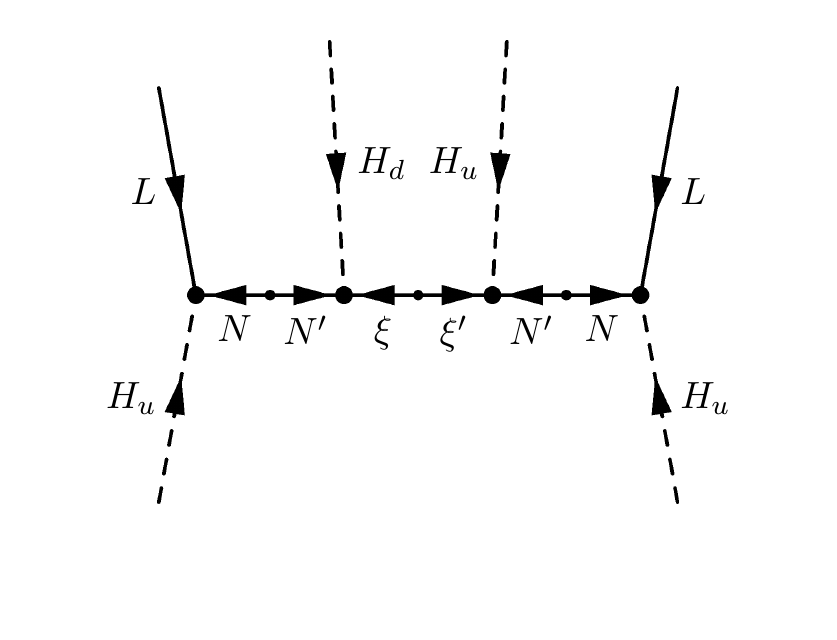}
\caption{Decomposition of the $d=7$ operator $LLH_uH_uH_dH_u$ with two SM singlets $N$ and $N'$ and two SU(2) doublets $\xi$ and $\xi'$ as mediators. Taken from Ref.~\cite{Krauss:2011ur}.}
\label{fig:decomp1}
\end{center}
\end{figure}

We choose an example with two additional SM singlets $\hat N$ and $\hat N'$ and two SU(2) doublets $\hat \xi$ and $\hat \xi'$ as mediators. The corresponding Feynman diagram is shown in Fig.~\ref{fig:decomp1}.
This model is specified by the superpotential
\begin{align}
 \begin{split}
 W = \ & W_\text{quarks} + Y_e \hat{e}^c \hat L \cdot \hat H_d
 - Y_N \hat N \hat L \cdot \hat H_u
 + \kappa_1 \hat N' \hat \xi \cdot \hat H_d
 - \kappa_2 \hat N' \hat \xi' \cdot \hat H_u
+ m_N \hat N \hat N' \\ &
+ m_\xi \hat \xi' \cdot \hat \xi
+ \mu \hat H_u \cdot \hat H_d\,,
\end{split}
\end{align}
The mass matrix for the neutral fermions reads
\begin{align}
\begin{split}
 &M^0_f = \begin{small}
\left(\begin{array}{ccccc}
	0			& Y_N v_u	        & 0		        & 0			& 0	\\
	Y_N^{\sf T} v_u		& 0			& m_N^{\sf T}	        & 0			& 0	\\
	0			& m_N		        & 0		        & \kappa_1 v_d	        & \kappa_2 v_u	\\
	0			& 0			& \kappa_1^{\sf T} v_d	& 0			& -m_\xi	\\
	0			& 0			& \kappa_2^{\sf T} v_u	& -m_\xi		        & 0			
\end{array}\right)\,\end{small}.
\end{split}
\label{equ:mass0}
\end{align}
in the basis $f^0 = (\nu, N, N',\xi^0, {\xi'}^0)$ after the Higgs fields have obtained a VEV. The charged fermion mass terms read $- v_d e^c Y_e e_L - m_\xi \xi^+ \xi'^-$.

To obtain the effective mass matrix for the light (SM) neutrinos, one has now to integrate out the mediator fields. If the hierarchy of the additional fields is $m_\xi > m_N$, we first have to integrate out the doublets and thus obtain a mass matrix similar to the inverse seesaw in the basis $(\nu,N,N')$
\begin{align}
 {M^0_f}' = \begin{small}\begin{pmatrix}
0	& Y_N \, v_u	& 0 \\
Y_N \, v_u	& 0		& m_N \\
0	& m_N	& \hat \mu \\
\end{pmatrix}\end{small}
\end{align}
where $\hat \mu = v_u v_d \, (2 \kappa_1 \kappa_2)/m_\xi $. In our model $\hat\mu$ is suppressed by the mass of the heavy doublets. If on the other hand $m_N > m_\xi$, we will obtain a linear seesaw structure in an intermediate step.
After integrating out the remaining mediators one finally arrives in both cases at an effective neutrino mass\footnote{For the sake of simplicity we have ignored the neutrino flavor so far.}
\begin{align}
  m_\nu = v_u^3 v_d Y_N^2 \frac{\kappa_1 \kappa_2}{m_\xi m_N^2} \, .
\end{align}
In order to obtain neutrino masses in agreement with the experimental bounds, the heavy mass scale can be of $\mathcal O (\text{TeV})$ if we assume couplings of $\mathcal O (10^{-3})$, which is in the range of the SM Yukawa couplings.

\section{Phenomenological implications at the LHC}
The production of the singlets $N$ and $N'$ will be suppressed due to the smallness of the couplings involved. The SU(2) doublets, however, couple to the weak gauge bosons and therefore can be produced in Drell-Yan processes, similar to charginos and neutralinos in the MSSM~\cite{Eichten:1984eu}. Numerical simulations with the program WHIZARD~\cite{Kilian:2007gr} using the SARAH package~\cite{Staub:2008uz,Staub:2009bi} give for $m_\xi = 200\,\text{GeV}$ cross sections $\sigma(pp\to \xi^\pm
\xi^0)$ of about 122 fb (417 fb) for 7 (14) TeV.

The dominant decay modes of the heavy particles are
\begin{align}
\xi^+ \to W^+ \nu_k \,,\,\, H^+ \nu_k
\end{align}
with $\Gamma(\xi^+) = 1.42 \cdot 10^{-5}\,\text{keV}$. The smallness of this decay width is due to the fact that the heavy particles decay via the mixing between light and heavy neutrinos, which is small in order to obtain light neutrino masses. The decay channels of the neutral mass eigenstates $n_i$ are
\begin{align}
 n_i &\to  W^\pm l^\mp_j \,,\,\, H^\pm l^\mp_j\,, &
n_i &\to  Z \nu_k  \,,\,\,h^0 \nu_k  \,,\,\,H^0 \nu_k  \,,\,\,A^0 \nu_k 
\end{align}
Here we obtain branching ratios of $\mathcal{O}(10^{-5}\dotsc1)\,\text{keV}$. Due to the small values of their decay widths these particles will have sizable decay lengths of up to between $100\,\mu\text{m}$ and several millimeters. This feature is useful for identifying the particles of our model and will be helpful to suppress background. The decay of the mass eigenstates $n_i$ into $W^\pm l^\mp$ will furthermore give a connection to neutrino physics, since it proves that these states carry lepton number. Another hint in this direction would be the test of their Majorana nature.
Therefore one can study the following processes which violate lepton number by two units:
\begin{align}
u \bar{d} &\to l^+ l'{}^+ W^-\label{equ:Wll} \\
u \bar{d} &\to l^+ l'{}^+ W^- Z\label{equ:ZWll}\\
q \bar{q} &\to l^+ l'{}^+ W^-  W^-,  l^- l'{}^- W^+  W^+ \label{equ:WWll}
\end{align}
The numerical values for these processes can also be found in Ref.~\cite{Krauss:2011ur}. We obtain results of several fb for the lepton number conserving (LNC) processes corresponding to (\ref{equ:Wll}), which are flavor mixed. For high enough luminosity they are therefore possible to be observed. In the lepton number violating (LNV) case, however, we have strong suppression, since the neutral mass eigenstates $n_4/n_5$ form a pseudo-Dirac pair and as a consequence the cross section, which is proportional to $m^2_{n_5}-m^2_{n_4} \simeq O(m^2_\nu)$, will be tiny.
For the LNV processes \ref{equ:WWll} we find on the other hand cross sections larger than one might expect. This is due to the fact that $\xi$ and $\xi'$ can be seen as vector-like representation of SU(2), which leads to some non-vanishing contributions to the LNV cross section. These are, however, at the limit of observability ($<\mathcal{O}(10^{-2}\,\text{fb})$).

\section{Conclusion}
Higher dimensional effective operators are possible extensions of the seesaw mechanism that allow for phenomenology in the TeV range. We have shown that these operators can be implemented in a SUSY framework such as the MSSM or the NMSSM. One specific example of a decomposition of a $d=7$ operator has been discussed, which can be tested at the LHC due to the appearance of displaced vertices. There are also non vanishing LNV processes possible, but their observation might be difficult.

%% file: Author/Vinzenz_Maurer.tex
{\bf Abstract}\\
\vskip5.mm
If supersymmetry (SUSY) will be discovered, successful models of flavour not only have to provide an explanation of the flavour structure of the Standard Model fermions, but also of the flavour structure of their scalar superpartners. We discuss aspects of such ``SUSY flavour'' models, towards predicting both flavour structures, in the context of supergravity (SUGRA). We point out the importance of carefully taking into account SUSY-specific effects, such as 1-loop SUSY threshold corrections and canonical normalization, when fitting the model to the data for fermion masses and mixings. This entangles the flavour model with the SUSY parameters and leads to interesting predictions for the sparticle spectrum. We demonstrate these effects by analyzing an example class of flavour models in the framework of an SU(5) Grand Unified Theory with a family symmetry with real triplet representations. For flavour violation through the SUSY soft breaking terms, the class of models realizes a scheme, where flavour violation effects are dominantly induced by the trilinear terms and in the reach of future experiments. 

\vskip5.mm

\section{Introduction}

The flavour puzzle is one of the biggest open questions in particle physics. Considering the standard model (SM), the flavour puzzle consists of finding some structure in the mixings and masses of the charged leptons and quarks. For example, it is tantalizing that down-type quarks and charged leptons show a similar hierarchical pattern, which differs substantially from the pattern of the up-type quark masses. Even more motivation to find structures in flavour physics arises when we extend the SM with neutrino masses and look at the large mixing angles needed to correctly describe their oscillations.

If supersymmetry (SUSY) or any other kind of new physics will be discovered at the LHC, this would add another ``dimension'' to the flavour puzzle. For one, SUSY breaking of the minimally supersymmetric SM (MSSM) has to have a flavour structure of its own, which is already at this point severely constrained from flavour physics, e.g.\ lepton flavour violation (LFV) or electric dipole moments (EDMs). This is often referred to as the ``SUSY flavour puzzle''.

In many cases these two puzzles are treated or solved independently. In this study, however, we focus on possible connections and relations between solutions to both puzzles.

\section{An Example Class of SUSY Flavour Models}
As an example, we first devise an example class of flavour models that can be brought into good agreement with experimental data. Important aspects of defining it in a supersymmetric context will be treated and it will be shown what consequences this can have on the flavour structure.

\subsection{Basic Structure}
Our starting point is a Grand Unified Theory (GUT) with gauge group $SU(5)$ plus a non-Abelian family symmetry group $G_F$ with real triplet representations, e.g.\ $A_4$ or $SO(3)$. We will assume that the three fields $F_i$ transforming as $\bar{\mathbf{ 5}}$ representations under $SU(5)$ are components of a triplet $F$ under the family symmetry $G_F$. The other three fields $T_i$, which transform as $\mathbf{10}$ representations under $SU(5)$, remain singlets under $G_F$. In addition, we add two right-handed neutrinos $N_1, N_2$ to the spectrum, which are singlets under both symmetry groups. These will be responsible for the masses of two light neutrinos via the seesaw mechanism.

\subsection{Pattern of Symmetry Breaking}

To break the full symmetry group to the SM, we introduce two types of fields. First, the usual $\mathbf{24}$-plet $H_{24}$ of $SU(5)$, which acquires a vacuum expectation value (VEV) $\langle H_{24} \rangle = v_{24} \: \text{diag} (1,1,1,-3/2,-3/2)$, second a group of 4 so-called ``flavon'' fields, that are triplets under $G_F$ break the family symmetry along the following directions in flavour space:
\begin{gather} 
\langle \phi_{1} \rangle \propto \begin{pmatrix}  0 \\  1\\  - 1  \end{pmatrix}  \;, \quad
\langle \phi_{2} \rangle \propto \begin{pmatrix} 1 \\  1 \\ 1  \end{pmatrix} \;, \quad
\langle \phi_{3} \rangle   \propto \begin{pmatrix}  0 \\  0\\  1  \end{pmatrix} \;, \quad
\langle \tilde\phi_{2} \rangle  \propto \begin{pmatrix}  0 \\  - \mathrm{i} \\  w  \end{pmatrix} \;.
\end{gather}
Beyond that, the field $H_{24}$ is assumed to be a singlet under $G_F$, while the flavons are singlets under $SU(5)$. This setup implies that these flavon VEVs generate the rows of $Y_d$ and the columns of $Y_e$ after family symmetry breakdown via effective operators with the index specifying which row/column. Moreover, involvement of $H_{24}$ in the generation of the Yukawa couplings can introduce distinct ratios between the different sub-multiplets of each $SU(5)$ multiplet \cite{arXiv:0902.4644}. 

To ensure that these ratios are left unperturbed by other operators, one has to resort to generation of the correct operators by integrating out appropriate messengers in a renormalisable underlying model \cite{arXiv:0902.4644}. 
However, in supersymmetry this introduces effective operators in the K\"ahler potential as well as in in the superpotential. So when we assume the operators generating the Yukawa couplings for charged leptons and down type quarks to be of the form
\begin{equation}
    W_{\text{eff}} \sim \frac{H_{24} \phi_i}{M^2} F T_i H_{\bar{5}} \;,
\end{equation}
we arrive at a K\"ahler potential 
\begin{equation}
    K \sim F^\dagger F + T_i^\dagger T_i + \frac{\phi_i^\dagger \phi_i}{M^2} F^\dagger F + \frac{\phi_i^\dagger \phi_i}{M^2} T_i^\dagger T_i \;.
\end{equation}
Assuming the hierarchy to reproduce the correct Yukawa couplings $\phi_3 / M \sim y_b \gg y_{d,s} \sim \phi_{1,2,\tilde{2}} / M$ also here, we find the K\"ahler metric of the field $F$ receives non-trivial and non-negligible corrections:
\begin{equation}
    \tilde{K}_{F F^\dagger} \approx \text{diag}(1, 1, 1 + \zeta^2) \;,
\end{equation}
with $\zeta^2 \sim |\phi_3|^2 / M^2$. As these imply non-canonical kinetic terms, we have to rescale $F$ using the transformation $F \to \text{diag}(1, 1, 1 - \frac{1}{2} \zeta^2) F$, see e.g.\ \cite{arXiv:0708.1282,arXiv:1104.3040}.

In summary, the textures of the relevant matrices in family space (in the convention used by \cite{Amsler:2008zzb}) will have the form:
\begin{align*}
    M_N &\phantom{=} \enspace\text{diagonal} \;, & Y_u &\phantom{=} \enspace\text{diagonal} \;, \\
    Y_\nu &= \begin{pmatrix}
            0   & y_2 \\
            y_1 & y_2 \\
            -y_1 k & y_2 k
        \end{pmatrix} \;, &
    Y_d &= \begin{pmatrix} 
        0 & \epsilon_{1} & -\epsilon_{1} k\\
        \epsilon_{2} & \epsilon_{2} + \text{i}\, \tilde{\epsilon}_{2} & (\epsilon_{2} + w\, \tilde{\epsilon}_{2}) k \\
        0 & 0 & \epsilon_3 k
        \end{pmatrix} \;,\\
    & &
    Y_e^T &= \begin{pmatrix} 
        0 & c_1\, \epsilon_{1} & - c_1\, \epsilon_{1} k\\
        c_2\, \epsilon_{2} & c_2\, \epsilon_{2} + \text{i}\, \tilde{c}_2\, \tilde{\epsilon}_{2} & (c_2\, \epsilon_{2} + w\, \tilde{c}_2\, \tilde{\epsilon}_{2}) k\\
        0 & 0 & c_3\, \epsilon_3 k
        \end{pmatrix} \;,
\end{align*}
with the rescaling factor $k = 1 - \tfrac{1}{2}\zeta^2$ and the Clebsch factors $c_1 = c_2 = c_3 = -\frac{3}{2}$, $\tilde{c}_2 = 6$ induced by the $SU(5)$ breaking effects of $H_{24}$ and giving rise to GUT scale Yukawa coupling ratios $y_\mu/y_s \approx 6$ and $y_\tau/y_b = 3/2$, see \cite{arXiv:0902.4644}. Since $\epsilon_1$ enters both in the submatrix of the first two generations of charged lepton and in the 13 down quark mixing, it is obvious that there is a  connection between $\theta_{13}^{\text{CKM}}$ and $y_{e, \mu, b}$ (and $\theta_{ij}^{\text{MNS}}$ via $\zeta$), which is made definite by the small uncertainty of the involved quantities. Moreover, we can also see features of the lepton sector, namely tribimaximal mixing with small corrections \cite{arXiv:0712.3759}
%
\begin{equation}
    \sin \theta^{\text{MNS}}_{12} \approx \frac{1}{\sqrt{3}} \left(1 + \frac{1}{6} \zeta^2 \right) \,,\; 
    \sin \theta^{\text{MNS}}_{23} \approx \frac{1}{\sqrt{2}} \left(1 + \frac{1}{4} \zeta^2 \right) \,,\;
         \theta^{\text{MNS}}_{13} \approx \frac{1}{\sqrt{2}} \left(1 + \frac{1}{4} \zeta^2 \right) \frac{1}{3}\theta_{12}^{\text{CKM}} \,,
\end{equation}
a normal neutrino mass hierarchy $0 = m_1 < m_2 < m_3$ and maximal CP violation $\delta_{\text{MNS}} = -90^\circ$.

\subsection{Soft Supersymmetry Breaking Terms}

One of the most popular ways to mediate the breaking of supersymmetry from some hidden sector to the visible sector is mediation via gravity. However, one additional effect of supergravity is that usually superfields that obtain VEVs in their scalar component also usually \cite{arXiv:0807.5047,arXiv:0910.4058} receive an ``irreducible'' contribution \cite{hep-ph/0211279} to their F-terms of the form
\begin{equation}\label{eq:fterm}
    F_\phi = \mathcal{O}(1) m_{3/2} \langle \phi \rangle \;.
\end{equation}
In the case of the flavons, these F-terms modify the soft terms from the usual flavour-universal via the formula for the soft squared mass matrices $\tilde{m}^2$ and the trilinear terms $A$ \cite{hep-ph/0312378}
\begin{align}
    \tilde{m}^2_{ij} &= m_{3/2}^2 \tilde{K}_{ij} - F_{\bar{n}}\, F_m\, \partial_{\bar{n}}\, \partial_m\, \tilde{K} \;, & A_{ijk} &= A_0\, Y_{ijk} + F_m\, \partial_m\, Y_{ijk} \;,
\end{align}
with the K\"ahler metric $\tilde{K}$ an index $m$ running over all flavons. In the considered class of models, thanks to sequestering, this surmounts to only small deviations from the constrained MSSM (CMSSM)
\begin{equation}
    m_{\tilde{F}}^2 = m_0^2\, \text{diag}(1, 1, 1 - \hat{x}_3^2 \, \zeta^2) \;,
\end{equation}
and representative for all the other $A$-terms 
\begin{equation}
    A_e^T =\begin{pmatrix}
            0 & x_1 \, c_1 \epsilon_{1} & -x_1\, c_1 \epsilon_{1}\, (1 - \frac{1}{2} \zeta^2) \\
            x_2 \, c_2 \epsilon_{2} & x_2\, \epsilon_{2} + \text{i}\, \tilde{x}_2\, \tilde{c}_2 \tilde{\epsilon}_{2} & (x_2\, c_2 \epsilon_{2} + \tilde{x}_2 \, w\, \tilde{c}_2 \tilde{\epsilon}_{2})\, (1 - \frac{1}{2} \zeta^2) \\
            0 & 0 & x_3\, \tilde{x}_3 \epsilon_{3}\, (1 - \frac{1}{2} \zeta^2)
        \end{pmatrix} \;, \label{eq:softterm_Ae}
\end{equation} 
where $x_i$ parametrises the $\mathcal{O}(1)$ factor between the flavon $\phi_i$ and its F-Term in Eq.\ \ref{eq:fterm}. As long as these $x_i$ are not all equal, the trilinear terms will not be diagonal in the same basis as the corresponding Yukawa matrices. As we can see the most prominent deviation from the CMSSM comes in the form of trilinear soft terms, so we expect additional flavour violating effects coming from there.

\section{Tests for SUSY Flavour Model}

\subsection{SUSY Threshold Corrections and Matrix Textures}

So far we have discussed the impact of a flavour model on the structure of the soft SUSY breaking sector at very high energy scales. However, to fit a SUSY flavour model to the available data, there is a certain class of supersymmetric corrections at the low scale which is mandatory to include, in particular in the large (or moderate) $\tan \beta$ region. These are the well known SUSY threshold corrections \cite{SUSYthresholds}.

\begin{figure}[ht]
    \begin{minipage}[t]{0.522\linewidth}
        \centering
        \includegraphics[width=\linewidth]{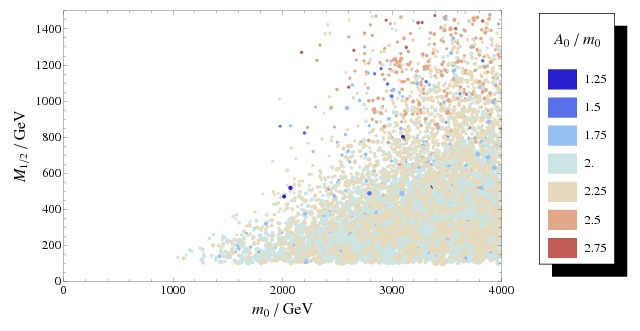}
        \caption{Points found by the Markov chains with $\mathcal{P} \ge 5\%$, ordered with respect to their $\chi^2$ such that the points which are fitting the experimental values better are drawn on top. In addition, point size scales with $\chi^2$, making points that agree better with experimental data larger.}
        \label{fig:planeplot}
    \end{minipage}
    \hspace{0.15cm}
    \begin{minipage}[t]{0.422\linewidth}
        \centering
        \includegraphics[width=\linewidth]{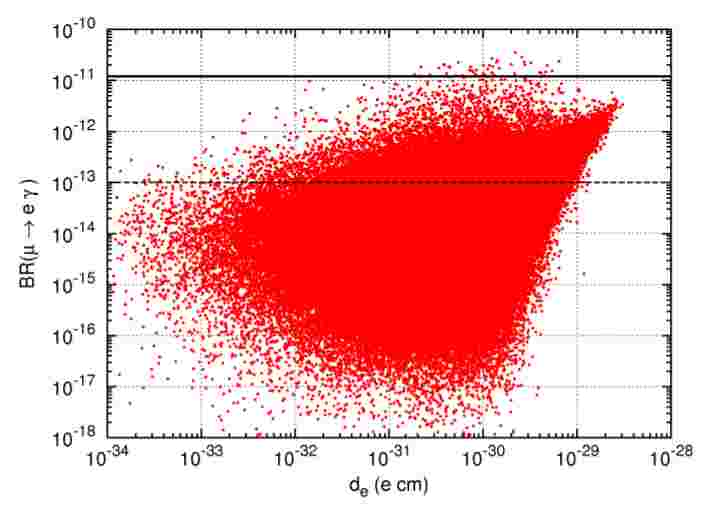}
        \caption{BR$(\mu\to e\gamma)$ versus the electron EDM, $d_e$, found by varying $x_1$, $x_2$, $x_{\tilde{2}}$ around points shown in Fig.\ \ref{fig:planeplot}. The solid (dashed) line corresponds to current (final) sensitivity of the MEG experiment \cite{meg}.}
        \label{fig:meg_edm}
    \end{minipage}
\end{figure}

Demanding that these corrections bring the considered class of models in agreement with experiment puts requirements on these corrections and provides a valuable connection between the flavour structure of the Yukawa matrices and the SUSY spectrum. To estimate where this is satisfied we performed a Monte Carlo Markov Chain analysis with $\tan\beta = 30$ and $\mu > 0$ to get a sample of the points which best fit the experimental data. The points obtained this way with p-value of more than $5\%$ are shown in Fig.\ \ref{fig:planeplot}. One can see that quite large trilinear couplings are needed and also only a quite massive spectrum suffices to meet the requirements. For further details see \cite{arXiv:1104.3040}.

\subsection{Lepton Flavour Violation and CP Violation}

The other aspect pertaining the SUSY spectrum lies in the deviations from the CMSSM in the considered class of models. Looking back at Eq.\ \ref{eq:softterm_Ae}, we see that only the three parameters $x_1$, $x_2$, $x_{\tilde{2}}$ determine the subsector of the first two generation charged sleptons. We can thus expect correlations between the lepton family number violating decay $\mu \to e \gamma$ and the CP violating flavour conserving electric dipole moment (EDM) of the electron $d_e$. To be specific, based on the numbers obtained from the fit shown in Fig.\ \ref{fig:planeplot}, one can make the following estimates 
\begin{align}
    \text{BR}(\mu\to e\gamma) &\simeq 3.5\times 10^{-13} \left(x_{1\tilde{2}}^2 + x_{2\tilde{2}}^2\right) \left(\frac{1~\text{TeV}}{m_0}\right)^4 \left(\frac{M_1}{100~\text{GeV}}\right)^2 \;, \\
    \frac{d_e}{e} &\simeq  2.2\times 10^{-29} \left(\frac{M_1}{100~\text{GeV}}\right)\left(\frac{1~\text{TeV}}{m_0}\right)^3 x_{2\tilde{2}}  ~{\rm cm} \;, 
\end{align}
with $x_{ij} = x_i - x_j$. Varying those parameters around the points obtained from the fit, one finds the values shown in Fig.\ \ref{fig:meg_edm}. One can see that current limits (being $d_e < 1.4\times 10^{-27}$ $e$ cm \cite{584532} and BR$(\mu\to e\gamma) < 1.2\times 10^{-11}$ \cite{meg}) do not put a constraint on the parameters. Given that the future experiments should reach $d_e \sim 10^{-30}$ $e$ cm, or below \cite{hep-ph/0504231} and that the final sensitivity of the MEG experiment is BR$(\mu\to e\gamma) \sim \mathcal{O}(10^{-13})$, it is interesting to notice that a large part of the parameter space should be in the reach both of MEG and the eEDM experiments.  In fact, evidence of $\mu\to e\gamma$ at MEG, at the level of, say, BR$(\mu\to e\gamma)\sim \mathcal{O}(10^{-12})$, would predict a lower bound on $d_e$ in the reach of future experiments. 

\section{Summary and Conclusions}

In this study, we have investigated aspects of ``SUSY flavour'' models, towards predicting both the SM and SUSY flavour structure, in the context of supergravity.
We highlighted the importance of including carefully all the SUSY-specific effects such as one-loop SUSY threshold corrections and canonical normalization effects when fitting the model to the low energy data for the fermion masses and mixing angles.
These effects entangle the flavour model with the SUSY parameters and leads to interesting predictions for the sparticle spectrum. In addition, family symmetries introduced to explain the flavour structure of the Standard Model fermions can modify the structure of soft terms with respect to the constrained MSSM in a specific way and can thus also make predictions testable in future flavour experiments.

%% file: Author/Panotopoulos.tex


\begin{center} \begin{large} Abstract \end{large} \end{center}

String theory constructions using D-brane physics offer a framework where ingredients like 
extra abelian factors in the gauge group, more than one Higgs 
doublet and a generalized Green-Schwarz mechanism appear at the same time.
Motivated by works towards the direction of obtaining the Standard Model in 
orientifold constructions, we study in the present work a Stueckelberg extension of 
the two-Higgs-doublet model. The distinctive features of our model are i) a sharp decay 
width for the heavy gauge boson, and ii) a charged Higgs boson having two main decay 
channels at tree level with equal branching ratios.
\section{Introduction}
%
The Standard Model of Particle Physics (SM) (for a review see e.g.~\cite{Pich:2007vu})
has been extremely successful in
describing all low energy phenomena, being in excellent agreement with
a vast amount of experimental data. The only missing part of the SM today is the
Higgs boson that gives masses to fermions and to $W^{\pm}$ and $Z$ bosons. The Stueckelberg
mechanism~\cite{Stueckelberg:1938zz} gives mass to abelian vector bosons without
breaking gauge invariance on the Lagrangian, and thus provides an alternative
to the Higgs mechanism. Most of the well motivated extensions of the SM, which have been developed to
address its open issues, involve an extra $U(1)$ in the gauge group. A new heavy
gauge boson, $Z^\prime$, is predicted which would have profound implications for
particle physics and cosmology. Another famous minimal extension of the SM consists in the 
addition of one scalar doublet to the theory \cite{Branco:2011iw}. This idea has been particularly 
successful for its simplicity and the rich phenomenology that generates, being able to introduce 
new dynamical possibilities, like different sources of CP violation or dark matter candidates, 
helps to solve some of the SM problems. In the most general version of the two-Higgs-doublet model 
(2HDM), the fermionic couplings of the neutral scalars are not diagonal in flavour, which generates 
dangerous flavour-changing neutral current (FCNC) phenomena. Since these are tightly constrained 
by the experimental data, it is necessary to implement ad-hoc dynamical restrictions to guarantee 
their absence at the required level. Many attempts have been made in order to embed the
SM in open string theory, with some success~\cite{Kiritsis:2003mc}. They consider the SM particles as open 
string states attached on different stacks of D-branes. $N$
coincident D-branes typically generate a unitary group $U(N) \sim SU(N) \times U(1)$.
Therefore, every stack of branes supplies the model with an extra abelian factor in the gauge group.
Such $U(1)$ fields have generically four-dimensional anomalies~\cite{Anastasopoulos:2003aj,Anastasopoulos:2004ga}. These anomalies are cancelled via the Green-Schwarz mechanism~\cite{Green:1984sg,Green:1984qs,Sagnotti:1992qw,Ibanez:1998qp} where a scalar axionic field is responsible for the anomaly 
cancellation. This mechanism gives a mass to the anomalous $U(1)$ fields and breaks the associated
gauge symmetry. This class of models is characterized by i) the existence of two Higgs doublets 
necessary to give masses to all fermions, and ii) the massive
gauge bosons acquire their mass from two sources, namely the usual Higgs mechanism, as well
as the stringy mechanism related to the generalized Green-Schwarz mechanism, which is
very similar to the Stueckelberg mechanism.
In the light of these developments, it becomes clear that it is natural to study the 2HDM
with additional $U(1)$s and the Stueckelberg mechanism together with the Higgs mechanism.
In the present work we wish to study the phenomenology of a simple four-dimensional, non-GUT,
non-supersymmetric model with an additional Higgs doublet, and just one extra $U(1)$ factor
in the gauge group for simplicity.

\section{$Z^ \prime$ searches}
%

The gauge group of the model is the SM gauge group times an extra abelian
factor $U(1)_X$, with a coupling constant $g_X$ and a gauge boson $C_\mu$
associated with it. We have three generations of quarks and leptons
with the usual quantum numbers under the SM gauge group, and they are
assumed to be neutral under the extra $U(1)$. This is a simple choice that ensures
that there are no anomalies in the model. We consider the presence
of two Higgs
doublets, $H_1$ and $H_2$,
with the same quantum numbers under the SM gauge group, the only difference
being is that $H_1$ is assumed to be neutral under
$U(1)_X$, while $H_2$ is charged
under the
additional abelian factor with charge $Y_X= \pm 1$. Finally, the Stueckelberg contribution is~\cite{Kors:2004dx}
\begin{eqnarray}
{\cal L}_{\rm St} = -\frac{1}{4} C_{\mu\nu}C^{\mu\nu}
- \frac{1}{2} (\partial_{\mu}\sigma  + M_1 C_{\mu} + M_2 B_{\mu})^2 \; ,
\end{eqnarray}
where $C_\mu$ is the gauge boson associated with the $U(1)_X$, $C_{\mu \nu}$ is the
corresponding field strength, $\sigma$ is the scalar axionic field which is assumed to couple
both to $B_\mu$ and $C_\mu$, and $M_1$ and $M_2$ are two mass scales which serve as two extra
parameters of the model. The details regarding the Higgs potential, the electroweak symmetry breaking
as well as the new interaction vertices can be found in~\cite{Panotopoulos:2011xb}.

The LHC is designed to collide protons with a center-of-mass energy
14 TeV. Since the center-of-mass energy of proton-proton
collisions at LHC is 14 TeV, the particle cascades
coming from the collisions might contain $Z^\prime$ if its mass
is of the order of 1 TeV. Therefore a heavy gauge boson can be discovered at LHC, and in fact
new gauge bosons are perhaps the next best motivated new physics, after the Higgs
and supersymmetric particles, to be searched for at future experiments.
The mass, total decay width as well as branching ratios for various decay modes are some
of the properties of  $Z^\prime$ that should be accurately measurable, and could be used to distinquish
between various models at colliders. Thus, in this section we discuss the phenomenology of the model as
far as the physics of the new gauge boson is concerned.

Our results are summarized in the figures below. We have fixed the Higgs boson masses (Set 1 and
Set 2 as can be seen in the table below), as well as the coupling constant $g_X$ considering two cases,
one in which the coupling is small, $g_X=0.001$, and one in which the coupling is comparable to the SM couplings, $g_X=0.1$. Then the only free parameter left is the heavy gauge boson mass. Therefore, in the figures shown below the
independent variable is the mass of $Z^\prime$. First we focus on the case where $g_X=0.001$.
Figures 1 and 2 show the total decay width of $Z^\prime$ (in GeV)
as a function of its mass for Set 1, with $M_2/M_1=$ 0.03 and 0.05, respectively. In the rest of the figures the impact of changing the value for the
ratio $M_2/M_1$ is negligible, so it is fixed at 0.03. Figures 3 and 4 show all branching ratios as a
function of $M_{Z^\prime}$ (for Set 1 and Set 2 respectively). All the decay channels into quarks have been
considered together as a single quark channel. However, we have checked that $Z^\prime$ decay into quarks
is dominated by the up quark contributions. The straight vertical
lines correspond to the thresholds, one for the top quark ($\sim 346$ GeV), one for the neutral Higgs bosons
(600 GeV for Set 2 only) and one for the charged Higgs bosons (400 GeV for Set 1 and 1000 GeV for Set 2). We
remind the reader that in the SM, the branching ratio of the $Z$ boson to electrons or muons
or tau leptons is 0.034 for each of them, to all neutrino species (invisible channel) is 0.2, and to
hadrons is 0.7. In the model with one Higgs doublet there are no decay channels to inert Higgs bosons, and for a large
enough $M_{Z^\prime}$, where the branching ratios of $Z^\prime$ to the inert Higgs bosons become
significant, the decay widths in the two models tend to differ. However, the difference is small 
since the dominant contribution to the decay width is from $Z^\prime$ to fermions, which scales as
$M_{Z^\prime}g_Y^2(M_2/M_1)^2$. Furthermore, in the model with
one Higgs doublet only, there is just the SM neutral Higgs boson, while in the model with
two Higgs doublets there are both neutral and charged Higgs bosons.
\begin{table}[h!]
\begin{center}
\begin{tabular}{|c|c|c|}
\hline
&Set 1& Set 2 \\
\hline
$M_{H^{\pm}}$ (GeV) &200& 500\\
$M_{H,A}$ (GeV) &100& 300\\
$M_{h}$ (GeV) &100& 250\\
\hline
\end{tabular}
\caption{The two sets of Higgs boson masses used in the analysis.}
\end{center}
\end{table}

\begin{figure}[h!]
\begin{center}
\includegraphics[width=6cm]{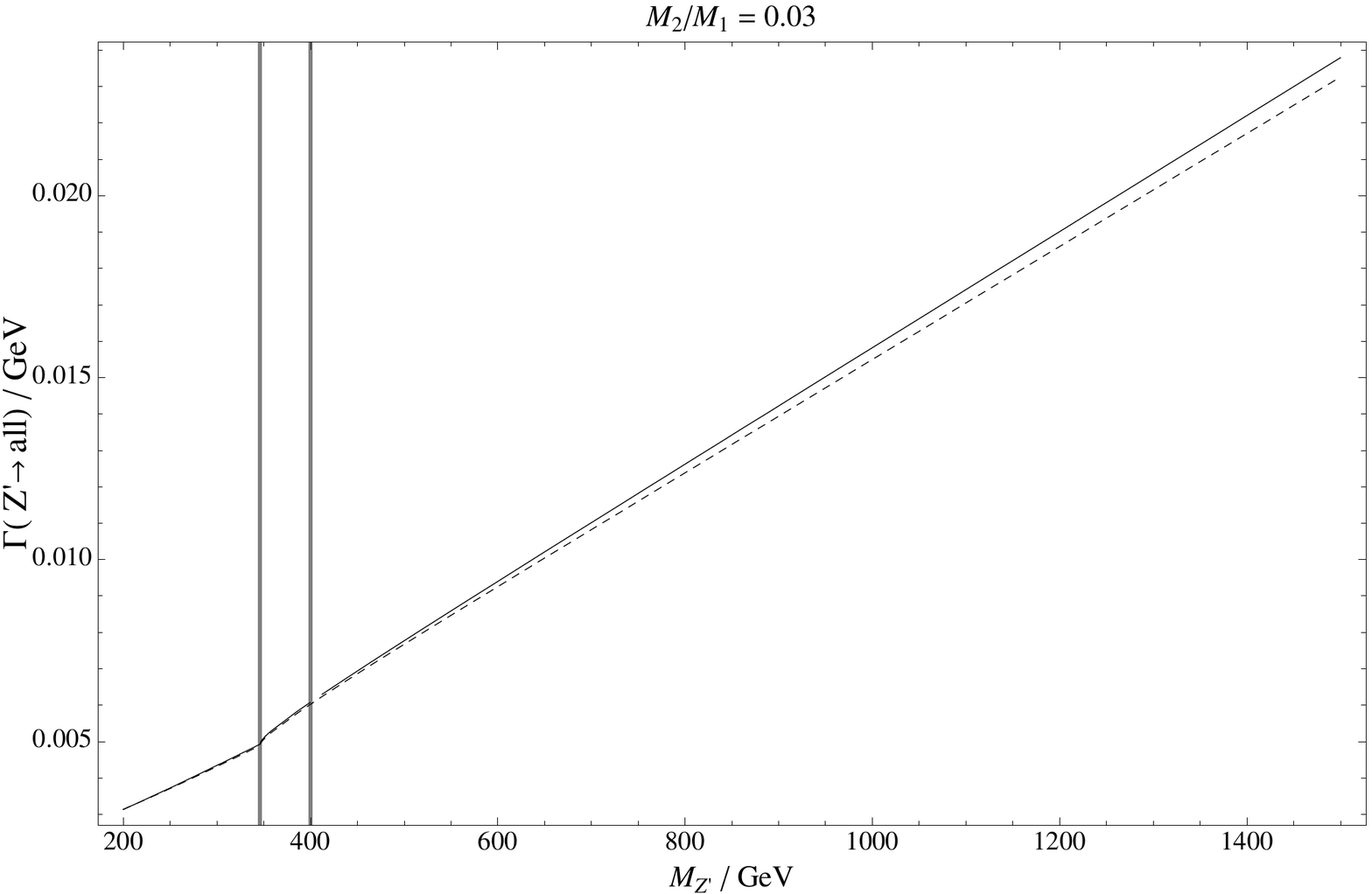}
\includegraphics[width=6cm]{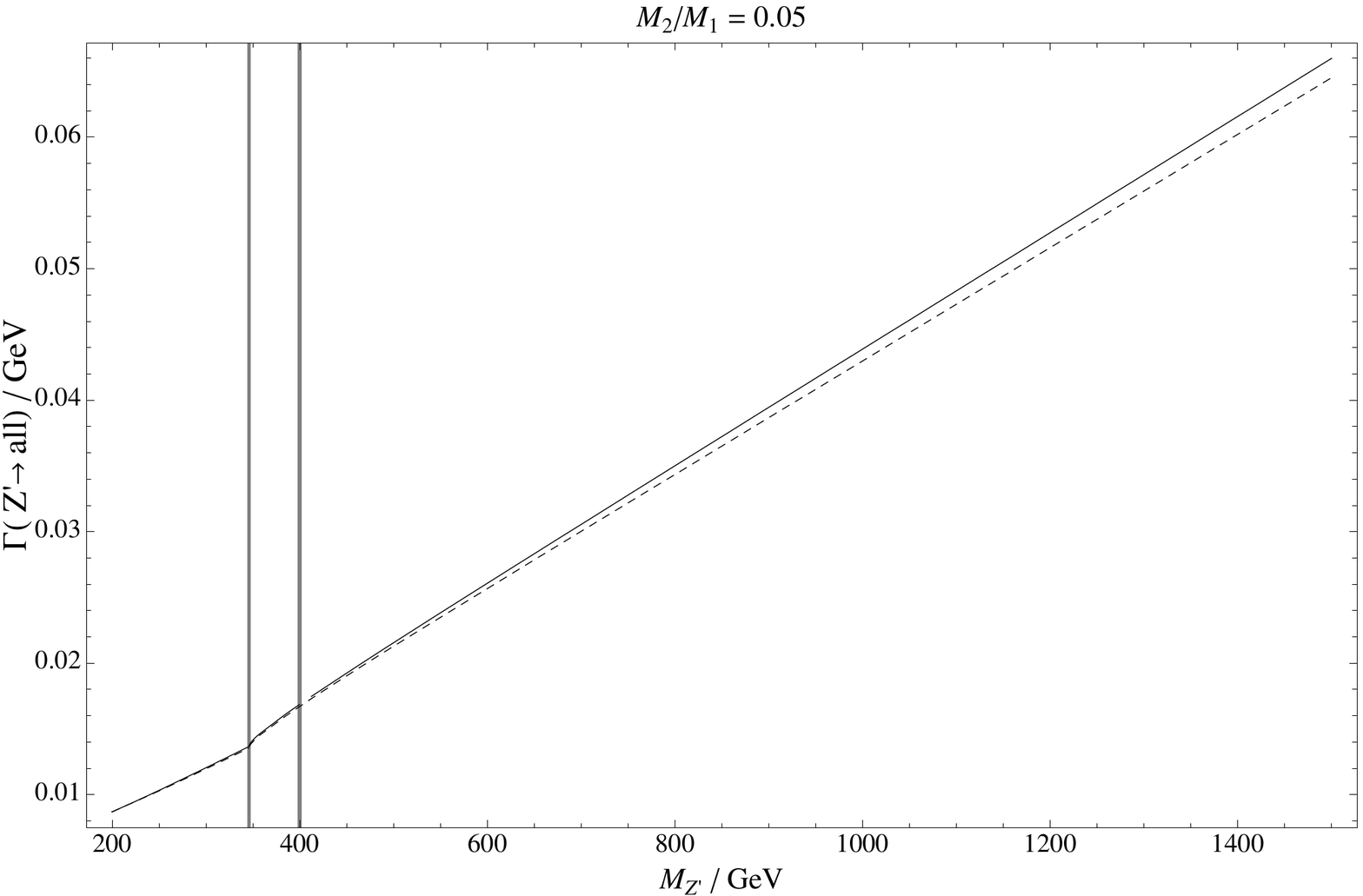}
\end{center}
\end{figure}

\vskip-5.mm
\begin{figure}[h!]
\begin{center}
\includegraphics[width=6cm]{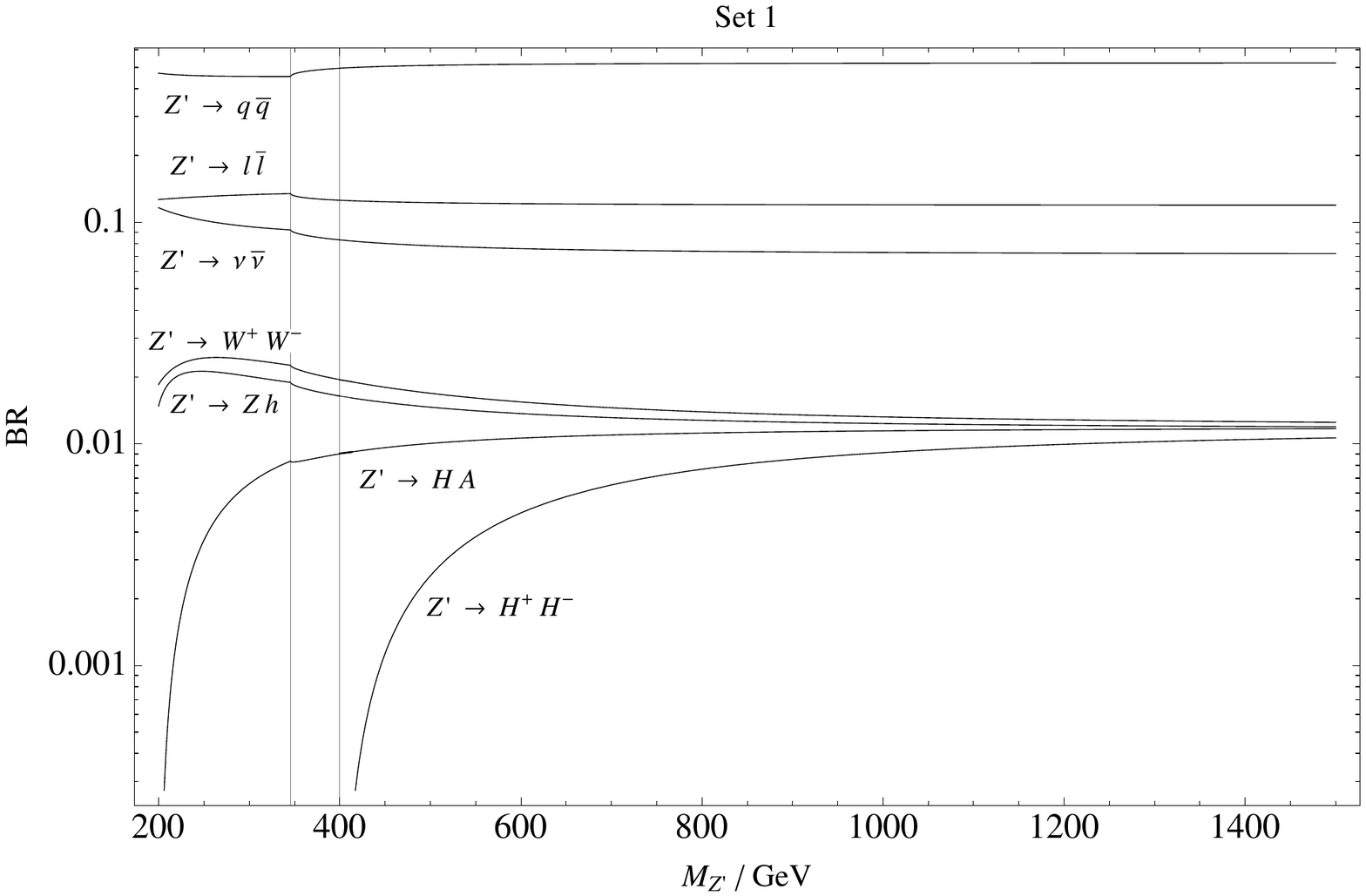}
\includegraphics[width=6cm]{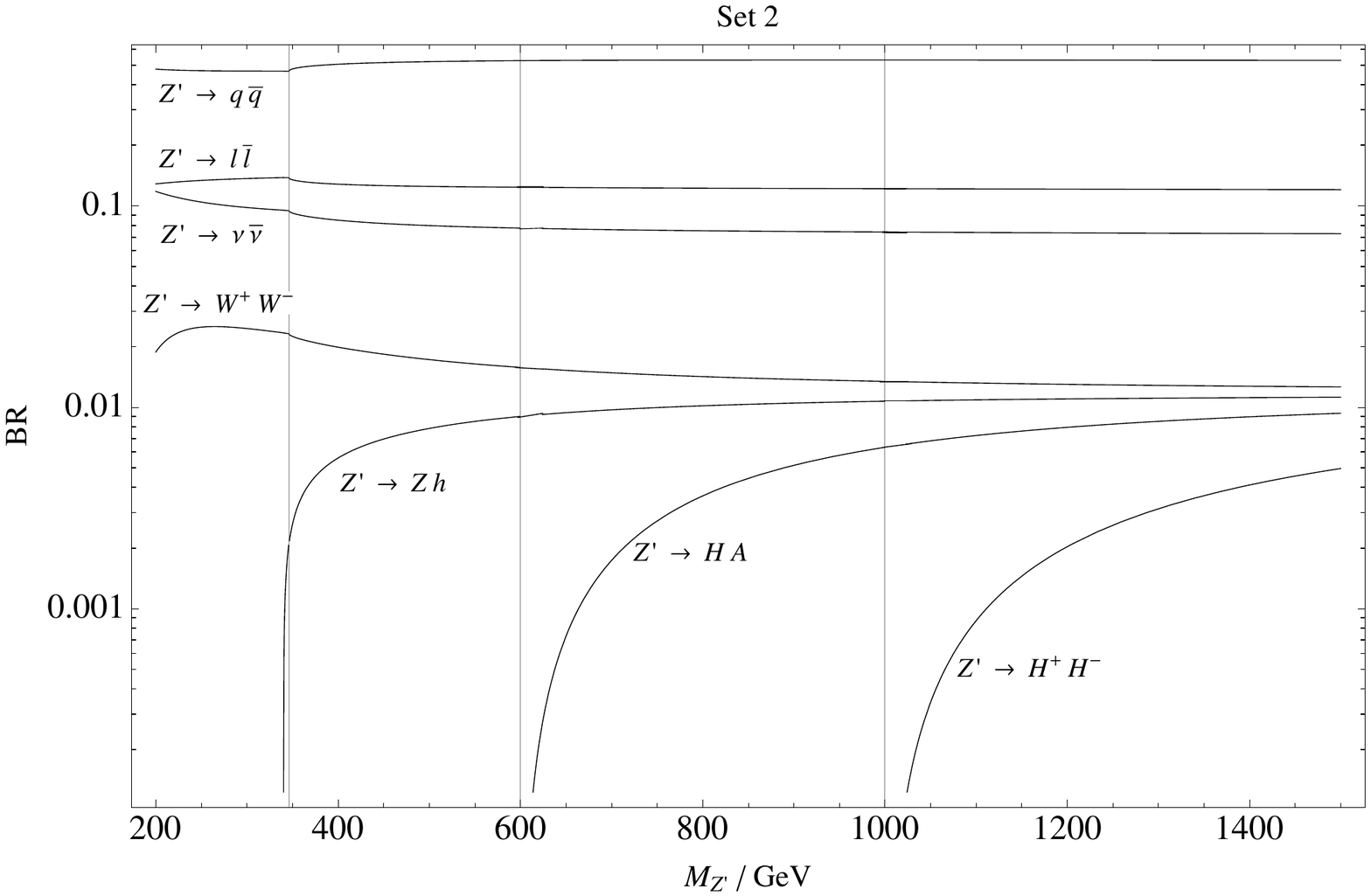}
\end{center}
\end{figure}

\vskip-5.mm
\begin{figure}[h!]
\begin{center}
\includegraphics[width=6cm]{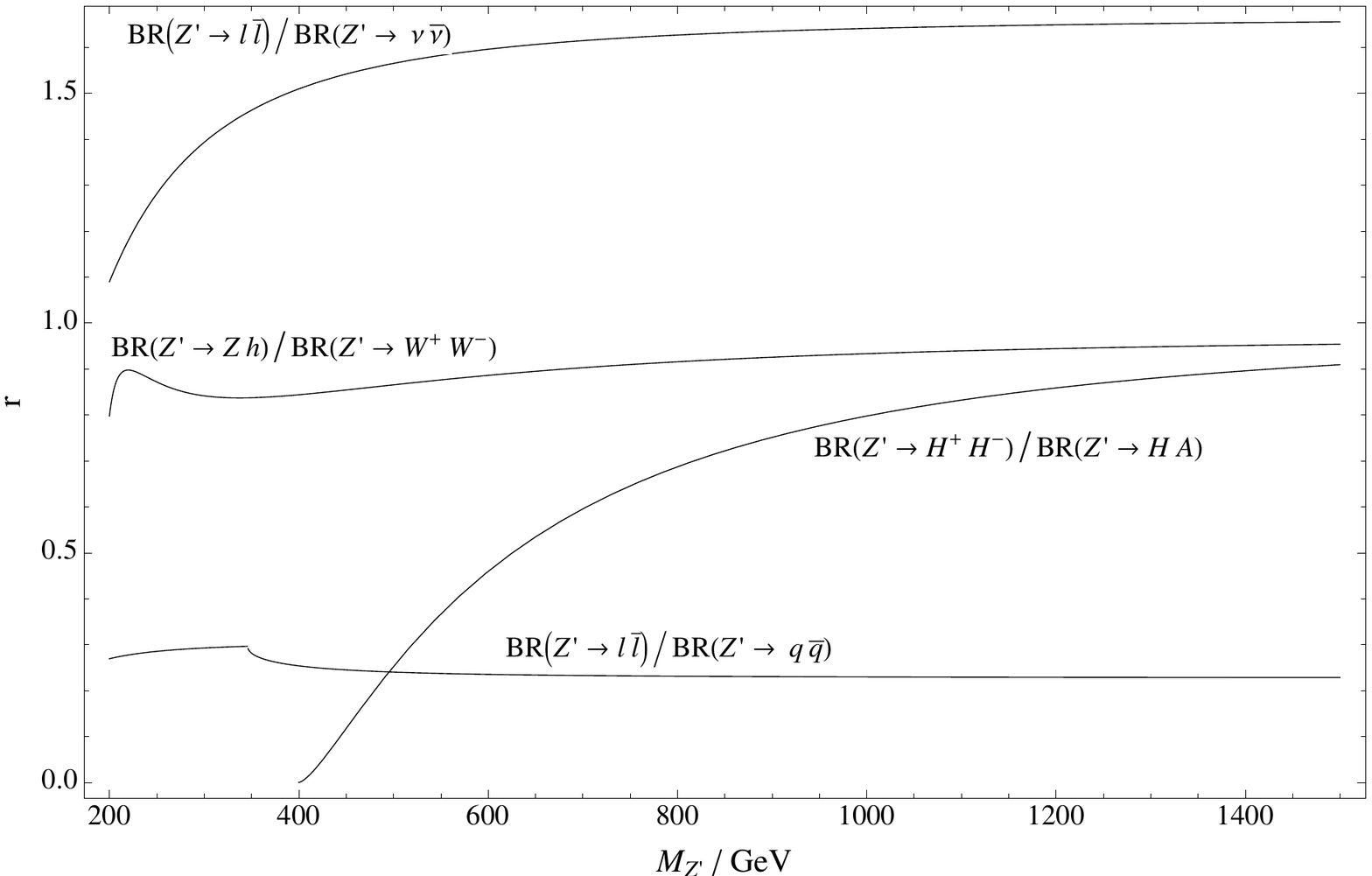}
\includegraphics[width=6cm]{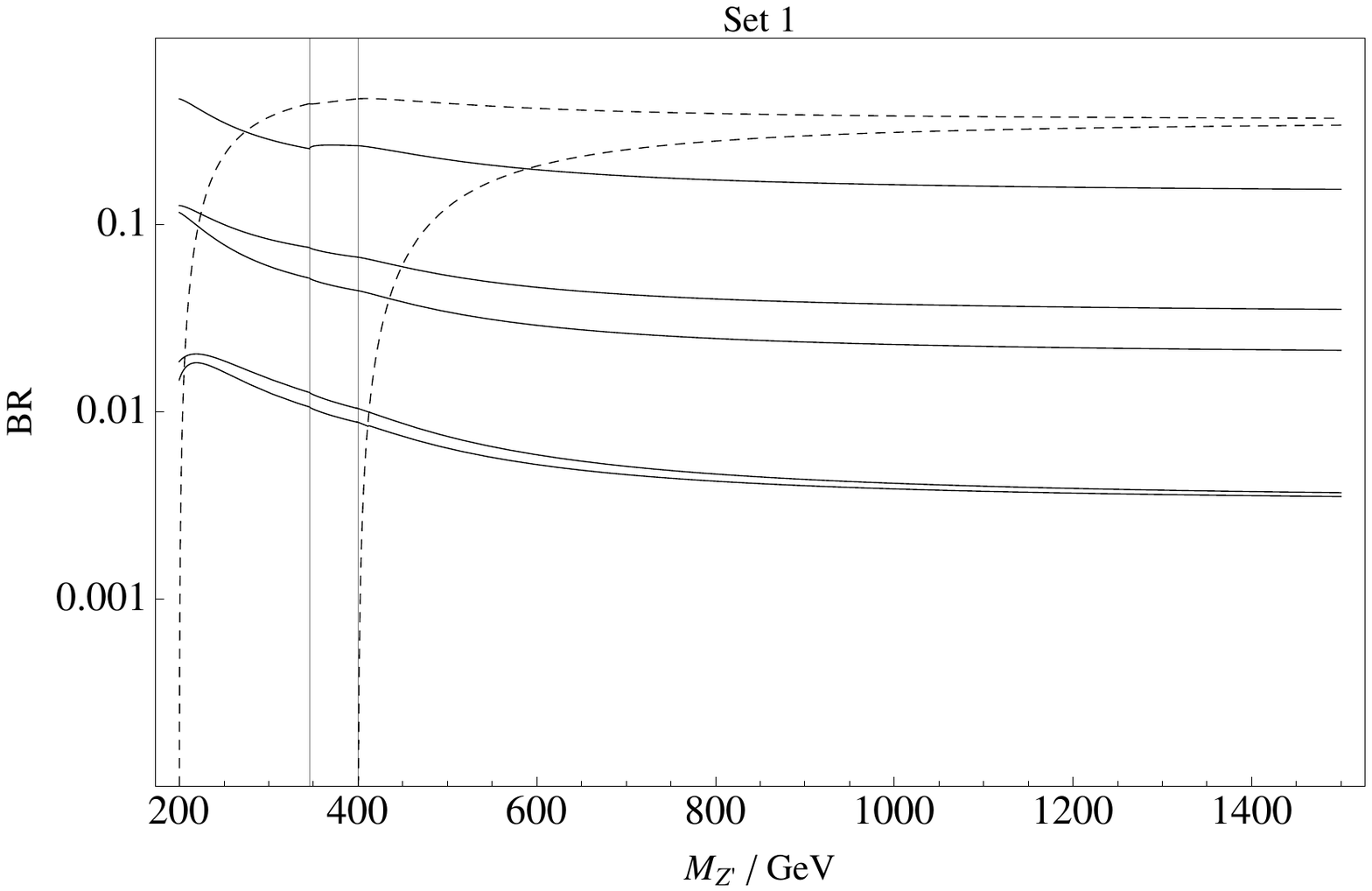}
\end{center}\end{figure}
Clearly, if a charged Higgs
boson is seen at colliders, this would be a direct evidence of physics beyond the SM.
Without Yukawa couplings the
charged Higgs bosons cannot directly decay into fermions, and therefore
the dominant decay channels of the charged Higgs bosons are just two, $H^{\pm} \rightarrow W^{\pm} \: H$
and $H^{\pm} \rightarrow W^{\pm} \: A$. Taking into account that $H$ and $A$ are degenerate in mass,
the model discussed here predicts that there are two main decay channels for $H^{\pm}$ with the two branching
ratios being equal to $1/2$. A detailed discussion of the Higgs phenomenology is postponed to a future work. 
We can also mention here in passing that if the decay channel $h \rightarrow ZZ$ is kinematically
allowed, the SM Higgs boson can be easily found through the so-called four-lepton golden Higgs channel,
$h \rightarrow ZZ \rightarrow l^+ l^- l^+ l^-$.
As in the case with one
Higgs doublet, the total decay width is much smaller than in other models~\cite{Anoka:2004vf,Aydemir:2009zz},
and therefore a heavy gauge boson is expected to show up at colliders as a sharp resonance. Finally,
in Figure 5 we show ratios of decay widths of two channels as a function of $M_{Z^\prime}$, and
in particular we have chosen to show the
following ratios: Leptons to hadrons, leptons to neutrinos, charged Higgs to neutral Higgs, and
$W^{\pm}$ bosons to SM Higgs and $Z$ boson. Recall that in the SM the ratio of leptons to neutrinos is
0.17, and the ratio of leptons to hadrons is 0.05.

Finally, notice that in Figures 1 and 2, although they look very
similar, the scale is different. When the ratio $M_2/M_1$ is increased from 0.03 to
0.05, the total decay width also increases by a factor $\sim 3$, because the couplings of the new
gauge boson are now larger. We have also checked that the plot with larger mass ratio showing the branching fractions
cannot be distinguished from the one with smaller mass ratio.

We now consider the case where $g_X=0.1$ for Set 1 and $M_2/M_1=0.03$. Most of the decay modes remain the same, apart from the ones into the inert Higgs bosons, for which the coupling now is larger, leading to
larger partial decay widths. Figure 6 shows the effect on the branching ratios. The curves 
corresponding to the decays
into the inert Higgs bosons preserve their shape, but now they are above the rest. The sign of $Y_X$
has been taken to be positive. If we change the sign of $Y_X$ we obtain a similar plot where
the branching ratios for the inert Higgs bosons are slightly larger.

\section{Conclusion}

A model with an extra $U(1)$ and a second Higgs doublet has been investigated.
It is assumed that the fermions and the SM Higgs are neutral under the extra
$U(1)$, while the dark Higgs is charged. Thus, Yukawa couplings for the additional
Higgs are not allowed, and the FCNC problem is avoided. From this point
of view the model is similar to the inert 2HDM, although the gauge symmetry
is more restrictive than the $\mathcal Z_2$ discrete symmetry. The massive gauge bosons
obtain their masses from two separate mechanisms, namely from the usual Higgs mechanism,
as well as from the Stueckelberg mechanism. The interplay between
the heavy gauge boson and the extended Higgs sector makes the phenomenology
of this model very rich. We have computed the total decay width and all the branching
ratios of $Z'$ as a function of its mass for two different sets of the Higgs bosons masses.
We find that two distinct features of the model are a) a sharp decay width for the heavy
gauge boson, characteristic of the Stueckelberg mechanism like in the corresponding model
with just one Higgs doublet, and b) a pair of charged Higgs bosons with no Yukawa couplings
decaying dominantly into a $W^{\pm}$ boson and a neutral Higgs boson $H$ or $A$, with the two branching
ratios being equal to $1/2$ each.

%% file: Author/Papa.tex
{\bf Abstract}\\
\vskip5.mm
A search for the decay $\mu^+ \to e^+ \gamma$ is going on at PSI. The 2009 collected data from the initial three months of operation of the MEG experiment yields an upper limit BR$(\mu^+ \to e^+ \gamma)\le2.8 \times 10^{-11}$ (90$\%$ C.L.). The analysis of the combined 2009 and 2010 data sample gives a 90$\%$ C.L. upper limit of $2.4 \times 10^{-12}$, constituting the most stringent limit on the existence of this decay to date. 
\vskip5.mm
\section{Introduction}
Lepton flavour violation (LFV) research is presently one of the most exciting branches of particle physics. Flavour violating processes, such as $\mu^+ \to e^+ \gamma$, which are not predicted by the Minimal Standard  Model (SM), are very sensitive to the physics beyond it. Neutrino oscillations are now an established fact, which can be accomodated in the SM by including right-handed massive neutrinos and mixing. This modified SM predicts unmeasureable branching ratios (BR) for lepton violating decays. Supersymmetric GUT theories naturally house finite neutrino masses and predict rather large and measurable branching ratios for LFV decays. The $\mu^+ \to e^+ \gamma$ process is therefore a powerful tool to investigate physics beyond the SM, since: $a$) the present experimental  upper limit is $\rm{BR = 1.2\cdot10^{-11}}$ at $90\%$~C.L. (by the MEGA collaboration~\cite{MEGA}) and $b$) supersymmetric GUT models, such as SO(10) SUSY-GUT or SU(5) SUSY-GUT, predict $\rm{BR \approx 10^{-14}-10^{-11}}$~\cite{barbieri1,barbieri2,kuno,raidal,masiero}.\\

The aim of the MEG experiment is to measure the branching ratio of the rare muon decay $\rm{BR = \frac{\mu^+ \to e^+ \gamma}{\mu^+ \to e^+\nu_e \overline{\nu}_{\mu}}}$ at a sensitivity of $\approx 10^{-13}$,  two orders of magnitude better than the present experimental limit and within the region of theoretical predictions~\cite{MEG}.\\

To reach this goal, the experiment must use the most intense continuous muon beam available ($\rm{\approx 10^8 \mu/s}$) and obtain the highest energy, time and space resolutions, today reachable. MEG started to collect data at the end of 2008. During 2009 a large part of the data taking time was devoted to calibration measurements and detector performance optimizations; a new physics data sample was collected at the end of this year in 1.5 months of acquisition time. We have continued to take data during 2010 and the final analysis of this sample is going on. Our final result will include both 2009 and 2010 data samples. A description of the main features of each subdetector and of the measured resolutions are given and the preliminary results of the search for $\mu^+ \to e^+ \gamma$ decay based on the 2009 data sample are presented.\\

The event signature of the $\mu^+ \to e^+\gamma$ decay at rest is a positron and a photon in timing coincidence, moving collinearly back-to-back with their energies equal to half the muon mass ($m_{\mu}/2 = 52.8$ MeV). Positive muons are used because they can be stopped in a target without being captured by a nucleus. 

Two kinds of background are present: $a$) the prompt background from the radiative muon decay, $\mu^+ \to e^+\nu_e \overline{\nu}_{\mu} \gamma$ and $b$) the uncorrelated background due to an accidental coincidence between a positron from the normal muon decay $\mu^+ \to e^+\nu_e \overline{\nu}_{\mu}$ and a high energy photon from radiative muon decay or positron annihilation in flight. The latter is the main background source and we evaluated its effective branching ratio at level of $\leq 10^{-13}$.

$\approx 10^{14}$ muons were stopped in the target. A data reduction was performed and the pre-selected events were further processed. The events falling into a pre-defined window in the  $E_{\gamma}$ vs   $t_{e\gamma}$ plane (``blinding-box"), which includes the signal region, were saved in separate hidden files; the $\mu \to e \gamma$ decay was searched within this sample. The other events (``side-bands") were used for optimizing the analysis parameters and for studying the background.\\
A likelihood analysis was performed to determine the  B.R. upper limit on $\mu \to e \gamma$ decay.

\section{Experimental set-up}
\begin{figure}
\begin{center} 
\includegraphics[scale=0.5]{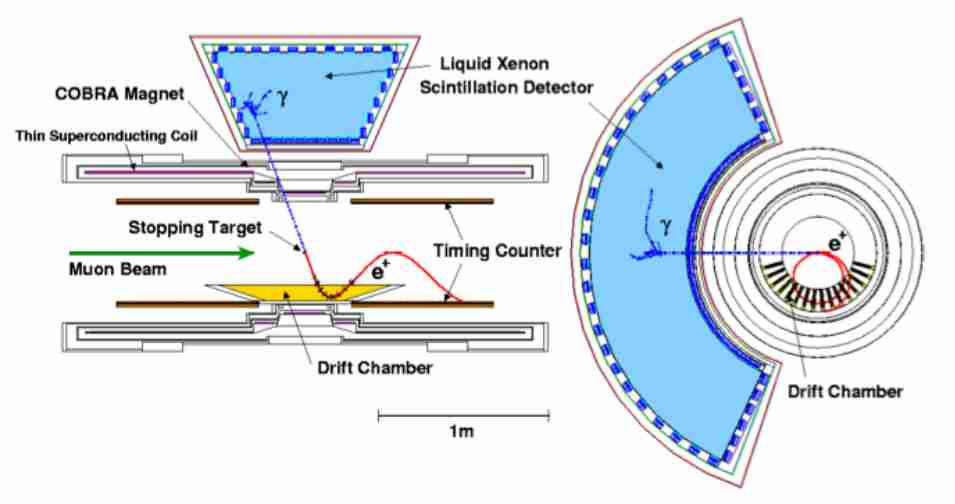}
\end{center}
\caption{The MEG experiment layout.}
\label{fig:MEGlayout}
\end{figure}

A schematic layout of the MEG detector is shown in the Fig.~\ref{fig:MEGlayout}.\\
Abundant low energy muons can be produced at high intensity proton accelerators. The so-called surface muons are obtained by bombarding protons in a thick production target and come out from the two-body decays of positive pions that stop near the surface of the target, with a  sharp momentum of approximately 29 MeV/c. Because of their low momentum and narrow momentum spread (typically 8$\%$ FWHM), a thin target ($\approx 10 mg\cdot cm^{-2}$) can be employed, with several advantages: good identification of decay region (thanks also to the small beam spot size $\sigma_x = \sigma_y \approx 10 mm$), low positron momentum degradation, minimum positron annihilation (a source of $\gamma$ background). The dc beam structure is also extremely important because the accidental background increases quadratically with the beam rate and this feature gives a lowest instantaneous background compared to a pulsed beam. 

The positron momentum, direction and time are measured by means of a novel spectrometer made by an array of 16 modules of low-mass drift chambers, inserted in a superconducting magnet, wich supplies a gradient field along the beam direction (Z-coordinate) with a maximum intensity at the center of 1.25 Tesla. The non-uniformity of B-field produces a constant projected radius of the positron trajectory, indipendently of the emitted positron angle and function only of the positron energy. Moreover low momentum positrons are swept away, without hitting the chambers, avoiding to busy the detector and to have not useful hits in the recostruction. 

Two sectors of 15 scintillating bars, mounted at each end of the spectometer and equipped with PMTs working in a high magnetic field region, provides the best timing measurement at this energy.

An innovative homogeneous liquid xenon calorimeter performes a precise measurement of the convertion point, timing and energy of the $\gamma$ ray. 846 photomultipliers are fully immersed in the xenon and only the VUV scintillating light is collect to preserve the rapity of detector based on the short scintillation time constants (22 and 45 ns).

The stability of the all detectors is continuously monitored and the performances are measured by means of complementary calibration methods. The position, energy and timing resolutions of the calorimeter at an energy closer to the signal one  are determined selecting 55 MeV $\gamma$ ray from the $\pi^{0}$ decay, produced from the pion charge exchange reaction $\pi^{-} p \to \pi^{0} n$ at rest. Energy calibration and linearity, stability, uniformity and purity are frequently measured illuminating  the inner face of the calorimeter with a sharp gamma line at 17.6 MeV from the resonant reaction  $Li(p,\gamma)Be $. The optical properties of the Xenon and the PMT's parameters (gain and QE) are extracted using Point-like Am-$\alpha$ source and LED. A 9 MeV line from the capture in nickel of neutrons from a pulsed and triggerable deuteron-deuteron neutron generator allows one to check the stability of the LXe detector even during data taking. The relative time between the TC and LXe detector is monitored using radiative muon decay events and 2 coincident $\gamma$ from the $B(p,2\gamma)C$ reaction. The absolute scale of the spectrometer is fixed by the Michel energy spectrum edge and a new method based on monochromatic Mott scattered positrons has being optimized to be used as standard tool for a deeper understanding of the spectrometer. 

An efficienty and flexible trigger which use only the fast detectors was developed to select with high efficiency the signal and to accomodate the different calibration methods. The DAQ is designed to digitize all waveforms, with an excellent capability to reject the pile-up. It is based on  the multi-GHz domino ring sampler chip (DRS), which can sample ten analogue input channels into 1024 sampling cells each at speeds of up to 4.5 GHz. The sampling speed for the drift chamber anode and cathode signals is 500 MHz, while that of the PMT signals from the photon detector and timing counters is 1.6 GHz.

A detailed GEANT 3.21 based Monte Carlo simulation of the full apparatus (transport system and detector) was developed and used throughout the experiment, from the design and optimization of all sub-systems to the calculation of acceptances and efficiencies. 

\section{Resuls from 2009 data sample} 
The data sample analyzed here was collected between September and December 2008 and corresponds to $\approx 9.5 \times 10^{13}$ muons stopping in the target. The collected sample is saved and a preliminary data reduction is performed. These pre-selected events are further processed. The events falling into a pre-defined window (``blinding-box"), containing the signal region on the $E_{\gamma}$-ray and on the $t_{e\gamma}$, were saved in separate hidden files; the $\mu \to e \gamma$ decay is searched within this sample. The other events (``side-bands") are used for optimizing the analysis parameters and for studying the background. The blinding-box is opened after completing the optimization of the analysis algorithms and the background studies.\\

A candidate $\mu \to e \gamma$ event is characterized by the measurement of five kinematical parameters: positron energy $E_e$, gamma energy $E_{\gamma}$, relative time between positron and gamma $t_{e\gamma}$ and the opening angles between the two particles $\theta_{e\gamma}$ and $\phi_{e\gamma}$.

A likelihood function is built in terms of the signal and the two kinds of background: the radiative Michel decay and the accidental background. A probability density function (PDF), depending on the five kinematical parameters, is associated to each component; the likelihood has the following expression:
\begin{equation} 
\mathcal{L}(N_{sig},N_{RMD},N_{BG}) = \frac{N^{N_{obs}}e^{-N}}{N_{obs}!}\prod_{i=1}^{N_{obs}}\left[ \frac{N_{sig}}{N}S + \frac{N_{RMD}}{N}R + \frac{N_{BG}}{N}B \right].
\end{equation}

The signal PDF $S$ is obtained as the product of the PDFs for the five observables ($E_{\gamma}$,  $E_e$, $t_{e\gamma}$,  $\theta_{e\gamma}$ and $\phi_{e\gamma}$). The radiative Michel decay PDF $R$ is the product of the theoretical PDF (in terms of the correlated $E_{\gamma}$,  $E_e$, $\theta_{e\gamma}$ and $\phi_{e\gamma}$), folded with the detector response, and the measured $t_{e\gamma}$ PDF (the same of the signal one); the PDF $B$ is the product of the background spectra for the five observables, which are precisely measured in the data sample in the side-bands.

\begin{figure}
\begin{center} 
\includegraphics[scale=0.55]{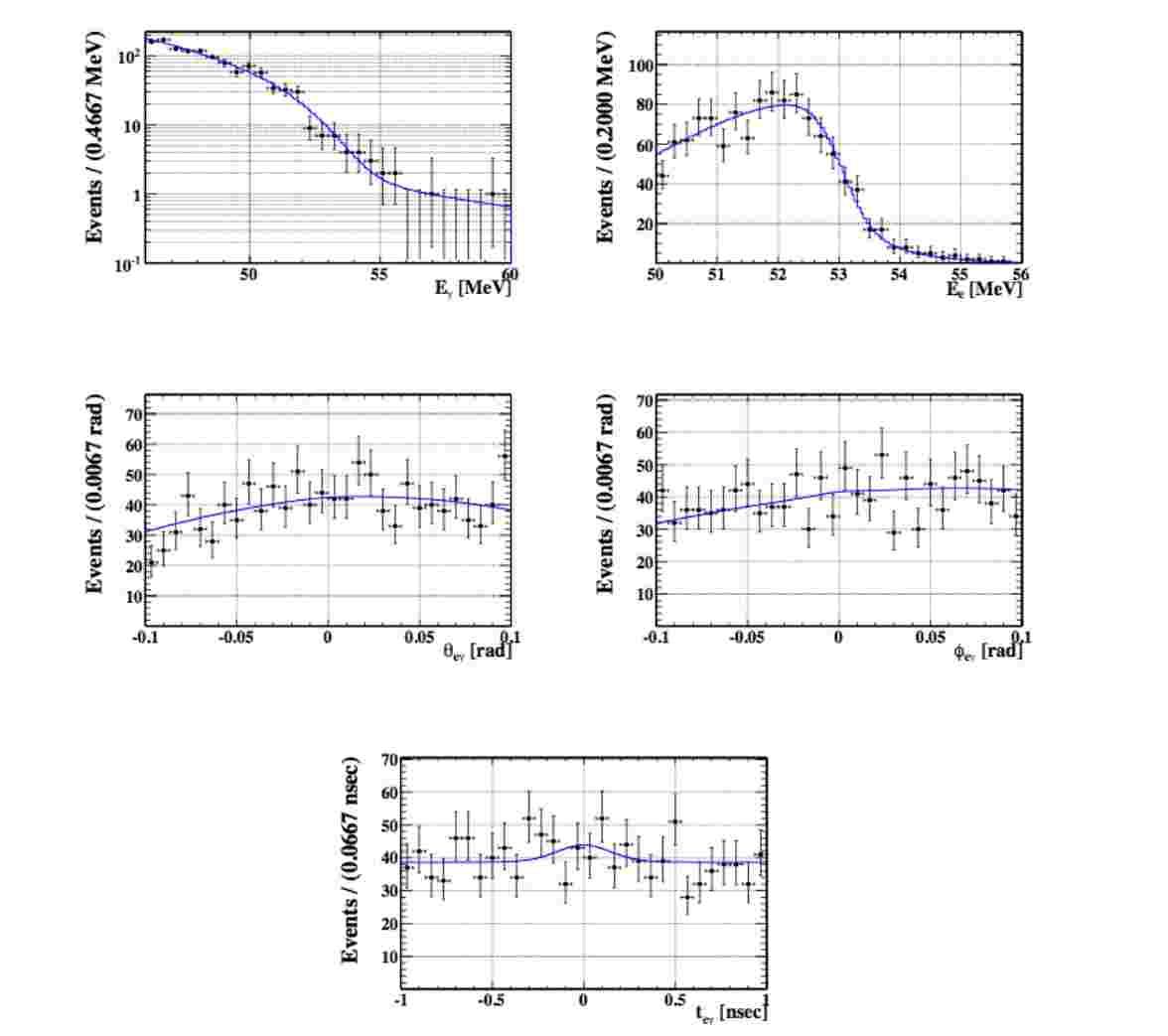}
\end{center}
\caption{The event distributions of the five observables for all events in the analysis window. The blu line shows the projections of the fitted likelihood function.}
\label{fig:Result2009}
\end{figure}

The event distributions of the five observables for all events in the analysis window are shown in Fig.~\ref{fig:Result2009}, togheter with the projections of the fitted likelihood function. The $90\%$ confidence intervals on $N_{sig}$ and $N_{RMD}$ are determined by the Feldman-Cousins approach~\cite{feldman}. A contuour of $90\%$ C.L. on the ($N_{sig}, N_{RMD}$)-plane is constructed by means of a toy Monte Carlo simulation. On each point on the contour, $90\%$ of the simulated experiments give a likelihood ratio ($\mathcal{L}/\mathcal{L}_{max}$) larger than that of the ratio calculated for the data. The limit for $N_{sig}$ is obtained from the projection of the contour on the $N_{sig}$-axis. The obtained upper limit at $90\%$ C.L. is $N_{sig} < 14.7$, where the systematic error is included. The largest contributions to the systematic error are from the uncertainty of the selection of photon pile-up events, the photon energy scale, the response function of the positron energy and the positron angular resolution.

The upper limit on BR($\mu^+ \to e^+ \gamma$) was calculated by the C.L. intervals normalizing the upper limit on $N_{sig}$ to the Michel positrons counted simultaneously with the signal, with the same analysis cuts, assuming  BR($\mu^+ \to e^+ \nu_e \overline{\nu}_{\mu}) \approx 1$. This method has the advantage of being indipendent of the instantaneous beam rate and is nearly insensitive  to the positron acceptance and efficiency factors associated with the DCH and TC, which differ only for small momentum dependent effects between the signal and the normalization sample.

The branching ratio can be written as:
\begin{equation}
\rm{BR}(\mu^+ \to e^+ \gamma) = \frac{N_{sig}}{N_{e\nu\overline{\nu}}} \times \frac{f^{E}_{e\nu\overline{\nu}}}{P}\times\frac{\epsilon^{trg}_{e\nu\overline{\nu}}}{\epsilon^{trg}_{e\gamma}}\times\frac{A^{TC}_{e\nu\overline{\nu}}}{A^{TC}_{e\gamma}}\times\frac{\epsilon^{DCH}_{e\nu\overline{\nu}}}{\epsilon^{DCH}_{e\gamma}}\times\frac{1}{A^{g}_{e\gamma}}\times\frac{1}{\epsilon_{e\gamma}}
\end{equation}
where $N_{e\nu\overline{\nu}}$ is the number of detected Michel positrons with $50\leq E_{e}\leq56$~MeV; $P$ is the prescale factor in the trigger used to select Michel positrons; $f^{E}_{e\nu\overline{\nu}}$ is the fraction of Michel positron spectrum above 50 MeV; $\epsilon^{trg}_{e\gamma}/\epsilon^{trg}_{e\nu\overline{\nu}}$ is the ratio of signal-to-Michel trigger efficiencies; $A^{TC}_{e\gamma}/A^{TC}_{e\nu\overline{\nu}}$ is the ratio of the signal-to-Michel DCH-TC matching efficiency; $\epsilon^{DCH}_{e\gamma}/\epsilon^{DCH}_{e\nu\overline{\nu}}$ is the ratio signal-to-Michel DCH recostruction efficiency and acceptance; $A^g_{e\gamma}$ is the geometrical acceptance for gamma signal given an accepted signal positron; $\epsilon_{e\gamma}$ is the efficiency of gamma recostruction and selection criteria.\\

The quoted limit on the branching ratio of the $\mu^+ \to e^+ \gamma$ decay is therefore:
\begin{equation}
\rm{BR}(\mu^+ \to e^+ \gamma) \leq 2.8 \times 10^{-11}  \; (90\% C.L.)
\end{equation}
where the systemetic uncertainty on the normalization is taken into account~\cite{FirstMEGResult}.

The analysis of the 2010 data is now finished and the results of both 2009 and 2010 data are available. The likelihood analysis of the combined data sample, which corresponds to a total of $1.8 \times10^{14}$ muon decays, gives an upper limit of:  
\begin{equation}
\rm{BR}(\mu^+ \to e^+ \gamma) \leq 2.4 \times 10^{-12}  \; (90\% C.L.)
\end{equation}
This is the most stringent limit on the existence of this decay to date~\cite{Results2010}.

%% file: Author/Ketan_M_Patel.tex
\begin{center}
{\bf Abstract}\\
\vskip5.mm
We derive the most general structures of the charged lepton and the neutrino mixing matrices which
lead to tri-bimaximal leptonic mixing. By integrating them into an \10 model, we show that one can
obtain excellent fits to all the fermion masses and quark mixing angles keeping tri-bimaximal
leptonic mixing intact. We also consider different perturbations to the basic structure which
can/cannot account for the recent T2K and MINOS results on the reactor mixing angle $\theta^l_{13}$.
\end{center}

\vskip5.mm

\section{Introduction}
The Tri-bimaximal (TBM) mixing in the lepton sector \cite{hs} provides a very important clue in
search of possible flavour structure \cite{af} which governs the leptonic masses and mixing angles.
The exact TBM predicts vanishing reactor mixing angle $\theta^l_{13}$ which can be reconciled with
the recent T2K \cite{t2k} (MINOS \cite{minos}) results at 2.5$\sigma$ (1.6$\sigma$) and with the
global analysis \cite{fogli,valle} at about 3$\sigma$. This suggests that the TBM may be good zeroth
order approximation which needs perturbations affecting mainly the reactor mixing angle. While such
perturbations may arise from some underlying flavor symmetry, it would be more appropriate if some
independent mechanism like grand unified theory (GUT) governs these perturbations. On the other
hand, incorporation of the exact TBM mixing into GUTs is a nontrivial task. This becomes more
challenging in the case of \10 based GUTs since all fermions in a given generation are completely
unified into a 16 dimensional irreducible representation of \10 and imposition of the TBM structure
on the leptonic mass matrices also constrains the quark mass matrices. It is not known whether the
requirement of the exact TBM mixing among leptons is consistent with the quark masses and mixing.
In this talk, we discuss this issue and present the detailed analysis of the exact TBM structure
and perturbations to it within a grand unified model based on SO(10) gauge symmetry.

\section{The exact TBM leptonic mixing in \10 Model}
The most general structures of the neutrino mixing matrix $M_{\nu}$ and the left handed charged
lepton mixing matrix $U_l$ which lead to the exact TBM leptonic mixing are derived in \cite{ak}. We
briefly discuss them here. It is well known that the neutrino mass matrix in the flavour
basis ${\cal M}_{\nu f}$ leads to the exact TBM mixing if it is invariant under a $Z_2\times Z_2$
symmetry whose elements are 
\be \label{s2s3} \begin{array}{cc}
S_2=\dfrac{1}{3}\left( \begin{array}{ccc} -1&2&2\\
		  2&-1&2\\
                  2&2&-1 \\ \end{array} \right) ~~~{\rm and}~~~
&
S_3=\left( \begin{array}{ccc} 1&0&0\\
		  0&0&1\\
                  0&1&0 \\ \end{array} \right),\end{array} \ee
{\it i.e.} ${\cal M}_{\nu f}$ must satisfy
\be \label{symmetrynuf}
S_{2,3}^{T}{\cal M}_{\nu f}S_{2,3}={\cal M}_{\nu f}. \ee
Let us now find out the circumstances under which the above condition is satisfied. One can always
choose a specific basis in which the original neutrino mass matrix $M_{\nu}$ exhibits the TBM
structure and thus satisfies $ S_{2,3}^{T}M_\nu S_{2,3}= M_\nu~$. In this basis, if $U_l$ (which
denote the mixing matrix among the left handed charged leptons) itself is  $Z_2\times Z_2$
symmetric, {\it i.e.} satisfies
\be \label{symmetryul}
S_{2,3}^{T}U_lS_{2,3}=U_l~ \ee
then ${\cal M}_{\nu f}$ will also satisfy Eq.~(\ref{symmetrynuf}) and thus would exhibit the TBM
structure. It is shown in \cite{ak} that such $U_l$ can be parameterized as 
\be \label{ulp}
U_l=e^{i \alpha} P_l\tilde{U_l} P_l~~ {\rm and}~~ \tilde{U_l}=\left(
\begin{array}{ccc}
 c_\theta & \frac{s_\theta}{\sqrt{2}}  & \frac{s_\theta}{\sqrt{2}}  \\
 \frac{s_\theta}{\sqrt{2}}& -\frac{1}{2} (c_\theta + e^{i \delta})
& -\frac{1}{2} (c_\theta - e^{i \delta}) \\
 \frac{s_\theta}{\sqrt{2}}& -\frac{1}{2} (c_\theta - e^{i \delta}) & -\frac{1}{2} (c_\theta + e^{i
\delta}) \end{array}
\right),\ee
where $P_l={\rm diag.}(1,e^{i \beta},e^{i \beta})$ is a diagonal phase matrix
and $\tan\theta=-2\sqrt{2}\cos \beta$. $U_l$ is thus fully determined by three phase angles
$\alpha, \beta$ and $\delta$.

Let us now integrate the above leptonic structures into an \10 model. For the simplification that
allow us to obtain quantitative description, we assume: (1) a supersymmetric \10 model with Higgs
transforming as $10, \overline{126}, 120$ representations of \10, (2) the generalized parity
\cite{jp} leading to Hermitian mass matrices and (3) the type-II seesaw dominance. The fermion
mass relations in this case after electroweak symmetry breaking can be  written in
their most general forms as \cite{jp}:
\beqa \label{genmass}
M_d= H+F+i G; &~& M_u=r (H+s F+i t_u~ G~); \nonumber \\
M_l= H-3 F+~it_l~ G; &~& M_D=r (H-3s F+i t_D~ G~);\nonumber\\
M_L= r_L F; &~& M_R= r_R^{-1} F.\eeqa
where ($G$) $H$, $F$ are real (anti)symmetric matrices. $r,s,t_l,t_u,t_D,r_L,r_R$ are
dimensionless real parameters. The effective neutrino mass matrix for three light neutrinos
resulting after the seesaw mechanism can be written as
\be \label{mnu}
M_\nu=r_LF-r_RM_DF^{-1}M_D^T\equiv  M_\nu^{II}+M_\nu^{I}
~.\ee
The first term proportional to $F$ denotes type-II seesaw contribution. We shall assume that $M_\nu$
is entirely given by this term and subsequently analyze the effect of a small type-I corrections on
the numerical solution found. 

One can always rotate the 16-plet fermions in generation space in such a way that $M_\nu \propto F$
is diagonalized by the TBM matrix. $F\rightarrow R^TFR=F_{TBM}\equiv O_{TBM}~{\rm
Diag.}(f_1,f_2,f_3)~O_{TBM}^T$ where $f_i$ are now real eigenvalues of $F$ and the $O_{TBM}$ is
tri-bimaximal orthogonal matrix. The matrix ($G$)$H$ maintains its (anti)symmetric form in such
basis. The charged lepton mass matrix in Eq.~(\ref{genmass}) can be rewritten as 
\be \label{mlfixing}
H+it_lG=V_l D_l V_l^\dagger+3 F_{TBM}~,\ee
where $D_l$ is a diagonal charged lepton mass matrix. $V_l$ is a unitary matrix that diagonalizes
$M_l$ and contains nine free parameters in the most general situation.

We perform the $\chi^2$ fitting to study the viability of Eq.~(\ref{genmass}) with the
experimentally observed values of fermion masses and mixing angles. The details of numerical
analysis are given in \cite{ak}. We shall present numerical analysis in two different cases. (A)
Corresponding to the most general $V_l$ and (B) with $V_l=U_l$ given as in
Eqs.~(\ref{ulp}). The case (A) has already been studied numerically in
\cite{jp,ab}. We refine this analysis taking into account the results of the most recent
global fits \cite{fogli} to neutrino data. This also serves as a benchmark with which to compare the
case (B) which leads to the exact TBM at $M_{GUT}$. The results of numerical analysis are displayed
in Table \ref{table:op}.
\begin{table}[ht]
\begin{center}
\begin{scriptsize}
\begin{math}
\begin{array}{|c||c|c||c|c|}
\hline
  & \multicolumn{2}{|c||} {\text{\bf Case A}} & \multicolumn{2}{|c|} {\text{\bf Case B}}  \\
\hline
 \text{Observables} & \text{Fitted value} & \text{Pull} & \text{Fitted value} &
\text{Pull}  \\
\hline
 m_d [{\rm MeV}] & 1.2339 & -0.0148738 & 1.22098 & -0.0463899 \\
 m_s [{\rm MeV}] & 21.7214 & 0.00411949 & 21.9922 & 0.0561874 \\
 m_b [{\rm GeV}] & 1.06614 & 0.0438763 & 1.16345 & 0.738942 \\
 m_u [{\rm MeV}] & 0.550018 & 0.000073755 & 0.550234 & 0.000936368\\
 m_c [{\rm GeV}] & 0.209977 & -0.00111886 & 0.209952 & -0.00230315\\
 m_t [{\rm GeV}] & 82.5278 & 0.00421748 & 82.5855 & 0.00612198 \\
 m_e [{\rm MeV}] & 0.3585 & - & 0.3585 & - \\
 m_{\mu } [{\rm MeV}] & 75.672 & - & 75.672 & -\\
 m_{\tau }[{\rm GeV}] & 1.2922 & - & 1.2922 & - \\
 \left( \dfrac{\dms}{\dma}\right) & 0.0323 & - & 0.031875 & - \\
 \sin  \theta _{12}^{q} & 0.224299 & -0.000688878 & 0.2243 & 0.0002182 \\
 \sin  \theta _{23}^{q} & 0.0351032 & 0.00246952 & 0.0350951 & -0.0038047 \\
 \sin  \theta _{13}^{q} & 0.00320513 & 0.0102511 & 0.00319436 & -0.0112796 \\
 \sin ^2 \theta _{12}^{l} & 0.306119& 0.00660722 & \textbf{0.3333} & - \\
 \sin ^2 \theta _{23}^{l} & 0.418475 & -0.0508353 & \textbf{0.5} & - \\
 \sin ^2 \theta _{13}^{l} & 0.0207708 & -0.0286467 & \textbf{0} & - \\
 J_{CP} & 2.19 \times 10^{-5} & -0.0183401 & 2.21\times 10^{-5} & 0.0194165 \\
 \delta _{MNS} & \textbf{282.396} & - & - & - \\
\hline
 \chi^2_{min} &   & \textbf{0.0061} &    & \textbf{0.5519} \\
\hline
\end{array}
\end{math}
\end{scriptsize}
\caption{Best fit solutions for fermion masses and mixing obtained in the SUSY \10 model with
$10+\overline{126}+120$ Higgs assuming (A) the general (non TBM) leptonic mixing and (B) exact TBM
leptonic mixing with $U_l$ of Eqs.~(\ref{ulp}). The predictions of different
approaches are shown in boldface.}
\label{table:op}\end{center}
\end{table}
\vspace*{-0.5cm}

We obtain an excellent fit corresponding to $\chi^2_{min}=0.552$ ($\chi^2_{min}/{\rm
d.o.f.}=0.276$) in the case of exact TBM leptonic mixing. Only the fitted value of $m_b$ deviates
slightly from the central value with a $0.74\sigma$ pull. All the remaining observables are fitted
within $0.06\sigma$. The obtained fit is not significantly different from the general case (A), in
which we obtain $\chi^2_{min}/{\rm d.o.f.}=0.0061$, showing that all the fermion masses and mixing
angles can be nicely reproduced along with the exact TBM within the \10 framework discussed here.

\section{Perturbations to the exact TBM mixing}
The TBM is an ideal situation and various perturbations to this can arise in the model. A deviation
from tri-bimaximality can arise in the model due to the following reasons.
\begin{enumerate}
 \item The quantum corrections due to running from $M_{GUT}$ to $M_Z$.
 \item Contribution from the sub dominant type-I seesaw term in Eq.~(\ref{mnu}).
 \item Corrections from the charged leptons due to the the breaking of the $Z_2\times Z_2$ symmetry
in $U_l$.
\end{enumerate}
The effect of (1) is known to be negligible \cite{rg} in case of the hierarchical neutrino mass
spectrum which we obtain here. We quantitatively discuss the implications of the other two
scenarios via detailed numerical analysis.

\subsection{Perturbation from type-I seesaw}
Depending on the GUT symmetry breaking pattern and parameters in the superpotential of the theory, a
type-I seesaw contribution can be dominant or sub dominant compared to type-II but it is always
present and can generate deviations in an exact TBM mixing pattern in general. In the approach
pursued here it is assumed that such contribution remains sub dominant and generates a small
perturbation in dominant type-II spectrum. Eq.~(\ref{mnu}) can be rewritten as
\be \label{mnu1}
M_\nu=r_L (F-\xi M_DF^{-1}M_D^T)
~\ee
where $\xi = r_R/r_L$ determine the relative contribution of type-I term in the neutrino mass
matrix.

The second term in Eq.~(\ref{mnu}) brings in two new parameters $\xi$ and $t_D$ present in the
definition of $M_D$ in Eq.~(\ref{genmass}) which generate departure from the exact TBM. For the
quantitative analysis of this deviation, we randomly vary the parameters $\xi$ and $t_D$ and
evaluate the neutrino masses and mixing angles. The correlations between different leptonic mixing
angles found from such analysis are shown in Fig.~(1) in \cite{ak}.

It is found that the perturbation induced by type-I term cannot generate considerable deviation in
the reactor angle. In particular, requiring that $\sin^2\theta_{12}^l$ remains within the 3$\sigma$
range puts an upper bound $\sin^2\theta_{13}^l \le 0.0002$ which does not agree with the latest
results from T2K and MINOS showing that a small perturbation from type-I term cannot be consistent
with data when the type-II term displays exact TBM.

\subsection{Perturbation from the charged lepton mixing}
\begin{figure}[ht]
 \centering
 \includegraphics[width=12cm]{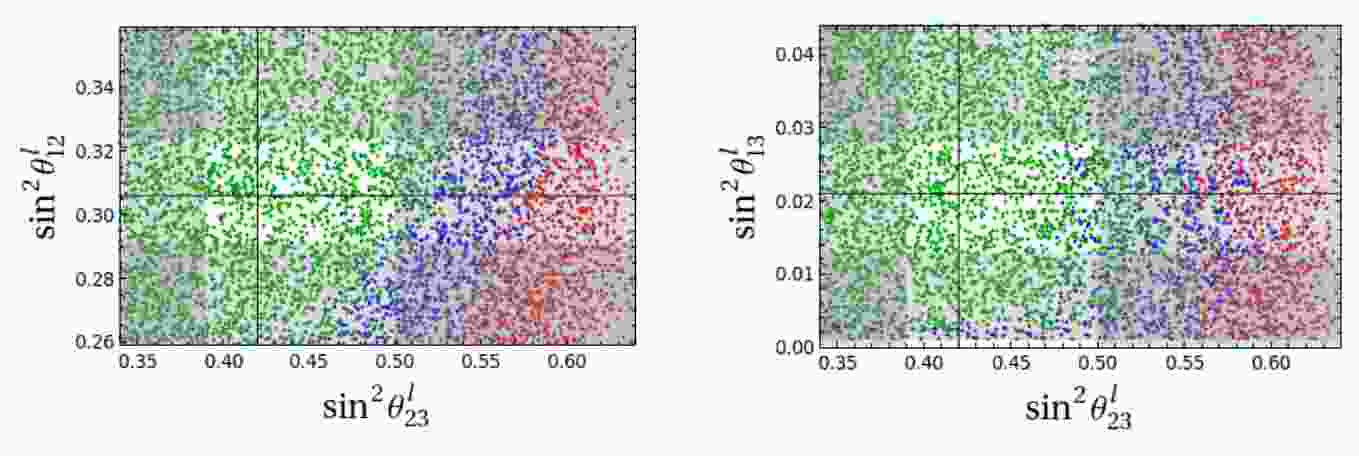}
\caption{Correlations among the lepton mixing angles in  case of the most general
charged lepton mixing matrix $U_l$. The points with different colors correspond to
$\bar{\chi}^2_{min} < 1$ (green), $1 \le \bar{\chi}^2_{min} < 4$ (blue) and $\bar{\chi}^2_{min} \ge
4$ (red).}
\label{fig:1}
\end{figure}

Perturbation to TBM arise when $U_l$ deviates from its $Z_2\times Z_2$ symmetric form given in
Eq.~(\ref{ulp}). In this case, the neutrino mass matrix has TBM structure but the charged lepton
mixing leads to departure from it. This case has been considered in the general context
\cite{leptonic} as well as in $SO(10)$ context \cite{ab}. Within our approach, we simultaneously
perturb all three mixing angles and look at the quality of fit compared to the exact TBM case. For
this we choose $U_l$ to be a general unitary matrix and repeat the analysis in case (A). There, we
have fitted the solar and the atmospheric mixing angles to their low energy values given in
\cite{fogli}. Here, we pin down specific values of the lepton mixing angles $p_i$ by modifying
the definition of $\chi^2$ function. Further details of such an analysis are given in \cite{ak}. The
results are displayed in Fig.~(\ref{fig:1}).

The green points represent very good fit in which all the observables are fitted within 1$\sigma$.
The obtained fit shown by the blue points is not as good as the previous one but it is statistically
acceptable. The red points represent poor fit which can be ruled out at 95\% confidence level.
Fig.~(\ref{fig:1}) shows definite correlations between $\theta_{23}^l$ and $\theta_{12}^l$. It is
also seen that the entire range $0.001\leq\sin^2\theta_{13}^l\leq 0.044$ is consistent with
statistically acceptable fits to fermion spectrum. This is to be contrasted with the previous case
where perturbation from type-I seesaw term led to an upper bound. The bounds obtained numerically
allows us to clearly distinguish the case of the exact TBM at $M_{GUT}$ in comparison to the one in
which the charged leptons lead to departures from the tri-bimaximality. 

\section{Conclusion}
In conclusion, we have analyzed the viability of the exact TBM lepton mixing in the context of the
grand unified $SO(10)$ theory taking a specific model as an example. It is shown that excellent
fits to all the fermion masses and quark mixing angles can be obtained keeping tri-bimaximal
leptonic mixing intact. The existence of TBM at the GUT scale may be inferred by considering its
breaking which can arise in the model and the reactor mixing angle is found to be a good pointer to
this. It is found that the quantum corrections and corrections coming from type-I seesaw are
disfavored by the recent T2K and MINOS results while the corrections coming from the charged lepton
mixing matrix can account for the T2K and MINOS results being consistent with the
detailed description of all the fermion masses and mixing angles.

%% file: Author/UJSaldanaSalazar.tex
{\bf Abstract}\\
\vskip5.mm
The Minimal $S_3$-Invariant Extension of the Standard Model (M$S_3$IESM) is formulated by introducing in the theory three Higgs fields that are $SU(2)_L$ doublets and a flavour permutational symmetry, $S_3$, in addition to the Majorana nature of massive neutrinos. In this way, the concept of flavour is extended to the Higgs sector and, hence, taken to a more fundamental level. The state of the art of the present model is first discussed and then, inspired in the already solved lepton sector, a $Z_2$ symmetry is introduced in the quark sector to achieve a further reduction in the number of free parameters. The study of the latter leads to conclude that a $Z_2$ symmetry is too constrictive and that a further analysis without it is needed. I end up making some meaningful remarks about the most general $S_3$-invariant Higgs potential that need to be taken into account.

\vskip5.mm

\section{Introduction}		
	The Standard Model (SM) has been succesful in describing the fundamental interactions between elementary particles. It is based upon the gauge symmetry group: $G_{SM} = SU(3)_C \otimes SU(2)_L \otimes U(1)_Y$. When the experimental precision improved it brought about a better understanding of the theory from the particle physicist point of view. Since then, the SM has been tested at every kind of imaginable experiment. Given its vast range of achieved predictions, it has become a very prestigious theory, but the story didn't have a happy ending, because when the massive nature of neutrinos in 1998 was finally confirmed the SM got its biggest hole. Which of course was rapidly covered by the introduction of the Majorana massive nature for neutral fermions, considering that Dirac neutrinos couldn't provide a natural explanation of the smallness of their masses.
	
	Other kind of open problems in the SM, just to mention some of them, are: the large number of free parameters, the mass hierarchy of fermions, the origin of the quark and lepton mixing patterns, that there are three and only three generations of fundamental particles, etc. The latter is called \textit{the flavour problem}.	
	
\section{The Minimal $S_3$-Invariant Extension of the Standard Model}
	Since the late 60's, a lot of work on how to relate the different families has been carried on, most of it motivated by the fact that the Cabibbo angle can be expressed as a quark mass ratio \cite{Gatto:1968ss,Cabibbo:1968vn,Pagels:1974qg,Weinberg:1977hb,Wilczek:1977uh,Fritzsch:1977za,Ebrahim:1978vv,Mohapatra:1977rj}, $\theta_c \approx \sqrt{m_d / m_s}$.
	
	In the following subsections, I will try to give a straightforward reasoning to show, how by demanding simplicity to a flavour extension of the SM we are inevitably taken to the construction of the Minimal $S_3$-Invariant Extension of the Standard Model (M$S_3$IESM), and, as an extraordinary consequence, we are brought to a \textit{natural correspondence} between experiment and theory.
	
	\subsection{A bottom-up approach}	
		Following a bottom-up approach, the simplest way to extend the SM, in order to give an explanation to some of the problems stated above, is the introduction of a \textit{flavour symmetry}. By doing this, we are taking into account that families are indistinguishable before introducing the Yukawa interactions. But, what kind of symmetry are we going to use? If we wish to avoid the appearance of Goldstone bosons or flavons then we need to focus only on discrete flavour symmetries.
		
		In 1977, F. Wilczek et al \cite{Wilczek:1977kr} stated, that mixing angles and masses, are possibly generated by the same dynamical mechanism and therefore, if this possibility is a fact, they should be somehow related. In the majority of the flavour models, it has been found, that a flavoured Higgs mechanism serves for this purpose, by relating mass ratios to mixing angles. Thus, the need to find a symmetry group that is capable of doing this task in the most natural and simplest way, is of major importance and pressing. 
		
		In 1978, Barbieri et al \cite{Barbieri:1978qh} showed that under the gauge group $SU(2)_{L} \otimes U(1)_{Y}$ of the SM, it is not posible to reproduce the Cabibbo angle in terms of quark mass ratios under the assumption of any abelian discrete symmetry and an arbitrary number of Higgs and families. Then, the only possibility left is a non-abelian discrete symmetry. In 1979, D. Wyler \cite{Wyler:1978fj} showed that by using a non-abelian discrete symmetry and by introducing at least three Higgs weak doublets, it becomes possible to express the Cabibbo angle in terms of quark mass ratios. Rephrasing the results stated before: Barbieri et al showed that when Flavor Changing Neutral Currents (FCNC's) are strictly forbidden, and therefore an abelian discrete symmetry is used, then, it becomes impossible to express the mixing angles in terms of mass ratios even with an arbitrary number of Higgs bosons and families; but, if we allow FCNC's to occur, then, we can use a non-abelian discrete symmetry, and, when conservation of the flavour symmetry is assumed, D. Wyler showed, that then, we need  to introduce at least three Higgs weak doublets.
		
		Hence, by demanding simplicity in the theory we are forced to introduce the simplest non-abelian discrete flavour symmetry, that is, the permutational symmetry $S_3$. And, since we want to test a model with its flavour symmetry conserved, we minimally extend the scalar sector by adding two more Higgs weak doublets. As an extraordinary consequence of having now three Higgs weak doublets, \textit{the concept of flavour is taken to a more fundamental level}, because now, the Higgs sector is flavoured in the same way as all fermions.
				
	\subsection{The permutational symmetry $S_3$}
		$S_3$ is the symmetry group of permutations of three objects. It has six elements, the smallest number of elements in non-abelian discrete groups. It has three irreducible representations (irreps): a doublet $\textbf{2}$, and two singlets, $\textbf{1}_S$ and $\textbf{1}_A$, symmetric and antisymmetric, respectively. Choosing a three dimensional real representation of the group, in accordance with the dimension of the mass matrices, we are led to have only two irreps, whose projectors satisfy: $\cal{P}_{\textbf{1}_S} + \cal{P}_{\textbf{2}}$ $= {1}_{3\times3}$.
		
		The direct products of irreps, including the antisymmetric singlet, are: $\textbf{1}_{S} \otimes \textbf{1}_{S} = \textbf{1}_{S}$, $\textbf{1}_{A} \otimes \textbf{1}_{A}= \textbf{1}_{S}$, $\textbf{1}_{A} \otimes \textbf{1}_{S}= \textbf{1}_{A}$, $\textbf{1}_{S} \otimes \textbf{2}= \textbf{2}$, $\textbf{1}_{A} \otimes \textbf{2}= \textbf{2}$, and $\textbf{2} \otimes \textbf{2} =  \textbf{1}_{A} \oplus \textbf{1}_{S} \oplus \textbf{2}$.
		
	\subsection{Assignment of families to $S_3$ irreps}		
\begin{figure}[h!]
			\centering
				\vspace{-15pt}
				\includegraphics[scale=.8]{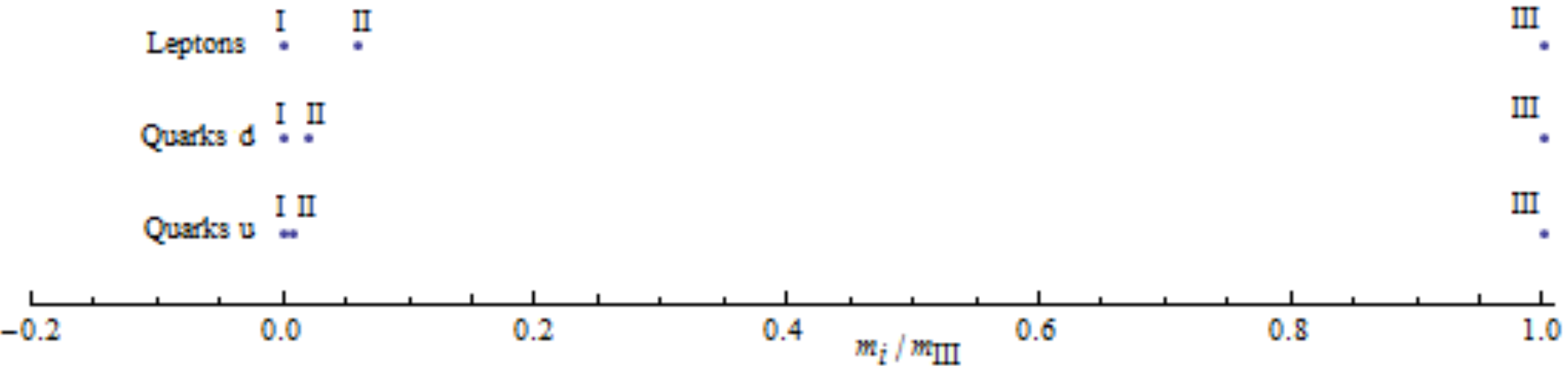}
				\vspace{-25pt}
				\caption{The flavour structure unveiled from the charged fermions mass spectrum. The indices I, II and III refer to the family index.}
				\vspace{-10pt}
				\label{Espectro}
		\end{figure}

		Mass values individually can not show us any relation between different families.  A direct comparison between the known fermion masses can only give us information about the minimum energy scale in which the particles may be created. To unveil the family or flavour structure in each fermion sector, it is only necessary to consider the mass ratios obtained by dividing masses over the largest of each sector. The fermion mass spectrum takes a form in which the flavour structure becomes readible, see Figure \ref{Espectro}.	
		
		It's now clear how the assignment should be done: the first two families in the doublet representation, $\textbf{2}$, and the third family in the symmetric singlet representation, $\textbf{1}_S$. 
		
		The natural correspondence between the $S_3$ irreps in the three dimensional real representation and the flavour structure may also be seen as \textit{a posteriori} argument on why $S_3$ should be chosen as the flavour symmetry.
		
	\subsection{Flavoured Yukawa interactions and the generic mass matrix for Dirac fermions}
		Once we have gathered all the necessary group symmetry ingredients into our theory, then we are ready for the construction of the two physical flavour theoretical elements: the Yukawa Lagrangian and the Higgs potential. In this subsection I will only discuss the Yukawa Lagrangian. 
		
		The most general $S_3$-invariant renormalizable Yukawa Lagrangian is constructed by asking that the product of left and right fermionic fields with a Higgs field is invariant under $S_3 \otimes G_{SM}$. In this way, the Yukawa Lagrangian is given by \cite{Kubo:2003iw}: 	
		
\noindent	\begin{center}$
			{\cal L}_{Y} = {\cal L}_{Y_{d}} + {\cal L}_{Y_{u}} + {\cal L}_{Y_{e}} + {\cal L}_{Y_{\nu}},		
 		$\end{center}
 		where,

\noindent	\begin{flushleft}$
	{\cal L}_{Y_d} = -Y^{d}_{1}\overline{Q}_I {H}_S d_{IR} -Y^{d}_3 \overline{Q}_3 {H}_S d_{3R} -Y^{d}_{2}[\overline{Q}_{I}\kappa_{IJ}H_1  d_{JR} + \overline{Q}_{I} \eta_{IJ} H_2  d_{JR}] -Y^{d}_{4}\overline{Q}_3 {H}_I d_{IR} - Y^{d}_{5} \overline{Q}_I {H}_I d_{3R} + h.c.$,

\noindent ${\cal L}_{Y_u} = -Y^{U}_1\overline{Q}_I (i \sigma_2)H_S^* u_{IR} -Y^u_3 \overline{Q}_3(i\sigma_2) H_S^* u_{3R}  - Y^{u}_{2}[\overline{Q}_{I}\kappa_{IJ}(i \sigma_2)H_1^* u_{JR} + \overline{Q}_{I} \eta_{IJ}(i\sigma_2) H_2^*  u_{JR}] -$ 

\indent$Y^u_{4}\overline{Q}_3(i \sigma_2) H_I^* u_{IR} -Y^u_{5} \overline{Q}_I (i\sigma_2)H_I^* u_{3R} + h.c.$,

\noindent ${\cal L}_{Y_e} = -Y^e_1\overline{L}_I H_S e_{IR} - Y^e_3 \overline{L}_3 H_S e	_{3R} - Y^{e}_{2}[\overline{ L}_{I}\kappa_{IJ}H_1  e_{JR} + \overline{ L}_{I} \eta_{IJ} H_2  e_{JR}] - Y^e_{4}\overline{ L}_3 H_I e_{IR} - Y^e_{5} \overline{ L}_I H_I e_{3R} + h.c.$,
		
\noindent ${\cal L}_{Y_\nu} = -Y^{\nu}_1\overline{ L}_I (i \sigma_2)H_S^* \nu_{IR} -Y^\nu_3 \overline{ L}_3(i\sigma_2) H_S^* \nu_{3R} -   Y^{\nu}_{2}[\overline{ L}_{I}\kappa_{IJ}(i \sigma_2)H_1^*  \nu_{JR} + \overline{ L}_{I} \eta_{IJ}(i\sigma_2) H_2^*  \nu_{JR}] - Y^\nu_{4}\overline{ L}_3(i \sigma_2) H_I^* \nu_{IR}	- Y^\nu_{5} \overline{ L}_I (i\sigma_2)H_I^* \nu_{3R} + h.c.$,
		\end{flushleft}
		with:
		\begin{eqnarray*}
			\kappa = \begin{pmatrix}
			0 & 1 \\
			1 & 0
			\end{pmatrix};
			\hspace{1.5cm}\hspace{1.5cm} 
			\eta = \begin{pmatrix}
			1 & 0 \\
			0 & -1
			\end{pmatrix}
			\hspace{1.5cm} I, J = 1,2.
		\end{eqnarray*}		
		
		After the spontaneous electroweak-symmetry breaking, the Higgses acquire a vacuum expectation value (vev) and with this, the Yukawa terms become mass terms for all fermions. The generic mass matrix for Dirac fermions is \cite{Derman:1978rx,Yahalom:1983kf}: 
		
\noindent	\begin{center}$
				{\bf M}_f = \begin{pmatrix}
	 -2 Y_1 v_s -2 Y_2 v_2 & -2 Y_2 v_1 & -2 Y_4 v_1 \\
	   	-2 Y_2 v_1 & -2 Y_1 v_s + 2 Y_2 v_2 & -2 Y_4 v_2 \\
		-2 Y_5 v_1 & -2 Y_5 v_2 & -2 Y_3 v_s
				\end{pmatrix},
	  $	\end{center}	
		where the vev's satisfy the SM relation: ${v_1}^{2} + {v_2}^{2} + {v_s}^{2} \approx (\frac{246}{\sqrt{2}} GeV)^{2}$.
			
		In the following sections it will be assumed a Majorana nature for massive neutrinos and, in order to achieve a further reduction in the number of free parameters, we will take $v_1 = v_2 \neq 0$ \cite{Pakvasa:1977in}. 		
			
	\subsection{Some highlights of the published work}
		The $S_3$ flavour symmetry has been analitical and numerically tested in different types of models \cite{Wyler:1978fj,Kubo:2003iw,Derman:1978rx,Yahalom:1983kf,Pakvasa:1977in,Teshima:2011wg,Frere:1978ds,Mondragon:1998gy,Kubo:2004ps,Mondragon:2007af,Meloni:2010aw,Bhattacharyya:2010hp,Koide:1999mx,Chen:2004rr,Beltran:2009zz,Morisi:2006pf,Araki:2005ec,Kimura:2005sx,Kaneko:2007ea,Mohapatra:2006pu,Xing:2010iu,Barranco:2010we,Mondragon:2007jx,Mondragon:2007nk,Koide:2005ep}. The following published results come from a similar extension as the one described here, and are intended to give a sample of how well works the model:
		\begin{itemize}
	\item Numerical analysis of the quark mixing matrix, without introducing a $Z_2$ symmetry, is in good agreement with experimental data \cite{Kubo:2003iw,Teshima:2011wg}.
	\item By adding a $Z_2$ symmetry to the lepton sector that forbade some Yukawa couplings, the mass and mixing flavour structure of the lepton sector was succesfully reproduced \cite{Mondragon:2007jx,Mondragon:2007nk,Mondragon:2009zza}:
	\begin{itemize}
	\item The leptonic mixing angles were successfully expressed in terms of lepton mass ratios.
	\item It predicts an inverted hierarchy for massive neutrinos. 
	\item A predicted value for the lepton mixing angle $\theta_{13}^{th} \approx 0.0034$, which now, with the new T2K and MINOS updated values, is well below the global fit lower bound $0.01 \leq \theta_{13}^{exp} \leq 0.154$ at $\pm 1\sigma$ \cite{Schwetz:2011zk}.
	\item The FCNC's contribution, $\delta {a}_\mu$, to the anomaly of the muon's magnetic moment is smaller than or of the order of 6\% of the discrepancy  $\Delta {a}_\mu$, between the experimental value and the SM prediction, $\frac{\delta {a}_\mu}{\Delta {a}_\mu} \approx 0.06$.
	\item Branching ratios were computed for leptonic processes via FCNC's as $\mu\rightarrow e\gamma$ and $\mu\rightarrow 3e$ and it gave for these particular processes $2$.$42\times10^{-20}$ and $2$.$53\times10^{-16}$, respectively.
	\end{itemize}
		\end{itemize}
		
	\subsection{Some new results}
	In the first subsection, the stated purpose of this work was to do a similar treatment of the number of quark free parameters as in the lepton sector, and, for the second subsection, to remade the calculation of the most general $S_3$-invariant Higgs potential.
	
	\subsection{A $Z_2$ symmetry in the quark sector}
	The lepton sector was analitically solved by introducing a $Z_2$ symmetry which forbade some of the Yukawa couplings. This reduction in the number of the free parameters allowed us to reparametrize the lepton mass matrices in terms of lepton mass ratios. Could this restriction have the same effect in the quark sector? We studied the interdependence between different Yukawa couplings, in such a manner that whenever we forbid one coupling as a consequence another one could be necessarily zero.
	
	Studying the particular $v_1 = v_2 \neq 0$ case, all the different allowed combinations of zeroes, in the quark mass matrices, gave us all the possible invariant under $S_3 \otimes Z_2$ matrix forms that we could get. Considering the same matrix form for both quark sectors, we found, both analitical and numerically, that a $Z_2$ symmetry in the quark sector is too constrictive,  because the predicted Cabibbo-Kobayashi-Maskawa (CKM) mixing matrix can not reproduce the experimental values  and moreover \textit{does not depend dominantly on the quark mass ratios}.
	
	An important remark on these results comes from the fact that the predicted Pontecorvo-Maki-Nakagawa-Sakata PMNS matrix depends poorly on the lepton mass ratios, but this is not seen as a bad feature because we know that the TriBiMaximal (TBM) form shares this behavior, but, in the case of the CKM mixing matrix, from the Wolfenstein Parametrization we know that mostly all of the dependence comes from a single mass ratio, $\theta_c \approx \sqrt{m_d / m_s}$, hence we expect to see this dominant nature of the Cabibbo angle in any reparametrization faithful to the flavour structure.
	
	\subsection{The most general $S_3$-invariant Higgs potential}
	In past references \cite{Wyler:1978fj,Derman:1978rx,Yahalom:1983kf,Pakvasa:1977in,Kubo:2004ps,Meloni:2010aw,Bhattacharyya:2010hp,Koide:1999mx,Chen:2004rr,Beltran:2009zz,Kimura:2005sx,Koide:2005ep} the expression written by each one of these authors and claimed as the most general $S_3$-invariant Higgs potential, differs when compared with the others. We believe the origin of this confusion comes from the fact that the Higgs fields carry flavour and weak indexes, and these features were not taken correctly into account. Because the Higgs phenomenology of this model depends on the Higgs potential, we repeated the calculation and we found some meaningful remarks. We briefly sketch how we did it and we give one example of an $S_3$ invariant term.
	
	The first step consists in finding all the different linearly independent product combinations of irreps, that according to the flavour symmetry group make an invariant term. We assign a different coupling for each linearly independent term formed. After that, the second step consists in contracting all the different and independent combinations of the $SU(2)_L$ indices which makes an invariant for each flavour singlet found. In this way, \textit{the flavour symmetry is preserved at the highest level of symmetry}. The next example should clarify our point.
	
	Each $S_3$ invariant term, for example: $(\bf{2} \otimes \bf{2})_{S} \otimes (\bf{2} \otimes \bf{2})_{S} = {\bf{1}}_{S_3}$, has its own coupling; is obtained first as a tensorial product of two or four irreps, followed by all the appropriate contractions of weak indexes. In terms of fields, it's clear how by contracting different combinations of weak indexes each term may give rise to different structures:
	
\noindent \begin{center}$
 \frac{1}{2}(H^{\dagger}_{1w} H_{1w} + H^{\dagger}_{2w} H_{2w})^{2},$

\noindent $\frac{1}{2}[(H^{\dagger}_{1w} H_{1w})^{2} + (H^{\dagger}_{2w} H_{2w})^{2} + (H^{\dagger}_{1w} H_{2w})^{2} + (H^{\dagger}_{2w} H_{1w})^{2}],$\\
$\frac{1}{2}[(H^{\dagger}_{1w} H_{1w})^{2} + (H^{\dagger}_{2w} H_{2w})^{2} + (H^{\dagger}_{1w} H_{2w})^{2} + (H^{\dagger}_{2w} H_{1w})^{2}]$.
\end{center}

	The most general $S_3$-invariant Higgs potential is found to have only \textit{six independent couplings}:	
\noindent \begin{center}$V_H = {\mu}^{2}_{S} ({H}^{\dagger}_{S} {H}_{S}) + {\mu}^{2}_{D} (H^{\dagger}_{1} {H}_{1} + {H}^{\dagger}_{2} {H}_{2}) + a({H}^{\dagger}_{S} {H}_{S})^{2} + b{f}_{ijk}[({H}^{\dagger}_{S} {H}_{i})({H}^{\dagger}_{j} {H}_{k}) + h.c.] + c[({H}^{\dagger}_{S} {H}_{1})({H}^{\dagger}_{1} {H}_{S})	+ ({H}^{\dagger}_{S} {H}_{2})({H}^{\dagger}_{2} {H}_{S}) + ({H}^{\dagger}_{S} {H}_{1})^2 + ({H}^{\dagger}_{S} {H}_{2})^2 + ({H}^{\dagger}_{1} {H}_{S})^2 + ({H}^{\dagger}_{2} {H}_{S})^2 + ({H}^{\dagger}_{S} {H}_{S})({H}^{\dagger}_{1} {H}_{1} + {H}^{\dagger}_{2} {H}_{2})] + (d+f)[({H}^{\dagger}_{1} {H}_{1} - {H}^{\dagger}_{2} {H}_{2})^{2} + ({H}^{\dagger}_{1} {H}_{2} + {H}^{\dagger}_{2} {H}_{1})^{2}] + (d+2f)({H}^{\dagger}_{1} {H}_{1} + {H}^{\dagger}_{2} {H}_{2})^{2} + (d+e-2f)({H}^{\dagger}_{1} {H}_{2} - {H}^{\dagger}_{2} {H}_{1})^{2}$,
\end{center}	
where $a,b,c,...,f$ are constants; $1$, $2$, and $S$ are flavour indexes for the doublet components and the symmetric singlet irrep of $S_3$, respectively; $f_{112}=f_{121}=f_{211}=-f_{222}=1$, whereas all the rest are zero.
	\subsection{Conclusions and remarks}
	The M${S_3}$IESM was constructed by asking simplicity in a flavour extension of the SM, this meant an $S_3$ non-abelian flavour symmetry and the introduction of two more Higgs weak doublets, and, by the latter addition, the concept of flavour was taken to a more fundamental level, namely the Higgs sector. It has been found to be a simple model that reflects the flavour structure of the fundamental particles in good agreement with experiment. 
	
	From the results highlighted in section 2.2.5 we may conclude that the state of the art of the M${S_3}$IESM shows interesting features, such as how well suppresed are the FCNC's in the model, as well as others. The only point that now it's not in accordance with experiment, is the leptonic mixing angle $\theta_{13}$ which is well below the latest global fit lower bound.
	
	Finally, some new results were discussed: an introduction of a $Z_2$ symmetry in the quark sector was tried, inspired in the solution of the leptonic sector, but, it was found that it is too constrictive; the steps to construct the most general $S_3$-invariant Higgs potential were remarked and when using them the Higgs potential was found to have only six independent couplings.

%% file: Author/HSerodio.tex
{\bf Abstract}\\
\vskip5.mm

Based on symmetry arguments, it is shown that in type-I seesaw models the Dirac-neutrino Yukawa coupling combination relevant for leptogenesis is diagonal in the physical basis where the charged leptons and heavy Majorana neutrinos are diagonal. This will lead to vanishing leptogenesis CP asymmetries in leading order. Type-II seesaw flavor models are not so restrictive and in general will allow for leptogenesis.

\vskip5.mm

\section{Introduction}
The fact that neutrinos oscillate, and therefore have mass, was one of the first evidences for physics beyond the Standard Model (SM). The smallness of neutrino masses and the existence of two large mixing angles, the solar $\theta_{12}\simeq 34^\circ$ and atmospheric $\theta_{23}\simeq 45^\circ$ (for detailed analysis see ref.~\cite{arXiv:1001.4524}), put the leptonic and quark sectors in a quite different foot. The SM fails in explaining these mixing patterns and mass spectrum. Another intriguing fact is the observed asymmetry between matter and antimatter in our Universe. From the analysis of the Wilkinson Microwave Anisotropy Probe (WMAP) seven-year data combined with the baryon acoustic oscillations it is inferred that the baryon to photon asymmetry is~\cite{arXiv:1001.4538}
\begin{equation}
\eta_B\equiv \frac{n_B-n_{\overline{B}}}{n_\gamma}=\left(6.20\pm0.15\right)\times 10^{-10}\,,
\end{equation}
with $n_B$ ($n_{\overline{B}}$) the number density of baryons (antibaryons). Once more, the SM fails in explaining this value. 

Among the mechanisms to explain both neutrino masses and the baryon asymmetry in the Universe, the seesaw became very popular since it can provide a natural solution to both problems (for recent reviews see refs.~\cite{arXiv:0802.2962,arXiv:1111.5332}). In the type-I seesaw, right-handed neutrinos $\nu_{R_i}$ are introduced in the SM, leading to the extended Lagrangian
\begin{equation}\label{lag}
\mathcal{L}=\mathcal{L}_{SM}+i\overline{\nu_{Ri}}\gamma_\mu\partial^\mu\nu_{Ri}
-\left[\mathbf{Y}_{\alpha i}\overline{\ell_\alpha}\tilde{\phi}\nu_{Ri}+\frac{1}{2}\overline{\nu_{Ri}^c}\left(\mathbf{M_R}\right)_{ij}\nu_{Rj}+\text{H.c.}\right]\,,
\end{equation}
where the masses of the right-handed neutrinos, $M_i$, are expected to be very large, such that they decouple and at low-energies the Lagrangian becomes, after spontaneous symmetry breaking,
\begin{equation}\label{seesaw}
\mathcal{L}_{eff}=\mathcal{L}_{SM}+\frac{1}{2}\overline{\nu_{Li}}\left(\mathbf{m_\nu}\right)_{ij}\nu_{Lj}^c+\text{H.c.}\quad\text{with}\quad \mathbf{m_\nu}=\mathbf{m_D}\mathbf{M_R}^{-1}\mathbf{m_D}^T\,,
\end{equation} 
where $\mathbf{m_D}=v\mathbf{Y}$. We then get a scale for the light neutrinos of the order $v^2/M$, leading to very small neutrino masses for large $M$. Within this framework, the three Sakharov conditions necessary for a dynamical generation of a baryon asymmetry are naturally fulfilled: (i) lepton number violation is introduced by the Majorana mass term, and nonperturbative (B+L)-violating sphaleron processes will then partially convert it into a baryon asymmetry; (ii) complex Yukawa couplings in the neutrino sector provide the necessary source of CP violation; (iii) departure from thermal equilibrium is achieved through the decay of the heavy right-handed neutrinos above the electroweak scale. 

\vspace{-0.7cm}
\begin{figure}[h!]
\begin{center}
\begin{tabular}{cc}
Type-I&Type-II\\
\includegraphics[width=6cm]{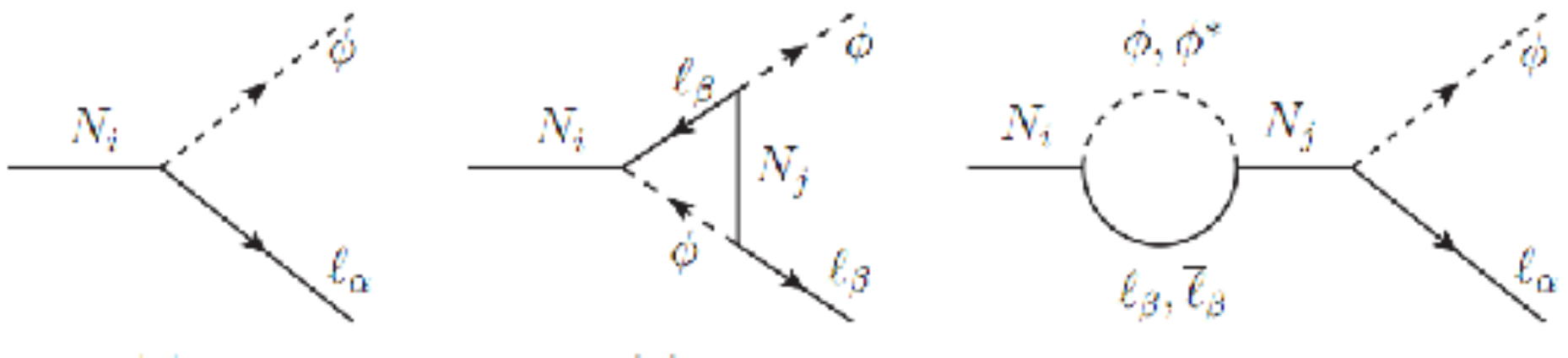}&
\includegraphics[width=7.5cm]{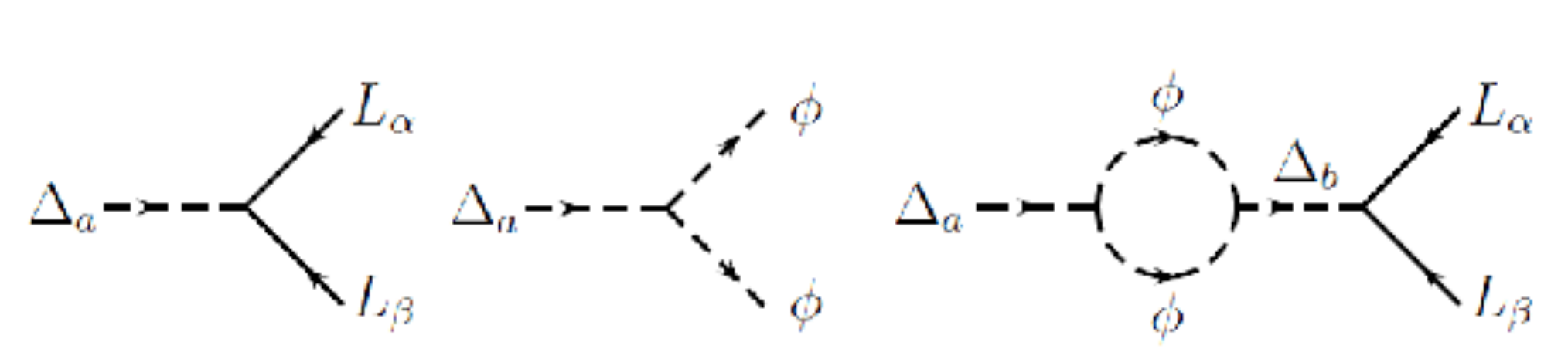}\\
(a)&(b)
\end{tabular}
\vspace{-0.6cm}
\caption{\label{fig1} (a) Type-I seesaw leptogenesis diagrams: Decay of the heavy neutrinos at tree level and one loop; (b) Type-II seesaw leptogenesis diagrams. Decay of the heavy scalar triplets at tree level and one loop.}
\end{center}
\end{figure}
\vspace{-1cm}

The CP asymmetry generated in the decay of the heavy mass eigenstates, $N_i$, is given by the interference between tree-level and one-loop diagrams, as presented in Fig.~\ref{fig1}a,
\begin{align}\label{CPflavorI}
\begin{split}
\epsilon_i^\alpha\equiv&\frac{\Gamma\left(N_i\rightarrow \phi\ell_\alpha\right)-\Gamma\left(N_i\rightarrow \phi^\dagger\overline{\ell}_\alpha\right)}{\sum_\beta \Gamma\left(N_i\rightarrow \phi\ell_\beta\right)+\Gamma\left(N_i\rightarrow \phi^\dagger\overline{\ell}_\beta\right)}\\
=&\frac{1}{8\pi \mathbf{H}_{ii}}\sum_{j\neq i}
\left\{\text{Im}\left[\mathbf{Y}_{\alpha i}^
\ast \mathbf{H}_{ij}\mathbf{Y}_{\alpha j}\right]\left(f(x)+g(x)\right)+\text{Im}\left[\mathbf{Y}_{\alpha i}^\ast \mathbf{H}_{ji}\mathbf{Y}_{\alpha j}\right]g^\prime(x)\right\}\,,
\end{split}
\end{align}
where $\mathbf{H}\equiv \mathbf{Y}^\dagger \mathbf{Y}$, $x\equiv M_j^2/M_i^2$,
\begin{equation}\label{fg}
f(x)=\sqrt{x}\left[1-(1+x)\ln(1+x^{-1})\right]\,,\quad g(x)=\frac{\sqrt{x}(1-x)}{(x-1)^2+\Gamma_{N_j}^2/M_i^2}=g^\prime (x)\sqrt{x},
\end{equation}
and $\Gamma_{N_i}=\mathbf{H}_{ii}M_i/(8\pi)$ is the total tree-level decay rate of $N_i$. Summing over the charged-lepton flavors, as usually done for temperatures above $10^{12}$ GeV when flavors are indistinguishable, one obtains the unflavored asymmetry
\begin{equation}
\epsilon_i=\sum_\alpha\epsilon_i^\alpha=\frac{1}{8\pi\mathbf{H}_{ii}}\sum_{j\neq i}\text{Im}\left[\mathbf{H}^2_{ij}\right]\left(f(x)+g(x)\right)\,.
\end{equation}

Another very popular seesaw is the type-II, where heavy scalar triplets are added to the SM. In this case, the relevant Lagrangian terms are 
\begin{equation}
\mathcal{L}=\mathcal{L}_{SM}-\mathbf{Y}_{\alpha\beta}^{a}\overline{\ell^c_\alpha}\Delta_a\ell_\beta-M_a^2\text{Tr}\left[\Delta_a^\dagger\Delta_a\right]-\mu_aM_a\tilde{\phi}^T\Delta_a\tilde{\phi}+\text{H.c.}+...\,,\quad{\small\Delta_a=
\begin{pmatrix}
\Delta_a^0&-\Delta_a^+/\sqrt{2}\\
-\Delta_a^+/\sqrt{2}&\Delta_a^{++}
\end{pmatrix}}\,,
\end{equation}
where the dots denote other terms in the scalar potential. After the scalar heavy fields decouple we get again the Lagrangian in Eq.~\eqref{seesaw} with the effective neutrino mass matrix given now by $\mathbf{m_\nu}=2\sum_a\mu_av^2/M_a\mathbf{Y}^{\dagger a}$. The three Sakharov conditions are also fulfilled in this seesaw framework, with the only difference that in this case the departure from thermal equilibrium is achieved through the decay of the heavy scalar fields. The CP asymmetry generated from the interference of the diagrams depicted in Fig.~\ref{fig1}b for each triplet $(\Delta_a^0,\Delta_a^+,\Delta_a^{++})$ is
\begin{align}\label{flavoured}
\begin{split}
\epsilon_a^{\alpha\beta}=&2\frac{\Gamma\left(\Delta_a^\ast\rightarrow\ell_\alpha+\ell_\beta\right)
-\Gamma\left(\Delta_a\rightarrow\overline{\ell}_\alpha+\overline{\ell}_\beta\right)}{\sum_{\alpha\beta} \Gamma\left(\Delta_a^\ast\rightarrow\ell_\alpha+\ell_\beta\right)
+\Gamma\left(\Delta_a\rightarrow\overline{\ell}_\alpha+\overline{\ell}_\beta\right)}=-\frac{g(x_b)}{2\pi}\frac{c_{\alpha\beta}\text{Im}\left[\mu_a^\ast\mu_b
\mathbf{Y}_{\alpha\beta}^{a}\mathbf{Y}_{\alpha\beta}^{b\ast}\right]}{\text{Tr}\left[\mathbf{Y}^{a\dagger}\mathbf{Y}^a\right]+|\mu_a|^2}\,,\quad(b\neq a)\,,
\end{split}
\end{align}
where $c_{\alpha\beta}=2-\delta_{\alpha\beta}$ for $\delta_a^0$ and $\Delta_a^{++}$, $c_{\alpha\beta}=1$ for $\Delta_{a}^+$; $x_b=M_b^2/M_a^2$ and the one-loop self-energy function $g(x_b)$ is given by Eq.~\eqref{fg} with the replacement $\Gamma^2_{N_j}/M_i^2\rightarrow \Gamma_b^2/M_a^2$, being $\Gamma_a$ the total triplet tree-level decay. Again, summing over the flavor states we obtain the unflavored CP-asymmetry
\begin{equation}\label{unflavoured}
\epsilon_a=-\frac{g(x_b)}{2\pi}\frac{c_{\alpha\beta}\text{Im}\left\{\mu_a^\ast\mu_b
\text{Tr}\left[\mathbf{Y}^{a}\mathbf{Y}^{b\dagger}\right]\right\}}{\text{Tr}\left[\mathbf{Y}^{a\dagger}\mathbf{Y}^a\right]+|\mu_a|^2}\,,\quad(b\neq a)\,.
\end{equation}
If the Universe reheats to a thermal bath after inflation, the baryon-to-photon number ratio $\eta_B$ can then be estimated for any of the seesaw types as the product of three suppression factors:
\begin{align}
\nonumber \eta_B=&\text{(leptonic CP-asymmetry $\epsilon$)}\times \text{(washout processes)}\times\text{(chemical equilibrium)}\,.
\end{align}

\section{Leptogenesis in type-I seesaw flavor models} 
In flavor models the breaking of the flavor symmetry usually leads to some residual symmetry and/or some very particular mass-independent textures. Let us discuss the consequences for leptogenesis due to the presence of these two features in flavor models, following ref.~\cite{arXiv:0908.2947}. In what follows, since $\mathbf{m_D}\propto \mathbf{Y}$, there will be no distinction between both matrices (when this is not the case our conclusions for leptogenesis may not be valid, see for instance ref.~\cite{arXiv:1110.3781}).
 
\subsection{Lagrangian residual symmetry}
We start by resuming some results concerning the algebra of symmetric matrices. Any symmetric matrix and, in particular, the $3\times3$ effective Majorana mass matrix for light neutrinos $m_\nu$ (and the heavy Majorana mass matrix $M_R$), has the symmetry
\begin{align}\label{GL}
\mathbf{\mathcal{G}_L}^\dagger \mathbf{m_\nu}\mathbf{\mathcal{G}_L}^\ast=\mathbf{m_\nu}\,,
\end{align}
where $\mathbf{\mathcal{G}_L}$ is a unitary matrix. The matrix $\mathbf{m_\nu}$ can be diagonalized through a unitary matrix $\mathbf{U_\nu}$, such that $\mathbf{U^\dagger_\nu m_\nu U_\nu^\ast}=\mathbf{d_\nu}$, where $\mathbf{d_\nu}=\text{diag}(m_1,m_2,m_3)$, with $m_i$ real and positive. One, therefore, obtains
\begin{align}\label{GLp}
\mathbf{\mathcal{G}^{\prime \dagger}_Ld_\nu\mathcal{G}^{\prime \ast}_L}=\mathbf{d_\nu}\,,\quad \text{where}\quad \mathbf{\mathcal{G}^\prime_L} = \mathbf{U_\nu^\dagger \mathcal{G}_L U_\nu}\,.
\end{align}
It is clear that after the diagonalization of the Majorana mass matrix $\mathbf{m_\nu}$, there is always a freedom to redefine the Majorana fields $\nu_{Li} \rightarrow \pm\, \nu_{Li}\,$.  Obviously, this transformation corresponds to the $Z_2\times Z_2\times Z_2$ symmetry group and leaves $\mathbf{d_\nu}$ diagonal, real and positive. If $\mathbf{\mathcal{G}^\prime_L}$ belongs to $SU(3)$ then the symmetry is reduced to $Z_2\times Z_2$. Note also that the symmetry group of $\mathbf{\mathcal{G}_L^\prime}$ is also the symmetry group of $\mathbf{\mathcal{G}_L}$, since they are connected by a similarity transformation. The symmetry group could be enlarge if some degeneracy is present.

We shall explore the consequences of having the symmetry of mass matrices as a residual symmetry of the Lagrangian. This condition requires the Eq.~(\ref{lag}) to be invariant under the transformations $\nu_L\rightarrow \mathbf{\mathcal{G}_L}\nu_L,\, \nu_R\rightarrow \mathbf{\mathcal{G}_R}\nu_R$, leading to the symmetry relations
\begin{align}\label{GLR}
\mathbf{\mathcal{G}_L}^\dagger \mathbf{m_D}\mathbf{\mathcal{G}_R}=\mathbf{m_D}\,,\quad
\mathbf{\mathcal{G}_R}^T \mathbf{M_R}\mathbf{\mathcal{G}_R}=\mathbf{M_R}\,.
\end{align}

In the basis where the heavy Majorana neutrinos are diagonal one can rewrite the symmetry equations as
\begin{align}\label{GH}
\mathbf{\mathcal{G}_R}^{\prime \dagger} \mathbf{H}\mathbf{\mathcal{G}_R}^\prime=\mathbf{H}\,,\quad\mathbf{\mathcal{G}_R}^{\prime T} \mathbf{d_R}\,\mathbf{\mathcal{G}_R}^\prime=\mathbf{d_R},\quad\text{with} \quad \mathbf{\mathcal{G}_R}^\prime = \mathbf{U_R}^\dagger\mathbf{\mathcal{G}_R} \mathbf{U_R}\,,
\end{align}
where $\mathbf{d_R}=\text{diag}(M_1,M_2,M_3)$ and $\mathbf{U_R}$ is the matrix that diagonalizes $\mathbf{M_R}$. The consequences of these relations for leptogenesis can be stated in two cases:
\begin{itemize}
\item \underline{No degeneracy in the right-handed Majorana sector.}

The second equation in Eq.~(\ref{GH}) requires the $Z_2\times Z_2\times Z_2$ symmetry generators $\mathbf{\mathcal{G}_R}^\prime$ to be diagonal. The latter are given by $\mathbf{\mathcal{G}_{R1}}^\prime = \text{diag}(1,1,-1)$, $\mathbf{\mathcal{G}_{R2}}^\prime = \text{diag}(1,-1,1)$ and $\mathbf{\mathcal{G}_{R3}}^\prime = \text{diag}(-1,1,1)$. The action of any two of these matrices in the first relation of Eq.~(\ref{GH}) would then enforce $\mathbf{H}$ to be diagonal.  This is turn implies a vanishing leptogenesis asymmetry, as can be seen from Eq.~(\ref{CPflavorI}).
Clearly, if one imposes a single $Z_2$ symmetry as the residual symmetry of the Lagrangian (\ref{lag}), the above conclusions do not necessarily hold. 

\item \underline{Some degeneracy in the right-handed Majorana sector.}

To be specific, let us assume a completely degenerate mass spectrum, i.e. $\mathbf{d_R}=M \text{diag}\,(1,1,1)$. The case with double degeneracy trivially follows from this analysis. Noticing that in this case
\begin{align}\label{dH}
\mathbf{\mathcal{G}_R}^{\prime T}\mathbf{\mathcal{G}_R}^\prime=\mathbb{I}\quad\Rightarrow\quad
\mathbf{\mathcal{G}_R}^{\prime\prime\dagger} \mathbf{d_H} \mathbf{\mathcal{G}_R}^{\prime\prime}=\mathbf{d_H}, \quad \mathbf{\mathcal{G}_R}^{\prime\prime}=\mathbf{V_H}^\dagger\mathbf{\mathcal{G}_R}^\prime \mathbf{V_H},
\end{align}
with $\mathbf{d_H}=|\mathbf{d_D}|^2$ and $\mathbf{V_H}= \mathbf{U_R^\dagger U_R^D}$, from the orthogonality condition of $\mathbf{\mathcal{G}_R}^\prime$ we get
\begin{align}\label{VH}
\mathbf{\mathcal{G}_R}^{\prime\prime T}\mathbf{V_H}^T\mathbf{V_H}\mathbf{\mathcal{G}_R}^{\prime\prime}
=\mathbf{V_H}^T\mathbf{V_H}\,.
\end{align}
This means that $\mathbf{V_H}^T\mathbf{V_H}$ is diagonal in accordance with Eq.~(\ref{dH}). One can conveniently parametrize it as $\mathbf{V_H}=\mathbf{O_1 K^\prime O_2}$, where $\mathbf{O_1}$ and $\mathbf{O_2}$ are two real orthogonal matrices with 3 rotation angles each, and $\mathbf{K^\prime}$ is a phase diagonal matrix with 3 independent phases. This parametrization allow us to easily shown that the matrix $\mathbf{H}$ is always real and, in general, non-diagonal. Yet, due to the heavy neutrino spectrum degeneracy there is always a freedom to redefine the right-handed fields by an orthogonal transformation, so that all the real off-diagonal entries in $\mathbf{H}$ are put to zero and the matrix $\mathbf{H}$ is rendered diagonal.
Therefore, even though leptogenesis is not possible with a fully degenerated spectrum of the heavy sector, this result already shows that a viable leptogenesis requires not only the breaking of such degeneracy, but also the need of non-zero off-diagonal elements in $\mathbf{H}$~\cite{arXiv:0904.3076}.
\end{itemize}

As a final remark we note that, when $\mathbf{m_D}$ is not Hermitian, Eqs.~(\ref{seesaw}) and (\ref{GLR}) imply that the unitary matrices $\mathbf{U_{L}^{D}}$ and $\mathbf{U_R^{D}}$ that diagonalize $\mathbf{m_D}$ have the form
\begin{align}\label{ULR}
\mathbf{U_L^{D}}=\mathbf{U_\nu\mathcal{P} K}, \quad \mathbf{U_R^{D}} =\mathbf{U_R \mathcal{P}^\prime K},
\end{align}
where $\mathcal{P}$ and $\mathcal{P}^\prime$ are two arbitrary permutation matrices and $\mathbf{K}$ is a phase diagonal matrix with 3 independent phases. For $\mathbf{m_D}$ Hermitian, Eq.~(\ref{GLR}) implies $\mathbf{\mathcal{G}_L}=\mathbf{\mathcal{G}_R}$ leading to $\mathbf{U_R}=\mathbf{U_\nu \mathcal{P} d}$. The case when the heavy neutrino sector has some degeneracy can be analogously analyzed since it simply corresponds to the replacement $\mathbf{U_R}\rightarrow \mathbf{U_R\,O}$, with $\mathbf{O}$ a real orthogonal matrix.

It is worth emphasizing that the above results are just a consequence of imposing the symmetries in the mass matrices to be the residual symmetry of the Lagrangian. Similar conclusions have been obtained in ref.~\cite{arXiv:0908.0161}.

\subsection{Mass-independent textures}

A common feature in many flavor models is that the matrices $\mathbf{m_D}$, $\mathbf{M_R}$ and $\mathbf{m_\nu}$ exhibit mass-independent textures, i.e. textures where the diagonalization is independent of mass parameters. It is worth studying this case and its implications for leptogenesis~\cite{arXiv:0908.2947,arXiv:0908.0907,arXiv:1011.6167,arXiv:1110.5676}. Let us we rewrite the seesaw formula, Eq.~(\ref{seesaw}), as
\begin{align}\label{Adef}
\mathbf{d_\nu} = \mathbf{A}\, \mathbf{d_R}^{-1} \mathbf{A}^T,\quad \mathbf{A} = \mathbf{U_\nu}^\dagger \mathbf{U_L^{D}}\, \mathbf{d_D}\, \mathbf{U_R^{D^\dagger} U_R}
\end{align}
where $\mathbf{A}$ is a real matrix, which is independent of the light $m_i$ and the heavy $M_i$ neutrino masses. One can then show that $\mathbf{A}$ has at least 6 zero entries. The solutions Eqs.~(\ref{Adef}), can be divided into two classes: 

\begin{itemize}
\item \underline{$\det\,\mathbf{m_\nu} \neq 0$}: Since the matrix $\mathbf{A}$ has at least 6 zero entries and $\det\,\mathbf{A} = \det\,\mathbf{d_D} \neq 0$, its possible textures are of the form of a permutation matrix. This in turn implies that the matrices $\mathbf{A} \mathbf{A}^\dagger$ and $\mathbf{A}^\dagger \mathbf{A}$ should be diagonal. From the definition of $\mathbf{A}$ given in Eq.~(\ref{Adef}), we then conclude that
\begin{equation}
\mathbf{U_\nu}^\dagger \mathbf{U_L^D} = \mathbf{\mathcal{P} K}, \quad  \mathbf{U_R^{D\dagger} U_R} = \mathbf{K^\ast \mathcal{P}^\prime},
\end{equation}
which are equivalent to the relations given in Eq.~(\ref{ULR}). Thus the assumption of mass-independent textures leads to the symmetry relation given in Eq.~(\ref{GLR}). In other words, the symmetry of the mass matrices is also the residual symmetry of the Lagrangian (\ref{lag}). This result holds for any mass-independent texture model.

\item \underline{$\det\,\mathbf{m_\nu} = 0$}: In this case one of the light neutrino masses is exactly zero. In the relevant basis for leptogenesis $\mathbf{m_D}^\prime=\mathbf{U_\nu A}$, the combination $\mathbf{H}=\mathbf{m_D^{\prime\dagger} m_D^\prime}=\mathbf{A}^T\mathbf{A}$ is always real, thus forbidding unflavored leptogenesis. The relevant combination for flavored leptogenesis is also real due to the particular textures of $\mathbf{A}$. Again, no CP asymmetry is generated at leading order.

\end{itemize}

\section{Leptogenesis in type-II seesaw}
The result presented in the previous section does not necessarily hold in type-II seesaw models~\cite{arXiv:1101.0602}. In order to show this, let us consider the particular case of tribimaximal leptonic mixing. A convenient parametrization of the effective neutrino mass matrix is
\begin{equation}
\mathbf{m}_\nu=x\mathbf{C}+y\mathbf{P}+z\mathbf{D}\,,\quad \mathbf{C}=\frac{1}{3}
\begin{pmatrix}
2&-1&-1\\
-1&2&-1\\
-1&-1&2
\end{pmatrix}\,,\quad
\mathbf{P}=
\begin{pmatrix}
1&0&0\\
0&0&1\\
0&1&0
\end{pmatrix}\,,\quad
\mathbf{D}=\frac{1}{3}
\begin{pmatrix}
1&1&1\\
1&1&1\\
1&1&1
\end{pmatrix}\,
\end{equation} 
where $x$, $y$ and $z$ are arbitrary complex parameters. 

For leptogenesis to be viable, at least two scalar $SU(2)$ triplets are needed. If both are family singlets, then one of them can be associated to the $\mathbf{P}$ contribution, and the other one to the $\mathbf{C}$ contribution. If a third scalar triplet is available, it may be associated to the democratic component $\mathbf{D}$. In this minimal setup, unless a democratic contribution is present, the unflavored asymmetry (\ref{unflavoured}) is zero, because $\text{Tr}\left[\mathbf{C}\mathbf{P}\right]=0$. Notice however that if each scalar triplet is simultaneously associated to  $\mathbf{C}$ and $\mathbf{P}$ contributions, the unflavored asymmetry (\ref{unflavoured}) is in general nonvanishing.
On the other hand, the flavored leptogenesis asymmetries in Eq.~(\ref{flavoured}) do not necessarily vanish, even when the democratic component is absent. 

If $\Delta$ are family triplets, there must be at least one extra singlet or triplet. Otherwise, it is not possible to generate a mass-independent mixing in agreement with low-energy neutrino data. It can be shown that any contributions to Eq.~\eqref{flavoured} that involve components of the same triplet cancel out, but a nonvanishing asymmetry can result from the interaction of a given component of the triplet with the extra singlet or triplet~\cite{arXiv:1101.0602}.

%% file: Author/shimizu.tex
{\bf Abstract}

\vspace{3mm}
In combination with supersymmetry, flavor symmetry may relate quarks with leptons, even in absence of a grand-unification group. 
We propose an $SU(3)\times SU(2)\times U(1)$ model where both supersymmetry and assumed 
$A_4$ flavor symmetries are softly broken, reproducing well observed fermion mass hierarchies and predicting 
a relation between down-type quarks and charged lepton masses, 
and a correlation between Cabibbo angle in quark sector, 
and reactor angle $\theta_{13}$ characterizing CP violation in neutrino oscillations.

		
\section{Introduction}

Understanding the observed pattern of quark and lepton masses and
mixing~\cite{nakamura2010review,Schwetz:2008er} constitutes one of the
deepest challenges in particle physics.
Flavor symmetries provide a very useful approach towards reducing the
number of free parameters describing the fermion
sector~\cite{Ishimori:2010au}.
It has long been advocated that grand unification offers a suitable
framework to describe flavor. In what follows we will adopt the
alternative approach, assuming that flavor is implemented directly at the
$SU(3)\times SU(2)\times U(1)$ level. 
Typically this requires several $SU(2)$ doublet scalars in order to
break spontaneously the flavor symmetry so as to obtain an acceptable
structure for the masses and mixing matrices.
(One may alternatively introduce ``flavons'' instead of additional
Higgs doublets, but in this case one would have to give up
renormalizability).

In order to construct a ``realistic'' extension of the Standard Model
(SM) with flavor symmetry one needs a suitable alignment of the scalar
vacuum expectation values (vevs) in the
theory~\cite{ma:2001dn,babu:2002dz,zee:2005ut,altarelli:2005yp}.
There are several multi-doublet extensions of the SM with flavor in
the market, but renormalizable supersymmetric extensions of the SM
with a flavor symmetry are only a few~\cite{PhysRevD.71.056006},
usually because the existence of additional Higgs doublets spoils the
unification of the coupling constants.

Here we choose to renounce to this theoretical argument, noting that
gauge coupling unification may happen in multi-doublet schemes due to
other effects.
What we now present is a supersymmetric extension of the SM based on
the $A_4$ group~\cite{Morisi:2011pt}, where all the matter fields as well as the Higgs
doublets belong to the same $A_4$ representation, namely, the
triplet. This leads us to two theoretical predictions. The first a
mass relation
\begin{equation}
\label{eq:ours}
\frac{m_{\tau}}{\sqrt{m_em_\mu}}\approx\frac{m_b}{\sqrt{m_d m_s}}~,
\end{equation}
involving down-type quarks and charged lepton mass ratios. Such
relation can be obtained by a suitable combination of the three
Georgi-Jarlskog (GJ) mass relations~\cite{Georgi:1979df},
\begin{equation}\label{GJ}
\begin{array}{lll}
m_b=m_\tau ,& m_s=1/3 m_\mu ,& m_d=3 m_e,
\end{array}
\end{equation}
which arise within a particular ansatz for the SU(5) model and hold at
the unification scale.  In contrast to eq. (\ref{GJ}), our relation
requires no unification group and holds at the electroweak scale. It
would, in any case, be rather robust against renormalization effects
as it involves only mass ratios.

The second prediction obtained in our flavor model is a correlation
between the Cabibbo angle for the quarks and the so-called ``reactor
angle'' $\theta_{13}$ characterizing the strength of CP violation in
neutrino oscillations~\cite{nunokawa:2007qh,bandyopadhyay:2007kx}.
Within a reasonable approximation we find
\begin{equation}
\label{eq:12q-13l}
\lambda_C\approx \frac{1}{\sqrt{2}} \frac{m_\mu m_b}{m_\tau m_s}
\sqrt{\sin^22 \theta_{13}}-\sqrt{\frac{m_u}{m_c}}\,.
\end{equation}
which arises mainly from the down-type quark
sector~\cite{Gatto:1968ss} with a correction coming from the up
isospin diagonalization matrix. This is a very interesting relation,
discussed below in more detail.\\[-.5cm]

\section{The Model}
\label{sec:themodel}

Here we review a supersymmetric model based on an $A_4$ flavor
symmetry studied in Ref.~\cite{Morisi:2011pt}.
The field representation content is given in
Table~\ref{tab:Multiplet1}. Note that all quarks and leptons transform
as $A_4$ triplets.  Similarly the Higgs superfields with opposite
hypercharge characteristic of the MSSM are now upgraded into two sets,
also transforming as $A_4$ triplets. Note that since all matter fields
transform in the same way under the flavor symmetry one may in
principle embed the model into a grand-unified scheme. However, given
the large number of scalar doublets, gauge coupling unification must
proceed differently, see, for example, Ref.~\cite{Munoz:2001yj}.
\begin{table}[h!]
\begin{center}
\begin{tabular}{|l||lllll||ll|}
\hline
fields & $\hat{L}$ & $\hat{E}^c$ & $\hat{Q}$& $\hat{U}^c$ & $\hat{D}^c$ & $\hat{H}^u $ & $\hat{H}^{d}$  \\
\hline
$SU(2)_L$ & $2$ & $1$ & $2$ & $1$ & $1$ & $2$ & $2$ \\
$A_4$ & ${\bf 3}$ & ${\bf 3}$ & ${\bf 3}$ & ${\bf 3}$ & ${\bf 3}$ & ${\bf 3}$ & ${\bf 3}$ \\
\hline
\end{tabular}
\caption{Basic multiplet assignments of the model}
\label{tab:Multiplet1}
\end{center}
\end{table}
\vspace{-10mm}

The most general renormalizable Yukawa Lagrangian for the charged
fermions in the model is~\cite{Morisi:2009sc}
\begin{equation}\label{y}
L_{\text{Yuk}}
= y^l_{ijk}\hat{L}_i\hat{H}^d_{j}\hat{E}^c_k
+ y^d_{ijk}\hat{Q}_i\hat{H}^d_{j}\hat{D}^c_k
+ y^u_{ijk}\hat{Q}_i\hat{H}^u_{j}\hat{U}^c_k~,
\end{equation}
where $y_{ijk}^{u,d,l}$ are $A_4$-tensors, assumed real at this
stage.

The Higgs scalar potential invariant under $A_4$ is
\begin{equation}\label{eq:V}
\begin{array}{lll}
V&=&
 (|\mu |^2+m_{H_u}^2)(|H_1^u|^2+|H_2^u|^2+|H_3^u|^2)+(|\mu |^2+m_{H_d}^2)(|H_1^d|^2+|H_2^d|^2+|H_3^d|^2) \\
&-&[b(H_1^uH_1^d+H_2^uH_2^d+H_3^uH_3^d)+\text{c.c.}] \\
&+&\frac{1}{8}(g^2+{g^\prime }^2)(|H_1^u|^2+|H_2^u|^2+ |H_3^u|^2-|H_1^d|^2-|H_2^d|^2-|H_3^d|^2)^2~. \\
\end{array}
\end{equation}

Assuming that the Higgs doublet scalars take real vevs $\vev{
  H_i^{u,d}} = v_i^{u,d}$ one can show that the minimization of the
potential $V$ gives as possible local minima the alignments
$\vev{{H^{0}}^{u,d}} \sim (1,0,0)$ and $(1,1,1)$.  Only the first is
viable and we verify that minimization leads to this solution within a
wide region of parameters.  By adding $A_4$ soft breaking terms to the
$A_4$-invariant scalar potential in eq.~(\ref{eq:V})
\begin{equation}\label{align0}
\begin{array}{lll}
V_{soft}&=&\sum_{ij}\left(\mu^u_{ij}H_i^{u*}H_j^u+\mu^d_{ij}H_i^{d*}H_j^d\right)+\sum_{ij}b_{ij}H_i^{d}H_j^u, \nonumber
\end{array}
\end{equation}
one finds that
\begin{eqnarray}
\label{eq:minima}
\vev{H^u}=(v^u,\varepsilon _1^u,\varepsilon _2^u),\quad
\vev{H^d}=(v^d,\varepsilon _1^d,\varepsilon _2^d)~,
\end{eqnarray}
where $\varepsilon_{1,2}^u\ll v^u$ and $\varepsilon_{1,2}^d\ll v^d.$\\[-.5cm]

\subsection{Charged fermions}
\label{sec:charged-fermions}

By using $A_4$ product rules it is straightforward to show that the
charged fermion mass matrix takes the following universal
structure~\cite{Morisi:2009sc}
\begin{equation}
M_{f}=
\left(
\begin{array}{ccc}
0 & a^f \alpha^f  & b^f  \\
b^f\alpha^f  & 0 & a^f r^f \\
a^f  & b^f r^f & 0
\end{array}
\right),
\label{Mf}
\end{equation}
where $f$ denotes any charged lepton, up- or down-type quarks, and 
$a^f=y_1^f\varepsilon_1^f$, $b^f=y_2^f\varepsilon_1^f$, with
$y_{1,2}^f$ denoting the only two couplings arising from the
$A_4$-tensor in eq.~(\ref{y}), $r^f=v^f/\varepsilon_1^f$ and
$\alpha^f=\varepsilon_2^f/\varepsilon_1^f$.  Thanks to the fact that
the same Higgs doublet $H^d$ couples to the lepton and to the
down-type quarks one has, in addition, the following relations
\begin{equation}\label{rel}
r^l=r^d,\qquad \alpha^l=\alpha^d,
\end{equation}
involving down-type quarks and charged leptons.

\begin{figure}[h!]
\begin{minipage}[]{0.45\linewidth}
\includegraphics[width=7cm]{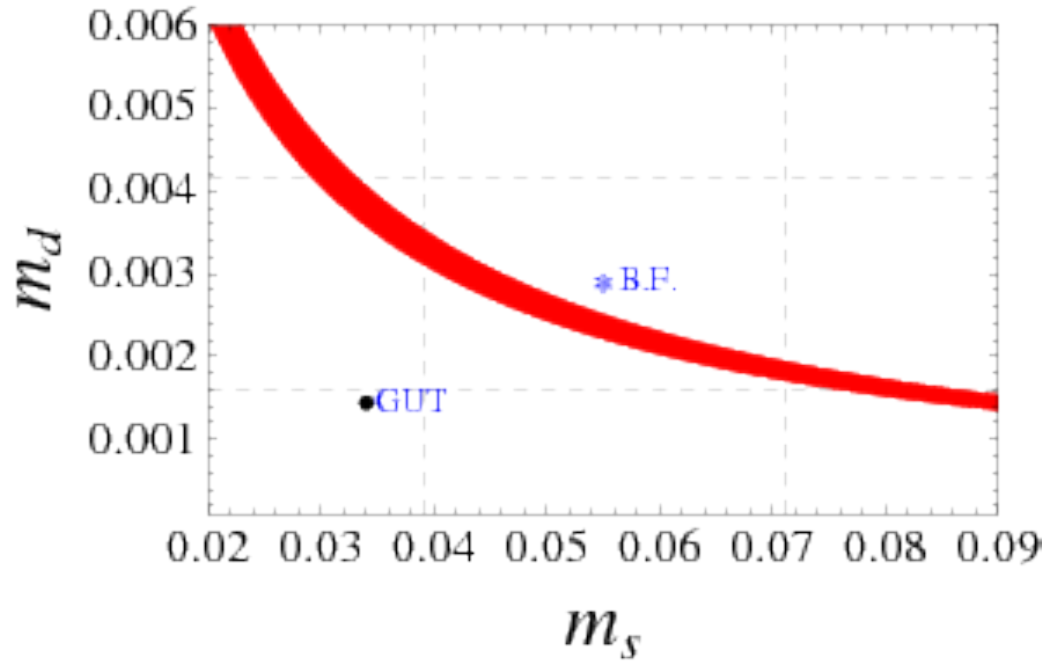}
\caption{The shaded band gives our prediction for the down-strange
  quark masses at $M_z$ scale, eq.~(\ref{massrel2}), vertical and
  horizontal lines are 1$\sigma$ experimental ranges from
  Ref.~\cite{Xing:2007fb}.}
\label{figmass}
\end{minipage}
\hspace{5mm}
\begin{minipage}[]{0.45\linewidth}
\includegraphics[width=7cm]{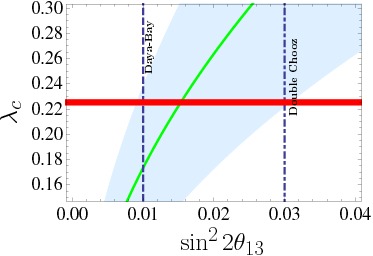}
\vspace{-3.5mm}
\caption{The shaded band gives our predicted 1~$\sigma$ correlation
  between Cabibbo angle and reactor angle, as above. Vertical
  lines give the expected sensitivities on
  $\theta_{13}$~\cite{Ardellier:2006mn,Guo:2007ug}.}
\label{fig2}
\end{minipage}
\end{figure}

It is straightforward to obtain analytical expressions for $a^f$,
$b^f$ and $r^f$ from eq.~(\ref{Mf}) in terms of the charged fermion
masses and $\alpha^f$,
\begin{equation}
\label{eq:rel2}
\frac{r^f}{\sqrt{\alpha^f}}\approx \frac{m_{3}^f}{\sqrt{m_{1}^fm_{2}^f}},\quad
a^f\approx\frac{m_{2}^f}{m_{3}^f}\frac{\sqrt{m_{1}^f{m_{2}^f}}}{\sqrt{\alpha^f}},\quad
b^f\approx \frac{\sqrt{m_{1}^f{m_{2}^f}}}{\sqrt{\alpha^f}}.\quad
\end{equation}
From eq.~(\ref{rel}) and eq.~(\ref{eq:rel2}) it follows that 
\[
 \frac{m_{\tau}}{\sqrt{m_{e}\, m_{\mu}}}\approx  \frac{m_{b}}{\sqrt{m_{d}\, m_{s}}},
\]
a formula relating quark and lepton mass ratios (to a very good
approximation this formula also holds for complex Yukawa couplings).
This relation is a strict prediction of our model, and appears in a
way similar to the celebrated SU(5) mass relation, despite the fact
that we have not assumed any unified group, but just the 
$SU(3)\times SU(2)\times U(1)$ gauge structure. 
It allows us to compute the down quark mass in terms of the
charged fermion masses and the $s$ and $b$ quarks, as
\begin{equation}\label{massrel2}
 m_d\approx m_{e}  \frac{m_{\mu}}{ m_{s}}\left(\frac{m_{b}}{m_{\tau}}\right)^2.
\end{equation}
This mass formula predicts the down quark mass at the scale of the $Z$
boson mass, to lie in the region
\begin{equation}
\begin{array}{c}
1.71~ MeV <m_d^{th}<3.35 ~MeV,\quad 1.71~ MeV <m_d< 4.14 ~MeV~,
\end{array}\end{equation}
at 1$\sigma$~\cite{Xing:2007fb}. This is illustrated in
Fig.~\ref{figmass} where, to guide the eye, we have also included the
1$\sigma$ experimental ranges from Ref.~\cite{Xing:2007fb}, as
well as the best fit point and the GJ prediction. 
Note also that, thanks to supersymmetry, we obtain a relation only
among the charged lepton and down-type quark mass ratios, avoiding the
unwanted relation found by Wilczek and Zee in
Ref.~\cite{Wilczek:1978xi}.
\vspace{-5mm}

\subsection{Neutrinos}
\label{sec:neutrinos}

To the renormalizable model we have so far we now add an effective
dimension-five $A_4$-preserving lepton-number violating operator
\begin{equation}\label{nudim5}
\mathcal{L}_{5d}=
\frac{f_{ijlm}}{\Lambda} \hat{L}_i \hat{L}_j \hat{H}^u_l \hat{H}^u_m ~,
\end{equation}
where the $A_4$-tensor $f_{ijlm}$ takes into account all the possible
contractions of the product of four $A_4$ triplets~\footnote{ Specific
  realizations of $\mathcal{L}_{5d}$ within various seesaw
  schemes~\cite{Valle:2006vb} can, of course, be envisaged.}.

Neutrino masses are induced after electroweak symmetry breaking from
the operator in eq.\,(\ref{nudim5}). In order to determine the flavor
structure of the resulting mass matrix we take the limit where the vev
hierarchy $\vev{H_1^u} \gg \vev{H_2^u}, \vev{H_3^u}$ holds, leading
to~\cite{Morisi:2009sc}
\begin{equation}
M_\nu=\left(\begin{array}{ccc}
    x {r^u}^2  & \kappa r^u  & \kappa r^u \alpha^u\\
    \kappa r^u  & y {r^u}^2  & 0\\
    \kappa r^u \alpha^u& 0 & z {r^u}^2
\end{array}
\right),
\label{mnu20}
\end{equation}
where $x,\,y,\,z$ and $\kappa$ are coupling constants, while $r^u$ and
$\alpha^u$ already been introduced above in the up quark sector.

The best fit of neutrino oscillation data~\cite{Schwetz:2008er} yields
maximally mixed $\mu$ and $\tau$ neutrinos. This is possible, in the
basis where charged lepton is diagonal, if and only if the
light-neutrino mass matrix is approximately $\mu-\tau$ invariant. In
turn this holds true if $y \approx z$ and $\alpha^u \approx
1$~\cite{Morisi:2009sc}\footnote{The charged lepton mass matrix is
  mainly diagonalized by a rotation in the 12 plane.}. When
$\alpha^u<1$ the ``atmospheric angle'' deviates from the
maximality. We have verified that for $\alpha^u\gtrsim 0.5$
the atmospheric angle is within its 3~$\sigma$ allowed range.\\[-.5cm]

\section{Relating the Cabibbo angle to $\theta_{13}$}
\label{sec:relating}

In the CP conserving limit we have taken so far we have in total 14
free parameters to describe the fermion sector: six $a^f$ and $b^f$
parameters (three for each charged fermion-type), plus four $r^f$ and
$\alpha^f$ (here only down-type are counted, in view of
eq.~(\ref{rel})), plus four parameters describing the neutrino mass
induced by the dimension-5 operator: $x, y, z, \kappa$.
These parameters describe 18 observables, which may be taken as the 9
charged fermion masses, the two neutrino squared mass differences
describing neutrino oscillations, the three neutrino mixing angles,
the neutrinoless double beta decay effective mass parameter, the
Cabbibo angle, in addition to $V_{ub}$ and $V_{cb}$. Hence we have
four relations.

The first of these we have already seen, namely the mass relation in
eq.~(\ref{eq:ours}) and Fig.~\ref{figmass}.
The second is a quark-lepton mixing angle relation concerning the
Cabibbo angle $\lambda_C$ and the ``reactor angle'' $\theta_{13}$
describing neutrino oscillations.  To derive it note first that the
matrix in eq.\,(\ref{Mf}) is diagonalized on the left by a rotation in
the 12 plane, namely $\sin\theta^f_{12}\approx\sqrt{\frac{m_1^f}{m_2^f}} \frac{1}{\sqrt{\alpha^f}}$.

In order to give an analytical expression for the relation between
Cabbibo and reactor angles, we neglect mixing of the third family of
quarks and go in the limit where our neutrino mass matrix, eq. (15) is
$\mu-\tau$ invariant, that is $\alpha^u=1$ and $y=z$. In this
approximation, the reactor mixing angle is given by
\begin{equation}
\sin \theta_{13}=\frac{1}{\sqrt{2}} \sin \theta_{12}^l=\frac{1}{\sqrt{2}}\sqrt{\frac{m_e}{m_\mu}}\frac{1}{\sqrt{\alpha^l}},
\label{t13-1}
\end{equation}
using our mass relation in eq.~(\ref{eq:ours}) one finds that the
Cabbibo angle may be written as
\begin{equation}\label{tc-1}
\lambda_C=\frac{m_b}{m_s}\frac{\sqrt{m_e m_\mu}}{m_\tau}\frac{1}{\sqrt{\alpha^d}}-\sqrt{\frac{m_u}{m_c}}.
\end{equation}
Comparing eq. (\ref{t13-1}) with eq. (\ref{tc-1}) leads immediately to
equation (\ref{eq:12q-13l}). 
In order to display this prediction graphically we take the quark
masses at 1~$\sigma$, obtaining the curved band shown in
fig.~\ref{fig2}.
The narrow horizontal band indicates current determination of the
Cabbibo angle, while the two vertical dashed lines represent the
expected sensitivities of the Double-Chooz~\cite{Ardellier:2006mn} and
Daya-Bay~\cite{Guo:2007ug} experiments on the ``reactor mixing angle''
$\theta_{13}$. The curved line corresponds to the analytical
approximation for the best fit value of the quark masses in
eq.~(\ref{eq:12q-13l}). Clearly the width of the curved band
characterizing our prediction is dominated by quark mass determination
uncertainties.
%

Finally note that mixing parameters of the third family of quarks $
U_{13}^q\approx \frac{m_2^q}{m_3^q}\frac{\sqrt{m_1^q
    m_2^q}}{m_3^q}\frac{1}{\sqrt{\alpha^q}}$ and $U_{23}^q\approx
\frac{m_1^q (m_2^q)^2}{(m_3^q)^3}\frac{1}{\alpha^q}$ ($q=u,d$)
are negligible, and can not account for the measured values of
$V_{ub}$ and $V_{cb}$. The predicted values obtained for these are too
small so that in its simplest presentation described above our model
can not describe the CP violation found in the decays of neutral
kaons.  However there is a simple solution which maintains the good
predictions described above, namely, adding colored vector-like
$SU(2)_L$ singlet states. In this case acceptable values for $V_{ub}$
and $V_{cb}$, leading to adequate CP violation can arise solely from
non-unitarity effects
of the quark mixing matrix.\\[-.5cm]

\section{Outlook}
\label{conc}

We proposed a supersymmetric extension of the standard model with an
$A_4$ flavor symmetry, where all matter fields in the model transform
as triplets of the flavor group.  Charged fermion masses arise from
renormalizable Yukawa couplings while neutrino masses are treated in
an effective way. 
The scheme illustrates how, in combination with supersymmetry, flavor
symmetry may relate quarks with leptons, even in the absence of a
grand-unification group.
Two good predictions emerge: (i) a relation between down-type quarks
and charged lepton masses, and (ii) a correlation between the Cabibbo
angle in the quark sector, and the reactor angle $\theta_{13}$
characterizing CP violation in neutrino oscillations, which lies
within the sensitivities of upcoming experiments.
Although the predicted values for the other mixing parameters $V_{uc}$
and $V_{cb}$ of the Cabibbo-Kobayashi-Maskawa matrix are too small, we
mentioned a simple way to circumvent this, making the scheme fully
realistic. 
Finally note that, with few exceptions such as those in
Refs.~\cite{King:2003rf,Dermisek:2005ij}, grand-unified flavor models
are not more predictive than the novel idea proposed here and
illustrate through this simple scheme.
As it stands the model fits well with the idea that gauge coupling
unification may be an effect of the presence of extra dimensions
rather than of grand-unified interactions~\cite{Munoz:2001yj}.
Notwithstanding, we wish to stress that our model is manifestfly
embeddable into a standard Grand-Unified scenario, which would result
in further predictive power.  A detailed study of this particular
model lies outside the scope of this letter and will be taken up
elsewhere.

%% file: Author/spinrath.tex
\chaptermark{Right Unitarity Triangles and Tri-Bimaximal Mixing}

{\bf Abstract}\\
\vskip5.mm
We discuss a recently proposed new class of flavour models
which predicts both close to tri-bimaximal lepton mixing (TBM) and
a right-angled Cabibbo-Kobayashi-Maskawa (CKM) unitarity
triangle, $\alpha \approx 90^\circ$. The ingredients of
the models include a supersymmetric (SUSY) unified gauge
group such as $SU(5)$, a discrete family symmetry such as
$A_4$ or $S_4$, a shaping symmetry including products of
$Z_2$ and $Z_4$ groups as well as spontaneous CP violation.
The vacuum alignment in such models allows a simple
explanation of $\alpha \approx 90^\circ$ by a combination
of purely real or purely imaginary vacuum expectation
values (vevs) of the flavon fields responsible for family
symmetry breaking.

\vskip5.mm

\section{Motivation}

Albeit the great success of the Standard Model (SM) of particle
physics, its flavour sector is still puzzling.
The SM flavour puzzle can be roughly divided into three aspects,
which are first the hierarchies of the observed fermion masses,
second the pattern of the observed mixing angles, and third
the origin of CP violation.

Here we are concerned mainly with two of those aspects.
The first one concerns the mixing angles. The fact that the
leptonic mixing angles turned out to be close to TBM \cite{HPS}
has led to increasing interest in non-Abelian discrete family
symmetries for flavour model building. Nevertheless, in many
realistic models another shaping symmetry has to be invoked to forbid
unwanted operators in the (super-)potential. These shaping symmetries
can shed some light on the second aspect of
the flavour puzzle we are concerned with, the origin
of CP violation, as was recently shown in
\cite{Antusch:2011sx}.

Experimental results point towards a right-angled CKM unitarity
triangle with $\alpha = (89.0^{+4.4}_{-4.2})^\circ$ \cite{PDG}.
This can be understood in terms of a simple phase sum rule
\cite{Antusch:2009hq}. As we will
revise later it becomes clear from this sum rule, that mass
matrices with purely real and purely imaginary elements can
lead to a right-angled CKM unitarity triangle, see also \cite{Masina:2006pe}.
These special phases in turn can be the result of a spontaneously
broken discrete symmetry \cite{Antusch:2011sx}.

In combination with a unified gauge group this proliferates an
attractive framework to describe mixing angles and CP violation
in the quark and the lepton sector as a result of spontaneously
broken discrete family and shaping symmetries.

\section{The Quark Mixing Phase Sum Rule}

First we revise the phase sum rule from \cite{Antusch:2009hq}.
For the mass matrices $M_u$ and $M_d$ in the Lagrangian we use the convention
\begin{equation}
\mathcal{L}_{Y}=-\overline{u^i_L} (M_u)_{ij} u^j_R - \overline{d^i_L}
(M_d)_{ij} d^j_R + H.c. \;.
\end{equation}
They are diagonalised by bi-unitary transformations
\begin{equation}
V_{u_L} M_u V_{u_R}^\dagger = \mbox{diag}(m_u, m_c, m_t) \quad \text{and} \quad V_{d_L} M_d V_{d_R}^\dagger = \mbox{diag}(m_d, m_s, m_b) \;,
\end{equation}
where $V_{u_{L}}$, $V_{u_{R}}$, $V_{d_{L}}$ and $V_{d_{R}}$ are
unitary $3\times 3$ matrices. The CKM matrix $V_{\text{CKM}}$ is given by
\begin{equation}\label{eq:UCKM_VuVd}
V_{\text{CKM}} =V_{u_L} V_{d_L}^\dagger = {U^{u_L}_{12}}^\dagger
{U^{u_L}_{13}}^\dagger {U^{u_L}_{23}}^\dagger
U^{d_L}_{23}U^{d_L}_{13} U^{d_L}_{12} \;,
\end{equation}
where the $U_{ij}$ matrices are unitary rotation matrices in the i-j plane, for instance,
\begin{equation}
 U_{12}= \begin{pmatrix}
  c_{12} & s_{12}e^{- \text{i} \, \delta_{12}} & 0\\
  -s_{12}e^{\text{i} \, \delta_{12}}&c_{12} & 0\\
  0&0&1\end{pmatrix} \;.
\end{equation}
For hierarchical quark mass matrices with a texture zero
in the 1-3 element it is straightforward to derive the following
approximate expressions for the quark mixing angles (for more details
see \cite{Antusch:2009hq})
\begin{align}
\label{F1} {\theta_{23}}e^{-\text{i} \, \delta_{23}} &=
{\theta_{23}^{d}}e^{-\text{i} \, \delta_{23}^{d}}
-{\theta_{23}^{u}}e^{-\text{i} \, \delta_{23}^{u}}\;,
\\
\label{F2} {\theta_{13}}e^{-\text{i} \, \delta_{13}} &=
-{\theta_{12}^{u}}e^{-\text{i} \, \delta_{12}^{u}}
({\theta_{23}^{d}}e^{-\text{i} \, \delta_{23}^{d}} - {\theta_{23}^{u}}e^{-\text{i} \, \delta_{23}^{u}})
\;,\\
\label{F3} {\theta_{12}}e^{-\text{i} \, \delta_{12}} &=
{\theta_{12}^{d}}e^{-\text{i} \, \delta_{12}^{d}}
-{\theta_{12}^{u}}e^{-\text{i} \, \delta_{12}^{u}} \;.
\end{align}
From these formulas we obtain for $\alpha$
\begin{equation} \label{eq:sumrule}
 90^\circ \approx \alpha = \arg \left( - \frac{V_{td} V_{tb}^*}{V_{ud} V_{ub}^*} \right) \approx \delta_{12}^d - \delta_{12}^u \quad \text{with} \quad \delta^{d/u}_{12} = \arg \left( \frac{M^{d/u}_{12}}{M^{d/u}_{22}} \right) \;.
\end{equation}
As a direct consequence it becomes obvious, that a relative phase
difference of $90^\circ$ in the 1-2 mixing is enough to
describe the CP violation in the quark sector, see also \cite{Masina:2006pe}.
The simplest realisation of this would be mass matrices with
purely real and purely imaginary elements.

In the following
we discuss a recent idea, how this can be accomodated in the
context of flavour models with discrete family and shaping
symmetries.

\section{The Method: Discrete Vacuum Alignment}

The class of models, we discuss here, is based on the method of discrete
vacuum alignment~\cite{Antusch:2011sx}, which has as its ingredients a
discrete family (like $A_4$ or $S_4$) and shaping
symmetry (like a product of $Z_n$'s), spontaneous CP violation and
a SUSY unified gauge group. The unified gauge group is not strictly necessary,
but it is very powerful, because it
relates the mixing and the CP violation in the quark and the lepton sector to each other.

The method can be described in a simple algorithm. First, use the
family symmetry to align the flavon vevs, so that only one complex
parameter $x$ is left undetermined, e.g.\ $\langle \phi \rangle \propto (0,0,x)^T$
or $\langle \phi \rangle \propto (x,x,x)^T$.
Then add for each flavon $\phi$ the following type of terms
to the superpotential
\begin{equation}\label{eq:flavonpotentialZn}
P \left( \frac{\phi^n}{\Lambda^{n-2}} \mp M^2 \right) \ ,
\end{equation} 
which are allowed by the discrete $Z_n$ shaping symmetries, and where
$M$ and $\Lambda$ are real mass parameters. By solving the $F$-term
condition, $F_P = 0$, the phase of the flavon vev is fixed to be
\begin{equation}\label{eq:phaseswithZn}
\arg(\langle \phi \rangle) = \arg(x) = \left\{ \begin{array}{ll}
\frac{2 \pi}{n}q \;, & q = 1, \dots , n \quad \mbox{\vphantom{$\frac{f}{f}$} for ``$-$'' in Eq.~\eqref{eq:flavonpotentialZn} ,}\\
\frac{2 \pi}{n} q +\frac{\pi}{n} \;,\quad & q = 1, \dots , n \quad  \mbox{\vphantom{$\frac{f}{f}$} for ``$+$'' in Eq.~\eqref{eq:flavonpotentialZn} .}
\end{array}
\right.
\end{equation}
If the shaping symmetries are only $Z_2$ or $Z_4$ symmetries
the phases can easily be arranged to fulfill the phase sum rule
in Eq.~\eqref{eq:sumrule}.

\section[One Example Model: $SU(5) \times A_4$]{One Example Model: $\boldsymbol{SU(5) \times A_4}$}

As an example we sketch now the $A_4$ model from \cite{Antusch:2011sx},
where an $S_4$ model is given as well.
The $A_4$ model has the symmetry
$SU(5) \times A_4 \times Z_4^4 \times Z_2^2 \times U(1)_R$
and five flavons with the alignments
 \begin{equation} \label{eq:A4Alignment}
\langle \phi_1 \rangle \propto \begin{pmatrix} 1 \\ 0 \\ 0 \end{pmatrix} , \; \langle \phi_2 \rangle \propto \begin{pmatrix} 0 \\ -\mathrm{i} \\ 0 \end{pmatrix} ,\; \langle \phi_3 \rangle \propto \begin{pmatrix} 0 \\ 0 \\ 1 \end{pmatrix} , \;
\langle \phi_{23} \rangle \propto \begin{pmatrix} 0 \\ 1 \\ -1 \end{pmatrix} , \; \langle \phi_{123} \rangle \propto \begin{pmatrix} 1 \\ 1 \\ 1 \end{pmatrix} \; .
\end{equation}
Note that only $\langle \phi_2\rangle$ has a purely imaginary vev, while
all other vevs are real. To demonstrate the method of discrete vacuum alignment
we discuss the simple alignment superpotential for $\phi_{1,2,3}$ (for the others see \cite{Antusch:2011sx}):
\begin{equation}
W = \; P_1  \left( \frac{(\phi_1 \cdot \phi_1)^2}{M_{\Upsilon_{1;1}}^2}  - M_1^2 \right) + P_2  \left( \frac{(\phi_2 \cdot \phi_2)^2}{M_{\Upsilon_{2;2}}^2}  - M_2^2 \right) + P_3 (\phi_3 \cdot \phi_3 - M_3^2) + A_i (\phi_i \star \phi_i) + O_{ij} ( \phi_i \cdot \phi_j )\;,  \label{eq:A4PhiAlignment} 
\end{equation}
where $M_{\Upsilon}$ labels messenger masses. We use the standard ``$SO(3)$ basis'' for which ``$\cdot$'' is the usual $SO(3)$ inner product and the symmetric ``$\star$'' product is defined analogous to the cross product but with a relative plus sign instead of a minus sign.

The $F$-term conditions $F_{A_i} = F_{O_{ij}} = 0$ give the directions of the flavon vevs and their mutual orthogonality. The vev of $\phi_3$ (charged only under a $Z_2$) is fixed to be real while the vevs of $\phi_2$ and $\phi_3$, which are charged under $Z_4$'s can be chosen to be either real or imaginary and we pick the phases from Eq.\ \eqref{eq:A4Alignment}.

The five-dimensional matter fields are organised in triplets under $A_4$ and the tenplets are $A_4$ singlets. Therefore in our conventions the flavon vevs form rows of the down-type quark Yukawa matrix. The up-type quark Yukawa matrix is given by the inner product of two flavon vevs apart from the 3-3 element, which is generated on the renormalisable level to account for the large top mass. With the symmetries and the field content (for details see \cite{Antusch:2011sx}) we obtain in the quark sector
\begin{equation}
Y_d = \begin{pmatrix} 0 & \mathrm{i} \, \epsilon_2 & 0 \\ \epsilon_{123} & \epsilon_{23} + \epsilon_{123} & -\epsilon_{23} + \epsilon_{123} \\ 0 & 0 & \epsilon_3  \end{pmatrix}  \quad \text{and} \quad
Y_u = \begin{pmatrix} a_{11} & a_{12} & 0 \\ a_{12} & a_{22} & a_{23} \\ 0 & a_{23} & a_{33} \end{pmatrix}  \;,
\end{equation}
where the $\epsilon_i$ and $a_{ij}$ are real coefficients.
First note that $\delta_{12}^d = \arg( (Y_d)_{12}/(Y_d)_{22} ) = 90^\circ$,
due to the purely imaginary 1-2 element of $Y_d$,
and $\delta_{12}^u = 0^\circ$, because $Y_u$ is real.
The 1-3 elements in $Y_d$ and $Y_u$
vanish and the sum rule from Eq.\ \eqref{eq:sumrule} can be applied
successfully.

In the lepton sector we obtain for the Yukawa matrices and the right-handed neutrino mass matrix
\begin{align}
Y_e^T = -\frac{3}{2} \begin{pmatrix} 0 & \mathrm{i} \, \epsilon_2 & 0 \\
 \epsilon_{123} & -3 \epsilon_{23} + \epsilon_{123} & 3 \epsilon_{23} +
 \epsilon_{123} \\ 0 & 0 & \epsilon_3  \end{pmatrix}  \, , \;
Y_{\nu} = \begin{pmatrix} 0 & a_{\nu_2} \\ a_{\nu_1} & a_{\nu_2} \\ - a_{\nu_1} & a_{\nu_2} \end{pmatrix} \, , \;
M_R = \begin{pmatrix} M_{R_1} & 0 \\ 0 & M_{R_2} \end{pmatrix} \;.
\end{align}
The first thing to note here, is that we do not use standard GUT relations, but instead use $y_\tau/y_\mu = -3/2$ and $y_\mu/y_s \approx 9/2$, which fit much better to current data for the quark and lepton masses and a CMSSM like scenario with $\mu > 0$ \cite{Antusch:2008tf}.

In the neutrino sector only two of the three neutrinos are massive
by construction since we have introduced only two right-handed neutrinos
and the mass pattern is normal hierarchical. For the mixing we obtain
exact tri-bimaximal mixing in the neutrino sector, which is disturbed by corrections
coming from the charged lepton sector inducing, for instance, a non-vanishing
$\theta^{\text{PMNS}}_{13} \approx 3^\circ$. It is also interesting to note,
that we predict all CP phases in the lepton sector, which turn out to be
very close to $0^\circ$ or $180^\circ$.

\section[Another Example]{Another Example}

The $SU(5) \times A_4$ model in \cite{Antusch:2010es} can also be read as another example of this class of models, if the flavon $\tilde{\phi}_{23}$ is split into two flavons
 \begin{equation}
\langle \tilde{\phi}_{23} \rangle \rightarrow
\langle \tilde{\phi}_2 \rangle + \langle \tilde{\phi}_3 \rangle \quad\text{where}\quad 
\langle \tilde{\phi}_2 \rangle  = \begin{pmatrix} 0 \\ -\text{i}  \\ 0  \end{pmatrix} \tilde{\epsilon}_{23} \quad\text{and}\quad \langle \tilde{\phi}_3 \rangle  = \begin{pmatrix} 0 \\ 0 \\ w \end{pmatrix} \tilde{\epsilon}_{23} \; .
\end{equation}
In this model the sum rule, Eq.\ \eqref{eq:sumrule}, is not applicable, because there are no texture zeros in the 1-3 elements, but the agreement with the experimentally determined CKM phase is still very good, which is closely related to the use of the GUT relation $y_\mu/y_s \approx 9/2$. In fact, the CKM phase can be predicted in this model from the precisely known values for the electron mass, the muon mass and the Cabibbo angle and we obtain
\begin{equation}
 \delta_{\text{CKM}}^{\text{pred}} = 69.9^\circ \quad \text{while} \quad \delta_{\text{CKM}}^{\text{exp}} = (68.8^{+4.0}_{-2.3} )^\circ \;.
\end{equation}
The fit to the quark masses and mixing parameters and the charged lepton masses in this model is quite good with a $\chi^2$ per degree of freedom of about 1.6.

\begin{figure}
\centering
\includegraphics[width=8.5cm]{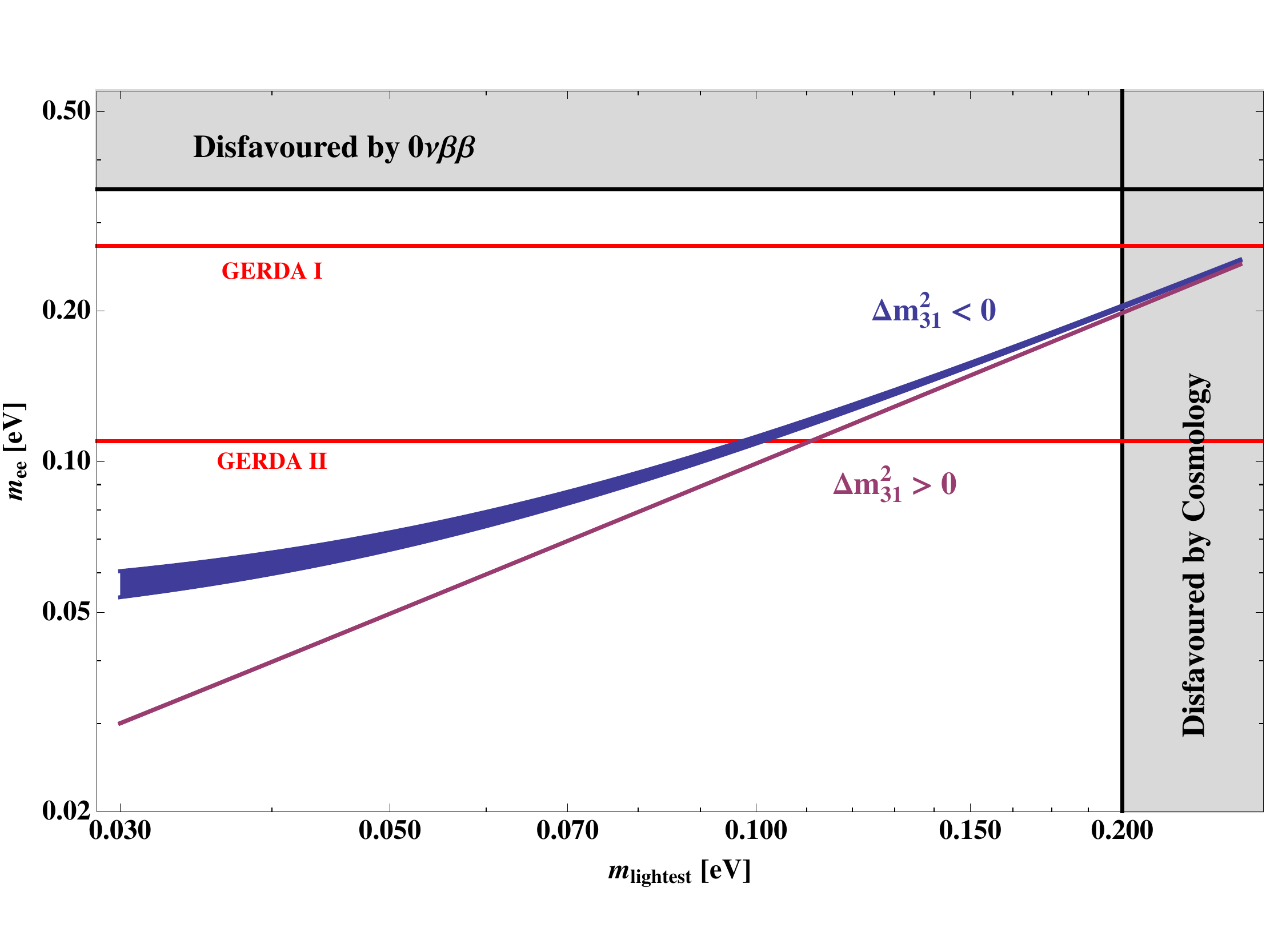}
\caption{The effective mass $m_{ee}$ in the setup from \cite{Antusch:2010es} relevant for neutrinoless double beta decay as a function of the mass $m_\mathrm{lightest}$ of the lightest neutrino, for an inverted neutrino mass ordering ($\Delta m_{31}^2 < 0$, upper line) and for a normal mass ordering ($\Delta m_{31}^2 > 0$, lower line). The bands represent the experimental uncertainties of the mass squared differences. The mass bounds from cosmology \protect\cite{Jarosik:2010iu} and from the Heidelberg-Moscow experiment \protect\cite{KlapdorKleingrothaus:2000sn} are displayed as grey shaded regions. The red lines show the expected sensitivities of the  GERDA experiment in phase I and II \protect\cite{Smolnikov:2008fu}.  \label{Fig:meePlot}}
\end{figure}

In the neutrino sector we have added a fifteen dimensional representation of $SU(5)$ giving a universal contribution to the neutrino masses, which can result in quasi-degenerate neutrino masses. All the mixing parameters are close to tri-bimaximal and the phases are fixed with $\delta_{\text{PMNS}} \approx 90^\circ$, $\alpha_1 \approx 9^\circ$ and $\alpha_2 \approx 0^\circ$. This has interesting phenomenological consequences. For example in Fig.\ \ref{Fig:meePlot} we have shown the prediction for neutrinoless double beta decay, which depends in this setup only on the neutrino mass scale and the sign of $\Delta m_{31}^2$.

\section{Summary and Conclusions}

Discrete symmetries are not only powerful in describing
leptonic mixing angles, but they can also be
used to predict the right-angled CKM unitarity triangle
by means of spontaneous CP violation.
In combination with a
unified gauge group this gives close relations between
the CP violation in the quark and the lepton sector.
In fact, in this new class of models all physical phases
can be predicted up to a discrete choice.
For example in the $A_4$ and $S_4$ model from
\cite{Antusch:2011sx} apart from $\alpha \approx 90^\circ$
in the quark sector, the leptonic Dirac and Majorana CP phases
are all close to $0^\circ$, $90^\circ$, $180^\circ$
or $270^\circ$. These predictions, especially for the
leptonic Dirac CP phase, can be tested at ongoing and
forthcoming neutrino experiments

%% file: Author/EmmanuelStamou.tex
{\bf Abstract}\\
\vskip5.mm
New neutral heavy gauge bosons appear automatically in many extensions
of the Standard Model with an extended gauge sector. Typical examples
are $Z'$ models and gauge-flavour models in which the flavour symmetry,
necessary to explain the Standard Model fermion masses and mixings, or
a part of it, is gauged. Often, additional heavy exotic fermions are
introduced to cancel anomalies of the new gauge sector. In
phenomenologically testable scenarios, the lightest heavy bosons and
fermions have masses around the TeV scale and may be directly produced
in current colliders. On the other hand, indirect bounds are present
since the neutral gauge bosons and exotic fermions affect the well-measured
branching ratio of $B\rightarrow X_s~\gamma$. We present the model-independent
constrains from $b\rightarrow s ~\gamma$ on the couplings of new neutral
gauge bosons to the standard model quarks, stressing the importance of QCD
mixing, and also discuss the contribution from the exotic down-type
quarks to the branching ratio. 

\vskip5.mm

\section{Introduction}\label{sec:introduction}
Extensions of the Standard Model (SM) with additional gauge symmetries are of particular interest
in view of current direct and indirect searches for physics beyond the SM. A prediction of such
theories is the existence of new gauge bosons, which may provide clear deviations from the SM
predictions and are being studied as possible discoveries at the LHC. Usually, the models
have a single additional $U(1)$ factor corresponding to an extra neutral gauge boson $Z'$
whose mass and couplings are strongly model-dependent \cite{Langacker:2008yv}. Recently, however,  in a 
series of papers \cite{Grinstein:2010ve,Albrecht:2010xh,Feldmann:2010yp,Guadagnoli:2011id}
it was suggested to  explain the SM fermion masses and mixings with a New Physics (NP) scale of a few TeV, by gauging flavour
symmetries. In such models, the gauge flavour symmetry is the product of non-abelian $U(3)$ factors.
For each broken gauge-group generator there is a new massive gauge boson, with no colour or 
electric charge, but with flavour violating couplings to fermions. In addition, new exotic fermions
are introduced to cancel anomalies from the new gauge sector.

The existence of one or more flavour-violating gauge bosons at the TeV scale
may have an impact on collider observables, but also on $\Delta F=2$ observables
and rare $B$ and $K$ decays, all of which have been studied more or less in detail in the literature.
The experimentally well-measured branching ratio of the inclusive decay
$\overline{B}\rightarrow X_s\gamma$ provides generally also a very strong constraint on extensions
of the SM due to the precision of its SM prediction. However, the impact of flavour-changing 
neutral gauge bosons on it
has only recently been considered in \cite{arXiv:1105.5146} and shall be discussed
here.

In Sec.~\ref{sec:thresholdcorrections} I present the threshold corrections due to
the presence of a neutral-gauge boson and possibly exotic fermions, while in
Sec.~\ref{sec:QCD} the QCD mixing effects, which are known from the SM to
be important for the $b\rightarrow s~\gamma$ transition. In Sec.~\ref{sec:pheno} I
give model-independent constraints and how they apply on representative toy-models,
to conclude in Sec.~\ref{sec:conclusions}.

\section{Threshold Corrections}\label{sec:thresholdcorrections}
We find the threshold corrections to the $b\rightarrow s~\gamma$ transition
originating from a heavy neutral gauge boson $A_H$ of mass $M_{A_H}$ by performing a
matching of the theory with $A_H$ and the effective field theory 
in which $A_H$ is integrated out. The matching is performed at the scale
$\mu_H\approx M_{A_H}$ and generates a Wilson coefficient $\Delta C_{7}(\mu_H)$ 
for the dipole operator $Q_{7\gamma}$, which is responsible for the 
$b\rightarrow s~\gamma$ transition. The $b\rightarrow s~\gamma$ transition is then
described by the effective Hamiltonian
\begin{equation}
  {\cal H}_{\rm eff}^{(b\to s\gamma)} = - \dfrac{4 G_{\rm F}}{\sqrt{2}} V_{ts}^*
  V_{tb} \Big[\Delta C_{7}(\mu_H) Q_{7\gamma} + \Delta  C_{8}(\mu_H) Q_{8G}
  +\Delta  C'_{7}(\mu_H) Q'_{7\gamma} +\Delta  C'_{8}(\mu_H) Q'_{8G}
  \Big]\,
  \label{eq:effectivehamiltonian}
\end{equation}
where we have already introduced the SM normalisation. The dipole operators read in our
conventions
\begin{equation}
    Q_{7\gamma}  =  \dfrac{e}{16\pi^2}\, m_b\, \bar{s}_\alpha\, \sigma^{\mu\nu}\,
    P_R\, b_\alpha\, F_{\mu\nu}
    \hfill\text{and}\hfill
    Q_{8G}     =  \dfrac{g_s}{16\pi^2}\, m_b\, \bar{s}_\alpha\, \sigma^{\mu\nu}\,
    P_R\, T^a_{\alpha\beta}\, b_\beta\, G^a_{\mu\nu} \,, 
  \label{eq:Q7Q8}
\end{equation}
while the primed operators $Q_{7\gamma}^\prime$ and $Q_8^{\prime}$ are obtained by
the interchange $L\leftrightarrow R$.

\begin{figure}[]
  \begin{minipage}{0.4\textwidth}
    \centering\includegraphics{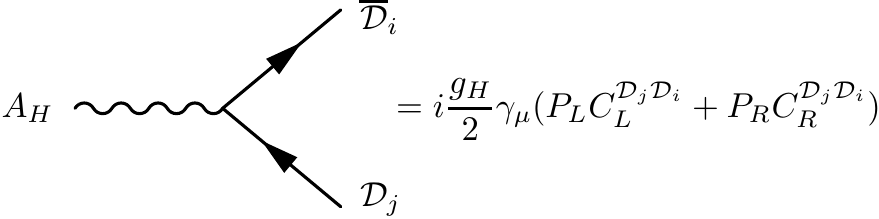}
  \end{minipage}
  \hfill
  \begin{minipage}{0.4\textwidth}
   \centering\includegraphics{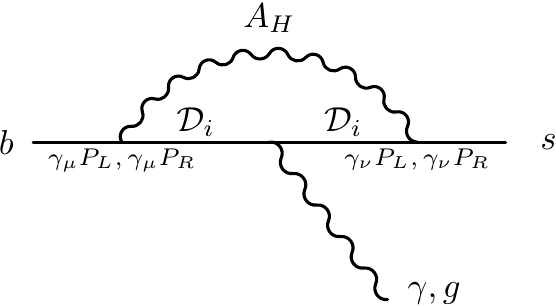}
  \end{minipage}
  \caption{Left: the general vertex for the $\mathcal{D}_j\rightarrow \mathcal{D}_i~A_H$
  transition. Right: all non-zero one-loop diagrams contributing to the quark-level transitions
  $b\rightarrow s~\gamma$ and $b\rightarrow s~g$ at the scale $\mu_H$.}
  \label{fig:bsgamma}
\end{figure}

We split the Wilson coefficients in Eq.~\eqref{eq:effectivehamiltonian} in two parts. The first part
involves only light quarks and is present in all models with a massive flavour
violating $A_H$. The second part is the impact of an exotic quark of mass $m_D$,
under the assumption that we may also integrate it out at $\mu_H$. The existence
of exotic quarks depends on the model under consideration.
The decomposition of the Wilson coefficient of $Q_7$ in terms of light- and
heavy-quark as well as SM-like (LL) and new (LR) contributions is then
\begin{equation}
  \Delta C_{7}(\mu_H)=\left(\Delta^{LL} C^\text{light}_{7}(\mu_H) +\Delta^{LR}C^\text{light}_{7}(\mu_H)\right)+
                          \left(\Delta^{LL} C^\text{heavy}_{7}(\mu_H) +\Delta^{LR}C^\text{heavy}_{7}(\mu_H)\right)
\label{eq:wilsonatmh}
\end{equation}
The general Feynman rule for the down type quark transition $D_j\rightarrow
D_i+A_H$ is given, in our notation, in Fig.~\ref{fig:bsgamma}. 
$\mathcal{D}_i,~\mathcal{D}_j$ are mass eigenstates of down-type quarks; we
denote with $d_i$ and $D_i$ light and exotic quarks, respectively and with $A_H$
the mass eigenstate of the colour- and electric-neutral gauge boson. With
the Feynman rule at hand we calculate all Wilson coefficients at the one-loop order, see
Fig.~\ref{fig:bsgamma}. 
They can all be expressed in terms of two known loop-functions, namely
\begin{equation}
  C_{7}(x)= \frac{3 x^3-2 x^2}{4(x-1)^4}\ln x -
             \frac{8 x^3 + 5 x^2 - 7 x}{24(x-1)^3}
             \hfill\text{and}\hfill
  C^{LR}_{8}(x)=\dfrac{-3x}{2(1-x)^3}\ln{x}+ \dfrac{3 x (x-3)}{4(x-1)^2} -1\,.
  \label{eq:C7C8LR}
\end{equation}
with $x$ the mass of the fermion in the loop over the mass of the gauge-boson
mass squared. $C_7(x)$ is known from the SM \cite{UT-KOMABA-80-8}, while
$C^{LR}_8$ from \cite{Bobeth:1999ww}. 

The individual contributions presented in \cite{arXiv:1105.5146} read:
\begin{align}
\Delta^{LL}C^\text{light}_{7}(\mu_H) &=-\dfrac{1}{6}\,\dfrac{g_H^2}{g_2^2}\,
                                                \dfrac{M_W^2}{M_{A_H}^2}\,\,\,
                                                \sum_{i=1}^3\dfrac{C_L^{sd_i*}\,C_L^{bd_i}}{V_{ts}^*\,V_{tb}}\,
                                                \left(\dfrac{1}{3}\right)
\label{eq:lightLL}\\
\Delta^{LR}C^\text{light}_{7}(\mu_H) &=-\dfrac{1}{6}\,\dfrac{g_H^2}{g_2^2}\,
                                                \dfrac{M_W^2}{ M_{A_H}^2}\,\,\,
						\sum_{i=1}^3\dfrac{m_{d_i}}{m_b}
                                                \dfrac{C_L^{sd_i*}\,C_R^{bd_i}}{V_{ts}^*\,V_{tb}}\,
                                                \left( -1\right)
\label{eq:lightLR}\\
\Delta^{LL}C^\text{heavy}_{7}(\mu_H) &=-\dfrac{1}{6}\,\dfrac{g_H^2}{g_2^2}\,\dfrac{M_W^2}{M_{A_H}^2}\,
                                         \dfrac{C_L^{sD*}\,C_L^{bD}}{V_{ts}^*\,V_{tb}}\,
					 \left(C_{8}(x)+\dfrac{1}{3}\right)
\label{eq:heavyLL}\\
\Delta^{LR}C^\text{heavy}_{7}(\mu_H) &=-\dfrac{1}{6}\,\dfrac{g_H^2}{g_2^2}\,\dfrac{M_W^2}{
						M_{A_H}^2}\,\dfrac{m_D}{m_b}\,\dfrac{C_L^{sD*}\,
						C_R^{bD}}{V_{ts}^*\,V_{tb}}\,C^{LR}_{8}(x)\,.
\label{eq:heavyLR}
\end{align}
The corresponding primed Wilson coefficients are obtained by the interchange
$L\leftrightarrow R$ and an additional suppression factor $m_b/m_{\mathcal{D}}$
depending on the down-type quark in the loop. Similar expressions also hold  for
$\Delta C_{8}(\mu_H)$, also to be found in \cite{arXiv:1105.5146}.

Eqs.~\eqref{eq:lightLL} to \eqref{eq:heavyLR} highlight the quadratic,
$M_W^2/M_{A_H}^2$, suppression of the threshold corrections with respect to
the SM contributions. When more than one gauge-boson is present in the theory, this
suppression factor renders the contribution of the lightest one the most relevant
one.
Also, in contrast to the SM no GIM mechanism is at work to
cancel mass-independent terms; this would for example completely cancel the sum of
all light-quark contributions. At last, $\Delta^{LR}C^\text{heavy}_{7}(\mu_H)$ is
strongly enhanced by $m_D/m_b$ if there is no extra suppression in the flavour
violating couplings of Eq.~\eqref{eq:heavyLR}. We return to this issue in
Sec.~\ref{sec:pheno}.

The connection to the SM is done by evolving the Wilson coefficients 
down to electroweak scale using the QCD Renormalisation Group Equations (RGE)
and subsequently integrating out $W$-bosons and the heavy top-quark, to which 
we turn our attention to now.

\mathversion{bold}
\section{Extended Operator Basis and QCD Mixing}\label{sec:QCD}
\mathversion{normal}
QCD corrections are very important role for the precise determination of the
$\overline{B}\rightarrow X_s~\gamma$ branching ratio. Within the SM they enhance
the rate by factor of $2-3$ \cite{Misiak:2006zs}, mainly from mixing of {\it charged}
current-current operators $Q^{cc}$ into the dipole operators $Q_{7\gamma}$ and $Q_{8G}$. These charged
operators originate from integrating out the $W$-bosons at the tree level.  

A similar situation occurs in the effective theory described in 
Sec.~\ref{sec:thresholdcorrections}; integrating out $A_H$ at the tree-level
generates $48$ new {\it neutral} current-current operators \cite{arXiv:1105.5146}: 
\begin{equation}
\begin{split}
  Q^{f}_1 &= (\overline s_\alpha \gamma_\mu P_A b_\beta)(\overline{f}_\beta\gamma^\mu P_B f_\alpha)\,,\qquad&
  Q^{f}_2 &= (\overline s_\alpha \gamma_\mu P_A b_\alpha)(\overline{f}_\beta\gamma^\mu P_B f_\beta)\,,&\\
  \hat{Q}^{d}_1 &= (\overline s_\alpha \gamma_\mu P_A d_\beta)(\overline{d}_\beta\gamma^\mu P_B b_\alpha)\,,\qquad&
  \hat{Q}^{d}_2 &= (\overline s_\alpha \gamma_\mu P_A d_\alpha)(\overline{d}_\beta\gamma^\mu P_B b_\beta)\,,&
\end{split}
  \label{eq:basisneutralcurrentcurent}
\end{equation}
where $f=\{u,c,t,d,s,b\}$, $A,B=\{L,R\}$ and $\alpha,\beta$ the usual colour indices.
We denote by $Q^{nn}$ and $Q^{nn\prime}$ all operators with $A=L$ and $A=R$,
respectively, since the former mix into $Q_{7\gamma}$ ($Q_{8G}$), while the latter into
$Q_{7\gamma}^{\prime}$ ($Q_{8G}^\prime)$.
\footnote{The operator basis of Eq.~\eqref{eq:basisneutralcurrentcurent}
is reducible under Fierz transformations. However, we can and shall use this
basis to best keep track of the mixing pattern between operators and only after
the RG evolution apply Fierz transformations if necessary \cite{MPI-PAE/PTh-63/89}.}

The full operator basis for the $b\rightarrow s~\gamma$ transition at scales
$\mu_b\leq\mu\leq\mu_H$ is given by the charged current-current, QCD-penguin, dipole,
and neutral current-current operators and the corresponding primed operators:
$Q^{cc}\,,\quad Q_P       \,,\quad Q_D       \,,\quad Q^{nn}\,,
          \quad Q_P^\prime\,,\quad Q_D^\prime\,,\quad Q^{nn\prime}$.
The anomalous dimension matrix describing the mixing of the operators into one another is
then at then leading order
\begin{equation}
\begin{tabular}{l|c|c|c|c|c|c|c|l}
\multicolumn{1}{r}{}& \multicolumn{1}{c}{$Q^{cc}$} & \multicolumn{1}{c}{$Q_P$}  & \multicolumn{1}{c}{$Q_D $}    & \multicolumn{1}{c}{$Q^{nn}$}
                                                   & \multicolumn{1}{c}{}       & \multicolumn{1}{c}{}          & \multicolumn{1}{c}{} \\
\cline{2-5}
$Q^{cc}$        & $X_1$         & $X_2$         & $X_3$         & 0     &\multicolumn{1}{c}{}   &\multicolumn{1}{c}{}   &\multicolumn{1}{c}{}\\
\cline{2-5}
$Q_P$           & 0             & $X_4$         & $X_5$         & 0     &\multicolumn{1}{c}{}   &\multicolumn{1}{c}{}   &\multicolumn{1}{c}{}\\
\cline{2-5}
$Q_D$           & 0             & 0             & $X_6$         & 0     &\multicolumn{1}{c}{}   &\multicolumn{1}{c}{}   &\multicolumn{1}{c}{}\\
\cline{2-5}
$Q^{nn}$        & 0             & $Y_1$         & $Y_2$         & $Y_3$ &\multicolumn{1}{c}{}   &\multicolumn{1}{c}{}   &\multicolumn{1}{c}{}\\
\cline{2-8}
\multicolumn{1}{c}{}            &\multicolumn{1}{c}{}&\multicolumn{1}{c}{}&\multicolumn{1}{c}{}&        & $X_4$ & $X_5$ & 0     & \multicolumn{1}{l}{$Q_P'$}\\
\cline{6-8}
\multicolumn{1}{c}{}            &\multicolumn{1}{c}{}&\multicolumn{1}{c}{}&\multicolumn{1}{c}{}&        & 0     & $X_6$ & 0     & \multicolumn{1}{l}{$Q_D'$}\\
\cline{6-8}
\multicolumn{1}{c}{}            &\multicolumn{1}{c}{}&\multicolumn{1}{c}{}&\multicolumn{1}{c}{}&        & $Y_1$ & $Y_2$ & $Y_3$ & \multicolumn{1}{l}{$Q^{nn\,\prime}$}\\
\cline{6-8}
\multicolumn{1}{r}{}            &\multicolumn{1}{r}{}&\multicolumn{1}{r}{}&\multicolumn{1}{r}{}&\multicolumn{1}{r}{}& \multicolumn{1}{c}{$Q_P'$}  & \multicolumn{1}{c}{$Q_D' $} & \multicolumn{1}{c}{$Q^{nn\,\prime}$}
\end{tabular}
\label{eq:AnomalousDimension}
\end{equation}
The matrices $X_i$  are known from the SM analysis \cite{Ciuchini:1993fk,Buchalla:1995vs}. The
$Y_i$ entries describe mixings related to the {\it neutral} current-current operators and
are found in \cite{arXiv:1105.5146}.
Using Eq.~\eqref{eq:AnomalousDimension} it is straightforward to evolve both
$\Delta C_{7}$ and $\Delta C_{7}^{\prime}$ from $\mu_H$ down to $\mu_W$, switching
on the SM and subsequently running down to $\mu_b$ to obtain $\Delta C_7(\mu_b)$
and $\Delta C_7^\prime(\mu_b)$ including the SM, threshold corrections from $\mu_H$
and QCD-mixing effects; more details in \cite{arXiv:1105.5146}.

\section{Phenomenological Implications}\label{sec:pheno}
The SM prediction for the $\overline{B}\to X_s\gamma$ branching ratio at
NNLO \cite{Gambino:2001ew,Misiak:2006ab} reads,
\begin{equation}
  Br(\overline{B}\to X_s \gamma)^{\text{SM}}
                                = R \left(|C^{\text{SM}}_{7}(\mu_b)|^2+|C^{\text{SM}\prime}_{7}(\mu_b)|^2+N(E_\gamma)\right)
                                =(3.15\pm0.23)\times 10^{-4}\,,
\label{eq:BRsm}
\end{equation}
and has been calculated for a photon-energy cut-off $E_\gamma>1.6$ GeV in the $\overline{B}$-meson rest frame 
($R=2.47\times10^{-3}$ and $N(E_\gamma)=(3.6\pm0.6)\times10^{-3}$). The SM prediction is to be compared with
the current experimental value \cite{FERMILAB-PUB-10-665-PPD}, 
\begin{equation}\label{eq:BRexp}
  Br(\overline{B}\to X_s \gamma)^{\text{exp}}=(3.55\pm 0.24 \pm 0.09)\times 10^{-4}
\end{equation}
for the same energy cut-off $E_\gamma$.

In the presence of NP, Eq.~\eqref{eq:BRsm} holds if we make the substitutions:
\begin{equation}
  C_7(\mu_b)^{\text{SM}}\longrightarrow \Delta C_7(\mu_b) 
  \,,\qquad
  C_7(\mu_b)^{\text{SM}\prime}\longrightarrow \Delta C_7^\prime(\mu_b)\,,
  \label{eq:C7NP}
\end{equation}
\begin{equation}  
\begin{split}
 \Delta C_{7}(\mu_b)\simeq C_7^{\text{SM}}(\mu_b)&+\quad\kappa_7~\Delta C_{7}(\mu_H) +\kappa_8~\Delta C_{8}(\mu_H)+\\[2mm]
                               &+\!\!\!\!\!\sum_{\substack{A=L,R\\f=u,c,t,d,s,b}}\!\!\!\!\! \kappa^{f}_{LA}~\Delta ^{LA} C_2^f(\mu_H)
                                +\!\!\!\sum_{A=L,R}\!\!\!\!\hat{\kappa}^{d}_{LA}~\Delta ^{LA} \hat{C}_2^d(\mu_H).
\end{split}
\label{eq:DeltaC7eff}
\end{equation}
The  $\kappa$-factors are the so-called ``magic'' numbers and summarise the RG
effects from $\mu_H$ down to $\mu_b$. They are independent of the model under
consideration and are given for different values of $\mu_H$ in \cite{arXiv:1105.5146}.
$\Delta ^{LA} C_2^f(\mu_H)$ and  $\Delta ^{LA} \hat{C}_2^d(\mu_H)$
are the Wilson coefficients of the {\it neutral} current-current operators in
Eq.~\eqref{eq:basisneutralcurrentcurent} $Q^f_2(L,A)$ and
$\hat{Q}_2^d(L,A)$, respectively \cite{arXiv:1105.5146}. Similar expressions hold
for the primed contributions.

The Wilson coefficients depend on the model. Once they are fixed the full NP
branching ratio prediction follows immediately from Eqs.~\eqref{eq:DeltaC7eff} and
\eqref{eq:C7NP} and gives the model-independent constraint
  \begin{equation}
      -\Delta C_{7\gamma}(\mu_b)+1.4 \left(\left|\Delta
      C_{7\gamma}(\mu_b)|^2+|\Delta C'_{7\gamma}(\mu_b)\right|^2\right)\lesssim4.2
      (6.1) \times 10^{-2},
  \end{equation}
corresponding to the $1\,\sigma$ ($2\,\sigma$) departure from the
{\it upper} experimental value of Eq.~\eqref{eq:BRexp}.

\begin{figure}[t]
  \begin{minipage}[]{0.4\textwidth}
    \centering
    \hspace*{3em}\mbox{Toy-Model 1}
    \centering\includegraphics[scale=0.9]{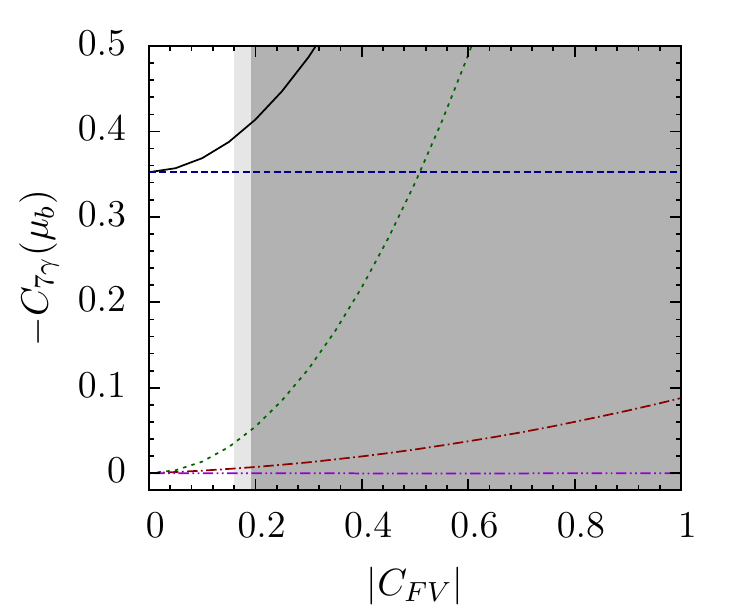}
  \end{minipage}
  \hspace*{5em}
  \begin{minipage}[]{0.4\textwidth}
    \hspace*{3em}\mbox{Toy-Model 2}
    \centering\includegraphics[scale=0.9]{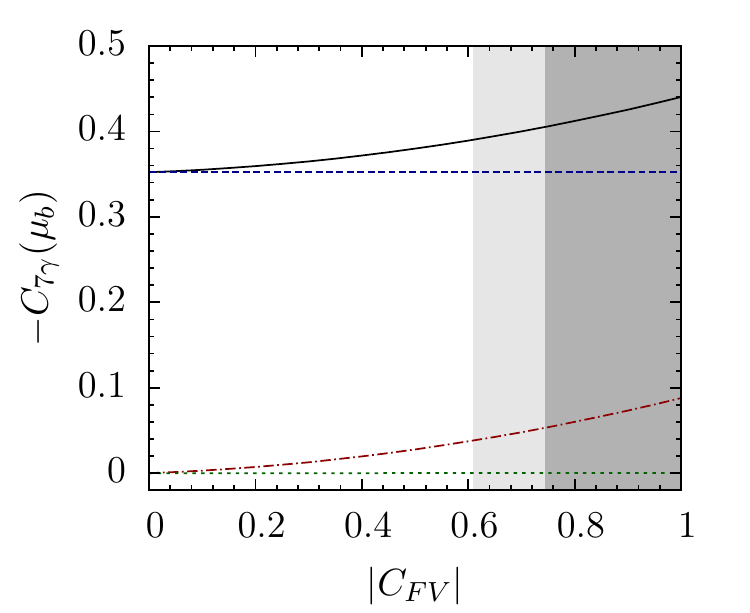}
  \end{minipage}
\caption{
Different contributions to $C_{7\gamma}(\mu_b)$ (solid line) are plotted as functions
of $C_{FV}$: in blue the SM , in red the neutral current-current ,
in green the exotic-fermion and in purple the light-fermion contribution. In
Toy-Model 2 exotic- and light-fermion contribution coincide. The shadowed
regions signal the departure of the predicted branching ratio from the $1\sigma$ (lighter)
and $2\sigma$ (darker) {\it upper} experimental value.
\label{fig:toymodels}}
\end{figure}
To illustrate the numerical importance of individual contributions we consider two
toy-models:
\begin{description}
  \item[Toy-Model 1 :] a model with one exotic fermion in which the coupling of SM fermions to $A_H$
    is independent of the mechanism of their mass generation. The mass of the
    exotic fermion is chosen to be $m_D=10$ TeV and $M_{A_H}=1$ TeV, while all
    flavour-violating couplings are identified to be $C_{FV}$ and all flavour
    conserving to be $1$. From Fig.~\ref{fig:toymodels} we see how the ($LR$)
    exotic-fermion renders all other contribution negligible and saturates the
    upper experimental bound even for small couplings.
  \item[Toy-Model 2 :] the SM fermion masses are governed by exotic
    fermions through a see-saw mechanism. In this case extra suppression factors
    enter in the coupling of  Eqs.~\eqref{eq:heavyLL} and \eqref{eq:heavyLR},
    which exactly cancel the enhancement of 
    $m_D/m_b$ in \eqref{eq:heavyLR} \cite{arXiv:1105.5146}.
    Again the mass of the  exotic fermion is chosen to be $m_D=10$ TeV and 
    $M_{A_H}=1$ TeV, while all flavour-violating couplings are identified to
    be $C_{FV}$ and all flavour conserving to be $1$. As seen in
    Fig.~\ref{fig:toymodels} the major effects originate from the QCD mixing of 
    neutral current-current operators into $Q_{7\gamma}$.
\end{description}

\section{Conclusions}\label{sec:conclusions}
Extensions of the SM with an enlarged gauge sector are attractive, since
they may provide an explanation of the flavour puzzle and have
testable predictions at colliders. In such models the flavour-violating neutral gauge bosons
often come with heavy exotic fermions to cancel anomalies and to justify the
observed SM fermion spectrum through a see-saw-like mechanism. The presence of flavour-violating
neutral gauge bosons translates into NP contributions to FCNC processes
and more specifically to the $\overline{B}\to X_s\gamma$ decay.
We have studied the impact of these new particles on the prediction of 
$Br(\overline{B}\to X_s\gamma)$ for the first time in a model-independent way including QCD-corrections,
and presented  expressions in a manner easily applied to test a specific model.
Our analysis shows once more how FCNC processes can put constraints on beyond-SM
constructions even before the discovery of new particles in direct searches.

%% file: Author/Sugiyama.tex
{\bf Abstract}\\
\vskip5.mm
 I will discuss phenomenology of
doubly charged scalars of $SU(2)_L$-triplet fields
in the simplest extension of the Higgs Triplet Model
with the $A_4$ symmetry.
 It is shown that their decays into a pair of leptons
have unique flavor structures which can be tested at the LHC
if some of their masses are below the TeV scale.
 Sizable decay rates for
$\tau \to \overline{\mu} e e$ and
$\tau \to \overline{e} \mu \mu$
can be obtained naturally
while other lepton flavor violating decays of charged leptons
are almost forbidden in this model,
which can be tested at the MEG experiment and future B factories.
 This talk is based on ref.~\cite{Fukuyama:2010mz}.

\vskip5.mm

\section{Introduction}

 Neutrino oscillation measurements
declared that neutrinos have masses
although they are regarded as massless particles
in the standard model of particle physics~(SM).
 The experiments also uncovered
the structure of the lepton flavor mixing matrix,
the Maki-Nakagawa-Sakata~(MNS) matrix $U_{\text{MNS}}$,
which can be parametrized as
\begin{eqnarray}
U_{\text{MNS}}
=
 \begin{pmatrix}
  1
  & 0
  & 0
\\
  0
  & c_{23}
  & s_{23}
\\
  0
  & -s_{23}
  & c_{23}
 \end{pmatrix}
 \begin{pmatrix}
  1
  & 0
  & 0
\\
  0
  & 1
  & 0
\\
  0
  & 0
  & 1
 \end{pmatrix}
 \begin{pmatrix}
  c_{12}
  & s_{12}
  & 0
\\
  -s_{12}
  & c_{12}
  & 0
\\
  0
  & 0
  & 1
 \end{pmatrix} ,
\label{HS-eq:MNS}
\end{eqnarray}
where $c_{ij}$ and $s_{ij}$
mean $\cos\theta_{ij}$ and $\sin\theta_{ij}$, respectively.
 Experimental results%
\footnote{
 If all experimental data is used
in 1 degree of freedom~(d.o.f.) analysis
where $\Delta\chi^2 = 9$ corresponds to 99.73\%~C.L.\ contour,
we can have the strongest constraint on a single parameter.
 However, as the cost for the strong constraint,
other parameters are not constrained at all.
 Therefore,
even if there are constraints on several parameters
(each of which is obtained in 1~d.o.f.\ analysis with all data),
we must use only one of them
in order to avoid multiple use of experimental data.
}
for mixing angles are
$\sin^2\!{\theta_{23}} \simeq 0.5$,
$\sin^2\!{\theta_{13}} \simeq 0$,
and $\sin^2\!{\theta_{12}} \simeq 0.3$.

 The most naive extension of the SM
to accommodate the neutrino mass
is the introduction of the right-handed neutrino $\nu_R^{}$
which is a singlet under the SM gauge group.
 Then
the Dirac mass of the neutrino can be obtained from
$y_\nu^{} \overline{L}\,i\sigma_2\,\Phi^\ast \nu_R^{}$,
where $\sigma_i (i=1\mbox{-}3)$ are the Pauli matrices,
$L = (\nu_L^{}, \ell)^T$ is a lepton doublet of $SU(2)_L$,
and $\Phi = (\phi^+, \phi^0)^T$ is the SM Higgs doublet.
 If the neutrino mass is given solely by the term
in the same way as the generation of other fermion masses,
it seems unnatural
because the Yukawa coupling constant $y_\nu^{}$ must be extremely small.
 We may expect that the neutrino mass
is produced in a different way.
 If we accept the lepton number non-conservation,
one possibility is
the Majorana mass term $1/2\,m_\nu^{} \overline{(\nu_L^{})^c}\,\nu_L^{}$,
where the superscript $c$ denotes the charge conjugation.
 The mass term is allowed only for the neutrino among the SM fermions
in order to keep $U(1)_{EM}$ gauge symmetry.
 Therefore
the neutrino mass can naturally be very different from other fermion masses.

 Before the breaking of $SU(2)_L\times U(1)_Y$ gauge symmetry
to $U(1)_{EM}$ gauge symmetry,
the weak-isospin $I_3$ and hypercharge $Y$ of the Majorana mass term
($I_3 = 1, Y = -2$)
should be compensated by those of scalar fields.
 If we do not introduce new scalar fields
which have their vacuum expectation values~(vev),
the compensation is achieved by the SM Higgs doublet $\Phi$
as a dimension-5 operator
$(\overline{L^c}\,i\sigma_2\,\Phi) (\Phi^T\,i\sigma_2\,L)$
or higher-dimensional ones.
 If we accept new scalar fields,
the simplest way of the compensation
is given by the Higgs Triplet Model~(HTM)~\cite{HS-HTM}
where the Majorana mass is provided
by a dimension-4 operator
$h_{\ell\HSpell}^{}\,\overline{L_\ell^c}\,i\sigma_2\,\Delta\,L_\HSpell$
with an $SU(2)_L$-triplet scalar field $\Delta$ of $Y=2$.
 The new Yukawa coupling constants
$h_{\ell\HSpell} (\ell,\HSpell = e, \mu, \tau)$
satisfy $h_{\ell\HSpell} = h_{\HSpell\ell}$.
 The triplet scalar field can be expressed as
\begin{eqnarray}
\Delta
=
 \begin{pmatrix}
  \Delta^+/\sqrt{2} & \Delta^{++}\\
  \Delta^0 & -\Delta^+/\sqrt{2}
 \end{pmatrix} .
\end{eqnarray}
 The Majorana mass matrix $(m_\nu^{})_{\ell\HSpell}$ for neutrinos
is obtained as
$(m_\nu^{})_{\ell\HSpell} = \sqrt{2}\, v_\Delta^{} h_{\ell\HSpell}^{}$,
where the triplet vev
$v_\Delta^{} (= \sqrt{2}\, \langle \Delta^0 \rangle)$
breaks the lepton number by 2 units.
 Since the HTM does not introduce new fermions to the SM,
neutrinos have no lepton number violating mixing%
~(ex.\ mixing between $\nu_L^{}$ and $(\nu_R^{})^c$)
which is the key in the seesaw mechanism.
 Even if $v_\Delta^{}$ is suppressed by a large mass scale,
it is just a consequence of the soft-breaking%
~(of the lepton number conservation)
rather than the seesaw mechanism.

 A doubly charged scalar $H^{++}$~($=\Delta^{++}$)
is the characteristic particle in the HTM\@.
 Its decay into a pair of same-signed charged leptons%
~($H^{++} \to \overline{\ell}\,\overline{\HSpell}$)
will give a clear signal even in hadron colliders,
and the flavor structure of the decay
can give direct information on $(m_\nu)_{\ell\HSpell}$%
~\cite{HS-HTMLHC,Petcov:2009zr}.
 The doubly charged scalar can also contribute to
flavor-violating decays of charged leptons (ex.\ $\tau \to \bar{\mu}ee$)
at the tree level~\cite{HS-HTMLFV}.

 On the other hand,
it seems interesting that
the lepton flavor mixing
has a nontrivial structure with two large mixings
while the quark mixing structure is rather simple with only small mixings.
 There might be some underlying physics for the lepton flavor.
 A candidate for that is the $A_4$ symmetry
which is a non-Abelian discrete group.
 The $A_4$ group is made from
twelve elements of even permutations of four letters.
 The group has three 1-dimensional representations%
~($\bf\underline{1}$, $\bf\underline{1}^\prime$, $\bf\underline{1}^\HSdprime$)
and one 3-dimensional representation ($\bf{\underline{3}}$).
 Only $\bf{\underline{1}}$ is the $A_4$-invariant.
 The $\bf{\underline{3}}$ seems to be fit the tree flavors of leptons,
and $A_4$ is the minimal one which has $\bf{\underline{3}}$.
 Some simple models based on the $A_4$ symmetry
can be found in e.g.,
refs.~\cite{HS-Ma,HS-A4model,HS-Altarelli}.
 Throughout this talk,
I will use $\bf{\underline{3}}$ etc.\ for $A_4$-representations
and "triplet" etc.\ for $SU(2)_L$-representations
in order to avoid confusions.

 The lepton mixing structure becomes
the tribimaximal mixing form~\cite{Harrison:2002er}%
~($\sin\theta_{23}=1/\sqrt{2}$, $\sin\theta_{13}=0$,
and $\sin\theta_{12}=1/\sqrt{3}$,
which agree reasonably with neutrino oscillation data)
without tuning Yukawa coupling constants
if $A_4$ is broken to $Z_3$ and $Z_2$
in the charged lepton and neutrino sectors,
respectively~\cite{HS-Altarelli}.
 It seems attractive that
the realization of the tribimaximal mixing
can be expressed simply in terms of the symmetry breaking pattern.
 If the lepton flavor mixing structure is reproduced
by a free fitting of parameters without such a guideline,
there would be no worth to deal with symmetries~($A_4$ etc.)
because such a fitting is also possible in the SM\@.

 In this talk,
I will present an extension of the HTM
by using the $A_4$ group~(we call the model as the A4HTM)
and discuss phenomenology of doubly charged scalars in the model.
 We will see that
the A4HTM has clear predictions
which can be tested experimentally in near future.

\section{Model}

\begin{table}[t]
\begin{tabular}{c||c|c|c|c}
 {}
 & $\psi_{1R}^-$
 & $\psi_{2R}^-$
 & $\psi_{3R}^-$
 & $\Psi_{AL}
    = \left(
       \begin{array}{c}
	\psi_{AL}^0\\
	\psi_{AL}^-
       \end{array}
      \right)$
\\[3mm]\hline\hline
 $A_4$
 & ${\bf\underline{1}}$
 & ${\bf\underline{1}}^\prime$
 & ${\bf\underline{1}}^\HSdprime$
 & ${\bf\underline{3}}$
\\\hline
 $SU(2)_L$
  & Singlet
  & Singlet
  & Singlet
  & Doublet
\\\hline
 $U(1)_Y$
 & $-2$
 & $-2$
 & $-2$
 & $-1$
\end{tabular}
\\[3mm]
\hspace*{25mm}
\begin{tabular}{c|c|c}
 $\Phi_A
    = \left(
       \begin{array}{c}
	\phi_A^+\\
	\phi_A^0
       \end{array}
      \right)$
 & $\delta
    = \left(
       \begin{array}{cc}
	\delta^+/\sqrt{2}
	 & \delta^{++}\\
	\delta^0
	 & -\delta^+/\sqrt{2}
       \end{array}
      \right)$
 & $\Delta_A
    = \left(
       \begin{array}{cc}
	\Delta_A^+/\sqrt{2}
	 & \Delta_A^{++}\\
	\Delta_A^0
	 & -\Delta_A^+/\sqrt{2}
       \end{array}
      \right)$
\\[3mm]\hline\hline
 ${\bf\underline{3}}$
 & ${\bf\underline{1}}$
 & ${\bf\underline{3}}$
\\\hline
 Doublet
  & Triplet
  & Triplet
\\\hline
 $1$
 & $2$
 & $2$
\end{tabular}
\vspace*{3mm}
\caption{
 The leptons and the Higgs bosons in the A4HTM\@.
 The subscript $A = x, y, z$ denotes
the index for ${\bf\underline{3}}$ of $A_4$;
 for example,
$(\Psi_{xL}, \Psi_{yL}, \Psi_{zL})$ belongs
to ${\bf\underline{3}}$
while each $\Psi_{AL}$ is an $SU(2)_L$-doublet field.
}
\label{HS-tab:particle}
\end{table}

 Table~\ref{HS-tab:particle}
shows particle contents in the A4HTM\@.
 No new fermion (ex.\ $\nu_R^{}$)
is added to the SM,
and only the scalar sector is extended. 
 This model has
three $SU(2)_L$-doublet
and four $SU(2)_L$-triplet scalars.
 For realization of appropriate flavor structure
of Yukawa coupling matrices,
we do not rely on singlet scalars 
under the SM gauge group~(the so-called flavons)
in order to respect renormalizability
which is preferred for predictability.
 For example,
renormalizable Yukawa interactions of triplet scalars with the $A_4$ symmetry
are expressed as
\begin{eqnarray}
\begin{pmatrix}\,
 \overline{(\Psi_{xL})^c} , \
 \overline{(\Psi_{yL})^c} , \
 \overline{(\Psi_{zL})^c}\,
\end{pmatrix}
 \begin{pmatrix}
  h_\delta i\sigma_2 \delta
   & h_\Delta i\sigma_2 \Delta_z
   & h_\Delta i\sigma_2 \Delta_y\\
  h_\Delta i\sigma_2 \Delta_z
   & h_\delta i\sigma_2 \delta
   & h_\Delta i\sigma_2 \Delta_x\\
  h_\Delta i\sigma_2 \Delta_y
   & h_\Delta i\sigma_2 \Delta_x
   & h_\delta i\sigma_2 \delta
 \end{pmatrix}
 \begin{pmatrix}
  \Psi_{xL}\\[1mm]
  \Psi_{yL}\\[1mm]
  \Psi_{zL}
 \end{pmatrix}
+ \text{h.c.} ,
\label{HS-eq:T-Yukawa}
\end{eqnarray}
where $h_\delta$ and $h_\Delta$ are Yukawa coupling constants.

 Let us just accept the following vev's
without analyzing the scalar potential%
\footnote{
 In order to reduce the number of parameters in scalar potentials
(not only in the A4HTM but also,
for example, in extensions of two-Higgs-doublet-model with $A_4$),
it is useful to notice
relations of rearrangements of $A_4$-invariant combinations,
which are similar to the Fierz transformation for the four-fermions.
 See Appendix~B in ref.~\cite{Fukuyama:2010mz}.
}%
~(See Sec.~III-A in ref.~\cite{Fukuyama:2010mz} for the detail):
\begin{eqnarray}
&&
\langle \phi_x^0 \rangle
= \langle \phi_y^0 \rangle
= \langle \phi_z^0 \rangle
= \frac{v}{\sqrt{6}},
\label{HS-eq:vevD}
\\
%
&&
\langle \delta^0 \rangle = \frac{v_\delta^{}}{\sqrt{2}} , \quad
\langle \Delta_x^0 \rangle
= \frac{v_\Delta^{}}{\sqrt{2}} , \quad
\langle \Delta_y^0 \rangle
= \langle \Delta_z^0 \rangle
= 0 .
\label{HS-eq:vevT}
\end{eqnarray}
 Masses of charged leptons and neutrinos are given
by the vev's in eqs.~\eqref{HS-eq:vevD} and \eqref{HS-eq:vevD},
respectively.
 In our convention of $A_4$-representations,
vev's in eq.~\eqref{HS-eq:vevD} break $A_4$ into $Z_3$
while ones in eq.~\eqref{HS-eq:vevT} do into $Z_2$.
 Then,
the tribimaximal mixing is obtained.
 However,
note that
this is just a mathematically beautiful reproduction
of known values~(lepton mixings).
 Here is the starting point of real physics
although the mathematical beauty can be a motivation.
 In the next section,
let us see predictions for phenomenology of doubly charged scalars
which have not been measured yet.
 See ref.~\cite{Fukuyama:2010mz}
for predictions on the mass of the lightest neutrino~(or a sum rule of masses)
and the Majorana phases which cannot be determined
by oscillation measurements.

\section{Phenomenology of doubly charged scalars}

 At first,
we must obtain mass eigenstates of relevant particles
to our discussion.
 Although we take vev's in eqs.~\eqref{HS-eq:vevD} and \eqref{HS-eq:vevT}
motivated by the lepton flavor mixing,
we can ignore triplet vev's
because the tree-level constraint from the $\rho$ parameter
results in $(v_\delta^2 + v_\Delta^2)/v^2 \lesssim 0.01$.
 Thus
$Z_3$ symmetry remains approximately in the A4HTM,
and this makes everything simple.
 Physical particles~(mass eigenstates)
should be classified by their $Z_3$-charges.
 Since triplet vev's are ignored hereafter,
we use flavor eigenstates for massless neutrinos.
 Table~\ref{HS-tab:t-ev} shows
$Z_3$-charges of charged leptons, neutrinos,
and four doubly charged scalars $H^{++}_i$~($i=1\mbox{-}4$)
made from four triplet fields.
 It is clear that
the flavor symmetry is not the original $A_4$
but the remaining $Z_3$.
 In that sense,
$\tau$ and $\overline{\mu}$ have the same flavor~(the same $Z_3$-charge).

\begin{table}[t]
\begin{tabular}{c||c|c|c}
 {}
  & \ $e$, $\nu_{eL}^{}$, $H_3^{++}$, $H_4^{++}$ \
  & \ $\mu$, $\nu_{\mu L}^{}$, $H_2^{++}$ \
  & \ $\tau$, $\nu_{\tau L}^{}$, $H_1^{++}$ \
\\[1mm]\hline\hline
 \ $Z_3$-charges \
  & $1$
  & $\omega$
  & $\omega^2$
\end{tabular}
\vspace*{3mm}
\caption{
 $Z_3$-charges of leptons and doubly charged scalars
where $\omega \equiv \exp(2\pi i/3)$.
}
\label{HS-tab:t-ev}
\end{table}

 Next,
let us investigate $H^{++}_i \to \overline{\ell}\,\overline{\HSpell}$.
 Yukawa interactions in eq.~\eqref{HS-eq:T-Yukawa}
are rewritten by using mass eigenstates.
 The Yukawa interactions of $H^{++}_i$ are
$(h_{i\pm\pm})_{\ell\ell^\prime}\,
\overline{(\ell_L)^c}\,\ell_L^\prime\,
H_i^{++}$.
 Yukawa coupling constants $(h_{i\pm\pm})_{\ell\ell^\prime}$ are given by
\begin{eqnarray}
&&
h_{1\pm\pm}
=
 \frac{1}{\sqrt{3}}\, h_\Delta
 \begin{pmatrix}
   0 & -1 & 0\\
  -1 &  0 & 0\\
   0 &  0 & 2
 \end{pmatrix} , \qquad
%
h_{2\pm\pm}
=
 \frac{1}{\sqrt{3}}\, h_\Delta
 \begin{pmatrix}
   0 &  0 & 1\\
   0 &  2 & 0\\
   1 &  0 & 0
 \end{pmatrix} ,
\nonumber\\
%
&&
h_{3\pm\pm}
=
 \frac{1}{\sqrt{3}}\,
 h_\Delta \cos\theta_{\pm\pm}
 \begin{pmatrix}
  2 & 0 & 0\\
  0 & 0 & 1\\
  0 & 1 & 0
 \end{pmatrix}
 +
 h_\delta
 e^{i\alpha_{\pm\pm}^{}}
 \sin\theta_{\pm\pm}
 \begin{pmatrix}
  1 &  0 & 0\\
  0 &  0 & -1\\
  0 & -1 & 0
 \end{pmatrix} ,
\nonumber\\
%
&&
h_{4\pm\pm}
=
 -\frac{1}{\sqrt{3}}\,
 h_\Delta
 \sin\theta_{\pm\pm}
 \begin{pmatrix}
  2 & 0 & 0\\
  0 & 0 & 1\\
  0 & 1 & 0
 \end{pmatrix}
 +
 h_\delta
 e^{i\alpha_{\pm\pm}^{}}
 \cos\theta_{\pm\pm}
 \begin{pmatrix}
  1 &  0 & 0\\
  0 &  0 & -1\\
  0 & -1 & 0
 \end{pmatrix} ,
\end{eqnarray}
where $\theta_{\pm\pm}$ and $\alpha_{\pm\pm}^{}$ are
mixing parameters of doubly charged scalars.
 These coupling constants
result in unique flavor structures
of $H^{++}_i$ decays into same-sign charged leptons
as listed in Table.~\ref{HS-tab:Hpp}.
 For example,
$H^{++}_1$ can decay only into
$\overline{e}\,\overline{\mu}$ and $\overline{\tau}\,\overline{\tau}$.
 Many zeros for $\HSBR(H^{++}_i \to \overline{\ell}\,\overline{\HSpell})$
is given by the conservation of $Z_3$-charges.
 Since ratios of nonzero parts
(ex.\
$\HSBR(H^{++}_1 \to \overline{\tau}\,\overline{\tau})
/\HSBR(H^{++}_1 \to \overline{e}\,\overline{\mu})=2$)
cannot be determined by $Z_3$ symmetry,
these are consequences of original $A_4$ symmetry.
 Therefore
both of $A_4$ and $Z_3$ can be tested
by measuring leptonic decays of $H^{\pm\pm}_i$ at the LHC
if they are right enough to be produced.

\begin{table}[t]
\begin{tabular}{c||ccccccccccc|c}
 {}
 & \multicolumn{11}{|c|}
   { $\HSBR(H^{++}_i \to \overline{\ell}\,\overline{\HSpell})$ }
 & \ LFV decays \
\\
 {}
 & $ee$ &:
  & $\mu\mu$ &:
  & $\tau\tau$ &:
  & $e\mu$ &:
  & $e\tau$ &:
  & $\mu\tau$ \
 & of charged leptons
\\\hline\hline
 $H_1^{\pm\pm}$
 & $0$ &:
 & $0$ &:
 & $2$ &:
 & $1$ &:
 & $0$ &:
 & $0$
 & None
\\\hline
 $H_2^{\pm\pm}$
 & $0$ &:
 & $2$ &:
 & $0$ &:
 & $0$ &:
 & $1$ &:
 & $0$
 & \ $\HStLeLmLmL$ \
\\\hline
 $H_3^{\pm\pm}$
 & $R^{\pm\pm}_3$ &:
 & $0$ &:
 & $0$ &:
 & $0$ &:
 & $0$ &:
 & $1$
 & \ $\HStLmLeLeL$ \
\\\hline
 $H_4^{\pm\pm}$
 & $R^{\pm\pm}_4$ &:
 & $0$ &:
 & $0$ &:
 & $0$ &:
 & $0$ &:
 & $1$
 & \ $\HStLmLeLeL$ \
\end{tabular}
\vspace*{3mm}
\caption{
 Ratios of decays of $H^{\pm\pm}_i$
into a pair of same-signed charged leptons
in the A4HTM\@.
 Here
$R^{\pm\pm}_3$ and $R^{\pm\pm}_4$
are combinations of model parameters.
 Contributions of $H^{\pm\pm}_i$
to $\HStlll$ at the tree level are also shown.
 Note that all of $H^{\pm\pm}_i$ does not
contribute to $\HSmeee$ and $\HSllg$
at the tree and 1-loop level, respectively.
}
\label{HS-tab:Hpp}
\end{table}

 Doubly charged scalars contribute also to
lepton flavor violating decays of charged leptons
at the tree level.
 However,
only $\tau \to \overline{e}\mu\mu$ and $\tau \to \overline{\mu}ee$
are allowed by the conservation of $Z_3$-charges
as shown in Table~\ref{HS-tab:Hpp}.
 Thus,
a stringent constraint
$\HSBR(\HSmeee)<1.0\times 10^{-12}$~\cite{Bellgardt:1987du}
is satisfied without fine tuning of parameters.
 The $Z_3$ symmetry also forbids $\ell \to \HSpell\gamma$
which look possible at the 1-loop level.
 Then
it is easy to expect sizable effects on $\tau$ decays.
 By virtue of these predictions,
the A4HTM can be tested in the MEG experiment and future B-factories
even if $H_i^{++}$ are too heavy to be produced at the LHC\@.
 Of course,
the A4HTM is excluded easily
if decays forbidden in the model are discovered.
 This is an excellent feature of the model
due to its high predictability.

\section{Conclusions}

 I have presented a renormalizable model, the A4HTM,
which is an extension of the HTM with $A_4$ symmetry.
 The A4HTM is compatible with the tribimaximal mixing.
 Phenomenology in the model is restricted
by an approximately remaining $Z_3$ symmetry.
 Then
sharp predictions have been obtained.
 It has been shown that
leptonic decays of $H^{\pm\pm}_i$ have
characteristic flavor structures
which would be tested at the LHC
if they are light enough to be produced.
 Even if $H^{\pm\pm}_i$ are too heavy to be produced,
they can affect on flavor violating decays of charged leptons.
 The $Z_3$ symmetry allows
only $\tau \to \overline{e}\mu\mu$ and $\tau \to \overline{\mu}ee$.
 The prediction would be tested at the MEG experiment and future B-factories.

%% file: Author/Toma.tex
\renewcommand{\thefootnote}{\fnsymbol{footnote}}
{\bf Abstract}\footnote[0]{This talk is based on ref.~\cite{Kajiyama:2011fe}.}\\
\vskip5.mm
Indirect and direct detection of Dark Matter is discussed in the $D_6$ flavor symmetric model. 
Dark Matter in this model is the $D_6$ singlet right handed neutrino $n_S$. 
The $D_6$ flavor symmetry predicts a certain mixings of leptons and also 
plays an important role in determining the final states of the Dark Matter annihilation.
A large annihilation cross section, which is required to explain the positron excess in cosmic ray 
observed by PAMELA experiment is obtained by the Breit-Wigner enhancement mechanism. 
Moreover, a certain elastic cross section with nucleon is derived by the mixing of Higgses which intermediate 
in the elastic scattering $n_Sq\to n_Sq$.

\vskip5.mm
\section{Introduction}
Many experimental evidences for the existence of Dark Matter (DM) are observed: 
for instance, rotation curves of spiral galaxy, CMB observation by WMAP, 
gravitational lensing and large scale structure of the Universe. 
DM candidate is often included as a stable particle due to a ${Z}_2$ symmetry in a particle physics model. 
A eligible DM has the thermally averaged annihilation cross section of $\left<\sigma v\right>\sim 10^{-9}~\mathrm{GeV^{-2}}$ 
in order to obtain the correct DM relic density. 

Several years ago, PAMELA reported excess of positron fraction in the cosmic ray~\cite{Adriani:2008zr}. 
This observation can be explained by annihilation and/or
decay of DM particles with mass of $\mathcal{O}(10^{2-3})~\mathrm{GeV}$. 
In this case, the required annihilation cross section is $\mathcal{O}(10^{-7})~\mathrm{GeV^{-2}}$ which is 
much larger than that for the relic DM density. 
Several ideas to overcome it are proposed such as the Sommerfeld enhancement, 
the Breit-Wigner enhancement~\cite{Ibe:2008ye}\cite{Feldman:2008xs}, 
non-thermal DM production and decaying DM. 
The PAMELA experiment searches antiproton as well in the cosmic ray, 
and it is consistent with the background~\cite{Adriani:2010rc}. 
Therefore, if these signals are from annihilation and/or decay processes of DM particles, 
this implies that the leptophilic DM is preferable. 
However, even if the DM is leptophilic, the resultant positron fraction depends on the
flavor of final state leptons. For instance, if the final state of annihilation and/or decay of the
DM is $\tau^{+}\tau^{-}$, it will overproduce gamma-rays as final state radiation~\cite{Meade:2009iu}\cite{Papucci:2009gd}. 
Therefore it is important to determine the flavor of final state leptons theoretically, and it could be possible 
by flavor symmetry of elementary particles which predicts the mixing of leptons. 

In this talk, we discuss the explanation of the positron excess in the cosmic ray observed by PAMELA in the model 
based on the $D_6$ flavor symmetry. 
The final states of the annihilation of DM are controlled by the $D_6$ flavor symmetry. 
The large annihilation cross section is obtained by the Breit-Wigner enhancement.
The elastic cross section for the direct detection of DM is also discussed briefly, and is obtained through the mixing of 
Higgses. The predicted elastic cross section is compared with the XENON100 and CDMS II results. 

\section{The Model}
We extend the SM by introducing three generations of right-handed neutrino $n_{S,I}$, 
Higgs doublets $\phi_{I,S}$, inert doublets $\eta_{I,S}$ which have no vacuum expectation values (VEVs), 
and one generation of inert singlet $\varphi$ where $I=1,2$ and $S$ denote $D_6$ doublet and singlet, respectively.
We also impose the additional discrete family symmery $\hat{Z}_2\times Z_2$ in order to suppress FCNC of 
the quark sector and forbid Dirac neutrino masses. In addition the imposed $Z_{2}$ symmetry stabilize a DM candidate. 
The $D_6\times\hat{Z}_2\times Z_2$ assignment is shown in Tab.\ref{a1} and \ref{a2}. 
\begin{table}[t]
\begin{center}
\begin{tabular}{|c|cccccc|} \hline
 & $L_S$ & $n_S$ & $e^c_S $&$L_I$&$n_I$&$e^c_I$ 
  \\ \hline
 $SU(2)_L\times U(1)_Y$ 
 & $({\bf 2}, -1/2)$  &  $({\bf 1}, 0)$  &  $({\bf 1}, 1)$
 & $({\bf 2}, -1/2)$&  $({\bf 1}, 0)$
 &  $({\bf 1}, 1)$
  \\ \hline
 $D_6$ & ${\bf 1}$  &  ${\bf 1}'''$  &  ${\bf 1}$
 & ${\bf 2}'$&  ${\bf 2}'$&  ${\bf 2}'$
 \\ \hline
 $\hat{Z}_2$
 & $+$ &$+$  & $-$  & $+$ 
 &$+$  & $-$
  \\ \hline
   $Z_2$
 & $+$ &$-$  & $+$  & $+$ &$-$  & $+$ 
   \\ \hline
\end{tabular}
\caption{The $D_6 \times \hat{Z}_2\times Z_2$ 
assignment for the leptons. 
$ L_{I,S}$ stands for the $SU(2)_L$ doublet leptons, and 
$e^c_{I,S}$ and $n_{I,S}$ are the $SU(2)_L$ singlet leptons.
}
  \label{a1}
\end{center}
\end{table}
\begin{table}[t]
\begin{center}
\begin{tabular}{|c|cccc|c|} \hline
 & $\phi_S$ &$\phi_I$ & $\eta_S$&$\eta_I $ & $\varphi$
  \\ \hline
   $SU(2)_L\times U(1)_Y$ 
 & $({\bf 2}, -1/2)$  &  $({\bf 2}, -1/2)$   &  $({\bf 2}, -1/2)$ 
 & $({\bf 2}, -1/2)$  & $({\bf 1}, 0)$ 
  \\ \hline
 $D_6$ & ${\bf 1}$ &${\bf 2}'$  &  ${\bf 1}'''$  & ${\bf 2}'$ & ${\bf 1}$
 \\ \hline
 $\hat{Z}_2$ &$+$ 
 & $-$ &$+$  & $+$& $+$ 
  \\ \hline
   $Z_2$
 & $+$ &$+$  & $-$  & $-$  & $+$
   \\ \hline
\end{tabular}
\caption{The $D_6 \times \hat{Z}_2\times Z_2$ 
assignment for the Higgs bosons.
}
\label{a2}
\end{center}
\end{table}
The invariant Lagrangian of the right handed neutrino sector under the imposed symmetry $D_6\times \hat{Z}_2\times Z_2$ is 
written as
\begin{eqnarray}
\mathcal{L}_{Y}\!\!\!&=&\!\!\!
\sum_{a,b,d=1,2,S}\left[Y^{ed}_{ab}L_{a}\phi_de^c_b+Y^{\nu d}_{ab}\eta^\dag_d L_{a}n_b\right]\nonumber\\
&&+\!\!\!\sum_{I=1,2}\frac{M_1}{2}n_In_I-\frac{M_S}{2}n_Sn_S-\sum_{I=1,2}\frac{\mathfrak{S}_1}{2}\varphi n_In_I
-\frac{\mathfrak{S}_S}{2}\varphi n_Sn_S+\mathrm{h.c}
\end{eqnarray}
where the couplings $\mathfrak{S}_1$ and $\mathfrak{S}_S$ are complex in general.
The following MNS (Maki-Nakagawa-Sakata) matrix is derived at leading order by the $D_6$ flavor symmetry.
\begin{equation}
V_{MNS}\simeq
\left(
\begin{array}{ccc}
\cos\theta_{12}&\sin\theta_{12}&0\\
-\frac{1}{\sqrt{2}}\sin\theta_{12}&\frac{1}{\sqrt{2}}\cos\theta_{12}&-\frac{1}{\sqrt{2}}\\
-\frac{1}{\sqrt{2}}\sin\theta_{12}&\frac{1}{\sqrt{2}}\cos\theta_{12}&\frac{1}{\sqrt{2}}\\
\end{array}
\right).
\end{equation}
The $D_6$ flavor symmetry gives two predictions. One is that the maximal mixing of atmospheric neutrino is derived. 
The other one is that inverted hierachy for the neutrino masses is only allowed~\cite{Kajiyama:2006ww}.

\section{DM Relic Density and $\mu\to e\gamma$ Constraint}
Several DM candidates which are $Z_2$ odd particles are included in the model. 
We assume that DM candidate is the $D_6$ singlet right handed neutrino $n_S$. 
The assumption is interesting since the Yukawa couplings are constrained by the $D_6$ symmetry and 
a few parameters which are relative with DM physics only remain in the model. 
We investigate whether the correct DM relic density can be satisfied by the DM $n_S$.
Due to the $D_6$ flavor symmetry, the neutrino Yukawa couplings 
$\eta^\dag_S\overline{\ell_i}Y^{\nu}_{ij}n_j$ are restricted as
\begin{equation}
Y^{\nu}_{ab}\simeq\left(
\begin{array}{ccc}
0&0&h\\
0&0&\frac{m_e}{m_\mu}h\\
0&0&0
\end{array}
\right)\qquad
\mbox{for charged leptons},
\end{equation}
\begin{equation}
Y^{\nu}_{ab}\simeq\left(
\begin{array}{ccc}
0&0&0\\
0&0&0\\
0&0&h
\end{array}
\right)\qquad
\quad\:\:\mbox{for neutrinos},
\end{equation}
where $m_{e}$, $m_{\mu}$ are electron and muon mass and $h$ the Yukawa coupling of $\mathcal{O}(1)$.
One can see that $e^{\pm}$ are dominantly generated as charged leptons due to the $D_6$ flavor symmetry. 
This point is crucial in order to explain the positron excess in the cosmic ray observed by PAMELA. 
The thermally averaged annihilation cross section of DM is calculated as
\begin{equation}
\left<\sigma_1 v\right>\simeq
\frac{|h|^4}{4\pi}\frac{M_S^2(M_S^4+M_\eta^4)}{(M_S^2+M_\eta^2)^2}\frac{T}{M_S}.
\end{equation}
where $M_\eta$ is $\eta$ mass which is included in the scalar potential $\mathcal{V}(\phi,\eta,\varphi)$ 
and $T$ is the temperature of the Universe.

\begin{figure}[h]
\begin{center}
\includegraphics[scale=0.5]{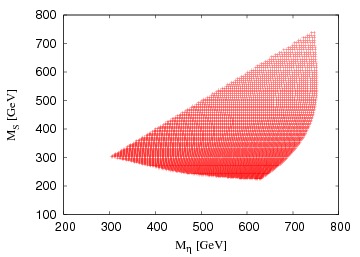}
\qquad
\includegraphics[scale=0.5]{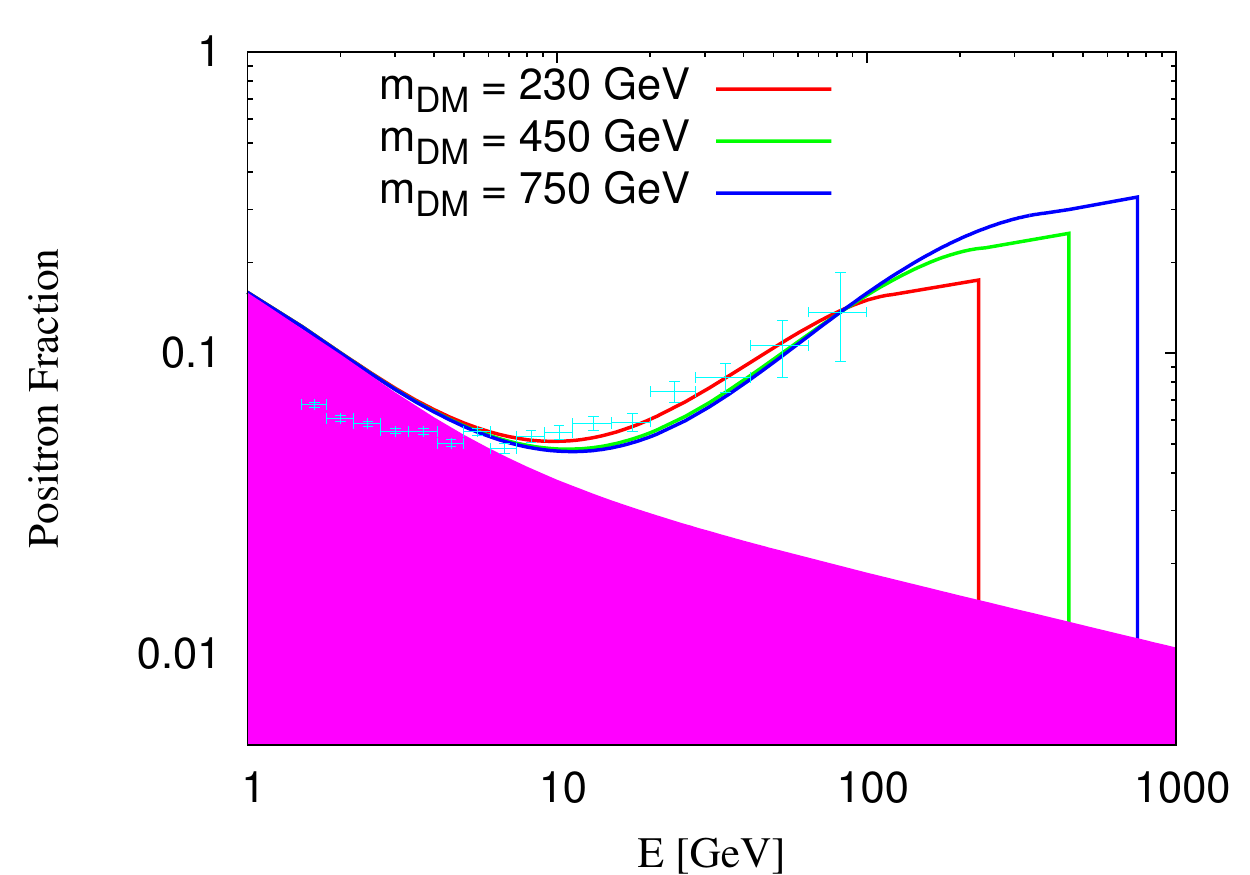}
\vspace{-0.7cm}
\caption{The allowed parameter region from the thermal DM relic density and $\mu\to e\gamma$ 
in the ($M_\eta$-$M_S$) plane (the left figure). 
The comparison of the positron flux with the PAMELA result for $M_S=230$, $450$, $750$ $\mathrm{GeV}$ 
(the right figure). The annihilation cross section $\left<\sigma_2v\right>$ is taken as $8.5\times10^{-8}$, 
$2.6\times10^{-7}$ and $6.8\times10^{-7}~\mathrm{GeV^{-2}}$ respectively.}
\label{fig:ms-ms}
\end{center}
\end{figure}

We also must take into account the constraint from Lepton Flavor Violation. 
In particular, $\mu\to e\gamma$ gives a severe constraint.
We explore allowed parameter region from the DM relic density, Lepton Flavor Violation, 
the perturbativity of the model $|h|<1.5$ and the condition of DM $M_S<M_\eta$. 
The left hand side of Fig.~\ref{fig:ms-ms} shows the allowed parameter region from these constraints. 
One can find that the allowed mass region of $M_S$ is $230~\mathrm{GeV}\lesssim M_S\lesssim 750~\mathrm{GeV}$ 
from the figure.

\section{Indirect Detection of DM}
\begin{figure}[t]
\begin{center}
\includegraphics[scale=0.8]{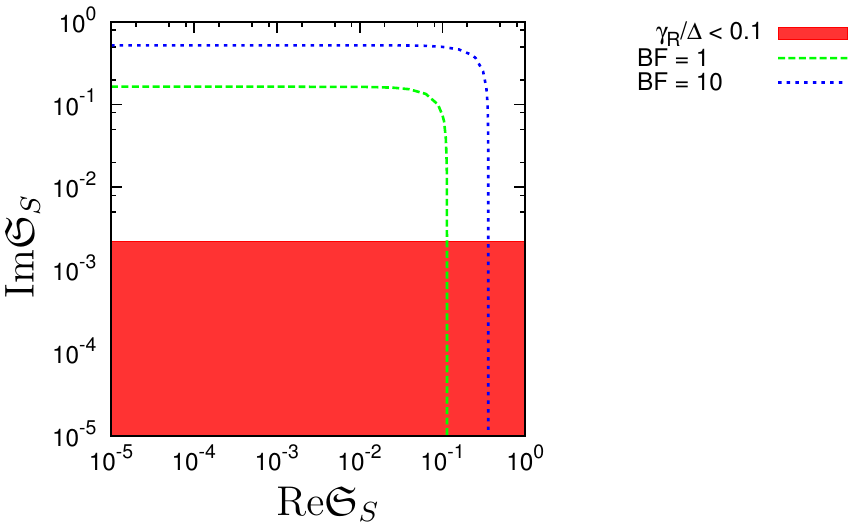}
\qquad
\includegraphics[scale=0.8]{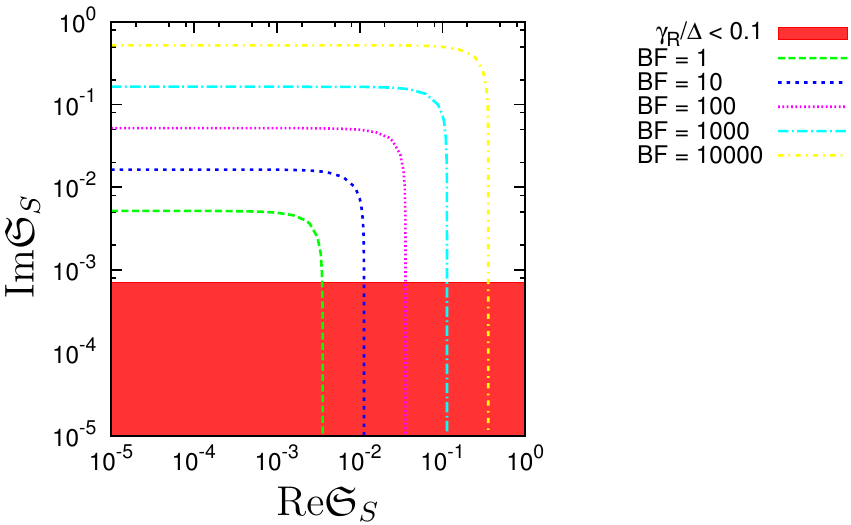}
\vspace{-0.7cm}
\caption{The contours of the boost facotor $BF$ defined in the text. 
$\gamma_R$ is defined as $\gamma_R=\Gamma_R/M_R$ where $\Gamma_R$ is the decay width of the resonance particle $R$.}
\label{fig:D}
\end{center}
\end{figure}

The positron excess in the cosmic ray is explained by the annihilation channel 
$n_Sn_S\to \varphi \to e^+e^-$. This process is s-channel and enhanced by the 
Breit-Wigner enhancement mechanism when the relation $2M_S\simeq M_R$ 
is satisfied where $M_R$ is the resonance particle mass. 
A similar analysis is done in ref.~\cite{Suematsu:2010gv}. 
The resonance particle $R$ is a mass eigenstate of 
Higgses and expressed by using the mixing matrix $\mathcal{O}$ as follows
\begin{equation}
R=\mathcal{O}_I\phi_I+\mathcal{O}_S\phi_S+\mathcal{O}_\varphi\varphi.
\end{equation}
If the condition $\gamma_R/\Delta\ll1$ is satisfied, the annihilation cross section through 
the s-channel is calculated as 
\begin{equation}
\left<\sigma_2v\right>\simeq
\frac{\sqrt{\pi}}{10(4\pi)^4}|h|^4\mathcal{O}_\varphi^2\left(\mathrm{Re}\mathfrak{S}_S\right)^2
\frac{m_e^2}{M_\eta^4}\left(\frac{M_S}{T}\right)^{3/2}e^{-\Delta M_S/T}
\label{eq:sigma2v}
\end{equation}
where the dimensionless parameter $\gamma_R$ is defined as $\gamma_R\equiv\Gamma_R/M_R$ and the mass degeneracy $\Delta$ 
is $\Delta\equiv 1-4M_S^2/M_R^2$~\cite{Griest:1990kh}\cite{Gondolo:1990dk}.
The annihilation cross section $\left<\sigma_2v\right>$ severely changes by the relative velocity of DM $v$. 
Namely, This enhancement is only effective at the present universe, and neglected at the
early universe. 
As a result, the size discrepancy of the annihilation cross section 
between obtaining the correct DM relic density and explaining the positron excess is solved. 
The positron flux is calculated by solving the diffusion equation~\cite{Hisano:2005ec}. 
The flux calculated in the model is shown in the right hand side of Fig.\ref{fig:ms-ms} 
where Isothermal profile is assumed here as DM density profile.
The contours of the boost factor which is defined as $BF\equiv\left<\sigma v\right>/3.0\times 10^{-9}$ 
here is shown in Fig.\ref{fig:D} for $\sqrt{\Delta}=10^{-6}$ and $10^{-7}$. 
The red region stands for $\gamma_R/\Delta\ll1$ region. 
The analysis is valid for only in the red region. 
From the figure, one can see that the relation $\mathrm{Re}\mathfrak{S}_S\ll\mathrm{Im}\mathfrak{S}_S$ 
and $\sqrt{\Delta}\lesssim 10^{-6}$ must be satisfied in order to obtain a large boost factor $BF$. 
We must take into account the constraint from no excess of anti-proton flux.
Due to the constraint, $\left<\sigma_3v\right>/\left<\sigma_2v\right>\lesssim10^{-2}$ is 
required where $\left<\sigma_3v\right>$ is the annihilation cross section of $n_Sn_S\to q\bar{q}$. This constraint corresponds to 
$|\mathcal{O}_{S}|/|\mathfrak{S}_S|\lesssim10^{-12}$, and 
it is the very severe constraint.

\section{Direct Detection of DM}
The elastic cross section with nucleon is derived from the mixing of Higgses. 
In particular, the mixing $\varphi$-$\phi_S$ is important since $\phi_S$ only 
couples to quarks. The SM Higgs is a superposition of Higgses $\phi_I$, $\phi_S$ and $\varphi$,
\begin{equation}
\mathrm{SM\mbox{-}higgs}=\mathcal{U}_I\phi_I+\mathcal{U}_S\phi_S+\mathcal{U}_\varphi\varphi.
\end{equation}
The elastic cross section is proportional to the mixing 
$\sigma^{N}_{\mathrm{SI}}\propto |\mathcal{U}_S\mathcal{U}_\varphi\mathfrak{S}_SY^{q}|^2$ 
We compare the predicted elastic cross section with direct detection experiments such as XENON100 and CDMS II 
which give the most severe constraint on direct detection of DM. 
As a result, we obtain the predicted elastic cross section which can be verified by the next future direct 
detection experiment XENON1T for $\mathcal{U}_S\mathcal{U}_\varphi\mathfrak{S}_S\simeq 0.1$. 

\section{Summary}
Indirect and direct detection of DM have been discussed in $D_6$ flavor 
symmetric model. 
The $D_6$ flavor symmetry gives the predictions for the mixings of leptons.
The mass of DM $n_S$ is constrained to $230\lesssim M_S\lesssim750~\mathrm{GeV}$ by 
the thermal DM relic density and Lepton Flavor Violation.
The $e^{\pm}$ excess in the cosmic ray is explained by the
Breit-Wigner enhancement. 
The flavor of the final states of the DM annihilation is determined 
by the $D_6$ flavor symmetry and the flavor is almost $e^\pm$.

The elastic scattering between DM and quarks occurs
through the Higgs mixing. 
A certain parameter region of the Higgs mixing 
will be verified by the next future direct detection experiments such as XENON1T.